\documentclass[graybox,envcountchap,sectrefs]{svmono}

\usepackage{makeidx}
\usepackage{slashed}
\usepackage{graphicx}
\usepackage{appendix}
\usepackage{chngcntr}
\usepackage{etoolbox}
\usepackage{lipsum}
\usepackage[utf8]{inputenc}

\AtBeginEnvironment{subappendices}{\chapter*{Appendix}
\addcontentsline{toc}{chapter}{Appendices}
\counterwithin{figure}{section}
\counterwithin{table}{section}
} 

\usepackage{amsmath,amssymb}
\usepackage{slashed}
\usepackage{graphicx}
\usepackage{bm}
\usepackage[T1]{fontenc}
\usepackage[utf8]{inputenc}
\usepackage{gauss} 
\usepackage[top=1.0in,bottom=1.0in,left=1.0in,right=1.0in]{geometry}
\usepackage[colorlinks,linkcolor=blue,citecolor=blue]{hyperref}
\newcommand{\be}{\begin{equation}}
\newcommand{\ee}{\end{equation}}
\newcommand{\bea}{\begin{eqnarray}}
\newcommand{\eea}{\end{eqnarray}}
\newcommand{\ba}{\begin{eqnarray}}
\newcommand{\ea}{\end{eqnarray}}

\newcommand{\Tr}{\mbox{Tr}\;}

\newcommand{\ket}[1]{\left|#1\right\rangle}

\newcommand{\LF}{\mathrm{LF}}
\newcommand{\dd}{\mathrm{d}}
\newcommand{\ii}{\mathrm{i}}

\newcommand{\Nc}{N_c}

\newcommand{\qb}{\bar{q}}
\newcommand{\vev}[1]{\left\langle #1 \right\rangle}
\newcommand{\ord}[1]{\mathcal{O}\!\left(#1\right)}
\newcommand{\Dlr}{\overleftrightarrow{D}}

\begin{document}
 \begin{center}
{\bf \Large
The Hadron-Parton Bridge\\
From the QCD Vacuum to Partons}
\\
 \vspace{2cm}
{\bf Edward Shuryak and Ismail Zahed }\\
Center for Nuclear Theory, Department of Physics and Astronomy,
\\
Stony Brook University, Stony Brook NY 11794-3800, USA
 \end{center}
\vspace{5cm}

\begin{abstract}
\\
Quantum Chromodynamics (QCD) exhibits complementary descriptions of hadrons: a rest-frame picture based on confinement, chiral symmetry breaking and interquark forces, and a high-energy light-front picture expressed through parton distributions (PDFs,TMDs,GPDs) and form factors. This review develops a unified framework that connects these two domains. It is based mostly on multiple studies by the authors in the past few years. Using the Instanton Liquid Model (ILM) to capture essential nonperturbative features of the QCD vacuum, we derive effective interactions for mesons, baryons, and multiquark states, construct their wave functions in hyperspherical coordinates, and boost them to the light front. The resulting light-front Hamiltonians, incorporating both perturbative and instanton-induced dynamics in the Wilsonian spirit, provide realistic nonperturbative inputs for computing PDFs, DAs, GPDs, quasi-distributions, and gravitational form factors at a well-defined low scale. The connection to perturbative QCD is then established by matching gradient-flow-renormalized operators and LF wave functions to the 
standard $\overline{\rm MS}$ scheme.
Perturbative DGLAP and ERBL evolution then connects these predictions to experimentally accessible regimes.
This approach is applied to quarkonia, glueballs, light mesons, baryons, tetraquarks, pentaquarks, and higher multiquark hadrons, yielding consistent descriptions of both their spectra and partonic structure. Special emphasis is placed on the energy-momentum tensor and the mechanical properties of hadrons, which emerge naturally from the same dynamical ingredients. Overall, the framework demonstrates a clear continuity between hadronic spectroscopy and partonic observables, offering a coherent multiscale picture of hadron structure rooted in the underlying dynamics of QCD.
\end{abstract}

\newpage


\tableofcontents


\chapter{Preamble}

The dynamics of strongly interacting matter, as encoded in Quantum Chromodynamics (QCD), remains one of the most intricate problems in theoretical physics. Although the QCD Lagrangian is deceptively simple, the richness of its nonperturbative phenomena continues to challenge our understanding, and is primary reason for the development of new analytic and numerical tools. Confinement, spontaneous chiral symmetry breaking, the proliferation of hadronic excitations, and the emergence of constituent degrees of freedom all defy simple perturbative descriptions. At the same time, high-energy scattering experiments reveal a complementary picture of hadrons in which quarks and gluons behave as nearly free, pointlike constituents carrying longitudinal momentum fractions. Reconciling these two disparate faces of QCD - the bound-state, rest-frame description and the parton-based, infinite momentum frame formulation - remains a central pursuit in contemporary hadron physics.

The purpose of this review is to assemble and expand upon a body of multiple works
done in the last few years.
Their unifying objective is to build a consistent theoretical bridge between traditional hadronic spectroscopy and the modern description of hadrons in terms of parton distribution functions (PDFs), distribution amplitudes (DAs), generalized parton distributions (GPDs), and gravitational form factors. A natural language for this connection is provided by light-front quantization, which renders many aspects of partonic observables transparent but is notoriously subtle when incorporating confinement and chiral symmetry breaking. The central thrust of the approach developed here is that these two essential nonperturbative features of QCD can be imported into the light-front formalism through explicit dynamical mechanisms rooted in the topological structure of the QCD vacuum.

Key among these mechanisms is the Instanton Liquid Model (ILM), which captures crucial aspects of the QCD vacuum through a semi-classical ensemble of instantons and anti-instantons. The ILM successfully accounts for the spontaneous breaking of chiral symmetry and provides nonlocal, chirally nontrivial interactions that leave observable signatures in hadron structure. Instanton-induced interactions give rise to effective constituent quark masses, characteristic spin-dependent potentials, and highly specific Dirac structures in the effective light-front Hamiltonians for both mesons and baryons. They also yield concrete predictions for higher-twist matrix elements, quasi-parton distributions, and the spin-orbit forces necessary to understand excited baryons and multiquark configurations. One central motivation of the present work is to trace these instanton effects continuously from the rest-frame description of hadrons to the light-front representation used to define partonic observables.

The first part of the review focuses on the QCD vacuum, topological fluctuations, and their impact on confinement and chiral symmetry breaking. These developments set the stage for the second major part: recent progress in spectroscopy of mesons, baryons, and (most important) the multiquark states. The hyperspherical approximation, combined with symmetry principles from the permutation group and color-spin-flavor algebra, provides a highly efficient representation of hadronic wave functions across sectors with increasing quark number. These wave functions encode the spatial correlations, spin couplings, and kinetic structures that underlie the observed spectra. They will later serve as the initial conditions for constructing light-front wave functions through controlled boosts and matching procedures.

The transition from hadronic spectroscopy to partonic observables proceeds through the construction of effective observables
(DAs,PDFs,GPDs,TMDs etc) 
and light-front Hamiltonians. These Hamiltonians incorporate both perturbative one-gluon exchange and instanton-induced nonperturbative interactions. They allow one to describe mesons, baryons, and even pentaquarks on the light front in a manner directly comparable to the parton model. Once the light-front wave functions are obtained, the calculation of DAs, PDFs, TMDs, GPDs, and form factors becomes a straightforward exercise in applying their standard operator definitions. The resulting observables provide a direct link between nonperturbative hadronic inputs and experimentally measurable quantities.

A critical element of this program is the evolution of observables from low to high resolution. Starting from the nonperturbative input scale determined by instanton physics, the DGLAP and ERBL equations govern the scale dependence of PDFs and DAs. The review emphasizes that this matching between low-scale models and high-scale observables is not merely a computational necessity but a conceptual synthesis: it explicitly connects the long-distance physics of the ILM with the short-distance partonic structure probed in deep inelastic scattering and hard exclusive reactions.

Beyond the traditional partonic observables, this work also incorporates gravitational form factors and the energy-momentum tensor (EMT), which encode information about the mechanical structure of hadrons, including pressure distributions, shear forces, and contributions to the mass and spin. The light-front representation of EMT matrix elements emerges naturally from the wave functions developed here and connects the mechanical interpretation of hadron structure with the same dynamical ingredients that control spectroscopy and PDFs.

Taken together, the developments presented in this review demonstrate that the hadronic and partonic pictures are not isolated but rather deeply interwoven. By constructing explicit maps between the confining, chiral-symmetry-breaking dynamics in the rest frame and the universal operator structures of the light front, we arrive at a unified, scale-resolving portrait of QCD bound states. The goal is not only to summarize a series of earlier works but to present them as a logically coherent program, highlighting the deep conceptual continuity that runs from the QCD vacuum to the partonic structure of hadrons.



\chapter{Introduction} 
This first chapter offers a broad and non-technical overview of the themes addressed in this book. The substantive material appears in the subsequent chapters, where we describe in more detail the current state of the phenomenology and theory pertinent to our subject. More technical discussions are deferred to Appendices placed at the end of individual chapters.

\section{The QCD vacuum, correlation functions and hadrons}
The QCD vacuum is the ground state of the theory, and hadrons constitute its lowest excitations. By examining how the masses and other properties of hadrons depend on their quantum numbers, we gain insight into the types of gauge and quark fields permeating the vacuum. The gauge fields composing the QCD vacuum, fall into two broad classes: the perturbative waves, or gluons, and the nonperturbative fields with amplitudes $A_\mu\sim O(1/g)$. The former resemble photons and arise from quantizing fluctuations around the  (empty) perturbative vacuum defined by the quadratic part of the action, whereas the latter encompass 
the various topological solitons 
composed of  glue. 
There is a distinction between gluons (small amplitude waves) and {\em glue} in general, including topological objects. The way to separate those is e.g. {\em gradient flow}  form of renormalization group.

Since the 1980s the primary tool for exploring the QCD vacuum has been large-scale numerical simulations on supercomputers. Wilson's lattice formulation of non-Abelian gauge fields, which preserves gauge symmetry at arbitrary lattice spacing, provides the foundation for these studies. Other symmetries, however, do not fare as well: in particular, most lattice formulations of quarks recover exact chiral symmetry only in the $m_q\rightarrow 0$ limit, which in practice requires challenging extrapolations.

In parallel with this numerical, first-principles lattice approach, a number of semiclassical models have been developed that describe the vacuum as an ensemble of solitonic configurations\footnote{Similarly, people who study weather and atmospheric dynamics use both numerical hydrodynamics and qualitative language such as  ``hurricane A moves from X to Y, while the anticyclone remains over the midwestern states."}. The earliest such objects discovered in non-Abelian gauge theories were magnetic monopoles \cite{tHooft:1974kcl,Polyakov:1974ek}, followed soon after by instantons \cite{Belavin:1975fg}. A more recent unifying picture, particularly relevant near QCD phase transitions, involves the so-called instanton-dyons (or instanton-monopoles) \cite{Kraan:1998sn}. While the technical details of these configurations are best sought in specialized references\footnote{E.g. *Nonperturbative topological phenomena in QCD and related theories*, E. Shuryak, Springer 2021.}, the next chapter provides an introductory discussion of the relevant phenomena.

A central tool uniting lattice practitioners, topologists, and researchers concerned with hadronic structure is the study of vacuum correlation functions. The simplest examples are of the form $\langle O_A(x)O_A(y)\rangle$, where $O_A$ is an operator built from fundamental fields with definite quantum numbers $A$. Inserting a complete set of intermediate states between the two operators yields the so-called QCD sum rules, which connect empirical hadronic information to vacuum correlators. For operators with a current structure, $O_A=\bar q\Gamma_A q$, the intermediate states include mesons carrying the corresponding quantum numbers.

\begin{figure} \centering\includegraphics[width=0.5\linewidth]{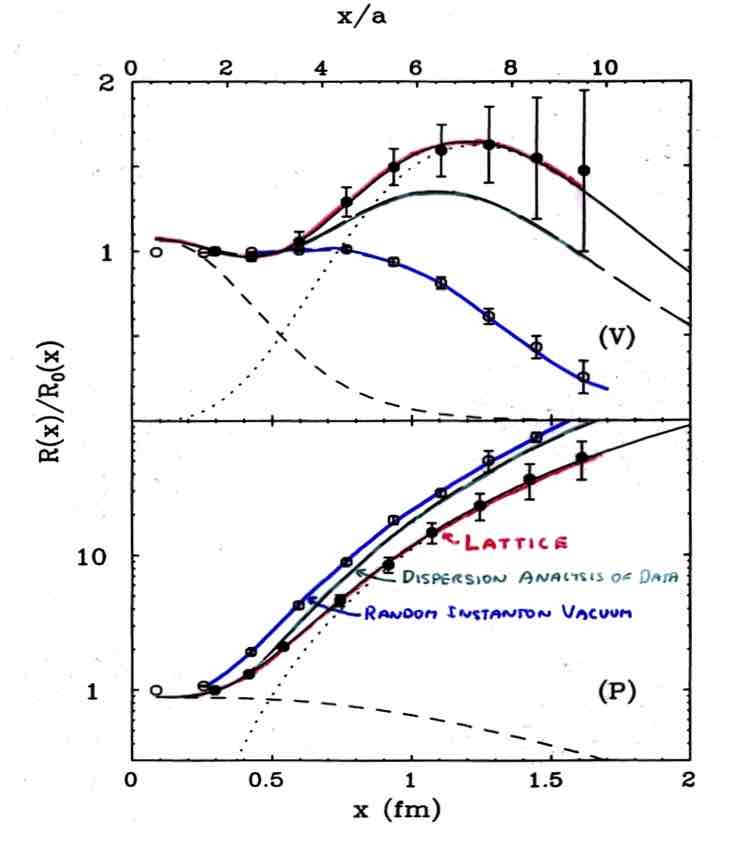}
    \caption{The   correlator of two local operators with the quantum numbers of the rho-meson (upper, V) and the pion  (lower, P), normalized to propagation of free massless quarks. 
    Note that the upper plot is linear and the lower logarithmic, as nonperturbative corrections
    in the pion channel are much larger. The dotted lines are contributions of the lowest state, the dashed ones those of non-resonance continuum.}
    \label{fig_pi_rho_correlators}
\end{figure}

The theory and phenomenology of correlation functions were reviewed in \cite{Shuryak:1993kg}, which emphasized that hadronic channels naturally fall into two groups with markedly different behaviors. Vector, axial, and many other mesonic channels exhibit a smooth transition from perturbative behavior at short distances to nonperturbative dynamics at longer distances. 
Spin zero channels (scalar and pseudoscalar) show earlier departure
from perturbative correlators.
This difference can be traced to single instanton contributions that strongly affect spin-zero correlators but are absent in the other channels. Figure~\ref{fig_pi_rho_correlators}, displaying lattice results from the MIT group \cite{Chu:1993cn}, illustrates this contrast and compares the findings to the instanton contribution.

\begin{figure}[h!]
    \centering \includegraphics[width=0.5\linewidth]{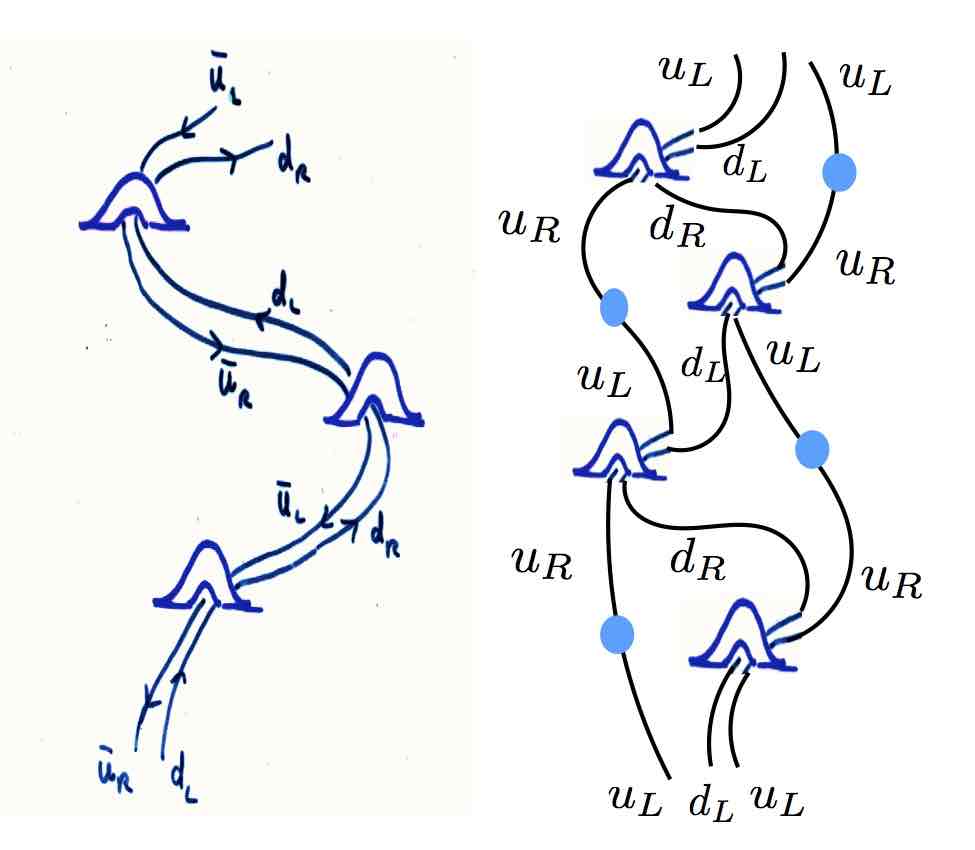}
    \caption{The left sketch shows a pion propagating through the instanton vacuum; the right shows a nucleon. In the latter, the $ud$ pair forms "good diquarks''. Blue dots mark quark mass insertions where chirality flips occur.}
    \label{fig_pion_proton_sketch}
\end{figure}

A broader discussion of the instanton liquid and correlators within it can be found in \cite{Schafer:1996wv}. Without delving into that extensive subject, Fig. \ref{fig_pion_proton_sketch} provides a qualitative 
picture (adapted from a talk by Negele)
 of how the pion and the nucleon propagate in the instanton ensemble.
 
  Instantons (or, more precisely, the associated 't Hooft vertices) are depicted as small hills with tunnels through which quark pairs must travel.

Correlation functions reveal not merely two but three distinct types of instanton-induced effects: attraction, repulsion, and neutrality (to first order in the instanton density). For quark-based correlators, channels such as $\sigma\pi$ are attractive, while $\delta$ and $\eta'$ are repulsive; many others, including vector channels such as those shown above, receive no leading-order correction. An analogous trichotomy appears even more strongly in gluonic correlators: scalar $J^P=0^+$ channels are attractive, pseudoscalar $0^-$ channels are repulsive, and tensor $2^+$ channels exhibit no first-order effect\footnote{This occurs because a strong instanton field nevertheless has zero stress tensor.}. One notable prediction made in \cite{Schafer:1994fd} was that the scalar glueball

has  much smaller size, with the radius near

 $r_{mass}\approx 0.2\,\mathrm{fm}$. After decades of hints, recent lattice works (e.g. \cite{Abbott:2025irb}) strongly support this prediction.

\section{Bridging hadronic spectroscopy to partonic observables}
Hadronic spectroscopy emerged in the 1950s with the discovery of numerous hadronic resonances, the excited states of mesons and baryons. The classification of these states was placed on firm footing in the 1960s through the flavor $SU(3)$ symmetry introduced by Gell-Mann and Zweig. The 1970s brought dramatic progress: asymptotic freedom and  QCD was discovered  \cite{Gross:1973id} and \cite{Politzer:1973fx}. Then  heavy-quark bound states ( quarkonia) were observed experimentally. Starting around 1980s, numerical simulation on 4d lattice has evolved into a highly successful tool, with simulations now reproducing hadronic spectra and properties with impressive accuracy. One might therefore think that most conceptual developments were concluded in the late twentieth century.

Yet the field has undergone a pronounced revival, 
as many new hadrons have been discovered.
In particularly tetraquarks and pentaquarks are in Particle Data.

Increasing attention is once again turning to six-quark (hexaquark) and twelve-quark configurations, which may play a role in the structure of the deuteron and $^4\mathrm{He}$.

The renewed theoretical efforts have reignited questions concerning the origin of the forces among quarks, both heavy and light, and have motivated deeper exploration of connections between these forces and the structure of the QCD vacuum. Relations between baryonic and mesonic forces have been generalized to multiquark systems, and puzzles associated with spin-dependent interactions, particularly spin-orbit couplings, have inspired new analyses. Later chapters will introduce the relevant quark symmetry group representations, methods for constructing spectra and wave functions using recent features of Mathematica, and approaches employing multi-component wave functions. Another powerful technique borrowed from nuclear physics involves the use of hyperdistance and spherical symmetry for three-, four-, and five-body systems.

Contemporary hadron spectroscopy relies heavily on two complementary fields. The first is lattice gauge theory, which has recently reached the point where sufficiently large volumes and small quark masses allow reliable chiral and infinite-volume extrapolations. The second consists of efforts to relate spectroscopy to partonic observables, chiefly through light-front (LF) quantization.

\begin{figure}
    \centering
    \includegraphics[width=0.75\linewidth]{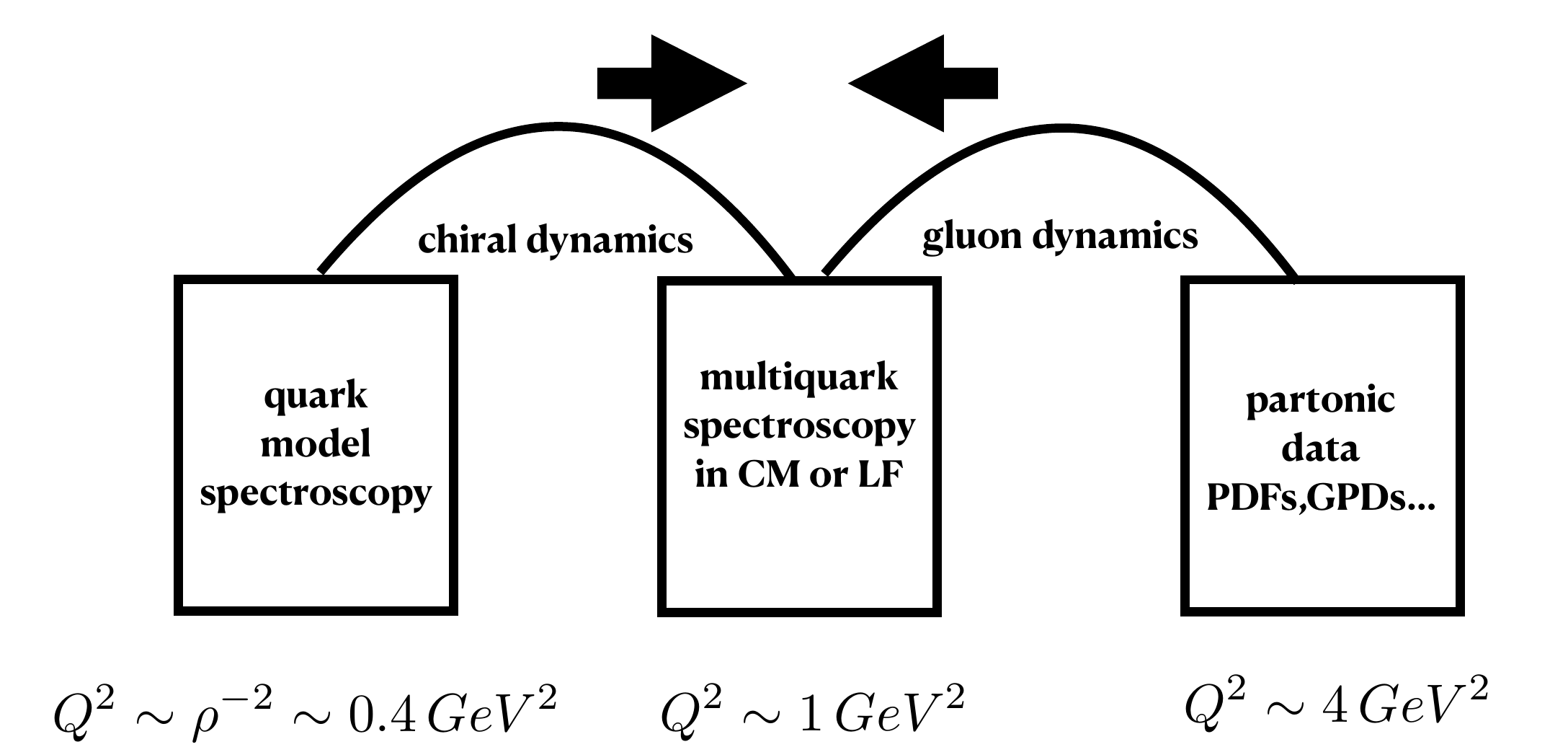}
    \caption{A sketch illustrating the "bridge'' to be discussed.}
    \label{fig:enter-label}
\end{figure}

The project described in this review  was motivated by the substantial conceptual gaps that still separate two major subfields. On one side lie models of mesons, baryons, tetraquarks, and pentaquarks formulated in terms of a fixed number of quarks. Their wave functions are eigenstates of appropriate Hamiltonians, mutually orthogonal, and built from color, flavor, and spin degrees of freedom, subject to Fermi statistics. On the other side lie partonic descriptions in which hadrons are composed of an indeterminate number of quarks, antiquarks, and gluons. These models, developed since the 1970s, rely on empirical parton distribution functions (PDFs) extracted from fits to hard-scattering data. Because PDFs 
include only partial information,
 as they are
summed over
   all variables but
    one parton momentum. 

Indeed, PDFs  are density matrices,
as as such they naturally possess  
nonzero entropy \cite{Kharzeev:2017qzs}.

Our attempt to bridge these two worlds required learning from both literatures and developing new elements as needed. These ideas are scattered through multiple papers, and the present review aims to assemble them in a coherent and pedagogical manner. We are aware that this approach is necessarily incomplete and may have omitted relevant contributions; we apologize to their authors for any such oversights. The review remains subjective and limited in scope, with the additional constraint that we deliberately chose not to involve students or postdocs except when absolutely necessary, in order to proceed quickly.

A crucial step in the program is the boosting of hadronic spectroscopy from the center-of-mass frame to the light front. This effort was pioneered by J. Vary and collaborators under what is now known as the BLFQ approach. To clarify the distinction between their program and ours, recall that solving a quantum-mechanical problem generally proceeds through three main steps: selecting the variables, choosing the Hamiltonian, and choosing the solution method. Our approach and BLFQ differ at each stage. We employ Jacobi coordinates to cleanly separate the center-of-mass motion; we adopt simple kinetic-plus-confinement Hamiltonians rather than those tailored to Brodsky's specific variable combinations such as $p_\perp^2/x$; and we emphasize that physical results must ultimately be independent of the solution method. The BLFQ basis is a technical tool that brings issues of truncation and convergence, and thus requires comparison to alternative methods. Such alternatives exist: for mesons and baryons one may directly solve the corresponding light-front Schrodinger equations numerically or adopt various variational methods, for instance parametrized basis functions or linear combinations of Gaussians.

\section{The multiquark hadrons and their admixtures}
Even in the simplest meson systems, such as charmonia $\bar cc $, are not so simple. In fact we learned recently that tetraquarks $\bar c c \bar q q$ are among those and can influence the spectrum through mixing, as those may  share the same quantum numbers. Over the past decade it has become increasingly clear that similar effects are ubiquitous. Baryons of the form $qqq$ may contain admixtures of pentaquark configurations $qqqq\bar q$, and such components may help clarify longstanding puzzles concerning the antiquark sea.

Systems containing even larger numbers of quarks are of considerable interest as well. On the theoretical side they pose obvious challenges, and on the phenomenological side they may appear in di-nucleon reactions, which seem to hint at hexaquark ($q^6$) states; even four-nucleon systems may contain significant twelve quarks $q^{12}$ admixtures.

Our discussion of multiquark hadrons rests on three broad ideas. First, we fix the center-of-mass by working in Jacobi coordinates. Second, when the relevant quarks are of equal mass, we invoke spherical symmetry, or the hyperdistance approximation. Third, we enforce Fermi statistics on the orbital, color, spin, and flavor degrees of freedom through the appropriate permutation-group representations of $S_n$. Although these concepts will be introduced pedagogically later, it is useful to recall their origins here. Jacobi introduced his coordinates in the mid-1800s for classical few-body problems. The hyperdistance approximation was developed in nuclear physics in the 1960s, originally applied to light nuclei such as tritium and $^4\mathrm{He}$. The permutation-group methods we use are of more recent vintage, developed in  \cite{Miesch:2024vjk}.

\section{The parton model} \label{sec_intro_partons}
The formulation and experimental observation of Bjorken scaling in deep-inelastic electron-nucleon scattering (DIS) led Feynman to propose the parton model \cite{Feynman:1969ej}. Its conceptual basis rests on three ideas: hadrons at high energies consist of many partons; in the infinite-momentum frame these partons are effectively frozen by time dilation and behave as if free from interactions; and photon-hadron inelastic scattering is predominantly incoherent. The first concrete application of this perspective to inelastic electron-proton scattering at high energies was made by Bjorken and Paschos \cite{Bjorken:1969ja}. Denoting by $q^\mu$ the space-like electron momentum transfer and by $\nu$ the virtual-photon energy in the nucleon rest frame, they found that inclusive electron-hadron cross sections depend only on the Bjorken variable $x_B=Q^2/(2m_N\nu)$ in the limit of large $Q^2=-q^2$.

This naive picture of hadrons at high energies predates QCD, whose formulation by 
\cite{Fritzsch:1973pi} provided the underlying quantum field theory. Remarkably, asymptotic freedom of QCD \cite{Gross:1973id,Politzer:1973fx} soon gave  theoretical support to the parton model. The perturbative description of scaling violation, based on anomalous dimensions and the perturbative evolution of parton distribution functions, emerges from logarithmically enhanced collinear gluon emissions and leads to the DGLAP evolution equations in the resolution scale $Q^2$. A related but distinct evolution in $x$ is described by the BFKL equation.

Feynman's parton model is rooted in the infinite-momentum frame, where a hadron with large $P^z$ is viewed as a collection of $N$ partons (quarks and gluons) with momenta
\bea
p_i=x_iP^z+k_{\perp i}\qquad \sum_{i=1}^Nx_i=1,\qquad\sum_{i=1}^N k_{\perp i}=0
\eea
and longitudinal momentum fractions bounded by $0\leq x_i\leq 1$, since all partons propagate along the hadron's direction of motion. Backward-moving partons are suppressed in the infinite-momentum frame, as first emphasized by \cite{Weinberg:1966jm}. The probability distribution for the longitudinal momentum fraction, or parton distribution function (PDF), became a central object of high-energy phenomenology. Global fits to experimental data, such as those produced by the CTEQ collaboration \cite{Hou:2019efy}, continue to refine these distributions.

The relation between Weinberg's time-ordered perturbation theory in the infinite-momentum frame \cite{Weinberg:1966jm} and Dirac's front-form quantization \cite{Dirac:1949cp} was first exploited by \cite{Chang:1968bh} in their analysis of QED in the large-momentum limit . The front-form formulation identifies time and space with the light-cone coordinates $x^\pm=(x^0\pm x^3)/\sqrt2$, with the corresponding Hamiltonian and momentum given by $p^\mp=(p^0\mp p^3)/\sqrt2$. For a free massive particle the front-form Hamiltonian is
\bea
p^-=\frac{p^0-p^z}{\sqrt2}=\frac{p_\perp^2+m^2}{2p^+}
\eea
which avoids sign ambiguities and eliminates the need for antiparticles, but introduces an infrared singularity at $p^+=0$. The Hamiltonian is non-local in the conjugate light-cone coordinate $x^-$, endowing the initial quantization surface with an infinite correlation length; this raises questions about the consistency of the front form in perturbative field theory.

Despite these challenges, the front form provides a natural pathway to the parton model. Consider the quantization of a free Dirac field by introducing the light-cone gamma matrices
\bea
\gamma^\pm=\frac{\gamma^0+\gamma^z}{\sqrt2}
\qquad \gamma^{+2}=\gamma^{-2}=0\qquad\qquad
\mathbb P_\pm=\frac12 \gamma^\mp\gamma^\pm \qquad
\mathbb P_++\mathbb P_-=1
\eea
and decomposing the Dirac field into $\psi=\psi_++\psi_-=\mathbb P_+\psi+\mathbb P_-\psi$. The component $\psi_-$ is a constrained variable, as follows from
\bea
\mathbb P_+(i\slashed{D}-m)\psi=
iD^+\gamma^-\psi_-+(i\slashed{D}^\perp-m)\psi_+=0 .
\eea
Since $D^+=\partial^{++} igA^+$ does not have derivative over time
$\partial^-$ ,
  $\psi_-$ can be eliminated:

\bea
\psi_-=-\frac{(i\slashed{D}^\perp-m)\psi_+}{iD^+\gamma^-}
=-\frac{\gamma^+}{iD^+}(i\slashed{D}^\perp-m)\psi_+
\eea
up to infrared poles at $D^+=0$. In light-cone gauge $A^+=0$, the field $A^-$ is also constrained and can be removed.

\section{The $bridge$ strategy and the resolution scales}
\label{sec_matching_point}
Before constructing a bridge, one must understand the shores it aims to connect. We therefore begin by briefly reviewing the theoretical situations at the two extremes. Since the 1970s it has been clear that hard processes at large momentum transfers, $Q^2\gg 1\,\mathrm{GeV}^2$, can be described in terms of nearly free partons: gluons, quarks, and antiquarks. The probabilities to find these partons in a hadron are encoded in the PDFs $q(x,Q^2)$. At sufficiently high resolution the pointlike constituents emit each other according to splitting functions derived directly from the QCD Lagrangian, and PDFs defined at different $Q^2$ are connected by perturbative DGLAP evolution. The combination of perturbative QCD and global fits to hard-scattering data has grown into a vast program; see, for example, CTEQ's recent review \cite{1912.10053}.

While the partonic description is a robust endpoint of the bridge, it does rely on certain approximations. In particular, the incoherence assumption underlying DGLAP evolution renders the equations probabilistic kinetic equations. As emphasized by \cite{Kharzeev:2017qzs}, this assumption leads to an entanglement entropy. Yet hadrons are pure quantum states, and a fully consistent description requires their complete light-front wave functions, without such entanglement and entropy.

Evolving PDFs downward in $Q^2$ reveals a scale at which gluon emission effectively ceases. The lightest glueballs have masses around $M_{\mathrm{glueball}}\sim 2\,\mathrm{GeV}$, suggesting a gluon effective mass of order $M_{\mathrm{eff}}(\mathrm{gluon})\sim1\,\mathrm{GeV}$ and placing the lower threshold of perturbative evolution at $Q^2\sim M_{\mathrm{eff}}^2$.

At lower $Q^2$ the dynamics are governed not by quarks and gluons but by the phenomena of chiral symmetry breaking, which foreground the Nambu-Goldstone pions and the scalar condensate $\sigma$. In this domain, the QCD Lagrangian is replaced by a chiral effective Lagrangian with an upper cutoff scale associated with the original NJL model, $\Lambda_\chi\sim1\,\mathrm{GeV}$. The mechanism was later related to instantons \cite{Shuryak:1982hk}, with the cutoff scale tied to the typical instanton size $\rho\sim1/3\,\mathrm{fm}$.

Thus we are led to a natural strategy: the two  arcs of the bridge, the chiral and perturbative domains, might meet smoothly at common  $1\,\mathrm{GeV}$ scale. In this section we examine whether PDFs computed from both sides indeed join at that point. Of course many observables require more refined approaches than an abrupt transition between effective theories. We note, for instance, various efforts to combine perturbative and nonperturbative contributions, including our own analyses of meson form factors \cite{Shuryak:2020ktq} and spin-dependent forces \cite{Shuryak:2021fsu}.

\begin{figure}[t]
\begin{center}
\includegraphics[width=16cm]{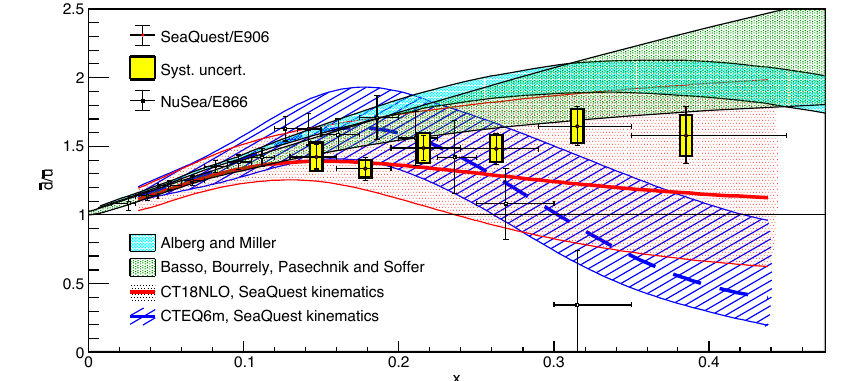}
\caption{Flavor asymmetry of the antiquark sea as the ratio $\bar d/\bar u$ versus momentum fraction $x$, from \cite{2103.04024}.}
\label{fig_flavor_asymmetry}
\end{center}
\end{figure}

It is not a place to discuss the whole set of well known standard PDFs, so let us jump now
to one of them we will be discussing most: the PDFs of the antiquark "sea". Before addressing the data at Fig.\ref{fig_flavor_asymmetry}, consider processes which 
may add 
 additional $\bar q q$ pair to a nucleon. Pairs which arise through gluon splitting would have $\bar uu$ and $\bar dd$ in equal amounts, since gluons are "flavor blind". By contrast, the instanton-induced 't Hooft four-fermion interaction generates sea quarks in a maximally flavor-antisymmetric pattern due to the Pauli principle governing instanton zero modes. 
 The predicted processes are $u\rightarrow u\bar d d$ and 
 $d\rightarrow d\bar u u$, but not $u\rightarrow u\bar u u$.
 Because the proton contains two valence $u$ quarks and one valence $d$ quark, the simplest first-order estimate based
 on instanton vertex predicts the ratio
\be {\bar d\over \bar u}\rightarrow 2 , \ee
while for the $\Delta^{++}$, which has valence content $uuu$, the sea should consist solely of $\bar d$ quarks, with no $\bar u$ component at all.
Last but not least, one may think of virtual process $N\rightarrow N+\pi$ creating a "pion cloud" around a nucleon.
Here one may have several flavor options.

The experimental data for the ratio $\bar d/\bar u$ PDFs are shown in Fig.\ref{fig_flavor_asymmetry}. The data points indicate this ratio to be $\sim 3/2$, somewhat below pion-based predictions (dark and light gray regions, from authors mentions). It is just in between one and two (as instanton vertices would predict.)
The flavor asymmetry of the sea is thus a powerful diagnostic for disentangling the contributions of chiral dynamics (step 2) and perturbative QCD (step 3) to the pentaquark sector of the light-front wave function of the nucleon. Only the former can generate 
full coherent description, the LF wave functions, from which
PDFs and much more would consistently follow.


\section{Overview of Related Approaches to Hadron Structure}
\label{sec:related_overview}

The problem of connecting hadronic spectroscopy at low energies with partonic
descriptions at high resolution has been a central theme of strong-interaction
physics since the inception of QCD. Over the past several decades, a variety of
theoretical frameworks have been developed to address different aspects of this
"hadron-parton duality'', often emphasizing complementary limits, degrees of
freedom, and organizing principles.

Historically, early insights were provided by current algebra and the parton
model, culminating in the operator product expansion and perturbative QCD
evolution equations for deep inelastic scattering
\cite{Wilson:1969zs,Altarelli:1977zs,Dokshitzer:1977sg}.
At the same time, constituent quark models and potential models successfully
accounted for large portions of the hadron spectrum
\cite{Isgur:1978xj,Godfrey:1985xj}, albeit without a direct link to partonic
observables.

More recent developments have focused on building explicit bridges between these
descriptions. These include lattice QCD calculations of hadron structure,
continuum functional methods, holographic approaches inspired by gauge/gravity
duality, and Hamiltonian light-front formulations. The framework developed in
this work belongs to this broader effort, sharing goals with these approaches
while differing in methodology and underlying assumptions.

In the following, we briefly summarize the main complementary lines of research, highlighting commonality  and distinction. For completeness, we will  refer to some reviews and their detailed references, where parts of these complementary lines of research are developed. 

\subsection{Nonperturbative QCD Frameworks}
\label{sec:nonperturbative_frameworks}

\subsection*{Lattice QCD}

Lattice QCD provides a systematically improvable, first-principles framework for
studying the nonperturbative regime of QCD by formulating the theory on an
Euclidean spacetime lattice. By discretizing spacetime and evaluating the QCD
path integral numerically, lattice methods enable controlled calculations of
hadron masses, decay constants, and matrix elements of local operators, with
quantifiable uncertainties arising from finite lattice spacing, finite volume,
and unphysical quark masses
\cite{Aoki:2019cca}.

Over the past decade, lattice simulations have reached a level of maturity where
precision calculations of hadron spectroscopy and low-energy structure observables
are routinely performed at or near the physical pion mass. In addition to static
quantities, lattice QCD has provided essential input for effective field theories
and phenomenological models, helping to constrain low-energy constants and
validate assumptions about confinement and chiral dynamics.

A major conceptual challenge for lattice QCD has been the extraction of
light-cone dominated observables, such as parton distribution functions, which
are intrinsically Minkowskian and defined through real-time correlations.
Recent methodological advances have addressed this limitation by relating
Euclidean correlation functions to light-cone observables through factorization
and matching procedures. Prominent among these is Large-Momentum Effective Theory
(LaMET), which enables the computation of quasi-parton distribution functions
(quasi-PDFs) from hadrons boosted to large spatial momenta
\cite{Ji:2013dva}. Closely related approaches include pseudo-PDFs and lattice
cross-section methods, which exploit short-distance operator expansions
\cite{Radyushkin:2017cyf,Lin:2017snn}.

These developments have opened a new avenue for direct comparison between lattice
QCD, phenomenological extractions, and model-based calculations at low
renormalization scales. At the same time, they highlight important practical
and conceptual limitations. Large hadron momenta amplify discretization effects
and statistical noise, while the interpretation of quasi-distributions requires
perturbative matching and careful control of higher-twist corrections.

From a broader perspective, lattice QCD excels at providing numerically reliable
results for well-defined observables, but offers limited transparency regarding
the underlying dynamical mechanisms responsible for confinement and dynamical
chiral symmetry breaking. In particular, the effective degrees of freedom that
emerge at low energies are encoded implicitly in the gauge-field ensembles rather
than appearing explicitly in a Hamiltonian description.

The approach pursued in the present work is complementary in spirit. Rather than
sampling the full gauge-field path integral numerically, it emphasizes explicit
nonperturbative vacuum structures and their effects on hadronic wave functions.
This allows for a more direct connection between confinement dynamics, chiral effective
interactions, and the structure of light-front wave functions at a defined low
scale, while retaining compatibility with lattice results through matching and
evolution.

\subsection*{Dyson-Schwinger and Bethe-Salpeter Approaches}

Continuum functional approaches based on Dyson-Schwinger equations (DSEs) and
Bethe-Salpeter equations (BSEs) provide a nonperturbative framework for QCD that
retains full Poincar\'e covariance and is formulated directly in Minkowski or
Euclidean momentum space. In this approach, hadrons emerge as bound states of
dressed quarks and gluons whose propagators and interaction vertices are obtained
from an infinite tower of coupled integral equations
\cite{Roberts:1994dr}.

A key achievement of the DSE/BSE program is its unified description of
dynamical chiral symmetry breaking and hadron structure. The momentum-dependent
quark mass function generated by the quark DSE encodes dynamical mass generation,
linking the infrared behavior of QCD to the emergence of constituent-like degrees
of freedom at low energies. When combined with Bethe-Salpeter equations for
mesons and covariant Faddeev equations for baryons, this framework yields
quantitative descriptions of hadron spectra, decay constants, and
electromagnetic form factors
\cite{Maris:2003vk,Eichmann:2016yit}.

Beyond spectroscopy, DSE/BSE methods have been extended to partonic observables,
including parton distribution amplitudes, generalized parton distributions, and
PDFs. These quantities are obtained by projecting covariant bound-state
amplitudes onto light-front kinematics, thereby establishing a connection between
Euclidean correlation functions and light-cone physics
\cite{Chang:2014lva,Gao:2017mmp}. Such studies have provided valuable insight into
the role of dynamical chiral symmetry breaking in shaping valence-quark
distributions and distribution amplitudes.

Despite these successes, the DSE/BSE framework relies on truncation schemes whose
systematic improvement is challenging. While symmetry-preserving truncations
exist and are well understood in certain channels, the treatment of multi-parton
correlations and higher Fock components remains indirect. Moreover, confinement
is implemented through analytic properties of propagators rather than through an
explicit Hamiltonian or potential. In addition, the spontaneous breaking of chiral in this approach has yet to be related to the topological character of the gauge fields, a key insight from lattice QCD.

Conceptually, the DSE/BSE program shares important common ground with the present
work, particularly in its emphasis on dynamical mass generation and the central
role of nonperturbative quark interactions. However, it differs in its treatment by ignoring the central role played by 
the QCD vacuum, and the importance played by its topological constituents  in setting up the emergent effective  interactions. In addition, the formulation lacks  an explicit Hamiltonian formulation.
The present approach instead constructs effective light-front Hamiltonians with explicit interactions that 
encode nonperturbative vacuum physics directly, facilitating a transparent
connection between bound-state dynamics, Fock-space structure, and partonic
observables.

\subsection*{Instantons and Topological Models}

As we noted earlier, topological gauge-field configurations play a fundamental role in the
nonperturbative dynamics of QCD. Among these, instantons-finite-action
solutions of the classical Yang-Mills equations in Euclidean spacetime-have
long been recognized as essential for understanding anomalous symmetries and
chiral dynamics
\cite{tHooft:1976snw}. Instantons generate effective multi-fermion interactions
that explicitly break the $U(1)_A$ symmetry and provide a microscopic mechanism
for strong correlations among light quarks.

Building on these insights, instanton-based models have been developed to
describe a wide range of hadronic phenomena. In the Instanton Liquid Model,
the QCD vacuum is approximated as an ensemble of instantons and anti-instantons
with a characteristic density and size distribution
\cite{Shuryak1982,Diakonov:1995ea,Schafer:1996wv,Nowak:1996aj}. This picture naturally accounts for
dynamical chiral symmetry breaking through the delocalization of quark zero
modes, leading to the emergence of a momentum-dependent constituent quark mass
and a nontrivial quark condensate.

Instanton models have been successfully applied to hadron spectroscopy,
two-point and three-point correlation functions, and static properties such as
magnetic moments and axial charges. Extensions of these ideas have also been
used to study parton distribution functions and transverse-momentum dependent
distributions, particularly at low normalization scales where nonperturbative
effects dominate
\cite{Kock:2020frx}.

While instantons provide a compelling mechanism for chiral symmetry breaking,
their role in confinement remains less direct. Dilute instanton ensembles do not
generate a linearly rising potential, although  dense instanton ensembles generate a considerable part of the confining and spin forces~\cite{Shuryak:2021fsu}.

The present work builds explicitly on the Instanton Liquid Model as a starting
point for constructing hadronic wave functions. Its distinctive feature is the
use of these wave functions as input for a light-front Hamiltonian formulation,
thereby translating topological vacuum structure into partonic degrees of
freedom. In this way, instanton-induced dynamics are not confined to static or
Euclidean observables, but are directly linked to light-front wave functions,
Fock-space structure, and parton distributions at a well-defined low scale.


\subsection{Light-Front Hamiltonian Approaches}
\label{sec:lf_hamiltonians_extended}

Light-front (LF) quantization offers a Hamiltonian formulation of relativistic
quantum field theory in which boosts along the longitudinal direction are
kinematical and hadronic states admit a Fock-space expansion in terms of
partonic degrees of freedom. This feature makes LF dynamics particularly
well suited for describing parton distribution functions, generalized parton
distributions, and distribution amplitudes, while simultaneously retaining a
connection to bound-state spectroscopy.

Early applications of light-front dynamics to QCD emphasized conceptual
foundations and formal developments, including the structure of LF wave
functions, vacuum triviality, and factorization
\cite{Dirac:1949cp,Brodsky:1997de}.
These studies established the basic language used in modern partonic analyses,
but left open the problem of constructing realistic, renormalized LF
Hamiltonians capable of describing both low-energy spectroscopy and high-energy
observables. More importantly, their relationship to the QCD vacuum as revealed by lattice QCD remains elusive.

\subsection*{Effective Light-Front Potentials and Semiclassical Models}

A major line of development has focused on constructing effective LF
Hamiltonians using phenomenological or semiclassical inputs. Light-front
holography, inspired by gauge/gravity duality, provides an effective Schrodinger
equation in a transverse variable that reproduces Regge trajectories and
captures aspects of confinement
\cite{Brodsky:2006uqa,Brodsky:2014yha}.

These models yield analytic or semi-analytic LF wave functions that have been
widely used to compute meson and baryon form factors, PDFs, and distribution
amplitudes. While highly economical and successful phenomenologically, such
approaches rely on guessed effective potentials whose relation to the microscopic QCD
vacuum remains indirect.

In contrast, the present framework constructs LF Hamiltonians by explicitly
boosting hadronic wave functions derived from nonperturbative vacuum dynamics,
thereby preserving a direct link between confinement, chiral symmetry breaking,
and partonic observables at a defined low scale.

\subsection*{Discretized Light-Cone Quantization}

Discretized light-cone quantization (DLCQ) represents one of the earliest
systematic numerical approaches to LF Hamiltonian field theory
\cite{Pauli:1985ps}.
By imposing periodic boundary conditions in the longitudinal direction, DLCQ
renders the LF momentum spectrum discrete and allows for numerical diagonalization
of truncated Hamiltonians.

DLCQ has been applied successfully in lower-dimensional gauge theories and as a
testing ground for nonperturbative renormalization ideas. However, its extension
to realistic (3+1)-dimensional QCD is hindered by severe truncation effects and
the complexity of renormalization in a discrete momentum basis.

\subsection*{Basis Light-Front Quantization (BLFQ)}

Basis Light-Front Quantization (BLFQ) represents a significant evolution of the
LF Hamiltonian program
\cite{Vary:2009gt,Li:2015zda}.
Instead of discretizing momentum space directly, BLFQ expands LF wave functions
in a particular basis, typically combining longitudinal momentum modes
with two-dimensional harmonic oscillator functions in the transverse plane.

This choice is motivated both by mathematical convenience and by its close
connection to confinement-inspired dynamics. The resulting Hamiltonian matrix
can be systematically improved by enlarging the basis, enabling controlled
studies of convergence and truncation effects.

BLFQ has been applied to a wide range of systems, including positronium,
heavy quarkonia, light mesons, and baryons
\cite{Zhao:2014cma,Wiecki:2014ola,Li:2017mlw}.
Extensions to time-dependent BLFQ have further enabled studies of scattering and
real-time evolution in strong external fields
\cite{Zhao:2013cma}.

\subsection*{Renormalization and Fock-Space Truncation}

A central challenge in all LF Hamiltonian approaches is renormalization in the
presence of Fock-space truncation. Several strategies have been developed,
including sector-dependent renormalization, similarity renormalization group
methods, and Hamiltonian flow equations
\cite{Perry:1990mz,Glazek:1993rc}.

In BLFQ, renormalization is typically implemented through effective interactions
matched to known physical observables or perturbative limits. While systematic
and increasingly sophisticated, these procedures highlight the tension between
maintaining a manageable Fock space and preserving the full dynamics of QCD.

The approach developed in this book addresses this issue from a complementary
angle: rather than relying solely on basis truncation, it incorporates
nonperturbative vacuum effects into the effective LF Hamiltonian itself. This
reduces the burden placed on higher Fock components at the matching scale and
clarifies the division of roles between nonperturbative input and perturbative
evolution.

\subsection*{Parton Distributions from Light-Front Wave Functions}

A major strength of LF Hamiltonian methods lies in the direct relation between
LF wave functions and partonic observables. PDFs, GPDs, and transverse-momentum
dependent distributions can be expressed as overlaps of LF wave functions,
making the connection between hadron structure and parton dynamics explicit
\cite{Diehl:2003ny}.

BLFQ and related approaches have produced explicit predictions for PDFs and
distribution amplitudes, particularly for heavy quarkonia and mesons
\cite{Li:2016mah}.
These results provide valuable benchmarks for lattice QCD and phenomenological
fits, especially in kinematic regions where experimental constraints are
limited.

In the present framework, LF wave functions are not postulated or fitted but
derived by boosting rest-frame wave functions obtained from a dynamical model
of the QCD vacuum. This ensures internal consistency between spectroscopy,
form factors, and partonic observables. Indeed, the results to be developed in the present work show that the partonic distribution functions at low resolution, are profiling of the emergent constituents from the QCD vacuum, e.g. quark zero modes and constitutive gluons as they emerge from the underlying instanton and anti-instanton dynamics.

\subsection*{Reviews of Light-Cone and Light-Front QCD}
\label{sec:key_lc_reviews}

A more comprehensive presentation of light-cone and light-front QCD and their applications to hadronic structure, can be found in a number of classic reviews of which we now cite a few and refer to their detailed references for a more complete list of references. 

An early and influential synthesis of light-cone quantization and its
application to relativistic bound states is provided by Brodsky, Pauli, and
Pinsky \cite{Brodsky:1997de}. This review established the conceptual foundations
of light-front wave functions, vacuum structure, and factorization, and remains
a standard reference.

A complementary pedagogical treatment emphasizing both formal aspects and
applications to parton physics is given by Heinzl \cite{Heinzl:2000ht}, which
clarifies the role of constrained fields, zero modes, and renormalization in
light-front dynamics.

The extensive review by Carbonell et al.\ \cite{Carbonell:1998rj} focuses on
relativistic few-body systems and bound-state equations on the light front,
providing a detailed discussion of practical implementations and numerical
methods.

For generalized parton distributions and their representation in terms of
light-front wave-function overlaps, the reviews by Diehl \cite{Diehl:2003ny} and by Belitski and Radyushkin~\cite{Belitsky:2005qn}
offer a comprehensive and widely used references.

Finally, Brodsky et al.\ \cite{Brodsky:2014yha} review modern developments in
light-front dynamics, including holographic approaches and connections to
conformal symmetry, illustrating how semiclassical light-front methods can be postulated to organize hadron spectroscopy and structure.

\section*{Relation to the Present Work}

The developments summarized above underline the breadth and diversity of
light-front Hamiltonian approaches to QCD. Basis-based methods such as BLFQ
offer systematic numerical control, while semiclassical and holographic models
provide analytic insight and phenomenological efficiency.

The framework developed in this book is aligned with these efforts in its use
of LF dynamics and Hamiltonian methods, but differs fundamentally in its
starting point. By anchoring the LF description in a nonperturbative model of
the QCD vacuum and hadronic wave functions, it seeks to unify spectroscopy and
parton physics within a single dynamical picture, rather than treating them as
separate regimes connected only by evolution equations.

Last but not least is different choice of coordinates for Hamiltonian quantization. We always use Jacobi coordinates, in CM spectroscopy and on the LF, eliminating spurious problem of center-of-mass motion.


\section{Light-front criticality and Wilsonian resolution.}
\label{sec_wilsonian}
Before closing this section, we would like to stress a key conceptual point that deserves explicit emphasis in reviewing light-front (LF) Hamiltonian approaches, and that  is the infrared (IR) nature of LF QCD itself. Canonical LF formulations correspond to quantizing QCD directly at a critical point, where the LF vacuum is trivial but long-distance dynamics are encoded in constrained fields and zero modes. As a consequence, LF QCD at infinite resolution is intrinsically afflicted by severe IR singularities. A canonical example is the appearance of inverse longitudinal derivatives in constrained fields, such as
\(
\psi_- \sim 1/D^-,
\)
as well as instantaneous gluon interactions proportional to \(1/(k^+)^2\), which diverge in the \(k^+ \to 0\) limit. These singularities are not accidental; they reflect the fact that the LF formulation, taken literally, has no intrinsic IR scale and therefore sits precisely at a critical point.

Many LF Hamiltonian approaches discussed in Sec.~\ref{sec:lf_hamiltonians_extended} attempt to manage this problem through explicit regulators, zero-mode subtractions, or basis truncations. While practically useful, such procedures obscure the separation between genuine nonperturbative physics and artifacts of the LF critical formulation. The approach developed in this work is conceptually different and explicitly Wilsonian. As detailed in Secs.~\ref{sec_flow_0} and \ref{sec_flow_1}, the LF theory is constructed at a low but finite resolution scale, where the relevant gauge fields are smooth and infrared-safe. This is achieved by defining effective LF degrees of freedom through gauge-covariant gradient flow, which suppresses short-distance fluctuations and introduces a natural separation of scales. At this stage, the LF Hamiltonian and wave functions are well-defined and free of spurious LF IR divergences.

The connection to perturbative QCD is then established by matching gradient-flow-renormalized operators and LF wave functions to the \(\overline{\mathrm{MS}}\) scheme at a fixed scale \(\mu_0\), followed by perturbative evolution, as discussed in Secs.~\ref{sec_flow_1} and \ref{sec_flow_3}. In this way, the LF formulation is treated as an effective theory at low resolution rather than as a fundamental critical description, allowing for a controlled bridge between hadronic spectroscopy and partonic observables.

We emphasize that the Wilsonian viewpoint advocated here does not preclude practical light-front Hamiltonian constructions. Rather, it implies that infrared-sensitive sectors associated with longitudinal zero modes or light-cone criticality must be treated nonperturbatively and absorbed into effective interactions. In concrete implementations, this may take the form of resumming classes of light-front-enhanced diagrams, identified by power counting, into dressed vertices or operators that encode the relevant vacuum structure, as detailed for example in section~\ref{sec_good}.

Later sections provide explicit examples of how such reorganizations arise in practice. Although the resulting Hamiltonians are written in light-front variables, they should be understood as effective descriptions in which the dominant infrared physics has already been accounted for, consistent with the Wilsonian philosophy outlined above.

\section{Comparison to the lattice LaMET approach}
It is instructive to
parallel

our strategy

with the Large Momentum Effective Theory 
(LaMET) approach.
 In LaMET, partonic physics is accessed through equal-time correlation functions evaluated at large but finite hadron momentum. Crucially, these calculations are performed away from the LF critical point: the finite hadron momentum provides an intrinsic IR regulator, and the theory does not suffer from the singular \(k^+ \to 0\) structure characteristic of canonical LF quantization. The matching to light-cone PDFs is achieved only after taking the large-momentum limit through a systematic factorization and perturbative matching procedure.

From this perspective, LaMET and the present LF formulation share an important conceptual similarity. In both cases, one deliberately avoids formulating QCD directly at the critical LF point. Instead, one works with an effective theory defined at a finite  momentum in LaMET, 
finite gradient flow scale in our approach are
 connecting 
to  the light cone theory through 
 perturbative matching and evolution. 
 This shared Wilsonian logic clarifies 
 why both approaches yield well defined 
 nonperturbative inputs for parton physics,
  while bypassing the intrinsic IR 
  pathologies of LF QCD formulated at infinite resolution.

The conceptual parallel between the present formulation and the LaMET can be sharpened by identifying explicitly how each framework avoids the light-front (LF) critical point. In canonical LF quantization, the criticality arises from enforcing exact light-like kinematics at the operator level, which removes any intrinsic longitudinal scale and leads to singular structures such as \(1/D^-\) and \(1/(k^+)^2\). Both LaMET and the present approach bypass this problem by introducing a finite control parameter that measures the distance from the LF critical point and by restoring the light-cone limit only through controlled perturbative matching.

In LaMET, the regulating parameter is the hadron momentum \(P^z\). Equal-time correlation functions are computed at finite but large \(P^z\), where the theory is infrared safe and free of LF zero-mode singularities. The limit \(P^z \to \infty\) is never taken at the nonperturbative level; instead, quasi-distributions are factorized and matched to light-cone PDFs in the \(\overline{\mathrm{MS}}\) scheme through an expansion in powers of \(1/P^z\). In this sense, \(P^z\) plays the role of a Wilsonian control scale that keeps the theory away from the LF critical surface.\footnote{As in any Wilsonian construction, power-suppressed corrections in \(1/P^z\) encode genuine long-distance physics and may exhibit ambiguities analogous to renormalons, which are systematically absorbed into higher-twist matrix elements rather than signaling a breakdown of the formalism.}

In the present formulation, the analogous control parameter is the gradient-flow scale \(\sqrt{t}\), or equivalently the matching scale \(\mu_0 \sim 1/\sqrt{t}\), as developed in Secs.~\ref{sec_flow_1}-\ref{sec_flow_3}. The LF Hamiltonian and wave functions are constructed at finite resolution using flowed gauge fields, for which constrained LF components are smooth and infrared safe. The LF critical point is not imposed as a starting point but is approached only indirectly, through perturbative matching and renormalization-group evolution.

This parallel may be summarized schematically as
\[
\text{LaMET:}\qquad P^z < \infty \;\;\Longrightarrow\;\; \text{matching} \;\;\Longrightarrow\;\; P^z \to \infty ,
\]
\[
\text{This work:}\qquad \sqrt{t} > 0 \;\;\Longrightarrow\;\; \text{matching} \;\;\Longrightarrow\;\; \sqrt{t} \to 0 .
\]
In both cases, it is essential that the limits \(P^z \to \infty\) and \(\sqrt{t} \to 0\) are \emph{never} taken at the nonperturbative level. The light-cone theory is instead viewed as an emergent description reached only after perturbative matching. This shared Wilsonian logic explains why both LaMET and the present LF formulation yield controlled nonperturbative inputs for parton physics, while bypassing the intrinsic infrared pathologies of canonical LF QCD formulated directly at the critical point.

\section{Perspective and Scope}

The approach developed in this book should be viewed as complementary to the frameworks summarized above. Its distinctive feature is the explicit dynamical continuity it establishes between the QCD vacuum, hadronic bound states, and partonic observables at a common low matching scale. By emphasizing mechanisms rather than parametrizations, it aims to provide both quantitative predictions and qualitative insight into the multiscale structure of hadrons.


\part{The QCD vacuum}
\chapter{The QCD vacuum, chiral symmetries and confinement}
\section{Modern theory of tunneling in quantum mechanics}

One of the fundamental differences between classical and quantum mechanics (QM) is that QM allows for the phenomenon of {\em tunneling} between regions where classical motion is permitted, 
separated by {\em potential
barriers}.
If  
the particle energy 
is smaller than
 the potential
$E<V(x)$ 
 the kinetic energy is negative,
\[
K=E-V(x)<0,
\]
and therefore the momentum becomes imaginary. Motion under the barrier can therefore be described using "Euclidean'' (imaginary) time $\tau=it$.

In the language of path integrals, tunneling corresponds to paths defined on the complex time plane, consisting of segments running along real and imaginary directions, separated by specific "turning points.''

Perhaps the most spectacular example of tunneling is the {\em $\alpha$ decay of heavy nuclei} (Gamow 1928), in which an $\alpha=p^2n^2$ cluster tunnels from the interior of the nucleus to the outside world through the Coulomb barrier. We will not discuss this celebrated example here; instead, we illustrate the main ideas using a simpler setting, the {\em double-well potential} (DWP).

We consider the nonrelativistic Hamiltonian with the potential
\be
V_{DWP}=\lambda (x^2-f^2)^2 ,
\ee
which has two minima at $x=\pm f$, often referred to as "classical vacua.'' Ignoring tunneling, one would expect the motion near each minimum to be independent, leading to a doubly degenerate spectrum. In particular, one expects two ground states with energy
\[
E=\omega/2=f\sqrt{2\lambda}.
\]

To make contact with QCD-inspired notation, we express all dimensional quantities in terms of the frequency $\omega$ and a dimensionless coupling
\[
\alpha=\lambda/\omega^3 .
\]
For example, the height of the barrier is
\be
V_{max}=V(x=0)=\omega {\omega^3 \over 64\lambda}= \omega {1 \over 64 \alpha}.
\ee
We define the "weak coupling'' regime as $\alpha\ll 1$, in which the barrier is very high, $V_{max}\gg \omega$. In this regime, one can develop a {\em perturbative theory} and compute the vacuum energy as a series
\be
E_{pert}={\omega \over 2}\Big[1+\sum_n C_n \alpha^n\Big],
\ee
using Feynman diagrams.\footnote{Here one must use diagrams without external legs, whose combinatorial coefficients are more subtle than those of standard diagrams with external lines.}

The double degeneracy of the vacuum (and of other states) is lifted by tunneling. The true eigenstates are symmetric and antisymmetric combinations of the classical vacua, with energies
\be \label{eqn_DWP_spilit}
E_\pm=E_{pert}\Big[1\mp \sqrt{2\over \pi \alpha}\,
\exp\!\Big(-{1 \over 12\alpha}\Big)+\ldots\Big],
\ee
separated by an exponentially small tunneling contribution.

Using Euclidean time $\tau$, one introduces the Euclidean action\footnote{Note that the potential appears with a plus sign here, unlike in Minkowski time. The dot denotes differentiation with respect to $\tau$, not real time.}
\be
S_E=\int d\tau \Big[{1 \over 2}\dot x^2+V(x)\Big].
\ee
Tunneling amplitudes are determined by minimizing this action over paths connecting the two vacua. The solution is easy to find: motion in the inverted potential $-V$ turns minima into maxima, and the trajectory "slides'' from one to the other,
\be
x_{inst}(\tau)/f=\tanh\!\Big[\omega(\tau-\tau_0)/2\Big].
\ee
This trajectory is called the {\em instanton}. Its action,
\[
S_{inst}=\omega^3/12\lambda ,
\]
is precisely the exponent appearing in Eq.~(\ref{eqn_DWP_spilit}). The existence of instanton-like paths removes the classical degeneracy and makes the quantum vacuum unique. 

So, quantum paths in vacuum can be considered a superposition of "zero-point" perturbative oscillations near one of the minima, and nonperturbative tunneling events, the instantons. The same general idea we will discuss below for the QCD vacuum as well.

For an in-depth presentation of modern semiclassical theory, one may consult, for example, the book \cite{shuryak2020lecturesnonperturbativeqcd}. The traditional WKB method discussed in QM textbooks can be replaced by the {\em flucton} calculus, in which the wave function is systematically constructed as a semiclassical series to arbitrary order. The perturbative and semiclassical expansions mentioned above are unified into a single {\bf transseries}, which cures the defects of each series individually.\footnote{Divergent series can often be summed using methods such as Borel regularization. These methods relate the ambiguities of the perturbative series to those arising in instanton calculus.}

In general, a transseries takes the form of a triple sum over integers $p,k,l$,
\[
\sum C_{pkl}\,\alpha^{2p}\,
\big[\exp(-\text{const}/\alpha)\big]^k\,
\big[\log(\pm 1/\alpha)\big]^l .
\]
The coefficients $C_{pkl}$ satisfy highly nontrivial relations due to the phenomenon known as {\em resurgence}; for a pedagogical review see, for example, \cite{Behtash:2015loa}. Flucton and instanton contributions have been computed diagrammatically up to three loops.

\begin{figure}[b]
    \centering
    \includegraphics[width=0.5\linewidth]{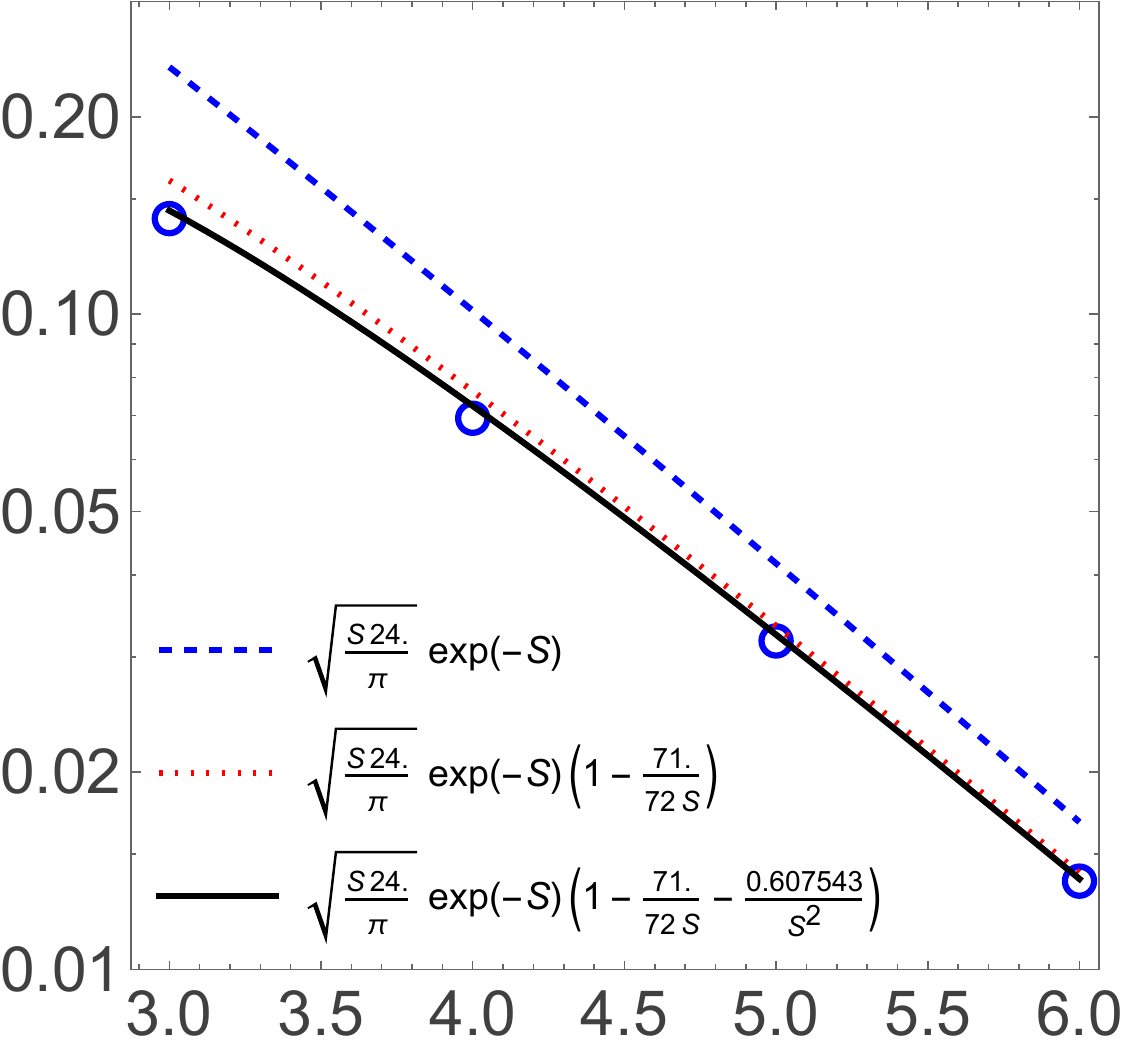}
    \caption{The splitting of the negative-parity state from the vacuum,
    $gap=E_- - E_+$, shown as data points. The three curves correspond to
    one-, two-, and three-loop semiclassical approximations.}
    \label{fig_DWP_gap_semi}
\end{figure}

To demonstrate that semiclassical theory is not merely a collection of complicated formulae, let us examine how well it works in practice. We focus on the $gap$ (splitting of negative-to-positive parity states) \footnote{In this sense, it is analogous to a pion mass though for the double-well potential.}
\[
gap=E_- - E_+ ,
\]
which can be computed numerically as a function of the instanton action $S$. The numerical results are shown as data points in Fig.~\ref{fig_DWP_gap_semi} and are compared with the one-, two-, and three-loop semiclassical predictions.

The one-loop result is based on evaluating the determinant of the fluctuation operator (see, for example, the pedagogical review \cite{Vainshtein:1981wh}). The two-loop correction was computed in \cite{Wohler:1994pg}; at this order, new diagrams appear that originate not from the action itself but from enforcing orthogonality to the zero mode. The three-loop correction was obtained in \cite{Escobar-Ruiz:2015nsa}, where many diagrams are involved, some evaluated analytically and others numerically. The explicit expressions are shown in the figure.

As the plot demonstrates, even for actions in the range
$S=3\ldots 6$,\footnote{The action is measured in units of $\hbar$, which is set to unity here and throughout.}
the three-loop approximation works remarkably well. The (unshown) residual discrepancy is consistent with an $O(S^{-3})$ correction, as expected from the yet-unknown four-loop contribution.

Looking ahead, QCD instantons have actions $S_{inst}\sim 10$-$12$, larger than those shown in the figure. Instanton constituents, known as instanton-dyons, have actions $S_{inst}/N_c\sim 4$, corresponding roughly to the middle of the plotted range. One might therefore expect semiclassical theory to work with comparable accuracy in this context. Unfortunately, for gauge theories the only explicit result currently available is the one-loop instanton density computed by 't~Hooft in the 1970. Despite half a century of effort  higher-loop results have not yet been obtained.\footnote{As the dates of the cited references suggest, each step in this program has taken on the order of two decades.}

Instantons correspond to extrema of the path integral, as they minimize the action. Another obvious extremum is the "lazy path'' $x(\tau)=0$. One may also search for complex paths with finite action, motivated by the principles of complex analysis, which suggest that all nearby extrema should be included.

Paths describing "incomplete tunneling'' can be represented by instanton-anti-instanton pairs, or "molecules.'' In the DWP, these configurations were first studied in \cite{Shuryak:1982dp} using numerical simulations of the path ensemble. The instanton-anti-instanton set was defined via "gradient flow''; in complex analysis, the corresponding objects are known as {\em Lefschetz thimbles}. For a pedagogical introduction, see \cite{Witten:2010zr}.

The basic idea, dating back to Cauchy, is to deform the integration contour from the real axis into the complex plane. This deformation is most naturally achieved by decomposing the integration domain into a set of thimbles connecting different extrema. On each thimble the phase of the integrand is constant, allowing the original complex integral to be rewritten as a sum of real integrals (with complex coefficients) that may be separately amenable to standard numerical methods, like Monte-Carlo sampling.

This opens an intriguing possibility for the {\em complexification} of path integrals, both in quantum mechanics and in quantum field theory. While this approach is not yet fully developed, it may ultimately provide a nontrivial bridge between perturbative and nonperturbative physics. Here, however, we stop and move on to instantons in quantum field theory, and in particular in QCD.

\section{The topological landscape of gauge fields}

\begin{figure}
    \centering
    \includegraphics[width=0.85\linewidth]{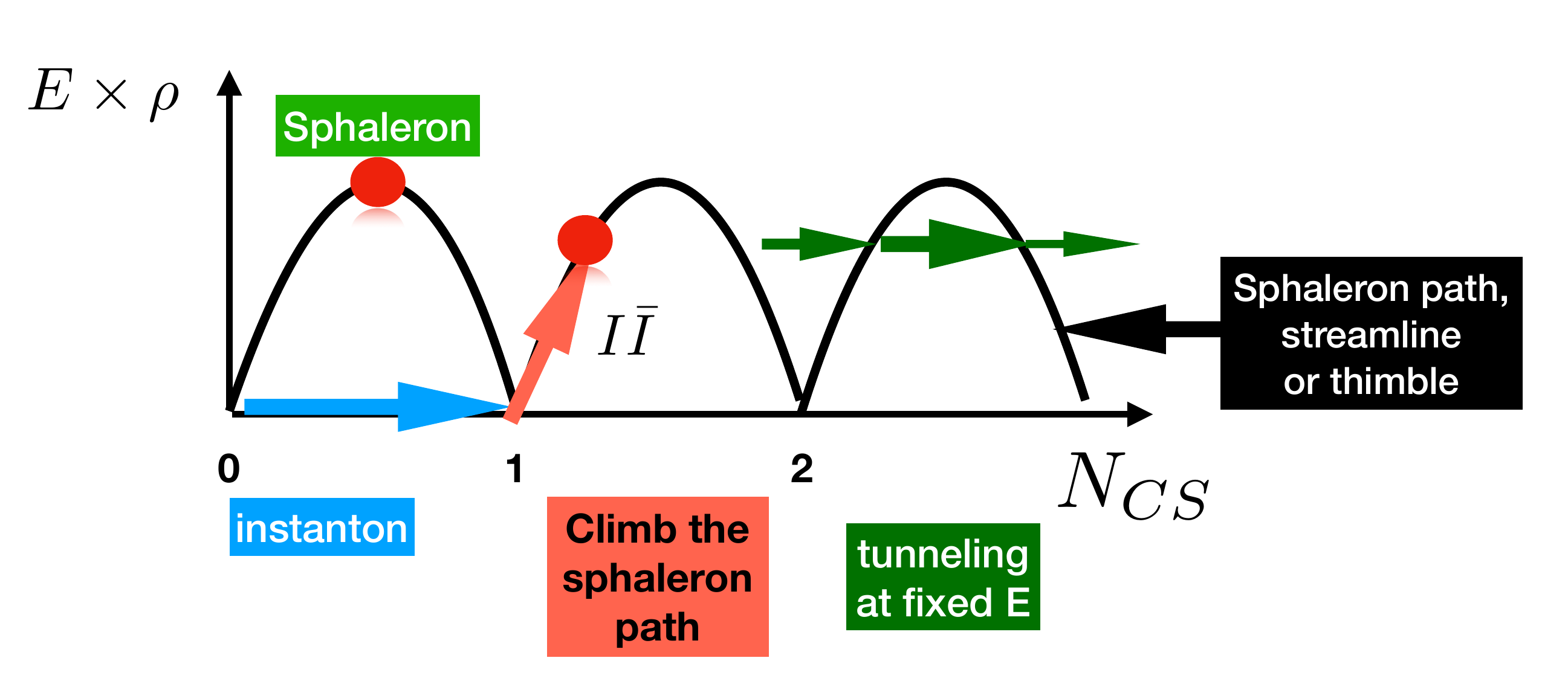}
    \caption{Topological landscape, shown as the gauge-field energy multiplied by the r.m.s. size versus the Chern-Simons number.}
    \label{fig_landscape}
\end{figure}

The $landscape$ refers to the set of {\em minimal-energy} gauge-field configurations as a function of two main variables.  
The first is the topological Chern-Simons number,
\be 
N_{CS}\equiv { \epsilon^{\alpha\beta\gamma} \over 16\pi^2}\int d^3x  
\left( A^a_\alpha \partial_\beta A^a_\gamma 
+{1\over 3}\epsilon^{abc}A^a_\alpha A^b_\beta A^c_\gamma \right),
\label{eqn_Ncs}
\ee
while the second variable is the r.m.s. size of the configuration, related to the spatial distribution of the squared field strength,
\be 
\rho_{r.m.s.}={\int r^2 G^2_{\mu\nu} d^3 r \over \int G^2_{\mu\nu} d^3 r}.
\ee
For fixed $\rho$, the landscape is shown in Fig.~\ref{fig_landscape}.  
It was defined in \cite{Ostrovsky:2002cg} in the parametric form
\begin{eqnarray}
U_{\rm min}(k, \rho)&=&(1-k^2)^2{3\pi^2\over g^2\rho}, \\ \nonumber
N_{CS}(k)&=&\frac 14 {\rm sign}(k)(1-|k|)^2(2+|k|),
\end{eqnarray}
with parameter $k$. At $k=\pm 1$ the minimal energy vanishes, while $k=0$ corresponds to a maximum of the energy. This point was called
sphaleron
 in electroweak theory 
 \cite{Klinkhamer:1984di}.

The set of configurations at arbitrary $k$ is known as the {\em sphaleron path}. These are static three-dimensional magnetic field configurations, also referred to as {\em turning points} (by analogy with quantum mechanics, where the semiclassical momentum vanishes). Such configurations are local energy minima in all directions except the size and $N_{CS}$ directions, which are held fixed.

The (anti)instanton corresponds to a tunneling path (blue arrow) connecting the bottoms of two neighboring valleys at zero energy.  
Since these configurations carry topological charge $Q=\pm 1$, they induce a change in the Chern-Simons number $\Delta N_{CS}=\pm 1$.

However, instanton paths are {\em not the only form of topological fluctuations} that may occur in this landscape. Indeed, we will discuss two additional classes of paths:
(i) those that move along the landscape following the sphaleron path (or $streamline$);
(ii) those that traverse the landscape at fixed energy, including tunneling.
All of these are illustrated in Fig.~\ref{fig_landscape}.

The first class is described by a constrained Yang-Mills equation with a nonzero right-hand side, interpreted as an external current that drags the configuration up (or down) the potential along its gradient. In the mathematical literature, it is known as
 Lefschetz timble,
  a special path connecting two extrema of a function by following its gradient. 

The second class is described by the Yang-Mills equations with a {\em zero} right-hand side and therefore occurs at fixed energy. The history of such a path passes through the turning points twice, with a Euclidean-time solution in between, a process known as {\em tunneling at nonzero energy}. In general, these paths should be supplemented by Minkowski-time solutions before and after the turning points. The technical term for this construction is a {\em zigzag path}, indicating a transition from real to imaginary time and back again.

\section{Chiral symmetries of QCD and their breaking}

The study of nonperturbative physics in strong interactions began even before the advent of QCD or the introduction of quarks. Already in the early 1960s, Nambu and Jona-Lasinio (NJL) \cite{Nambu:1961tp}, inspired by the BCS theory of superconductivity, investigated the vacuum structure of a theory with {\em massless} (but strongly interacting) fermions. They showed that an attractive interaction can lead to the formation of a gap at the surface of the Dirac sea. In other words, fermions can acquire a mass dynamically, rather than through an explicit mass term in the Lagrangian.  

A key difference from the case of superconductivity is that this phenomenon does not occur in the weak-coupling limit, but instead requires strong coupling.

It is instructive to compare perturbative and nonperturbative forces, using the NJL model as an example of the latter. The first parameter is the coupling constant. The second important parameter is the ultraviolet cutoff $\Lambda_{NJL}\sim 1\,{\rm GeV}$, below which the hypothetical attractive four-fermion interaction operates. Typical values used in NJL phenomenology are
\begin{equation}
G_{NJL}=19\,{\rm GeV}^{-2},\qquad 
\Lambda_{IR}=0.24\,{\rm GeV},\qquad  
\Lambda_{UV}=0.645\,{\rm GeV},
\end{equation}
for which the constituent quark mass is found to be $M\approx 0.4\,{\rm GeV}$.  

\begin{figure}[h]
\begin{center}
\includegraphics[width=7cm]{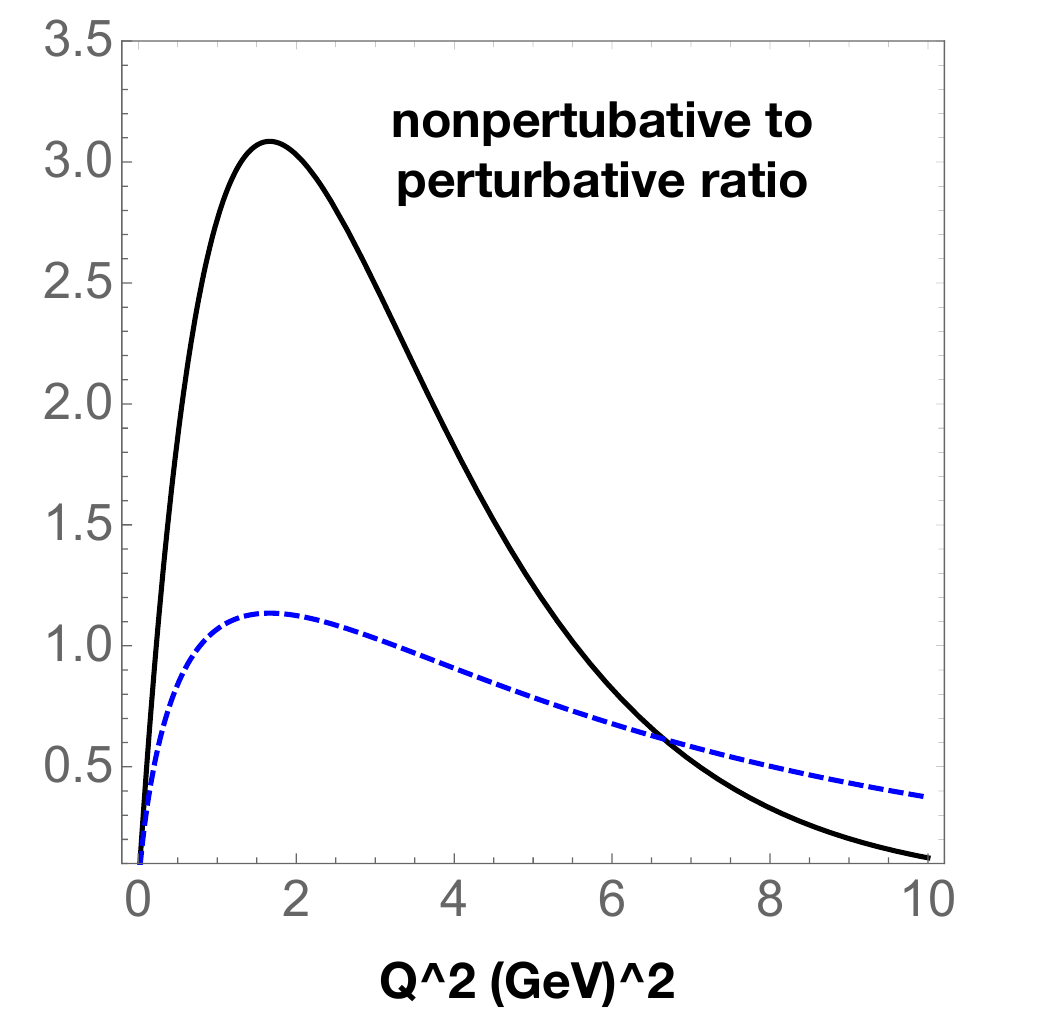}
\caption{The ratio of the nonperturbative-to-perturbative four-fermion effective vertex (\ref{nonpert_to_pert}), shown for a Gaussian form factor (solid) and an exponential form factor (dashed), as a function of the momentum transfer squared $Q^2\,({\rm GeV}^2)$.}
\label{fig_nonpert_to_pert}
\end{center}
\end{figure}

Let us compare this to the force arising from one-gluon exchange,
$F_{\rm gluon}(k^2)=g^2/k^2$. For a typical momentum exchange inside a meson,
\begin{equation}
k^2=x\bar x\, Q^2 \approx Q^2/4.
\end{equation}
The ratio of the NJL interaction to the gluon-exchange force can then be estimated as
\begin{equation}
{ G_{NJL} \over F_{\rm gluon} }\,
\exp\!\left(-{k^2 \over \Lambda_{UV}^2}\right).
\label{nonpert_to_pert}
\end{equation}
This ratio is shown in Fig.~\ref{fig_nonpert_to_pert}. While it decreases at large momenta due to the form factor, it remains larger than unity over a wide range of momentum transfers.

Thus, at a qualitative level, the puzzlingly large values of hadronic form factors at intermediate momenta, $Q^2\sim 1$-$7\,{\rm GeV}^2$, may be understood if nonperturbative contributions are added to the perturbative diagrams.

An important consequence of the NJL construction was the introduction of {\em chiral symmetries}, their {\em spontaneous breaking} (specifically of the $SU(N_f)_A$ subgroup), the emergence of a nonzero quark condensate $\langle\bar q q\rangle\neq 0$, and the appearance of the {\em Nambu-Goldstone modes}: pions, kaons, and related mesons. NJL suggested that the interquark interaction can be modeled by an attractive local four-fermion operator of the form\footnote{Boldface denotes isovectors.}
\be 
L_{NJL}=G_{NJL}(\vec\pi^{\,2}+\sigma^2),
\ee
where $\vec\pi=i\bar q\gamma_5\vec\tau q$ and $\sigma=\bar q q$. It is implicitly assumed that the NJL interaction is restricted to momentum transfers $Q^2<\Lambda_{NJL}^2\sim 1\,{\rm GeV}^2$. Thus, the NJL model contains two parameters: the coupling $G_{NJL}$ and the cutoff $\Lambda_{NJL}$.

Many major developments of the 1960s and 1970s, such as the rise of perturbative QCD and its phenomenological applications, temporarily shifted attention away from these issues. Practical discussions of the nonperturbative QCD vacuum structure resumed with the introduction of the {\em QCD sum rule} approach by \cite{Shifman:1978bx}. In this framework, perturbative short-distance descriptions of Euclidean point-to-point correlation functions were complemented by the {\em operator product expansion} (OPE), which introduced vacuum condensates such as the quark condensate $\langle\bar q q\rangle\neq 0$, the {\em gluon condensate} $\langle G^2\rangle$, and many higher-dimensional operators.

By matching the long-distance behavior of correlators, expressed in terms of the lightest hadronic states, with their OPE expansions at short distances, it became possible to extract the values of these condensates and to observe a remarkable consistency between the two regimes. Only several spin-zero (scalar and pseudoscalar) quark and gluon channels resisted such a description, indicating much stronger nonperturbative effects \cite{Novikov:1981xi}.

Extensive phenomenological studies of Euclidean point-to-point correlation functions in many hadronic and glueball channels have confirmed that correlators of operators with different quantum numbers behave strikingly differently (see the review \cite{Shuryak:1993kg}). While correlators of vector currents $j_\mu=\bar q\gamma_\mu q$, built from light quarks $q=u,d,s$, show only minor deviations from free-quark propagation over distances from zero to about $1\,{\rm fm}$, the correlators of spin-zero scalar and pseudoscalar operators ($\gamma_\mu\to 1,\gamma_5$) exhibit strong deviations from perturbative QCD already at relatively short distances, $x\gtrsim 0.2\,{\rm fm}$. These observations demonstrate that the dominant quark interactions at short distances cannot be reduced to perturbative gluon exchange alone.
(Attempts to explain these effects using, for example, rainbow diagrams at strong coupling continue (see e.g. \cite{Raya:2024ejx}). While such approaches may generate a nonzero quark condensate, they do not explain why some correlation functions remain essentially unaffected while others  like the $\pi$ and $\eta-prime$ channels exhibit strong effects of opposite sign. In particular, it is difficult to understand how gluon exchange could produce repulsive forces in a meson channel.)

An alternative picture, developed since the early 1980s and summarized in \cite{Shuryak:1993kg}, attributes the strong nonperturbative forces responsible for splittings in spin-zero channels to the {\bf collectivization} of instanton zero modes. Equivalently, one may say that nonperturbative interquark interactions are described by the multi-fermion 't~Hooft Lagrangian \cite{tHooft:1976snw}. These effects are induced by topological fluctuations of the gluon fields, which can be described semiclassically by topological solitons known as {\em instantons} \cite{Belavin:1975fg}.  

For two quark flavors ($u,d$), the interaction has the schematic four-fermion form\footnote{Letters in boldface are isovectors.}
\be 
L_{'t\,Hooft}\sim \big(\vec\pi^{\,2}+\sigma^2-\eta'^2-\vec s^{\,2}\big),
\ee
where the first two terms reproduce the NJL structure, while the latter two correspond to repulsion in the channels $\eta'=\bar q i\gamma_5 q$ and $\vec s=\bar q\vec\tau q$. The negative sign in front of these terms reflects the explicit breaking of $U(1)_A$ symmetry by instantons, a key consequence of the axial anomaly.


\section{The QCD vacuum as an {\em instanton liquid} (ILM)}

Models of the QCD vacuum that view it as an ensemble of semiclassical solitons -instantons and anti-instantons  - are collectively known as instanton liquid models (ILM)\footnote{The key word here is $liquid$, as opposed to an {\em ideal gas}. It emphasizes that while solitons retain their individual identity, their interactions are strong.}. 
Depending on the theoretical tools and assumptions employed, these models can be classified into the following categories:\\
(i) In the original model~\cite{Shuryak:1981fza}, instanton parameters were inferred from empirical values of vacuum observables, such as the {\em quark and gluon condensates} and the {\em topological susceptibility}.\\
(ii) Mean-field analyses of the instanton partition function based on the sum ansatz~\cite{Diakonov:1985eg,Nowak:1996aj}.\\
(iii) Numerical simulations of the partition function, based on the common quark zero-mode determinant; see~\cite{Shuryak:1992jz} and the review~\cite{Schafer:1996wv}.\\
(iv) The "dense ensemble'' (DILM), which includes both isolated instantons and {\em instanton-anti-instanton molecules}~\cite{Shuryak:2019zhv}.

Here we cannot discuss these approaches in detail and instead limit ourselves to comments on their main assumptions and results. For illustrative figures and quantitative lattice results, see Section~\ref{sec_top_lat}.

The typical instanton size and density in ILM are 

\begin{equation} \label{eqn_rho_n}
\rho_{ILM}\sim \frac 13\, {\rm fm} , \,\,\, n_{ILM}\approx 1\, {\rm fm}^{-4} .
\end{equation}

Because $\rho$ is relatively small compared to the sizes of most hadrons\footnote{With the exception of the lowest $\Upsilon$ states and the pions.}, it is often convenient to employ a {\em quasi-local approximation}, in which $\rho$ is taken to zero. In this limit, the ILM density appears as a coupling constant multiplying the local 't~Hooft Lagrangian. Such "quasi-local'' multi-quark vertices are reminiscent of those proposed earlier by Nambu and Jona-Lasinio. 

Indeed, the two NJL parameters, the coupling $G_{NJL}$ and the cutoff $\Lambda_{NJL}\sim 1\,{\rm GeV}$,  can be directly related to the two ILM parameters quoted above. One can then calculate various diagrams involving these vertices. In contrast to the NJL model, the ILM contains no ultraviolet divergences, since instanton zero modes are not truly pointlike.

The most important property shared by all versions of the ILM is the {\em spontaneous breaking of the $SU(N_f)_a$ chiral symmetry}. In particular, this mechanism leads to massless pions as Nambu-Goldstone modes.

The well-known {\em axial anomaly} states that, unlike the vector isoscalar current, the axial current is not conserved. Its divergence is proportional to $\tilde G G$, the topological charge density. Tunneling events contribute to this quantity with properly normalized positive or negative integers, and therefore must also change the chirality of each sufficiently light fermion. The microscopic mechanism underlying this phenomenon was discovered in~\cite{'tHooft:1977hy}: massless fermions (quarks in QCD, and quarks and leptons in the electroweak theory) possess zero modes in the instanton background. As a result, instantons induce novel fermionic interactions described by the {\em 't~Hooft effective Lagrangian} $L_{tHooft}$, an operator with $2N_f$ fermion legs. In QCD with $u$ and $d$ quarks, this reduces to a four-fermion operator resembling the hypothetical NJL Lagrangian~\cite{Nambu:1961fr}. Unlike the NJL interaction, however, it explicitly violates the $U(1)_a$ chiral symmetry.

Returning to the ILM in the early 1980s, the central question was whether the instanton density and sizes are {\em large enough} to generate spontaneous breaking of the $SU(N_f)_a$ chiral symmetry. In other words, can $L_{tHooft}$ play the same dynamical role as the NJL Lagrangian proposed two decades earlier? If so, does it quantitatively reproduce the parameters of chiral perturbation theory, such as the pion mass and couplings? In~\cite{Shuryak:1982hk} and subsequent work, it was shown that this is indeed the case. The most direct way to understand this result is through a mean-field analysis, discussed in Section~\ref{sec_app_chiral}.

An important qualitative feature of the instanton vacuum is the prediction that vacuum gauge fields are extremely inhomogeneous. Near instanton centers ($r\rightarrow 0$), the fields are very strong,
\be
G_{a\mu\nu}^2(r)={192\rho^4 \over (\rho^2+r^2)^4},
\ee
reaching scales of several ${\rm GeV}^2$. These strong fields can capture quarks and confine them into quasi-bound states, the zero modes. At the same time, the fields are nearly zero in the regions between instantons. Clearly, quark propagators behave very differently in these two extreme environments\footnote{As an analogy, consider a town experiencing a storm with multiple tornadoes. The behavior of people far from the tornadoes and those caught inside them would be drastically different. Consequently, a {\em mean propagator} averaged over the entire vacuum ( as often used, for example, in the Schwinger Dyson community~\cite{Raya:2024ejx})  has limited physical meaning.}.

An {\em alternative view of chiral symmetry breaking} emphasizes quark $hopping$ between instantons, which is possible because their zero modes form degenerate bound states. The resulting quark loops can be either short or long (infinite in the thermodynamic limit $V_4\rightarrow\infty$). The latter are responsible for the accumulation of Dirac eigenvalues near zero, $\lambda\sim 1/V_4$. If {\em collectivization} of instanton zero modes into the so-called {\em zero mode zone} (ZMZ) occurs, no gap appears in the Dirac eigenvalue spectrum. The near-zero Dirac eigenvalues are confined to a narrow strip,
\begin{equation}
|\lambda| \sim {\rm width\,(ZMZ)} \sim {\rho^2\over R^3} \sim 20\,{\rm MeV},
\end{equation}
as predicted by the original ILM using the parameters quoted above. In~\cite{Glozman:2012fj}, meson and baryon spectroscopy was studied after removing all Dirac eigenstates within a given strip $|\lambda|<\Delta$. A strong restructuring of the light-hadron spectrum was observed when $\Delta > {\rm width\,(ZMZ)}$. In particular, the Nambu-Goldstone modes (pions) disappeared entirely from the spectrum, as expected. It would be interesting to extend this analysis to heavy-light systems and to investigate more generally the fate of various spin-dependent forces.

Further statistical descriptions of interacting instanton ensembles were developed using mean-field techniques and numerical simulations; for reviews see~\cite{Diakonov:1995ea,Schafer:1996wv,Nowak:1996aj}. Related lattice studies include~\cite{Chu:1994vi,Faccioli:2003qz}. Effective models employing the quasi-local approximation or direct numerical simulations of instanton ensembles were actively developed throughout the 1990s.


\section{Hadrons in the instanton vacuum}

The important role of topology and instantons is technically understood through the inclusion of the so-called 't~Hooft effective Lagrangian. Since its explicit form is rather technical, we discuss it in the Appendix to this chapter and here only summarize its main properties.

The 't~Hooft effective Lagrangian generally includes all fermion flavors in a single nonlocal vertex. In QCD, these are only the light quarks ($u,d,s$), so the interaction takes the form of a six-fermion operator%
\footnote{In electroweak theory, the instanton-induced Lagrangian involves 12 fermions: 9 quarks and 3 leptons. Which fermions appear depends on the $SU(2)$ gauge orientation of the instanton. An extreme possibility is all up-type fermions: $u,c,t$ (each with three colors) together with $e,\mu,\tau$. As required in electroweak theory, 
all of them must be left-handed.}.

The chiralities of quark and antiquark lines are always opposite, so any attempt to close these lines into loops vanishes for massless quarks. This interaction explicitly violates the $U_A(1)$ chiral symmetry, since it changes the chiral charge $Q_5$ of a state%
\footnote{Instanton-induced production of axial charge is, of course, a specific example of a sphaleron process : whether the motion in the topological landscape proceeds via real or virtual fields is irrelevant, and the general relation between the Chern-Simons number $N_{CS}$ and $Q_5$ holds for any such process.}.

For temperatures $T<T_c$, the $SU(N_f)$ chiral symmetry is spontaneously broken, and therefore one (or more) quark-antiquark pairs can be replaced by their condensate. This leads to effective four- (or two-) quark operators. Both are extremely important for understanding hadronic physics. The two-quark operator generates a nonzero constituent quark mass, while the resulting four-fermion operators lead to a variety of effects that we discuss below.

As already mentioned, the instanton liquid model (ILM) provides not only qualitative insights, but also a rather good quantitative description of {\em chiral physics}. Without going into technical details, let us enumerate some of its implications for hadronic spectroscopy. First, it establishes a quantitative connection between the quark condensate and the instanton density. Furthermore, the lightest chiral multiplet, the $\sigma$ and $\pi$ mesons, is dominated by the so-called "instanton chain"  diagrams involving the 't~Hooft interaction. These diagrams can be summed using the Bethe-Salpeter equation, allowing one to relate all chiral parameters%
\footnote{These include $\langle\bar q q\rangle$, $f_\pi$, $m_\pi^2$, as well as higher-order couplings of the effective chiral Lagrangian.}
to the key vacuum properties given in Eq.~(\ref{eqn_rho_n}).

\begin{figure*}
  \includegraphics[height=5.5cm,width=.46\linewidth]{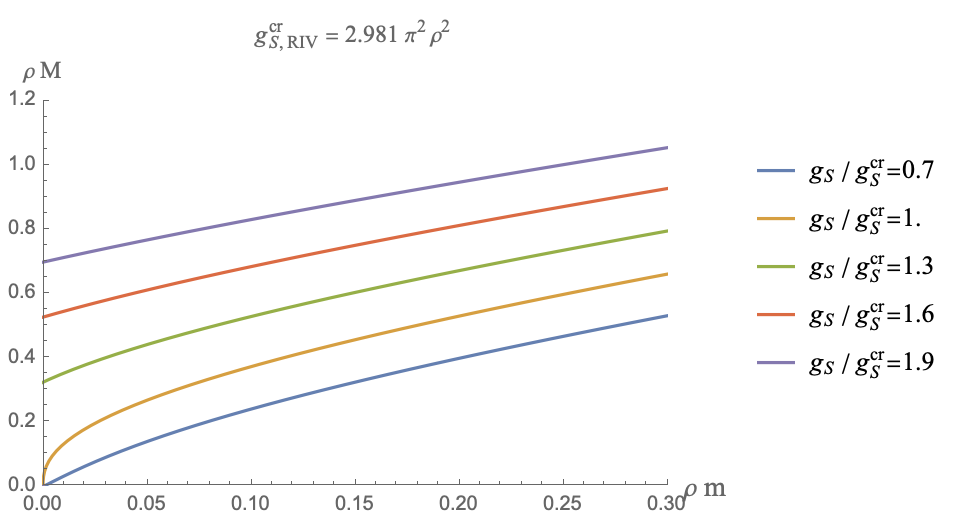}%
  \includegraphics[height=5.5cm,width=.46\linewidth]{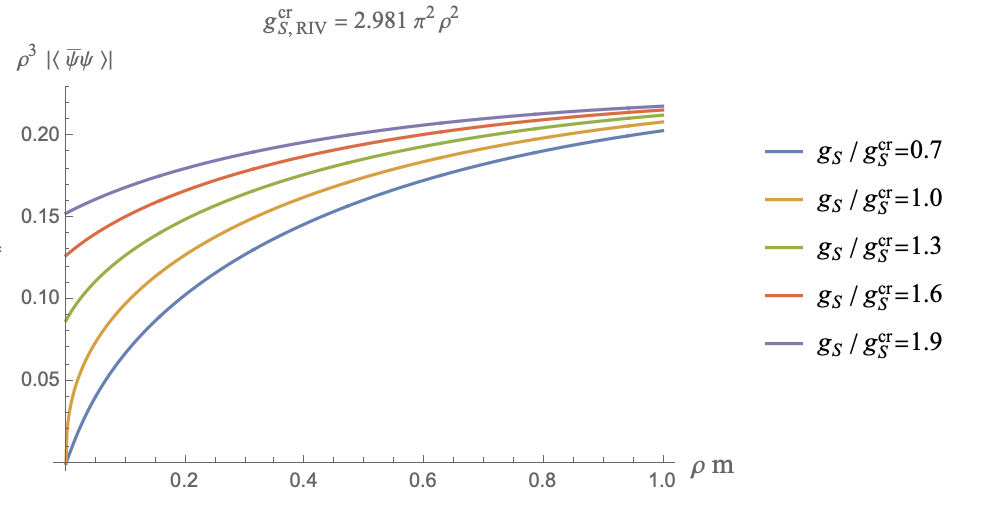}%
\caption{(a) Quark constituent mass $M$ versus the current quark mass $m$ in the ILM, for increasing strength of the 't~Hooft coupling $g_S$ (from bottom to top), with the critical coupling $g_{S,\mathrm{RIV}}^{\rm cr}=2.981\,\pi^2\rho^2$ and fixed instanton size $\rho$. 
(b) Quark condensate $\langle \bar\psi\psi\rangle$ versus the current quark mass in the ILM~\cite{Liu:2023feu}.}
\label{fig_Gap}
\end{figure*}

Unlike superconductors or the Fermi surface in metals, the surface of the massless Dirac sea can be $gapped$ only if the attractive interaction exceeds a critical strength. This was already observed by Nambu and Jona-Lasinio and also applies to the 't~Hooft interaction. In Fig.~\ref{fig_Gap} we show representative results for the quark constituent mass (left panel) and the chiral condensate (right panel) for different values of the coupling indicated in the figure.

Other mesons-such as vector mesons $\rho,\omega$, tensor mesons, and higher-spin states-do not couple directly to instantons, as already indicated by analyses of empirical correlation functions \cite{Shuryak:1993kg}. Chiral channels of the $LL+RR$ type, however, are sensitive to forces induced by $\bar I I$ molecules. We will discuss these effects in the next section.

The simplest baryons in the flavor $SU(3)_f$ decuplet are $\Delta(qqq),\ldots,\Omega$. These states can be viewed as three constituent quarks bound together primarily by confinement. The corresponding nonrelativistic potentials, generated by averaging three Wilson lines in the background field of $\bar I I$ molecules, have been used extensively%
\footnote{Unfortunately, purely heavy states such as $ccc$ or $bbb$ have not yet been observed. However, the same methods can be applied to $cc\bar c\bar c$ tetraquarks, as discussed below.}.

Finally, nucleons $p,n$ (and the other members of the $SU(3)_f$ octet) differ from the decuplet baryons because the $ud$ pairs in this case are in the so-called $good$ diquark. These constitute attractive channels, both through perturbative Coulomb forces \cite{Jaffe:2003sg} and through instanton-induced interactions%
\footnote{In a sense, diquarks are {\em half-brothers  of the pions} and can also be described by instanton-induced loop diagrams. In fact, in two-color QCD ($N_c=2$), they are true twins and become massless in the chiral limit \cite{Rapp:1997zu}.}
In Section~\ref{sec_Delta_Nucleon}, we will show how accounting for such pairing effects leads to the observed differences between the nucleon and $\Delta$ baryons.

\begin{subappendices}


\section{Chiral symmetry breaking: brief derivations}
\label{sec_app_chiral}

Historically, the concept of effective quark masses generated by dynamical chiral symmetry breaking was introduced by Nambu and Jona-Lasinio \cite{Nambu:1961fr}. The central idea is that an attractive interaction between a quark and an antiquark can be sufficiently strong to open a gap at the surface of the massless Dirac sea, in close analogy with electron-electron attraction producing a spectral gap at the Fermi surface in superconductors. As discussed in the Introduction and in the chapter on the QCD vacuum, the modern formulation \cite{Shuryak:1981ff,Diakonov:1985eg} of the NJL four-fermion operator is nothing but the instanton-induced 't~Hooft vertex.

The mechanism can be demonstrated most simply in the mean-field approximation, following the NJL approach. The emergent constituent quark mass in the rest frame appears as a solution of the {\em gap equation}~(\ref{MCONST}),
\begin{equation}
\label{MKKX}
    M(k)=m+2g_S\mathcal{F}(k)\int\frac{d^4q}{(2\pi)^4}\frac{4M(q)}{q^2+M^2(q)}\mathcal{F}(q).
\end{equation}
The last term represents a dressed quark loop generated by the four-fermion interaction, with $g_S=G_S/N_c$ denoting the rescaled 't~Hooft coupling. In contrast to the local NJL model, instanton semiclassical theory introduces the form factor $\mathcal{F}(q)$, which reflects the finite size of instantons.

A formal solution of Eq.~(\ref{MKKX}) can be written as
\begin{equation}
\label{MFK}
    M(k)=m\left[1-\mathcal{F}(k)\right]+M\,\mathcal{F}(k)\sim M\,\mathcal{F}(k),
\end{equation}
which reduces Eq.~(\ref{MKKX}) to the standard {\em gap equation}  for the constituent mass $M$,
\begin{equation}
\label{MASSGAP}
    \frac{m}{M}=1-8g_S\int\frac{d^4k}{(2\pi)^4}\frac{\mathcal{F}^2(k)}{k^2+M^2}.
\end{equation}

After neglecting the momentum dependence of the running mass in the denominator of Eq.~(\ref{MASSGAP}), as discussed in \cite{Kock:2021spt}, one finds that solutions for $M$ exist only for couplings exceeding a critical value. In the instanton liquid model (ILM), this critical coupling is determined by the instanton-antiinstanton density.

The chiral condensate is obtained in a similar manner,
\bea
\label{PSIPSI}
    \langle\bar{\psi}\psi\rangle
    =-\int\frac{d^4k}{(2\pi)^4}\mathrm{Tr}\,S(k)
    =-8N_cM\int\frac{d^4k}{(2\pi)^4}\frac{\mathcal{F}(k)}{k^2+M^2}.
\eea
Note that this expression reduces to the scalar mean-field expectation value $\langle\sigma\rangle=\langle\bar\psi\psi\rangle/N_c$ only in the local  limit of $vanishing$ instanton size. In that limit, both Eqs.~(\ref{MASSGAP}) and (\ref{PSIPSI}) diverge logarithmically in the infrared and ultraviolet. This problem is resolved in the instanton vacuum, where the quark zero modes generate the form factor $\mathcal{F}(k)$, which is essential for obtaining finite results.

In Fig.~\ref{fig_Gap}a we show the dependence of the constituent quark mass $M$ on the current quark mass for different values of the multi-fermion coupling $g_S$ in the ILM of the QCD vacuum. In Fig.~\ref{fig_Gap}b we show the corresponding dependence of the scalar quark condensate on these parameters.

\section{Condensate on the light front}

On the light front, the gap equation is modified; we now explain how this occurs.
After eliminating the bad fermionic component in the mean-field approximation,
only the physical modes of the good component $\psi_+$ remain.
The running quark mass on the light front behaves as
$M(k^-)\sim M{\mathcal F}(k^-)$, together with the on-shell condition
$2k^-k^+=k_\perp^2+M^2$, where $M$ is determined by the scalar gap equation
(\ref{MASSGAP}). Expressed in terms of the physical modes of $\psi_+$,
Eq.~(\ref{MASSGAP}) can be rewritten as
\begin{equation}
\label{gap_eq}
    \frac{m}{M}=1-2g_S\int \frac{dk^+d^2k_\perp}{(2\pi)^3}\frac{\epsilon(k^+)}{k^+}\mathcal{F}^2(k)\bigg|_{k^-=\frac{k_\perp^2+M^2}{2k^+}} \, .
\end{equation}
Here we assume that ${\mathcal F}(k^2)$ is free of physical poles.
This is the case for (\ref{FKK}) after analytic continuation, up to spurious
branch points whose contributions can be neglected at leading order in the
diluteness factor~\cite{Kock:2021spt}.

To express the chiral condensate on the light front, we first note that the quark
propagator takes the form
\bea
\label{eq:prop}
    S(k)\rightarrow\left[\frac{i[\slashed{k}+M(k^2)]}{k^2-M(k^2)^2}-\frac{i\gamma^+}{2k^+}\right]
    \rightarrow \left[\frac{i[\slashed{k}+M(k^2)]}{k^2-M^2}-\frac{i\gamma^+}{2k^+}\right] \, ,
\eea
following the elimination of the bad component in favor of the good one.
The condensate is then given by
\begin{equation}
\label{cond_eq}
    \langle\bar{\psi}\psi\rangle=-2N_cM\int\frac{dk^+d^2k_\perp}{(2\pi)^3}\frac{\epsilon(k^+)}{k^+}\mathcal{F}(k^-) \, .
\end{equation}

Using the explicit form of the instanton-induced form factor ${\cal F}(k)$,
the gap equation becomes
\bea
    \frac{m}{M}
    =1-\frac{4g_S}{\pi^2\rho^2}\int_0^{\infty} dz \, z\frac{z^3}{z^2+\frac{\rho^2M^2}{4}}
    |z(I_0(z)K_0(z)-I_1(z)K_1(z))'|^4 \, ,
\eea
and similarly the quark condensate reads
\begin{equation}
    \rho^3\langle\bar{\psi}\psi\rangle=-\frac{4N_c}{\pi^2}\rho M
    \int_0^\infty dz \frac{z^3}{z^2+\frac{\rho^2M^2}{4}}
    |z(I_0(z)K_0(z)-I_1(z)K_1(z))'|^2 \, .
\end{equation}

In the chiral limit, the constituent mass is nonzero only when the
't~Hooft coupling exceeds a critical value $g^{\mathrm{cr}}_{S,\mathrm{RIV}}$,
defined by
\bea
    g^{\mathrm{cr}}_{S,\mathrm{RIV}}=
    2\pi^2\rho^2\left[8\int_0^{\infty} dz \, z
    |z(I_0(z)K_0(z)-I_1(z)K_1(z))'|^4\right]^{-1}
    \approx 2.981\pi^2\rho^2 \, .
\eea
We note that in the limit ${\cal F}\rightarrow 1$ the above expressions become
divergent, clearly demonstrating the essential role of the instanton-induced
zero-mode profile ${\cal F}(k)$. In the absence of this structure, ad hoc
cutoffs are required. For vacuum loops (the 0-body sector), such a cutoff must
apply to on-shell momenta satisfying $k^2=2k^+k^-+k_\perp^2$ and should preserve
both boost and parity invariance, for example
$|\sqrt{2}k^\pm|\leq \Lambda\sim 1/\rho$.
Further discussion of this issue can be found in Ref.~\cite{Liu:2023fpj}.


\section{QCD topological objects observed on the lattice}
\label{sec_top_lat}

Over the last decades, lattice gauge theory has developed into a large and mature field, eventually reaching the capability to perform numerical simulations with (near-)realistic quark masses.
As a result, the main features of the hadronic spectrum have been successfully reproduced.
One may therefore ask why further efforts to derive and understand many aspects of hadronic structure,
as described in this review, are still justified.

There are essentially two reasons.
The first, discussed in the first half of this review, concerns a deeper understanding of the confining
and spin-dependent forces between quarks.
The second is the goal of {\em bridging the gap between spectroscopy and light-cone observables.}
Lattice results, used to the fullest extent possible, guide attempts to explain these phenomena
in terms of models that are significantly simpler than full-scale numerical simulations of
Euclidean gauge fields.

\begin{figure}[t!]
	\begin{center}
		\includegraphics[width=9cm]{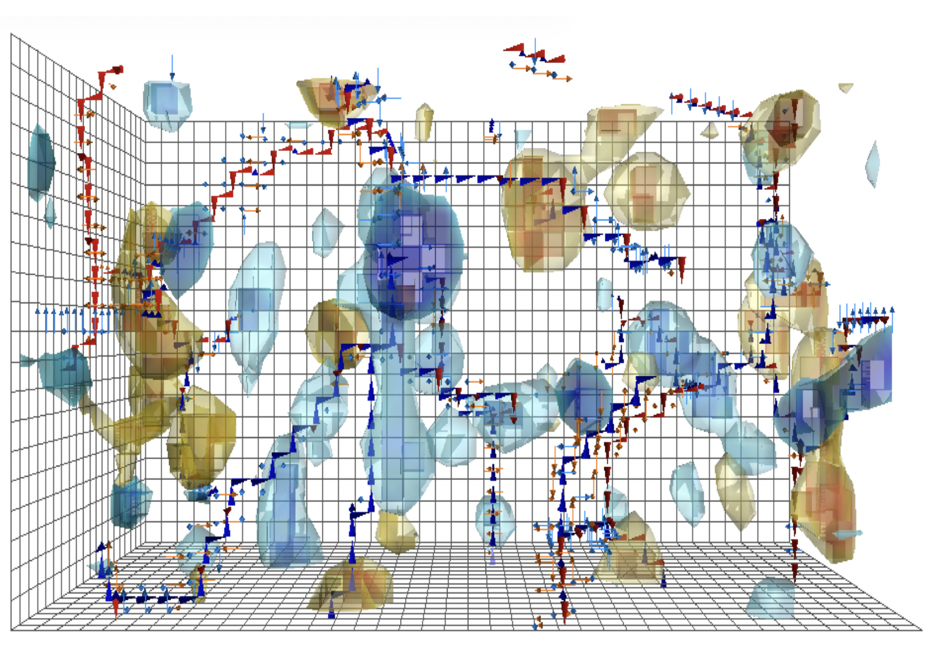}
		\caption{Instantons (yellow) and anti-instantons (blue) in the {\em deep-cooled Yang-Mills vacuum}.
		After center projection, they are threaded by center P-vortices~\cite{Biddle:2019gke,Biddle:2020eec}.
		These configurations constitute the primordial gluon epoxy (hard glue) at the origin of
		light hadron masses~\cite{Zahed:2021fxk}.
		The center P-vortices are responsible for confinement. See text for details.}
		\label{fig_VAC}
	\end{center}
\end{figure}

Proceeding in this spirit, we present two relatively recent examples of visualizations of vacuum fields in QCD.
The Instanton Liquid Model discussed above describes these fields as highly inhomogeneous:
localized regions of very strong gauge fields at instanton centers are surrounded by
"empty'' space-time largely free of nonperturbative fields.
Since the mid-1990s, such field distributions have been extracted from lattice configurations
using various "cooling'' methods.
Since a picture is worth many words, we refer the reader to the visualization shown in
Fig.~\ref{fig_VAC}, which displays a representative topological charge distribution from
Refs.~\cite{Leinweber:1999cw,Biddle:2019gke,Biddle:2020eec}.
It reveals a number of instantons and anti-instantons populating the QCD vacuum.

Note that the topological clusters are also threaded by thin center vortices, or
Z$_3$-flux strings, which span world-sheet surfaces in four dimensions.
While center vortices are essential for enforcing confinement at long distances,
Fig.~\ref{fig_VAC} shows that they are, on average, largely decoupled from the strong and
inhomogeneous topological fields, except at the branching points of P-vortices.
Moreover, the field strength associated with these vortices in the vicinity of topological
objects is approximately $\sigma_T \bar{\rho} \approx 0.3\,{\rm GeV}$.
This is weaker than the typical chromo-electric or chromo-magnetic field strength at an
instanton center, $\sqrt{E}=\sqrt{B}\approx 2.5/\bar{\rho}\approx 1.5\,{\rm GeV}$,
yet it is crucial for long-distance color correlations.

\begin{figure}[t]
\begin{center}
\includegraphics[width=7.cm]{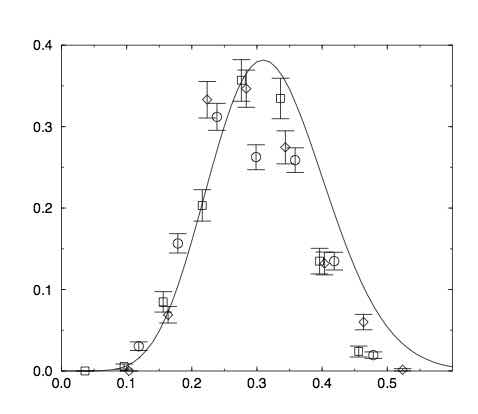}
\caption{Left: Size distribution of instantons from lattice configurations~\cite{Hasenfratz:1999ng}.
The curve represents the theoretical prediction in Eq.~(\ref{eqn_inst_distr}).
Right: Schematic paths of quarks and antiquarks for pions and nucleons.
Instantons are depicted as small hills with tunnels.
Note that chirality flips $L\leftrightarrow R$, as well as at blue dots indicating quark mass insertions.}
\label{fig_inst_sizes1}
\end{center}
\end{figure}

Many years later, lattice simulations have quantified instanton parameters in a systematic way,
see for example the size distribution shown in Fig.~\ref{fig_inst_sizes1}.
Technically, this is obtained from lattice gauge fields that have been {\em deeply cooled,}
meaning that the action is minimized in a way that removes perturbative gluons while preserving
the gauge topology.
The curve shown corresponds to the distribution
\be
\label{eqn_inst_distr}
{dn \over d\rho}= {dn \over d\rho}\bigg|_{semiclassical} \cdot e^{-2\pi \sigma \rho^2} \, ,
\ee
proposed in Ref.~\cite{Shuryak:1999fe}, where the suppression of large instanton sizes arises
from a dual monopole condensate inferred from the flux-tube tension $\sigma$.
The semiclassical contribution is known only to one-loop order~\cite{tHo_76b};
unfortunately, higher-loop corrections, analogous to those studied in quantum mechanics,
have not yet been calculated.

Note that the four-dimensional ball volume is $\pi^2 \rho^4/2$, and the diluteness parameter
$n_{I+\bar{I}} \pi^2 \rho^4/2 \sim 1/20 \ll 1$ is indeed small.
Nevertheless, instantons interact strongly with each other, which motivates the term
$liquid$ in the Instanton Liquid Model.
The theoretical description began with a mean-field approximation~\cite{Diakonov:1985eg},
followed by statistical numerical simulations; for a review, see Ref.~\cite{Schafer:1996wv}.
An important qualitative feature is that the QCD vacuum is highly inhomogeneous%
\footnote{Attempts to describe the vacuum using $averaged$ propagators or vertices are therefore
doomed to fail. Instead, one should compute entire processes (such as form factors or scattering
amplitudes) in the background of a single instanton and subsequently average over the ensemble.},
with most regions essentially free of strong fields, while at instanton centers the field
strengths are large, $G \sim 2$-$5\,{\rm GeV}^2$.

The broader implication of this analysis is that the topological structure of the QCD vacuum is not merely a theoretical curiosity but the organizing principle behind hadron formation. Instantons break chiral symmetry, generate dynamical masses, produce multiplet structures in the diquark sector, and ultimately give rise to the bound-state structure of the low-lying baryons. The instanton liquid model therefore provides a unified, predictive, and physically transparent framework for understanding low-energy hadron physics.

Confinement is not as essential for ground state hadrons, as it is for the high lying ones and their manifest Reggeization. In the Euclidean formulation, the pion and nucleon worldlines are kept bunched by the 't Hooft interactions, which makes them less sensitive to flux piercing by P-vortices in the dual description. This is not the case for the Delta and high-lying hadrons, where the worldlines are  wider and more prone to flux piercing.  

In the LF descrition (to which we turn in the last part of this review) the Hamiltonian
actually produces square masses of hadrons,
and Regge trajectories would naturally appear.

 \section{The {\em gradient cooling}  hunting for the $I\bar I$ molecules}

The general notion of the renormalization group (RG) is arguably the most important concept in quantum field theory.
In partonic physics it appears in the form of DGLAP evolution: as the resolution scale $\mu$ changes,
one must add or subtract gluonic degrees of freedom.

Nowadays, $cooling$ via the gradient flow can be consistently related to the
renormalization group flow~\cite{Luscher:2011bx}, placing the scale dependence of the
"molecular component'' on a firm theoretical footing.
Skipping several decades of developments, we consider as an example the recent work
of Ref.~\cite{Athenodorou:2018jwu}, which studied three- and four-point gluon field correlators
and related their evolution under cooling to topology.
The original motivation was to extract the gluon coupling $\alpha_s(k)$, leading to the
definition of the observable
\begin{equation}
\alpha_{MOM}(k)={k^6 \over 4\pi}
{\langle G^{(3)}(k^2) \rangle^2 \over \langle G^{(2)}(k^2) \rangle^3} \, ,
\label{ratioG3toG2}
\end{equation}
given by the ratio of three-point to two-point Green functions.

In the uncooled quantum vacuum, where perturbative gluons are present,
the effective coupling decreases at large momenta $k>1\,{\rm GeV}$, as expected from
asymptotic freedom.
However, in the low-momentum limit $k\rightarrow 0$, a positive power-law behavior persists,
with a slope that exactly matches the prediction from an instanton ensemble~\cite{Boucaud:2002fx},
\begin{equation}
\alpha_{MOM}(k)\rightarrow \frac{k^4}{18\pi n_{I+\bar I}} \, .
\end{equation}
Furthermore, it was observed that as the cooling time $\tau$ increases, this same power-law
behavior extends to higher momenta, $k>1\,{\rm GeV}$.
Cooling was found to eliminate not only perturbative gluons, but also closely bound
instanton-anti-instanton pairs.

\begin{figure}[htbp]
\begin{center}
\includegraphics[width=7cm]{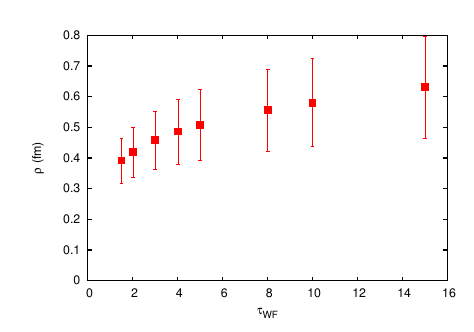}
\includegraphics[width=8cm]{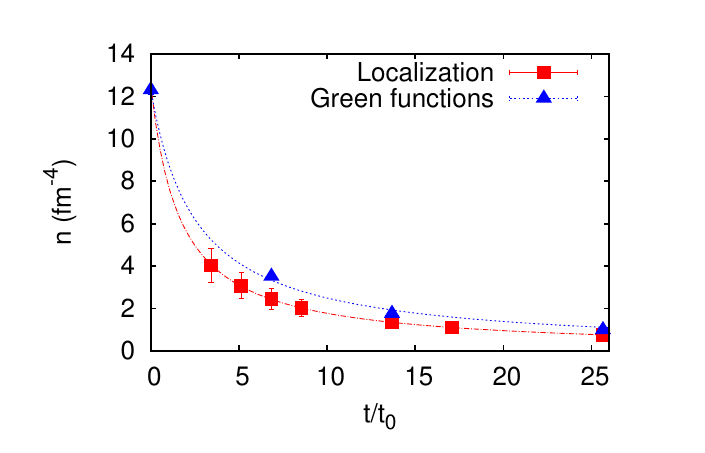}
\caption{Dependence of the mean instanton size (upper panel) and density (lower panel)
on the gradient-flow cooling time ($\tau$ or $t/t_0$, arbitrary units).
The quantum vacuum corresponds to the extrapolation $\tau\rightarrow 0$.}
\label{fig_cooling}
\end{center}
\end{figure}

The dependence of the mean instanton size and density on the gradient-flow cooling time
is shown in Fig.~\ref{fig_cooling}, taken from Ref.~\cite{Athenodorou:2018jwu}.
The main conclusion of that analysis is that, when extrapolated to zero cooling time
($\tau\rightarrow 0$), the characteristic values are
\be
\rho \rightarrow {1 \over 3}\,{\rm fm}, \qquad n \rightarrow 10\,{\rm fm}^{-4} \, .
\ee
Thus, the density of nonperturbative fields is significantly larger than previously
estimated in the Instanton Liquid Model.
While the results agree with the ILM in the deep-cooling limit (large $\tau$),
where $n\sim 1\,{\rm fm}^{-4}$, in the physically relevant limit $\tau\rightarrow 0$
the density is larger by roughly an order of magnitude%
\footnote{The density may in fact be even larger, since the instanton definition employed
in Ref.~\cite{Athenodorou:2018jwu} excludes closely bound instanton-anti-instanton pairs
($molecules$).}.
We will use these values as  inputs for a {\em dense instanton liquid model},
which explicitly includes instanton-anti-instanton molecules.

\section{Poisson duality: instantons versus monopoles}

Among the various nonperturbative mechanisms proposed to explain infrared QCD phenomena, instantons-topologically nontrivial tunneling configurations of the gauge field-occupy a  central role. The instantons induce chirality-changing fermionic zero modes and generate  't~Hooft multi-fermion interaction, explaining the mechanism of spontaneous  chiral symmetry breaking, the $U_{A}(1)$ anomaly, and the mass splitting in the pseudoscalar meson sector. When embedded into an ensemble of instantons and anti-instantons, this mechanism forms the basis of the instanton liquid model (ILM), which captures the dominant infrared degrees of freedom of the gluonic vacuum. Lattice gauge theory, particularly with the aid of gradient flow, now provides strong evidence that cooled gauge configurations reveal a dilute but strongly correlated ensemble of instantons and anti-instantons, supporting the ILM picture.

The same gradient flow techniques using the P-vortex projection as we have discussed earlier, have also shown a direct correlation between the location of the emergent branching points, sources of monopoles and anti-monopoles and the location of the instantons and anti-instantons. This is not surprising,  given the geometrical origin of both configurations, and their intimate relationship at finite temperature~\cite{Kraan:1998pm,Lee:1998bb}. What emerges is the intimate relationship between chiral symmetry and confinement, where the instantons and anti-instantons are dominant in the former and in particular in the low-lying hadrons (relatively bunched worldlines not prone to flux piercing), and where monopoles and anti-monopoles through center vortices are dominant for the latter, especially for the high-lying hadrons (relatively broad worldlines prone to flux piercing). The two descriptions are likely dual in the vacuum,  extending the initial observation in the thermal state~\cite{Ramamurti:2018evz}.

\section{More on $I\bar I$ molecules}
In Euclidean field ensembles generated on the lattice, an analogous procedure is known as
{\em gradient cooling}. As a function of the cooling time $\tau$, hard gluons above a certain scale are eliminated,
rendering gauge field configurations progressively less noisy.
Eventually, this procedure leads to a dilute ensemble of instantons.
While gluons (small-amplitude fluctuations) are removed, the "topological glue'' in the form of instantons remains.
Because instantons are local minima of the action, they remain stable under further action reduction.

What we wish to emphasize in this section is that strong gauge-field fluctuations also exist between these two limits:
the full quantum vacuum with gluons and the dilute instanton ensemble.
These configurations are neither self-dual nor do they possess (near-)zero Dirac eigenvalues.
As a result, they do not contribute to chiral symmetry breaking and were therefore not included in the original
Instanton Liquid Model (ILM).
Nevertheless, they represent genuine fluctuations of the vacuum gauge fields and do contribute to certain observables,
most notably Wilson lines.
Accordingly, we will investigate their contributions to central and spin-dependent forces between quarks.
The inclusion of such $molecules$  is the novel element of the vacuum model developed here.

Close instanton-anti-instanton pairs are, of course, well known and have been observed on the lattice.
However, they are not visible in Fig.~\ref{fig_VAC}, which was obtained using so-called deep cooling of gauge configurations.
During this process, closely spaced instanton-anti-instanton pairs are already annihilated.
The role of the {\em molecular component} of the vacuum was previously explored in connection with phase transitions
in hot and dense matter.
Indeed, neglecting quark masses, this component is the only one that survives at temperatures $T>T_c$,
where chiral symmetry is restored.
The combined treatment of "atomic'' and "molecular'' components dates back to Ref.~\cite{Ilgenfritz:1988dh}.
The molecular component was also shown to be important at high baryon density, where it contributes to quark pairing
and color superconductivity~\cite{Rapp:1997zu}.
More recently, we have explored its role through nonperturbative contributions to mesonic form factors~\cite{Shuryak:2020ktq}
and matching kernels~\cite{Liu:2021evw}.

The theory of sphaleron processes (schematically illustrated in the middle of Fig.~\ref{fig_landscape})
is closely related to the issue of $I\bar I$ interactions.
The numerical calculation of the $streamline$ along the action gradient was first performed for
quantum-mechanical instantons in Ref.~\cite{Shuryak:1987tr}.
For gauge theories, the streamline equation was derived in Ref.~\cite{Balitsky:1986qn},
solved approximately in Ref.~\cite{Yung:1987zp}, and computed numerically in Ref.~\cite{Verbaarschot:1991sq}.
A surprising finding of the latter work was that the Yung ansatz is remarkably accurate not only at large separations
$R\gg \rho$, where it was originally derived, but in fact down to zero separation.
Note that the last two works employed conformal inversion about the instanton center,
rendering the instanton and anti-instanton co-central.

Consider an instanton $I$ and an anti-instanton $\bar I$ of equal size $\rho$ and identical color orientation,
with their four-dimensional separation $R$ serving as the parameter of the configuration.
Their centers are located at $x_{1\mu}=(0,0,0,R/2)$ and $x_{2\mu}=(0,0,0,-R/2)$.
In the mathematical literature, such a family of configurations is known as a {\em Lefschetz thimble}.
It interpolates between one extremum at $R\rightarrow\infty$, corresponding to independent $I$ and $\bar I$,
and the perturbative vacuum with vanishing mean field at $R=0$.
For $y_4<0$, the gauge fields are approximately anti-self-dual, $\vec E\approx -\vec B$,
while for $y_4>0$ they are approximately self-dual, $\vec E\approx \vec B$.
At $y_4=0$, the electric field vanishes.
As shown in Ref.~\cite{Ostrovsky:2002cg}, the three-dimensional magnetic objects obtained using the Yung ansatz
closely resemble sphaleron-path configurations obtained via constrained energy minimization.

The instanton-anti-instanton streamline thus provides a semiclassical description of sphaleron production.
In the context of electroweak theory, it was first applied in Refs.~\cite{Khoze:1991sa,Shuryak:1991pn} more than three decades ago.
Recent interest in the production of QCD sphalerons at the LHC and RHIC is discussed in our work~\cite{Shuryak:2021iqu}.

We now turn to lattice observables to which the molecular component of the instanton ensemble contributes.
The simplest example is the gluon condensate $\langle G_{\mu\nu}^2 \rangle$, introduced within the framework of
QCD sum rules~\cite{Shifman:1978bx}.
The numerical value of this condensate has been subject to significant revisions, with later studies suggesting
a substantially larger magnitude.
Additional discrepancies were identified in lattice attempts to extract local or nonlocal observables involving
powers of the gauge-field strength $G_{\mu\nu}$.

As already mentioned, instantons ($I$) are solutions of the Yang-Mills equations in Euclidean time
$\tau=ix(4)$~\cite{Belavin:1975fg}, semiclassically describing tunneling between gauge-field configurations
with different topology (Chern-Simons number).
Anti-instantons ($\bar I$) describe tunneling in the opposite direction.
By $molecules$ we mean strongly overlapping $I\bar I$ configurations, corresponding to incomplete tunneling events.

Historically, overlapping $I\bar I$ configurations were first studied in the context of the well-known
quantum-mechanical double-well potential.
As discussed, for example, in Ref.~\cite{Shuryak:1987tr}, one can fix certain points along the path
and minimize the action, leading to the concept of {\em conditional minima}.
Action minimization along the direction of its gradient (known as gradient flow) produces
$streamline$ configurations. (The analogous objects in complex analysis are known as Lefschetz thimbles.)

In gauge theory, the first derivations of such configurations were performed in
Refs.~\cite{Balitsky:1986qn,Yung:1987zp} for large $I\bar I$ separations,
and in full generality in Ref.~\cite{Verbaarschot:1991sq}, using conformal transformations
to render $I$ and $\bar I$ co-central.
These configurations were later applied to finite-temperature QCD in Ref.~\cite{Ilgenfritz:1988dh},
and to the semiclassical theory of sphalerons at high-energy colliders in Ref.~\cite{Shuryak:1991pn},
although such objects have not yet been observed experimentally in either QCD or the electroweak sector.

Gauge-field configurations obtained from lattice simulations, after gradient-flow filtering of gluons,
reveal a rather dense ensemble of $I\bar I$ molecules.
With continued gradient flow, these molecules annihilate, leaving behind a dilute ensemble of
isolated instantons and anti-instantons that is stable under action minimization.
This ensemble is what is traditionally referred to as the Instanton Liquid Model~\cite{Shuryak:1981fza}.
Examples of recent lattice studies investigating molecular configurations and their annihilation
under gradient flow can be found in Ref.~\cite{Athenodorou:2018jwu}.

In later work, however, we did not employ either Verbaarschot's or Yung's streamline configurations.
Instead, we adopted a variant of the so-called ratio ansatz
\be
\label{eqn_ansatz}
A^{\mu a}(x)=
\frac{ \bar\eta^{a\mu\nu} y_I^\nu \rho^2/ Y_I^2
      + \eta^{a\mu\nu} y_A^\nu \rho^2 /Y_A^2 }
{1+\rho^2/Y_A +\rho^2/ Y_I } \, ,
\ee
where $I$ and $A$ denote the instanton and anti-instanton, respectively.
The centers are located at $y_{I,A}^m=x^m$ for $m=1,2,3$, with
$y^4_I=x^4-R/2$ and $y^4_A=x^4+R/2$.
Capital $Y$ denotes the squared distances, for example $Y_A=y_A^\mu y_A^\mu$.
Near one of the centers (e.g.\ $Y_I$ small and $Y_A$ large), the first term dominates,
and cancellation of $Y_A$ between numerator and denominator yields the usual singular-gauge
fields of an instanton or anti-instanton.

\begin{figure}
    \centering
    \includegraphics[width=7cm]{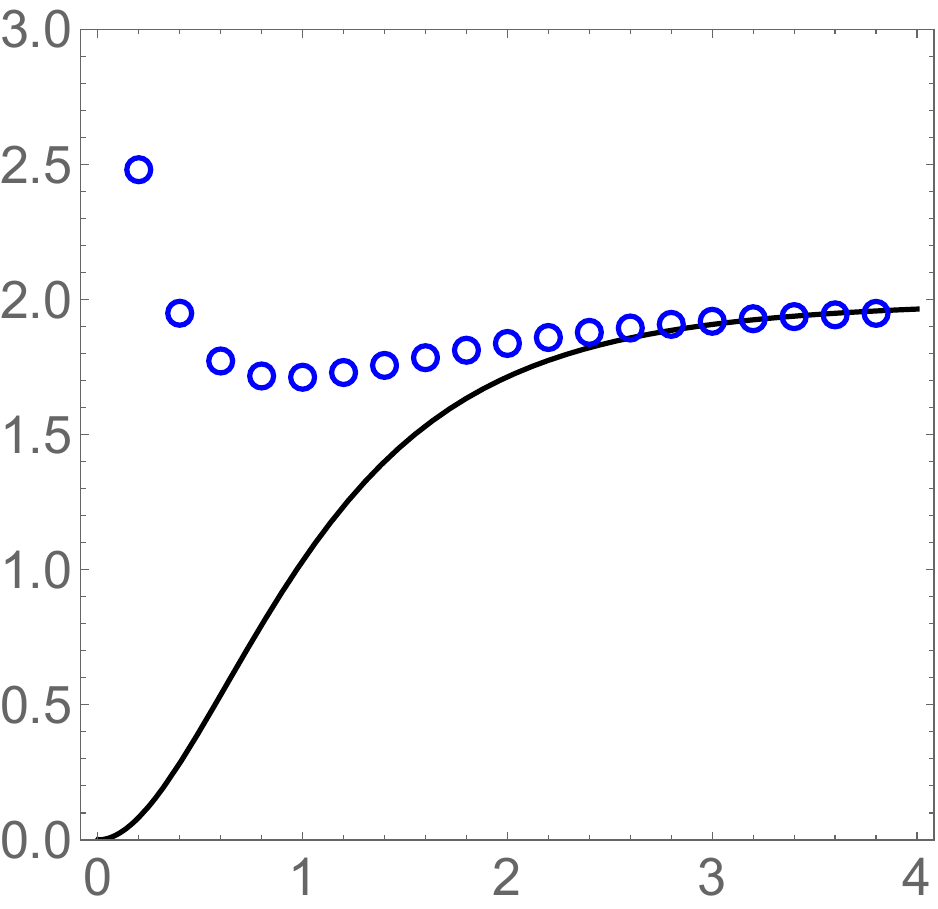}
    \caption{Action of the $I\bar I$ configurations (in units of the single-instanton action $8\pi^2$)
    as a function of the separation $R/\rho$.
    The solid line corresponds to the streamline configuration~\cite{Verbaarschot:1991sq},
    while the points show the result obtained using the ansatz~(\ref{eqn_ansatz}).}
    \label{fig_action}
\end{figure}

\begin{figure}
    \centering
    \includegraphics[width=7cm]{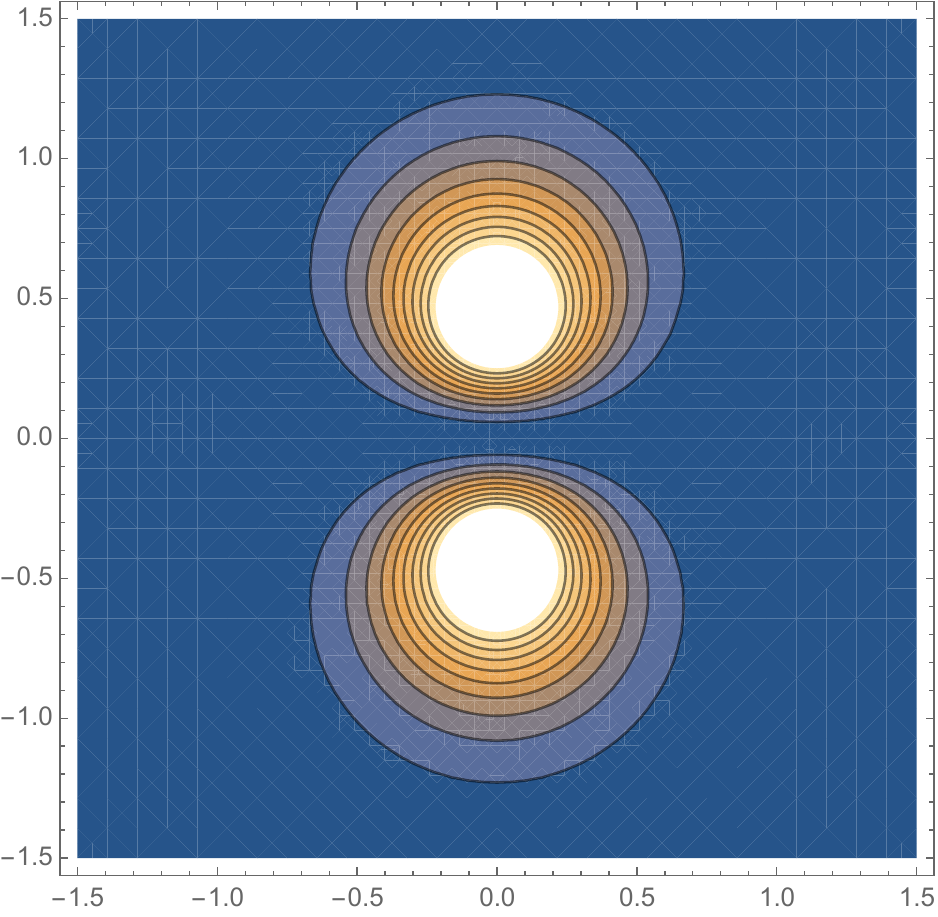}
    \includegraphics[width=7cm]{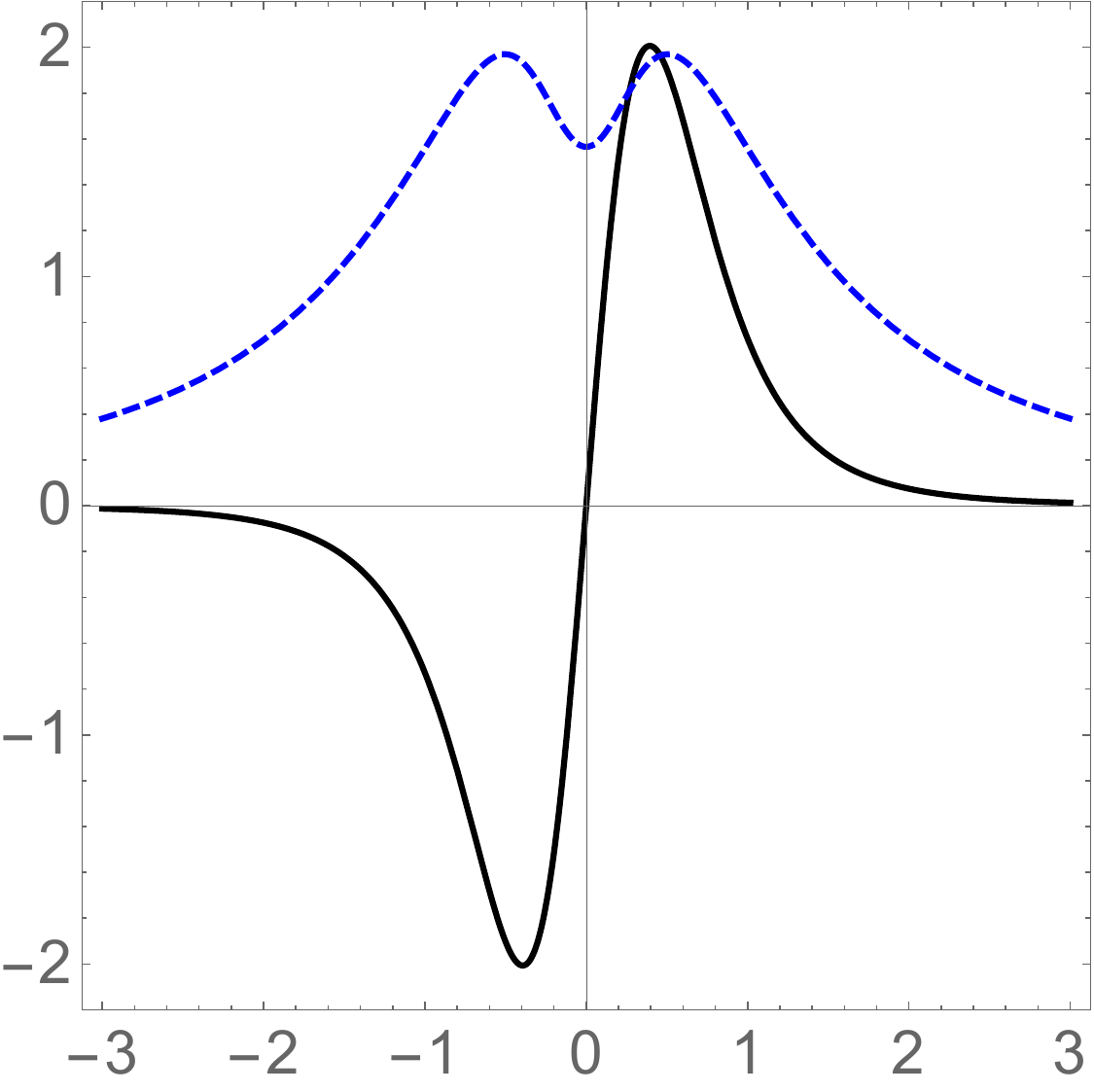}
    \caption{Left: Contour plot of the action density as a function of $x^1$ and $x^4$,
    in units of the instanton size $\rho$.
    The separation between the centers is $R=\rho$, oriented along the Euclidean time direction $\tau=x^4$.
    Right: Electric (black solid) and magnetic (blue dashed) field components with $\mu=3,a=3$
    as functions of $x^4/\rho$.}
    \label{fig_molec_R1}
\end{figure}

It is straightforward for symbolic programs such as {\sc Mathematica} to generate explicit
(though lengthy) expressions for the field strengths and their squares.
The distribution of the gauge-invariant scalar combination $[G_{\mu\nu}^a]^2$ (the action density)
is shown in Fig.~\ref{fig_molec_R1}.
The integrated action, measured in units of the single-instanton action
$S_0=8\pi^2/g^2$, approaches 2 at large separation $R\rightarrow\infty$.
At small $R$, however, it does not vanish  but instead
exhibits a repulsive core for $R\lesssim 0.5$.
This feature is unimportant for our purposes, since in subsequent calculations we take
$I\bar I$ molecules with $R=\rho$ as representative of the vacuum.

The behavior of the gauge fields is illustrated in the right panel of Fig.~\ref{fig_molec_R1}.
The electric field is odd under $x^4\rightarrow -x^4$, while the magnetic field is even.
Similarly, $A^4$ is odd, implying that the configuration restricted to the three-dimensional
subspace at $x^4=0$ is purely magnetic and belongs to the so-called sphaleron path.
The molecular configurations can therefore be interpreted as tunneling amplitudes and anti-amplitudes,
separated by the analogue of a quantum-mechanical turning point, at which the electric field
(and hence the momentum) vanishes.

Individual instantons (anti-instantons) are self-dual (anti-self-dual), and as a result all components
of the stress tensor vanish.
This is not the case for molecular configurations.
In particular, the Poynting vector $\vec P=[\vec E^a\times\vec B^a]$, which characterizes the direction
and magnitude of energy flow, is nonzero.
Since the configuration is localized, the energy flow must form closed loops, which turn out to be circular.

 \section{'t Hooft effective Lagrangian}

The derivation is based on the zero-mode contribution to the quark propagator,
\be
S(x,y)=\frac{\psi_0(x)\psi_0^+(y)}{i m}
+\sum_{\lambda\neq 0}\frac{\psi_\lambda(x)\psi_\lambda^+(y)}{\lambda + i m} \, .
\ee
At large distances, the zero mode $\psi_0$ behaves as $x_\mu \gamma_\mu/x^4$,
corresponding to a free propagator in empty space.
Amputating these external legs leads to an effective quasi-local vertex with $2N_f$ fermion legs,
known as the 't~Hooft effective Lagrangian~\cite{tHooft:1976snw}.
It includes the zero modes of all light quark flavors ($u,d,s$) and therefore induces
flavor-mixing interactions,
\[
\bar u u \leftrightarrow \bar d d \leftrightarrow \bar s s \, .
\]

For $N_f=3$, in the local (zero-size instanton) approximation and in the chiral limit
($m_q=0$), the interaction takes the form of determinantal interactions among the
$u,d,s$ quarks~\cite{Chernyshev:1995gj},
\bea
	\label{DETUDS}
	{\cal V}^{L+R}_{qqq}=\frac {G}{N_c(N_c^2-1)}\bigg[
	\bigg(\frac{2N_c+1}{2(N_c+2)}\bigg)  {\rm det}(UDS)
	+\frac 1{2(N_c+1)}\,\bigg({\rm det}(U_{\mu\nu}D_{\mu\nu}S)
	+ {\rm cyclic}\bigg)\bigg] +(L\leftrightarrow R) \, ,
\eea	
where $U=\overline u_R u_L$ and $U_{\mu\nu}=\overline u_R\sigma_{\mu\nu} u_L$, with analogous
definitions for $D$ and $S$.
For a single instanton within the instanton liquid model, the coupling constant $G$
is related to the instanton density as
\be
\label{DETUDSX}
 G=\frac{n_{I+\bar I}}{2}
 \bigg(4\pi^2\rho^3 \bigg)^3
 \bigg(\frac{1}{m^*_u\rho}\bigg)
 \bigg(\frac{1}{m^*_d\rho}\bigg)
 \bigg(\frac{1}{m^*_s\rho}\bigg) \, ,
\ee
with the so-called {\em determinantal mass}
\bea
m_f^*=m_f+ \frac{n_{I+\bar I}}{2N_c}\frac{4\pi^2\rho^2}{m_f^*} \, ,
\eea
where $m_f$ is the Higgs-induced quark mass appearing in the fundamental QCD Lagrangian.

For most of the analyses in this work, we will specialize to $N_f=2$, with $u$ and $d$
quarks of nearly degenerate masses.
Strangeness will also be addressed in a similar spirit, using $U$- and $V$-spin.
The interaction~(\ref{DETUDS}) explicitly breaks the $U_A(1)$ axial symmetry,
while preserving flavor and left-right symmetry.
This structure becomes more transparent after a Fierz transformation,
\begin{eqnarray}
\label{FTH}
   \frac{G}{8(N^2_c-1)}\Bigg(
   \frac{2N_c-1}{2N_c}
   \big[(\bar{\Psi}\Psi)^2-(\bar{\Psi}\tau^a\Psi)^2
   -(\bar{\Psi}i\gamma^5\Psi)^2+(\bar{\Psi}i\gamma^5\tau^a\Psi)^2\big]
   -\frac{1}{4N_c}
   \big[(\bar{\Psi}\sigma_{\mu\nu}\Psi)^2
   -(\bar{\Psi}\sigma_{\mu\nu}\tau^a\Psi)^2\big]
   \Bigg) \, .\nonumber\\
\end{eqnarray}
This interaction is strongly attractive in the $\sigma$ and pion channels and repulsive
in the $\eta^\prime$ channel, thereby resolving the $U_A(1)$ problem.
Equation~(\ref{FTH}) is the QCD realization of the interaction originally proposed
in the pre-QCD era~\cite{Nambu:1961tp}.
Finite instanton-size effects can be restored by the replacement
$\psi(x)\rightarrow\sqrt{\mathcal{F}(i\partial)}\psi(x)$
in~(\ref{FTH}), with the universal profile $\mathcal{F}(k)$ in momentum space
($k=\sqrt{k^2}$),
\begin{equation}
\label{FKK}
    \mathcal{F}(k)=\left[(zF'(z))^2\right]\bigg|_{z=\frac{k\rho}{2}},
    \qquad\qquad
    F(z)=I_0(z)K_0(z)-I_1(z)K_1(z) \, .
\end{equation}

In addition to changing quark flavors, this Lagrangian also induces chirality
flips\footnote{In a talk, T.\,D.~Lee once compared instanton tunneling to driving
from Queens to Manhattan. Shuryak remarked that a better analogy is tunneling
from Britain to France, since it naturally incorporates the interchange of
left- and right-handed fermions.}.

This Lagrangian can be written in several equivalent forms, which may initially
appear confusing.
Any multi-fermion operator can be identically rewritten using Fierz identities,
which regroup fermions into different pairs.
For four-fermion operators there are three possible pairings, corresponding to
pairing a given fermion with any of the remaining three.
As a result, the $N_f=2$ 't~Hooft Lagrangian admits three equivalent representations.
Different forms may be more convenient for different applications, but their coexistence
in the literature can lead to confusion.
Following Ref.~\cite{Rapp:1999qa}, we present all three forms here.
In what follows, $t^a$ denote color generators (Gell-Mann matrices for $N_c=3$),
$\tau^- = (\vec \tau,i)$ with $\vec \tau$ the isospin matrices, and the
momentum-dependent form factor $F(k)$ (a Dirac matrix) is related to the zero-mode
solution.

\begin{eqnarray}
L_1&=&\frac{G}{4 (N_c^2-1)} \bigg[
\frac{2N_c -1}{2 N_c}  (\bar q F^+ \tau_\alpha^- F q)^2
+ \frac{2N_c -1}{2 N_c} (\bar q F^+ \gamma_5 \tau_\alpha^- F q)^2
+ \frac{1}{4N_c} (\bar q F^+ \sigma_{\mu\nu} \tau_\alpha^- F q)^2
\bigg]\nonumber\\
L_2&=&\frac{G}{8 N_c^2} \bigg[
(\bar q F^+ \tau_\alpha^- F q)^2
+(\bar q F^+ \gamma_5 \tau_\alpha^- F q)^2
+\frac{N_c-2}{2(N_c^2-1)}
\big((\bar q F^+ \tau_\alpha^- t^a F q)^2
+(\bar q F^+ \gamma_5 \tau_\alpha^- t^a F q)^2\big)\nonumber\\
&-&\frac{N_c}{4 (N_c^2-1)} (\bar q F^+ \sigma_{\mu\nu} \tau_\alpha^- t^a F q)^2
\bigg]
\end{eqnarray}

\begin{eqnarray}
&&L_3=\frac{G}{8 N_c^2}  \bigg[
-\frac{1}{N_c-1}(q^T F^T C \tau_2 t^a_A F q)
(\bar q F^+ \tau_2 t^a_A C F^* \bar q^T)
\nonumber\\
&&-\frac{1}{N_c-1}(q^T F^T C \tau_2 \gamma_5 t^a_A F q)
(\bar q F^+ \tau_2 \gamma_5 t^a_A C F^* \bar q^T)
+\frac{1}{2(N_c+1)}(q^T F^T C \tau_2 \sigma_{\mu\nu} t^a_A F q)
(\bar q F^+ \tau_2 \sigma_{\mu\nu} t^a_A C F^* \bar q^T)
\bigg] \, .\nonumber\\
\end{eqnarray}

The first two forms, $L_1$ and $L_2$, are often referred to as the mesonic form,
since they involve squares of colorless bilinears $(\bar q \dots q)$,
which can be interpreted as mesonic currents.
These include four species: scalar and pseudoscalar mesons with isospin zero
($\sigma,\eta^\prime$) and one ($\pi,\delta$).
To first order in $L_i$, the masses of these mesons are shifted upward or downward
according to the signs of the corresponding terms:
the $\pi$ and $\sigma$ masses decrease, while those of the $\delta$ and $\eta^\prime$
increase.

The Lagrangian $L_3$ is known as the {\em diquark form,} since it contains $(qq)$ and
$(\bar q\bar q)$ pairs.
Here $T$ denotes transposition, $C$ is the charge-conjugation matrix
(equivalent here to the antisymmetric isospin matrix $\tau_2$),
and the color generators $t^a_A$ are restricted to those antisymmetric in color indices,
as indicated by the subscript $A$.
As in the mesonic case, the first two terms have a sign opposite to that of the last one,
implying that, to first order, scalar diquarks receive attractive mass corrections
while vector diquarks are repelled (spin-0 down, spin-1 up).

 \end{subappendices}



\part{Hadronic spectroscopy}
\chapter{Quarkonia}

\section{Quarkonia and tetraquarks}

Quarkonia is the collective name for mesons composed of heavy quarks, namely
charmonia $\bar c c$ and bottomonia $\bar b b$.
Discovered in the 1970s, they immediately became central pillars of hadronic spectroscopy.
Hundreds of experimental and theoretical studies have been devoted to their properties,
regularly summarized in the Reviews of Particle Properties (RPP).
This review is not the place to recount the historical developments or enter into detailed discussions;
it suffices to recall that the mean (spin-averaged) spectra of quarkonia are well reproduced
by the nonrelativistic Schrodinger equation with a simple Cornell potential%
\footnote{We use this as an exercise in Appendix~\ref{sec_Math_Sch}, where we show how to solve it using {\sc Wolfram Mathematica}.}.

\begin{figure}
    \centering
    \includegraphics[width=0.5\linewidth]{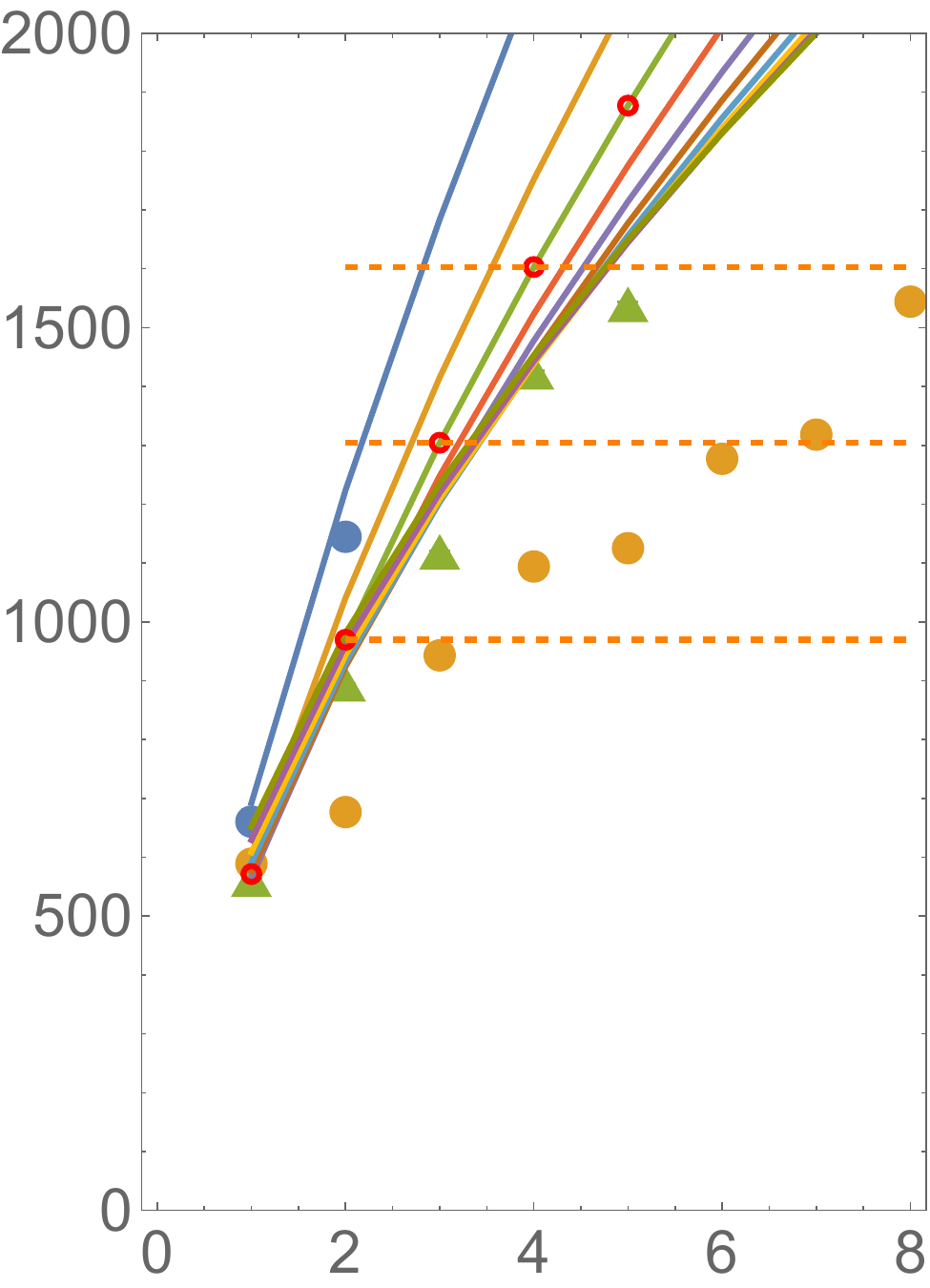}
    \caption{Gaps between levels $gap_n=M_{n+1}-M_1$ versus $n$.
    Blue disks, orange disks, and green triangles correspond to
    $\bar s s$, $\bar c c$, and $\bar b b$ quarkonia
    ($\phi$, $\psi$, and $\Upsilon$ families, $J^{PC}=1^{-}$), respectively.
    The curves show gaps calculated using the Cornell potential for quark masses
    $0.5, 1.0, 1.5,\dots,5\,\mathrm{GeV}$ from left to right.
    Horizontal dashed lines and small red circles indicate calculated values for
    successive charmonium states.}
    \label{fig_gaps_quarkonia}
\end{figure}

Beyond their foundational role half a century ago, quarkonium spectroscopy has also played
a key role in the recent revolution in hadronic spectroscopy over the past decade.
To set the stage, let us briefly explain what has emerged from a closer scrutiny of charmonium states.
A plot that clearly illustrates the current situation is shown in Fig.~\ref{fig_gaps_quarkonia},
which displays the $gaps$
$gap_n=M_{n+1}-M_1$, i.e.\ quarkonium masses measured relative to the ground state.
This representation allows one to compare charmonium and bottomonium data on a single plot%
\footnote{It also removes the ambiguity associated with an additive constant in the interquark potential and quark masses.}.

The first observation concerns a well-known but striking fact:
the lowest gaps $gap_1=M_2-M_1$ for $\bar s s$, $\bar c c$, and $\bar b b$ quarkonia
are nearly identical, even though the corresponding quark masses span more than
an order of magnitude%
\footnote{Remarkably, the higher gaps also show only a weak dependence on the quark mass.
This behavior is a consequence of specific features of the Cornell potential, in which
the Coulombic $1/r$ and confining $r$ terms partially cancel.}.

Calculations using the Cornell potential are shown as multiple curves corresponding to
different quark masses, spanning the range from $\bar s s$ to $\bar b b$ systems.
However, if one naively plots all $\psi$ states listed in the RPP,
the charmonium data appear to deviate strongly from these curves
(and from the small red circles indicating theoretical expectations).
The reason is well known:
out of the nine listed states, only five (the ground state, the first excitation,
and three additional states lying near the horizontal dashed lines)
are genuine $\bar c c$ quarkonia.
The remaining four states are the {\bf tetraquarks} of the composition
$\bar c c \bar q q$.
This striking difference between charmonium and bottomonium arises from the very different
locations of two-meson thresholds in the two systems.

We emphasize these points at the outset because they   illustrate
 dramatic changes  taken place in the field.
For nearly 50 years, hadrons were classified almost exclusively as mesons or baryons,
while all other candidates were relegated to the category of $exotics$,
mostly with questionable experimental status.
Today, it is evident that roughly half of the observed states in the vector
($J^{P}=1^{-}$) and axial-vector ($J^{P}=1^{+}$) charmonium channels are in fact
tetraquarks $\bar c c \bar q q$. 

While their experimental status is excellent, let us
add a brief comment on nomenclature further highlights the fluid state of the field.
Axial tetraquarks have now acquired their own designation in the RPP, $T_{\bar c c}$,
whereas vector states are still commonly labeled as $\psi$, $\eta_c$, and so on.
This reflects the fact that in the axial channel the $\bar q q$ pair forms an
isovector ($I=1$, with quantum numbers analogous to the pion),
whereas in the vector channel it corresponds to its chiral partner, the $\sigma$.

The theoretical literature on tetraquark structure suggests several possible interpretations:
(i) deuteron-like states, bound weakly near threshold;
(ii) diquark-antidiquark configurations with two characteristic spatial scales; or
(iii) genuinely compact tetraquarks.
We will not review these possibilities in detail here, as our primary interest lies
in compact multiquark states composed of quarks with equal masses,
such as compact tetraquarks, pentaquarks, and hexaquarks.
In such systems, there are good reasons to expect that the lowest states are
spherically symmetric in the corresponding multidimensional configuration spaces.
Accordingly, the only charm-related tetraquark we will discuss explicitly is
$T_{\bar c\bar c c c}$, for which there are strong indications that it belongs
to the compact tetraquark category.

\section{From Wilson lines to effective potentials}
\label{sec_Wilson_mesons}
Static potentials, introduced by K.~Wilson, are defined via the so-called Wilson loops,
constructed from Wilson lines
\be
\label{eqn_W}
W\equiv P\exp\left[i \int A^a_4 (\lambda^a/2)\, dx^4\right]
\ee
of the gauge field taken along the Euclidean time direction $\tau=x^4$.
Here $\lambda^a$ are the Gell-Mann matrices, and the factor $1/2$ reflects the fact
that quarks transform in the fundamental\footnote{In the chapter discussing glueballs those will be substituted by adjoint color generators.} representation of $SU(3)$.
Provided the gauge fields are periodic in $\tau$, $W$ is a gauge-invariant
holonomy operator\footnote{The average of $W$, known as the Polyakov line, plays an important role
in finite-temperature field theory, which will not be discussed in this review.}.
By construction, $W$ is a unitary matrix describing the rotation of the quark color
degree of freedom.

The static potential for a quark-antiquark pair is obtained from the correlator
of two Wilson lines, one of them Hermitian conjugated, separated by a spatial distance $r$,
\be
\exp[-V(r)\tau]
=
\left\langle
1-\frac{1}{N_c}\,
\mathrm{Tr}\, W(0)\,W^\dagger(r)
\right\rangle .
\ee
Expressions for the static potentials involving three quarks will be discussed in
Sec.~\ref{sec_baryons}, and for tetraquarks in Sec.~\ref{sec_tetra}.

At lowest order in perturbation theory, the two Wilson lines are connected by a single
gluon propagator, leading to the Coulomb interaction.
Unlike in QED, higher-order diagrams in QCD include gluon self-interactions, which
generate the running coupling in the Coulomb potential,
\be
V_C(r)=-\frac{4}{3}\frac{\alpha_s(r)}{r} \, ,
\ee
dominant at short distances.

Nonperturbative calculations of the static potential have been the subject of extensive
lattice studies since the 1980s.
These works established that at large distances the potential exhibits
{\em linear confinement},
\be
V_{\rm string}(r)=\sigma \, r \, ,
\ee
generated by a flux tube (or confining QCD string) composed predominantly of longitudinal
electric field, stabilized by circulating magnetic currents.

By combining the short-distance Coulomb interaction with the long-distance linear term,
one arrives at the so-called Cornell potential.
Solving the nonrelativistic Schrodinger equation with this potential yields spectra
for charmonium and bottomonium that are already in qualitative agreement with experimental data.

While these results are well known, a more detailed examination reveals several open issues.
One of them concerns the degree to which the effective potential is known at intermediate
distances.

The first issue is that a quantum QCD string can vibrate, implying that the linear
potential must be supplemented by so-called L\"uscher terms,
which scale as $\sim 1/r$ and $1/r^3$~\cite{Luscher:1980ac}.
Disentangling the L\"uscher term from the Coulomb contribution is nontrivial.
Moreover, the Nambu-Goto geometric string action leads to the Arvis potential~\cite{Arvis:1983fp},
\be
V_A=\sigma \, r \sqrt{1-\frac{\pi}{6}\frac{1}{\sigma r^2}} \, ,
\ee
which reproduces the universal L\"uscher term but cannot be extended to small $r$.

(Effective models that reproduce confinement, and the QCD string structure in particular,
are often based on variants of dual superconductor models.
Unfortunately, these rely on effective Lagrangians containing scalar fields that do not
appear in the fundamental QCD Lagrangian.)

A second issue arises in QCD with light quarks, where QCD strings can break through the
creation of light quark-antiquark pairs.
As a result, the static linear potential should eventually saturate to an
$r$-independent constant related to the $\bar D D$ or $\bar B B$ thresholds,
typically for $r\gtrsim 1\,\mathrm{fm}$.
Dynamical level-crossing arguments suggest that the effective potential in this regime
is a superposition of linear and constant contributions.

We approach these issues from two complementary perspectives.
First, quarkonium spectroscopy can be used to extract more precise information about
effective potentials.
Second, we consider explicit models of vacuum gauge fields, substitute them into the
definition of the static potential, and test whether the resulting potentials are
consistent with spectroscopic data (or lattice results).
Guided by visualizations such as Fig.~\ref{fig_VAC}, we explored in Ref.~\cite{Shuryak:2021fsu}
the consequences of modeling the vacuum as a gas of instantons or as a gas of closely bound
instanton-anti-instanton pairs (molecules).
Below, we extend this analysis to systems of three and four quarks, and proceed to calculate
the associated spin-dependent potentials.
As the reader will see, in nearly all cases the resulting potentials are in reasonable
agreement with known phenomenology.

\subsection{Central potential and instanton-anti-instanton molecules}

The definition of the Wilson line~(\ref{eqn_W}) involves a path-ordered exponential
because gauge fields at different points generally correspond to color rotations
that do not commute.
Fortunately\footnote{As already noted in 1978 by the Princeton group~\cite{Callan:1978ye}.},
for instanton gauge fields of the form
\be
A_\mu^a=\frac{2}{g}
\frac{\bar\eta_{\mu\nu}^a \rho^2 x^\nu}{x^2(x^2+\rho^2)} \, ,
\ee
the color rotations $A_4^a\,d\tau$ along a straight line
$x^\mu=(\vec z,\tau)$ occur around a fixed direction in color space,
namely $z^a/|\vec z|$, for all $\tau$.
This property eliminates the need to explicitly evaluate the path ordering in $W$
and allows for an analytic expression.
(In singular gauge, the angle of color rotation is
$\theta=\pi-\pi z/\sqrt{z^2+\rho^2}$.)%
\footnote{Owing to the $hedgehog$ structure of the instanton field, this property also
holds for lines not aligned with the Euclidean time direction, a fact we will later
exploit when approaching the light cone.}

\begin{figure}[h]
\begin{center}
\includegraphics[width=6cm]{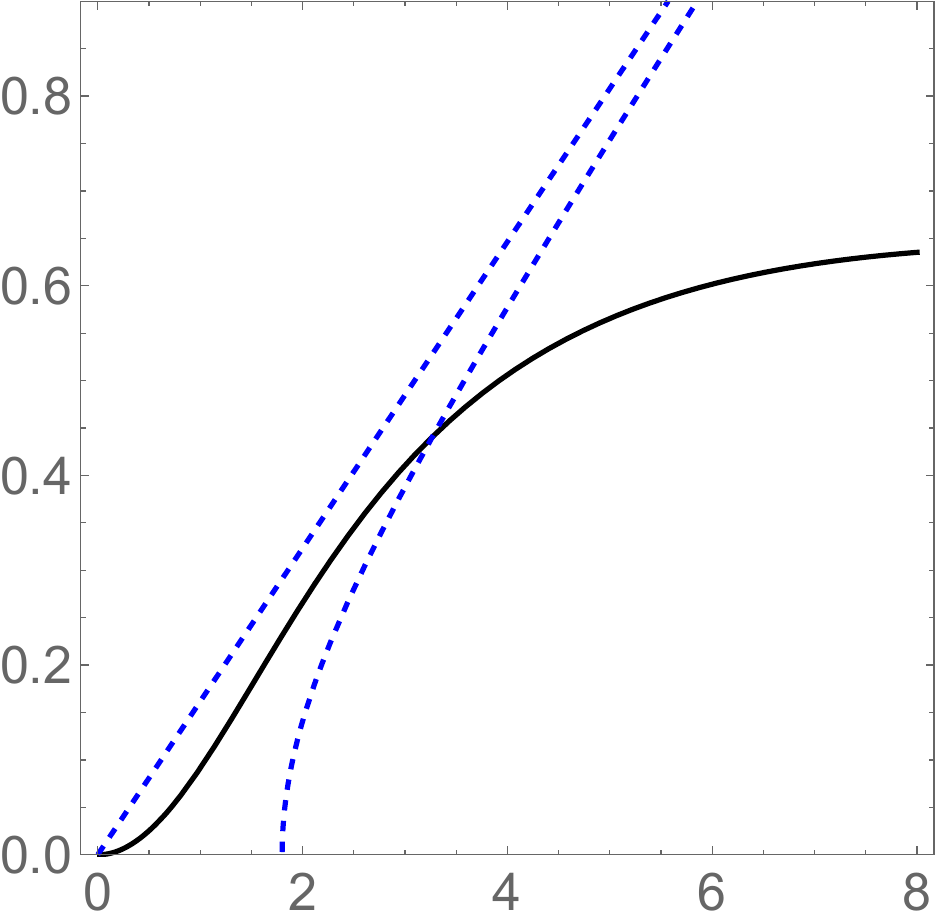}
\caption{Comparison of the central potentials $V_C(r)$ as a function of the interquark
distance $r$ in ${\rm GeV}^{-1}$ (with $1\,{\rm fm}\approx 5\,{\rm GeV}^{-1}$).
The solid line shows the potential derived from the dense instanton ensemble.
The confining flux-tube contribution is illustrated by two blue dashed lines:
the upper line corresponds to the classical linear string,
while the lower line represents the Arvis potential,
which accounts for quantum string vibrations~\cite{Arvis:1983fp}.}
\label{fig_two_potentials}
\end{center}
\end{figure}

Assuming an ensemble of uncorrelated instantons with fixed density $N/V_4$,
the Wilson lines exponentiate, and the static potential reduces to a correlator
over a single instanton,
\bea
\label{Ww}
V_{\rm inst}
=
\frac{N\rho^3}{V_4}
\int \frac{d^3 z}{\rho^3}
\big[
1-\cos(\theta_1)\cos(\theta_2)
-\sin(\theta_1)\sin(\theta_2)(\vec n_1\cdot\vec n_2)
\big] .
\eea
The integral over the instanton position converges because at large distances both
angles vanish and the integrand tends to zero.
In Fig.~\ref{fig_two_potentials} we compare the resulting potential with the linear
part of the Cornell potential.
The instanton-induced contribution to $V_C(r)$ is shown as the solid line for the density
\begin{equation}
n\equiv n_{\rm mol}+n_{\rm ILM}=7\,{\rm fm}^{-4} \, .
\end{equation}
This density is approximately a factor of two smaller than that suggested by the
triangles and extrapolation lines in Fig.~\ref{fig_cooling}(b).

The instanton-induced central potential comes remarkably close to the phenomenological
potential in the phenomenologically important intermediate region
$r\lesssim 3\,{\rm GeV}^{-1}$.
Assuming this agreement is meaningful, we proceed below to compute spin-dependent effects,
and will show that these results are likewise reasonable.

Note that for this choice of parameters the diluteness parameter is $\kappa\approx 1$.
In other words, the ensemble of $I\bar I$ molecules is rather dense, with a mean
interparticle spacing
\[
R_{\rm dense}\equiv n^{-1/4}=0.61\,{\rm fm}\approx 2\rho \, .
\]
While this may appear surprisingly dense, one should recall that an $I\bar I$ molecule
does not consist of two independent instantons.
Their gauge fields and actions partially cancel, and it is precisely this cancellation
that allows for a relatively large density.

Before proceeding further, let us consider a well-studied example, the bottomonium system,
to assess how differences between the linear and instanton-induced potentials at
$r\gtrsim 1\,{\rm fm}$ are reflected in the mass spectrum.

\begin{figure}[h]
\begin{center}
\includegraphics[width=5cm]{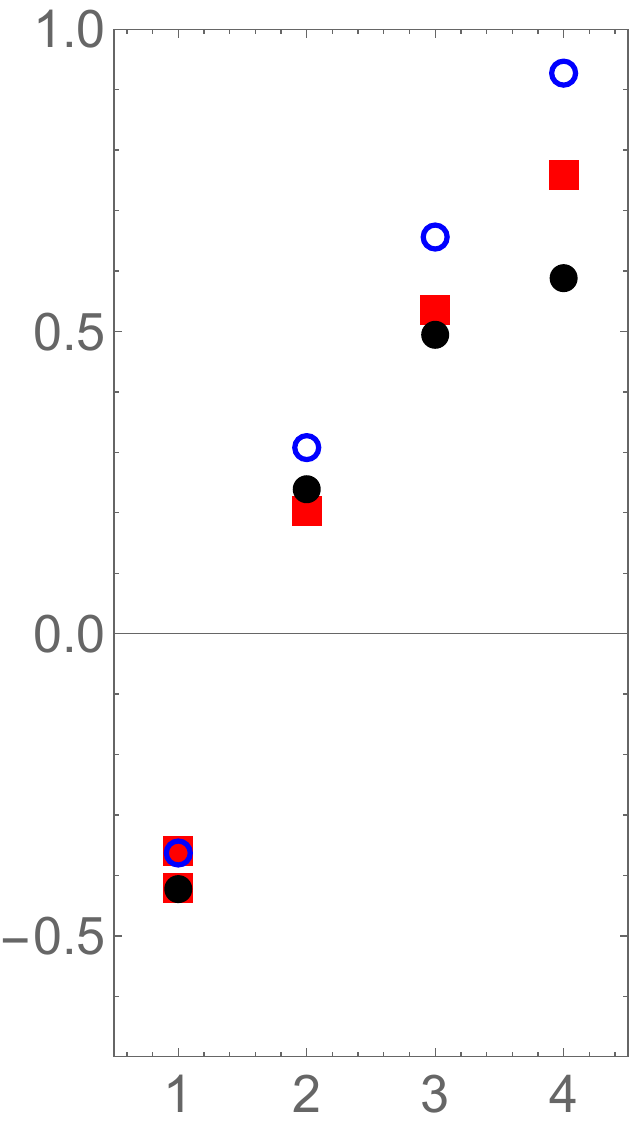}
\caption{Nonrelativistic energies $M_i-2M_b$ for spin-averaged $\bar b b$ states
as a function of the principal quantum number $n$.
The five red squares indicate the experimentally measured masses of
$\Upsilon(1S)$, $\eta_b(1S)$, $\Upsilon(2S)$, $\Upsilon(3S)$, and $\Upsilon(4S)$.
Blue circles correspond to the standard Cornell potential,
while black filled circles are obtained using the instanton-induced potential
shown in Fig.~\ref{fig_two_potentials}.}
\label{fig_four_upsilons}
\end{center}
\end{figure}

The levels of $\bar b b$ states obtained using the two potentials are shown in
Fig.~\ref{fig_four_upsilons}.
We display the nonrelativistic energies of four radial excitations in the $S$-wave
($L=0$) channel.
Since spin-dependent interactions are not included at this stage
(they will be discussed in the next subsections),
the calculation corresponds to spin-averaged combinations of the
$J=1$ $\Upsilon$ and $J=0$ $\eta_b$ states.
The difference between the two potentials at large distances,
$r\gtrsim 5\,{\rm GeV}^{-1}\approx 1\,{\rm fm}$,
indeed leads to distinct predictions for highly excited radial states.
At large $n$, the instanton-induced potential does not reproduce the expected
Regge behavior $m_n^2\sim n$.
However, for the specific bottomonium states considered here,
the practical difference between the two potentials is small.
(In the matrix elements for lighter quark systems discussed below,
we will nevertheless continue to use wave functions obtained with the standard
Cornell potential.)

\subsection{Spin-dependent potentials}

Spin-dependent interactions of heavy quarks, formulated using temporal Wilson lines with additional field insertions, were first derived in Ref.~\cite{Eichten:1980mw}.
They can be expressed in terms of five a priori independent potentials
(see Appendix~A for definitions),
which contribute as
\bea
\label{SD}
V_{SD}=&&\bigg(\frac{S_{ Q}\cdot L_Q}{2m^2_Q}-\frac{S_{\bar Q}\cdot L_{\bar Q}}{2m_{\bar Q}^2}\bigg)
\bigg(\frac 1r\frac d{dr}\big(V(r)+2V_1(r)\big)\bigg)\nonumber\\
+&&\bigg(\frac{S_{\bar Q}\cdot L_Q}{m_Q m_{\bar Q}}-\frac{S_{Q}\cdot L_{\bar Q}}{m_{\bar Q} m_{\bar Q}}\bigg)
\bigg(\frac 1r\frac d{dr}V_2(r)\bigg)\nonumber\\
+&&\frac {(3S_{Q}\cdot \hat r\, S_{\bar Q}\cdot \hat r-S_Q\cdot S_{\bar Q})}{3m_Qm_{\bar Q}}V_3(r)
+\frac 13\frac{S_Q\cdot S_{\bar Q}}{m_Qm_{\bar Q}}V_4(r) \, .
\eea
Here $\vec S_{Q,\bar Q}$ and $\vec L_{Q,\bar Q}$ denote the spin and orbital angular momenta of the $\bar Q Q$ pair.
$V(r)$ is the central static potential, while $V_1(r)$ and $V_2(r)$ arise from inserting a chromo-electric or chromo-magnetic field,
respectively, into the temporal Wilson loop.
The spinspin and tensor contributions, $V_3(r)$ and $V_4(r)$, follow from the insertion of two chromo-magnetic fields.
Equation~(\ref{SD}) is exact up to order $1/m_Q^2$.
The spinspin potential is further constrained by
$V_4(r)=2\nabla^2 V_2(r)$~\cite{Eichten:1980mw}.

Lorentz invariance relates the central potential $V(r)$ to the spinorbit potentials $V_1(r)$ and $V_2(r)$
through the Gromes relation~\cite{Gromes:1984ma},
\be
\label{V12}
V(r)=V_2(r)-V_1(r) \, .
\ee
Spin-dependent contributions generated by the confining string were discussed by
Buchm\"uller~\cite{Buchmuller:1981fr} and others~\cite{Pisarski:1987jb,Gromes:1984ma}.
Since spinspin interactions are short-ranged, only the spinorbit terms survive at
large transverse separation $b_\perp$.
This is evident from Eq.~(\ref{V12}), where asymptotically $V_2(r)\rightarrow 0$ and
$V(r)\rightarrow -V_1(r)$.
Consequently~\cite{Buchmuller:1981fr,Pisarski:1987jb,Gromes:1984ma},
\be
V_{SL,{\rm string}}(b_\perp)
=-\bigg(\frac{S_Q\cdot L_Q}{2m_Q^2}-\frac{S_{\bar Q}\cdot L_{\bar Q}}{2m_{\bar Q}^2}\bigg)
\frac {\sigma _T}{b_\perp}
\rightarrow -\frac {\sigma_T}{2m_Q^2b_\perp}\,S\cdot L \, ,
\ee
where $V(r)=\sigma_T r$, $\vec S=\vec S_Q+\vec S_{\bar Q}$, and
$\vec L=\vec L_Q=-\vec L_{\bar Q}$.
Because the electric flux tube is confined within the string, the spinorbit interaction
arises solely from Thomas precession.
Its sign is opposite and its magnitude is half that of the spinorbit contribution
generated by the central potential.
This behavior follows directly from Eqs.~(\ref{SD}) and~(\ref{V12}) if one assumes
that $V_2(r)$ is short-ranged, implying $V(r)=-V_1(r)$.
The resulting string-induced spinorbit interaction is therefore termed
{\em scalar-like}, in contrast to the {\em vector-like} Coulomb-induced spinorbit interaction.

\begin{figure}[htbp]
\begin{center}
\includegraphics[width=6cm]{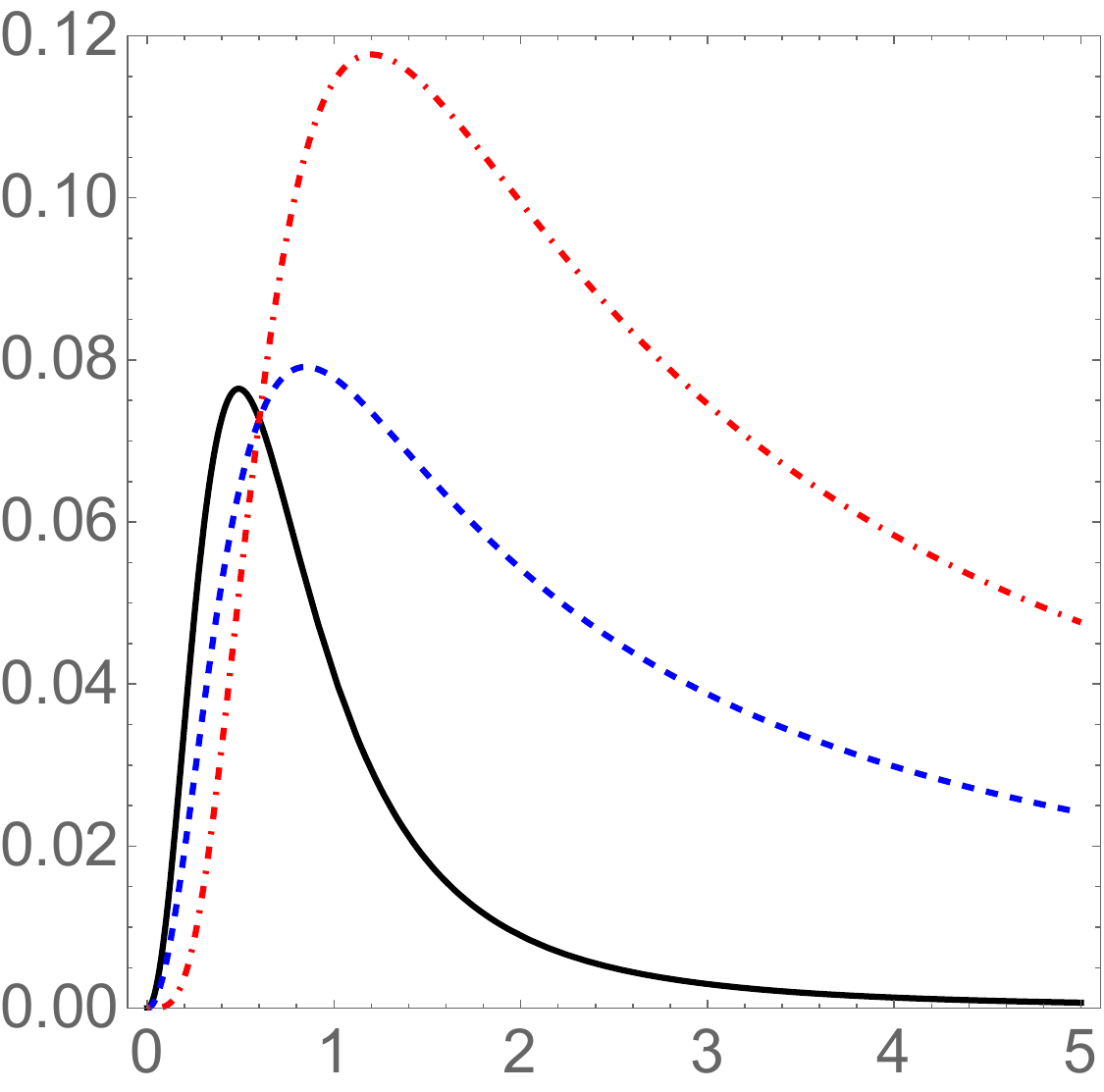} 
\includegraphics[width=6cm]{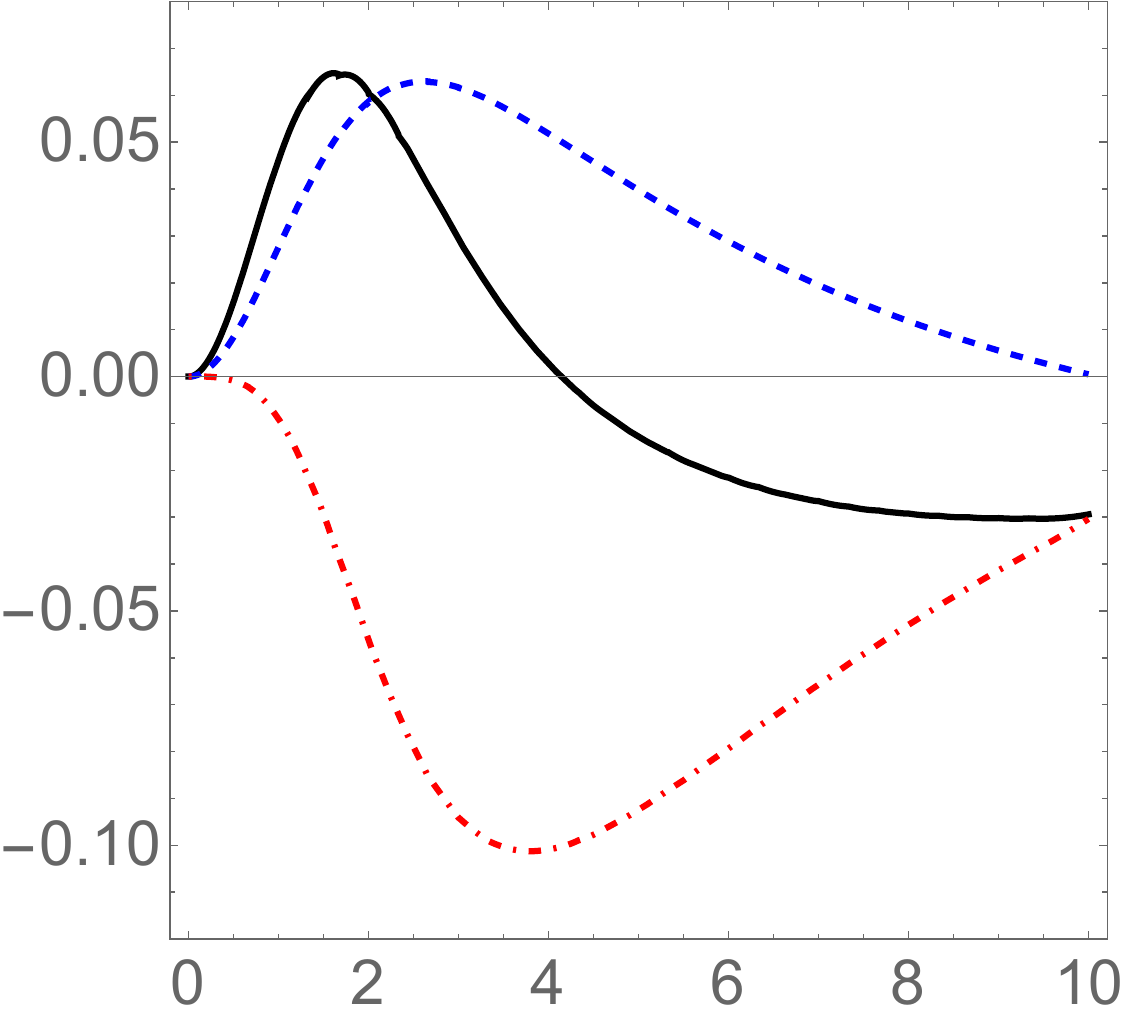}
\caption{Perturbative (left) and instanton-induced (right) spin-dependent potentials
for charmonium.
The black solid, blue dashed, and red dash-dotted curves correspond to
$r^2 V_{SS}$, $r^2 V_{SL}$, and $r^2 V_{T}$ (in ${\rm GeV}^{-1}$), respectively,
plotted as functions of $r$ in ${\rm GeV}^{-1}$.}
\label{fig_spin_V}
\end{center}
\end{figure}

Four decades after the discovery of quarkonia and the formulation of perturbative
quarkonium theory, static spinspin potentials were evaluated on the lattice using
correlators of Wilson lines with explicit field-strength insertions.
As shown in Ref.~\cite{Koma:2006fw}, while $V_{SS}$ is indeed short-ranged,
it does not fit the simple vector-plus-scalar exchange paradigm.
Instead, pseudoscalar glueball exchange was proposed as an explanation.
A decade later, Ref.~\cite{Kawanai:2015tga} not only extracted the central and spinspin
potentials for the $\bar c c$ and $\bar c s$ systems with high precision,
but also employed them in effective Schrodinger equations, achieving good agreement
with spectroscopic data.
In Fig.~\ref{fig_spin_V}, we display (as a solid line) their exponential fit for $\bar c c$,
\begin{equation}
V_{SS}^{\bar c c}\equiv \alpha e^{-\beta r}, \qquad
\alpha=2.15\,{\rm GeV}, \quad \beta=2.93\,{\rm GeV} \, .
\label{eqn_VSS_lat}
\end{equation}

In the instanton vacuum, all potentials appearing in Eq.~(\ref{SD}) are directly related
to the central potential $V(r)=V_C(r)$~\cite{Callan:1978ye,Eichten:1980mw,Turimov:2016adx},
\be
V_C(r)+2V_1(r)\rightarrow 0 , \qquad
V_2(r)\rightarrow \frac 12 V_C(r) , \qquad
V_3(r)\rightarrow -\bigg(\frac 1rV_C^\prime-V_C^{\prime\prime}\bigg) \, .
\ee
These relations lead to the simplified spinorbit interaction
\be
\label{SD1}
\bigg[\frac 1{2m_Q^2}\frac 1r \frac d{dr} V_C(r)\bigg]
(S_{Q}+S_{\bar Q})\cdot L_{Q}
\equiv
\bigg[\frac 1{2m_Q^2}\frac 1r \frac d{dr} V_C(r)\bigg] S\cdot L \, .
\ee
This simplification follows from the self-duality of the gauge fields,
$\vec E^a=\pm \vec B^a$, in Euclidean space.
Schematically, the central electric correlator
$\langle E_i^a(x)[x,0]^{ab}E_i^b(0)\rangle$
is then directly related to the spinorbit correlator
$\langle E_i^a(x)[x,0]^{ab}B_i^b(0)\rangle$
and to the spinspin correlator
$\langle B_i^a(x)[x,0]^{ab}B_i^b(0)\rangle$.

The derivation of the corresponding potentials for a sloped Wilson loop on the light front
follows the same reasoning as in Refs.~\cite{Callan:1978ye,Turimov:2016adx}.
The only difference is that the relevant integration becomes cylindrical rather than spherical.
The resulting central potential carries over to the light cone as
\bea
V_C(r\rightarrow b_\perp)
=\bigg(\frac{4\kappa}{N_c\rho}\bigg)\,
{\bf I}\bigg(\frac{b_\perp}{\rho}\bigg) \, .
\eea
We note the sign flip of the spinorbit contribution between the confining
(string-induced) potential and the instanton-induced potential, as well as the sign flip
of the instanton tensor interaction relative to the perturbative one-gluon
spintensor contribution.

\section{Bridging together the worlds of heavy and light quarks}

Let us begin by introducing the concept through the notion of
approximate universality of heavy-quark spectra, namely their
{\em unexpectedly weak} dependence on the quark mass.

For purely Coulombic systems, such as the hydrogen atom, the Coulomb
potential $\alpha/r$ contains no intrinsic dimensional parameter.
As a result, the bound-state energies $E_n$ scale linearly with the
electron mass $m$.
In the asymptotic heavy-quark limit, quarkonia are expected to be dominated
by an approximately Coulombic interaction, and one would therefore
anticipate a similar scaling, $E_n\sim m$.

This expectation is not borne out by real-world $\bar c c$ and
$\bar b b$ quarkonia.
In Fig.~\ref{fig_wfN31} we show the (spin-weighted) spectra of charmonia
and bottomonia, {\em counted from their respective ground states},
$J/\psi$ and $\Upsilon$.
Despite the quark masses differing by roughly a factor of three, the
excitation spectra are nearly coincident.
In particular, the $2S-1S$ gaps are $0.605$ and $0.572\,{\rm GeV}$,
respectively, differing by only a few percent.
Why should this be the case?

A traditional explanation invokes the Cornell potential, which combines
a Coulombic $-1/r$ term with a confining $\sim r$ contribution.
One might speculate that their effects push the level spacings in opposite
directions, leading to an accidental cancellation.
This possibility is easy to test%
\footnote{Note that in doing so we remove relativistic corrections present
in the experimental masses.}.
Adopting the Cornell potential
\be
\label{eqn_cornell}
V=-\frac{0.641}{r}+0.113\, r \, ,
\ee
(with all energies in GeV and $r$ measured in ${\rm GeV}^{-1}$),
we computed the $1S$, $2S$, $3S$, and $4S$ masses and the corresponding gaps,
shown as open blue points in Fig.~\ref{fig_wfN31}.
One finds that the higher gaps actually agree even better than the first.
Thus, the observed universality is not the result of a single accidental
cancellation; it persists and even improves for higher excitations.

This observation naturally raises the question: what form of potential
would lead to exact universality?
If the quark mass is rescaled by a factor $\xi$ and distances are rescaled
by $\xi^{-1/2}$, the kinetic energy remains unchanged.
For a power-law potential $V\sim r^\alpha$, the potential energy rescales
as $\xi^{-\alpha/2}$.
The so-called {\em Martin potential}, introduced for charmonium,
\be
\label{eqn_Martin}
V_{\rm Martin}=-9.054+6.870\, r^{0.1} \, ,
\ee
rescales as $\xi^{-0.05}$.
Does this explain the observed universality?
Our calculations 
 show that
it does not: the agreement is far less accurate than the empirical
universality of the spectra.

For a logarithmic potential, $V\sim \log r$, a rescaling of the mass
simply shifts all energy levels by an additive constant,
$-(1/2)\log\xi$, while leaving the level spacings invariant.
Thus, the quarkonium spectra suggest that the effective interquark
potential is, to a good approximation, logarithmic.
Why should this be so?

\begin{figure}[h]
\begin{center}
\includegraphics[width=8cm]{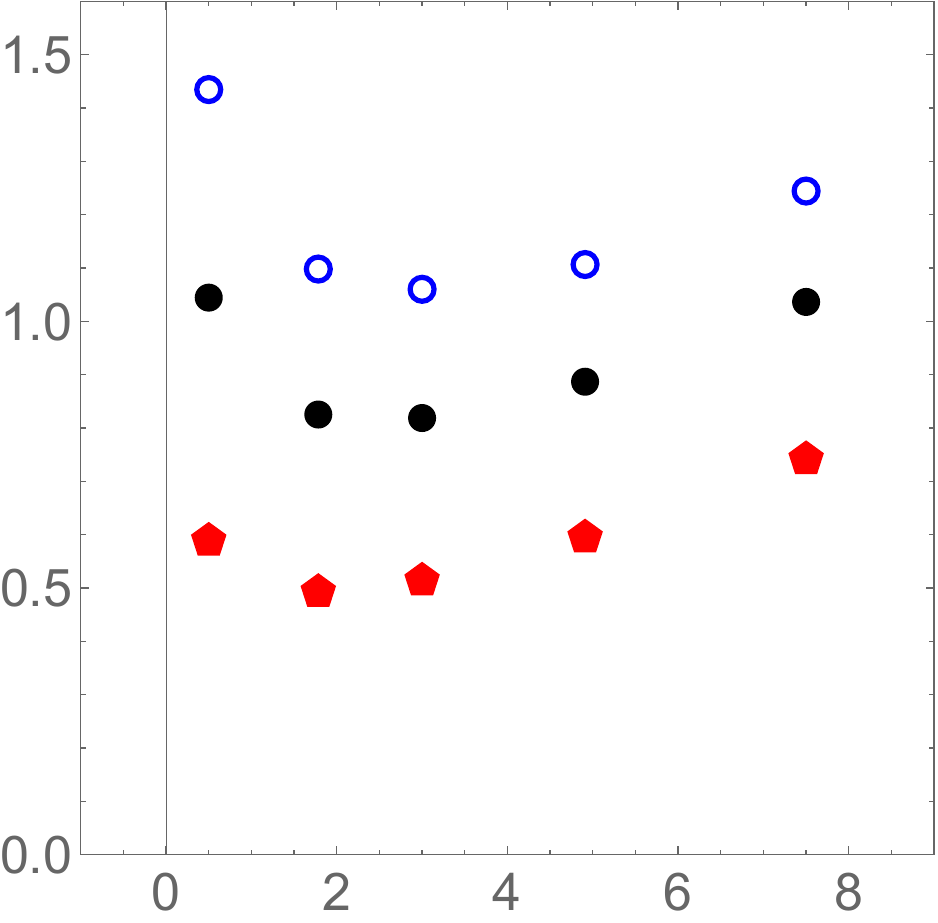}
\includegraphics[width=8cm]{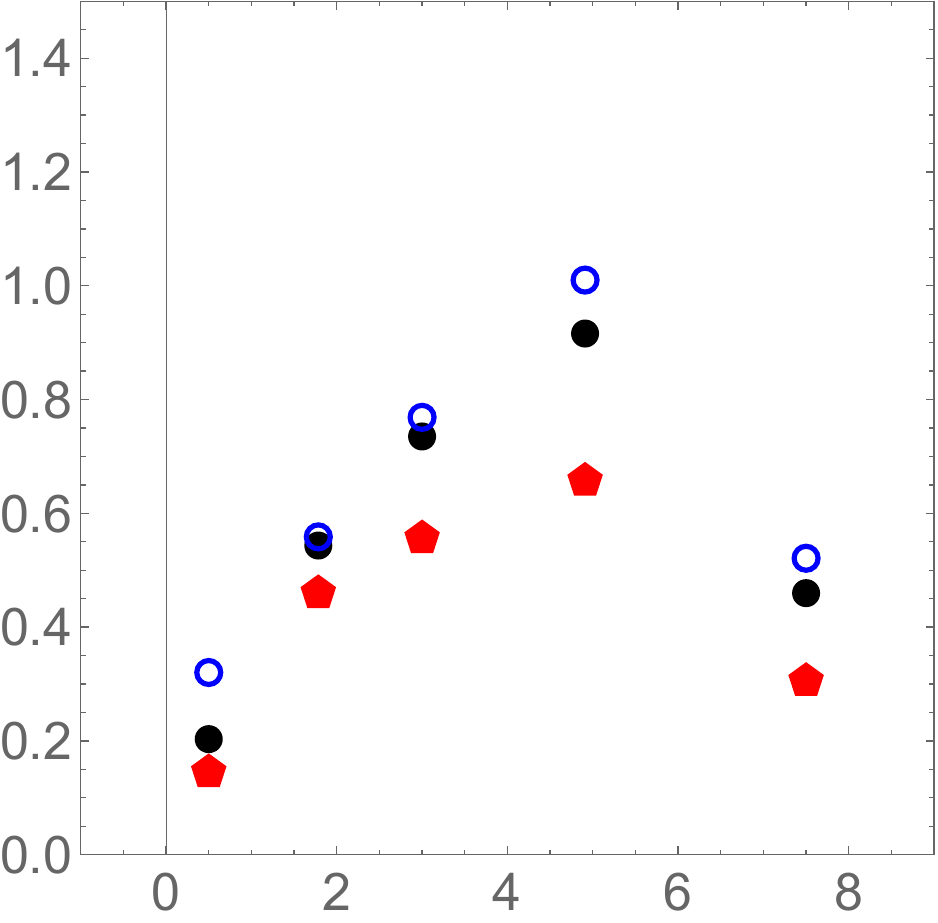}
\caption{Spectral gaps $E(2S)-E(1S)$, $E(3S)-E(1S)$, and $E(4S)-E(1S)$
(in ${\rm GeV}$, from bottom to top) as functions of the quark mass
$m_q$ (in GeV).
The chosen values are $0.5$ (strange), $1.7$ (charm), $3$, $4.7$ (bottom),
and $7.5\,{\rm GeV}$.
The left panel corresponds to the Cornell potential~(\ref{eqn_cornell}),
while the right panel shows the instanton-induced potential.}
\label{fig_gaps}
\end{center}
\end{figure}

To address this point, we recalculated the dependence of the spectra
on the quark mass and present the results in Fig.~\ref{fig_gaps}.
The left panel shows the results for the Cornell potential~(\ref{eqn_cornell}),
while the right panel corresponds to the instanton-induced potential.
In both cases, as the quark mass varies by more than an order of magnitude,
from $1$ to $15\,{\rm GeV}$, the gaps between the spherical $S$-wave states
change only modestly, by about $0.1\,{\rm GeV}$ in the Cornell case.

The key point illustrated by these plots is that the universality between
charmonium and bottomonium arises because the spectral gaps depend only weakly
on the quark mass.
However, the detailed behavior differs qualitatively: for the Cornell
potential the mass dependence exhibits a minimum, whereas for the
instanton-induced potential it exhibits a maximum.
The true interaction law must therefore lie somewhere in between.

In principle, hadronic spectroscopy can thus provide information about
rather subtle features of the interquark potential.
In practice, however, the available data are limited:
for bottomonia we know the masses of four $S$ states, while for charmonia
only the $1S$ and $2S$ states (and hence a single gap) are well established.
We have not overlaid the experimental points on Fig.~\ref{fig_gaps} to avoid
cluttering the plot.
Doing so would not allow one to discriminate between the two cases, and we
are therefore forced to conclude that present data do not reveal whether
the observed universality corresponds to a minimum or a maximum in the
mass dependence.
At present, we cannot decisively determine which potential is favored.

Finally, let us address the broader idea of {\em bridging the worlds of light
and heavy quarks.}
The vacuum paths of heavy quarks are well approximated by straight lines
along the time direction, namely Wilson lines.
In contrast, the vacuum paths of nearly massless quarks are completely
different: instanton zero modes correspond to relativistic motion of light
quarks bound within instantons.
At first sight, these two pictures appear impossible to reconcile.

However, suppose we are not interested in the detailed dynamics of quarks
inside instantons, but only in how they propagate {\em between} them.
After appropriate resummations, one arrives at the notion of
constituent quarks with effective masses
$M_{u,d}\sim 400\,{\rm MeV}$, generated by the chiral condensate.

Now imagine temporarily setting aside chiral dynamics and using the
$\bar Q Q$ potential derived from Wilson lines in an $I\bar I$ background.
At large distances the potential saturates,
$V(r\gtrsim 0.8\,{\rm fm})\sim 800\,{\rm MeV}$.
Per quark, this corresponds to a mass shift
$\Delta M\sim V/2\approx 400\,{\rm MeV}$.
Thus, somewhat unexpectedly, two frameworks that appear radically
different merge in a rather smooth way.

This $bridging$ idea is particularly appealing in the strange sector.
The effective strange constituent quark mass is $M_s\sim 500\,{\rm MeV}$,
roughly three times smaller than for charm and an order of magnitude
smaller than for bottom.
Yet the spectral gap for the $\phi=\bar s s$ vector mesons shown in
Fig.~\ref{fig_gaps_quarkonia} is quite similar to those of heavier quarkonia.
Once again, quarkonium-like nonrelativistic Schrodinger equations and
chiral BetheSalpeter equations yield nearly identical results in this case.

\begin{subappendices}
\section{Solving Schrodinger equations in Wolfram Mathematica} 
\label{sec_Math_Sch}
Standard quantum mechanics courses include well-known examples of analytic
solutions, such as the harmonic oscillator and the hydrogen atom.
In practice, however, one often has to solve the Schrodinger equation
numerically, in a variety of dimensions and physical settings.
In one dimension, several numerical solvers are available that start from
a prescribed value of the wave function and its first derivative at one end
of the domain, and then tune the energy eigenvalue to obtain acceptable
behavior at the other end.
In higher dimensions, variational methods have traditionally been used
most often.

Recent versions of {\sc Wolfram Mathematica} have significantly simplified
such calculations.
There is no longer a need to rely on numerical PDE solvers that propagate
solutions from one side of the domain to the other.
Instead, the command \texttt{NDEigensystem} provides a much more convenient
approach, directly yielding eigenvalues and eigenfunctions of a given
differential operator.
In this way, one can obtain wave functions and energies for as many states
as required.
Below we present an example that computes seven bottomonium states.
Note that \texttt{u[x]} corresponds to the radial wave function $\psi(r)$,
energies and masses are expressed in GeV, and the radial coordinate $r$ is
measured in inverse GeV units.

\begin{verbatim}
In: VCornell[r_] := -(4/3)*0.4/r + (.4^2)*r;
Mb = 4.8; 
\[ScriptCapitalL]b = (-u''[x] - 2/x*u'[x])/Mb + (2*Mb + VCornell[x])*
    u[x];
{valb, funb} = 
  NDEigensystem[\[ScriptCapitalL]b, u[x], {x, 0, 40}, 7, 
   Method -> {"Eigensystem" -> "Arnoldi", 
     "SpatialDiscretization" -> {"FiniteElement", {"MeshOptions" -> \
{MaxCellMeasure -> 0.001}}}}];
valb

Out: {9.42516, 9.98909, 10.315, 10.5758, 10.8034, 11.01, 11.2015}

In: Plot[{-funb[[1]], -funb[[2]]}, {x, 0, 10}, PlotRange -> {-1, 3}]   
\end{verbatim}
The experimentally measured masses are
$9.4604, 10.0234, 10.3551, 10.5794, 10.8852,$ and $11.\,{\rm GeV}$.

\begin{figure}
    \centering
    \includegraphics[width=0.5\linewidth]{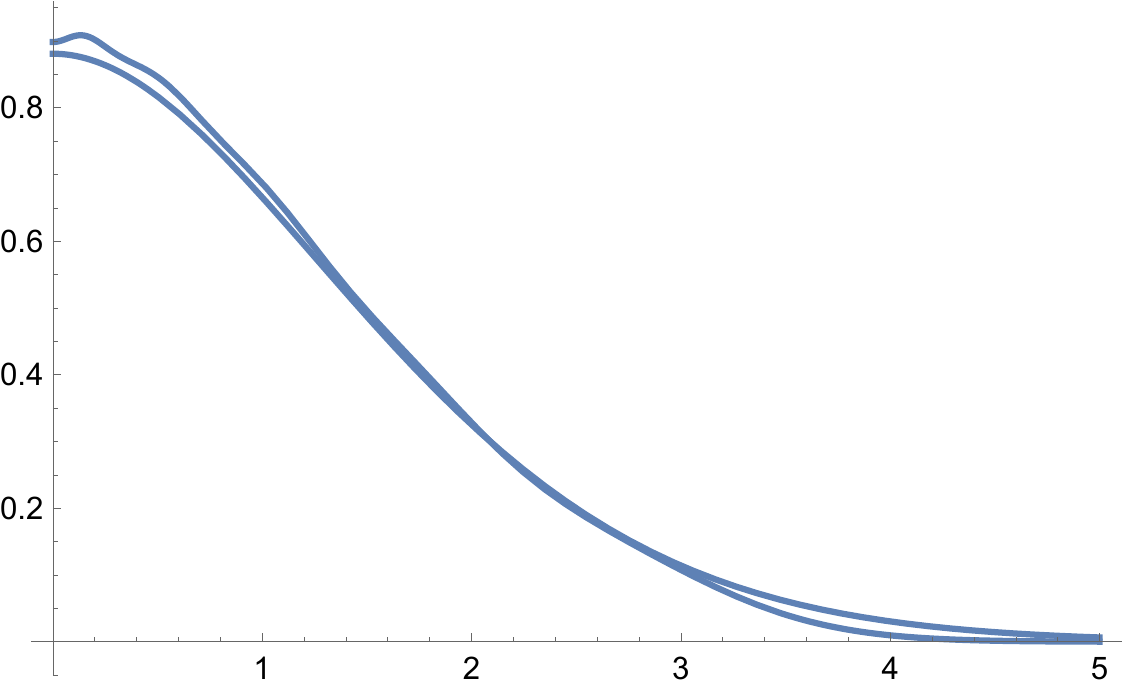}
        \includegraphics[width=0.5\linewidth]{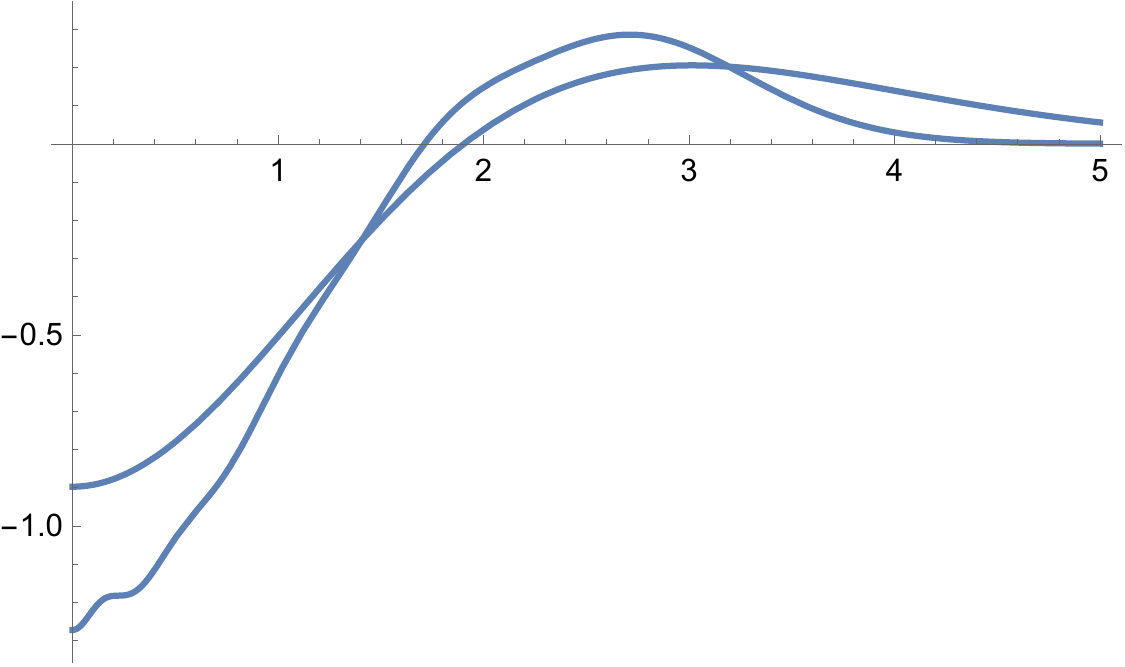}
    \caption{Comparison of the first and second WFs, between numerical and variational solutions.}
    \label{fig_Ritz}
\end{figure}

The final line of the code illustrates how to plot the wave functions;
the resulting smooth curves are shown in Fig.~\ref{fig_Ritz}.
Plotting wave functions is particularly useful, as it allows one to verify
that the number of nodes matches the expected ordering of the states.

In the present example this issue does not arise, but it is worth noting that
the states returned by \texttt{NDEigensystem} are not necessarily ordered from
lowest to highest energy, as one might expect.
Instead, they are listed according to the magnitude of the {\em modulus} of
the eigenvalues, which can sometimes lead to confusion in their ordering.

This method also works in two dimensions (we will use it for the forward part
of baryon wave functions defined on a triangular domain) and in three
dimensions.
Unfortunately, it does not extend to four dimensions, which will be required
for the pentaquark applications discussed below.

\section{Ritz variational method: simple examples}

Here we compare  direct numerical solutions with the widely used variational
method, e.g. employed 
nowadays in multidimensional applications such as tetraquarks.

The method, proposed by W.~Ritz in 1909 for general elliptic partial
differential equations, consists of minimizing the energy using a chosen set
of basis functions.
These basis functions need not form an orthogonal set, although orthogonality
can simplify the calculations.
In the example below, we compare the variational approach with numerically
obtained solutions for a linear potential in the bottomonium problem.
We use ten variational coefficients $c_i$ as parameters and minimize the ratio
of the expectation value of the Hamiltonian to the norm of the wave function.

\begin{verbatim}
In: nmax = 10; Radii = Table[0.1*n, {n, 1, nmax}];
Cset = Array[c, nmax]
uMin[x_] := Sum[c[n]*Exp[-x^2/2/Radii[[n]]^2], {n, 1, nmax}]; 
(* select WF as a set of Gaussians *)
\[Psi]H\[Psi] = Expand[uMin[x]*((-D[uMin[x], {x, 2}] - (2/x)*D[uMin[x], x])/Mc +  VcK[x]*uMin[x])];
\[Psi]H\[Psi]av = Integrate[\[Psi]H\[Psi]*x^2, {x, 0., Infinity}];
norm = Integrate[uMin[x]^2*x^2, {x, 0., Infinity}];
cond := c[1]^2 < 1 && c[2]^2 < 1 && c[3]^2 < 1 && c[4]^2 < 1 && 
  c[5]^2 < 1 && c[6]^2 < 1 && c[7]^2 < 1 && c[8]^2 < 1 && c[9]^2 < 1 &&
   c[10]^2 < 1
In: NMinimize[{\[Psi]H\[Psi]av/norm, cond},
 {c[1], c[2], c[3], c[4], c[5], c[6], c[7], c[8], c[9], c[10]}]
Out: {0.415895, {c[1] -> 2.30567*10^-8, c[2] -> -1.53966*10^-7,  c[3] -> 9.77366*10^-7, 
c[4] -> -4.90733*10^-6, c[5] -> 0.0000181519, c[6] -> -0.0000475488, c[7] -> 0.0000847055, c[8] -> -0.000097003,  c[9] -> 0.0000642188, c[10] ->-0.000018777}}
Show[Plot[(-uMin[x]/Sqrt[norm]) /. {c[1] -> 2.3056657172291844`*^-8, c[2] -> -1.5396634565426175`*^-7, c[3] -> 9.773662975575236`*^-7, 
c[4] -> -4.907333728277542`*^-6, c[5] -> 0.000018151944935911747`,c[6] -> -0.00004754881284624645`, c[7] -> 0.00008470554538661096`, c[8] -> -0.00009700295113376238`, c[9] -> 0.00006421875804716321`,  c[10] -> -0.00001877695480986375`}, {x, 0, 5}],
\end{verbatim}

The upper figure compares smooth curve obtained via $NDEigensystem$
to (slightly wiggling) variational result. The error in WF and energy is about 1\%.

In order to repeat it for the next level, one has to calculate
scalar product of the ground state to the next, and put it as a
condition into $NMinimize$: in this case it generates the lower
plot. Note that the agreement between curves is fair, but accuracy is significantly degraded, already
for the second state it is about 10\%. 
Perhaps playing more with radii can get a bit more accurate results.

   In general, NMinimize is not very stable and in some cases it
   get confused, working only if the number of parameters is 
   reduced. Also  restricting the minimization domain by extra conditions
   sometimes helps to get to the right minimum.

\end{subappendices}

\chapter{Glueballs}
\section{Introduction}
Glueballs are the simplest color-singlet bound states of non-Abelian gauge theory, composed entirely of gluonic degrees of freedom. As such, they provide a unique window into confinement, dynamical mass generation, and the role of topology in quantum chromodynamics (QCD). In the pure Yang-Mills theory, where quark degrees of freedom are absent, the glueball spectrum is well defined and has been studied extensively using lattice gauge theory, which remains the most reliable nonperturbative approach to this problem.

One of the most obvious question in QCD is why the observed hadrons are made of quarks and not of gluons. It appears that glueballs are much heavier than typical quark-made hadrons, and therefore they have large widths and/or complicated decay patterns in the real world, making them difficult to identify. But why are glueballs heavier? What are their masses, radii and other parameters in a purely gluonic world, and how do they change if one includes light quarks?

Three decades ago \cite{Schafer:1994fd} have shown that instantons lead to a strong attractive force in the $J^{PC}= 0^{++}$ channel, which results in the scalar glueball being much smaller in size than other hadrons (including the pions). In the pseudoscalar $0^{-+}$ channel the corresponding force is repulsive, and in the $2^{++}$ case the instanton effect is absent. Furthermore, due to the strong classical field of the instantons, these forces are even stronger as compared to the ones for mesons made of light quarks. 


Early high-precision lattice calculations, particularly those employing anisotropic lattices, established the ordering of the lowest-lying glueball states and set quantitative benchmarks for their masses. These studies  identified the scalar $0^{++}$ glueball as the lightest state, followed by the tensor $2^{++}$, with the pseudoscalar $0^{-+}$ and higher-spin excitations appearing at significantly larger masses \cite{Morningstar:1999rf,Chen:2006dv}. Subsequent work with improved actions, larger operator bases, and careful continuum extrapolations has refined this picture and confirmed its stability across lattice spacings and volumes. In particular, recent high-statistics calculations have provided a  reference spectrum for pure $\mathrm{SU}(3)$ Yang-Mills theory, covering a wide range of $J^{PC}$ channels \cite{Athenodorou:2020ani}.

While the quenched glueball spectrum is now comparatively well established, several important questions have gained renewed attention in recent years. One concerns the {\em internal structure} of glueballs in hadrons, including their spatial extent and form factors. Lattice studies of energy-momentum tensor matrix elements and related gravitational form factors have begun to probe the size and mass distribution of glueballs, with emerging evidence that the scalar $0^{++}$ state may be unusually compact compared to higher-spin excitations. Such results provide new, nontrivial constraints on phenomenological and microscopic models of glueballs. A recent review summarizes the current status of lattice calculations, experimental searches, and theoretical interpretations \cite{Morningstar:2025glueballreview}.

Another major direction concerns the fate of glueballs in full QCD. When light quarks are dynamical, pure-glue operators can mix with flavor-singlet $q\bar q$ states, particularly in the scalar and pseudoscalar channels. This mixing complicates the identification of experimental candidates and blurs the connection between quenched lattice spectra and observed isoscalar mesons. Lattice calculations incorporating dynamical quarks have begun to address these issues directly, including studies of pseudoscalar glueball-$\eta$ and $\eta'$ mixing, highlighting the importance of gluonic components even in channels traditionally associated with quark model states \cite{Jiang:2022ffl}.

In parallel with lattice developments, a variety of theoretical approaches have been pursued. Functional methods based on Dyson-Schwinger and Bethe-Salpeter equations reproduce many qualitative features of the glueball spectrum and offer insight into the role of gluon mass generation and effective interaction kernels. Phenomenological models, including constituent-gluon Hamiltonians, flux-tube and bag models, and holographic constructions, provide intuitive pictures for the organization of glueball states and their Regge behavior. Among these, instanton-based approaches are particularly suggestive: topological fluctuations of the gauge field generate a momentum-dependent dynamical gluon mass and induce channel-dependent short-range interactions that are strongest in the scalar channel. Calculations in the instanton vacuum yield a dynamical gluon mass of a few hundred MeV in the infrared and motivate enhanced attraction for the $0^{++}$ glueball at distances comparable to the instanton size \cite{Musakhanov:2017erp,Shuryak:2021xkq}.

Motivated by these developments, we construct a constituent two-gluon description of the  glueball states in pure Yang-Mills theory. Our framework combines adjoint Coulomb interactions fixed by group theory, Casimir-scaled confinement with gluonic  screening, and instanton-induced central and spin-dependent forces.

\begin{table}[h!]
    \centering
    \begin{tabular}{|c|c|c|c|} \hline
J= 0 &  1.475 &(30) &(65) \\
  0 & 2.755 &(70) &(120)  \\
  0 & 3.370 & (100)& (150) \\
   0& 3.990 &(210) &(180) \\
  2 & 2.150 &(30) &(100)  \\
  2 & 2.880 &(100) &(130) \\
  3 & 3.385 & (90) &(150) \\
  4 & 3.640 &(90) &(160)  \\
  6 & 4.360 & (260) &(200) \\ \hline
 \end{tabular}
    \caption{Masses (GeV) with their error bars of 
    glueballs, for all $P=C=+$ glueballs, from \cite{Meyer:2004gx}}
    \label{tab_all_pp}
\end{table}

A popular point of view among spectroscopists
(which we also shared until recently) was that,
unlike hadrons made of quarks, glueballs cannot be
described by {\em additive-type} models, in which glueball
masses are built predominantly from some effective
gluons, with interactions playing only a secondary
role. This view was largely shaped by the traditional
focus on the lowest states in the scalar, pseudoscalar,
and tensor sectors.

In this paper we reverse this logic, and instead
consider first a broader set of $normal$ states,
including radial excitations in the scalar channel
and states with higher angular momentum, up to
$J=6$. As a basis we use the lattice study
\cite{Meyer:2004gx}, which, although not recent,
addresses the largest number of states; from it we
retain only the $P=C=+$ subset listed in
Table~\ref{tab_all_pp}.

What we find is that these states can in fact be
described by a Schrodinger equation, without
invoking any extreme assumptions. Moreover, the
spectrum of $normal$ glueballs turns out to resemble that of strange $\bar s s$ and charmed
$\bar c c$ mesons, although it is true that
the role of potential is enhanced, roughly by
the factor 9/4 expected from color algebra.

Particular emphasis is placed on the scalar and tensor channels. While the lowest $0^{++}$ state is dominated by the short-distance dynamics and is very compact, the other $0^{++}$ states 
are not. The $J^{++}$ glueballs with larger angular momentum are affected by angular momentum barriers and significant $S$-$D$ wave mixing. We solve  the resulting coupled-channel problem and discuss the  mass hierarchies and spatial structure.

This chapter is organized as follows. We introduce the constituent two-gluon Hamiltonian and specify the confining, Coulomb, instanton-induced, and spin-dependent interactions that define the model. In Sec.~\ref{sec:schrodinger} we solve the resulting Schrodinger equation and present the glueball spectrum in the $C=+$ sector, including radial excitations in the scalar channel and higher-$J$ states up to $J=6$, together with the corresponding wave functions and radii. The lowest $0^{++}$ state, whose short-distance dynamics is not captured reliably in the nonrelativistic treatment, is discussed separately and motivates the relativistic Bethe-Salpeter treatment .  We complement the numerical spectrum with a semiclassical WKB analysis in appendix, which clarifies the systematics of excitations and the emergence of Regge behavior, and motivates a smooth $J$-dependent parametrization of short-distance mass shifts. Vector glueball channels and the implications of Bose symmetry and the Landau-Yang selection rule.  A number of appendices provide technical details, including the density scaling of the gluon mass, the construction and numerical evaluation of the effective potentials, the formalism and explicit calculations for tensor mixing, analytic expressions for radial integrals, parameter choices, and additional semiclassical and instanton-based derivations supporting the main text.

\section{The Hamiltonian}
\label{sec:hamiltonian}
\subsection{Confining potential}
We model glueballs as two effective constituent gluons in their center-of-mass frame. The relative motion Hamiltonian is
\begin{eqnarray}
H=2m_g(\eta)+\frac{\bm p^2}{m_g(\eta)} +V_0
+V_{\rm conf}(r)+V_C(r)+V_{\rm inst}(r)+V_{SS}(r)+V_T(r).\nonumber\\
\label{eq:Hfull}
\end{eqnarray}

The confining term is taken as a screened adjoint (double) string,
\begin{eqnarray}
V_{\rm conf}(r)=\sigma_8\,r\,e^{-r/r_{\rm scr}}+V_0,\quad
\sigma_8=\frac{C_A}{C_F}\sigma_3=\frac94\,\sigma_3,\quad
\sigma_3=0.18~\mathrm{GeV}^2,
\label{eq:Vconf}
\end{eqnarray}
The overall shift by a constant $V_0$
is not known \emph{a priori}, but fits to the spectrum
yield a rather modest negative value. This shift can
be avoided altogether if only level spacings, rather
than absolute masses, are used in the fits.

\begin{figure}
    \centering  \includegraphics[width=0.75\linewidth]{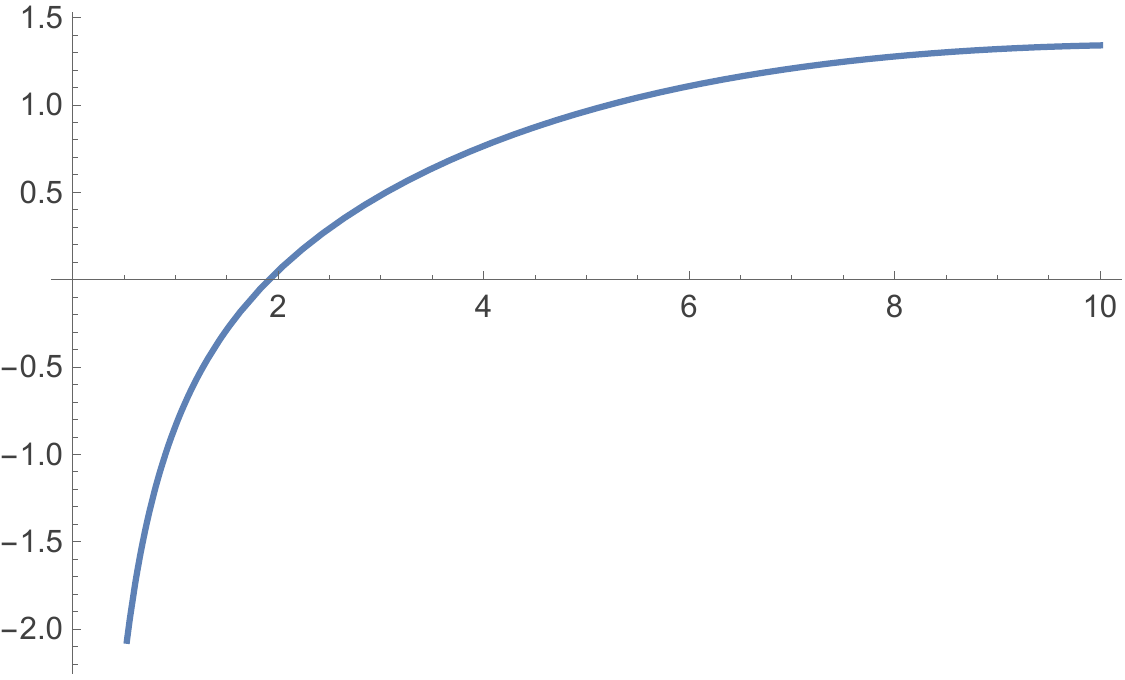}
    \caption{Central potential $V_{conf}(r)\, (GeV)$ used versus $r \,(GeV^{-1})$}
    \label{fig_pot}
\end{figure}

The choice of the adjoint screening length $r_{\rm scr}$
is a rather delicate issue. If $r_{\rm scr}$ is smaller
than the typical size of the states, the potential
acts merely as a barrier and is unable to support
bound states of larger spatial extent, producing at
most a few quasibound states. To avoid this
situation we place the states in a box of radius
$R = 2\,\mathrm{fm}$ and take $r_{\rm scr}\sim R$, so that
the potential inside the box does not decrease but
instead saturates, as illustrated in
Fig.~\ref{fig_pot}. A more detailed discussion, as
well as the relation of this potential shape to our
Monte-Carlo calculations with a single instanton or
$\bar I I$ molecule using numerically generated Wilson
lines, is given in Appendix~\ref{sec_pote_numeric}.

\subsection{The short-distance effects}
The  Coulomb term follows from one-gluon exchange. Using
\begin{equation}
V_C(r)=\frac{\alpha_s}{r}\,T_1\!\cdot T_2,
\qquad
T_1\!\cdot T_2=\frac12\big(C_R-C_{R_1}-C_{R_2}\big),
\label{eq:VCgeneral}
\end{equation}
one obtains the Casimir form
\begin{equation}
V_C(r)=-\frac{\alpha_s}{2r}\big(C_{R_1}+C_{R_2}-C_R\big).
\label{eq:VCcasimir}
\end{equation}
For two gluons $R_1=R_2=A$ with $C_A=3$ coupled to a singlet $R=1$ with $C_R=0$,
\begin{equation}
V_C^{(gg,1)}(r)=-\frac{3\alpha_s^{\rm eff}}{r},
\label{eq:VCgg}
\end{equation}
which is attractive and $9/4$ times stronger than the $q\bar q$ singlet Coulomb coefficient at the same $\alpha_s^{\rm eff}$.

The  instanton-induced interactions
are large, but $not$ parametrically larger than 
that for quarks. Naively, writing the field as a sum $A_{class}+A_{gluon}$ one can get their
product O(1/g) which however vanishes due to
random instanton orientation. Correct interaction estimate comes e.g. from the quartic term 
$$ g^2 A_{class}^2 A_{gluon}^2\sim O(g^0)A_{gluon}^2  $$
contributing to effective gluon mass.
Note that for light quarks the instanton-induced Lagrangian is built from fermion zero modes which also
do not have $g$. Yet numerically it is large and that is why the scalar glueball's size is only $\sim 0.2\, fm$, while the smallest of all mesons - the pion - have radius $\sim 0.5\, fm$ .

\begin{figure}[h] 
\centering  \includegraphics[width=0.45\linewidth]{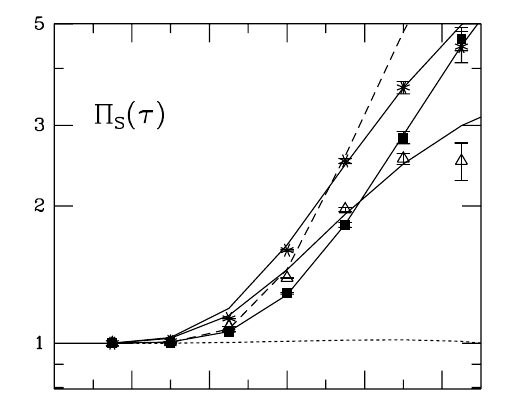}
\includegraphics[width=0.45\linewidth]{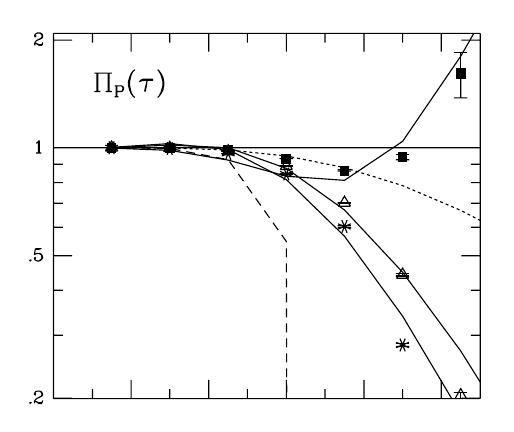}
\includegraphics[width=0.45\linewidth]{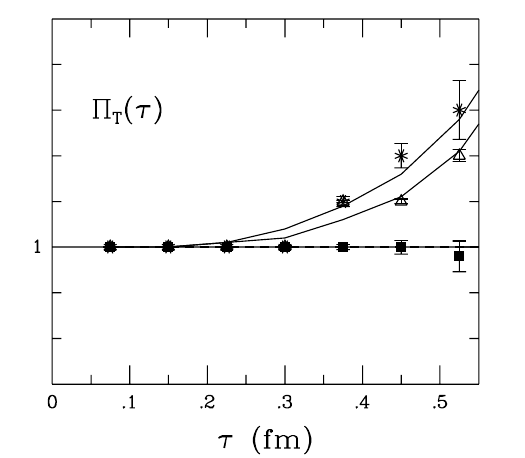}
    \caption{
    Scalar, pseudoscalar and tensor glueball vacuum correlation functions normalized to the corresponding free correlators,
    as a function of Euclidean time between operators, from \cite{Schafer:1994fd}. The results in the random, quenched and full ensembles are denoted by stars, open triangles and solid squares, respectively. The solid lines show the parametrization described in the text, the dashed line the dilute instanton gas approximation, and the dotted line the QCD sum rule calculation. The horizontal line in the second figure was added to guide the eye, the vertical scale in the third figure is $10^{-4}$.
    }
    \label{fig_corr}
\end{figure}

Correlation functions of scalar, pseudoscalar and tensor operators calculated in \cite{Schafer:1994fd} 
in ILM are reproduced in Fig.\ref{fig_corr}. They clearly show that interaction in these three channels are attractive, repulsive, and small, respectively.

Predicted values of the lowest glueball masses and radii (from Bethe-Salpeter amplitudes) are  given in
the Table \ref{tab_corr}. 

\begin{table}[]
    \centering
    \begin{tabular}{|c|c|c|c|} \hline
       & scalar & pseudoscalar & tensor \\ \hline
    $ M (GeV)$   & 1.4$\pm$ 0.2 & > 3. & ? \\ \hline
     $r_{r.m.s}$ (fm) &   0.21 & ? & 0.61\\
     \hline
    \end{tabular}
    \caption{Masses and sizes of glueballs from fits to correlation functions and Bethe-Salpeter amplitudes, predicted by \cite{Schafer:1994fd}.}
    \label{tab_corr}
\end{table}

The main modification which these old
calculations now would require would be in the tensor channel. True, instantons have zero stress tensor. Yet instanton-antiinstanton
molecules have comparable electric and magnetic fields, and nonzero
stress tensor. Therefore the 
effects in the tensor channel is expected to be enhanced by the ratio
of their density to that of dilute instantons, denoted by  
parameter $\eta\sim 7$.

The perturbative spin-spin
interaction is local $\sim \delta^3(r)$. The instanton-induced
attraction is introduced as a Gaussian centered at the origin, with range $\rho$,
\begin{equation}
V_{\rm inst}(r)=-G(\eta)\,e^{-r^2/2\rho^2},
\qquad
G(\eta)=G_0\,\eta,
\label{eq:Vinst}
\end{equation}
with $\rho\simeq \frac13~\mathrm{fm},$
and is assumed to act dominantly in the scalar channel, encoding the strong instanton enhancement of $0^{++}$.
 Because the (anti)instanton field is (anti)self-dual, the induced coupling in~\cite{Liu:2024glue} (see Eq.~(45) ) carries the 't~Hooft symbols $\eta,\bar\eta$ and therefore projects most strongly onto the parity-even scalar combination $G^{a}_{\mu\nu}G^{a}_{\mu\nu}$ (enhancing the $0^{++}$ channel) while the Levi-Civita piece controls the pseudoscalar $G\tilde G$ channel. 

The spin-dependent terms are taken in the standard form
\begin{eqnarray}
V_{SS}(r)&=&\frac{C_{SS}(\eta)}{m_g^2(\eta)}\,\delta^{(3)}_\Lambda(\bm r)\,\bm S_1\!\cdot\!\bm S_2,
\nonumber\\
V_T(r)&=&\frac{C_T(\eta)}{m_g^2(\eta)}\,\frac{1-e^{-r^2/\rho^2}}{r^3}\,S_{12},
\label{eq:Vsd}
\end{eqnarray}
with $S_{12}=3(\bm S_1\!\cdot\!\hat{\bm r})(\bm S_2\!\cdot\!\hat{\bm r})-\bm S_1\!\cdot\!\bm S_2$, and a Gaussian-smeared contact regulator
\begin{equation}
\delta^{(3)}_\Lambda(\bm r)=\left(\frac{\Lambda^2}{\pi}\right)^{3/2}e^{-\Lambda^2 r^2},
\qquad \Lambda\sim \rho^{-1}.
\label{eq:deltaReg}
\end{equation}
The dense-ILM scaling is implemented by
\begin{equation}
C_{SS}(\eta)=C_{SS}^{(0)}\eta,\qquad C_T(\eta)=C_T^{(0)}\eta,
\label{eq:Ceta}
\end{equation}
with $C_T^{(0)}$ allowed to be negative, consistent with the ILM result in~\cite{Shuryak:2021fsu}.
The dynamical constituent gluon mass is parametrized as
\begin{equation}
m_g(\eta)=m_{g0}\sqrt{\eta},\qquad m_{g0}\simeq 0.36~\mathrm{GeV},
\label{eq:mgEta}
\end{equation}

\section{Glueball spectrum from the Schrodinger equation}
\label{sec:schrodinger}
\subsection{The $S=0,\,J=L$ sector}

Since the expected effective gluon mass is
$O(1\,\mathrm{GeV})$, general validity of the nonrelativistic approximation
is expected to be comparable to that for charmonium. Of course, there
are important differences in accuracy arising from  larger potentials etc.

\begin{figure}
    \centering
    \includegraphics[width=0.55\linewidth]{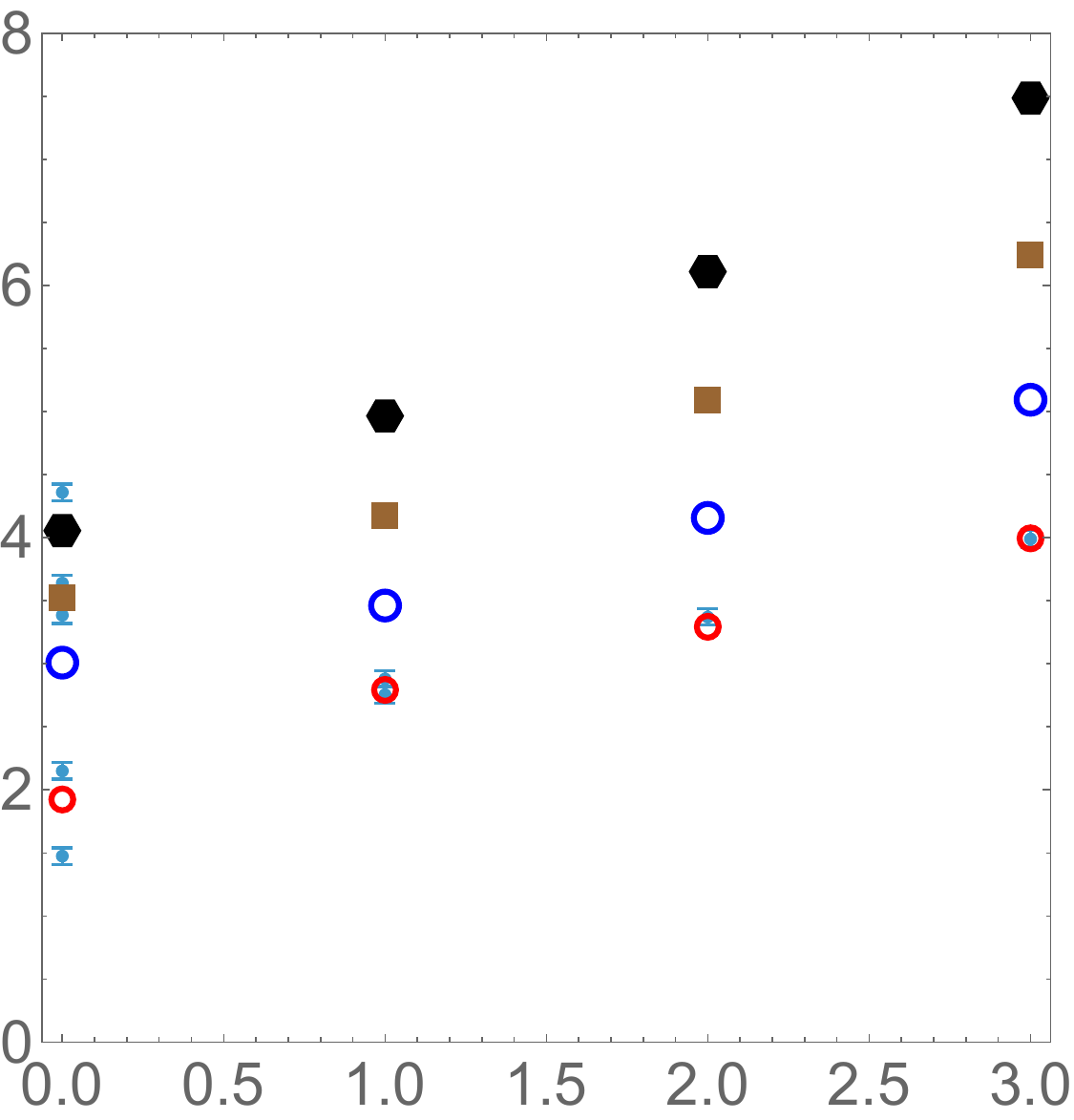}
    \caption{The colored points show the calculated energies $E_n$ (GeV) of the
    four lowest states $0^{++}$, $2^{++}$, $4^{++}$, and $6^{++}$ (from bottom
    to top) as a function of the principal quantum number $n$. The smaller
    points with error bars are from lattice simulations in pure SU(3) gauge
    theory \cite{Meyer:2004gx}.}
    \label{fig_gbpp}
\end{figure}

We begin with the simplest case, with total spin $S=0$, so that the total
and orbital angular momenta coincide, $J=L$. The calculated spectrum of
several $J^{PC}=J^{++}$ channels is shown in
Fig.~\ref{fig_gbpp}. At this stage only the Coulomb and confining
potentials are included, with no spin-dependent forces, as is commonly
done, for example, in charmonium studies. The model parameters are
chosen to reproduce the three higher states in the scalar channel,
which, unlike the lowest $n=0$ state, have normal hadronic sizes
(see the table of r.m.s.\ radii) and are therefore less sensitive to
short-distance forces. The most important parameter obtained in this
way is the effective gluon mass, fitted to be
\footnote{For comparison, the effective masses of the strange and charmed quarks
are $m_s \approx 0.5\,\mathrm{GeV}$ and $m_c \approx 1.5\,\mathrm{GeV}$,
respectively.}
\be
m_g = 0.90\,\mathrm{GeV}.
\ee

This implies a value of the parameter $\eta$ (the enhancement of the
gluon mass relative to the dilute ILM value due to instanton-antiinstanton
molecules) of $\eta = 6.25$. This is close to the value $\eta \approx 7$
used in \cite{Shuryak:2021fsu} to reproduce the central charmonium
potential. That value was originally motivated by {\em gradient flow
cooling} of lattice gauge configurations \cite{Athenodorou:2018jwu}.

\begin{figure}
    \centering
    \includegraphics[width=0.55\linewidth]{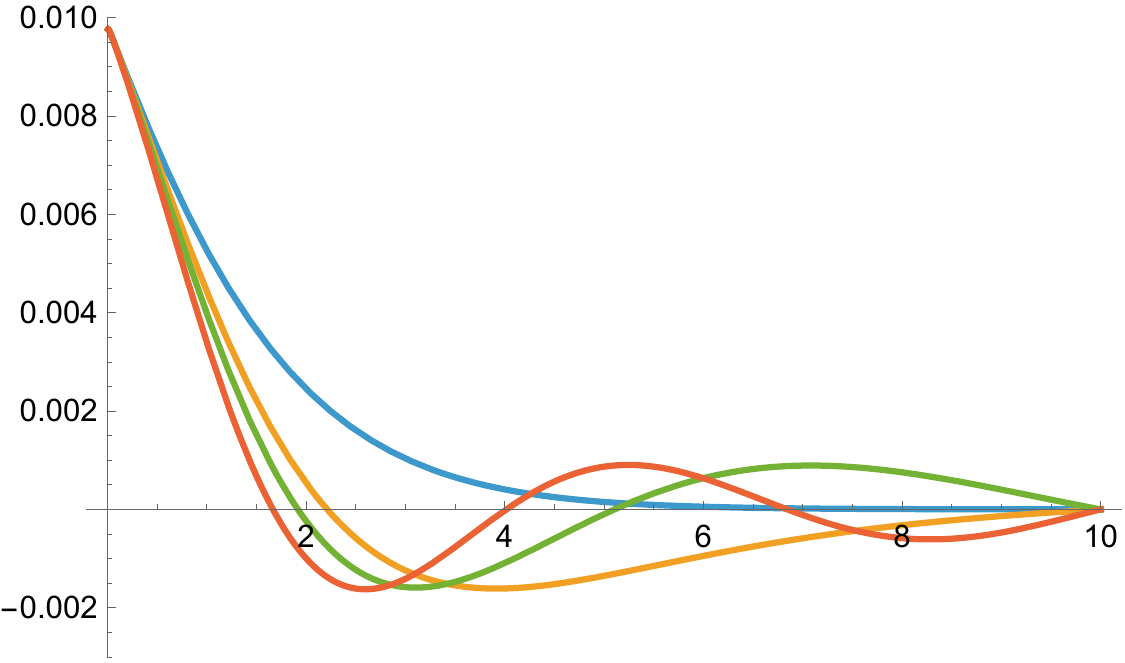}
    \caption{Wave functions (unnormalized) $\psi_n(r)$ for $n=0,1,2,3$ as
    functions of $r$ ($\mathrm{GeV}^{-1}$).}
    \label{fig_GB-func}
\end{figure}

To quantify the sizes and shapes of these states, we display the
corresponding wave functions in Fig.~\ref{fig_GB-func} and list the
root-mean-square radii,
\be
r_{\rm r.m.s.}
=
\left[
\frac{\int dr\,\psi_n^2(r)\,r^4}
     {\int dr\,\psi_n^2(r)\,r^2}
\right]^{1/2},
\ee
in Table~\ref{tab_0++}.

\begin{table}[h!]
    \centering
    \begin{tabular}{|c|c|c|c|c|} \hline
      $E_n$ (GeV) & 1.92 & 2.79 & 3.29 & 4.0 \\
      $r_{\rm r.m.s.}$ ($\mathrm{GeV}^{-1}$) & 2.03 & 5.0 & 6.26 & 6.25 \\ \hline
    \end{tabular}
    \caption{Energies and r.m.s.\ radii of scalar glueballs obtained from the
    Schrodinger equation with the parameters defined in the text.}
    \label{tab_0++}
\end{table}

Fixing the model parameters from the masses of the $normal$
$0^{++}$ states with $n=1,2,3$, which have sizes
$r_{\rm r.m.s.}\sim 5\,\mathrm{GeV}^{-1}\sim 1\,\mathrm{fm}$, we can then
consider other channels. The predicted masses for the higher-$J$ states
with $J=4,6$ turn out to be close to the lattice values. This is expected,
since the large centrifugal barrier increases their spatial extent and
further suppresses sensitivity to short-distance forces. When discussing
these states one should keep in mind that, unlike for mesons, the
confining potential for glueballs is expected to be screened at large
distances, rendering sufficiently large states unstable, formally
corresponding to complex energies. This issue is avoided here by placing
the system inside a spherical cavity: in our calculations we take a
radius $R = 10\,\mathrm{GeV}^{-1}\approx 2\,\mathrm{fm}$ and impose
Dirichlet boundary conditions.

\subsection{The lowest $0^{++}$ scalar glueball}

Finally, we turn to the lowest $0^{++}$ state with $n=0$. As discussed in
the Introduction, this state has a long history, with an r.m.s.\ radius
predicted in \cite{Schafer:1994fd} to be
$r_{\rm r.m.s.}(0^{++},n=0)\approx 1\,\mathrm{GeV}^{-1}\approx 0.2\,\mathrm{fm}$,
a result confirmed three decades later in \cite{Abbott:2025irb}. As is
clear from Table~\ref{tab_0++}, neither the calculated mass nor the size
of this state agrees with the lattice results. This discrepancy is not
unexpected, since the present calculation does not yet include
short-distance attractive spin-spin forces, whether perturbative or
instanton-induced. To estimate their effect, we add a local interaction
\be
V_{\rm local}(r)=G\,\delta^{3}(r).
\ee
After doing so, we find that reproducing Meyer's value
$M_{0^{++}}=1.475\,\mathrm{GeV}$ requires
\[
G \approx 38\,\mathrm{GeV}^{-2}.
\]
The corresponding r.m.s.\ radius is reduced, but only to
$r_{\rm r.m.s.}(0^{++},n=0)\approx 1.6\,\mathrm{GeV}^{-1}$, which is still
significantly larger than expected. We therefore conclude that the
lowest scalar glueball is unlikely to be described reliably within the
Schrodinger framework, and we will instead treat it below using a
relativistic Bethe-Salpeter approach.

\subsection{$0^{-+}$ glueballs}

The pseudoscalar glueball channel requires separate discussion, since its
quantum numbers forbid a direct continuation of the leading even-$J$
adjoint-string-like trajectories. In the constituent-gluon picture,
$C=+$ glueballs are composed of two transverse gluons bound by an adjoint
flux tube. Parity and charge conjugation are determined by the orbital
angular momentum $L$ and total gluon spin $S$ according to
\begin{equation}
P = (-1)^{L+1},
\qquad
C = (-1)^{L+S}.
\end{equation}
For $J=0$ and $C=+$, the pseudoscalar quantum numbers $0^{-+}$ require
$L=1$ and $S=1$, corresponding to a ${}^3P_0$ configuration. Thus, unlike
the scalar $0^{++}$ glueball, the $0^{-+}$ state is intrinsically an
orbital excitation and cannot lie on the same trajectory as the
$0^{++}$, $2^{++}$, and higher even-$J$ states.

According to \cite{Meyer:2004gx}, the masses of the two lowest $0^{-+}$
states are
\[
M_{0^{-+}}^{0}=2.250\,(60)(100),
\qquad
M_{0^{-+}}^{1}=3.370\,(150)(150).
\]
Solving the Schrodinger equation with $L=1$ and the shifted potential
$V_{\rm conf}+V_0$ (as done for the $0^{++}$ channel), we obtain
\[
M_{0^{-+}}^{0}=2.65,
\qquad
M_{0^{-+}}^{1}=3.13.
\]
Once again, the excited state is reproduced reasonably well, while the
lowest state shows a noticeable discrepancy
which
 should originate from short-distance effects. The
situation in this channel is quite subtle. The perturbative spin-spin
interaction in the $S_{\rm tot}=1$ channel is attractive, since
$(\vec S_1\!\cdot\!\vec S_2)=-1$, unlike the value
$(\vec S_1\!\cdot\!\vec S_2)=+1/2$ in $L=S=1$ charmonium. In contrast, the
instanton-induced interaction is repulsive (see, for example,
Fig.~\ref{fig_corr}). Moreover, for nonzero $L$ one has $\psi(0)=0$, so
the perturbative contribution depends sensitively on the smearing of the
delta function. For these reasons, we leave this issue unresolved.

\section{Relativistic treatment of short-distance dynamics}
\label{sec:relativistic}
\subsection{Emergent 4-gluon interaction}

At fixed pseudoparticle moduli, the low-mode fermion determinant generates the one-(anti)instanton induced vertices (for each instanton $I$ and anti-instanton $A$) dressed with perturbative gluons of the form~\cite{Liu:2024glue}
\begin{align}
\Theta_{I}
&=\prod_{f}\Bigg[
\frac{m_{f}}{4\pi^{2}\rho^{2}}
+i\,\psi_f^{\dagger}(x)\,U_{I}\,
\frac{1}{2}\Big(1+\frac{1}{4}\tau^{a}\,\bar\eta^{a}_{\mu\nu}\sigma_{\mu\nu}\Big)\,
U_{I}^{\dagger}\,\frac{1-\gamma_{5}}{2}\,\psi_f(x)
\Bigg]\,
\exp\!\Big[-\kappa\,\rho^{2}\,R^{ab}(U_{I})\,\bar\eta^{b}_{\mu\nu}\,G^{a}_{\mu\nu}(x)\Big],\nonumber\\
\Theta_{A}
&=\prod_{f}\Bigg[
\frac{m_{f}}{4\pi^{2}\rho^{2}}
+i\,\psi_f^{\dagger}(x)\,U_{A}\,
\frac{1}{2}\Big(1+\frac{1}{4}\tau^{a}\,\eta^{a}_{\mu\nu}\sigma_{\mu\nu}\Big)\,
U_{A}^{\dagger}\,\frac{1+\gamma_{5}}{2}\,\psi_f(x)
\Bigg]\,
\exp\!\Big[-\kappa\,\rho^{2}\,R^{ab}(U_{A})\,\eta^{b}_{\mu\nu}\,G^{a}_{\mu\nu}(x)\Big],
\label{eq:Theta45_rewrite}
\end{align}
in the local (zero size) approximation, with $\kappa\equiv 2\pi^{2}/g$, and 
$$R^{ab}(U)=\frac{1}{2}\Tr(\tau^{a}U\tau^{b}U^{\dagger})$$.
The exponential is the (LSZ-reduced) gluonic emission from the pseudoparticle, written compactly as a coupling to the field strength $G^a_{\mu\nu}$.


For the present purpose we isolate the purely gluonic source factor in \eqref{eq:Theta45_rewrite}, and restaure the
finite size form factor
\begin{eqnarray}
\mathcal{E}_I[G]=
\exp\!\left[
-\kappa\rho^2
\int\frac{d^4q}{(2\pi)^4}e^{iq\cdot x}
\beta_{2g}(\rho|q|)
R^{ab}(U)\bar\eta^b_{\mu\nu}G^a_{\mu\nu}(q)
\right],
\nonumber\\
\mathcal{E}_A[G]=
\exp\!\left[
-\kappa\rho^2
\int\frac{d^4q}{(2\pi)^4}e^{iq\cdot x}
\beta_{2g}(\rho|q|)
R^{ab}(U)\eta^b_{\mu\nu}G^a_{\mu\nu}(q)
\right].
\nonumber\\
\label{eq:E_def}
\end{eqnarray}
For a single (anti)-instanton of size $\rho$ centered at the origin, the field strength is
\begin{eqnarray}
    {\mathbb G}_{\mu\nu}^a(x)=\frac 4g\,\bar{\eta}^a_{\mu\nu}\frac{\rho^2}{(x^2+\rho^2)^2}
\end{eqnarray}
and the field-strength Fourier transform or form factor, is
\begin{equation}
{\mathbb G}_{\mu\nu}^a(q)=\frac{4\pi^2\rho^2}g\bar\eta_{\mu\nu}^a\,\beta_{2g}(t)
\equiv \frac{4\pi^2\rho^2}g\bar\eta_{\mu\nu}^a\,\bigg(\frac{t^2}{2}K_2(t)\bigg)\,.
\end{equation}
It is gauge invariant, satisfies $\beta_{2g}(0)=1$ and decays exponentially for $t\gg1$. An analogous expression holds for anti-instantons with $\bar\eta\to\eta$. The extra $\frac 12$ that appears in the exponential arises from the  coupling to the background field
\begin{eqnarray}
    \frac 14\int d^4x\,2\,{\mathbb G}_{\mu\nu}^a(x)G_{\mu\nu}^a(x)
    \rightarrow \frac 12
    G^a_{\mu\nu}(q)\int d^4x\,{\mathbb G}^a_{\mu\nu}(x)e^{-iq\cdot x}
\end{eqnarray}
where the rightmost equation follows by LSZ reduction of the perturbative gluon. Note that this reconstruction of the exponent is fully gauge invariant.

The instanton-induced four-gluon operator arises from the fourth-order term in the expansion of the tail-emission functional. Writing
\begin{equation}
X_I[G]=\kappa\rho^2\int\frac{d^4q}{(2\pi)^4}e^{iq\cdot x}
\beta_{2g}(\rho|q|)
R^{ab}(U)\bar\eta^b_{\mu\nu}G^a_{\mu\nu}(q),
\end{equation}
and similarly for $X_A[G]$, 
the quartic contribution is proportional to
\begin{equation}
n\left\langle \frac{X_I^4}{4!}+\frac{X_A^4}{4!}\right\rangle_U,
\end{equation}
where $n$ is the instanton density and $\langle\cdots\rangle_U$ denotes averaging over color orientations. 
Averaging over color orientation with the Haar measure projects onto singlets. For the adjoint dimension $d_A=N_c^2-1$, the fourth group average yields the standard contraction structure
\begin{equation}
\Big\langle R_{a_1 b_1}R_{a_2 b_2}R_{a_3 b_3}R_{a_4 b_4}\Big\rangle_U
\;\Rightarrow\;
\frac{3}{d_A(d_A+1)}\,
\Big(\delta_{a_1 a_2}\delta_{b_1 b_2}\Big)
\Big(\delta_{a_3 a_4}\delta_{b_3 b_4}\Big)
+\text{perm.},
\label{eq:R4_avg_tensor_new}
\end{equation}
To proceed, we recall the identity
\begin{equation}
\bar\eta^{\,b}_{\mu\nu}\bar\eta^{\,b}_{\rho\sigma}
=
\delta_{\mu\rho}\delta_{\nu\sigma}
-
\delta_{\mu\sigma}\delta_{\nu\rho}
-
\epsilon_{\mu\nu\rho\sigma},
\label{eq:eta_id_new}
\end{equation}
which yields to the contractions
\begin{equation}
\bar\eta^{\,b}_{\mu\nu}\bar\eta^{\,b}_{\rho\sigma}
G^{a}_{\mu\nu}G^{a}_{\rho\sigma}
=
2\,G^{a}_{\mu\nu}G^{a}_{\mu\nu}
-
2\,G^{a}_{\mu\nu}\tilde G^{a}_{\mu\nu}.
\label{eq:G2_decomp_new}
\end{equation}
Projecting onto the parity-even scalar channel retains only the $(G^a_{\mu\nu}G^a_{\mu\nu})^2$ part, leading the induced and non-local scalar four gluon interaction
\begin{equation}
\Delta\mathcal{L}_{4g}
=
n\,
\frac{\kappa^4\rho^8}{2\,d_A(d_A+1)}
\int\prod_{i=1}^4\frac{d^4q_i}{(2\pi)^4}\,
\beta_{2g}(\rho|q_i|)
(2\pi)^4\delta^{(4)}\!\left(\sum_i q_i\right)
\big[G^a_{\mu\nu}(q_1)G^a_{\mu\nu}(q_2)\big]
\big[G^c_{\rho\sigma}(q_3)G^c_{\rho\sigma}(q_4)\big]
+\text{perm.}
\label{eq:DeltaL4g_final_newx}
\end{equation}
Each external gluon momentum carries its own form factor $\beta_{2g}(\rho|q_i|)$. Note the leg-by-leg factorization which will be key for the bound state equation to follow. In the local approximation (zero size) the 4-gluon vertex is
\begin{equation}
\Delta{\cal L}^{(0^{++})}_{4g}(x)
=
n\,
\frac{\kappa^{4}\rho^{8}}{2\,d_A(d_A+1)}\,
\Big(G^{a}_{\mu\nu}(x)G^{a}_{\mu\nu}(x)\Big)^2.
\label{eq:DeltaL4g_final_new}
\end{equation}
Eq.~\eqref{eq:DeltaL4g_final_newx} is the microscopic one-(anti)instanton contact vertex in the $0^{++}$ channel that we now reduce to an instantaneous two-gluon potential.

\subsection{Emergent 2-gluon $0^{++}$-potential}
To extract the gluon two-body  potential in the scalar $0^{++}$ channel, we will use on-shell
in-out gluonic external states and the non-local operator \eqref{eq:DeltaL4g_final_newx}. This procedure parallels the one discussed by us~\cite{Shuryak:2021fsu} for the constituent quark pair interactions. 

For the  on-shell one-gluon states in the CM frame, we use the normalization
\begin{equation}
\langle \vec k,\lambda,a | \vec k',\lambda',a'\rangle
=
(2\pi)^3\,2\omega\,
\delta^{(3)}(\vec k-\vec k')\,
\delta_{\lambda\lambda'}\,\delta_{aa'}.
\label{eq:gluon_norm_new}
\end{equation}
For a physical transverse polarization vector $\epsilon^\mu(k,\lambda)$, the field-strength matrix element is
\begin{equation}
\langle 0|\,G^a_{\mu\nu}(0)\,|g^b(k,\lambda)\rangle
=
i\,\delta^{ab}\,
\big(k_\mu\epsilon_\nu(k,\lambda)-k_\nu\epsilon_\mu(k,\lambda)\big),
\label{eq:G_ext_new}
\end{equation}
and similarly for outgoing legs with $\epsilon\to\epsilon^{*}$. Define the gauge-invariant two-gluon contraction induced by $G^a_{\mu\nu}G^a_{\mu\nu}$,
\begin{equation}
\mathcal{F}(i,j)\equiv
G^a_{\mu\nu}(i)\,G^a_{\mu\nu}(j)
\;\Rightarrow\;
-\,2\,\delta^{a_i a_j}\Big[(k_i\!\cdot\!k_j)(\epsilon_i\!\cdot\!\epsilon_j)
-(k_i\!\cdot\!\epsilon_j)(k_j\!\cdot\!\epsilon_i)\Big].
\label{eq:Fij_new}
\end{equation}
At Born level, \eqref{eq:DeltaL4g_final_new} yields
\begin{align}
i\mathcal{M}^{(4)}(1,2\to3,4)
=
i\,n\,
\frac{\kappa^4\rho^8}{2\,d_A(d_A+1)}\prod_{i=1}^4\beta_{2g}(\rho|k_i|)
\Big[
\mathcal{F}(1,3)\mathcal{F}(2,4)
+
\mathcal{F}(1,4)\mathcal{F}(2,3)
\Big]\,.
\label{eq:M4_corrected}
\end{align}
Projecting onto the two-gluon color singlet state $|1\rangle_{\rm color}=\delta^{ab}|ab\rangle/\sqrt{d_A}$ sets the color factors in \eqref{eq:M4_corrected} to unity for both exchange structures.

The scalar $0^{++}$ projection is implemented by taking the $J_z=0$ helicity combination of two transverse gluons,
\begin{equation}
|0^{++}\rangle_{\rm pol}
=
\frac{1}{\sqrt{2}}
\Big(|\lambda_1{=}{+},\lambda_2{=}{-}\rangle
+
|\lambda_1{=}{-},\lambda_2{=}{+}\rangle\Big),
\label{eq:scalar_pol_state_new}
\end{equation}
and similarly for the outgoing state. In CM kinematics with scattering angle $\theta$ between $\vec k_1$ and $\vec k_3$, the detail contractions of the transverse helicities for massive on-shell gluons yield
\begin{equation}
\langle 0^{++}|\,
\Big[\mathcal{F}(1,3)\mathcal{F}(2,4)+\mathcal{F}(1,4)\mathcal{F}(2,3)\Big]
\,|0^{++}\rangle
=2\,(\omega^2+k^2)^2\,\Big(1+\cos^{2}\theta\Big),
\label{eq:pol_factor_theta_new}
\end{equation}

Using the Legendre polynomials,
\begin{equation}
1+\cos^2\theta
=
\frac{4}{3}\,P_0(\cos\theta)+\frac{2}{3}\,P_2(\cos\theta),
\label{eq:L_Pdecomp}
\end{equation}
the decomposition apparently shows both $J=0,2$. 
If we regard \eqref{eq:pol_factor_theta_new} as a helicity-0 object and performs a
spinless partial-wave projection,
\begin{equation}
\mathcal M^{(4)}_{J}
=
\frac{2J+1}{2}\int_{-1}^{1}dx\,
P_J(x)\,\mathcal M^{(4)}(x),
\qquad x=\cos\theta,
\label{eq:L_spinlessPW}
\end{equation}
then \eqref{eq:L_Pdecomp} implies the  weights and ratio
\begin{equation}
\mathcal K_{0^{++}}=\frac{4}{3},
\qquad
\mathcal K_{2^{++}}=\frac{2}{3}, \quad \frac{\mathcal K_{2^{++}}}{\mathcal K_{0^{++}}}=\frac{1}{2}.
\label{eq:K00_new}
\end{equation}
However, this ratio  reflects only the relative size of the $P_2$ and $P_0$ components in
\eqref{eq:L_Pdecomp}, with the latter  essentially an S-channel amplitude. While  $\mathcal K_{0^{++}}$ is the proper weight in the iteration  of the $0^{++}$, $\mathcal K_{2^{++}}$ does not carry the correct
weight in the $2^{++}$.  Although the emergent instanton  vertex contains a $P_2$ component with relative weight $1/2$ in the helicity-0 decomposition, the \emph{physical} $2^{++}$ interaction built from transverse gluons follows from   helicity projection.

With this in mind, the in-out legs form factors in the S-channel can be restaured as follows. In the glueball rest frame, the incoming legs carry relative momentum $\bm p$ and the outgoing legs carry $\bm k$. The four form factors combine as
\begin{eqnarray}
\prod_{i=1}^4\beta_{2g}(\rho|k_i|)
=\beta_{2g}^2(\rho p)\beta_{2g}^2(\rho k).
\end{eqnarray}
Since the exchange is taking place inside an instanton (anti-instanton) this is appropriately described by an {\it instantaneous} scalar potential, hence  
\begin{equation}
V(\bm p,\bm k)
=
-\,V_0\,
\beta_{2g}^2(\rho p)\,
\beta_{2g}^2(\rho k).
\end{equation}
which is separable.
The strength $V_0$ is obtained by combining the quartic coefficient, the scalar projection factor, and the normalization required to convert an invariant amplitude into a three-dimensional potential. The latter introduces reduced-state factors $(2\omega)^{-1/2}$ per external leg. Matching these factors at the dominant instanton scale $p,k\sim\rho^{-1}$ yields
\begin{equation}
\omega_\rho=\sqrt{\rho^{-2}+m_g^2},
\end{equation}
and produces a factor $(2\omega_\rho)^{-2}=1/(4\omega_\rho^2)$ in the effective coupling. Collecting all contributions gives
\bea
V_0
=
n\,
\frac{\kappa^4}{2\,d_A(d_A+1)}
\;\frac{16{\mathcal K}_{0^{++}}}{4\omega_\rho^2}\,
\rho^4.\\\nonumber
\eea

\subsection{Reduced Bethe-Salpeter equation}

Let $\Gamma(p;P)$ be the amputated Bethe-Salpeter (BS) vertex for two equal-mass constituent gluons of mass $m_g$, with total four-momentum $P$ and relative four-momentum $p$. In the ladder approximation with an instantaneous kernel $V(\bm p,\bm k)$, the reduced BS (Salpeter) equation for the scalar $0^{++}$ glueball is
\begin{equation}
\Gamma(p;P)
=
\int\frac{d^4k}{(2\pi)^4}\;
V(\bm p,\bm k)\;
D\!\left(k+\frac{P}{2}\right)\;
D\!\left(-k+\frac{P}{2}\right)\;
\Gamma(k;P),
\label{eq:BS_general}
\end{equation}
where $-iD=1/(q^2-m_g^2+i0)$ is the constituent-gluon propagator. .

In Eq.~\eqref{eq:BS_general} the interaction kernel is constructed using on-shell transverse gluons.
This reflects the instantaneous (Salpeter) reduction to follow: after integrating
over the relative energy, the dominant contributions arise from the poles of the gluon
propagators, and the bound state below threshold is governed by physical transverse
degrees of freedom. Off-shell and gauge components are usually implicitly encoded in the effective kernel of the Salpeter reduction, and do not affect the mass eigenvalue. In our case the instanton induced pair interaction is
gauge invariant by construction.

With this in mind and in the glueball rest frame
\begin{equation}
P^\mu=(M,\bm 0),\qquad p^\mu=(p_0,\bm p),\qquad k^\mu=(k_0,\bm k),
\end{equation}
we define the equal-time Salpeter amplitude
\begin{equation}
\phi(\bm p)\equiv \int\frac{dp_0}{2\pi}\;
D\!\left(p+\frac{P}{2}\right)\;
D\!\left(-p+\frac{P}{2}\right)\;
\Gamma(p;P).
\label{eq:Phi_def}
\end{equation}
Since $V(\bm p,\bm k)$ does not depend on $p_0$ or $k_0$, we can multiply both sides 
of \eqref{eq:BS_general}  by the product of propagators and integrate over $p_0$ as in \eqref{eq:Phi_def}. This yields
\begin{equation}
\phi(\bm p)
=
\int\frac{d^3k}{(2\pi)^3}\;
V(\bm p,\bm k)\;
\left[
\int\frac{dp_0}{2\pi}\,
D\!\left(p+\frac{P}{2}\right)\;
D\!\left(-p+\frac{P}{2}\right)
\right]\;
\phi(\bm k)
\label{eq:Phi_intermediate}
\end{equation}
The remaining $p_0$ integral is elementary,
\begin{equation}
\int\frac{dp_0}{2\pi}\,D\!\left(p+\frac{P}{2}\right)D\!\left(-p+\frac{P}{2}\right)
=\int\frac{dp_0}{2\pi}\,
\prod_{s=\pm}\frac{i}{(p_0+sM/2)^2-\omega_p^2+i0}
=\frac{1}{2\omega_p}\;
\frac{1}{M^2-4\omega_p^2+i0},
\end{equation}
 by contour integration, hence 
the reduced Bethe-Salpeter integral equation
\begin{equation}
\phi(\bm p)
=
\frac{1}{2\omega_p}\;
\frac{1}{M^2-4\omega_p^2}\;
\int\frac{d^3k}{(2\pi)^3}\;
V(\bm p,\bm k)\;
\phi(\bm k),
\label{eq:Salpeter_3D}
\end{equation}

\subsection{$0^{++}$ mass equation}
Substituting the separable kernel yields the rank-1 integral equation for the scalar glueball $0^{++}$ wavefunction in the CM frame
\begin{equation}
\phi(\bm p)=-
\,V_0
\frac{\beta_{2g}^2(\rho p)}{2\omega_p
(M^2-4\omega_p^2+i0)}
\int\frac{d^3k}{(2\pi)^3}
\beta_{2g}^2(\rho k)\phi(\bm k).
\end{equation}
Eliminating the normalization constant,  leads to the  root equation
\begin{equation}
1=
V_0
\,{\bf PP}\,\int_0^\infty\frac{p^2\,dp}{(2\pi)^2}
\frac{\beta_{2g}^4(\rho p)}{2\omega_p(4\omega_p^2-M^2)}.
\end{equation}
with the Principal Part (${\bf PP}$) retained for a bound state. 
Introducing $x=\rho p$, $\mu=m_g\rho$, and $\mathcal{M}=M\rho$ yields the dimensionless form
\begin{equation}
1=
\lambda_M\,{\bf PP}\,
\int_0^\infty dx
\frac{x^2}{\sqrt{x^2+\mu^2}}
\frac{\left[\frac{x^2}{2}K_2(x)\right]^4}{4(x^2+\mu^2)-\mathcal{M}^2},
\label{eq:root_mass}
\end{equation}
with
$$\lambda_M=\frac{V_0\rho^2}{(2\pi)^2}.$$
Eq.~\eqref{eq:root_mass} is the explicit rank-1 root mass equation for the $0^{++}$ glueball rescaled mass.

Finally, note that the bound-state mass is fixed by the pole condition of the two-body Green function,
$1-V_0\Pi(P^2)=0$, which is analytic below threshold. Hence, the Minkowski equation
evaluated at $P^2=M^2$ and the Euclidean equation evaluated at $P_E^2=-M^2$, yield
the same root mass \eqref{eq:root_mass} by analytical continuation, which in Euclidean signature reads
\begin{equation}
1=
\lambda_M
\int_0^\infty dx
\frac{x^2}{\sqrt{x^2+\mu^2}}
\frac{\left[\frac{x^2}{2}K_2(x)\right]^4}{4(x^2+\mu^2)+\mathcal{M}^2},
\label{eq:root_massE}
\end{equation}

\subsection{$0^{++}$ wavefunction}

In contrast, the reduced momentum-space wavefunctions in Minkowski and Euclidean space are different, yet related to each other. Indeed, 
in the equal-time (Minkowski) reduction the wavefunction in momentum space is explicitly  given by
\begin{equation}
\phi_M(\mathbf p)={\cal N}_M
\frac{\beta_{2g}^2(\rho p)}
{2\omega_p\,(M^2-4\omega_p^2+i0)} ,
\label{eq:wf_momentum_CM}
\end{equation}
In contrast, the Euclidean reduced Bethe-Salpeter amplitude is a 4-dimensional function of the form 
\begin{equation}
\chi_E(p_4,\mathbf p)
\;\sim\;
\frac{\beta_{2g}^2(\rho p)}
{\big[(p_4+\tfrac{P_4}{2})^2+\omega_p^2\big]
 \big[(p_4-\tfrac{P_4}{2})^2+\omega_p^2\big]},
\end{equation}
originating from the product of the two Euclidean propagators in the iterating kernel,
with no additional factor $1/(2\omega_p)$. The equal-time (Salpeter) wavefunction follows by slicing over the relative energy,
\begin{equation}
\phi_M(\mathbf p)
=
\int_{-\infty}^{\infty}\frac{dp_4}{2\pi}\,
\chi_E(p_4,\mathbf p)
\;\propto\;
\frac{\beta_{2g}^2(\rho p)}
{2\omega_p\,(M^2-4\omega_p^2+i0)} .
\end{equation}
The extra factor of $1/(2\omega_p)$ arises from the relative-energy integration, and is
absent in the 4-dimensional Euclidean amplitude.

\subsection{Numerical results}
\begin{figure}
    \centering
    \includegraphics[width=0.55\linewidth]{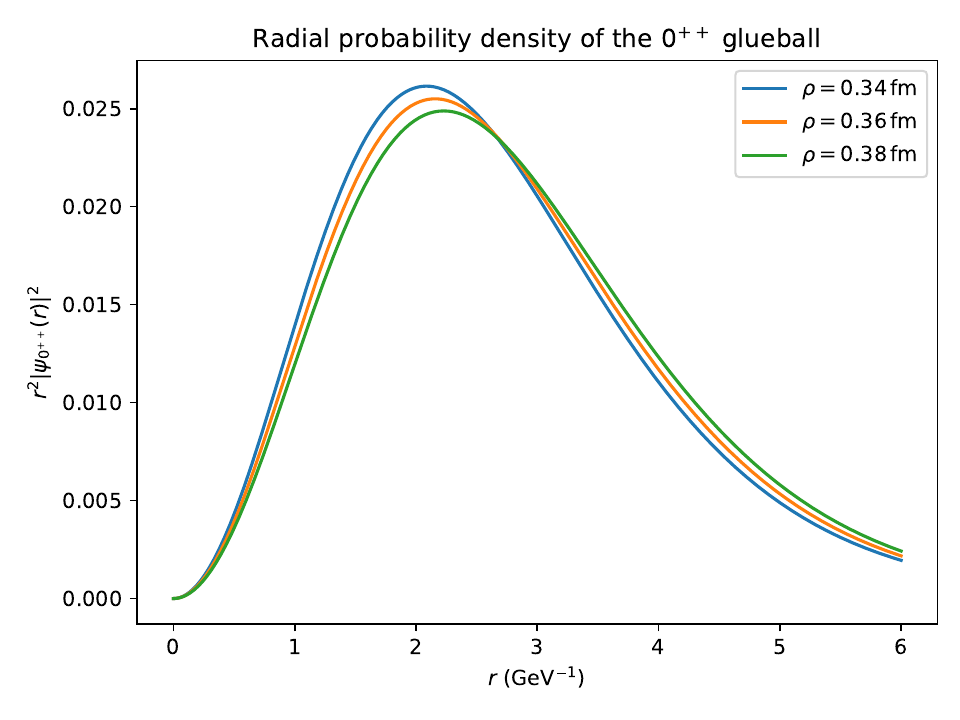}
    \caption{$0^{++}$ Glueball radial probability  $r^2|\psi_{0^{++}}(r)|^2$ versus $r(\rm GeV^{-1})$ from the reduced Bethe-Salpeter equation, for different (anti)instanton size $\rho=0.34, 0.36, 0.38$ fm.}
\label{fig:glueball_radial}
\end{figure}

For sufficiently strong (anti)instanton coupling $\lambda_M$ a bound state below 2-constitutive gluon treshold can form.
Since $\lambda_M\sim\rho^{\,p+2}$ with $p+2\gtrsim6$, this coupling is very sensitive to the (anti)instanton size $\rho$, e.g.
\begin{equation}
\left(\frac{\rho}{0.32~\mathrm{fm}}\right)^6
\sim 10\text{-}15
\end{equation}
for $\rho\simeq0.36\text{-}0.38~\mathrm{fm}.$
With this in mind, for a dense ILM, with
$\eta=6-7$ and
\begin{eqnarray}
m_g(\eta)=m_{g0}\sqrt{\eta},\quad m_{g0}=0.36~\mathrm{GeV},\quad
\alpha=\frac{g^2}{4\pi}=0.3\text{-}0.5,
\end{eqnarray}
 the scalar $0^{++}$ glueball mass is
\begin{equation}
M_{0^{++}} \;\simeq\; 1.4\text{-}1.5~\mathrm{GeV},
\end{equation}
well  below the two-gluon threshold
$2m_g(\eta)=1.8$-$1.9~\mathrm{GeV}$.
The $0^{++}$ rms radius follows from the exact momentum-space eigenstate
\eqref{eq:wf_momentum_CM}, using
\begin{equation}
\langle r^2\rangle
=
\int\frac{d^3p}{(2\pi)^3}\,
\phi_M^*(\mathbf p)
\left(-\nabla_{\mathbf p}^2\right)
\phi_M(\mathbf p).
\end{equation}
with the result
\begin{equation}
\langle r^2\rangle^{1/2}_{0^{++}}
\simeq
0.25\text{-}0.32~\mathrm{fm},
\end{equation}
This shows a low-lying and compact scalar glueball, dominated by instanton-scale dynamics, as illusttrated by the radial probability distribution shown in Fig.~\ref{fig:glueball_radial}

\section{Glueball summary}
\label{subsec:spectrum_PC}

The constituent two-gluon Hamiltonian organizes the glueball spectrum naturally
into radial and orbital families, closely mirroring the familiar structure of
quarkonium spectroscopy.
This separation becomes particularly transparent when the instanton density
parameter is fixed at $\eta=1$, or within the instanton liquid model (ILM),
where the scalar and tensor ground states are fitted to their lattice values.

Radial excitations correspond to states with fixed total spin and parity but an
increasing number of nodes in the relative wavefunction.
In the scalar channel, the ground state $0^{++}_0$ is strongly shifted downward by
the combined effects of Coulomb and instanton-induced attraction.
As a result, it is a compact state with a characteristic size of order the
instanton radius.
The first radial excitation, $0^{++}_1$, is far less sensitive to short-distance
dynamics: its wavefunction extends to larger radii, reducing its overlap with the
instanton core.
Consequently, its mass lies much closer to the confinement-dominated WKB
prediction.
This hierarchy illustrates the rapid decoupling of instanton physics with
increasing radial quantum number.

Orbital excitations, by contrast, form Regge-like towers at fixed radial quantum
number.
The tensor family $2^{++},4^{++},6^{++},\ldots$ provides the clearest example.
In this case, the centrifugal barrier already suppresses short-distance overlap at
the ground state, and ${}^5S_2$-${}^5D_2$ mixing further shifts probability density
toward larger radii.
As a result, the tensor ground state is only moderately displaced from the WKB
baseline, while higher-$J$ states align along an approximately linear Regge
trajectory with a slope determined by the adjoint string tension.

\begin{figure}[t]
\centering
\includegraphics[width=1.0\linewidth]{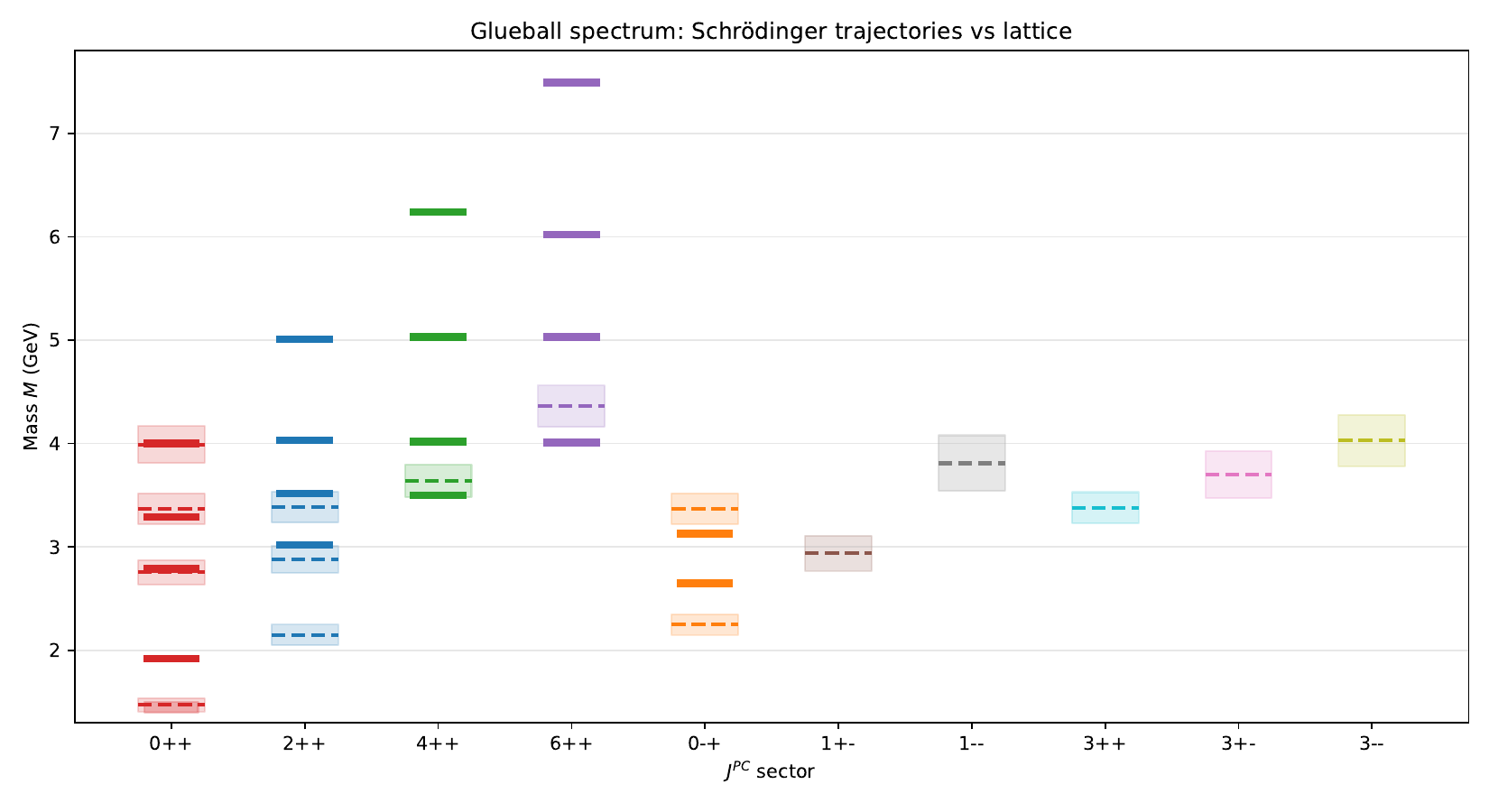}
\caption{
Glueball mass spectrum organized by $J^{PC}$ sectors, as listed in 
Table~\ref{tab:glueball_twocol_JPC}. Solid markers (this work) and
dashed (mean)  and spread (uncertainty) markers from quenched SU(3) lattice~\cite{Meyer:2004gx,Morningstar:1999rf}.
}
\label{fig:glueball_spectrum_PC}
\end{figure}

Figure~\ref{fig:glueball_spectrum_PC} displays the resulting glueball spectrum
organized by parity and charge conjugation, following the standard layout used in
lattice spectroscopy.
The comparison emphasizes the selective impact of instanton-induced dynamics.
In the $C=+$ scalar channel, coherent Coulomb and instanton attraction generate a
deeply bound and compact $0^{++}$ ground state, well below the confinement
baseline.
The tensor $2^{++}$ channel, on the other hand, exemplifies a Reggeized orbital
sequence.
The residual discrepancy in its lowest state arises from treating it purely as a
D-wave; in reality, it is better described as an S-wave dominated state with a
significant D-wave admixture, as discussed in the semiclassical analysis.
By contrast, vector channels appear only at substantially higher masses and do not
support compact two-gluon configurations, in agreement with symmetry constraints
and lattice results~\cite{Morningstar:1999rf,Morningstar:2025glueballreview}.


\begin{table}[t]
\centering
\begin{tabular}{|c|c|c|}
\hline
\bf $J^{PC}$ & \bf This work & \bf Lattice \cite{Meyer:2004gx,Morningstar:1999rf} \\
\hline
$0^{++}$ & 1.92 \quad (BS: 1.4-1.5) & 1.475(30)(65) \\
$0^{++}$ & 2.79  & 2.755(70)(120) \\
$0^{++}$ & 3.29  & 3.370(100)(150) \\
$0^{++}$ & 4.00  & 3.990(210)(180) \\
\hline
$2^{++}$ & 3.01  & 2.150(30)(100) \\
$2^{++}$ & 3.46  & 2.880(100)(130) \\
$2^{++}$ & 4.16  & 3.385(90)(150) \\
$2^{++}$ & 5.10  &  \\
\hline
$4^{++}$ & 3.52  & 3.640(90)(160) \\
$4^{++}$ & 4.18  &  \\
$4^{++}$ & 5.10  &  \\
$4^{++}$ & 6.24  &  \\
\hline
$6^{++}$ & 4.06  & 4.360(260)(200) \\
$6^{++}$ & 4.97  &  \\
$6^{++}$ & 6.11  &  \\
$6^{++}$ & 7.49  &  \\
\hline
$0^{-+}$ & 2.65  & 2.250(60)(100) \\
$0^{-+}$ & 3.13  & 3.370(150)(150) \\
\hline
$1^{+-}$ &     & 2.94(17) \\
$1^{}$ &     & 3.81(27) \\
\hline
$3^{++}$ &     & 3.38(15) \\
$3^{+-}$ &     & 3.70(23) \\
$3^{}$ &     & 4.03(25) \\
\hline
\end{tabular}
\caption{
Glueball masses in pure SU(3) gauge theory: comparison of this work as in Fig.~\ref{fig_gbpp}, with lattice results from Refs.~\cite{Morningstar:1999rf,Meyer:2004gx}.
}
\label{tab:glueball_twocol_JPC}
\end{table}

\begin{subappendices}
\section{WKB spectrum}
A complementary analytic description follows from the semiclassical quantization. For the linear potential $V(r)=\sigma_8 r$ with reduced mass $\mu=m_g/2$, the WKB spectrum with the Langer modification yields
\begin{eqnarray}
E_{nL}^{\rm WKB}(\eta)&=&
\left[\frac{3\pi}{2}\left(n+\frac{L}{2}+\frac{3}{4}\right)\right]^{2/3}
\frac{\sigma_8^{2/3}}{m_g^{1/3}(\eta)},
\nonumber\\
M_{nL}^{\rm WKB}(\eta)&=&2m_g(\eta)+E_{nL}^{\rm WKB}(\eta),
\label{eq:WKBspec}
\end{eqnarray}
which captures the universal scaling $$M-2m_g\propto \sigma_8^{2/3}m_g^{-1/3}$$.

A convenient WKB size measure is the outer turning point $r_+=E/\sigma_8$. For $L=0$, the WKB radial probability density implies closed expressions for moments. With $p(r)=\sqrt{2\mu(E-\sigma_8 r)}$ and 
$$|\psi|^2 d^3r\propto dr/p(r)\,,$$ we have
\begin{eqnarray}
\langle r\rangle_{L=0}^{\rm WKB}=\frac{2}{3}r_+,\quad
\langle r^2\rangle_{L=0}^{\rm WKB}=\frac{8}{15}r_+^2,\quad
r_{{\rm rms},\,L=0}^{\rm WKB}&=&\sqrt{\frac{8}{15}}\,r_+.
\label{eq:WKBrms}
\end{eqnarray}
For $L>0$ we quote $r_+$ and use $r_{\rm rms}\approx \sqrt{8/15}\,r_+$ as a uniform semiclassical estimate, which is adequate for trajectory comparisons.

Table~\ref{tab:wkb} provides the WKB masses and radii at $\eta=1$ for low-lying $L=0$ (scalar-like) and $L=2$ (tensor-like orbital) levels, including the first few excitations. The conversion $1~\mathrm{GeV}^{-1}=0.197~\mathrm{fm}$ is used.

\section{Gluon mass and density scaling}
\label{app:ME}

Musakhanov and Egamberdiev~\cite{Musakhanov:2017erp} derived the gluon polarization operator in the instanton medium and extracted a momentum-dependent dynamical mass. In their ILM setup the {\em scalar gluon mass} $M_s(q)$ is generated by rescattering on instantons, and the physical transverse gluon mass satisfies $M_g^2(q)=2M_s^2(q)$, implying at $q=0$ the standard-ILM value $M_g(0)\simeq 0.36~\mathrm{GeV}$ \cite{Musakhanov:2017erp}. The key scaling with density follows from the structure of the self-energy in a random instanton background: to leading order in the density expansion the polarization operator is proportional to the instanton density $n$, and the mass is obtained from the infrared limit of the propagator denominator, schematically $D^{-1}(q)\sim q^2+\Pi(q)$ with $\Pi(0)\propto n$. This implies $m_g^2\propto n$ and therefore
\begin{equation}
m_g(\eta)=m_{g0}\sqrt{\eta},
\qquad \eta\equiv \frac{n_{\rm eff}}{n_0},
\label{eq:mgScalingApp}
\end{equation}
where $n_0\simeq 1~\mathrm{fm}^{-4}$ is the traditional ILM density used to define $m_{g0}$. In the dense-ILM-ensemble~\cite{Shuryak:1982ilm}, $\eta$ is naturally taken as a flow- or resolution-dependent effective density, rather than a fixed vacuum number \cite{Schafer:1996wv}.


\begin{table}[ht]
\centering
\begin{tabular}{cccccc}
\hline\hline
$n$ & $L$ & $J$ & $M^{\rm WKB}_{nL}$ (GeV) & $r_+$ (fm) & $r^{\rm WKB}_{\rm rms}$ (fm) \\
\hline
0 & 0 & 0 & 3.196 & 0.628 & 0.459 \\
1 & 0 & 0 & 4.176 & 1.105 & 0.807 \\
2 & 0 & 0 & 4.974 & 1.493 & 1.090 \\
0 & 2 & 2 & 4.176 & 1.105 & 0.807 \\
1 & 2 & 2 & 4.974 & 1.493 & 1.090 \\
0 & 4 & 4 & 4.974 & 1.493 & 1.090 \\
0 & 6 & 6 & 5.680 & 1.836 & 1.341 \\
\hline\hline
\end{tabular}
\caption{WKB masses and radii at $\eta=7$ for the linear adjoint potential. Here $J$ is identified with $L$ for Regge-trajectory purposes. These semiclassical radii are larger than the variational scalar radius because Coulomb and instanton attractions are not included in the baseline WKB table; they can be incorporated as short-distance energy shifts.} 
\label{tab:wkb}
\end{table}

\section{Static potentials for fundamental and adjoint charges from instantons and $I\bar I$ molecules}
\label{sec_pote_numeric}

Wilson loops involve path-ordered exponents
\be
W_C = P \exp\!\left[i \sum \Delta x_\mu\, A_\mu^a\, T^a \right],
\ee
taken over closed contours $C$, usually of rectangular
shape. The sum runs over infinitesimal elements
$\Delta x_\mu$ along the loop, and $A_\mu^a$ are vacuum
gauge fields. Since the contour is closed, $W_C$ is
gauge invariant.

\begin{figure}[t!]
    \centering   \includegraphics[width=0.45\linewidth]{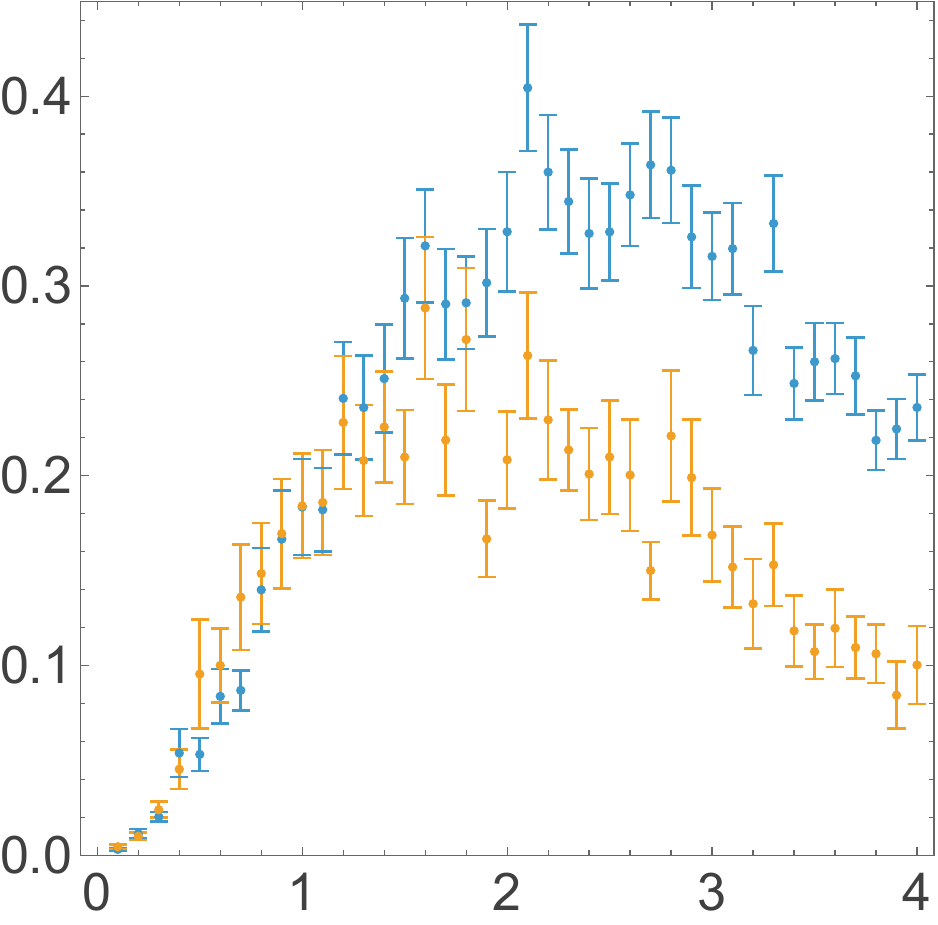}
    \includegraphics[width=0.45\linewidth]{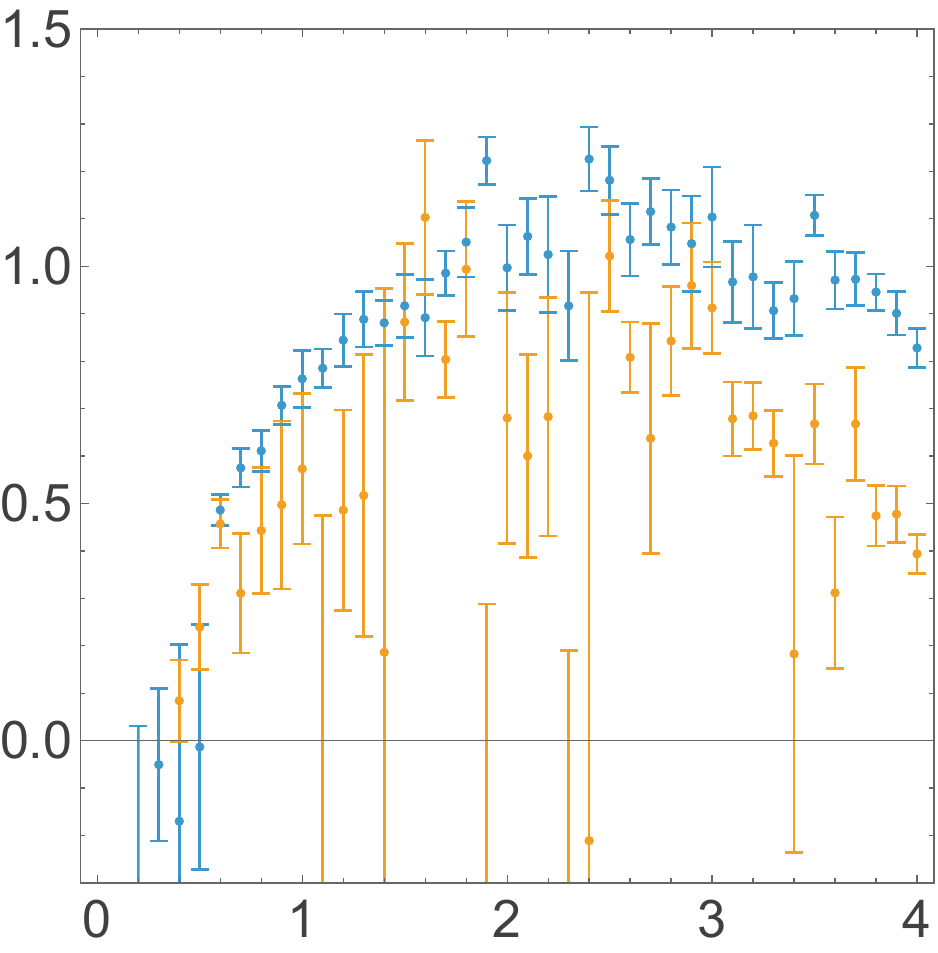}
    \caption{Static potentials $V_{conf}(r)\, (GeV)$ versus $r\, (GeV^{-1})$ calculated from fundamental (left) and adjoint (right)
    Wilson lines. Blue dots are for a  $\bar I I$ molecule, red are for a single instanton. Potentials still needs to be rescaled proportional to their density.}
\label{fig_W_fund_adj}
\end{figure}

For static color charges corresponding to quarks,
the color generators are in the fundamental
$SU(N_c)$ representation, $T^a=\lambda^a/2$, with
Pauli or Gell-Mann matrices. In several of our
earlier publications we have numerically calculated
the corresponding static confining potentials from
ensembles of instantons or $I\bar I$ molecules, and
applied them to quarkonia, baryons, and tetraquarks
\cite{Miesch:2023hjt}. The upper panel of Fig.~\ref{fig_W_fund_adj}
shows an example of such a Monte-Carlo simulation.
(The Wilson line locations and orientations are
randomized.)

Since a {\em constituent gluon} belongs to the adjoint
color representation, the only modification in the
Wilson loop is that the generators are now adjoint,
\be
(T^a)_{bc} = -i\,\epsilon^{a}_{\ bc}.
\ee
Unlike the case of Pauli matrices, for which a
compact closed form for the exponent exists, no such
expression is available here. The practical method
we employ is based on a Taylor expansion of the
exponent to second order in $\Delta x_\mu$, assumed
to be small. The adjoint potential shown in the lower
panel of Fig.~\ref{fig_W_fund_adj} was calculated in
this way, using the same gauge-field configurations
and the same set of Wilson lines as in the upper
panel.

First, note that the vertical scales of the two plots
are different, and that, crudely, the ratio of the
two potentials is approximately $9/4$, as suggested
by the ratio of the corresponding color Casimir
operators.

Second, since the Wilson lines and the resulting
potentials contain nonlinear terms (higher powers of
the gauge field), such scaling is not expected to
hold exactly. Indeed, the shapes of the two
potentials differ somewhat. Nevertheless, all
potentials reach a maximum at
$r \sim 2.5\,\mathrm{GeV}^{-1} \sim 0.5\,\mathrm{fm}$
and then decrease slightly. This feature is likely
an artifact of the setup used here, in which a single
instanton (or molecule) is surrounded by a Wilson
loop. In an ensemble of such objects, the potential
is expected to saturate at large $r$ to a constant,
equal to twice the effective mass of the static
charges.

For heavy quarks interacting via instantons, this
asymptotic value is
\be
V_{\rm conf}(r \rightarrow \infty)
\approx 2\,\eta \times (0.070\,\mathrm{GeV})
\approx 1\,\mathrm{GeV},
\ee
for $\eta \sim 7$, corresponding to a rescaling from
the dilute instanton liquid model to a dense ensemble
of molecules. In physical QCD such an energy is
sufficient to produce an additional $\bar q q$ pair
and split quarkonium $\bar Q Q$ into two $\bar Q q$
mesons.

For {\em constituent gluons} in pure gauge theory, the
large-distance splitting instead corresponds to
$gg \rightarrow (gg)(gg)$, so that
$V_{\rm conf}(r \rightarrow \infty)$ must be at least
of order $M_{0^{++}} \sim 1.5\,\mathrm{GeV}$. Guided by
these considerations, we model our potential in the
form shown in Fig.~\ref{fig_pot}.

\end{subappendices}

\chapter{Baryons}
\label{sec_baryons}
\section{Brief history}

Before starting our brief historical account, let us note that baryon spectroscopy has long been at the center of hadronic physics, both theoretically and experimentally. Consequently, there is no shortage of extensive reviews on the subject. One comprehensive recent source is the collection {\em 50 years of QCD} \cite{Gross:2022hyw}. The particular approximation we will employ is known as the $hypercentral$ approximation; for a comprehensive review of its use in baryon studies, see \cite{Giannini:2015dca}.

The nonrelativistic quark model originated in the 1960s. With the underlying spin-flavor $SU(6)$ symmetry, researchers were able to fix baryon wave functions and calculate important properties, such as magnetic moments. After the discovery of QCD, significant progress followed in the 1970s. Using quadratic (oscillatory) confinement and perturbative spin-dependent forces, Isgur and Karl derived the wave functions of positive-parity baryons \cite{Isgur:1978wd,Isgur:1979be} and of $P$-wave baryons \cite{Isgur:1978xj}, achieving considerable success for the lowest-lying states.

Further refinements, including in particular the introduction of relativized quarks, were made by Capstick and Isgur \cite{Capstick:1985xss}. With time, the issue of missing resonances faded away, as states in the second and third shells were nearly all observed, leading to the impression that most of the outstanding problems in this field had been resolved.

The extension of this framework to excited baryons, especially to the negative-parity $P$-shell \cite{Isgur:1978xj}, nevertheless revealed a number of conceptual and technical issues. To simplify some of the symmetry aspects for three-light-quark states $qqq$, the authors focused instead on mixed $qqs$ states ($\Sigma,\Lambda$), and then discussed the light-quark limit $m_s\rightarrow m_{u,d}$.

They showed that perturbative predictions based on one-gluon exchange worked qualitatively well for spin-spin and tensor forces, whereas the spin-orbit interaction, calculated within the same approximation, was essentially absent. A major shortcoming at the time was the prediction of $D$-shell states that were not observed experimentally, giving rise to the so-called {\em missing resonances problem.} As already mentioned, later refinements, particularly the inclusion of relativized quarks by Capstick and Isgur \cite{Capstick:1985xss}, and subsequent experimental progress largely alleviated this issue.

In our (mostly methodological) paper \cite{our_negative_parity}, we elaborated on several technical aspects of baryon spectroscopy, taking as an example excited nucleon states with negative parity. Some of the novel features of that work will be presented in this section.

Textbook discussions of baryons usually begin with the observation that, since the color wave function $\sim \epsilon_{ijk}$ is antisymmetric, Fermi statistics require the remaining part of the wave function-coordinates $\times$ spins $\times$ isospins-to be symmetric under quark interchange. Isgur and Karl \cite{Isgur:1978wd,Isgur:1979be} avoided constructing such fully symmetric wave functions explicitly for negative-parity baryons by focusing on strange baryons $\Lambda,\Sigma$, where 1-2 quark symmetry is enforced, followed by a limiting procedure $m_s\rightarrow m_d$. However, there is no fundamental reason to retrace this route repeatedly: the issue can be addressed directly, and the required symmetric wave functions can be constructed explicitly.


Mesons, as two-body systems, have two obvious coordinates: the relative distance $r_{12}=|\vec x_2-\vec x_1|$ and the center-of-mass (CM) position $\vec X_{CM}=M_1\vec x_1+M_2\vec x_2$. Only the former is physically relevant, while the latter is set to zero.

When dealing with few-body problems, one must proceed analogously and employ variables in which the center-of-mass motion is removed\footnote{If calculations are performed in momentum space, this corresponds to setting the total momentum to zero, $\vec P=0$.}. While this may appear to be an obvious first step, it is worth emphasizing that the complete elimination of center-of-mass motion is essential. The general procedure for an $N$-body system was developed by Jacobi in the mid-19th century, defining each new coordinate as the displacement between the center of mass of a subsystem and the next particle.

For a three-body system with equal masses, the standard Jacobi coordinates are
\bea
\vec\rho=(\vec r_1-\vec r_2)/\sqrt{2} , \,\,\,\,
\vec \lambda=(\vec r_1+\vec r_2-2\vec r_3)/\sqrt{6}
\eea
and have been traditionally used in baryon studies since the 1970s, for example in \cite{Isgur:1978xj}.

It is instructive to contrast this with alternative approaches in which the center-of-mass motion is not removed. In nuclear physics, for instance, oscillator-type mean-field potentials have often been employed. On the one hand, oscillator-based algebras of $a,a^+$ operators provide a convenient way to enforce Fermi statistics. On the other hand, the center-of-mass coordinate remains dynamical and introduces spurious degrees of freedom. While this may be relatively harmless for large particle numbers $N\gg1$, it becomes problematic for small systems. Attempts to {\em subtract center-of-mass effects (such as the CM energy)} are fundamentally flawed, since quantum entanglement between physical and spurious degrees of freedom persists and contaminates the calculation\footnote{This is an important distinction between our approach and that of J.~Vary and collaborators, who employed such methods even for baryons with $N=3$.}.

\begin{figure}[t!]
    \centering
    \includegraphics[width=0.25\linewidth]{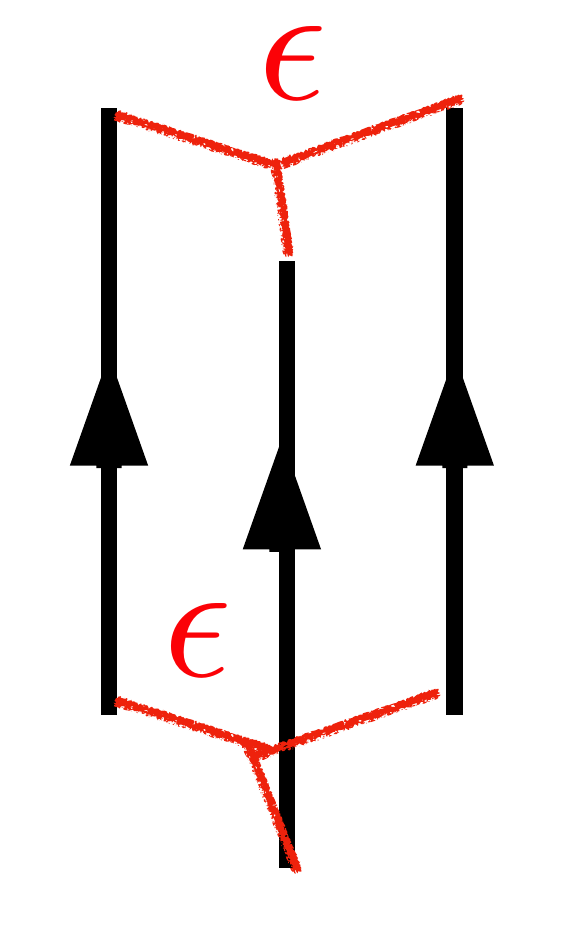}
    \includegraphics[width=0.55\linewidth]{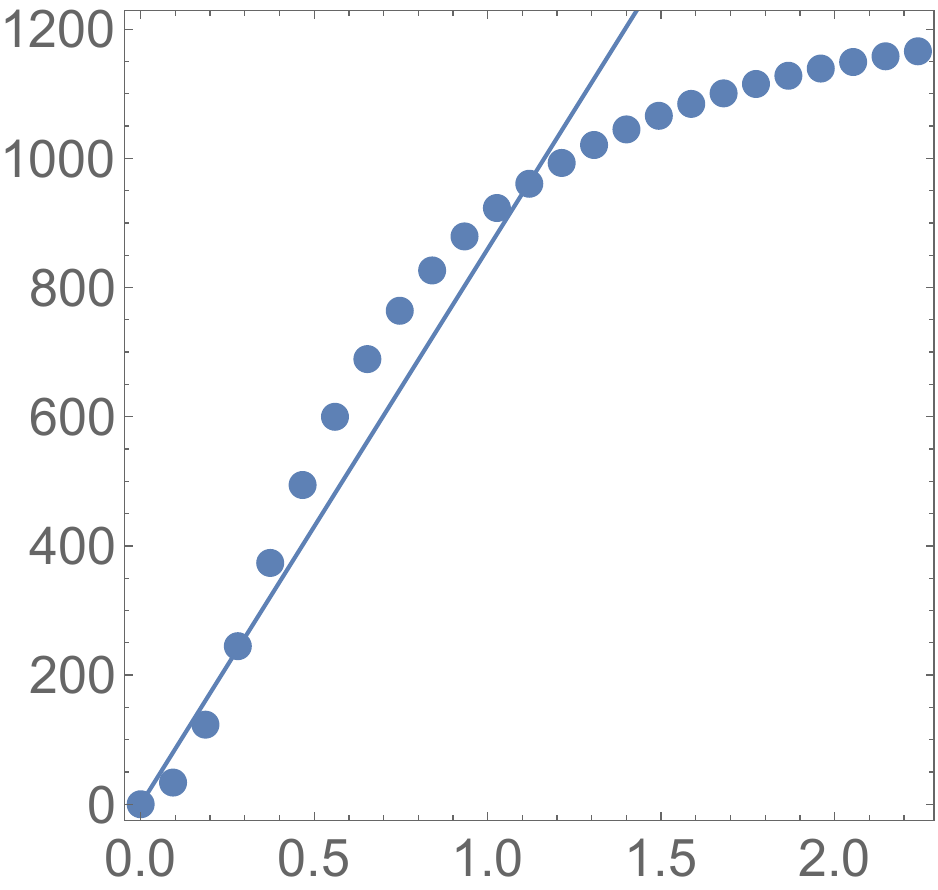}
    \caption{Left: a sketch of three Wilson lines connected with Levi-Civita symbols. 
    Right: Effective $QQ$ potential (MeV) versus distance (fm), calculated from instanton-antiinstanton molecules, from \protect\cite{Miesch:2023hjt} (points).
    The solid line shows the slope of the standard quarkonium potential for comparison.
    }
    \label{fig_WWW}
\end{figure}

\section{Interquark forces in baryons}
\label{sec_forces_baryons}

\subsection{The central $qq$ potential}

For static (infinitely heavy) quarks, interquark forces are described by Wilson lines, either without or with additional gauge-field insertions. The same logic applies to systems with any number of quarks (and antiquarks), where $N$ Wilson lines are contracted with the appropriate color indices of the color wave functions. For baryons, this structure is schematically shown in Fig.~\ref{fig_WWW} (left), with the red $\epsilon$ denoting the three-index Levi-Civita symbol.

The simplest case arises when the interaction is generated by gluon exchanges between Wilson lines. At lowest order, one readily finds that the binary $QQ$ Coulomb interaction is $2\pi\alpha_s/(3r)$, i.e.\ one half of the corresponding $Q\bar Q$ force \footnote{A natural explanation is geometric: the color vectors form angles of $2\pi/3=120^\circ$, with $\cos(2\pi/3)=1/2$.}.

Empirical observations discussed in \cite{Karliner:2016zzc} suggest that the same ratio,
$V_{QQ}/V_{Q\bar Q}=1/2$, persists also for nonperturbative effects. A fully nonperturbative derivation can be carried out on the lattice; as an example we mention \cite{deForcrand:2005vv}.

\subsection{Flavor dependence of spin forces}

As is well known, perturbative spin-spin forces are proportional to Dirac magnetic moments, which involve effective quark masses,
$$
\mu_q \sim e/m_q^{\rm eff}.
$$
This immediately implies that the magnetic moment of the strange quark is smaller than that of light quarks by the ratio $m_q^{\rm eff}/m_s^{\rm eff}$. Indeed, empirical $sq$ and $ss$ spin-spin forces follow this pattern, which has often been taken as evidence for the validity of the perturbative description.

An alternative perspective is provided by nonperturbative QCD vacuum models, such as instanton-based approaches. Here we recall the early work \cite{Shuryak:1988bf} on instanton-induced spin forces. It was observed that two-body forces arise from the reduction of the six-fermion 't~Hooft interaction by contracting two of its legs with vacuum condensates. For a light-quark pair $qq$, the resulting interaction is proportional to $\langle\bar s s\rangle$, whereas for a strange-quark pair it is proportional to $\langle\bar q q\rangle$. The ratio of these condensates coincides precisely with the ratio of effective constituent masses. As a result, both perturbative and instanton-induced spin-spin forces exhibit the same flavor dependence.

Instanton-induced spin forces were applied to baryon spectroscopy in \cite{Loring:2001ky}\footnote{This paper and subsequent works are commonly referred to as the {\em Bonn baryon model}. It is formulated with relativistic quarks and includes not only binary but also three-body forces.}.
Further contributions to this topic are presented in \cite{Miesch:2023hjt}, from which the right panel of Fig.~\ref{fig_WWW} is taken. A key conceptual issue addressed there is that, for $N$ Wilson lines piercing an instanton field, it is not {\em a priori} obvious that the resulting interaction can be reduced to a sum of binary forces. Nevertheless, for the baryon case ($N=3$), this reduction turns out to be valid; see the explicit derivation in \cite{Miesch:2023hjt}.

A second, more recently appreciated issue is that strongly correlated instanton-antiinstanton pairs ($molecules$) are more abundant and generate stronger effects than the dilute instanton ensemble of the original instanton liquid model. Their contributions to various Wilson-line correlators were calculated in the same work.

\section{The orbital-color-spin-flavor wave function and the permutation group $S_3$}
\label{sec_WF_S3}

The wave function of spherical ($L=0$) light-quark baryons contains color, spin, and flavor indices. The total number of monoms is
$$
N_{\rm monoms}=3^3\times 2^3\times 2^3,
$$
so the full wave function must be expressed as a linear combination of these basis elements with appropriate coefficients.

The simplest example is the $\Delta^{3/2,3/2}$ state. Its spin and isospin structure, e.g.\ $\uparrow\uparrow\uparrow$, is fully symmetric under particle permutations. Consequently, the color wave function must be antisymmetric, $\sim \epsilon_{c_1 c_2 c_3}$. In this case, the construction is straightforward.

Next come the nucleons, for which the color part factorizes in the same antisymmetric form. One must then construct a symmetric spin-isospin wave function. As an example, consider the proton with spin and isospin $S=1/2$, $I=1/2$, and projections $S_z=1/2$, $I_z=1/2$. A simple preliminary observation is that one of the three spins (and one of the three isospins) must be $down$, while the other two are $up$. One therefore expects $3\times3=9$ terms in the spin-isospin wave function. Since the total number of spin-isospin monoms is $2^3\times2^3=64$, the proton wave function occupies a nine-dimensional subspace of this 64-dimensional monom space.

In the literature, this wave function is often written in two equivalent forms: a symmetric representation and an explicit expansion in spin-flavor monoms,
\bea \label{eqn_N_monoms}
\bigg|0 \frac 12\frac 12 \frac 12 \bigg\rangle_{p_M^+}&=&{1 \over \sqrt{18}}\big[(\uparrow \downarrow \uparrow-\downarrow \uparrow \uparrow) (udu-duu) + (\uparrow \uparrow \downarrow-\uparrow \downarrow\uparrow)(uud-udu) + ( \uparrow \uparrow \downarrow-\downarrow \uparrow \uparrow)(uud-duu)\big] \nonumber  \\
&=& {1 \over \sqrt{18}}\big[ 2 (u^\uparrow d^\downarrow u^\uparrow)+2(d^\downarrow  u^\uparrow  u^\uparrow) +2(u^\uparrow  u^\uparrow d^\downarrow)-(d^\uparrow u^\downarrow u^\uparrow)-(u^\downarrow d^\uparrow u^\uparrow) \nonumber \\ &-&
(u^\uparrow d^\uparrow u^\downarrow) -(d^\uparrow u^\uparrow u^\downarrow)- (u^\downarrow u^\uparrow d^\uparrow)
-( u^\uparrow u^\downarrow d^\uparrow)
\big] 
\eea
After combining the terms that appear twice in the first expression, one indeed obtains nine distinct monoms.

To gain a deeper understanding of how this structure arises, let us start by adding three spin-$1/2$ representations:
$$
(1/2)\otimes(1/2)\otimes(1/2)=(3/2)\oplus(1/2)\oplus(1/2).
$$
Counting states on both sides confirms this decomposition,
$$
2\times2\times2=8=4+2+2.
$$
The key lesson is that there are {\em two} inequivalent ways to combine three spins into a total spin $S_{\rm total}=1/2$. When additional quantum numbers (orbital, color, flavor) are included, the number of possible combinations grows rapidly, and only a small subset is compatible with Fermi statistics.

A systematic tool for organizing these representations is provided by Young tableaux. Horizontal rows correspond to symmetrization, while vertical columns correspond to antisymmetrization. Although one might naively expect three possibilities for placing the lowest box, Young tableau rules exclude one of them, leaving precisely two allowed configurations-in agreement with the above counting. These two states form what we shall call a {\em good basis} for the three-spin problem.

We fix these two basis states as
\bea 
S_{\frac 12}^\rho=&&\frac 1{\sqrt{2}}(\uparrow\downarrow-\downarrow\uparrow)\uparrow\nonumber\\
S_{\frac 12}^\lambda=&&-\frac 1{\sqrt{6}}(\uparrow\downarrow\uparrow+\downarrow\uparrow\uparrow-2\uparrow\uparrow\downarrow)
\eea  \label{eqn_S12}
and define analogous isospin-$1/2$ states,
\bea
F_{\frac 12}^\rho=&&\frac 1{\sqrt{2}}(ud-du)u\nonumber\\
F_{\frac 12}^\lambda=&&-\frac 1{\sqrt{6}}(udu+duu-2uud)
\eea

The totally symmetric spin-$3/2$ states are
\bea \label{eqn_S_sym}
S_{\frac 32 m}^S= \bigg(\uparrow\uparrow\uparrow, \frac 1{\sqrt 3}(\uparrow\uparrow \downarrow + {\rm perm.}), 
\frac 1{\sqrt 3}(\uparrow\downarrow \downarrow + {\rm perm.}), \downarrow\downarrow\downarrow\bigg).
\eea

Note that these definitions parallel those of the Jacobi coordinates $\rho,\lambda$ for three-body systems. At this stage, we know that the proton wave function must be constructed as a specific combination of these components, and that it must be orthogonal to all $\Delta$ states built from the spin-$3/2$ wave functions.

To determine the correct combination, we use the fact that quarks are fermions, and therefore the full spin-isospin wave function must be symmetric under {\em all} permutations of quark labels. The relevant symmetry group is the permutation group of three elements, $S_3$, which consists of $3!=6$ elements,
\bea
\label{PERM}
\hat P_{i=1,\ldots,6}=I,\,(12),\,(13),\,(23),\,(123),\,(132).
\eea

Consider first the transposition $(12)$. The $\rho$-type wave functions are antisymmetric under $(12)$, while the $\lambda$-type are symmetric. Consequently, a fully symmetric spin-isospin wave function must be constructed as a combination of $S^\rho F^\rho$ and $S^\lambda F^\lambda$. The original four-dimensional space spanned by
$$
X^\rho X^\rho,\; X^\rho X^\lambda,\; X^\lambda X^\rho,\; X^\lambda X^\lambda
$$
is reduced to a two-dimensional subspace, since the mixed combinations $S^\rho F^\lambda$ and $S^\lambda F^\rho$ are antisymmetric under $(12)$ and must be discarded.

Remarkably, to enforce symmetry under {\em all} permutations, it is sufficient to check only one additional transposition, e.g.\ $(23)$. Using Jacobi coordinates, the action of $(12)$ and $(23)$ can be written in matrix form,
\bea
\label{eqn_perm_12}
{[\hat P_2=(12)]}
\begin{pmatrix}
\rho\\
\lambda
\end{pmatrix}
&=&
\begin{pmatrix}
-1&0\\
0 & 1
\end{pmatrix}
\begin{pmatrix}
\rho\\
\lambda
\end{pmatrix}\nonumber\\
{[\hat P_4=(23)]}
\begin{pmatrix}
\rho\\
\lambda
\end{pmatrix}
&=&
\begin{pmatrix}
\frac 12&\frac{\sqrt 3}2\\
\frac{\sqrt{3}}2& -\frac 12
\end{pmatrix}
\begin{pmatrix}
\rho\\
\lambda
\end{pmatrix}.
\ea  

The central idea is that spin and isospin indices must be combined using a tensor product. In {\sc Mathematica}, this is implemented by the command {\tt KroneckerProduct}, which combines two matrices of dimensions $M$ and $N$ into a single matrix of dimension $MN$. Applying this operation to the two $\hat P(23)$ matrices yields
\ba
\hat P(23)\bigotimes \hat P(23) = 
\begin{bmatrix} 
1/4 & \sqrt{3} /4&  \sqrt{3} /4& 3/4 \\
 \sqrt{3}/4 &   3/4 & -1/4 & - \sqrt{3}/4 \\ 
\sqrt{3/4}& -1/4&  3/4 & - \sqrt{3}/4 \\
   3/4  & - \sqrt{3/4}& - \sqrt{3}/4 & 1/4 
\end{bmatrix},
\ea
which acts in the four-dimensional {\em good basis} of spin-isospin space.

Diagonalizing this matrix (using the {\tt Eigensystem} command) yields three eigenvalues equal to $1$ and a single eigenvalue equal to $-1$. Performing the same analysis for $\hat P(12)\bigotimes \hat P(12)$, which is diagonal, gives two eigenvalues $+1$ and two eigenvalues $-1$.

The final step is to identify the {\em common eigenvector} with eigenvalue $+1$ for both operators. There is one and only one such vector, $(1,0,0,1)$, corresponding to
$$
\psi_{\rm sym}=(S^\rho I^\rho+S^\lambda I^\lambda)/\sqrt{2}.
$$
Thus, the symmetric proton spin-isospin wave function is
\bea \label{eqn_proton_wf}
\bigg|0 \frac 12\frac 12 m\bigg\rangle_{p_M^+}\sim \frac 1{\sqrt{2}}(S_{ \frac 12 m}^\rho F_{\frac 12}^\rho+S_{\frac 12 m}^\lambda F_{\frac 12}^\lambda).
\eea
The conversion of expressions of this type into vectors in the standard monom basis is explained in Appendix~\ref{sec_spin_tensor}.

The reader may feel that this derivation of a well-known result is unnecessarily complicated. However, following this procedure allows one to construct wave functions of multiquark hadrons that have eluded theorists for decades. In the problems discussed below, we will encounter matrices in the {\em good basis} with dimensions of hundreds or even thousands-large, but still manageable. The corresponding operators in the monom basis often have dimensions of millions by millions. As we will show, even in such cases one can perform practical calculations while preserving the familiar quantum-mechanical notation.

\section{Hyperspherical approximation for single-flavor baryons}

We now turn to the discussion of complicated wave functions of states containing
3, 4, 5, 6, or even 12 quarks, starting with states that possess the {\em highest symmetries}.
In the quantum mechanics of atoms, the discussion begins with the S shell, and
the principal quantum number $n$ is introduced. Only afterward are states with
nonzero orbital angular momentum $L\neq 0$ considered. Subsequently, the electron
spin $S$ is included and the total angular momentum $\vec J=\vec L+\vec S$ is defined.

Similarly, we will start with {\em spherically symmetric states}, for which there
exists a single (hyper-radial) wave function obtained from a single radial
Schrodinger equation.

Few-body systems composed of quarks with different flavors generally have
nontrivial kinetic energy operators, with coefficients depending on the
corresponding quark masses. However, if all quarks are identical, then in
modified Jacobi coordinates (see Appendix~\ref{sec_d_dimensions}) the kinetic
energy operator $\hat K$ takes a {\em spherically symmetric form of a Laplacian},
multiplied by the universal coefficient $1/2m$. After removing the motion of the
center of mass, one finds a six-dimensional hyperspace for baryons, a
nine-dimensional one for tetraquarks, and so on, reaching fifteen dimensions for
hexaquarks ($n=6$). The details for these cases are described in
Appendices~\ref{sec_d_dimensions} and~\ref{sec_jacobi_baryons}. 

The Laplacian in these higher dimensions differs from the familiar $d=3$ case
because of the appearance of a quasi-centrifugal term $\sim 1/mr^2$. Nevertheless,
its scaling behavior remains the same as that of the ordinary Laplacian. As we
have checked, this universality appears to persist across different systems.

The {\em hyperspherical approximation} has been successfully used in nuclear
physics since the 1970s. One of us (ES) has also applied this method to the $^4$He
wave function in Ref.~\cite{Shuryak:2019ikv}, not only reproducing the ground
$1S$-state binding energy $B\approx 28\,\mathrm{MeV}$, but also discovering a
second, shallow $2S$ bound state, which is known experimentally. This success
naturally increased our confidence in the method.

The next issue is that binary interactions $\sum_{i>j} V(\vec r_{ij})$ depend on
the interparticle distances $r_{ij}$ and therefore are {\em not} spherically
symmetric in hyperdimensional coordinates. Nevertheless, one can always define
their spherical projection by averaging over the solid angle (with 5, 8, etc.,
angular variables):
\be
V \rightarrow \bar V \equiv {1\over \Omega} \int V\, d\Omega \, .
\ee
The difference $\Delta V \equiv (V-\bar V)$ can affect a spherically symmetric
state only starting at second order, and is therefore expected to be small.

If the flavors of all three quarks are the same, the flavor wave function is
symmetric. Since the color wave function is necessarily antisymmetric, the
spin wave function must also be symmetric for an S-wave state, implying
$J=S=3/2$. Examples of such states are the $uuu=\Delta^{++}$ and
$sss=\Omega^-$ baryons. Unfortunately, analogous $ccc$ or $bbb$ baryons have not
yet been observed experimentally, so for these systems we can only make
predictions and compare them with lattice results.

The radial Schrodinger equation we use for the reduced wave function $u(R_6)$
is related to the full radial wave function by
\be
\psi(R_6)=u(R_6)/R_6^{5/2} \, .
\ee
Note that the factor $R_6^5$ is absorbed into the volume element, so that $u(R_6)$
has the usual one-dimensional normalization,
$\sim \int dR_6\, |u(R_6)|^2$. This reduction removes the first-derivative term
from the Laplacian, but introduces an additional quasi-centrifugal term. The
resulting equation has the form
\be
\label{RED6}
\bigg(-u''+{15\over 4 R_6^2 }u \bigg){1\over 2M}+(V_6-E_{Ns}) u=0 \, .
\ee
The radial projection of the potential onto $R_6$ is defined in
Appendix~\ref{sec_jacobi_baryons} and is worked out explicitly for a Cornell-type
binary potential.

\begin{figure}[h]
\begin{center}
\includegraphics[width=8cm]{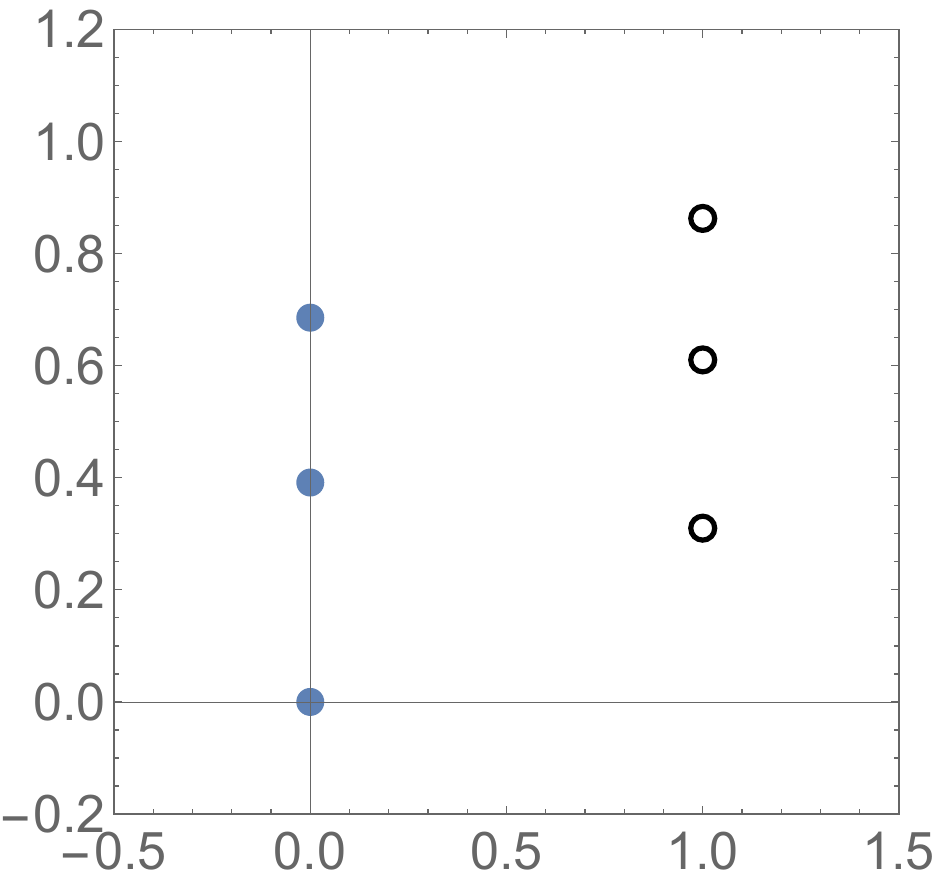}
\includegraphics[width=8cm]{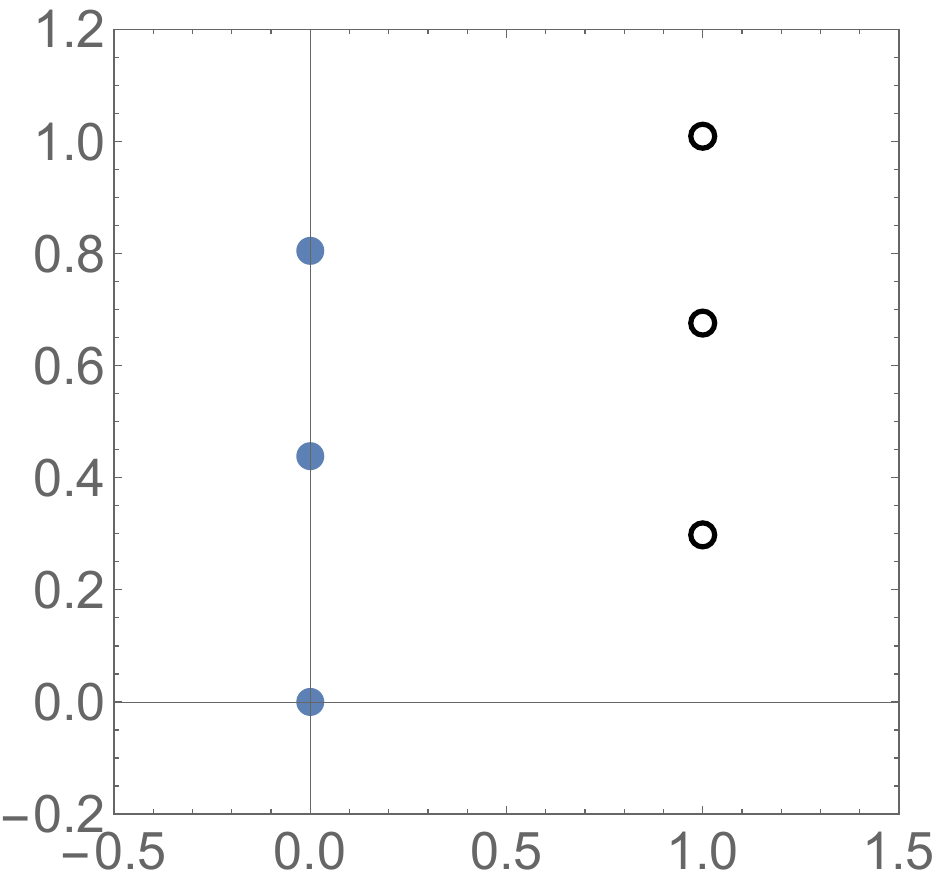}
\caption{Masses of $ccc$ baryons (upper plot) and $uuu=\Delta^{++}$ (lower plot)
$M(N,L)$ for $L=0$ and $L=1$ shells, measured relative to the corresponding ground
states.}
\label{fig_ccc}
\end{center}
\end{figure}

The mass splittings of the lowest s- and p-states for $ccc$ and $uuu$ baryons,
obtained numerically from Eq.~(\ref{RED6}), are shown in Fig.~\ref{fig_ccc}. The
binary $QQ$ potential is taken to be one half of the Cornell potential (used
above for the $\Upsilon$ family). We do not show the absolute masses, but instead
focus on the level splittings.

Experience shows that these splittings depend only weakly on the choice of quark
mass and are primarily controlled by the interaction potential. Indeed, as shown
in Fig.~\ref{fig_ccc}, the spectra of $ccc$ and $uuu$ baryons appear nearly
identical, despite the quark masses differing by more than a factor of three.
This mirrors the well-known similarity of splittings in the charmonium and
bottomonium families emphasized earlier, where the absolute mass scales also
differ by a comparable factor.

More specifically, for the s-shell $ccc$ baryons we find
\be
M_{ccc}^{2S}-M_{ccc}^{1S}\approx 391, \,\,M_{ccc}^{3S}-M_{ccc}^{1S}\approx  685 \,
\mathrm{MeV} \nonumber
\ee
while for the $uuu=\Delta^{++}$ states we obtain
\be
M_{uuu}^{2S}-M_{uuu}^{1S}\approx 438, \,\,M_{uuu}^{3S}-M_{uuu}^{1S}\approx 804 \,
\mathrm{MeV} \nonumber
\ee
These should be compared with the experimental values
\be
M_{uuu}^{2S}-M_{uuu}^{1S}\approx 278, \,\, M_{uuu}^{3S}-M_{uuu}^{1S}\approx 688 \,
\mathrm{MeV} \nonumber
\ee
(from the 2022 Particle Data Tables, using masses from Breit-Wigner fits).

The deviations between the calculated and experimental splittings are comparable
to the shifts expected from spin-spin and 't~Hooft interactions, which have so
far been neglected. However, their sign would be opposite, shifting the $1S$
state downward more strongly than the excited states. These effects are also
relativistic corrections of order $\sim 1/M^2$, and thus should be much less
important for charm quarks. We therefore conclude that the effective baryonic
potential used here (Ansatz~A) is somewhat stronger than in reality. We will
return to this point in the next section.

\begin{figure}[h]
\begin{center}
\includegraphics[width=8cm]{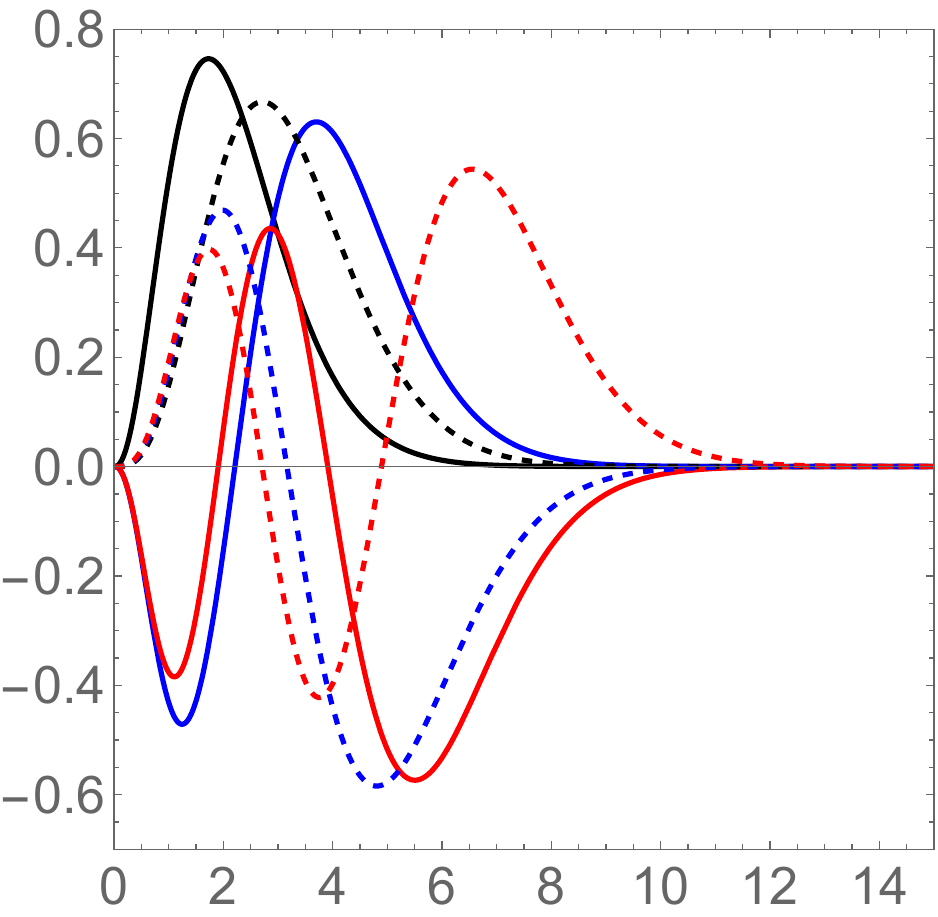}
\includegraphics[width=8cm]{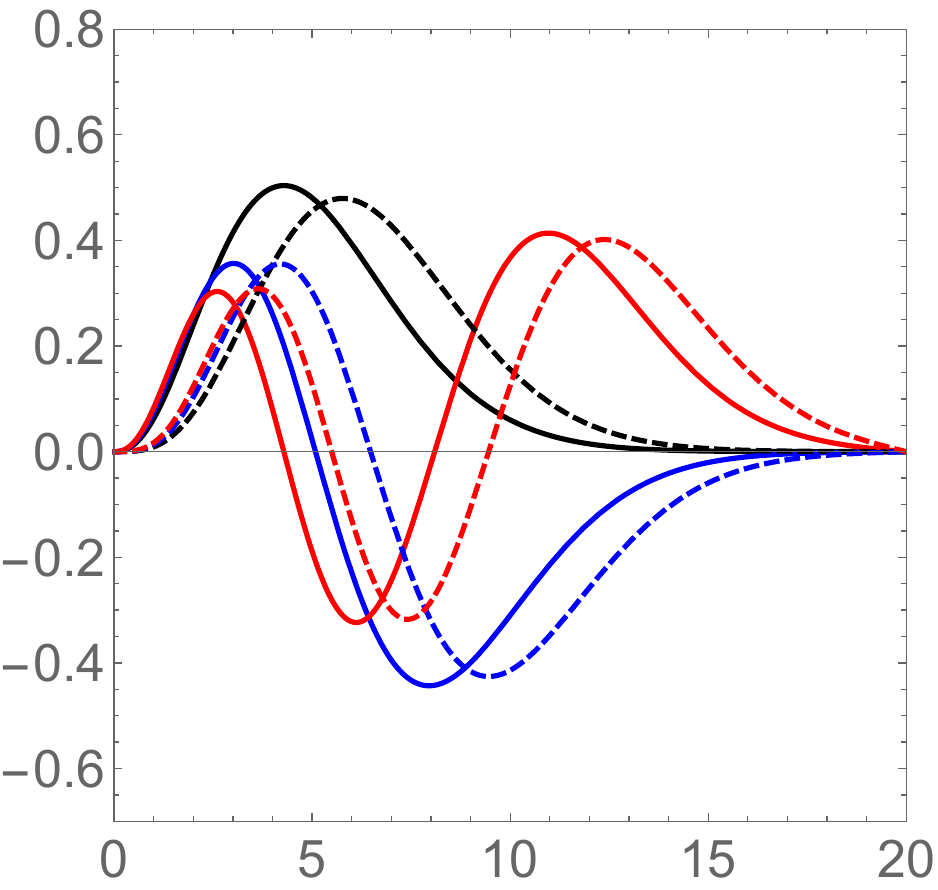}
\caption{Wave functions $u(R_6)$ of $ccc$ (upper) and $uuu$ baryons (lower) for
$L=0$ (solid lines) and $L=1$ (dashed lines), shown as functions of the
hyper-radial distance $R_6\,(\mathrm{GeV}^{-1})$.}
\label{fig_ccc_wfs}
\end{center}
\end{figure}

The corresponding wave functions are shown in Fig.~\ref{fig_ccc_wfs}. Unlike the
splittings, they exhibit dramatic differences between the $uuu$ and $ccc$
baryons. This reflects the fact that these hadrons have very different spatial
sizes, with the $ccc$ baryons being much more compact. Moreover, even the ground
states display somewhat unusual behavior: despite the enhanced Coulomb forces in
six-dimensional systems, the quasi-centrifugal potential leads to wave functions
that are strongly suppressed at small hyper-radial distances.

These wave functions can be used to compute matrix elements of various operators.
For example, the root-mean-square radii of the three s-states of the $\Delta$ are
\be
\sqrt{\langle R_6^2 \rangle}=4.97,\,7.53,\,9.72\, \mathrm{GeV}^{-1} \, .
\ee

To summarize, in this section we have tested the assumption that the approximate
universality of spectra observed for S- and P-shell states of $c$ and $b$
quarkonia can also extend to baryons. Our results indicate that this is indeed
the case.

Specifically, we modeled the nonperturbative forces between quarks in baryons as
a sum of Cornell-style binary interactions, reduced by the same factor of $1/2$
(as for perturbative forces) relative to the mesonic $\bar Q Q$ potential. We
find that this leads to s-p shell splittings that are somewhat larger than those
observed in the $uuu=\Delta^{++}$ baryon, indicating that the effective baryonic
potential is slightly too strong.


\section{Electric, magnetic and transitional formfactors}
\label{sec_FF}

\subsection{Brief history}

The most direct observables related to the spin structure are the hadronic
magnetic moments. Already in the 1960s it was shown that the proton and
neutron magnetic moments are reasonably well described by nonrelativistic
$qqq$ models, without any additional admixtures or orbital motion. If admixtures of states with
$L\neq 0$ in excited $qqq$ configurations, or of pentaquark $qqqq\bar q$ sectors,
are included, one must examine whether mechanisms exist that lead to
cancellations between spin and orbital contributions.

The next class of sensitive observables are transitional matrix elements
between on-shell states of relevant operators (such as electromagnetic currents),
either at zero or finite momentum transfer $Q$. In the latter case, the
corresponding quantities are referred to as (transitional) form factors. The
diagonal $pp$ and $nn$ form factors are further decomposed into electric and
magnetic Sachs form factors. The electromagnetic current itself contains
contributions proportional to the quark charges (Dirac term) and to the quark
spins (Pauli term). There is no need to reproduce the well-known standard
formulae here.

For many decades it was widely accepted that the ratio of electric to magnetic
form factors is independent of the momentum transfer $Q$. This belief was based
on the assumption that the spin and radial parts of the nucleon wave function
factorize. In addition, it was expected that the neutron electric form factor
vanishes, $G_E^n=0$.

However, experiments at JLAB
\cite{JeffersonLabHallA:1999epl,JeffersonLabHallA:2001qqe,Puckett:2011xg}
have demonstrated that neither of these expectations is supported by data. The
ratio $G_E^p/G_M^p$ decreases with increasing $Q$, with a possible sign change at
the upper end of the measured range, $Q^2\sim 8.5\,\rm GeV^2$. Moreover, the
electric form factor of the neutron is clearly nonzero, $G_E^n\neq 0$.

Before turning to theory, it is worth noting that the JLAB data lie in the range
$Q^2\leq 9\,\rm GeV^2$, corresponding to $Q/2\approx 1.5\,\rm GeV$. Thus, these
experiments probe neither the small-$Q$ regime, where a soft chiral cloud is
expected to dominate, nor the asymptotically large-$Q$ region, where hard
perturbative predictions should apply. Instead, they belong to an intermediate,
or {\em semi-hard}, regime in which form factors must be computed
nonperturbatively\footnote{See, for example, the discussion and calculations in
\cite{Shuryak:2020ktq}.}.

Clearly, any mixing between states will affect the form factors. As one example
from the literature, the mixing of ground-state nucleons (proton and neutron)
with D-shell states was studied in Ref.~\cite{Simonov:2020wql}. In that work, a
schematic model was shown to reproduce the JLAB data for the electric and magnetic
form factors of both the proton and neutron, but only at the price of rather
strong and specific mixing. This study inspired our work
\cite{Miesch:2025vas}, in which we analyzed the D-shell states, their mixing, and
their possible admixture into the nucleon wave function in detail. Some of these
results will be presented below in the chapter on the D-shell. Our analysis
indicates that the admixture of D-wave components can indeed be large enough to
explain the observed behavior of the electric form factors.

\subsection{Theoretical approaches and some results}

Before evaluating form factors explicitly, let us comment on several general
issues. At medium and large momentum transfer, $Q\sim{\rm few}\,\rm GeV >
M_N$, relativistic effects are clearly essential. It is therefore natural to
formulate the problem using light-front (LF) dynamics, as discussed in the later
parts of this review. At the very least, one should attempt to boost wave
functions from the center-of-mass frame to momenta of order $Q/2$. Examples for
the nucleon and the $\Delta$ are given in Ref.~\cite{Shuryak:2022wtk}, where the
corresponding form factors are also analyzed.

At small $Q$, one might naively attempt to use Fourier transforms of
nonrelativistic center-of-mass wave functions, as is common in atomic physics.
However, this approach fails. One reason is that constituent quarks, unlike
electrons, are not pointlike. In addition, baryon radii are affected by their
{\em chiral clouds.}

At intermediate $Q$, nonrelativistic wave functions may still be employed,
provided a  {\em boost prescription} is used (following e.g.
\cite{Simonov:2020wql}),
$$
\psi(\vec p_\perp, p_{l}) \rightarrow  {1\over \gamma}\,
\psi(\vec p_\perp, p_{l}/\gamma),
$$
which amounts to a Lorentz contraction of the longitudinal size by the gamma
factor $1/\gamma=\sqrt{1-v^2}$. Below, we will discuss to what extent this
procedure can be used to describe form factors in the intermediate
"semi-hard'' region.

In addition to elastic observables, {\em transitional form factors} from the
nucleon to $N^*$ and $\Delta$ resonances provide a rich set of probes of internal
structure. In particular, the Roper resonance $N^*(1440)$ exhibits a transitional
form factor that is in clear disagreement with a {\em minimal scenario} in which
it is interpreted as a simple radial $2S$ excitation. This provides further
motivation for future studies.

Having completed this general discussion, we now turn to explicit calculations.
In the nonrelativistic limit, form factors are evaluated directly from the wave
functions. More specifically, in the commonly used Breit frame\footnote{Also
known as the {\em brick-wall frame}, in which the momentum reverses sign after the
collision.}, elastic scattering corresponds to a transition from an initial
hadron with momentum $\vec Q/2$ to a final hadron with momentum $-\vec Q/2$ along
the $z$ axis, induced by the electromagnetic current operator
$\langle j|J_\mu^{em}|i\rangle$.

The first step in evaluating form factors is the computation of Fourier
transforms of the wave functions, followed by their convolution. In the
nonrelativistic approximation, the electric form factor can be inverted to
obtain the charge distribution, which in turn determines the wave function and
the underlying Schrodinger potential\footnote{For details, see, for example,
\cite{Giannini:2015dca}.}. The resulting potential is finite at infinity,
suggesting an apparent "no confinement'' behavior. In particular, the
hypercoulomb potential leads to an analytic form factor of the form
$\sim 1/(1+Q^2/\Lambda^2)^{7/2}$.

We found this situation rather confusing and therefore repeated these
calculations ourselves. The outcome was unambiguous: the form factors obtained
within the hypercentral approach do {\em not} agree\footnote{Recall, however,
that this model reproduces the energy spectrum quite well.} with experimental
data.

\begin{figure}
    \centering
    \includegraphics[width=0.45\linewidth]{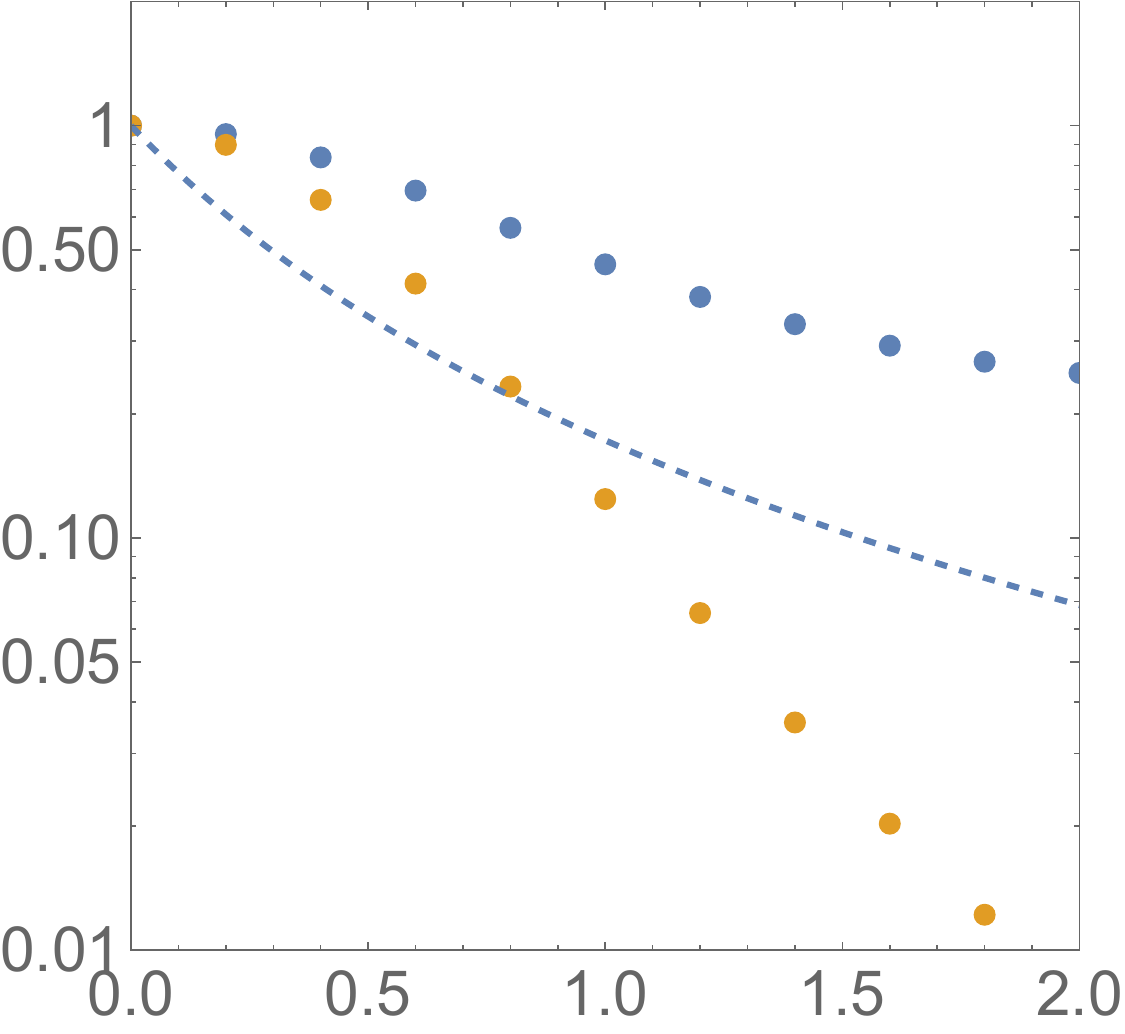}
    \includegraphics[width=0.45\linewidth]{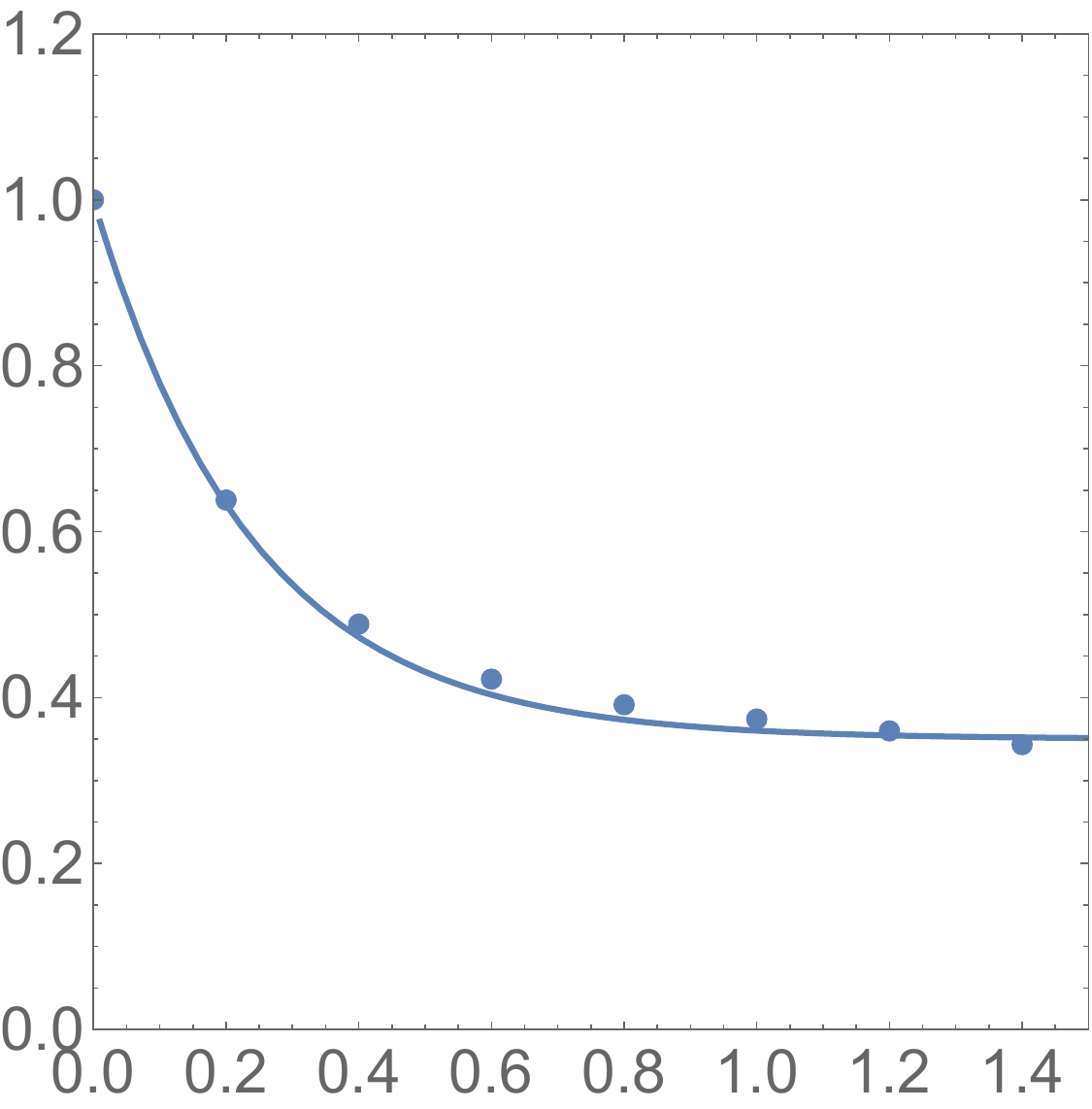}
    \caption{The left plot shows the ${\frac 12}^+$ form factor obtained from the
    1S (blue dots) and 1D (orange dots) wave functions as a function of
    $Q^2\,(\rm GeV^2)$. The dashed blue line represents the nucleon dipole fit
    $FF_{dipole}=1/(1+Q^2/0.71)^2$.
    The right plot shows the constituent quark or {\em pion cloud} form factor
    versus $Q^2\,(\rm GeV^2)$. The points correspond to the ratio of the
    calculated form factor to the fitted experimental curve in the left panel.
    The solid line is $0.35+0.65\,\exp(-Q^2R_{\rm cloud}^2/6)$ with
    $R_{\rm cloud}=1\,\rm fm\approx 5\,\rm GeV^{-1}$.
    }
    \label{fig_FF_1S1D}
\end{figure}

As noted earlier, relativistic effects can be partially included through Lorentz
contraction in the Fourier transforms. In the upper panel of
Fig.~\ref{fig_FF_1S1D} we show the form factors calculated using the 1S (blue dots)
and 1D (orange dots) wave functions obtained in the hyperdistance approximation.
As expected, the 1D state is significantly softer than the 1S state, since the
1S wave function is finite at the origin, while the 1D wave function is
quadratically suppressed by the centrifugal barrier.

These results should be compared with the empirical dipole fit to the proton
magnetic form factor (dashed line). Clearly, the 1S form factor obtained from the
hypercentral wave function (blue dots) fails to reproduce the data, even at small
$Q^2$. Nevertheless, its behavior at larger $Q^2\gtrsim 0.5\,\rm GeV^2$ is
qualitatively similar.

The fact that hypercentral wave functions significantly underestimate the
nucleon radius has been noted previously; see, for example,
\cite{Gallimore:2024fcz}. Indeed, standard radial wave functions are expected to
describe the spatial distributions of constituent quark {\em centers}, rather
than the distributions of charge, mass, or energy. We now briefly discuss two
ideas that may help explain this discrepancy.

First, constituent quarks are not pointlike objects; they possess intrinsic
spatial extent and internal charge and mass distributions, which introduce
additional form factors.

Second, the three-quark ground-state wave function represents only part of the
full nucleon wave function. There is mixing with excited states (to be discussed
in the chapter on the D-shell) and with higher Fock components (to be discussed
in the chapter on pentaquarks). In a different language, the nucleon is
surrounded by a pion cloud, as described, for example, in the {\em little bag
model } \cite{Brown:1979ui} or in {\em chiral soliton} models
\cite{Diakonov:1995qy}.

To make this discussion more quantitative, let us assume that the experimental
form factor can be written as the {\em product} of two contributions: one
associated with the global wave function described by the hyperradius, and one
associated with the constituent quark structure,
\be
\label{TIMES}
FF_{N}(Q^2)=FF_{\rm WF}(Q^2)\cdot FF_{\rm cloud}(Q^2).
\ee
The right panel of Fig.~\ref{fig_FF_1S1D} shows the extracted form factor
$FF_{\rm cloud}(Q^2)$. A pointlike contribution from the constituent quarks is
present, accounting for approximately one third of the total form factor.

At small momentum transfer, this contribution can be parametrized by a Gaussian
$\exp(-Q^2R^2/6)$ with a large radius
$R\approx 5\,\rm GeV^{-1}\approx 1\,fm$. Such a form factor is well described by
instanton-based semiclassical theories of chiral symmetry breaking. In order to
generate a nonzero quark condensate, virtual quarks bound to an instanton (zero
modes) must propagate to neighboring (anti)instantons in the ensemble, which
are typically separated by distances of order
$R\sim n_{\rm inst}^{-1/4}\sim 1\,\rm fm$. In other words, constituent quarks
carry a substantial chiral cloud.

\begin{figure}
    \centering
    \includegraphics[width=0.65\linewidth]{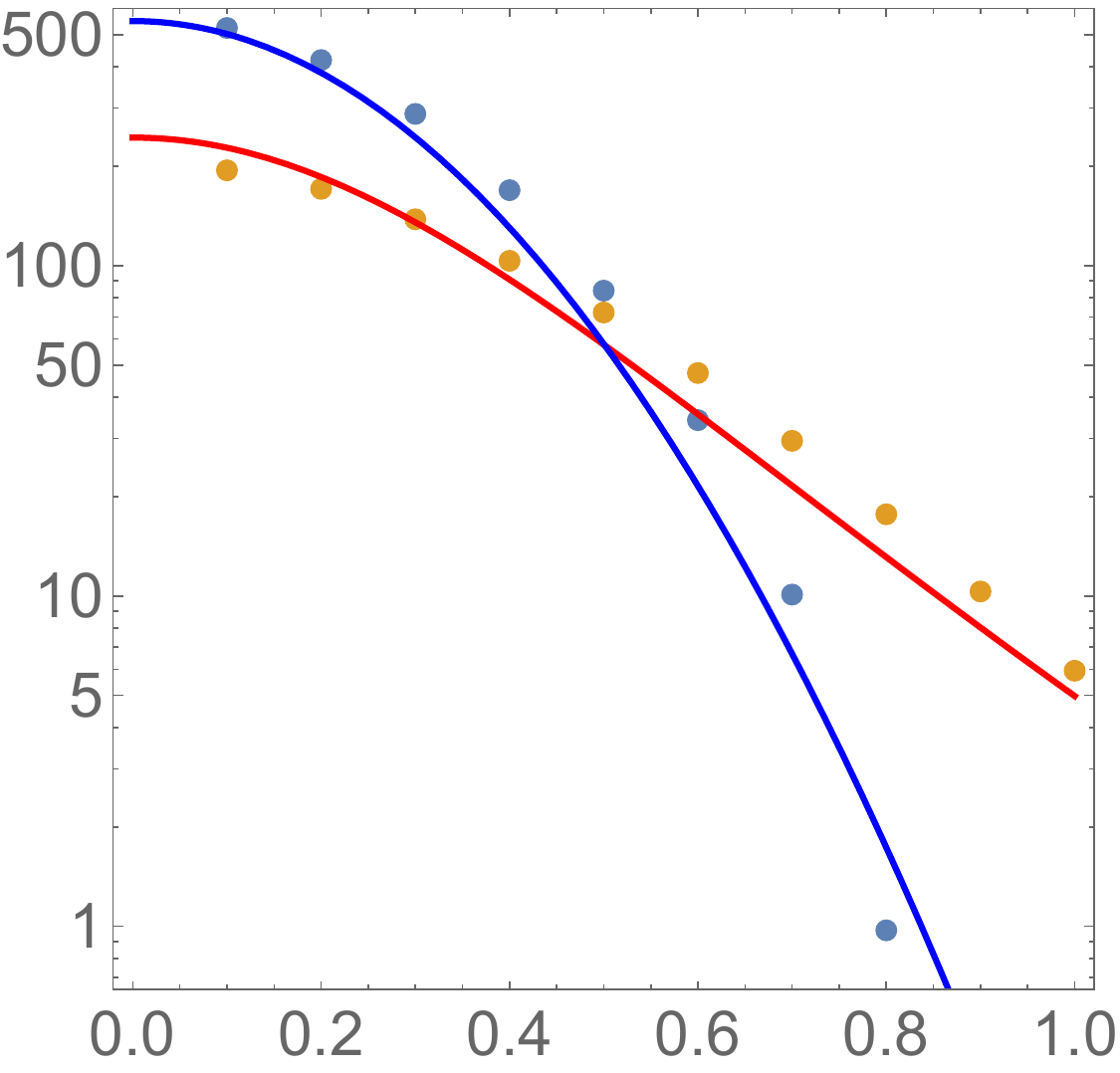}
    \caption{Blue and red points show the Fourier transforms of the 1S and 1D
    radial wave functions as functions of momentum $p\,(\rm GeV)$. The solid
    lines represent the analytical fits given in
    the text.}
    \label{fig_Fourier_1S1D}
\end{figure}

We now briefly outline the calculations underlying the discussion above. The
radial (hyperdistance) wave functions and their six-dimensional Fourier
transforms were evaluated numerically. For analytical parametrizations, it is
useful to recall the exact Fourier transforms of Gaussian and exponential wave
functions,
\ba
\rm exp\bigg(-{Y^2 \over 2 \Lambda^2} \bigg)
&\rightarrow&
\rm exp\bigg(-{P^2 \Lambda^2\over 2 } \bigg), \nonumber\\
\rm exp\bigg(-Y\cdot A\bigg)
&\rightarrow&
{1 \over (1+P^2/A^2)^{7/2}}.
\ea
With this in mind, the numerical form factors shown in
Fig.~\ref{fig_Fourier_1S1D} for the 1S (blue dots) and 1D (orange dots) states are
well fitted by simple analytical expressions.

In the Breit frame, the form factor is given by a convolution of wave functions
corresponding to total momenta $\vec Q/2$ and $-\vec Q/2$. The boost amounts to
dividing the longitudinal momentum components by the gamma factor
$$
1/\gamma=\sqrt{1-v^2}=\sqrt{1/(1+Q^2/4M^2)},
$$
so that the Lorentz-contracted wave functions are
$\phi=\psi(\vec p_\perp,p_{\rm long}/\gamma)$. The form factor includes
scattering off all three quarks, labeled by $i=1,2,3$, weighted by their charges
$e_i$,
\bea
\label{eqn_FF}
G_E(Q)=&&\sum_i {e_i \over e \gamma}
\int {d^3 p_\rho d^3 p_\lambda\over (2\pi)^6} \nonumber\\
&&\times \phi(p_\rho,p_\lambda)
\phi(\vec p_\rho+\nu^i_\rho \vec Q,
      \vec p_\Lambda+\nu^i_\lambda \vec Q),
\eea
where for the proton $e_i=(2/3,2/3,-1/3)$, and $\nu_\rho,\nu_\lambda$ denote the
fractions of the total momentum $\vec Q$ carried by the corresponding Jacobi
coordinates.

The form factors obtained from relativized versions of the hyperdistance model\footnote{See
 Fig.~24 of \cite{Giannini:2015dca}.} show better agreement with experimental
data. Rather than pursuing these models further, we will instead turn, in the
later parts of this review, to a derivation of form factors directly from
light-front wave functions.


\begin{subappendices}
\section{Schrodinger equations in $d$ dimensions}
\label{sec_d_dimensions}

The nonrelativistic kinetic energy of $A$ particles of identical mass $m$, written
in terms of $N=A-1$ modified Jacobi coordinates, takes the simple form of a
Laplacian (multiplied by $1/m$)~\cite{FABREDELARIPELLE1979185},
\be
-\frac 12 \sum_{i=1}^A\,\nabla^2_{i}=
-\frac 12 \bigg(\partial_\rho^2+\frac{3N-1}\rho\,\partial_\rho-\frac 1{\rho^2}K_N^2\bigg).
\ee
The particular cases of three and four particles will be discussed below.

The hyperspherical harmonics (HHs) are eigenstates of the grand-angular momentum
operator,
\be
K_N^2\,{\cal Y}_{[K]}^{KLM_L}(\Omega_N)
=(K(K+3N-2))\,{\cal Y}_{[K]}^{KLM_L}(\Omega_N).
\ee
The $2^N+N$ angular variables
$\Omega_N=(\hat{\xi}_1,\ldots,\hat{\xi}_N;\varphi_1,\ldots,\varphi_N)$
consist of the directions of the individual Jacobi coordinates $\hat{\xi}_i$ and
the hyperangles defined by ${\rm cos}\,\varphi_j=\xi_j/\rho$.
The quantum numbers $L,M_L$ are the usual eigenvalues of the total orbital
angular momentum operators $L^2$ and $L_z$. The explicit form of the HHs follows
from recoupling the individual angular momenta $l_i$.

The hyperspherical harmonics are normalized according to
\be
\int d\Omega_N\,
{\cal Y}_{[K]}^{KLM_L\,*}(\Omega_N)\,
{\cal Y}_{[K']}^{K'L'M'_L}(\Omega_N)
=\delta_{[K],[K']},
\ee
and their total degeneracy is given by~\cite{FABREDELARIPELLE1979185}
\be
d_K=(2K+3N-2)\frac{(K+3N-3)!}{K!(3N-2)!}.
\ee
For example, for $A=4$ particles ($N=3$), the $K=0$ HH has degeneracy $d_0=1$,
while the $K=1$ HHs have degeneracy $d_1=9$.

For the applications relevant to this paper, however, we mostly employ different
sets of angular coordinates. For instance, in the case of pentaquarks, the
original $5\times 3=15$ coordinates are reduced to four Jacobi vectors
$\vec\alpha,\vec\beta,\vec\gamma,\vec\delta$, corresponding to 12 coordinates
after fixing the center of mass at the origin. In this parametrization, there
are $4\times 2=8$ angles $\theta_i,\phi_i$ ($i=1,\ldots,4$) specifying directions
in four three-dimensional subspaces, and three additional angles, denoted
$\chi_2,\chi_1,\phi_\chi$, which parametrize directions in the four-dimensional
space spanned by the vector lengths $\alpha,\beta,\gamma,\delta$.

Using any such coordinate system, we follow the same general procedure. One
starts by computing the metric tensor $g_{\mu\nu}$ as the coefficients of the
differentials in the quadratic form $ds^2=g_{\mu\nu}dx^\mu dx^\nu$. All geometric
quantities then follow from standard expressions. In particular, the volume
element is given by $\sqrt{\det g}$. For the 12-dimensional example discussed
above, it takes the form
\[
\sqrt{\det g}\,\prod dx^i
=Y^{11} dY
\bigg[\prod_{i=1..4} \sin(\theta_i)\, d\theta_i\, d\phi_i \bigg]
\bigg[\sin^2(\chi_2)\,\sin(\chi_1)\, d\chi_2\, d\chi_1\, d\phi_\chi \bigg].
\]

If all quarks have the same mass, the kinetic energy operator is proportional to
the Laplace-Beltrami operator,
\be
-{L} = {1 \over \sqrt{\det g}}
{\partial \over \partial x^\mu}
\sqrt{\det g}\, g^{\mu\nu}
{\partial \over \partial x^\nu},
\ee
where $g^{\mu\nu}$ is the inverse metric tensor.


\section{Baryons in hyperspherical coordinates}
\label{sec_jacobi_baryons}

To describe the specifics of the hyperspherical approximation, we first consider
three quarks forming a baryon. The three coordinate vectors of the quarks are
compressed into two Jacobi vectors,
\bea
\vec\rho=\frac 1{\sqrt{2}} \vec r_{12}, \qquad
\vec \lambda=\frac{1}{\sqrt{6}}(\vec r_{13}+\vec r_{23}),
\eea
where $\vec r_{ij}=\vec r_i-\vec r_j$ are three-dimensional vectors. The square
root factors define the so-called modified Jacobi coordinates, which ensure that
the kinetic energy takes the form of a sum of second derivatives with the same
coefficient $1/2m$. As a result, the kinetic energy becomes spherically symmetric
in six dimensions.

The hyperdistance is the six-dimensional radial coordinate, denoted by $R_6$,
and is defined as
\be
R_6^2\equiv \vec\rho^2+\vec \lambda^2
={1\over 3} \big[(\vec r_1-\vec r_2)^2+(\vec r_1-\vec r_3)^2+(\vec r_3-\vec r_2)^2\big].
\ee
It is naturally related, in a fully symmetric manner, to the sum of all squared
interparticle distances.

In its most symmetric form, the hyperspherical approximation does not
distinguish between $\vec\rho$ and $\vec\lambda$, but instead treats all six
coordinates on an equal footing. Standard spherical coordinates can be
introduced using five angular variables
$\theta_1,\theta_2,\theta_3,\theta_4\in[0,\pi]$ and
$\phi\in[0,2\pi]$, leading to the six-dimensional solid angle
\ba
\Omega_6 &=& \int d\theta_1\, d\theta_2\, d\theta_3\, d\theta_4\, d\phi \,
\sin^4(\theta_1)\,\sin^3(\theta_2)\,\sin^2(\theta_3)\,
\sin(\theta_4)
=\pi^3 .
\ea
Alternatively, a six-dimensional coordinate parametrization with five angular
variables can be defined as
$$
\vec \rho = R_6 \cos(\Phi)\,\hat{\vec n}_1, \qquad
\vec \lambda = R_6 \sin(\Phi)\,\hat{\vec n}_2,
$$
where $\hat{\vec n}_1$ and $\hat{\vec n}_2$ are unit vectors with independent
solid-angle integrations.

The numerical coefficients in the definitions of these coordinates ensure that
the kinetic energy operator takes the form of a Laplacian in the six-dimensional
space $\{\vec\rho,\vec\lambda\}$, divided by $2m$. (The total center-of-mass
momentum carries a different coefficient, but it is assumed to vanish.) We
defer the explicit derivation to the section devoted to four-particle systems.

In general, the effective potential $V_6$ depends on all six coordinates. It can,
however, be projected onto the six-dimensional radial coordinate $R_6$ by
performing an angular average. If the binary interactions are of Cornell type,
consisting of linear and Coulomb-like terms, for example
$V_{12}=\kappa_2/r_{12}+\sigma_2 r_{12}$, their projections are obtained through
the solid-angle integrals
\ba
\langle r_{12} \rangle
= \sqrt{2}\,\langle |\vec \rho | \rangle
= \sqrt{2} R_6
{\int d\Omega_6\, d_{12} \over \Omega_6} 
\approx 0.960\, R_6 ,
\ea
where
\be
d_{12}
=\sqrt{C_1^2 + S_1^2 C_2^2 + S_1^2 S_2^2 C_3^2}
=\sqrt{1-S_1 S_2 S_3},
\ee
and we have introduced the shorthand notations
$C_i=\cos(\theta_i)$ and $S_i=\sin(\theta_i)$.

Dimensional considerations also imply that the Coulomb term $1/r$ in the binary
potential must generate a contribution $\sim 1/R_6$ in hyperspherical
coordinates. The angular averaging of the Coulomb interaction involves the
factor $1/(\sqrt{2}d_{12})$, leading to
\bea
\bigg\langle {1 \over r_{12}} \bigg\rangle
= {1 \over \Omega_6 }
\int d\Omega_6\, {1 \over \sqrt{2}\, d_{12}} 
\approx {1.20 \over R_6}.
\eea
The convergence of this integral follows from the fact that $d_{12}$ can vanish
only if all three angles $\theta_1,\theta_2,\theta_3\approx \pi/2$.

As a result, the effective potential in the hyperspherical coordinate $R_6$
takes the form
\be
\label{eqn_Cornell_6d}
V(R_6)= 3\big[0.96\, R_6\,(0.113/2)
- {1.2 \over R_6}\,(0.64/2)\big],
\ee
where the factor of 3 accounts for the interactions of the three quark pairs
$12$, $13$, and $23$. The numerical values $0.113$ and $0.64$ (in GeV$^2$ units)
are taken from our version of the Cornell potential.


\section{The spin-tensor and vector representations of the multi-index WFs}
\label{sec_spin_tensor}

Nucleon-like baryons have color$\times$spin$\times$flavor wave functions derived
earlier in this chapter, for example the spin-flavor part of the proton
wave function in Eq.~(\ref{eqn_proton_wf}). Multiquark hadrons possess much more
complicated wave functions of the form
(orbital)$\times$color$\times$spin$\times$flavor, often containing thousands of
components in the monomial space and, in some cases, even millions. Clearly, an
automatic procedure is required to record such wave functions in a form that
allows for efficient manipulation.

The procedure consists of two main steps. The first is to define the wave
functions as spin tensors. We illustrate this for the proton wave function
(\ref{eqn_proton_wf}), written schematically as
$S_\rho I_\rho + S_\lambda I_\lambda$.

We begin by specifying the basis states. For both spin and isospin, these are
\[
up=\{1,0\}, \qquad down=\{0,1\}.
\]
The spin-isospin wave function in spin-tensor form is then represented as a
$2^6$-dimensional table, defined by the Mathematica command
\ba
SIproton &=& \mathrm{Table}\Big[(up[[s1]]*down[[s2]] - up[[s2]]*down[[s1]])up[[s3]] \nonumber \\
&\times& (up[[i1]]*down[[i2]] - up[[i2]]*down[[i1]])up[[i3]])/2 \nonumber \\
&+& (up[[s1]]down[[s2]]up[[s3]]+down[[s1]]up[[s2]]up[[s3]]-2up[[s1]]up[[s2]]down[[s3]]) \nonumber \\
&\times& (up[[i1]]down[[i2]]up[[i3]]+down[[i1]]up[[i2]]up[[i3]]-2up[[i1]]up[[i2]]down[[i3]])/6 \nonumber \\
&,& \{s1, 1, 2\}, \{s2, 1, 2\}, \{s3, 1, 2\},
     \{i1, 1, 2\}, \{i2, 1, 2\}, \{i3, 1, 2\} \Big]
\ea
The output is
    $$=  \{\{\{\{\{\{0, 0\}, \{0, 0\}\}, \{\{0, 0\}, \{0, 0\}\}\}, \{\{\{0, \sqrt{2}/
      3\}, 
      \{-1/(3 \sqrt{2}), 0\}\}, \{\{-1/(3 \sqrt{2}), 0\}, $$
$$      
      \{0, 
      0\}\}\}\}, \{\{\{\{0, -1/(3 \sqrt{2})\}, \{\sqrt{2}/3, 
      0\}\}, \{\{-1/(3 \sqrt{2}), 0\}, \{0, 0\}\}\}, \{\{\{0, 0\}, \{0, 0\}\}, $$
$$      \{\{0, 
      0\}, \{0, 0\}\}\}\}\}, \{\{\{\{\{0, -1/(3 \sqrt{2})\}, \{-1/(3 \sqrt{2}), 
      0\}\}, \{\{\sqrt{2}/3, 0\}, \{0, 0\}\}\}, \{\{\{0, 0\}, \{0, 0\}\}, \{\{0, 0\}, $$
$$      \{0, 
      0\}\}\}\}, \{\{\{\{0, 0\}, \{0, 0\}\}, \{\{0, 0\}, \{0, 0\}\}\}, \{\{\{0, 0\}, \{0, 
      0\}\}, \{\{0, 0\}, \{0, 0\}\}\}\}\} $$
      

In the second step, a single 64-dimensional vector representation is obtained by
applying the \texttt{Flatten} command, which removes all internal brackets. In
this form, only nine of the 64 components of the proton wave function are
nonzero.

The normalization can be verified straightforwardly by computing the scalar
product $SIproton\cdot SIproton=1$. Expectation values of any operator $\hat O$
(for example, the magnetic moment) can then be obtained as
$SIproton\cdot \hat O \cdot SIproton$, and similarly for other observables.


\end{subappendices}


\chapter{Tetraquarks}
\label{sec_tetra}
Historically, the first in-depth study of tetraquarks dates back to
Ref.~\cite{Jaffe:1976ig}, where it was argued that tetraquarks composed of light
quarks could be rather light, form specific $SU(3)_f$ multiplets, and in
particular provide an explanation for some known scalar mesons. These
calculations were performed within the framework of the MIT bag model and were
received with considerable skepticism at the time, largely because of the
well-known limitations of that particular model.

It took roughly four decades for the community's attitude to shift from
skeptical to enthusiastic, following the discovery of
$\bar c c \bar q q$ tetraquark candidates, already mentioned in the charmonium
chapter. In this chapter, we begin with the kinematics of
the general case involving four different quark flavors,  and then turn to special limiting cases, in particular the
single-flavor $cc\bar c \bar c$ systems.

\section{Four-flavor tetraquarks}

When all four quark flavors are different, the identification of exotic states
is relatively straightforward. For example, in the $D^-K^+$ channel, any
resonance must have the minimal valence quark content
$\bar c d \bar s u$, and is therefore manifestly exotic. Resonances observed by
LHCb in this channel include
\[
X_0(2900):\quad M=2.866 \pm 0.007 \pm 0.002 \,\mathrm{GeV}, \qquad
\Gamma =57\pm 12 \pm 4 \,\mathrm{MeV},
\]
\[
X_1(2900):\quad M=2.904 \pm 0.005 \pm 0.001 \,\mathrm{GeV}, \qquad
\Gamma =110\pm 11 \pm 4 \,\mathrm{MeV}.
\]
At present, the exact quantum numbers of these states are not yet firmly
established, nor is it clear whether they represent the lowest-lying states in
this channel.

Among the early theoretical studies of these resonances, we focus here on
Ref.~\cite{Karliner:2020vsi}. In that work, the authors describe the states as
two diquarks connected by a QCD string stretched between two string junctions.
Using a schematic additive mass model, combined with earlier results for
baryons obtained within the same framework, they note that part of such a
tetraquark configuration (a $ud$ diquark bound to a $c$ quark) resembles the
structure of the $\Lambda_c$ baryon. The remaining contributions arise from the
$sc$ binding and from the additional mass associated with the string junction,
for which previous studies suggested a value
$S\approx 165\,\mathrm{MeV}$.

It should be emphasized that in such string-based models only the $ud$ and $cs$
pairwise interactions are included, while all other quark pairs are assumed to
be noninteracting. Clearly, more comprehensive approaches, incorporating
realistic interquark forces and fully quantum-mechanical treatments, are
required in order to obtain reliable wave functions and a complete
spectroscopy.

In four-flavor tetraquarks, all quarks are distinguishable, and therefore no
antisymmetrization of the wave function is required. As a result, such systems
can, in principle, be studied using direct numerical methods, such as path
integral techniques\footnote{One of us employed such methods more than forty
years ago to study two electrons in the helium atom and four nucleons in the
helium nucleus; see
E.~V.~Shuryak and O.~V.~Zhirov,
Testing Monte Carlo methods for path integrals in some quantum mechanical
problems,
Nucl.\ Phys.\ B \textbf{242}, 393 (1984).}. A closely related first-principles
approach is the solution of the Schrodinger equation in imaginary (Euclidean)
time, also known as diffusion Monte Carlo; see, for example,
Ref.~\cite{Gordillo:2025caj}.

In that work, the Hamiltonian includes $(\lambda\lambda)(\sigma\sigma)$
operators and is represented as a matrix in an appropriate monomial basis. The
operator $\exp(-\hat H \tau)$ is then implemented numerically as a relaxation
from an initial Ansatz toward the true ground state. The results indicate that
the experimentally observed resonances are close to isospin $I=1$ states, while
the corresponding $I=0$ states (not yet observed) are predicted to lie roughly
$400\,\mathrm{MeV}$ lower in mass.

An important feature of the resulting wave functions is that the distributions
of interparticle distances for all quark pairs are very similar in shape and
are quite compact, with $r^2|\psi|^2$ peaking at
$r\sim 0.5\,\mathrm{fm}$. If confirmed, this behavior would strongly disfavor interpretations, whether in terms of two mesons or two loosely
bound diquarks.


\section{Color-spin-flavor structure of tetraquarks in terms of diquarks}

Tetraquarks consist of two quarks and two antiquarks. The constraints imposed by
Fermi statistics are therefore even simpler than in the baryonic case discussed
above, and they can be analyzed without any special techniques.

We begin with diquarks, $qq$. For light quarks ($uu,ud,dd$), the combined
color-spin-flavor wave function must be antisymmetric under permutation of the
two quarks. The same applies to a pair of antiquarks. The relevant permutation
group is $S_2$, which is particularly simple, so all possibilities can be
enumerated explicitly.  

{\bf Color} admits an antisymmetric $\underline{3}$ and a symmetric $6$
representation.  
{\bf Spin} admits an antisymmetric singlet (spin 0) and a symmetric triplet
(spin 1), and the same structure applies to $SU(2)$ isospin.  

Altogether, there are four possible diquark configurations,
\[
d_1=\underline{3}\otimes 1 \otimes 1, \qquad
d_2=\underline{3}\otimes 3 \otimes 3, \qquad
d_3=6\otimes 3 \otimes 1, \qquad
d_4=6\otimes 1 \otimes 3 .
\]

Combining a diquark with an antidiquark is only possible if their color
representations match, ensuring that the resulting tetraquark is a color
singlet. This leads to eight possible combinations,
\[
d_1 \bar d_1,\;
d_1 \bar d_2,\;
d_2 \bar d_1,\;
d_2 \bar d_2,\;
d_3 \bar d_3,\;
d_3 \bar d_4,\;
d_4 \bar d_3,\;
d_4 \bar d_4 .
\]

As usual, we start with the extreme cases, which are the simplest. Suppose the
total spin and isospin of the tetraquark are maximal,
\[
S_{\rm tet}=I_{\rm tet}=2 .
\]
As an explicit example, consider
$T^{++}=u\uparrow u\uparrow \bar d\uparrow \bar d\uparrow$. In this case, the only
remaining degree of freedom is color, which must be antisymmetric,
$\underline{3}$. Consequently, only a single structure is allowed,
schematically $d_2\bar d_2=(333)-(333)$.

Next, suppose that $S_{\rm tet}$ is arbitrary, while $I_{\rm tet}=2$ remains
fixed. In this situation, both color and spin degrees of freedom contribute,
and the diquarks $d_2$ and $d_4$ are allowed. This leads to four possible
tetraquark configurations, and so on. As we shall see, these states can mix
through various interactions.

The first interaction we consider is the "relative color'' force,
\be
H_{\lambda\lambda}=\sum_{i>j} w_{ij}(\vec\lambda_i \vec\lambda_j),
\ee
where the sum runs over all six quark-quark and quark-antiquark pairs. The
explicit Mathematica implementation of this operator is given in
Appendix~\ref{sec_lambdas}.

The color wave function can be written as a superposition of two components,
\[
|\psi\rangle = F_1\,|\underline{3}\rangle + F_2\,|6\rangle .
\]
Sandwiching the operator $V_0$ defined above between diquark-antidiquark states
and performing the summation over all color indices yields
\be
H_{\lambda\lambda}^{66}
= \frac{2}{3}\,
(2 w_{12} - 5 w_{13} - 5 w_{14} - 5 w_{24} - 5 w_{32} + 2 w_{34}),
\ee
\be
H_{\lambda\lambda}^{\underline{3}\underline{3}}
= -\frac{4}{3}\,
(2 w_{12} + w_{13} + w_{14} + w_{24} + w_{32} + 2 w_{34}).
\ee
In the $66$ case, the contributions from $qq$ and $\bar q \bar q$ pairs are
repulsive, whereas in the $\underline{3}\underline{3}$ case all terms are
attractive. A naive application of the diquark model would therefore suggest
that the $\underline{3}\underline{3}$ configuration should have a lower energy
than the $66$ one. However, this conclusion is incorrect. If all interaction
strengths are equal, $w_{ij}=w$, both expressions reduce to the same value,
$-(32w/3)$.

This remarkable result is well known
\cite{Badalian:2023krq,Miesch:2023hjt}. We emphasize it here because it contradicts
a widespread intuition: starting from "good diquarks'' ($F_1$) rather than
"bad'' ones ($F_2$) does {\em not} automatically lead to lower tetraquark
energies. In this case, the repulsion within the diquarks is exactly compensated
by stronger attraction between them.

Further splittings and mixings arise once more complicated interactions are
included, such as the color-spin Hamiltonian,
\be
H_{\lambda\lambda\sigma\sigma}
=V \sum_{ij}(\vec \lambda_i \vec \lambda_j)(\vec \sigma_i \vec \sigma_j).
\ee
Since this operator does not change isospin, there are four states that split
into two pairs, $(311311,311631)$ and $(333333,333613)$. The corresponding
Hamiltonian matrices are
\be
H(311311-311631)= 
V\begin{pmatrix}
44 /3 & -6 \sqrt{2} \\  
-6 \sqrt{2} & -2/3 
\end{pmatrix},
\ee
with eigenvalues after mixing $V(18.43,-4.44)$, and
\be
H(613613-311631)= 
V\begin{pmatrix}
20 /3 & -6 \sqrt{2} \\ 
-6 \sqrt{2} & -34/3 
\end{pmatrix},
\ee
with eigenvalues after mixing $V(-17.80,-0.20)$.

Another possible interaction is pion exchange, proposed by Ripka and Glozman,
\be
H_{\sigma\sigma\tau\tau}
=V_{\sigma\sigma\tau\tau}
\sum_{i>j}(\vec\sigma_i\vec\sigma_j)(\vec\tau_i\vec\tau_j),
\ee
where $\vec\tau$ are Pauli matrices for $SU(2)$ flavor (or Gell-Mann matrices for
$SU(3)$). In this case, the color structure remains unchanged, while the isospin
is affected. The Hamiltonian then takes the form
\be
H(311311-311333)= 
V_2\begin{pmatrix}
18 & 3 \\ 
3  & 2
\end{pmatrix},
\ee
with eigenvalues after mixing $V_2(18.544,1.456)$. The color-sextet pair splits in
an analogous manner.

We conclude this technical subsection by explaining why we have chosen to
discuss tetraquarks in a somewhat old-fashioned way, using spin tensors and
explicit summation over all indices. In principle, all states can be represented
as vectors in a universal {\em monom basis} of dimension
$N_{\rm monoms}=3^4\cdot 2^4 \cdot 2^4=20736$. In \textit{Wolfram Mathematica},
this is easily implemented using the \texttt{TensorProduct} command. For
example, the first $d_1\bar d_1$ wave function can be written as
\be
\Psi_1=\mathrm{Flatten}\big[\mathrm{SparseArray}[
  \mathrm{TensorProduct}(color33, spin0, spin0, spin0, spin0)]\big],
\ee
which produces a vector of length $N_{\rm monoms}$, with only 192 nonzero
components\footnote{This is why the \texttt{SparseArray} format is convenient,
though optional.}. The individual factors can be prepared as spin tensors; for
example,
\[
spin0=\mathrm{Flatten}\big[\mathrm{Table}[
(up[[s1]]*down[[s2]]-up[[s2]]*down[[s1]])/\sqrt{2},
\{s1,1,2\},\{s2,1,2\}]\big],
\]
which returns the spin-zero diquark wave function
$\{0,1/\sqrt{2},-1/\sqrt{2},0\}$.

For the present discussion, where only eight wave functions are needed, this
formalism may appear excessive. Its real advantage becomes clear when various
Hamiltonians are represented as universal matrices of dimension
$N_{\rm monoms}\times N_{\rm monoms}$, enabling systematic and automated
calculations.

\section{Quantum mechanics of four quarks in hyper-spherical coordinates} \label{sec_hyper_tetra}
The modified  Jacobi coordinates  for four particles are
defined by
\ba  \label{eqn_Jacobi_4}
\vec\xi_1 &=& \sqrt{1 \over 2}(\vec r_1-\vec r_2) \\ 
\vec\xi_2 &=& \sqrt{1 \over 6}(\vec r_1+\vec r_2-2\vec r_3) \nonumber \\
\vec\xi_3 &=& {1 \over 2 \sqrt{3}}(\vec r_1+\vec r_2+\vec r_3-3\vec r_4) \nonumber \ea
plus the CM coordinate $\vec X=(\vec r_1+\vec r_2+\vec r_3+\vec r_4)/4 $, complleting it to 12 coordinates.
In this case the $hyperdistance$ is defined as 
\ba R_9^2& \equiv &\vec \xi_1^2+\vec \xi_2 ^2+\vec \xi_3^2 \\
&=& {1\over 4} [(\vec r_1-\vec r_2)^2+(\vec r_1-\vec r_3)^2+(\vec r_1-\vec r_4)^2  \nonumber \\
&+&(\vec r_3-\vec r_2)^2+(\vec r_4-\vec r_2)^2+(\vec r_3-\vec r_4)^2]
\nonumber \ea
 is connected to the sum of the squared distances  of all six pairs of quarks. When supplemented  by
 8 angles, it describes the 9-dimensional space in which quantum mechanics is performed.

 Calculating from (\ref{eqn_Jacobi_4}) the Jacobian matrix $\partial \xi_i / \partial r_j $ and using it twice, one can transform the kinetic energy
 into new coordinates. 
\be -2 m K=  {\partial^2 \over \partial \vec\xi_1^2} +  {\partial^2 \over \partial \vec\xi_2^2} +  {\partial^2 \over \partial \vec\xi_3^2} + {1 \over 4} {\partial^2 \over \partial \vec X^2}\ee
Keeping the CM coordinate $\vec X=0$ and ignoring the last term, we can use  the kinetic energy proportional to the 9-dimensional Laplacian.

The dependence on the wave function on the hyper-distance $only$ select the lowest spherically symmetric S-shell.  Only the radial Schrodinger equation needs to be solved. Note that
after changing to the reduced wave-function,
it differs from the familiar 3-dimensional case, by only the
quasi-centrifugal term $12/R_9^2$. 

The angular averaging is performed as in the previous section, except now
there is a different volume element, namely
\ba \Omega_9&=&\int (\prod_{i=1}^7 d\theta_i) d\phi \,
sin(\theta_1)^7*sin(\theta_2)^6*sin(\theta_3)^5*sin(\theta_4)^4 \,
  sin(\theta_5)^3*sin(\theta_6)^2*sin(\theta_7) ={32 \pi^4 \over 105} \nonumber \\
\ea
The angular integrations when averaging using the Cornell potential, are done as in the previous section, with obvious changes
\ba \langle r_{12} \rangle &=& \sqrt{2}\langle |\vec \rho | \rangle  
={\sqrt{2}R_9 \over \Omega_9}
\int d\Omega_9 d_{12}  
\approx  0.773 R_9  
\ea
\ba \bigg< {1 \over r_{12}} \bigg> &=&{1 \over R_9 \Omega_9 } \int { d\Omega_9 \over \sqrt{2} d_{12}}  
\approx  1.55/R_9 
  \ea

\section{Single-flavor tetraquarks}

The $cc\bar c\bar c$ tetraquarks were first observed by the LHCb Collaboration,
and subsequently by the CMS and ATLAS collaborations, in
$J/\psi\,J/\psi$ (and related) decay channels. Three interfering states have been
reported, all with the same quantum numbers, most likely $2^{++}$. The current
LHCb values (in MeV) are
\ba
M_1 &=& 6593^{+15}_{-14}, \qquad \Gamma_1 = 446^{+66}_{-54}, \nonumber\\
M_2 &=& 6847 \pm 10, \qquad\;\;\, \Gamma_2 = 135^{+16}_{-14}, \nonumber\\
M_3 &=& 7173^{+9}_{-10}, \qquad \Gamma_3 = 73^{+18}_{-15},
\ea
and these states are observed both in the $J/\psi\,J/\psi$ and
$J/\psi\,\psi'$ channels.

Theoretical investigations on the lattice began with studies of charmonium
scattering lengths, such as $\eta_c$-$\eta_c$ and $J/\psi$-$J/\psi$
scattering, reported in Ref.~\cite{Meng:2024czd}. While the lowest-order diagrams
naively suggest an attractive interaction between the two quarkonia, the lattice
calculations instead indicate that the low-energy interaction is in fact
{\em repulsive}. The corresponding scattering lengths are
\be
a^{0+}_{\eta_c-\eta_c}=-0.104(09)\,\mathrm{fm}, \qquad
a^{2+}_{J/\psi-J/\psi}=-0.165(16)\,\mathrm{fm}.
\ee

Since the absolute masses of the tetraquark states are sensitive to additive
constants in the potential (or, equivalently, to the quark mass), we will focus
instead on the mass splittings, or gaps
\be
M_2-M_1, \qquad M_3-M_1,
\ee
which should be compared with the gaps between successive radial excitations.

\begin{subappendices}



\section{Kinematics of four bodies with different masses}
\label{4bodies_different_masses}

The kinematics of four bodies with equal masses were discussed above in
Section~\ref{sec_hyper_tetra}. In this Appendix, we consider the more general
case of four bodies with different masses, as well as alternative choices of
coordinates that are sometimes used in the literature.

We begin with the original Jacobi construction, defining coordinates for four
objects with unequal masses. These coordinates are obtained through the
following transformation from the original particle coordinates $x[[i]]$,
\ba
\alpha &=& -x[[1]] + x[[2]],\\
\beta &=& (M_1 x[[1]] + M_2 x[[2]])/(M_1 + M_2) - x[[3]], \nonumber\\
\gamma &=& (M_1 x[[1]] + M_2 x[[2]] + M_3 x[[3]])/(M_1 + M_2 + M_3) - x[[4]], \nonumber\\
\delta &=& (M_1 x[[1]] + M_2 x[[2]] + M_3 x[[3]] + M_4 x[[4]])/(M_1 + M_2 + M_3 + M_4).
\nonumber
\ea
The coordinate $\delta$ corresponds to the center-of-mass (CM) position and, as
usual, is set to zero.

By computing the inverse transformation, one can express the original
coordinates $x[[i]]$ in terms of $\alpha,\beta,\gamma,\delta$ and then determine
the metric in the new coordinate system from the line element. The resulting
expression is
\ba
dl^2 &=&
\frac{(M_1^2 + M_2^2)}{(M_1 + M_2)^2}\, d\alpha^2
+ \frac{(M_1^2 + 2 M_1 M_2 + M_2^2 + 2 M_3^2)}{(M_1 + M_2 + M_3)^2}\, d\beta^2
\nonumber\\
&-&
\frac{2 (M_1 + M_2 - 2 M_3)\, d\beta
(M_4 d\gamma + (M_1 + M_2 + M_3 + M_4) d\delta)}
{(M_1 + M_2 + M_3)(M_1 + M_2 + M_3 + M_4)}
\nonumber\\
&+&
\frac{1}{(M_1 + M_2 + M_3 + M_4)^2}
\Big[(M_1^2 + M_2^2 + 2 M_2 M_3 + M_3^2
+ 2 M_1 (M_2 + M_3) + 3 M_4^2)\, d\gamma^2
\nonumber\\
&-&
2 (M_1^2 + M_2^2 + M_3^2 + 2 M_2 (M_3 - M_4)
+ 2 M_1 (M_2 + M_3 - M_4)
- 2 M_3 M_4 - 3 M_4^2)\, d\gamma\, d\delta
\nonumber\\
&+&
4 (M_1 + M_2 + M_3 + M_4)^2\, d\delta^2 \Big]
\nonumber\\
&+&
\frac{2 (M_1 - M_2)\, d\alpha}
{(M_1 + M_2)(M_1 + M_2 + M_3)(M_1 + M_2 + M_3 + M_4)}
\nonumber\\
&\times&
\Big[ M_3 (M_1 + M_2 + M_3 + M_4)\, d\beta
+ (M_1 + M_2 + M_3)
(M_4 d\gamma + (M_1 + M_2 + M_3 + M_4)\, d\delta) \Big].
\ea
From this metric one can derive the Laplacian involving the inverse metric
$g^{\mu\nu}\partial_\mu\partial_\nu$, which is likewise non-diagonal.

In the special case of equal masses,
$M=M_1=M_2=M_3=M_4$, both the metric and the Laplacian become diagonal,
\be
dl^2= (6\, d\alpha^2 + 8\, d\beta^2 + 9\, d\gamma^2 + 48\, d\delta^2)/12.
\ee
By rescaling the coordinates (denoted by bars) with appropriate factors and
neglecting the CM motion ($d\delta\rightarrow 0$), the line element reduces to
\be
d\bar\alpha^2 + d\bar\beta^2 + d\bar\gamma^2,
\ee
which exhibits manifest spherical symmetry.

In some studies of four-body systems (including tetraquarks), an alternative
set of coordinates is employed\footnote{These are also often referred to as
Jacobi coordinates, although this terminology is somewhat misleading. The
original idea of Jacobi was to define relative coordinates between the center of
mass of a subset of particles and the next particle.},
\ba
a &=& x[[2]] - x[[1]],\\
b &=& x[[3]] - x[[4]], \nonumber\\
c &=& (M_1 x[[1]] + M_2 x[[2]])/(M_1 + M_2)
- (M_3 x[[3]] + M_4 x[[4]])/(M_3 + M_4), \nonumber\\
d &=& (M_1 x[[1]] + M_2 x[[2]] + M_3 x[[3]] + M_4 x[[4]])
/(M_1 + M_2 + M_3 + M_4).
\nonumber
\ea
Here again, $d$ represents the center-of-mass coordinate and is set to zero. The
corresponding metric is, in general, non-diagonal,
\ba
dl^2 &=&
\frac{(M_1^2 + M_2^2)}{(M_1 + M_2)^2}\, da^2
+ \frac{(M_3^2 + M_4^2)}{(M_3 + M_4)^2}\, db^2
\nonumber\\
&-&
\frac{2 (M_3 - M_4)\, db
\big[-(M_1 + M_2) dc + (M_1 + M_2 + M_3 + M_4) dd\big]}
{(M_3 M_4)(M_1 + M_2 + M_3 + M_4)}
\nonumber\\
&+&
\frac{2 (M_1 - M_2)\, da
\big[(M_3 + M_4) dc + (M_1 + M_2 + M_3 + M_4) dd\big]}
{(M_1 + M_2)(M_1 + M_2 + M_3 + M_4)}
\nonumber\\
&+&
\frac{1}{(M_1 + M_2 + M_3 + M_4)^2}
\Big[2 (M_1^2 + 2 M_1 M_2 + M_2^2 + (M_3 + M_4)^2)\, dc^2
\nonumber\\
&-&
4 (M_1^2 + 2 M_1 M_2 + M_2^2 - (M_3 + M_4)^2)\, dc\, dd
+ 4 (M_1 + M_2 + M_3 + M_4)^2\, dd^2 \Big].
\ea
A non-diagonal metric again implies a non-diagonal Laplacian. However, when all
masses are equal ($M=M_1=M_2=M_3=M_4$), the metric simplifies to
\be
dl^2=(da^2 + db^2 + 2\, dc^2 + 8\, dd^2)/2,
\ee
which is diagonal.

Is there any preference between these two coordinate choices? We find that there
is. For tetraquarks of the type $qQ\bar q \bar Q$, with masses chosen as
$m=M_1=M_3$ and $M=M_2=M_4$, the alternative coordinates lead to a diagonal
Laplacian,
\ba
L=
{\partial \over \partial a^2}
+ \frac{(m + M)^2}{(m^2 + M^2)}\,{\partial \over \partial b^2}
+ {\partial \over \partial c^2}
+ \frac{1}{4}\,{\partial \over \partial d^2}.
\ea

For convenience, the main features of the two coordinate choices discussed above
are summarized in the appended table. It compares the original Jacobi construction
and the alternative pair-based coordinates for different mass patterns, focusing
on whether the resulting metric (and hence the Laplacian) is diagonal after
removal of the center-of-mass motion. In the generic case of four unequal masses,
both choices lead to non-diagonal kinetic terms and offer no particular
advantage. For equal masses, the two constructions become equivalent and reduce
to a spherically symmetric form. The most important case for applications,
however, is the $qQ\bar q\bar Q$ tetraquark with $m=M_1=M_3$ and $M=M_2=M_4$,
where the alternative coordinates yield a diagonal Laplacian even for unequal
masses. This makes them the optimal choice for practical calculations of
tetraquark spectra and wave functions.

\begin{center}
\begin{tabular}{|c|c|c|p{4.2cm}|}
\hline
Mass pattern 
& Coordinates 
& Laplacian 
& Comment \\
\hline
$M_1\neq M_2\neq M_3\neq M_4$ 
& Jacobi $(\alpha,\beta,\gamma)$ 
& Non-diagonal 
& Fully general, but algebraically cumbersome \\
\hline
$M_1\neq M_2\neq M_3\neq M_4$ 
& Alternative $(a,b,c)$ 
& Non-diagonal 
& No practical simplification \\
\hline
$M_1=M_2=M_3=M_4$ 
& Jacobi 
& Diagonal 
& Spherical symmetry after CM removal \\
\hline
$M_1=M_2=M_3=M_4$ 
& Alternative 
& Diagonal 
& Equivalent to Jacobi choice \\
\hline
$qQ\bar q\bar Q$ \newline
$m=M_1=M_3$ \newline
$M=M_2=M_4$
& Alternative 
& Diagonal 
& Optimal choice for tetraquarks \\
\hline
$qQ\bar q\bar Q$ \newline
$m=M_1=M_3$ \newline
$M=M_2=M_4$
& Jacobi 
& Non-diagonal 
& Less convenient \\
\hline
\end{tabular}
\end{center}


\section{Color matrices for quarks and antiquarks} \label{sec_lambdas}
Here 8 $\lambda$ matrices are Gell-Mann generators of $SU(3)$ for quarks, given in Mathematica notations\footnote{Note however that
in order to use it one has to rewrite it to their Mathematica: direct copy would lead to font problems, as they are different in Wolfram and PDF.}. Matrices $\lambda b=-\lambda^*$ are appropriate for antiquarks   
 \begin{verbatim}
\[Lambda][1] = {{0, 1, 0}, {1, 0, 0}, {0, 0, 0}}; \[Lambda][
  2] = {{0, -I, 0}, {I, 0, 0}, {0, 0, 0}};
 \[Lambda][3] = {{1, 0, 0}, {0, -1, 0}, {0, 0, 0}}; 
\[Lambda][4] = {{0, 0, 1}, {0, 0, 0}, {1, 0, 0}};
\[Lambda][5] = {{0, 0, -I}, {0, 0, 0}, {I, 0, 0}};  
\[Lambda][6] = {{0, 0, 0}, {0, 0, 1}, {0, 1, 0}}; \[Lambda][
  7] = {{0, 0, 0}, {0, 0, -I}, {0, I, 0}}; 
 \[Lambda][8] = {{1, 0, 0}, {0, 1, 0}, {0, 0, -2}}/Sqrt[3]; 
 \[Lambda]b[1] = -{{0, 1, 0}, {1, 0, 0}, {0, 0, 0}}; \[Lambda]b[
  2] = {{0, -I, 0}, {I, 0, 0}, {0, 0, 0}};
 \[Lambda]b[3] = -{{1, 0, 0}, {0, -1, 0}, {0, 0, 0}}; 
\[Lambda]b[4] = -{{0, 0, 1}, {0, 0, 0}, {1, 0, 0}};
\[Lambda]b[5] = {{0, 0, -I}, {0, 0, 0}, {I, 0, 0}};  
\[Lambda]b[6] = -{{0, 0, 0}, {0, 0, 1}, {0, 1, 0}}; \[Lambda]b[
  7] = {{0, 0, 0}, {0, 0, -I}, {0, I, 0}}; 
 \[Lambda]b[8] = -{{1, 0, 0}, {0, 1, 0}, {0, 0, -2}}/Sqrt[3]; 
 \end{verbatim}
 \begin{verbatim}
 V0 :=w12* Sum[\[Lambda][A][[c1p, c1]]*\[Lambda][A][[c2p, c2]],{A, 1, 8}]*\[Delta]3[[c3p, c3]]*\[Delta]3[[c4p,c4]]+
  w34*Sum[\[Lambda]b[A][[c3p, c3]]*\[Lambda]b[A][[c4p, c4]], {A, 1, 8}]*\[Delta]3[[c1p, c1]]*[Delta]3[[c2p, c2]] +w13*Sum[\[Lambda][A][[c1p, c1]]*\[Lambda]b[A][[c3p, c3]], {A, 1, 8}]*\[Delta]3[[c2p, c2]]*\[Delta]3[[c4p, c4]] + 
  w24*Sum[\[Lambda][A][[c2p, c2]]*\[Lambda]b[A][[c4p, c4]], {A, 1, 8}]*\[Delta]3[[c3p, c3]]*[Delta]3[[c1p,c1]] +
  w14*Sum[\[Lambda][A][[c1p, c1]]*\[Lambda]b[A][[c4p, c4]], {A, 1, 8}]*\[Delta]3[[c3p, c3]]*\[Delta]3[[c2p, c2]] +
  w32*Sum[\[Lambda]b[A][[c3p, c3]]*\[Lambda][A][[c2p, c2]], {A, 1, 8}]*\[Delta]3[[c1p, c1]]*\[Delta]3[[c4p, c4]]
\end{verbatim}

\end{subappendices}


\chapter{Excited baryons}

\section{The negative-parity P-shell baryons}

To reveal the role of various spin-dependent forces between quarks, it is {\em
not} sufficient to consider only the lowest S-shell baryons. Spin-spin
interactions generate essentially a single mass splitting between the $N$ and
$\Delta$. In contrast, spin-orbit and tensor forces become operative only in
higher baryonic shells, where the spectrum is richer and the phenomenology more
informative.

The pioneering study by Isgur and Karl (IK)~\cite{Isgur:1978wd} focused on the
negative-parity P-shell baryons. Their analysis concentrated primarily on
hyperons containing a strange quark, for which wave functions with 12-symmetry
were sufficient. The light-quark baryons were then obtained as appropriate
limits of these hyperon states, by reducing the strange-quark mass to that of
the light quarks.

IK employed two Jacobi vectors, $\vec\rho$ and $\vec\lambda$, which naturally
separate structures that are antisymmetric or symmetric under permutation of
quarks 1 and 2, $\hat P(12)$. In the P-shell, the wave functions factorize into
three parts: orbital, spin, and isospin. Using three building blocks with the
same $\rho,\lambda$ structure, one obtains a $2^3=8$-dimensional space, within
which the states symmetric under {\em all} permutations can be identified.
This observation played a central role in the work of IK and their successors.

A systematic treatment, without relying on educated guessing and with a clear
path toward generalization to multiquark systems, requires the explicit
construction of representations of the relevant permutation groups. We present
these constructions in the appendix to this chapter.

In contrast to the Isgur-Karl approach, we choose not to introduce additional
parameters associated with strangeness, and instead focus exclusively on
light-quark states. As is well known, the negative-parity $L=1$ (P-shell)
baryons consist of seven states: five $N^*$ and two $\Delta^*$. Restricting
attention to the $N^*$ sector, the five masses together with two mixing matrix
elements (for the $J=3/2$ and $J=1/2$ pairs) provide seven independent inputs.
In Table~\ref{tab_masses} we list the current experimental masses of the P-shell
states and compare them with the original Isgur-Karl predictions~\cite{Isgur:1978wd}.

We adopt here a strategy that inverts the usual logic of spectroscopy. Rather
than postulating a specific model and comparing its predictions with data, we
first extract {\em phenomenological values} of the matrix elements of the
relevant operators directly from experiment. Given the known structure of the
wave functions and the spin-orbit-isospin structure of the operators, the
masses can be expressed as linear combinations of these matrix elements. The
seven experimental inputs are sufficient to determine them uniquely.

\begin{table}[b]
\caption{The P-shell baryons composed of light quarks. The original Isgur-Karl
predictions~\cite{Isgur:1978wd} are compared with experimental masses from the
RPP. The last column lists the masses after unmixing (see text).}
\begin{center}
\begin{tabular}{|c|c|c|c|}
 \hline
 states $J^P$ & Isgur-Karl & experiment & unmixed \\ 
 \hline
 $ N^*_{1/2^-}$ & 1490 & 1535 & 1567.3, $S=1/2$\\
 $ N^*_{1/2^-}$ & 1655 & 1650 & 1617.7, $S=3/2$\\
 $ N^*_{3/2^-}$ & 1535 & 1520 & 1521.97, $S=1/2$ \\
 $ N^*_{3/2^-}$ & 1745 & 1700 & 1698.0, $S=3/2$\\
 $ N^*_{5/2^-}$ & 1670 & 1675 & $S=3/2$ \\
 \hline
\end{tabular}
\end{center}
\label{tab_masses}
\end{table}

The mixing angles of the $S=3/2$ and $S=1/2$ states are determined from decay
data and are listed in the RPP reviews,
\[
\theta_{S1/2}=-32^\circ, \qquad \theta_{S3/2}=6^\circ.
\]
The relation between the energies of the mixed states and the unmixed ones is
\begin{equation}
M_{\pm}= \frac 12 \bigg( M_1+M_2\pm
\sqrt{4 H_{mix}^2+(M_1-M_2)^2} \bigg),
\end{equation}
and the corresponding mixing angle is
\be
\tan\theta=\frac{2 H_{mix}}
{M_1-M_2-\sqrt{4 H_{mix}^2+(M_1-M_2)^2}}.
\ee
Using the observed masses $M_{\pm}$ and mixing angles, we extract the unmixed
masses $M_{1,2}$, which are listed in the final column of
Table~\ref{tab_masses}. The resulting mixing matrix elements are
\ba
\label{eqn_mixing_values}
H_{mix}(J=1/2)&=&51.7\,\mathrm{MeV}, \nonumber\\
H_{mix}(J=3/2)&=&-18.7\,\mathrm{MeV}.
\ea
Below, we will use the five masses to determine five matrix elements and compare
the value of the tensor matrix element with these two mixing matrix elements.

In the P-shell with $L=1$, one has either $L_\rho=1,L_\lambda=0$ or
$L_\lambda=1,L_\rho=0$. Accordingly, for the orbital wave functions
$\varphi_{LM}^{\rho,\lambda}$ we define
\bea
\label{ZMX}
\varphi^\rho_{1m}
&=&(\rho_-,\sqrt{2}\rho_z,-\rho_+)\varphi_{00}
\equiv z_m^\rho\varphi_{00}, \nonumber\\
\varphi^\lambda_{1m}
&=&(\lambda_-,\sqrt{2}\lambda_z,-\lambda_+)\varphi_{00}
\equiv z_m^\lambda\varphi_{00},
\eea
with $\rho_\pm=(\rho_1\pm i\rho_2)$ and
$\lambda_\pm=(\lambda_1\pm i\lambda_2)$. Up to the overall radial dependence,
these are the standard angular functions $Y_1^m$ with $m=-1,0,+1$.

Combining these orbital states with all possible spin configurations yields five
excited nucleon states,
\bea
J^P=L^P\oplus S^P
=1^-\otimes\bigg(\frac12^+,\frac32^+\bigg)
\rightarrow
2\times\bigg(\frac12^-,\frac32^-\bigg),\;\frac52^-.
\eea
That is, two $\frac12^-$ states, two $\frac32^-$ states, and one $\frac52^-$
state. Standard Clebsch-Gordan coefficients specify how to combine the orbital
and spin parts into states of definite $J,J_z$. However, to construct fully
antisymmetric baryon wave functions consistent with Fermi statistics, the
isospin structure must also be incorporated properly. The construction of
symmetric wave functions from the three $\rho,\lambda$ structures is described
in the appendix to this chapter; see Eq.~(\ref{eqn_sym_in_3}).

We now present the explicit results, beginning with the case $S=\frac32$ and
the corresponding $J=\frac52,\frac32,\frac12$ states. Starting from the maximal
spin state, one may use the familiar two-structure combination,
\bea
\bigg|1 \frac32 \frac52 \frac52\bigg\rangle_{p_M^-}
&=&
C_A\,S^S_{\frac32\frac32}\,
\frac1{\sqrt{2}}
\big(F^\rho_{\frac12}\varphi^\rho_{11}
+F^\lambda_{\frac12}\varphi^\lambda_{11}\big),
\nonumber
\eea
while the remaining states require combinations involving all three
$\rho,\lambda$ structures,
\bea
\bigg|1 \frac32 \frac32 \frac32\bigg\rangle_{p_M^-}
&=&C_A\bigg(
\sqrt{\frac35}\,S^S_{\frac32\frac32}\,
\frac1{\sqrt{2}}
\big(F^\rho_{\frac12}\varphi^\rho_{10}
+F^\lambda_{\frac12}\varphi^\lambda_{10}\big)
\nonumber\\
&&\quad
-\sqrt{\frac25}\,S^S_{\frac32\frac12}\,
\frac1{\sqrt{2}}
\big(F^\rho_{\frac12}\varphi^\rho_{11}
+F^\lambda_{\frac12}\varphi^\lambda_{11}\big)
\bigg),
\nonumber\\
\bigg|1 \frac32 \frac12 \frac12\bigg\rangle_{p_M^-}
&=&C_A\bigg(
-\frac1{\sqrt2}\,S^S_{\frac32\frac32}\,
\frac1{\sqrt{2}}
\big(F^\rho_{\frac12}\varphi^\rho_{1-1}
+F^\lambda_{\frac12}\varphi^\lambda_{1-1}\big)
\nonumber\\
&&\quad
+\frac1{\sqrt3}\,S^S_{\frac32\frac12}\,
\frac1{\sqrt{2}}
\big(F^\rho_{\frac12}\varphi^\rho_{10}
+F^\lambda_{\frac12}\varphi^\lambda_{10}\big)
\nonumber\\
&&\quad
-\frac1{\sqrt6}\,S^S_{\frac32-\frac12}\,
\frac1{\sqrt{2}}
\big(F^\rho_{\frac12}\varphi^\rho_{11}
+F^\lambda_{\frac12}\varphi^\lambda_{11}\big)
\bigg).
\eea
Here $S^S_{\frac32\frac32}=\uparrow\uparrow\uparrow$ denotes the totally symmetric
three-quark spin-$\frac32$ state.

In general, constructing a state of definite $J$ from orbital angular momentum
$L$ and spin $S$ requires summing over all kinematically allowed combinations
$m=J_z=m_L+m_S$, weighted by the standard Clebsch-Gordan coefficients,
\bea
\label{SPIN32}
\bigg|1 \frac32 J m\bigg\rangle_{p_M^-}
=
\sum_{m_S}
{\bf C}^{Jm}_{1m_L\,\frac32 m_S}
\bigg[
C_A\,S^S_{\frac32 m_S}\,
\frac1{\sqrt{2}}
\big(F^\rho_{\frac12}\varphi^\rho_{1m_L}
+F^\lambda_{\frac12}\varphi^\lambda_{1m_L}\big)
\bigg].
\eea
We use the convention
\bea
{\bf C}^{Jm}_{Lm_LSm_S}
=
(-1)^{S-L-m}\sqrt{2J+1}
\begin{pmatrix}
L & S & J\\
m_L & m_S & -m
\end{pmatrix}.
\eea

The expressions for spin $\frac12$ states with $J=\frac32,\frac12$ are more
involved. The maximal-$J$ states for fixed $L,S$ are
\bea
\label{SPIN12}
\bigg|1 \frac12 \frac32 \frac32\bigg\rangle_{p_M^-}
&=&C_A\,
\frac1{\sqrt{2}}
\bigg(
F^\rho_{\frac12}
\frac1{\sqrt{2}}
(\varphi^\rho_{11} S^\lambda_{\frac12\frac12}
+\varphi^\lambda_{11} S^\rho_{\frac12\frac12})
\nonumber\\
&&\quad
+F^\lambda_{\frac12}
\frac1{\sqrt{2}}
(\varphi^\rho_{11} S^\rho_{\frac12\frac12}
-\varphi^\lambda_{11} S^\lambda_{\frac12\frac12})
\bigg),
\nonumber\\
\bigg|1 \frac12 \frac12 \frac12\bigg\rangle_{p_M^-}
&=&C_A\bigg(
\sqrt{\frac23}
\frac1{\sqrt{2}}
\bigg(
F^\rho_{\frac12}
\frac1{\sqrt{2}}
(\varphi^\rho_{11} S^\lambda_{\frac12-\frac12}
+\varphi^\lambda_{11} S^\rho_{\frac12-\frac12})
\nonumber\\
&&\qquad
+F^\lambda_{\frac12}
\frac1{\sqrt{2}}
(\varphi^\rho_{11} S^\rho_{\frac12-\frac12}
-\varphi^\lambda_{11} S^\lambda_{\frac12-\frac12})
\bigg)
\nonumber\\
&&\quad
-\sqrt{\frac13}
\frac1{\sqrt{2}}
\bigg(
F^\rho_{\frac12}
\frac1{\sqrt{2}}
(\varphi^\rho_{10} S^\lambda_{\frac12-\frac12}
+\varphi^\lambda_{10} S^\rho_{\frac12-\frac12})
\nonumber\\
&&\qquad
+F^\lambda_{\frac12}
\frac1{\sqrt{2}}
(\varphi^\rho_{10} S^\rho_{\frac12-\frac12}
-\varphi^\lambda_{10} S^\lambda_{\frac12-\frac12})
\bigg)
\bigg).
\eea
For the isobar one finds
\bea
\bigg|1 \frac12 \frac32 \frac32\bigg\rangle_{\Delta_M^-}
=
C_A\,F^S_{\frac32}\,
\frac1{\sqrt{2}}
\big(S^\rho_{\frac12\frac12}\varphi^\rho_{11}
+S^\lambda_{\frac12\frac12}\varphi^\lambda_{11}\big).
\eea
The lower-$J$ states for fixed $L,S$ follow by standard Clebsch-Gordan
construction,
\bea
\label{MX1}
\bigg|1 \frac12 J m\bigg\rangle_{p_M^-}
=
\sum_{m_S}
{\bf C}^{Jm}_{1m_L\,\frac12 m_S}
\bigg[
C_A\,
\frac1{\sqrt{2}}
\bigg(
F^\rho_{\frac12}
\frac1{\sqrt{2}}
(\varphi^\rho_{1m_L} S^\lambda_{\frac12 m_S}
+\varphi^\lambda_{1m_L} S^\rho_{\frac12 m_S})
\nonumber\\
\qquad
+F^\lambda_{\frac12}
\frac1{\sqrt{2}}
(\varphi^\rho_{1m_L} S^\rho_{\frac12 m_S}
-\varphi^\lambda_{1m_L} S^\lambda_{\frac12 m_S})
\bigg)
\bigg],
\eea
and similarly for the odd-parity $\Delta$ states,
\bea
\bigg|1 \frac12 J m\bigg\rangle_{\Delta_M^-}
=
\sum_{m_S}
{\bf C}^{Jm}_{1m_L\,\frac12 m_S}
\bigg[
C_A\,F^S_{\frac32}\,
\frac1{\sqrt{2}}
\big(S^\rho_{\frac12 m_S}\varphi^\rho_{1m_L}
+S^\lambda_{\frac12 m_S}\varphi^\lambda_{1m_L}\big)
\bigg].
\eea



\subsection{Spin forces from P-shell states}

Knowledge of the coefficients of the relevant operators for the five (unmixed)
negative-parity $N^*$ masses allows one to determine the corresponding matrix
elements of the spin-spin, spin-orbit, and tensor forces.

Without going into technical details, let us recall one striking feature first
noted by Isgur and Karl~\cite{Isgur:1979be}: the contribution of the spin-orbit
force is strongly suppressed relative to the spin-spin and tensor forces, and
its matrix elements are consistent with zero within errors. Our fit fully
confirms this observation. We therefore adopt what we refer to as the
{\em optimized IK model,} in which the spin-orbit interaction is neglected and
only the spin-spin and tensor forces ( considered to be more reliable) are
retained,
\ba
\label{eqn_optimized}
\langle H_0 \rangle &=& 1607\,\mathrm{MeV}, \nonumber\\
{\langle \rho^2 V_{SS}\rangle \over \langle \rho^2+\lambda^2 \rangle }
&=& 83.2\,\mathrm{MeV}, \nonumber\\
{\langle \rho^4 V_{\rm tensor}\rangle \over \langle \rho^2+\lambda^2 \rangle }
&=& 43.7\,\mathrm{MeV}.
\ea
Here $H_0$ denotes the spin-independent part of the Hamiltonian, which is common
to all five resonances. Our only assumptions are that the spin-dependent forces
are treated to first order, and that near-threshold effects are neglected%
\footnote{Isgur and Karl made additional assumptions, such as adopting a Gaussian
radial wave function.}.

\begin{figure}[h]
\begin{center}
\includegraphics[width=8cm]{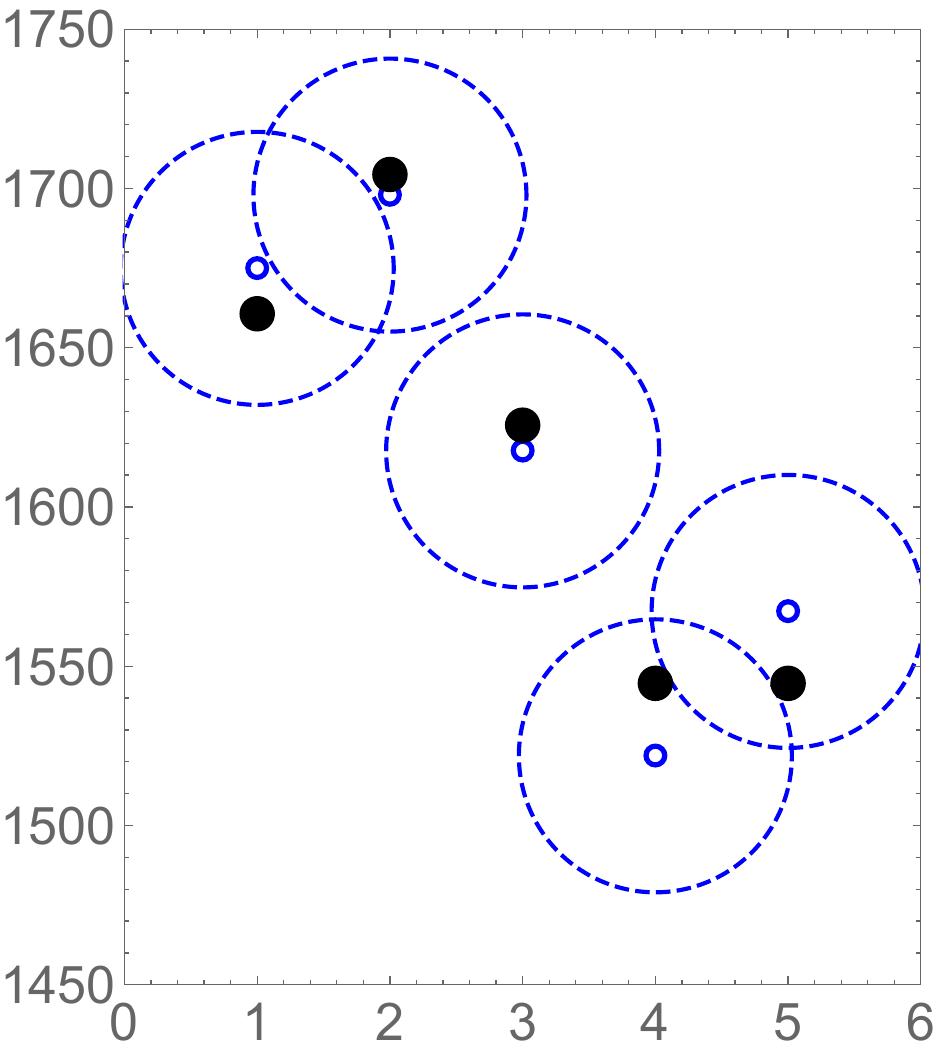}
\caption{The black dots show the masses of the negative-parity P-shell baryons
obtained from the optimized IK model with fitted radial matrix elements
(\ref{eqn_optimized}), ordered as
$|J=\frac52\rangle$, $|J=\frac32,S=\frac32\rangle$,
$|J=\frac12,S=\frac32\rangle$, $|J=\frac32,S=\frac12\rangle$,
$|J=\frac12,S=\frac12\rangle$.
The small blue circles indicate the experimental pole positions. They are
surrounded by large dashed blue circles whose radii are of order the typical
half-width $\Gamma/2$.}
\label{fig_5nucs}
\end{center}
\end{figure}

The overall quality of the description of the masses is illustrated in
Fig.~\ref{fig_5nucs}. In all cases, the deviations are significantly smaller than
the typical half-widths of the resonances%
\footnote{These half-widths provide a natural scale for the threshold effects
that are ignored in the present analysis.}.

As independent "external tests'' of the accuracy of the model, we use the
empirical mixing matrix elements in
Eq.~(\ref{eqn_mixing_values}). The observed ratio of the magnitudes of the mixing
matrix elements in the tensor channel is
\be
{H_{mix}(J=1/2)\over H_{mix}(J=3/2)} \approx -2.76.
\ee
If only the tensor operator is retained, the ratio of the corresponding
coefficients is $-\sqrt{10}=-3.16$, which is reasonably close.

Using the fitted tensor matrix element extracted from the splittings shown in
Fig.~\ref{fig_5nucs}, we obtain the value given in
Eq.~(\ref{eqn_optimized}), namely $43.7\,\mathrm{MeV}$. This should be compared
with $51.6\,\mathrm{MeV}$ inferred from the mixing matrix element in the
$J=\frac12$ P-shell pair discussed above.

In summary, the spectra of P-shell baryons are well described by the combined
effects of spin-spin and tensor forces. Any additional contributions not
explicitly included here can affect the masses only at the level of
$\sim 8\,\mathrm{MeV}$. This uncertainty is small compared both to the typical
splittings, $\sim 100\,\mathrm{MeV}$, and to the characteristic scale
$\Gamma/2\sim 50\,\mathrm{MeV}$, which we use as an estimate of the maximal
possible shifts due to threshold effects.

\section{The wave functions and splittings of negative-parity $N^*$}

Having extracted empirical values of the matrix elements of the spin-dependent
forces, we now compare them with the corresponding theoretical values obtained
within the hyperdistance approximation. We begin by outlining the general
structure of the calculation.

All matrix elements involve six-dimensional integrals over the Jacobi
coordinates $\vec\rho$ and $\vec\lambda$, which can be separated into four
angular and two radial integrations,
\be
\int \bigg({d\Omega_\rho \over 4\pi}{d\Omega_\lambda \over 4\pi}\bigg)
(\rho^2 d\rho\,\lambda^2 d\lambda)\,\ldots\, .
\ee
The wave functions consist of a common six-dimensional spherically symmetric
radial function $\phi_{00}(R)$, with $R^2=\rho^2+\lambda^2$, multiplied by
appropriate orbital factors. For the P-wave baryons considered here, the orbital
parts are linear in $\vec\rho$ and $\vec\lambda$. The matrix elements receive
multiple contributions from different components of the fully symmetrized wave
functions, as detailed in the Appendix.

In the expressions below, all angular integrations are performed explicitly,
while the integrals over the moduli $\rho$ and $\lambda$ are left unevaluated,
since no specific assumption about the functional form of $\phi_{00}(R)$ is
made. We therefore introduce the shorthand notation
\begin{equation}
\langle \hat V(\rho,\lambda) \rangle \equiv
\int \int d\rho\, d\lambda \;\rho^2 \lambda^2
|\phi_{00}|^2\, V(\rho,\lambda),
\end{equation}
for such radial integrals. For example, the normalization integral
($\hat V=\hat 1$) averages the squared orbital wave function,
\be
\label{eqn_averaging}
\langle \rho^2+\lambda^2 \rangle =
\int \int d\rho\, d\lambda\;\rho^2 \lambda^2
|\phi_{00}|^2 (\rho^2+\lambda^2).
\ee
This quantity appears in the denominators of all subsequent expressions.

The masses of the excited nucleon states $N^*_J$ contain an overall constant
(not written explicitly below), supplemented by contributions from four
spin-dependent operators: spin-spin, spin-orbit, tensor, and 't~Hooft terms.
The resulting mixing matrices in the P-shell subspaces are
\ba
\label{eqn_results}
{\mathbb M}_{\frac52}'=
{3\over4}\,
{\langle (\rho^2+\lambda^2)V_S(\sqrt{2}\rho)\rangle
\over \langle (\rho^2+\lambda^2) \rangle}
+3\,
{\langle \rho^2 V_{LS}(\sqrt{2}\rho) \rangle
\over \langle (\rho^2+\lambda^2) \rangle}
-{1\over5}\,
{\langle \rho^4 V_T(\sqrt{2}\rho) \rangle
\over \langle (\rho^2+\lambda^2) \rangle},
\ea
\bea
{\mathbb M}_{\frac32}' &=&
\begin{pmatrix}
	3/4 & 0\\
	0 & -3/4
\end{pmatrix}
{\langle (\rho^2+\lambda^2)V_S(\sqrt{2}\rho)\rangle
\over \langle (\rho^2+\lambda^2) \rangle}
-
\begin{pmatrix}
	2 & -\sqrt{5/2}\\
	-\sqrt{5/2} & -1
\end{pmatrix}
{\langle \rho^2 V_{SL}(\sqrt{2}\rho)\rangle
\over \langle (\rho^2+\lambda^2) \rangle}
\nonumber\\
&+&
\begin{pmatrix}
	4/5 & 1/\sqrt{10}\\
	1/\sqrt{10} & 0
\end{pmatrix}
{\langle \rho^4 V_T(\sqrt{2}\rho)\rangle
\over \langle (\rho^2+\lambda^2) \rangle}
+
\begin{pmatrix}
	0 & 0\\
	0 & 1+3a
\end{pmatrix}
{G_{tH}\langle \lambda^2\rangle
\over 8\sqrt{2}\pi\langle (\rho^2+\lambda^2)\rangle},
\\
{\mathbb M}_{\frac12}' &=&
\begin{pmatrix}
	3/4 & 0\\
	0 & -3/4
\end{pmatrix}
{\langle (\rho^2+\lambda^2)V_S(\sqrt{2}\rho)\rangle
\over \langle (\rho^2+\lambda^2) \rangle}
-
\begin{pmatrix}
	5 & 1\\
	1 & -2
\end{pmatrix}
{\langle \rho^2 V_{SL}(\sqrt{2}\rho)\rangle
\over \langle (\rho^2+\lambda^2) \rangle}
\nonumber\\
&+&
\begin{pmatrix}
	-1 & -1\\
	-1 & 0
\end{pmatrix}
{\langle \rho^4 V_T(\sqrt{2}\rho)\rangle
\over \langle (\rho^2+\lambda^2) \rangle}
+
\begin{pmatrix}
	0 & 0\\
	0 & 1+3a
\end{pmatrix}
{G_{tH}\langle \lambda^2\rangle
\over 8\sqrt{2}\pi\langle (\rho^2+\lambda^2)\rangle}.
\eea
The rows and columns of the $2\times2$ matrices correspond to states with
$S=\frac32$ and $S=\frac12$, respectively,
\bea
\label{DEG1}
\mathbb M_J=
\begin{pmatrix}
\langle S=\frac32|\mathbb V_{S+T+SL+V_{TH}}|S=\frac32\rangle &
\langle S=\frac32|\mathbb V_{S+T+SL+V_{TH}}|S=\frac12\rangle \\
\langle S=\frac12|\mathbb V_{S+T+SL+V_{TH}}|S=\frac32\rangle &
\langle S=\frac12|\mathbb V_{S+T+SL+V_{TH}}|S=\frac12\rangle
\end{pmatrix}_J .
\eea
Some of these matrices contain non-diagonal elements, implying that the total
spin $S$ is not, in general, a good quantum number for the observed states.


\subsection{The shape of P-wave states}

The explicit wave functions for all P-wave states are derived in spin-tensor
form in the Appendix to this section. Once these expressions are available, it
is straightforward to evaluate matrix elements of arbitrary
spin-isospin-angular operators%
\footnote{Note, for example, that the orbital angular-momentum operator involves
derivatives with respect to angles or coordinates. \textsc{Mathematica} can
perform such operations directly on tables of functions, such as our
orbital-spin-tensor wave functions.}.

Although we do not present the detailed derivations here, it is instructive to
discuss the {\em shape} of the P-wave states. Except for the total
$J=\frac12$ case, these shapes are expected to be nontrivial.

The coordinate-space densities $|\psi|^2$, summed over all spin and isospin
indices, for the five negative-parity $N^*$ states with $J_z=\frac12$ are
\ba
| N^*_{J=5/2,J_z=1/2,S=3/2}|^2
&\sim&
\frac{2}{5}\big(
\lambda_1^2 + \lambda_2^2 + 3 \lambda_3^2
+ \rho_1^2 + \rho_2^2 + 3 \rho_3^2
\big),
\nonumber\\
| N^*_{J=3/2,J_z=1/2,S=3/2}|^2
&\sim&
\frac{2}{15}\big(
7\lambda_1^2 + 7\lambda_2^2 + \lambda_3^2
+ 7\rho_1^2 + 7\rho_2^2 + \rho_3^2
\big),
\nonumber\\
| N^*_{J=1/2,J_z=1/2,S=3/2}|^2
&\sim&
\frac{2}{3}\big(
\lambda_1^2 + \lambda_2^2 + \lambda_3^2
+ \rho_1^2 + \rho_2^2 + \rho_3^2
\big),
\nonumber\\
| N^*_{J=3/2,J_z=1/2,S=1/2}|^2
&\sim&
\frac{2}{15}\big(
7\lambda_1^2 + 7\lambda_2^2 + \lambda_3^2
+ 7\rho_1^2 + 7\rho_2^2 + \rho_3^2
\big),
\nonumber\\
| N^*_{J=1/2,J_z=1/2,S=1/2}|^2
&\sim&
\frac{2}{3}\big(
\lambda_1^2 + \lambda_2^2 + \lambda_3^2
+ \rho_1^2 + \rho_2^2 + \rho_3^2
\big).
\nonumber
\ea
We note that the $J=\frac52$ and $J=\frac32$ states exhibit elliptic
deformations of opposite sign, while both $J=\frac12$ states remain
spherically symmetric in this sense.

\begin{subappendices}

\section{Enforcing Fermi statistics by representations of the permutation group $S_3$}

Negative-parity baryons possess nontrivial {\em orbital} wave functions in
addition to spin and isospin. The orbital part can be constructed from either
the $\rho$ or the $\lambda$ Jacobi vector, so one must understand the action of
permutations on three objects of $\rho,\lambda$ type. There are $2^3=8$ such
possibilities, namely
$X^\rho_1 X^\rho_2 X^\rho_3,\ldots$, forming an eight-dimensional space.

Symmetry under the transposition $(12)$ is straightforward to analyze: of the
eight combinations, four containing an even number of $X^\rho$ factors are
symmetric, while the remaining four are antisymmetric. To proceed further, one
must determine how these basis states transform under the permutation
$\hat P(23)$. We follow the same strategy as before and evaluate this
transformation explicitly using \textsc{Mathematica}%
\footnote{The command used is \texttt{KroneckerProduct} with three arguments.}.

The resulting matrix representation of $\hat P(23)$ in this basis is
\ba
M_{23} &\bigotimes&  M_{23}\bigotimes  M_{23}=\\
&=&\begin{bmatrix} 
-1/8 & \sqrt{3}/8& \sqrt{3}/8& -3/8& \sqrt{3}/8& -3/8& -3/8&  3 \sqrt{3}/8 \\
  \sqrt{3}/8& 1/8& -3/8& -\sqrt{3}/8& -3/8& -\sqrt{3}/8& 3 \sqrt{3}/8& 3/8 \\ 
  \sqrt{3}/8& -3/8& 1/8& -\sqrt{3}/8& -3/8& 3 \sqrt{3}/8& -\sqrt{3}/8& 3/8 \\ 
  -3/8& -\sqrt{3}/8& -\sqrt{3}/8& -1/8& 3 \sqrt{3}/8& 3/8&
   3/8& \sqrt{3}/8 \\
   \sqrt{3}/8& -3/8& -3/8& 3 \sqrt{3}/8& 1/8& -\sqrt{3}/8& -\sqrt{3}/8& 3/8 \\
   -3/8& -\sqrt{3}/8&
  3 \sqrt{3}/8& 3/8& -\sqrt{3}/8& -1/8& 3/8& \sqrt{3}/8 \\
  -3/8& 3 \sqrt{3}/8& -\sqrt{3}/8& 3/8& -\sqrt{3}/8& 3/8& -1/8& \sqrt{3}/8 \\
  3 \sqrt{3}/8& 3/8& 3/8& \sqrt{3}/8& 3/8& \sqrt{3}/8& \sqrt{3}/8& 1/8 
\end{bmatrix}
\nonumber
\ea
written in the basis of the following eight monomials:
\[
X^\lambda_1 X^\lambda_2 X^\lambda_3,\;
X^\rho_1 X^\lambda_2 X^\lambda_3,\;
X^\lambda_1 X^\rho_2 X^\lambda_3,\;
X^\rho_1 X^\rho_2 X^\lambda_3,\;
X^\lambda_1 X^\lambda_2 X^\rho_3,\;
X^\rho_1 X^\lambda_2 X^\rho_3,\;
X^\lambda_1 X^\rho_2 X^\rho_3,\;
X^\rho_1 X^\rho_2 X^\rho_3.
\]

The determinant of this matrix is unity, with four eigenvalues equal to $-1$ and
four equal to $+1$. This spectrum is the same as for the (diagonal) matrix
representing the permutation $\hat P(12)$, although the corresponding
eigenvectors are different. One can then readily verify that the two symmetric
subspaces have {\em only one common} symmetric vector,
\[
(-1,\,0,\,0,\,1,\,0,\,1,\,1,\,0),
\]
in this basis. This vector corresponds to the unique orbital-spin-isospin
structure compatible with full $S_3$ symmetry,
\be
\label{eqn_sym_in_3}
- X^\lambda X^\lambda X^\lambda
+ X^\rho X^\rho X^\lambda
+ X^\rho X^\lambda X^\rho
+ X^\lambda X^\rho X^\rho .
\ee

Proceeding to the next excitation, the D-shell with $L=2$, the orbital wave
function is a tensor constructed from {\em two} coordinates. One convenient
approach is therefore to consider permutations of four objects of $\rho,\lambda$
type. In a suitable basis of $2^4=16$ combinations, eight are symmetric under
$\hat P(12)$ and eight are symmetric under $\hat P(23)$. Examining their
intersection, one finds two independent combinations that are symmetric under
both permutations. These will be discussed in the section devoted to the
$L=2$ D-shell baryons.

\begin{figure}[h]
\begin{center}
\includegraphics[width=14cm]{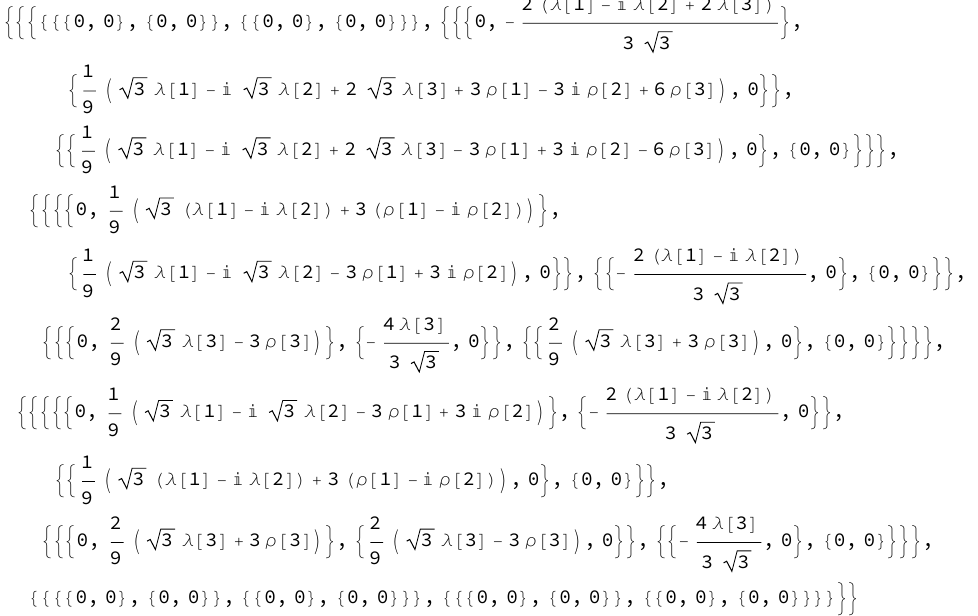}
\caption{ Spin-tensor wave function for $ | N^*, J=3/2,Jz=1/2,S=1/2 > $ .}
\label{fig_wfN31}
\end{center}
\end{figure}

\begin{figure}[h]
\begin{center}
\includegraphics[width=14cm]{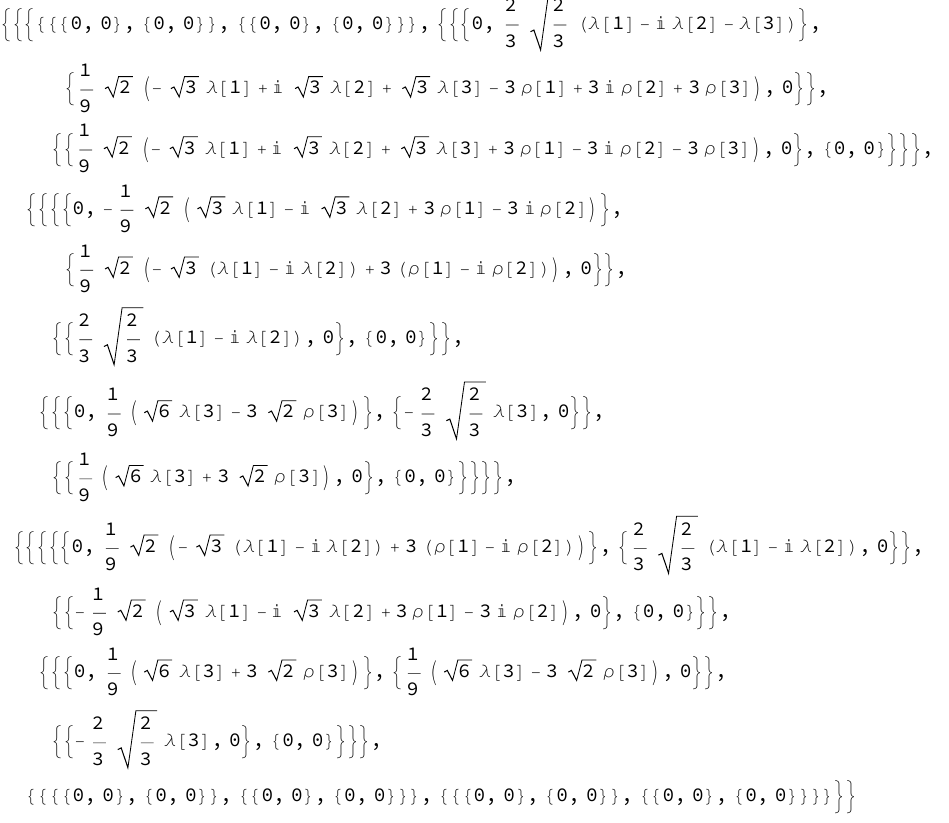}
\caption{Spin-tensor wave function for $ | N^*, J=1/2,Jz=1/2,S=1/2 > $}
\label{fig_wfN11}
\end{center}
\end{figure}

Finally, converting these symbolic expressions in the {\em good basis} into
explicit wave functions ( including orbital dependence, spin, and flavor
indices) proceeds in a standard way; see
Ref.~\cite{Miesch:2023hvl}. From a mathematical perspective, this amounts to
mapping the 8- (or 16-)dimensional good-basis space back to the full monomial
space. This is accomplished by systematic use of the spin-tensor representation
of the orbital-spin-isospin wave functions. The symbolic objects
$S_\rho,S_\lambda$ are replaced by their explicit forms in coordinate, spin, and
isospin variables, after which \textsc{Mathematica} generates a table with
indices $s_1,s_2,s_3,i_1,i_2,i_3$. The resulting coefficients are linear
functions of the $\vec\rho$ and $\vec\lambda$ coordinates; examples are shown in
Fig.~\ref{fig_wfN31} and Fig.~\ref{fig_wfN11}.

\section{The D-shell ($L=2$) baryons}

Representation theory of angular momentum tells us that one can construct
positive-parity excited nucleon states with the following $J^P$ assignments,
\[
L^\pi\otimes S
=
2^+\otimes \frac12,\frac32
=
\bigg(\frac32^+,\frac52^+\bigg),
\;
\bigg(\frac12^+,\frac32^+,\frac52^+,\frac72^+\bigg).
\]
However, the explicit construction of wave functions that are symmetric under
quark permutations requires additional care. Such wave functions have been
discussed, for example, in Ref.~\cite{Capstick:2000qj}. A fully systematic
treatment is most naturally achieved using representations of the permutation
group $S_3$~\cite{Miesch:2023hvl}.

Since the orbital part of the wave function for $L=2$ consists of symmetric
tensors constructed from the coordinate vectors $\rho^i$ and $\lambda^i$, there
are three independent tensor structures,
\[
\rho^i\rho^j,\qquad
\lambda^i\lambda^j,\qquad
(\rho^i\lambda^j+\lambda^i\rho^j).
\]
The first two are symmetric under the transposition $(12)$, while the last is
antisymmetric. Under the permutation $(23)$ these structures mix according to a
nontrivial matrix. To construct fully symmetric (physical) wave functions, these
orbital tensors must therefore be combined with spin and isospin wave functions
of appropriate symmetry.

For total spin $S=\frac32$, the spin wave functions are fully symmetric (for
example $\uparrow\uparrow\uparrow$). In this case one must construct the $S_3$
representation from three objects (two coordinates and one isospin factor) which
we denote symbolically as $X^{A1}X^{A2}X^{A3}$ with binary indices
$A=\{\rho,\lambda\}$. As shown earlier, this procedure yields a single fully
symmetric combination, given in Eq.~(\ref{eqn_sym_in_3}).

For total spin $S=\frac12$, there are two independent spin structures,
$S^\rho$ and $S^\lambda$. The most direct way to construct the corresponding
wave functions is to build symmetric representations of {\em four}
spinor-like objects,
\[
X^{A1}X^{A2}X^{A3}X^{A4},
\]
again with binary indices $A=\rho,\lambda$. There are $2^4=16$ such combinations,
half of which are symmetric and half antisymmetric under the permutation $(12)$.
The $16\times16$ matrix representing the permutation $(23)$ (not shown) has eight
symmetric and eight antisymmetric eigenvectors. As demonstrated in the Appendix,
there are precisely two combinations that are symmetric under {\em all}
permutations. These can be written as
\be
\lambda^i \lambda^j S_\lambda I_\lambda
+\rho^i \lambda^j S_\lambda I_\rho
+\lambda^i \rho^j S_\rho I_\lambda
+\rho^i \rho^j S_\rho I_\rho,
\ee
\be
\lambda^i \lambda^j S_\rho I_\rho
-\rho^i \lambda^j S_\lambda I_\rho
-\lambda^i \rho^j S_\rho I_\lambda
+\rho^i \rho^j S_\lambda I_\lambda.
\ee
Linear combinations of these two structures generate the tensor-excited states
with total spin $S=\frac12$. For states with definite $J$ and $J_z$, one then
proceeds in the standard way by evaluating the appropriate Clebsch-Gordan
coefficients.

\end{subappendices}

\chapter{Pentaquarks}
\section{Why are pentaquarks important?}

In the Introduction we mentioned the origins of the
recent revolution in hadronic spectroscopy, triggered by the
discovery of $c\bar c q\bar q$ tetraquarks hiding among charmonium states.
Today, such tetraquark states listed in the PDG are nearly {\em as numerous as}
the genuine charmonium levels themselves. These states may be viewed as the
addition of a $\bar q q$ pair (with $\sigma$- or $\pi$-like quantum numbers)
to a compact $\bar c c$ core. In close analogy, adding an extra $\bar q q$ pair
to a baryon naturally leads to a rich spectrum of pentaquark states.
The experimental and theoretical history of pentaquarks is full of puzzles,
which we will not attempt to review exhaustively here.

Although this may not be immediately apparent, there exists a substantial
"theoretical barrier'' between the spectroscopy of mesons, baryons, and
tetraquarks on the one hand, and that of pentaquarks on the other.
In the former cases, color indices can be factorized in the wave function,
whereas in pentaquarks (and larger multiquark systems) this is no longer possible.
As a result, the construction of color-spin-flavor wave functions leads to a
much larger monom space, and enforcing Fermi statistics becomes
significantly more challenging. Developing systematic tools to address this
problem is a central goal of the present work.

In our recent paper~\cite{Miesch:2025wro}, we studied light-quark pentaquarks in
detail and derived explicit antisymmetric wave functions for their S- and
P-shell states. The latter are essential for investigating baryon-pentaquark
mixing, a phenomenon that may help resolve several long-standing puzzles,
including the flavor asymmetry of the antiquark sea and the unexpectedly strong
orbital motion observed in nucleons. The latter issue emerged prominently in
connection with the so-called "spin puzzle.'' The naive expectation that the
nucleon spin is carried entirely by valence quarks was disproved by experiment,
and orbital angular momentum is now believed to provide a major contribution.

In a recent study~\cite{Miesch:2025vas}, we investigated whether such orbital
motion could be induced by a deuteron-like mechanism through mixing with
D-shell baryons. We concluded that this scenario is highly unlikely, even when
assuming unrealistically strong tensor forces. In contrast, in the present
work devoted to admixtures of pentaquark states, the available data on orbital
motion appear to be reproduced successfully.

A broader lesson is that pentaquarks are perhaps less important as isolated
states than as admixtures to ordinary baryons. Already forty years ago,
Isgur and collaborators~\cite{Kokoski:1985is,Geiger:1989yc} argued that adding an
extra $\bar q q$ pair with $\sigma$ (vacuum) quantum numbers requires it to be in
a $^3P_0$ state ($J=0$, $S=1$, $L=1$), following earlier ideas such as those in
Ref.~\cite{LeYaouanc:1972vsx}. This insight highlighted why some amount of orbital
motion in nucleons may be unavoidable.

In those early studies, pentaquarks were modeled effectively as multiple
baryon-meson states. Subsequent developments along these lines were pursued
in Refs.~\cite{Santopinto:2007aq,Bijker:2009up} and others, leading to what is now
known as the {\em unquenched constituent quark model} (UCQM). These works led to
the realization that all baryons, including the nucleon, contain a significant
admixture (of order $\sim 40\%$) of five-quark components. Remarkably, many
textbook results from the three-quark constituent model (such as magnetic
moments) remain intact due to cancellations within the five-quark sector, while
new phenomena, such as orbital angular momentum, can be naturally explained.

A different line of thought pursued the inclusion of a "pion cloud'' around
the nucleon. An extreme realization of this idea is the Skyrmion model, in which
the $qqq$ core is eliminated altogether and baryons emerge as solitons of
semiclassical pion fields. Less radical approaches introduce pions through
diagrams derived from effective chiral Lagrangians.

In the present work, we treat the addition of $\sigma$-like and $\pi$-like quark
pairs on equal footing, as required by chiral symmetry. The relations between
the corresponding admixture operators date back to the 1950s, such as the
Goldberger-Treiman relation and the Gell-Mann-Levy linear sigma model
\cite{Gell-Mann:1960mvl}.

For completeness, we note that several other groups continue to study the
five-quark Fock component of the nucleon with similar goals. Perhaps the closest
in spirit to our approach is Ref.~\cite{An:2019tld}, where states of the $L=1$
pentaquark shell are analyzed, including their mixing with the nucleon via the
$^3P_0$ model. As the reader will see, our construction is technically quite
different: we derive pentaquark wave functions directly from the requirements of
Fermi statistics, rather than from combinations of color-spin-flavor group
representations. The latter approach is nevertheless discussed in the Appendices
for comparison and completeness. More importantly, in contrast to
Ref.~\cite{An:2019tld}, our treatment of nucleon-pentaquark mixing explicitly
enforces chiral symmetry.

Experimental searches for light pentaquarks have been carried out for decades,
but so far no universally accepted signals have emerged. A notable attempt was
the 2003 report by the LEPS collaboration~\cite{LEPS:2003wug} of a
$\Theta=uudd\bar s$ resonance with
\[
M_\Theta=1.54\pm0.01~\mathrm{GeV},\qquad \Gamma_\Theta<25~\mathrm{MeV}.
\]
Although this signal was observed in a few other experiments, it was not
confirmed by subsequent measurements with higher statistics. For a recent
review of the experimental situation and future prospects, see for example
Ref.~\cite{Amaryan:2025muw}. The chiral soliton model~\cite{Diakonov:2012yi}
predicted such a light $\Theta$ as a member of an antidecuplet multiplet, and
many alternative models were proposed; see Ref.~\cite{Praszalowicz:2024mji} for a
recent discussion.

At that time, we proposed a schematic quark-diquark symmetry
\cite{Shuryak:2003zi}, in which the pentaquark was approximated as a three-body
system composed of two $good$ diquarks, $(ud)^2\bar s$, rotating in a P-wave.
Such a picture suggested near-degeneracy with P-wave excited decuplet baryons,
implying a higher mass scale, $M\sim1.9~\mathrm{GeV}$.

Needless to say, our present approach is far removed from such simplified
models. We now work with a complete set of wave functions derived explicitly
below. Although at present no firm identification of the resulting states with
known resonances can be made, we place confidence in the underlying principles:
Fermi statistics and the hyperdistance approximation. In this sense, we follow
the same strategy as in our previous studies of excited baryons, tetraquarks,
and hexaquarks.

The results obtained here serve as intermediate stepping stones toward a
quantitative understanding of five-quark Fock components in the nucleon.
As usual, the analysis of complicated multiquark wave functions begins with
states of extreme quantum numbers, such as the channel with maximal spin,
$J^P=5/2^+$. Before turning to its theoretical description, let us note that it
may correspond to the resonance $N^*(2000)\,5/2^+$, observed in channels such as
$N\pi$, $N\sigma$, $\Delta\pi$, and $\Lambda K^*$; see the multichannel analysis
of Ref.~\cite{Anisovich:2011fc}. This state is the second resonance with these
quantum numbers, lying well above the first $5/2^+$ state, $N^*(1680)$, which is
traditionally identified as a $qqq$, $L=2$ D-shell excitation. Moreover, it lies
close to the negative-parity state $N^*(2060)\,5/2^-$, which may be interpreted
as its chiral partner.

Finally, we comment briefly on pentaquarks containing charm quarks. Our use of
the hyperdistance approximation for fully charmed tetraquarks
\cite{Miesch:2024fhv} successfully reproduces the observed level splittings.
Unfortunately, this approach cannot be straightforwardly extended to the
$uudc\bar c$ pentaquarks observed by LHCb, and such states will therefore not be
discussed here.



\section{The five-body kinematics}

The coordinates of five bodies, with the center-of-mass (CM) motion removed, are
described by four three-dimensional Jacobi coordinates\footnote{A general
Mathematica statement generating Jacobi coordinates for arbitrary $n$ is given
in the Appendix to this chapter.}, denoted
$\vec\alpha,\vec\beta,\vec\gamma,\vec\delta$:
\ba \label{eqn_penta_Jacobi}
 \vec X_1 &=& ( 15 \sqrt{2} \vec\alpha + 5 \sqrt{6} \vec\beta +
    5 \sqrt{3} \vec\gamma + 3 \sqrt{5} \vec\delta)/30, \nonumber \\
 \vec X_2 &=&
  (- 15 \sqrt{2} \vec\alpha + 5 \sqrt{6} \vec\beta +
    5 \sqrt{3} \vec\gamma + 3 \sqrt{5} \vec\delta )/30,  \nonumber\\
 \vec X_3 &=&   ( - 10 \sqrt{6} \vec\beta + 5 \sqrt{3} \vec\gamma +
    3 \sqrt{5} \vec\delta)/30,  \nonumber \\
 \vec X_4 &=&   ( - 5 \sqrt{3} \vec\gamma + \sqrt{5} \vec\delta)/10,  \nonumber\\
 \vec X_5 &=&   - (2 /\sqrt{5}) \vec\delta
\ea
with the inverse relations
\ba
\vec\alpha &=& ( \vec X_1 - \vec X_2)/\sqrt{2}, \nonumber \\
\vec\beta &=& ( \vec X_1 + \vec X_2 - 2 \vec X_3)/\sqrt{6}, \nonumber \\
\vec \gamma &=&
 \sqrt{3}/6 (\vec X_1 + \vec X_2 + \vec X_3 -
    3 \vec X_4), \nonumber \\
\vec \delta &=&
 \sqrt{5}/2 ( \vec X_1 + \vec X_2 + \vec X_3 + \vec X_4)
\ea

The Jacobi coordinates are defined sequentially, with the first two coinciding
with the $\vec\rho$ and $\vec\lambda$ coordinates used in baryon spectroscopy.
Since the CM is placed at the origin, $\sum_{i=1}^5 \vec X_i=0$, the last
coordinate $\vec\delta$ can be re-expressed solely in terms of $\vec X_5$, which
we identify with the antiquark coordinate.

Each three-dimensional vector can be parameterized by its magnitude and two
polar angles $\theta_i,\phi_i$, with $i=1,2,3,4$. The four-dimensional space of
their magnitudes $|\alpha|,|\beta|,|\gamma|,|\delta|$ is described by the
hyperdistance in 12 dimensions,
\be
Y^2=\vec\alpha^2+\vec\beta^2+\vec\gamma^2+\vec\delta^2 ,
\ee
together with three additional angular variables
$\theta1_\chi,\theta2_\chi,\phi_\chi$. In total, the configuration space is
parameterized by $Y$, eight ordinary angles, and three $extraordinary$
ones.

The integration measure factorizes into a product of four three-dimensional and
one four-dimensional solid angles,
\be
d\Omega_{11}= d\Omega_\alpha d\Omega_\beta d\Omega_\gamma d\Omega_\delta d\Omega_\chi ,
\ee
where
$d\Omega_\chi=\rm sin^2(\theta2_\chi)\, sin(\theta1_\chi)\,
d\theta2_\chi\, d\theta1_\chi\, d \phi_\chi$,
and the usual
$d\Omega_i=\rm \sin(\theta_i)d\theta_i d\phi_i$.
The generic wave functions may then be written in a factorized form,
\be
\Psi= R_n(Y)\, F(\theta1_\chi,\theta2_\chi,\phi_\chi)
\prod_{i=1}^4 Y_{l_i,m_i}(\theta_i,\phi_i) .
\ee

\section{Building the pentaquark wave functions}
\subsection{Pentaquarks with maximal spin $S=5/2$}

Before discussing the pentaquark case $q^4\bar q$, let us briefly recall the
logic used in Ref.~\cite{Miesch:2024vjk} for the related system of hexaquarks
($q^6$). As usual, one starts from the simplest configurations, namely those
with maximal possible spin: $S=3$ for hexaquarks and $S=5/2$ for pentaquarks.

Recall that the $S=3$ hexaquark is the only case for which a fully antisymmetric
wave function has been derived analytically~\cite{Kim:2020rwn}. This construction
used traditional group-theoretical methods, combining available color and
flavor representations, and resulted in an explicit expression as a sum of five
terms with different color-flavor structures. This state is widely believed to
have been observed experimentally~\cite{WASA-at-COSY:2011bjg} as the resonance
$d^*(2380)$ in the reaction
\be
p+n \rightarrow d +\pi^0 +\pi^0 .
\ee
Its width, $\Gamma_{d^*}\approx 70\,\mathrm{MeV}$, is significantly smaller than
that of the $\Delta$ baryon, $\Gamma_\Delta\approx 115\,\mathrm{MeV}$, which was
one of the arguments against interpreting it as a loosely bound $\Delta\Delta$
state. Moreover, a deuteron-like $\Delta\Delta$ interpretation would require
a binding energy of order $\approx 84\,\mathrm{MeV}$, which appears uncomfortably
large.

By analogy, the simplest pentaquark configurations are those with maximal spin
$S=5/2$. If the isospin is also maximal, the only remaining nontrivial structure
is the color wave function, which cannot be fully antisymmetric because the
height of the color Young tableau is limited to $N_c=3$. Nevertheless,
antisymmetric pentaquark states with $S=5/2$ and $I=3/2$ or $I=1/2$ can still be
constructed, even using the conventional procedure of combining color and flavor
representations~\cite{Miesch:2023hvl}. 

The wave function for the $S=5/2$, $I=1/2$ state is given explicitly in
\cite{Miesch:2023hvl} using a monom basis using Wolfram
Mathematica,  and also obtaining a wave function with 144 nonzero components. Comparing permutation-group methods, with good agreement.
Several properties of the maximal-spin $S=5/2$ states, such as
\be
\langle \sum_{i>j}\vec\lambda_i\vec \lambda_j \rangle=-40/3
\ee
and
\be
\langle \sum_{i>j}(\vec\lambda_i\vec \lambda_j)(\vec S_i \vec S_J) \rangle =-10/3 ,
\ee
can be obtained analytically. Remarkably, these relations continue to hold for
other, much more complicated pentaquark states discussed below. Such numerical
checks proved to be very useful throughout our analysis.

The distributions of flavor and color between the four-quark core and the
antiquark are highly nontrivial. For example, for the $S=5/2$, $I=1/2$ state the
mean values of the isospin projections are
\be
\langle I_z(q) \rangle=1/4, \qquad
\langle I_z(\bar q)\rangle=-1/2 .
\ee
Note that the sum rule
$4\langle I_z(q)\rangle+\langle I_z(\bar q)\rangle=1/2$
is satisfied, as expected.

\subsection{Color-spin-flavor wave functions for S-shell ($L=0$) pentaquarks from the permutation group $S_4$}

A significant technical difficulty in constructing the theory of multiquark
hadrons is the requirement to enforce complete antisymmetry of the quark
wave functions dictated by Fermi statistics. The traditional approach proceeds
by successively adding quarks (or quark pairs) in all possible representations
of the relevant symmetry groups ( $SU(3)$ for color and $SU(2)$ for spin and
flavor) and combining them step by step. A systematic description of this
procedure is given in the Appendix to this chapter.

\begin{table}[b!]
    \centering
    \begin{tabular}{|c|ccc|}   \hline
         I \textbackslash \, S &0 & 1 & 2   \\
         \hline
         0&0& 1& 0 \\
         1& 1&1 & 1 \\
         2&0&1&0 \\ \hline
    \end{tabular}
    \caption{Number of states (per choice of $I_z$ and $S_z$) for each combination
    of total spin and isospin for four quarks. Adding an antiquark shifts each
    state's spin up or down by $1/2$.}
    \label{tab:4q}
\end{table}

The wave functions derived here are based on the novel technique developed in
Ref.~\cite{Miesch:2024vjk}. For pentaquarks, this method relies on constructing
antisymmetric representations of the four-quark core using the permutation
group $S_4$. We begin with Table~\ref{tab:4q}, which lists the number of core
states as a function of total isospin $I$ and spin $S$. As expected, the table
is symmetric under interchange of $I$ and $S$.

After adding the antiquark via the standard procedure, one obtains the full set
of pentaquark states summarized in Table~\ref{tab_L0}. Because the antiquark is
not subject to exchange symmetry with the quarks and may carry either spin or
isospin projection, the number of available states increases substantially.
The antisymmetric representations of the permutation group $S_4$ for four light
quarks are constructed following the procedure of Ref.~\cite{Miesch:2024vjk}.

The standard monom spacefor S-shell ($L=0$) pentaquarks has
color-spin-flavor dimension
\be \label{eqn_monoms}
d_{monoms}=3^6 \times 2^5 \times 2^5=746496 ,
\ee
where the antiquark corresponds to two boxes in the color Young tableau, giving
rise to the factor $3^6$.
For P-shell ($L=1$) states (which will be needed below for the study of
baryon-pentaquark mixing)  this dimension is multiplied by four, reflecting the
number of Jacobi coordinates that enter linearly. As a result, the monom space
dimension exceeds one million.

At first sight, writing operators as matrices in such a large space may appear
prohibitive, even with the aid of Mathematica. Fortunately, this is not the
case, for two reasons. First, in each of the four subspaces ( orbital, color,
flavor, and spin) one can define a {\em good basis} based on the corresponding
Young tableaux. In practice, this allows one to work with permutation matrices
of dimension much smaller than $d_{monoms}$,
\[
N_{GB}=N_{GB}^{\rm color}\times N_{GB}^{\rm spin}\times N_{GB}^{\rm flavor}.
\]
The two basic generators of the permutation group are constructed using the
\emph{KroneckerProduct} operation in Mathematica and subsequently diagonalized.
The desired wave functions correspond to antisymmetric eigenvectors that are
common to both generators. Since all $n!$ elements of $S_n$ can be generated from
these two, such eigenvectors are antisymmetric under all permutations.

For spherically symmetric S-shell ($L=0$) states, the relevant antisymmetric
color-spin-flavor subspaces are listed in Table~\ref{tab_L0}. The integers give
the multiplicities of the states, while the numbers in parentheses indicate the
dimensions of the corresponding good basis used in each $(S,I)$ sector.
Although the resulting permutation matrices are still too large to display
explicitly, their dimensions are far smaller than that of the full monom space
(\ref{eqn_monoms}), making their manipulation straightforward.

As an illustrative example, consider the pentaquarks with maximal spin
$S=5/2$. In this case, the only nontrivial Young tableaux are those associated
with color and isospin. Using standard $SU(2)$ spin addition, one finds five
distinct ways to combine the quarks into total isospin $I=1/2$, implying a
"good basis'' of dimension five in isospin space. The color Young tableau that
yields a color singlet consists of two closed columns,
$\epsilon_{abc}\epsilon_{def}$, as in the hexaquark case. Unlike the latter,
however, permutations act only on the four quarks, with indices $a,b,c,d$.
This leads to a color "good basis'' of dimension three, and hence a total
dimension of $3\times5=15$.

The two generators of $S_4$ are then constructed and diagonalized within this
space. A single common eigenvector with eigenvalue $-1$ is found, corresponding
to the fully antisymmetric state. Thus, there exists a unique pentaquark state
with $S=5/2$ and $I=1/2$, mirroring the situation encountered for maximal-spin
hexaquarks.

\begin{table}[h!]
    \centering
    \begin{tabular}{|c|c|c|c|} \hline
  I/S    & 1/2 & 3/2   & 5/2  \\  \hline
  1/2 & 3 (75) & 3 (60)   & 1 (15) \\
  3/2 & 3 (60) & 3 (48) &  1 (12)  \\
  5/2 & 1 (15)  & 1 (12) & 0 \\ \hline
    \end{tabular}
    \caption{Spin-isospin table of antisymmetric S-shell ($L=0$) pentaquark
    states. The integers denote state multiplicities, while the numbers in
    parentheses give the dimensions of the corresponding good basis.}
    \label{tab_L0}
\end{table}

Once all relevant antisymmetric states have been identified, the most efficient
strategy is to return to the universal description in the full monom basis of
dimension (\ref{eqn_monoms}). Although vectors and operators in this space
formally involve matrices of size
$d_{monoms}\times d_{monoms}$, {\em Wolfram Mathematica} allows one to handle
them efficiently using standard expressions such as
$vector^\ast\!\cdot matrix\cdot vector$\footnote{While the \texttt{SparseArray}
format can accelerate computations for large vectors and matrices, some general
commands (such as \texttt{Expand}, \texttt{Conjugate}, and \texttt{ComplexExpand})
do not function correctly in this format in version 14.1. For this reason, we
found it safer to work in the \texttt{Normal} format for algebraic manipulations.}.

In practice, the number of nonzero components in the wave functions is much
smaller than $d_{monoms}$, so these objects are highly sparse. While it is
clearly impractical to publish the full wave functions, it is nevertheless
instructive to compare the number of nonzero terms in their monom
representations. As anticipated, the smallest numbers occur for maximal spin
$S=5/2$: the state with $I=3/2$ has only 252 nonzero components, while the
$I=1/2$ state has 708. Moving to lower spin and then to P-shell ($L=1$) states
leads to a dramatic increase. For example, the 13 states with
$L=1$, $S=1/2$, $I=1/2$ used below for mixing with nucleons together contain
187\,200 nonzero components.

For $I=1/2$ and $S=1/2,3/2$, we find two triplets of states in each case, in
agreement with expectations based on Young tableaux. These triplet states are
not uniquely defined; only the eigenstates of specific physical operators are
physically meaningful. The sums over all ten quark-quark and quark-antiquark
pairs,
$(\lambda\lambda)=\sum_{i>j}\vec \lambda_i \vec\lambda_j$  and
$SS=\sum_{i>j}\vec \sigma_i \vec\sigma_j$, are related to the Casimir operators
of color and spin and are therefore diagonal in this basis.

To lift the remaining degeneracies, one needs a Hamiltonian. As an example, we
consider the one-gluon-exchange color-spin interaction summed over all ten
pairs,
\be \label{eqn_H}
H_{\lambda\sigma}=-C_{\lambda\sigma}\sum_{i>j}
(\vec\lambda_i \vec\lambda_j)(\vec S_i \vec S_j) .
\ee
We evaluated the full Hamiltonian matrix between all states and diagonalized it.
The resulting eigenvalues for the seven states with $I=1/2$ (the $S=1/2$ triplet,
the $S=3/2$ triplet, and the $S=5/2$ singlet)  are listed in
Table~\ref{tab_L0_properties}. As expected, the maximal-spin state $S=5/2$ has
the highest energy. If the resonance $N^*(2000)\,5/2^+$ is indeed identified
with this pentaquark, the remaining states should lie somewhat below
$2\,\mathrm{GeV}$.

(An alternative Hamiltonian, used for example in Ref.~\cite{An:2019tld}, is the
flavor-spin interaction associated with pion exchange, following Ripka and
Glozman,
\be
H_{\tau\sigma}=-C_{\tau\sigma}\sum_{i>j}
(\vec\tau_i \vec\tau_j)(\vec S_i \vec S_j) .
\ee
Here $\vec S=\vec\sigma/2$, whereas for color and flavor we use Gell-Mann and
Pauli matrices without the factor $1/2$. Using the same wave functions, we also
constructed and diagonalized this Hamiltonian. Since its spectrum is not
qualitatively different from that of the color-spin Hamiltonian, we do not
display it explicitly.)

By symmetry, the mean value of the $S_z$ component is the same for all four
quarks, but differs for the antiquark. By construction, the expectation values
satisfy the sum rule
\be \label{eqn_S_sumrule}
\sum_{i=1..5}\langle S_z(i) \rangle
=4\langle S_z^{1}\rangle+\langle S_z^{5}\rangle
= S_z^{\rm total} .
\ee

\begin{table}[h!]
    \centering
    \begin{tabular}{|c|c|c|c|c|c|c|c|} \hline
  S & 1/2  & 1/2 & 1/2 & 3/2 & 3/2 & 3/2  & 5/2 \\  \hline
$\lambda \lambda$ & -40/3  & -40/3 & -40/3 &  -40/3  &  -40/3  &  -40/3  & -40/3  \\
$SS$ & -3/2  & -3/2 & -3/2 & 0 & 0 & 0 & 5/2  \\
$H_{\lambda \sigma}/C_{\lambda \sigma}$ &
-4.66 & -1.44 & 2.77 & -3 & 1/3 & 10/3 & 10/3 \\
 \hline
    \end{tabular}
    \caption{Properties of the seven antisymmetric $L=0$, $I=1/2$ pentaquark
    states: three $S=1/2$, three $S=3/2$, and one $S=5/2$.}
    \label{tab_L0_properties}
\end{table}

The mean spin and isospin values of the quarks and the antiquark can be compared
with expectations from naive models. For instance, in the diquark-based picture
of Ref.~\cite{Shuryak:2003zi}, two spin-isospin zero diquarks imply
$\langle S_z^{q}\rangle\approx \langle I_z^{q}\rangle\approx 0$, with all spin
and isospin carried by the antiquark,
$\langle S_z^{\bar q}\rangle\approx \langle I_z^{\bar q}\rangle\approx 1/2$.
Inspection of Table~\ref{tab_L0_properties} shows that while some states exhibit
either vanishing quark spin or vanishing quark isospin, both never vanish
simultaneously. This provides yet another illustration of how oversimplified
such naive models can be.

An alternative naive picture, in which spin and isospin are shared equally among
all five constituents, is likewise not supported by the present, more detailed
analysis.

In summary, the explicit construction of fully antisymmetric S-shell pentaquark
wave functions reveals a spectrum and internal structure that are considerably
richer than suggested by simple diquark-based or equal-sharing models. Even in
the most constrained case of maximal spin, the allowed states are unique and
highly nontrivial, while for lower spins multiple states appear whose physical
distinction only emerges after diagonalization of realistic interaction
Hamiltonians. The resulting distributions of spin and isospin between the quark
core and the antiquark demonstrate that neither extreme localization on the
antiquark nor uniform sharing among constituents is realized. This underscores
the necessity of enforcing Fermi statistics exactly and treating color, spin,
and flavor dynamics on equal footing when analyzing pentaquark structure.

\subsection{Pentaquarks of the $L=1$ shell}

Because the $\sigma$ and $\pi$ operators that introduce the corresponding mesons
(to be used below for $unquenching$ of the nucleon)
necessarily carry nonzero orbital angular momentum,
P-shell pentaquark states can mix with the nucleon.

For $L=1$ pentaquarks, the dimension of the good basis is increases by
a factor of four relative to the S-shell case. This factor simply reflects the
number of Jacobi coordinate vectors
$\vec\alpha,\vec\beta,\vec\gamma,\vec\delta$ that can appear linearly in the
orbital wave function.
Correspondingly, one finds approximately four times as many fully antisymmetric
states; these are listed in Table~\ref{tab_L1}.

The pentaquark states that can mix with the nucleon must have isospin $I=1/2$
and spin $S=1/2$ or $S=3/2$.
The complete list of states compatible with Fermi statistics is given in
Table~\ref{tab_L1}, amounting to 24 states in total%
\footnote{In Ref.~\cite{An:2019tld} the number of $L=1$ compatible states is
quoted as 17, and they are not organized into fully Fermi-statistics-compliant
combinations. We also disagree with the way their energies are derived from the
Hamiltonian.}.

Even the largest good basis encountered here, of dimension 300, is still many
orders of magnitude smaller than the full monom basis of dimension 746496.
The corresponding vectors are sparse, though not extremely so: the 13 states
with $S=1/2$ contain 187\,200 nonzero components, while the 11 states with
$S=3/2$ contain 198\,000 nonzero components.
P-shell wave functions are therefore substantially more complicated than their
$L=0$ counterparts. In particular, the amplitudes are no longer simple numbers
but explicit functions of 12 variables: eight angles
$\theta_i,\phi_i$ ($i=1,2,3,4$) and four radial coordinates.
The angular-momentum operators involve derivatives with respect to these angles,
which can nevertheless be applied directly to our sparse vectors of very large
dimension.

\begin{table}[h!]
    \centering
    \begin{tabular}{|c|c|c|c|} \hline
  I/S    & 1/2 & 3/2   & 5/2  \\  \hline
  1/2 & 13 (300) & 11 (240)   & 3 (60) \\
  3/2 & 11 (240) & 10 (192) &  3 (48)  \\
  5/2 & 3 (60)  & 3 (48) & 1 (12) \\ \hline
    \end{tabular}
    \caption{Antisymmetric pentaquark states with $L=1$, total spin $S$ and
    isospin $I$. As in Table~\ref{tab_L0}, the first numbers indicate the number
    of states, while the numbers in parentheses give the dimension of the
    corresponding good basis.}
    \label{tab_L1}
\end{table}

Some operators, such as the color operator
\[
\big\langle\sum_{i>j}\vec\lambda_i\vec\lambda_j\big\rangle = -\frac{40}{3},
\]
have identical expectation values for all states. This follows from the fact
that the four-quark core always transforms according to the $[211]$ color Young
tableau.

\begin{figure}[h!]
    \centering
    \includegraphics[width=0.55\linewidth]{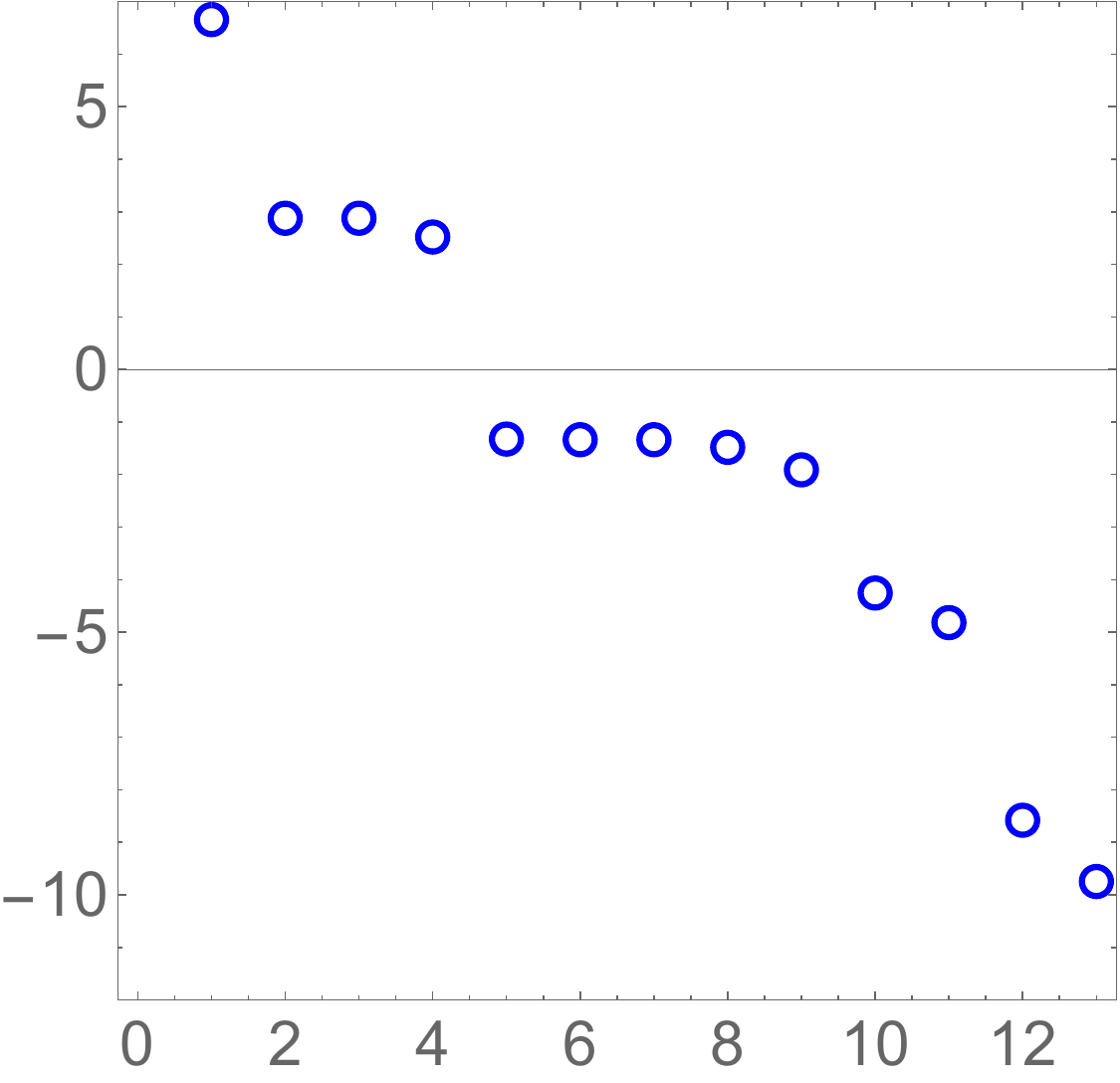}
    \caption{Eigenvalues of the color-spin Hamiltonian for $L=1$, $S=1/2$
    pentaquark states, normalized to the matrix element
    $H_{\lambda\sigma}/C_{\lambda\sigma}$.}
    \label{fig_pentas_L1}
\end{figure}

The eigenvalues of the $13\times13$ color-spin Hamiltonian for the $L=1$
pentaquark shell with $S=1/2$ are shown in Fig.~\ref{fig_pentas_L1}.
Although the resulting spectra are rather similar, the underlying states
themselves can differ substantially.
While it is unlikely that individual pentaquark states can be experimentally
resolved due to their overlapping widths, we will show below that specific
linear combinations of these states constitute the five-quark Fock component of
the nucleon.

As for the $L=0$ pentaquark shell, we evaluated the distributions of various
physical quantities over the four Jacobi solid angles and among the five
constituents.
To avoid excessive clutter, we present results only for three representative
states (1, 7, and 13 of the $S=1/2$ set)   listed in
Table~\ref{tab_3states}.

The four Jacobi coordinates are associated with four pairs of angles
$\theta_i,\phi_i$, and we computed the expectation values of the squared orbital
angular momenta in each corresponding solid angle,
\be
L^2_i f = -\frac{1}{\sin\theta_i}\frac{\partial}{\partial\theta_i}
\Bigl(\sin\theta_i\frac{\partial f}{\partial\theta_i}\Bigr)
-\frac{1}{\sin^2\theta_i}\frac{\partial^2 f}{\partial\phi_i^2},
\ee
with $i=1,2,3,4$.
Symmetry among the quarks in the core implies
$\langle L^2(1)\rangle=\langle L^2(2)\rangle=\langle L^2(3)\rangle$, while the
corresponding value associated with the antiquark (the fourth angle) is
different.
As shown in Table~\ref{tab_3states}, the three selected states exhibit markedly
different distributions.
The lowest-energy state (state~1) carries almost no orbital angular momentum on
the antiquark and almost no spin on the quarks.

We also computed expectation values of the $z$-components $L_z(i)$ for
$i=1,\dots,4$.
All results satisfy the expected sum rules,
\be
\sum_{i=1..4}\langle L^2(i) \rangle = 3L^2(1)+L^2(4)=2,
\ee
\be
\sum_{i=1..4}\langle L_z(i) \rangle = 3L_z(1)+L_z(4)=-1,
\ee
as required for the $L=1$ shell.
For completeness, we recall that the spin and isospin components satisfy the
analogous relations
\[
\sum_{i=1..5}\langle S_z(i) \rangle
=4S_z(1)+S_z(5)=\frac12,\qquad
\sum_{i=1..5}\langle I_z(i) \rangle
=4T_z(1)+T_z(5)=\frac12.
\]

\begin{table}[h!]
    \centering
    \begin{tabular}{|c|c|c|c|} \hline
  state number    & 1 & 7   & 13  \\  \hline
$L^2(1)$ & 0.4965 & 0.6652 & 0.6547 \\
$L^2(4)$ & 0.0258 & 0.0043 & 0.0358 \\
$S_z(1)$ & 0.0963 & 0.0562  & 0.0931 \\
$S_z(5)$ & 0.1146 & 0.2748 & 0.1272  \\
$I_z(1)$ & 0.1523 & 0.0516 & 0.0545 \\
$I_z(5)$ & -0.1094 & 0.2934 & 0.2817 \\ \hline
    \end{tabular}
    \caption{Distribution of selected observables over Fermi-antisymmetric
    pentaquark states with $L=1$, $S=1/2$, and $I=1/2$.
    The operator $L^2(i)$ denotes the squared orbital angular momentum associated
    with the $i$th Jacobi solid angle.
    Mean $z$-components of spin and isospin are shown for a representative quark
    ($i=1$) and for the antiquark ($i=5$).}
    \label{tab_3states}
\end{table}

\subsection{Where pentaquarks become chaotic?} \label{sec_chaotic}

Quantum few-body  systems are known to undergo a
transition to the so-called {\em quantum chaos} regime; see, for example,
Ref.~\cite{Zelevinsky:1996dg} for a review.
This phenomenon is well documented in atomic and nuclear systems with several
particles or holes near closed shells.
A historically important example is the cerium atom, which has four valence
electrons and thus 12 effective coordinates. It was shown in
Ref.~\cite{flambaum1997applystatisticallawssmall} that even its lowest-energy
states display chaotic behavior.

Since this phenomenon is generic, it is natural to expect it to arise in
multiquark systems as well.
Pentaquarks ( the subject of this section)  also involve 12 Jacobi coordinates, the same
number as the cerium atom.
One may therefore anticipate a similar transition to chaos in these systems.
One of us has previously discussed this issue for baryons, pentaquarks, and
their admixture within a light-cone framework, using different wave functions
and basis states~\cite{Shuryak:2019zhv}.

\begin{figure}[t]
    \centering
    \includegraphics[width=0.4\linewidth]{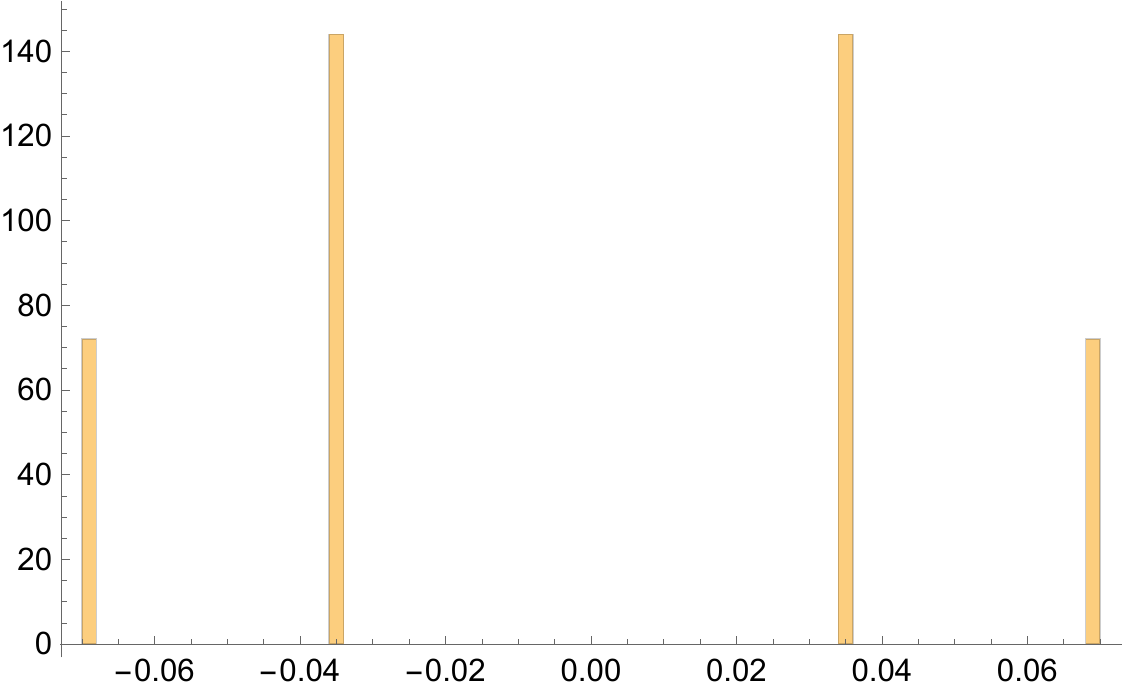}
    \includegraphics[width=0.4\linewidth]{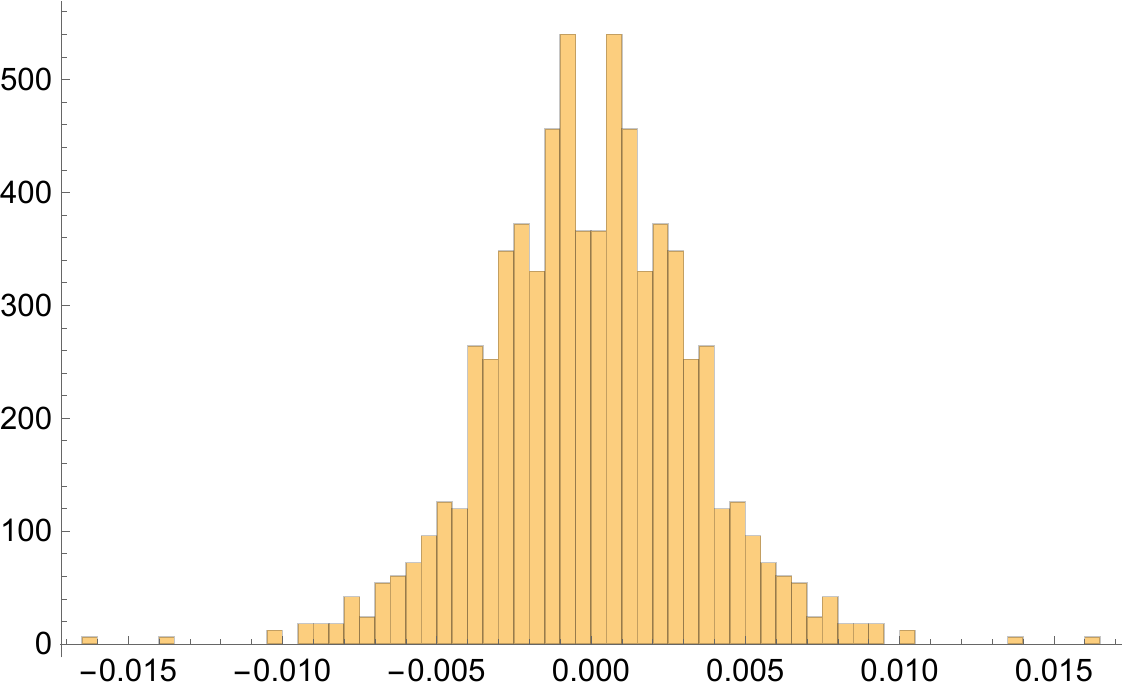}
    \caption{Distribution of monom coefficients for an S-shell state with
    $L=0$, $S=5/2$, $I=1/2$ (upper panel) and for a P-shell state with
    $L=1$, $S=1/2$, $I=1/2$ (lower panel).}
    \label{fig_Hist_L1}
\end{figure}

The simplest signature of quantum chaos is a random (Gaussian-like)
Porter-Thomas distribution of amplitudes of exact Hamiltonian eigenstates when
expanded in a natural basis, such as that of the single-particle Hamiltonian.
In Fig.~\ref{fig_Hist_L1} we show two such distributions for fixed-energy states,
plotted in terms of their monom coefficients.

The upper histogram corresponds to the S-shell state with
$L=0$, $S=5/2$, $I=1/2$, which we obtained both analytically and numerically.
Its wave function consists of three distinct structural terms, multiplied by
permutations.
The histogram confirms that only two amplitude values (up to an overall sign)
appear, each repeated many times.
Since there are $4!=24$ quark permutations and three structural terms, the total
occupancy of the histogram is $72=24\times3$.
This state is therefore highly regular and far from chaotic.

The lower histogram corresponds to one of the P-shell $L=1$ pentaquark states.
Its overall shape is indeed close to a Gaussian, although systematic deviations
are clearly visible.
The most prominent feature is a dip near zero amplitude.
Such dips are often associated with finite-size effects.  (For example in the
Dirac eigenvalue spectrum of lattice or instanton studies of the QCD vacuum.)
In the present case, however, the dip is well understood in chiral random-matrix
models~\cite{Shuryak:1992pi,Verbaarschot:1993pm} in terms of finite volume and
finite matrix size, and has been confirmed in numerous lattice simulations.
The precise origin of the dip in the present histogram remains unclear.
We only note that the total number of nonzero coefficients is about 14\,400 in a
monom basis of dimension 746\,496, so the wave function is still relatively
sparse.

Another notable feature of the lower histogram is the presence of regularly
spaced peaks.
This may signal a quantum system whose phase space supports both regular and
chaotic motion, with the chaotic component being dominant.

Quasi-random expansion coefficients in a given basis constitute only one of many
possible indicators of chaos.
One may also examine probability distributions of single-particle observables,
such as spatial densities or distributions over single-particle energies, and
compare them with predictions of statistical models (for instance, a Fermi gas
characterized by an effective temperature and entropy).
Indeed, even a single many-body eigenstate can admit an effective thermal
description at the level of single-particle observables.
We defer such investigations to future work.

In summary, we have presented evidence that a transition toward quantum chaos
may occur in pentaquarks already at the second excitation level, namely in the
$L=1$ shell.

The onset of chaotic behavior in the $L=1$ pentaquark sector has direct
consequences for baryon-pentaquark mixing.
As the density of antisymmetric five-quark states increases and their wave
functions become broadly delocalized in the underlying basis, the coupling of a
three-quark baryon to the five-quark sector ceases to be dominated by a small
number of well-defined configurations.
Instead, the mixing proceeds through a large set of near-degenerate pentaquark
states with statistically distributed amplitudes.
In this regime, individual mixing matrix elements lose predictive power, while
their ensemble-averaged properties constrained by symmetry, sum rules, and the
structure of the interaction operators,  become the relevant quantities.
Consequently, sizable baryon-pentaquark admixtures arise naturally, even in the
absence of narrow pentaquark resonances, providing a dynamical mechanism for
enhanced five-quark Fock components in low-lying baryons.

\section{Nucleon-pentaquark mixing}

The $unquenching$ of mesons and baryons (namely, attempts to quantify the
four-quark sector of mesons and the five-quark sector of baryons)  has a long
history and remains an open problem.
For several decades, from the 1960s until relatively recently, multiquark
hadrons were largely regarded as "exotica'', a category viewed with skepticism
due to the lack of firmly established experimental evidence.
This situation has changed dramatically in the last decade.

Although this shift has not yet fully permeated the broader community, its
impact can be clearly illustrated by examining one of the central pillars of
QCD spectroscopy: heavy quarkonia.
While bottomonium spectroscopy remains fully consistent with a simple
$\bar b b$ interpretation well described by a Cornell-type potential, states of  charmonium
already deviates significantly from this picture.
The first state unambiguously inconsistent with a pure $\bar c c$ assignment
was the $X(3872)$ with $J^P=1^+$.
Originally grouped under the generic $X,Y,Z$ labels reflecting their unknown
structure, such states are now separated in the PDG into distinct categories.

In particular, axial-vector states with quantum numbers $I=1$, $J^P=1^+$,
which necessarily carry isospin, are interpreted as pion-added systems and
are classified as tetraquarks, denoted $T_{\bar c c 1}$ in the 2024 PDG.
In contrast, vector ($J^P=1^-$) and pseudoscalar ($J^P=0^-$) states with
$I=0$  (which may be interpreted as "$\sigma$-added)  continue to be listed under
the traditional $\psi$ and $\eta_c$ labels.
This asymmetric treatment is unfortunate, particularly in view of the
empirical observation that axial and vector states often appear as
near-degenerate pairs, suggestive of chiral partner relationships.

(Some charmed tetraquarks are often interpreted as molecular, deuteron-like
two-meson bound states located near threshold, such as $DD^*$.
However, this description does not apply universally.
In recent years, a large number of tetraquark states have been observed,
including systems with $cc$ pairs and even fully charmed $\bar c\bar c c c$
tetraquarks at the LHC.
For details on such states and their mixing patterns we refer to
Ref.~\cite{Ferretti:2013faa} and references therein.)

The key point of this detour is the following: the number of known
pion-admixed and sigma-admixed charmonium-like states appears to be
roughly equal.
With four $T_{\bar c c 1}$ states and four additional $\psi$-like states that
cannot be accommodated within the $\bar c c$ spectrum, the observed pattern
suggests an approximate $\pi$-$\sigma$ symmetry, reminiscent of the original
chiral symmetry.
Chiral doubling in heavy-light systems was first discussed in
Refs.~\cite{Nowak:1992um,Bardeen:1993ae,Nowak:2003ra} and later confirmed
experimentally by the BaBar~\cite{BaBar:2003oey} and CLEO~\cite{CLEO:2003ggt}
collaborations.

Turning now to baryons, a similar pattern emerges.
Historically, the appearance of the $\sigma=f_0$ meson in hadronic reactions has
been systematically underestimated.
To emphasize the relevance of chiral mixing in the baryon sector, we list in
Table~\ref{tab_branchings} the branching ratios of well-established nucleon
resonances into $N\pi$ and $N\sigma$ final states.
Although the pion channels generally dominate, this primarily reflects the
larger available phase space for $N\pi$ decays.
When translated into coupling constants, the values of $g_{NN^*\pi}$ and
$g_{NN^*\sigma}$ turn out to be comparable within uncertainties, precisely as
expected from naive chiral symmetry considerations.

\begin{table}[t]
    \centering
    \begin{tabular}{|c|c|c|}
    \hline
     state    &  $N\pi$ & $N\sigma $ \\ \hline
    $N^*(1440)\, 1/2^+$ & 0.55-0.75 & 0.11-0.23 \\ 
   $N^*(1520) \,3/2^+$ & 0.55-0.65 &  $<0.1 $ \\
    $N^*(1535) \,1/2^-$ & 0.32-0.52 &  $0.02-0.1 $ \\
     $N^*(1650)\, 1/2^-$ & 0.5-0.7 &  $ 0.02-0.18$ \\
      $N^*(1675) \,5/2^-$ & 0.38-0.42 &  $0.03-0.07 $ \\ 
    $N^*(1675)\, 5/2^-$ & 0.38-0.42 &  $0.03-0.07 $ \\
        $N^*(1680)\, 5/2^+$ & 0.6-0.7 &  $0.09-0.19 $ \\
      $N^*(1700) \,3/2^-$ & 0.07-0.17 &  $0.02-0.14 $ \\
        $N^*(1710)\, 1/2^+$ & 0.05-0.20 &  $<0.16 $ \\
            $N^*(1720)\, 3/2^+$ & 0.08-0.14 &  $0.02-0.14 $ \\  
       $N^*(1875)\, 3/2^-$ & 0.03-0.11 &  $0.02-0.08 $ \\ 
          $N^*(1880)\, 1/2^+$ & 0.03-0.31 &  $0.08-0.40 $ \\ 
             $N^*(1895)\, 1/2^-$ & 0.02-0.18 &  $<0.13 $ \\
                    $N^*(1900)\, 3/2^+$ & 0.01-0.20 &  $0.01-0.07 $ \\ 
            $N^*(2060)\, 5/2^-$ & 0.07-0.12 &  $0.03-0.09 $ \\   
             $N^*(2100) \,1/2^+$ & 0.08-0.32 &  $0.14-0.35 $ \\
              $N^*(2120) \,3/2^-$ & 0.05-0.15 &  $0.04-0.14 $ \\
            $N^*(2190) \,7/2^-$ & 0.1-0.2 &  $0.03-0.09 $ \\
         \hline
    \end{tabular}
    \caption{Branching ratios of $N\pi$ and $N\sigma$ decays of $N^*(mass) J^P$ resonances (left column),  from PDG24}
    \label{tab_branchings}
\end{table}

After these phenomenological detours, we return to the theoretical framework.
The traditional strategy for estimating the five-quark admixture in baryons
proceeds in two steps:
(i) the formulation of {\em meson-admixture operators} that add a $\bar q q$ pair
to a baryon, followed by
(ii) the projection of the resulting states onto good pentaquark
configurations consistent with Fermi statistics and other symmetry constraints.

Although we have carried out a detailed calculation of nucleon-pentaquark
mixing in Ref.~\cite{Miesch:2025wro}, we do not reproduce those results here.
The reason is that a subsequent study reformulated the same physics on the
light front, yielding results that are more directly connected to experimental
observables such as antiquark parton distribution functions.
Readers interested in these applications are therefore referred to the final
part of this book, where this subject is discussed in detail.

Before turning to the explicit mechanisms of nucleon-pentaquark mixing, it is
useful to place the discussion in the broader dynamical context established in
the previous section.
The appearance of a dense spectrum and quasi-chaotic wave functions in the
$L=1$ pentaquark shell implies that five-quark states form a highly entangled
many-body sector rather than a small set of isolated configurations.
In such a regime, mixing between three-quark baryons and five-quark states is
not governed by proximity to individual resonances, but instead by collective
properties of the pentaquark spectrum and by symmetry-driven selection rules.
This perspective motivates treating baryon-pentaquark mixing as a generic and
robust feature of low-energy QCD, closely tied to chiral dynamics and the
ubiquity of $\bar q q$ excitations in hadronic wave functions.

\chapter{Hadrons made of six, nine, or twelve quarks}
\section{Some evidences for multiquark states}

In this chapter we consider quantum states composed of more quarks.
Dibaryons may be viewed as hexaquarks ($q^6$), three-baryon systems as $q^9$, and
four-baryon systems as $q^{12}$ objects.

The central issue is not merely one of nomenclature.
Rather, the key question is whether such multiquark configurations are
\emph{compact enough} to exist as objects qualitatively distinct from ordinary
nuclear states, either as bound states or as resonances.
An alternative possibility is that compact multiquark configurations appear
only as \emph{admixtures} to conventional nuclear wave functions, in close
analogy with the baryon-pentaquark mixing discussed earlier.

\subsection*{Phenomenological indications}

We begin with several phenomenological examples.
The experiment WASA-at-COSY~\cite{WASA-at-COSY:2011bjg} observed a dibaryon
resonance $d^*(2380)$ with spin $J=3$ in the reaction
\[
p+n \rightarrow d + \pi^0 + \pi^0 .
\]
This state lies approximately $84\,\mathrm{MeV}$ below the $\Delta\Delta$
threshold.
From a nonrelativistic perspective, the spin assignment $J=3$ could arise from
the vector addition $3/2 + 3/2$ of two $\Delta$ baryons, suggesting a molecular,
deuteron-like interpretation.\footnote{
If this were the case, one would expect the width to be dominated by the $\Delta$
width, $\Gamma_\Delta \approx 115\,\mathrm{MeV}$.
The observed width, $\Gamma_{d^*} \approx 70\,\mathrm{MeV}$, is significantly
smaller, which may indicate a more compact structure.
}
Equally well, the state may correspond to an S-shell hexaquark with maximal spin
$6 \times \tfrac12$.
In general, the physical state may be a mixture of both configurations.

Additional hints for multiquark components arise from hadron-nucleus scattering.
While the standard Glauber-type description based on multiple $pN$ scatterings
is highly successful in many applications, it fails for scattering on
$^4\mathrm{He}$.
As noted in~\cite{Dakhno:1984zd}, the total cross section, the diffraction slope,
and the position of the diffractive minimum all deviate from standard
expectations.
The inclusion of a compact 12-quark component was shown to resolve all three
discrepancies~\cite{Dakhno:1984zd,Mosallem:2002ck}.
The parameters extracted in these studies are mutually consistent and indicate
that the 12-quark configuration is not extremely small in size.

\subsection*{Relation to nuclear structure}

From a theoretical standpoint, three- and four-baryon systems naturally include
light nuclei, which have been extensively studied in nuclear physics.
Particular attention has long been paid to the $\alpha$ particle,
$^4\mathrm{He}=ppnn$, which is a "double-magic'' system with closed $1S$ shells
for both protons and neutrons.
Its exceptional binding and compactness have motivated decades of discussion on
whether heavier nuclei such as $^{12}\mathrm{C}$ or $^{16}\mathrm{O}$ exhibit
$\alpha$-cluster substructure.
In the same spirit, one may ask whether the $\alpha$ particle itself should be
described as containing a compact 12-quark $1S$ component
($3$ colors $\times\,2$ spins $\times\,2$ isospins).

Even at the nucleon level, achieving an accurate description of the $\alpha$
particle is challenging.
A naive sum of pairwise nuclear potentials yields a very deep minimum,
approximately $6 \times 50 \sim 300\,\mathrm{MeV}$, but only for a small fraction
of the available four-body configurations.
One way to resolve this tension is through the hyperdistance approximation,
which successfully reproduces the physical binding energy,
$B \approx 28\,\mathrm{MeV}$.
Another approach employs first-principles simulations based on path-integral
Monte Carlo methods.\footnote{
One of us pioneered numerical path-integral techniques over forty years
ago~\cite{Shuryak:1984xr}.
Recently, this approach has been revived to study $\alpha$ preclustering in
heavy-ion collisions~\cite{DeMartini:2020hka}.
}
These studies indicate that four-nucleon preclusters can survive at temperatures
$T \sim 100\,\mathrm{MeV}$, characteristic of hadronic freeze-out.

\subsection*{Limitations of nucleon-based descriptions}

The deeper conceptual issue, however, is not technical.
While modern renormalization-group methods allow low-energy nuclear forces to be
fixed uniquely from nucleon scattering data, this is not yet true for the
short-distance repulsive core.
Different potential models fitted to spectra and phase shifts exhibit a wide
spread in their short-range behavior.
Naively summing six such cores in a multi-nucleon system leads to repulsions of
order several GeV, an outcome that is both unrealistic and highly uncertain.
As has been emphasized repeatedly in the literature, this problem is more
naturally addressed at the quark level.

\subsection*{Beyond diquark models}

A related theoretical issue concerns "good diquark'' models.
As argued already in the Introduction, singling out a small subset of pairwise
interactions (six diquarks in a 12-quark system) while neglecting the remaining
pairs (66 in total) is logically inconsistent and leads to unobserved phenomena.
At the perturbative level, this can be demonstrated explicitly.

At the nonperturbative level, one must explain the origin of strong diquark
binding.
For light $u,d,s$ quarks, such binding is largely attributed to the
instanton-induced 't~Hooft interaction~\cite{Rapp:1997zu,Alford:1997zt}.
Achieving a binding of order $6 \cdot B_{qq}$ in a 12-quark cluster would require
six instantons within the same spatial region, which is problematic given the
diluteness of instantons in the QCD vacuum.
Thus, for light-quark 12-quark systems, the problem is naturally elevated to one
of instanton correlations in the nonperturbative vacuum,  a topic well beyond the
scope of the present work.

\subsection*{Lattice and symmetry considerations}

An alternative approach is provided by lattice QCD.
Although progress remains limited, recent studies of strange and charmed
dibaryons have yielded encouraging results~\cite{Dhindsa:2025gae}.

Finally, for light $u,d$ quarks, spin and isospin share the same $SU(2)$
structure and can be interchanged.
One therefore expects "mirror'' multiquark states-- for example, hexaquarks with
$S=3,I=0$ and $S=0,I=3$.\footnote{
These mirror states need not be degenerate if the Hamiltonian is of color-spin
type, but they would be nearly so in flavor-spin models of the
Riska-Glozman type, dominated by pion exchange.
}

\subsection*{Practical construction of multiquark wave functions}

Although the "monom space'' and spin-tensor notation are conceptually natural,
their size grows rapidly with the number of quarks.
Color, flavor, and spin spaces scale as $3^n$, $2^n$, and $2^n$, respectively,
yielding a total dimension $12^n$.
For hexaquarks ($n=6$), this already reaches the millions, making direct matrix
manipulations impractical.

Fortunately, these dimensions can be drastically reduced using the "best
basis'' approach introduced in~\cite{Miesch:2024vjk}.
This method constructs orthonormal bases corresponding to the required Young
tableaux in each monom subspace, rendering explicit calculations feasible even
for large multiquark systems.

In summary, existing experimental indications and theoretical considerations
suggest that multiquark configurations with six or more quarks cannot be dismissed
as purely academic constructs.
While in some cases they may appear as distinct resonances, in others they are
more naturally interpreted as compact components admixed into conventional
nuclear wave functions.
The limitations of nucleon-based descriptions at short distances, together with
symmetry arguments and nonperturbative dynamics at the quark level, motivate a
systematic construction of multiquark wave functions consistent with Fermi
statistics.
In the following sections we therefore turn to explicit realizations of
hexaquark, nona\-quark, and dodeca\-quark states, using permutation-group methods
and reduced bases that make such analyses tractable.


\section{Building of the multi-quark wave functions}

We will not present Jacobi coordinates explicitly here. Instead, in Appendix
\ref{sec_Jacobi_equal} we provide a Mathematica script that generates Jacobi
coordinates for an arbitrary number of particles $N$.
When all quark masses are equal, the kinetic energy is proportional to the
Laplacian, allowing one to employ the same hyperdistance approximation used in
previous chapters.

The general strategy for constructing multiquark wave functions was already
introduced in earlier chapters and is based on representations of the
permutation group $S_n$. The corresponding Mathematica implementation and
additional technical details are described in Ref.~\cite{Miesch:2024vjk}.
Here we briefly summarize the main elements of the procedure.

A single light quark carries $3\times2\times2=12$ color-spin-flavor states.
Consequently, an $N$-quark state\footnote{In contrast to the pentaquark case,
antiquarks are absent here, so the counting of color states is simpler.}
can be expanded in a "monom'' basis of dimension
$N_{\text{monoms}}=12^N$.
In spin-tensor notation, implemented as Mathematica \texttt{Table}s\footnote{Since
these tables are often used in flattened form, the ordering of indices
(color, spin, isospin) must be chosen consistently at construction time.},
a generic wave function takes the form
\[
\psi=\sum_{\text{all indices}} C_{\text{indices}}
\ket{c_1\ldots c_N\, s_1\ldots s_N\, i_1\ldots i_N},
\]
with explicit color, spin, and isospin indices.

The procedure for constructing wave functions consistent with Fermi statistics
can be summarized as follows:
\begin{enumerate}
\item Identify the \textbf{allowed tensor structures} in each sector (color,
spin, flavor, orbital), corresponding to the required Young tableaux.
For example, in the color sector a baryon is described by the antisymmetric
tensor $\epsilon_{c_1c_2c_3}$, while for hexaquarks the color structure generalizes
to $\epsilon_{c_1c_2c_3}\epsilon_{c_4c_5c_6}$ together with all index permutations.
In the presence of antiquarks, Kronecker symbols
$\delta^{c}_{\bar c}$ may also appear.

\item Generate \textbf{all permutations} of the $n$ indices under the generators
of $S_n$, and express the resulting $n!$ tensors as vectors in the monom space
$\mathbb{C}^{N^n}$.

\item \textbf{Orthogonalize} this set of vectors, for example using the
Gram-Schmidt procedure. This produces a significantly reduced orthonormal
"good basis''. The expected dimension of this basis can be estimated~\cite{Miesch:2024vjk}.

\item Construct the matrices representing the two independent
\textbf{permutation-group generators} acting on the original monom space, and
project them onto the reduced basis obtained in the previous step. This yields
two matrices of dimension $n'\times n'$.

\item After performing steps 1-4 for each sector (color, spin, flavor, orbital,
etc.), take the \textbf{Kronecker product}\footnote{Recall that the Kronecker
product combines matrices into a single matrix acting on the product space,
whereas the tensor product combines vectors into a higher-rank tensor.} of the
corresponding permutation generators. This produces two matrices acting in the
full reduced Hilbert space, of total dimension
$n'_{\text{total}}=n'_{\text{color}}\cdot n'_{\text{spin}}\cdot n'_{\text{flavor}}\cdots$.

\item \textbf{Diagonalize} the two generators simultaneously, searching for
common eigenvectors with the required symmetry (eigenvalues $\pm1$, depending on
the fermionic or bosonic character of the sector).
These eigenvectors are the desired wave
functions, which may be projected back to the full monom basis if needed.
\end{enumerate}


\subsection{The simplest case of maximal spin $S=3$}
We already used these procedures for baryons and pentaquarks, so now let us proceed directly to $N=6$ Hexaquark case. As usual,
we start with the simplest case of maximal spin\footnote{Or maximal isospin, but not both}.

Simplification in this case comes from  the total spin being at its maximal value  $S=3$, so that all quark spins
$\uparrow\uparrow\uparrow\uparrow\uparrow\uparrow$
point in the same direction. Thus  the spin
part of the WF is trivially symmetric and factorizes. What remains to deal with are
the intermixed  $color$ and $flavor$ WFs. The former make representations of the $SU(3)$ group, and the latter either $SU(2),u,d$ or $SU(3),u,d,s$ flavor groups.  

\begin{figure}
    \centering
    \includegraphics[width=6cm]{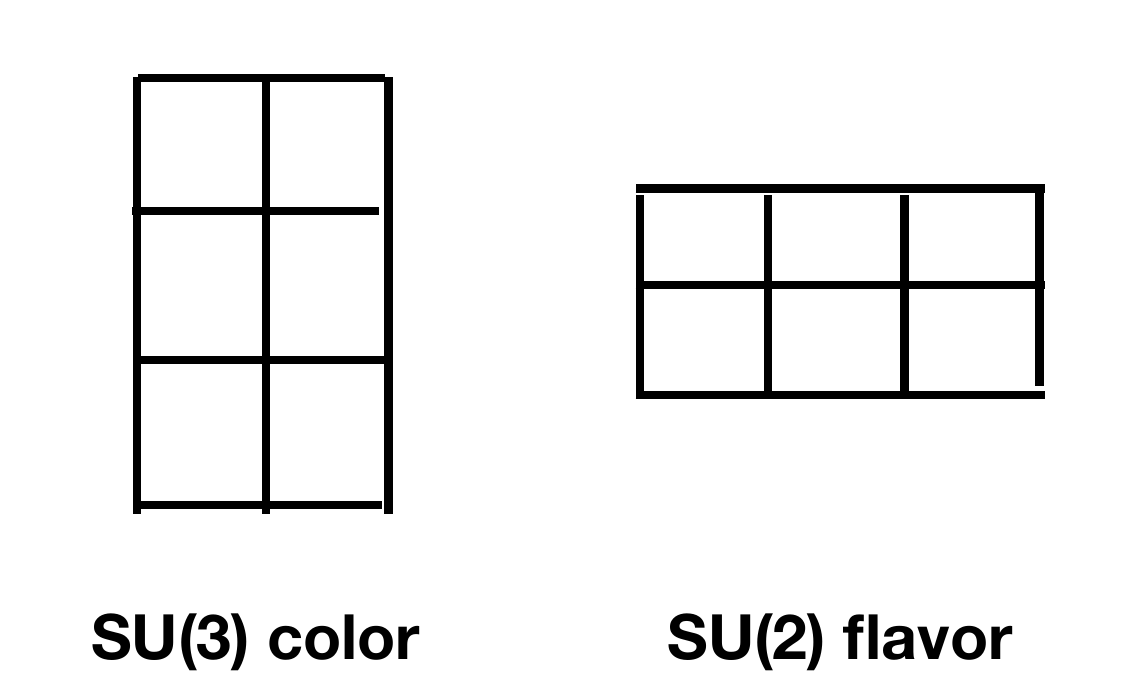}
    \caption{Color and flavor Young tableax for hexaquarks}
    \label{fig_young_6}
\end{figure}

The space of all color states have $3^6$ monoms. 
Color Young tableaux shown in Fig.\ref{fig_young_6}
should look like two complete vertical sets of squares. Those are antisymmetrized, or represented by 3-index Levi-Civita symbol. In our notations they correspond to all permutations of two of those
\begin{verbatim}
 TensorProduct[LeviCivitaTensor[3], LeviCivitaTensor[3]]  
\end{verbatim}
%
For step 2, we considered every possible one of these rearrangements and wrote them in as vectors in $\mathbb{C}^{3^6}$, using Mathematica's $Flatten$ function.  There are $6!$ permutations in symmetric $S_6$ group,  so that is the number of $3^6$-dimensional vectors in our list. (By a coincidence $6!=720$  happens to be close to $3^6=729$ .)

Orthogonalization of this set of vectors was accomplished through Mathematica's $Orthogonalize$ procedure, which by default uses Gram-Schmidt method and generates a set of independent orthonormal vectors.  From dimension 720 that reduces to "good basis" of only $5$ linearly independent combinations.  The notations below use $k=1$ is $P_{12}$,$k=2$ is $P_{cycle}$ permutation generators, defined in this basis $\{\mathbf{b}_i\}_{i=1}^5$ were then found by taking
$$
    (P^k_{color})_{ij}=\mathbf{b}_i^{T}\cdot P^k_{\mathbb{C}^{729}}\cdot \mathbf{b}_j,
$$
where $P^k_{\mathbb{C}^{729}}$ was generated as we did in smaller dimensions for baryons.

For the flavor sector we select $uuuddd$ quark composition, so the necessary tensor structure is three diquarks or three columns each being 2d Levi-Civita
\begin{verbatim}
 TensorProduct[LeviCivitaTensor[2], LeviCivitaTensor[2],LeviCivitaTensor[2]]  
\end{verbatim}
%
Just like for color, there are $6!$ quark permutations of the flavor indices $f_i$'s, also leading to just 5 unique linearly independent combinations.  We then
compute two $P^k_{flavor}$  $5\times5$ matrices for generators of $S_6$.

The next step is to perform $KronekerProduct$ of color and spin (or flavor) spaces\footnote{Perhaps it is useful to remind here the difference between $KronekerProduct$ and $TensorProduct$ commands. The former convert any number of matrices into one big matrix. The latter convert product of vectors into a large vector.}, and
representing two
 generators of the $S_6$ group, $P_{12}$ and $P_{cycle}$. Their diagonalization allows to search for common eigenstates with total eigenvalue $-1$ (antisymmetry for Fermions).
There is indeed  one such state found. 

Most of the wavefunctions we obtained are too large to list here, but for instance the color-flavor 25-dimensional wavefunction can be written in the ("flattened") basis $\{\mathbf{b}_{color}\}\otimes\{\mathbf{b}_{flavor}\}$
\bea
    (0, 0, 0, 0, \frac{1}{\sqrt{5}}, 0, 0, 0, -\frac{1}{\sqrt{5}}, 0, 0, 0, 
    \frac{1}{\sqrt{5}}, 0, 0, 
0, \frac{1}{\sqrt{5}}, 0, 0, 0, -\frac{1}{\sqrt{5}}, 0, 0, 0, 0)
\eea

Note that there are 3 positive terms and 2 negative terms with otherwise equal coefficient: we checked that they are exactly those  found by Kim, Kim, 
and Oka in \cite{Kim:2020rwn}.

\subsection{Generic spin cases}
\begin{table}[]
    \centering
    \begin{tabular}{|c|cccc|}
          $S$\textbackslash $S_z$ & 0 & 1 & 2 & 3   \\
         \hline
          0 & 0 &&&\\
          1 & $\{1,5\times 5\times 9\}$ & $\{1,5\times 5\times 9\}$ & & \\
          2 & 0 & 0 &  &\\
          3 & $\{1,5\times 5\times 1\}$ & $\{1,5\times 5\times 1\}$ & 
          $\{1,5\times 5\times 1\}$ & $\{1,5\times 5\times 1\}$ 
    \end{tabular}
    \caption{Hexaquarks with $uuuddd$ flavor content, with different
    values of spin $S$
    (rows) and 
    its projections $S_z$ (columns).
    When the number of states is 0, as it is for $S=0$ and 2, the row is all zeroes. When the number of antiperiodic wave functions found is one, we give  the dimension of the orthogonalized basis for tensor product of color-flavor-spin states. }
    \label{tab:udududTableSmS}
\end{table}


Using the power of the proposed method, we apply it for 
hexaquarks with other quantum numbers.  


For spins smaller than 3, the spin WF is no longer trivial.
For $S=1, S_z=0$, the general symmetry structure can be obtained from Young tableaux.  It is antisymmetry in 2 pairs of indices and symmetry in the other two.  The tensor structure to be permuted is therefore $[\uparrow\downarrow][\uparrow\downarrow] \{\uparrow\downarrow\}$, where 
symmetric and antisymmetric  combinations are
$$[\uparrow\downarrow]=(\uparrow\downarrow-\downarrow\uparrow)/\sqrt{2}$$
$$\{\uparrow\downarrow\}=(\uparrow\downarrow+\downarrow\uparrow)/\sqrt{2}$$
 with large number of permutations of indices. Yet, 
after $orthogonalization$ the basis of independent states  possesses only 9 linearly independent vectors. Therefore, two $S_6$ generators  $\sigma^1_{spin}$ and $\sigma^2_{spin}$  in this basis are $9\times 9$ matrices, given as examples in equation \ref{eq:9x9}.

    \bea
        P^1_{spin}&=&\left(
        \begin{array}{ccccccccc}
         -1 & 0 & 0 & 0 & 0 & 0 & 0 & 0 & 0 \\
         0 & -1 & 0 & 0 & 0 & 0 & 0 & 0 & 0 \\
         0 & 0 & -1 & 0 & 0 & 0 & 0 & 0 & 0 \\
         0 & 0 & 0 & 1 & 0 & 0 & 0 & 0 & 0 \\
         0 & 0 & 0 & 0 & 1 & 0 & 0 & 0 & 0 \\
         0 & 0 & 0 & 0 & 0 & 1 & 0 & 0 & 0 \\
         0 & 0 & 0 & 0 & 0 & 0 & 1 & 0 & 0 \\
         0 & 0 & 0 & 0 & 0 & 0 & 0 & 1 & 0 \\
         0 & 0 & 0 & 0 & 0 & 0 & 0 & 0 & 1 \\
        \end{array}
        \right),\nonumber\\
        P^2_{spin}&=&\left(
        \begin{array}{ccccccccc}
         \frac{1}{4} & \frac{1}{4 \sqrt{3}} &
           \frac{1}{\sqrt{6}} & -\frac{\sqrt{3}}{4} &
           -\frac{1}{4} & -\frac{1}{\sqrt{2}} & 0 & 0 & 0
           \\
         -\frac{\sqrt{3}}{4} & \frac{1}{12} & \frac{1}{3
           \sqrt{2}} & -\frac{1}{4} & \frac{1}{12
           \sqrt{3}} & \frac{1}{3 \sqrt{6}} &
           -\frac{\sqrt{\frac{2}{3}}}{3} & -\frac{4}{3
           \sqrt{3}} & 0 \\
         0 & -\frac{\sqrt{2}}{3} & \frac{1}{6} & 0 &
           -\frac{\sqrt{\frac{2}{3}}}{3} & \frac{1}{6
           \sqrt{3}} & -\frac{1}{3 \sqrt{3}} & \frac{1}{6
           \sqrt{6}} & -\frac{\sqrt{\frac{5}{2}}}{2} \\
         -\frac{\sqrt{3}}{4} & -\frac{1}{4} &
           -\frac{1}{\sqrt{2}} & -\frac{1}{4} &
           -\frac{1}{4 \sqrt{3}} & -\frac{1}{\sqrt{6}} &
           0 & 0 & 0 \\
         \frac{3}{4} & -\frac{1}{4 \sqrt{3}} &
           -\frac{1}{\sqrt{6}} & -\frac{1}{4 \sqrt{3}} &
           \frac{1}{36} & \frac{1}{9 \sqrt{2}} &
           -\frac{\sqrt{2}}{9} & -\frac{4}{9} & 0 \\
         0 & \sqrt{\frac{2}{3}} & -\frac{1}{2 \sqrt{3}} &
           0 & -\frac{\sqrt{2}}{9} & \frac{1}{18} &
           -\frac{1}{9} & \frac{1}{18 \sqrt{2}} &
           -\frac{\sqrt{\frac{5}{6}}}{2} \\
         0 & 0 & 0 & \sqrt{\frac{2}{3}} &
           -\frac{\sqrt{2}}{9} & -\frac{4}{9} &
           -\frac{1}{9} & -\frac{2 \sqrt{2}}{9} & 0 \\
         0 & 0 & 0 & 0 & \frac{8}{9} & -\frac{2
           \sqrt{2}}{9} & -\frac{1}{9 \sqrt{2}} &
           \frac{1}{36} & -\frac{\sqrt{\frac{5}{3}}}{4}
           \\
         0 & 0 & 0 & 0 & 0 & 0 &
           \frac{\sqrt{\frac{3}{10}}}{2}+\frac{7}{2
           \sqrt{30}} &
           \frac{\sqrt{\frac{3}{5}}}{4}-\frac{2}{\sqrt{15
           }} & -\frac{1}{4} \\
        \end{array}
        \right)
    \label{eq:9x9}
    \eea

The next step is to define these two generators ($k=1,2$ for $P_{12}$ and $P_{cycle}$)  written as tensor product incorporating every sector of the WF, e.g. 
\be
    P^k_{total}=P^k_{color}\otimes P^k_{spin}\otimes P^k_{flavor}.
\ee

In the previous subsection - hexaquarks with maximal spin $S=3$ - these were  matrices (in "good basis")  of dimension $5\cdot 5\cdot 1=25$. For other spin values and $uuuddd$ quarks, those we found to be matrices in the following minimal dimensions:
 for $S=2$ they are  in 125-d space, for $S=1$ in 225-d, and for spin 0 they are matrices in 125-dimensions again. 

While all of them are too large to be given in press, we still emphasize that these
dimensions are many times  smaller than that of the full  space of monoms, $12^6$.
Furthermore, in practice there is absolutely no problem to operate with them
inside Mathematica. In particularly, all are  generated in a second,  and 
diagonalized as  quickly. 

A  procedure to find common antisymmetric eigenstates one can e.g.
add two matrices with coefficients
being some random numbers, e.g. $e$ and $\pi$, and then look for eigenvalues $-e-\pi$.

 The particular number of solutions for each spin  depends only on the spin value $S$, and of course not on its projection $S_z$, as follows from rotational symmetry. Yet the calculation themselves are not technically identical, so 
 we did it for $all$ values of $S_z$ to check for their mutual consistency. 
 Some of the results are shown in Table \ref{tab:udududTableSmS}.
 
 At $S=0$ and $S=2$ we found no solutions were possible with the permutation antisymmetry desired.  At $S=1$ and $S=3$ however, we found a
 $single$ antiperiodic wave function for each value of $S_z$. 

\begin{table}[]
    \centering
    \begin{tabular}{|c|cccc|}
        \hline 
        $I=0$&&&&\\
      L \textbackslash \, S   &0 & 1 & 2 & 3   \\
         \hline
         0&0& 1& 0 & 1\\
         1& 1&1 & 2 &0\\
         2&4&9&5&2 \\ \hline
    \end{tabular}
    \begin{tabular}{|c|cccc|}
        \hline 
        $I=1$&&&&\\
      L \textbackslash \, S   &0 & 1 & 2 & 3   \\
         \hline
         0&1& 0& 1 & 0\\
         1& 1&4 & 2 &1\\
         2&9&15&10&2 \\ \hline
    \end{tabular}
    \begin{tabular}{|c|cccc|}
        \hline 
        $I=2$&&&&\\
      L \textbackslash \, S   &0 & 1 & 2 & 3   \\
         \hline
         0&0& 1& 0 & 0\\
         1& 2&2 & 1 &0\\
         2&5&10&5&1 \\ \hline
    \end{tabular}
    \caption{Number of antisymmetric 6 $q$ states (per choice of $m_s$ and $m_l$) at each combination of total orbital and spin angular momentum for the light quark hexaquark. For example, last in row 1 of the $I=0$ plot is the KKO state with S=3, L=0.  At $L=0$ there are 3 possible combinations of $S$ and $I$, with the ability to swap them identically as they are both $SU(2)$: (1,0), (3,0), and (2,1).  It is worth noting that these exactly match the allowed quantum number combinations for dibaryons made from nucleons and $\Delta$'s first calculated in 1964 by Dyson and Xuong~\cite{Dyson:1964xwa}.}
    \label{tab:ududud}
\end{table}
\begin{table}[]
    \centering
    \begin{tabular}{|c|cccc|}   \hline
         L \textbackslash \, S &0 & 1 & 2 & 3   \\
         \hline
         0&0& 1& 0 & 0\\
         1& 0&2 & 1 &0\\
         2&5&7&4&1 \\ \hline
    \end{tabular}
    \caption{Number of $udsuds$ antisymmetric states (per choice of $m_s$ and $m_l$) at each combination of total orbital and spin angular momentum for $udsuds$ hexaquark.  -}
    \label{tab:udsuds}
\end{table}
Let us now change the flavor content, adding two strange quarks to $uuddss$ hexaquark. The flavor becomes $SU(3)$ and its treatment is similar to 
that of color, if total adds to zero. The resulting antisymmetric states are reported in Table \ref{tab:udsuds}.

Completing the hexaquark discussion, let us consider another simplified case, of same-flavor quarks (e.g. $cccccc$). 
The number of good states is in the Table \ref{tab:cccccc}.

Let us explain some cases without solutions first.
If both flavor and spin is flat-symmetric, then Fermi statistics requirement falls on color
WF, which cannot be fulfilled for 6 quarks.

\begin{table}[]
    \centering
    
    \begin{tabular}{|c|cccc|}  \hline 
         L \textbackslash \, S&0 & 1 & 2 & 3   \\
         \hline
         0&1& 0& 0 & 0\\
         1& 0&1 & 0 &0 \\
         2& 2& 2& 1& 0 \\  \hline
    \end{tabular}
    \caption{Number of antisymmetric states (per choice of $m_s$ and $m_l$) at each combination of total orbital and spin angular momentum for $cccccc$ hexaquark.}
    \label{tab:cccccc}
\end{table}


\section{Matrix Elements of basic operators and hexaquark masses}
With the color-flavor-spin wave functions available, one can
attempt to calculate the average values of pertinent operators. The obvious step one is to do that
perturbatively, for a gluon exchanges. The lowest order
gluon exchange generates potentials proportional to "relative
color" operators made out of color generators $\langle\lambda_i^A\lambda_j^A/4\rangle$ where
$A=1..8$ and $i,j=1..n$. Relativistic corrections lead to
spin-spin, spin-orbit and tensor forces, as usual. Perturbative
one-gluon exchange require that those also are proportional to  colors, e.g. spin-spin is proportional to $$\langle\sum_{i>j}({1 \over 4}\lambda_i^A\lambda_j^A) (\vec S_i \vec S_j)\rangle$$
For $S-shell, L=0$ hadrons the spin-spin forces are the only relativistic corrections. 

The resulting masses are shown for light hexaquarks in table \ref{tab:matrixElements6q} and for all other $L=0$ states in table \ref{tab:matrixElements}.   Note that the ordering of the compact hexaquark states  in \ref{tab:matrixElements6q} (fourth column) is rather different from the ordering of the dibaryon molecules with the same quantum numbers (fifth column).  
The (0,3) state uniquely breaks the pattern:  not only  it is degenerate with the (1,2) state and lighter than the (2,1) state, it is lighter than (or at least close to) the dibaron molecule state with spin 3 and isospin 0, which was predicted to have a mass of 2350 MeV.  Every other channel show that a molecule is lighter than  compact hexaquark. Perhaps this is the only hexaquark state that has been experimentally observed.

\begin{figure}[h]
    \centering
    \includegraphics[width=6cm]{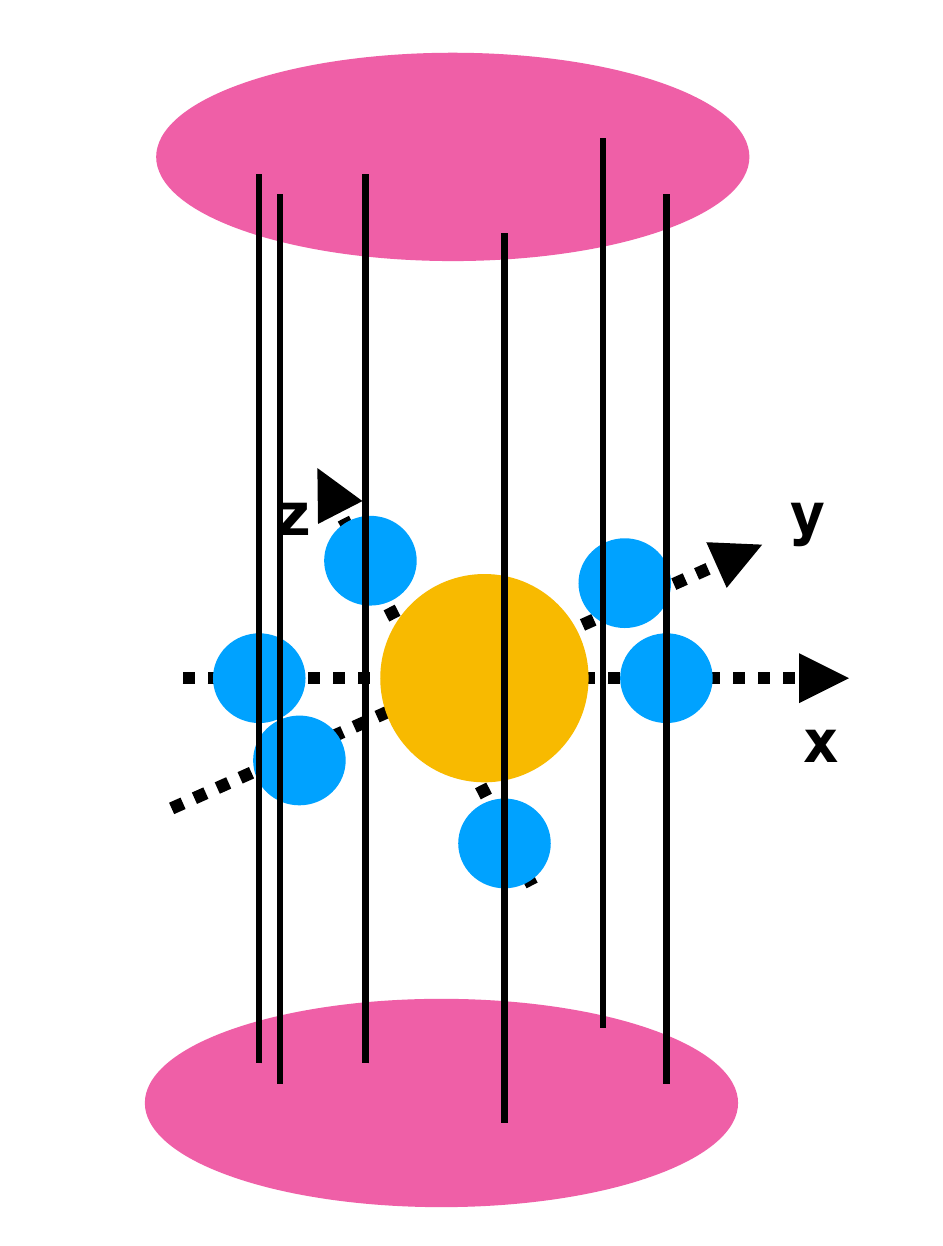}
    \caption{The setting for calculation of the effective potential for $n=6$ quarks, located at (blue) points in three dimensions. 
    The vertical direction is Euclidean time, six vertical lines are Wilson lines. Red ovals above and below indicate
    color wave functions to which Wilson lines are traced. Yellow circle indicate gauge fields of an instanton.}
    \label{fig_6W}
\end{figure}

Higher order gluon exchanges lead to operators with higher orders of Gell-Mann matrices. Those diagrams can best be obtained from an expansion of the set of Wilson lines
convoluted with color wave functions, see e.g. Fig.\ref{fig_6W} for hexaquarks.  For example, for hexaquarks one can either put two color epsilons with all $6!$  permutations or simplify it to just 5 "good basis" color convolutions.  As far as we know, next order
gluon exchanges were not yet used in spectroscopy.

    \begin{table}[]
        \centering
        \begin{tabular}{|c|c|c|c|c|c|}
            \hline
              $(I,S)$ & $\langle\lambda\lambda\rangle$ & $\langle\lambda\lambda S S\rangle$  & 6$q$ mass &Molec.model & Experiment\\
             \hline
              $(0,1)$&-16&-2/3&2098&1876&1876\\
              $(1,0)$&-16&-2&2196&1876&1878\\
              $(1,2)$&-16&-4&2342&2160&2160\\
              $(2,1)$&-16&-20/3&2536&2160&2160\\
              $(0,3)$&-16&-4&2342&2350&2380\\
              $(3,0)$&-16&-12&2926&2350&2464\\
              \hline
        \end{tabular}
        \caption{6$q$ matrix elements of color and color-spin operators for all of the $L=0$ light hexaquark states. The resulting masses (in MeV) obtained from simple additive model  \cite{Kim:2020rwn} are in the 4-th column (using $m_q=330$ MeV, $m_{s}=500$ MeV, and $m_c=1270$ MeV).  The fifth column show  prediction of molecular dibaryons model~\cite{Dyson:1964xwa}. The 6-th column show  experimentally measured values of the resonances with corresponding quantum numbers (taken from compilation \cite{Clement:2016vnl}).}
        \label{tab:matrixElements6q}
    \end{table}
\begin{table}[]
        \centering
        \begin{tabular}{|c|c|c|c|}
            \hline
              State & $\langle\lambda\lambda\rangle$ & $\langle\lambda\lambda S S\rangle$  & \text{Mass (MeV)}\\
              \hline
              $udsuds$, $S=0$ &-16&6&1611\\
              \hline
              $cccccc$, $S=0$ &-16&-12&7819\\
              \hline
              $udud\overline{q}$, $S=\frac32$ & $-\frac{20}{3}$&$7/4$&1623\\
              $udud\overline{q}$, $S=\frac12$&$-\frac{20}{3}$&$1/2$&1714\\
              \hline
              $udud\overline{c}$, $S=\frac32$&$-7.654$&$1.117$&2607\\
              $udud\overline{c}$, $S=\frac12$&$-7.654 $&$1.442$&2582\\
              \hline
              $udud\overline{Q}$, $m_Q \to \infty$, $S=\frac32$  & $-8$ & $4/3$ & 1319+$m_Q$ \\
              \hline
        \end{tabular}
        \caption{Matrix elements and masses for all other $L=0$ states found in this paper, using the same fit from \cite{Kim:2020rwn}.}
        \label{tab:matrixElements}
    \end{table}
The nonperturbative confining potentials are defined via correlators of $n$ Wilson lines, 
\be \langle \Psi |\big(\delta_{c1}^{c1'}...\delta_{cn}^{cn'}- W_{c1}^{c1'} ...W_{cn}^{c1n'}\big) | \Psi'  \rangle\ee
with path-ordered exponential of color generators
$$ W=Pexp[i (\lambda^a/2) \int g A_0^a dt ]_{c1}^{c1'} $$
The setting is shown schematically
in Fig.\ref{fig_6W}

Note that while total wave functions $\Psi$ is a spin-tensor with many different indices (color, flavor, spin etc) the main perturbative $\lambda\lambda$ operator has  only color indices. So, if $\Psi$
consists of several factorizable parts 
$$ W=\sum_A C^A\psi^A_{color}\psi^A_{flavor} \psi^A_{spin} $$
one can sum over non-color indices using normalization of those wave functions and put this operator as 
"sums of squares" of the color wave functions 
$$\sum_A C_A^2\psi^A_{color}(c)\big(\delta_{c1}^{c1'}...\delta_{cn}^{cn'}- W_{c1}^{c1'} ...W_{cn}^{c1n'}\big)\psi^A_{color}(c')$$
We have calculated it explicitly for variouse states. For hexaquarks, their five color wave functions
  are explicitly given in Appendix of the original paper. We also of course convoluted those
 with Wilson lines to get color potentials, but the expression
 is too long to be given here (can be obtained from the 
authors upon request).
This expression can be directly evaluated on the lattice,
or by any vacuum model (e.g., in the instanton model),
producing forces among quarks in all hexaquarks.


\section{Nine and twelve-quark S-shell states}
\label{sec_9_12}

Light $u,d$ quarks form an isospin-$1/2$ doublet and carry three color and two
spin degrees of freedom, resulting in a total of
$2\times3\times2=12$ internal states. This number therefore plays the role of a
"magic number'', completing the $1S$ ($L=0$) quark shell.

The corresponding quantum numbers naturally bring to mind the
$\alpha$ particle, $^4\mathrm{He}=ppnn$, the classic example of a "double magic''
nucleus, known to be deeply bound and compact.
A substantial nuclear-physics literature discusses the extent to which heavier
nuclei such as $^{12}\mathrm{C}$ and $^{16}\mathrm{O}$ may exhibit
$\alpha$-cluster correlations.
Indeed, nucleons in the $\alpha$ particle do not strongly overlap, making a
four-body nuclear description well justified.
Nevertheless, it remains a legitimate question whether the $\alpha$-particle
wave function may contain a sufficiently compact component that is more
naturally described in terms of twelve quarks.

The theoretical description of the $^4\mathrm{He}$ wave function has a long
history, dating back to the 1930s, and played a key role in the development of
the hyperdistance approximation that we employ for multiquark systems.
Alternatively, one may avoid approximations altogether and perform numerical
Path Integral Monte Carlo calculations~\cite{Shuryak:1984xr}. This approach has
recently been revived in Ref.~\cite{DeMartini:2020hka}, where evidence for
$\alpha$ preclustering was reported at temperatures $T\sim100\,\mathrm{MeV}$,
relevant for hadronic freezeout in heavy-ion collisions.

Before turning to explicit calculations, it is useful to comment on
diquark-based approaches to multiquark systems. Such models often argue that the
12-quark problem can be reduced to a system of six diquarks, thereby transforming
a fermionic many-body problem into a bosonic one.
As already discussed in the Introduction, this line of reasoning is fundamentally
flawed. There is no justification for singling out six pairs out of the total
66 possible quark-quark pairs, even if those six are assumed to experience the
strongest attraction\footnote{We have already seen explicitly that the "good
diquark'' ideology fails for tetraquarks.}. All pairwise interactions must be
treated on equal footing\footnote{This can be demonstrated already at the
perturbative level.}.

The origin of strong diquark correlations is commonly traced to the
instanton-induced 't~Hooft interaction, see
Refs.~\cite{Rapp:1997zu,Alford:1997zt}. To obtain a binding energy of
$6\,B_{qq}$ from six diquarks would require six instantons to be localized within
a single 12-quark cluster. This scenario is highly implausible, given the dilute
nature of instantons in the QCD vacuum.

As a theoretical starting point, one may instead consider a toy model in which
the quarks are sufficiently heavy to justify a nonrelativistic 
Schrodinger
description with perturbative Coulomb interactions.
We begin with the most symmetric case, involving two heavy flavors (e.g.\ $c$
and $b$) with a common mass $M$.

As shown above for systems of three, four, five, and six quarks, one can construct
wave functions obeying Fermi statistics by explicit construction of
color-spin-flavor tensor products and subsequent diagonalization of the
Hamiltonian within this space.
For 12-quark systems (and even for nine quarks), this approach rapidly becomes
intractable.
The full monom space has dimension
$N_{\mathrm{monoms}}=12^{12}\approx8.9\times10^{12}$, and even the initial
reduction to a "good basis'' poses severe difficulties.
Although the color space alone has dimension $3^{12}\sim10^6$, the number of
permutations of open-index tensors constructed from four Levi-Civita symbols is
prohibitively large, even with the aid of Wolfram Mathematica.

Nevertheless, by restricting to a reduced set of permutations and starting with
nine quarks, we were able to identify partial "good basis'' structures.
For $N=9$, the color basis has dimension 33, and the flavor basis dimension 42.
If the total spin is maximal, $S=9/2$, the spin sector factorizes, and the
combined color-flavor basis has dimension $33\times42=1386$.
The two generators of the permutation group $S_9$ can be constructed as matrices
in this space. Diagonalizing them leads to the negative result that no common
eigenvectors with eigenvalue $-1$ (corresponding to full antisymmetry) exist.
Extending the analysis to non-maximal spin introduces an additional factor of 42
from the spin sector, yielding matrices of dimension $33\times42\times42$, which
are beyond our current computational reach.

Although the explicit construction of wave functions becomes prohibitively
difficult for $N=9$ and $N=12$, it remains possible to determine whether
Fermi-statistics-compliant states exist at all.
This can be achieved by consulting tables of Kronecker coefficients available in
the mathematical literature.
For the lowest spin and isospin values ($S,I=0$ for even $N$, and $S,I=1/2$ for odd
$N$), such analyses indicate the existence of a single state for $N=9$, $N=12$,
and $N=24$, with no corresponding states for $N=15$, $18$, or $21$.

The possibility that such compact quark clusters exist as admixtures to
conventional nuclear states has been discussed for decades and remains an
important direction for future investigation.

\paragraph{Concluding remarks.}
In summary, the analysis of nine- and twelve-quark systems highlights both the
conceptual importance and the technical difficulty of extending multiquark
spectroscopy beyond the few-body domain. While the explicit construction of
fully antisymmetric wave functions becomes rapidly impractical as the number of
quarks increases, group-theoretical considerations indicate that such states do
exist at selected quark numbers, including $N=9$ and $N=12$. This strongly
suggests that compact multiquark configurations may arise not as isolated
exotic resonances, but rather as well-defined components admixed into ordinary
nuclear wave functions. Clarifying the dynamical role of these components, and
their interplay with conventional nuclear forces, remains a key challenge for
future theoretical and lattice studies.

\begin{subappendices}
\section{Modified Jacobi coordinates for N bodies } \label{sec_Jacobi_equal}
The general idea is to compute location of the CM of the first $n$ 
bodies and then define coordinate as distance to $n+1$th. 
For bodies of the same mass, let us give compact Mathematica commands generating those 
\begin{verbatim}
In:   Jacobi[n_] := Inverse[DiagonalMatrix[
    Table[Sqrt[i + 1]/Sqrt[i], {i, 1, n}]]] . (Table[
     Join[Table[1/i, i], Table[0, n - i]], {i, 1, n}] + 
    DiagonalMatrix[Table[-1, n - 1], 1]) 
\end{verbatim}

 Let us check what it will produce
 for 3,4 and 5 bodies, baryons, tetraquarks and pentaquarks
 as we have defined those coordinates above:
 \begin{verbatim}
  In:  Jacobi[3] 
Out: {{1/Sqrt[2], -(1/Sqrt[2]), 0}, 
{1/Sqrt[6], 1/Sqrt[ 6], -Sqrt[(2/3)]},
{1/(2 Sqrt[3]), 1/(2 Sqrt[3]), 1/(2 Sqrt[3])}}

In: Jacobi[4]
Out: {{1/Sqrt[2], -(1/Sqrt[2]), 0, 0},
{1/Sqrt[6], 1/Sqrt[6], -Sqrt[(2/3)], 0}, 
   {1/(2 Sqrt[3]), 1/(2 Sqrt[3]), 1/( 2 Sqrt[3]), -(Sqrt[3]/2)}, 
  {1/(2 Sqrt[5]), 1/(2 Sqrt[5]), 1/(2 Sqrt[5]), 1/(2 Sqrt[5])}}
  
In: Jacobi[5]
Out: {{1/Sqrt[2], -(1/Sqrt[2]), 0, 0, 0},
{1/Sqrt[6], 1/Sqrt[6], -Sqrt[(2/3)], 0, 0}, 
{1/(2 Sqrt[3]), 1/(2 Sqrt[3]), 1/(2 Sqrt[3]),-(Sqrt[3]/2), 0},
{1/(2 Sqrt[5]), 1/(2 Sqrt[5]), 1/(2 Sqrt[5]), 1/(2 Sqrt[5]), -(2/Sqrt[5])},
{1/Sqrt[30], 1/Sqrt[30],  1/Sqrt[30], 1/Sqrt[30], 1/Sqrt[30]}}

\end{verbatim}
The Jacobi coordinates we call $\alpha,\beta...$ are given as combination
of actual body coordinates $x_i, \, i=1 .. n$ with coefficients just obtained. For example $\vec\alpha=(\vec x_1-\vec x_2)/\sqrt{2}$ etc. Note that the last bracket always gives the location
of the Center of Mass, to be put to zero. Thus the number of Jacobi
coordinates is not $n$ but $n-1$.

Note that for baryons (n=3), the traditional names of Jacobi
coordinates are $\vec\rho,\vec \lambda$ rather than $\vec\alpha,\vec\beta $,
and in this case we will use traditional ones. 

Note that in order to calculate the metric tensor in new coordinates it is convenient
to start with the Cartesian form for old coordinates $ds^2=\sum dx_i^2$, and then, using the matrix $inverse$ to that obtained above, rewrite it as $ds^2= g_{11} d\alpha^2+... $.
Note further that the Laplacian includes $g^{ij}$, the inverse of the metric $g_{ij}$ obtained. The volume element is given by $\sqrt{det(g)}$.
In Mathematica use $Inverse$ and $Det$ commands.

For  the cases of bodies with different masses we discuss
 in tetraquark chapter. 
\end{subappendices}

\part{Parton model and the light front properties of hadrons}

\chapter{Deep Inelastic scattering, PDF moments and DGLAP evolution}

This part is partly introductory in context. We start with parton distributions (PDFs) and related theory. The Distribution Amplitudes (DAs) will be introduced in connection with pion in \ref{sec_pion_DA}. Generalizations of PDFs are
discussed in chapters on GPDs and TMDs later in this part. 

\section{Light front quantization}
We are not going to repeat here the standard lore related to
Deep Inelastic scattering and other hard processes: there is no shortage
of literature for that. And still, a number of basic definitions (with
commentaries) are perhaps needed for non-experts. 

To introduce the notations, we start with free fermions on the  light front. The Hamiltonian  is
obtained from the expanded relativistic kinetic energy, assuming the longitudinal momentum is the largest~\cite{Brodsky:1998hn}
\bea
P^-=\int dx^-d^2x_\perp~ T^{+-}=
\int [d^3k]_+\int [d^3q]_+\frac{k^2_\perp+M^2}{2k^+}\bar{\psi}_+(k)\gamma^+\psi_+(q)(2\pi)^3 \delta^3_+(k-q)
\eea
What we call a Hamiltonian gives the spectrum of mass squared $M^2=P^+P^-$.
The LF form of quantized fermionic field in momentum space is defined by
\begin{equation}
    \psi_+(x^-,x_\perp)=\int [d^3k]_+\psi(k)e^{-ik^+x^-+ik_\perp\cdot x_\perp}
\end{equation}
with the free LF spinors for the quarks and anti-quarks 
\begin{equation}
u_s(p)=\frac{1}{\sqrt{\sqrt{2}p^+}}\left(\slashed{p}+M\right)\begin{pmatrix}\chi_s \\ \chi_s \end{pmatrix}
\qquad\qquad
v_s(p)=\frac{1}{\sqrt{\sqrt{2}p^+}}\left(\slashed{p}-M\right)\begin{pmatrix}2s\chi_s \\ -2s\chi_s \end{pmatrix}
\end{equation}
respectively. Here  $\chi_s$  a 2-spinor with a spin pointing in the $z$-direction. Slashed momentum vectors mean, as usual, their convolution with gamma matrices.
$\psi(k)$  annihilates a particle in a  $u_s(k)$ mode,  or creates  an antiparticle in a $v_s(k)$ mode, i.e. 
\bea
\psi_+(k)=\sum_s u_s(k)b_s(k)\theta(k^+)+v_s(-k)c_s^\dagger(-k)\theta(-k^+)\nonumber\\
\eea
The measure in momentum space is defined as
\bea
[d^3k]_+=\frac{dk^+d^2k_\perp}{(2\pi)^32k^+}\epsilon(k^+)
\eea
which sums over the  positive $k^+$ region for  particle modes,  and over the negative $k^+$ region for antiparticle modes.

\section{Partons at the light front and PDFs}
The mass spectrum diagonalizes the light front Hamiltonian
\begin{equation}
\label{PMINUS2}
    P^-|X,P\rangle=\frac{m_X^2}{2P^+}|X,P\rangle
\end{equation}
with the states expanded in multi-component Fock representations. For example, a meson in the lowest Fock representation reads

\begin{equation}
\label{Meson_bound_state2}
    |\mathrm{Meson} ~X,P\rangle=\frac{1}{\sqrt{N_c}}\int_0^1 \frac{dx}{\sqrt{2x\bar{x}}}\int\frac{d^2k_\perp}{(2\pi)^3}\sum_{s_1,s_2}\Phi_X(x,k_\perp,s_1,s_2)b^\dagger_{s_1}(k) c^\dagger_{s_2}(P-k)|0\rangle
\end{equation}
with the light front normalization $\langle P|P'\rangle=(2\pi)^32P^+\delta^{3}(P-P')$
and
\begin{equation}
\label{normal2}
    \int_0^1dx\int\frac{d^2k_\perp}{(2\pi)^3}\sum_{s_1,s_2}\left|\Phi_X(x,k_\perp,s_1,s_2)\right|^2=1
\end{equation}

The parton distribution functions (PDFs) are pertinent matrix elements of the light front 
quark and gluon operators organized in a twist expansion (the twist of an operator is the dimension minus spin). In the large momentum limit $[D^+]$ factors out, with the space-time dimensions reduced by 1, e.g.
$[\psi_+]=1, [\psi_-]=2$. The twist-2 and twist-4 unpolarized PDF for a meson are respectively,
\begin{equation}
    \langle X,P|\bar{\psi}_+\frac{1}{2}\gamma^+\psi_+|X,P\rangle
    \qquad\qquad
    \langle X,P|\bar{\psi}_-\frac{1}{2}\gamma^+\psi_-|X,P\rangle
\end{equation}
in the light cone gauge. They count the number of partons in the hadron state, in the
infinite momentum frame.

\subsection{Partons from leading pQCD factorization}
The light front form of the leading twist unpolarized hadron PDFs, follow from the 
factorization of the inclusive photon-hadron cross section as illustrated by the hand-bag
diagram 
with
$d\sigma\sim l^{\mu\nu}(q)\,W_{\mu\nu}(k)$.
The hadronic kernel 
\bea
M_{\alpha\beta}(k)=\int d^4x\,e^{ik\cdot x}\,
\langle P|T^*(\overline\psi_\beta(0)\psi_\alpha(x)|P\rangle
\eea
which corresponds to removing a quark of 
momentum-$k$ and spin-$\alpha$ from the hadron at $x$, and returning as a  a quark of momentum-$k$ but spin-$\beta$ at point $0$, factors out  in the hand-bag diagram, defines the hadronic tensor
\bea
\label{WMUNU}
W^{\mu\nu}(k)=\int\, d^4k\, {\rm Tr}\bigg(M(k)\gamma^\mu\frac{i(\slashed{k}+\slashed{q})}{(k+q)^2+i0}\gamma^\nu\bigg)+[\rm cross]
\eea
which is frame covariant. Note that the leading contribution to the hadronic matrix element 
is from the photon transverse polarizations
$\mu,\nu=\perp, \perp$. In the nucleon rest frame, $P^\mu=(m_N,0^\perp, 0)$  and $q^\mu=(\nu,0^\perp, -\sqrt{\nu^2+Q^2})$. In the Bjorken limit with $\nu, Q^2\rightarrow\infty$
but fixed $x_B=Q^2/2\nu m_N$,  the photon 4-momentum simplifies $q^\mu\approx (\nu, 0^\perp, -(\nu+m_Nx_B))$, hence $q^-\approx \nu\gg q^+\approx m_Nx_B$ and (\ref{WMUNU}) reduces to
\bea
W^{\mu\nu}(k)=-g_\perp^{\mu\nu}\int\, d^4k\, {\rm Tr}\bigg(M(k)\frac{i\gamma^+}{2(k^++p^+)+i0}\bigg)+[\rm cross]
\eea
which simplifies to
\bea
W^{\mu\nu}(k)=-ig_\perp^{\mu\nu}\int\, \frac{dx}{2\pi}
\bigg(\frac {q(x)}{x-x_B+i0}+\frac {\bar q(\bar x)}{\bar x+x_B+i0}\bigg)=-g_\perp^{\mu\nu}\frac 12\big(q(x_B)+\bar q(-x_B)\big)
\eea
The light cone quark PDF is
\bea
\label{QX}
    q(x)=\int_{-\infty}^\infty\frac{dx^-}{2P^+}e^{ix P^+x^-}\langle P|\bar{\psi}(0)\gamma^+ W(0,x^-)\psi(x^-)|P\rangle\rightarrow \int\frac{d^2k_\perp}{(2\pi)^3}\left|\Phi_X(x,k_,s,s')\right|^2
\eea
and  the anti-quark PDF is
\bea
\label{QBX}
    \bar{q}(x)=\int_{-\infty}^\infty\frac{dx^-}{2P^+}e^{-ix P^+x^-}\langle P|\bar{\psi}(0)\gamma^+ W(0,x^-)\psi(x^-)|P\rangle\rightarrow \int\frac{d^2k_\perp}{(2\pi)^3}\left|\Phi_X(\bar{x},k_,s,s')\right|^2
\eea
We have restored the gauge link  $$W(x^-,0)=\mathrm{exp}\left[-ig\int_0^{x^-}d\eta^- A^+(\eta^-)\right]\rightarrow 1$$
which is 1 in the light cone gauge $A^+=0$.
The unpolarized inclusive cross section counts the number of partonic quarks in a hadron, in the leading twist aproximation
at parton $x=x_B$ in the Bjorken limit.
In the hadron rest frame, the twist-2 PDFs 
probe the quark correlation function along the $x^-$ light front direction. They are inherently non-perturbative.

\section{Operators and Twist classification}
\subsection{Operators of the leading twist}
"Twist"  is defined as operator's dimension minus its spin.
Quark operators of  spin n contain symmetrized product of covariant derivatives, and thus twist 2 
\begin{align}
& O^q_n(x) \equiv \bar\psi \nabla_{\{\alpha 1} ... \nabla_{\alpha n-1}\gamma_{\beta\}} \psi,
\end{align}
Similar definitions hold for  gluonic operators. For example, the  gluonic  stress tensor
\begin{align}
O^g_{\alpha\beta}(x) \equiv G^a_{\alpha\mu} G^a_{\mu\beta} - \textrm{trace},
\end{align}
has dimension 4 and spin 2, so its twist is dimension-spin 4-2=2.
Let us recall that the instanton field, although strong, has zero
stress tensor, and so there is no instanton contributions to it.
The $I\bar I$ molecules however, do contribute. The gluonic PDF's second moment
is related to its average in a hadron: it is known both from data
and lattice.

The correlator of the gluonic  stress tensors in $vacuum$
\begin{align}
& \Pi^{qg} (|x|) \sim \langle O^q_{\alpha\beta}(x) O^g_{\alpha\beta}(0)\rangle
\label{twist2_quark_gluon}
\end{align}
 describes the propagation of tensor glueballs. And indeed, unlike the scalar and pseudoscalar
 ones, it is  enhanced by instantons,  but remains relatively small.
 (Perhaps it can be used to see the contribution of the $I\bar I$ molecules .)

\subsection{Anomalous dimensions of operators}
Generally, perturbative renormalization of operators yields mixing of quark and gluon operators.
The ensuing expressions are unwieldy and we will not cover them here.
For twist 3 and 4 one can find these expressions in the
original work in~\cite{Shuryak:1981kj}. Instead,  we will 
give an  example of  the use of the anomalous dimensions relevant to the "spin crisis" 
as discussed in~\cite{Thomas:2008ga}. It is based on the RG form 
preserving chiral symmetry. Using standard  (one-loop) notations
\be
t=ln(Q^2/\Lambda_{QCD}^2), \,\,\, \alpha_s={4\pi\over \beta_0 ln(Q^2/\Lambda_{QCD}^2)},\,  \,\,\,\beta_0=11-{2\over 3}N_f
\ee
one can derive the RG evolution for the orbital momentum contributions
in closed forms.  For the flavor-singlet and isovector
combinations they are given by
\ba
L^{u+d+s}(t)+\Sigma &=& ({1\over 2}){3N_f \over 16+3N_f}+({t \over t_0})^{-{32+6N_f \over 9\beta_0}}\big(L^{u+d}(t_0)+\Sigma/2 - {3N_f \over 16+3N_f} \big) \\
L^{u-d}(t) &=&  -(\Delta u -\Delta d)/2+({t \over t_0})^{-{32 \over 9\beta_0}}\big( L^{u-d}(t_0)+(\Delta u -\Delta d)/2\big)
\ea
From which one can obtain the orbital momenta for each
flavor separately. 
At large $Q$ the values of  $L_u\approx -0.36,L_d\approx 0.28$ are of opposite sign,
but evolve to lower scale $\rm 1 \,GeV$ to be about the same. Further evolution down to the initial model scale~\footnote{In~\cite{Thomas:2008ga} the low scale was set to  $\rm 0.4\, GeV$, perhaps too low for the use of pQCD. Yet the results seem to make sense.}, yields vanishing $L_d\approx 0$ and a larger  $L_u\approx 0.3$. These results support the idea of a $d$ quark  always bound
into a scalar $ud$ diquark and  is decoupled from the nucleon's spin, while the $u$ quark  is
free with most of the nucleon spin.
 
\subsection{Examples of DGLAP evolution}

Using anomalous dimensions in practice is not very convenient,
because the operator averages are $moments$ of the PDFs, that is
their convolution with powers of $x$. Those naturally
require knowledge of the PDFs over all $x\in (0,1)$, which is not available
in practice.

Dokshitzer-Gribov-Lipatov, and Altarelli-Parisi (DGLAP)
worked out convenient "evolution equations" describing
parton splitting reactions, such as $q\rightarrow q g, g\rightarrow gg$ and so on. Those equations allows to 
"evolve" PDFs in resolution, in any direction $$PDF(x,Q_1^2)\leftrightarrow PDF(x,Q_2^2)$$
In particular, the standard normalization point used by the lattice community is
$Q^2=2\,\rm  GeV^2$, to which results from different lattices are evolved for comparison.

The
experiments are done at much larger resolutions, in a wide range
$Q^2=10-1000\, \rm GeV^2$, so a practical tool allowing simultaneous
fit to many different experiments was badly needed.
Large collaborations have been created for performing this task,
such as CTEQ collaboration. Its web page provides a lot of
textbook-like material and useful practical tools.

While there is no need to reproduce most of it here, let us still
mention a particularly useful source we used, for the purposes of this book.
We suggest ManeParse \footnote{ A Mathematica reader for Parton Distribution Functions
D.B.Clark, E.Godat, F.I.Olness
Published in : Comput.Phys.Commun .216 (2017) 126 - 137
e - Print : 1605.08012[hep - ph]} providing directly  the $PDF(x,Q)$
functions in Mathematica notebook form.

We now provide
three examples of DGLAP evolution from this source. 
One is the evolution of the density of (relatively soft) gluons.
It shows a rapid disappearance of the soft gluons at resolution at or below
    $Q\rightarrow 1\, \rm GeV$. So, at lower resolution
    one can assume that gluons are absent\footnote{Once again, we remind the reader that "gluons" (as small amplitude waves) are different from the  $glue$ which include e.g. semiclassical glue in the form of instantons.}.
    
\begin{figure}
    \centering
    \includegraphics[width=0.5\linewidth]{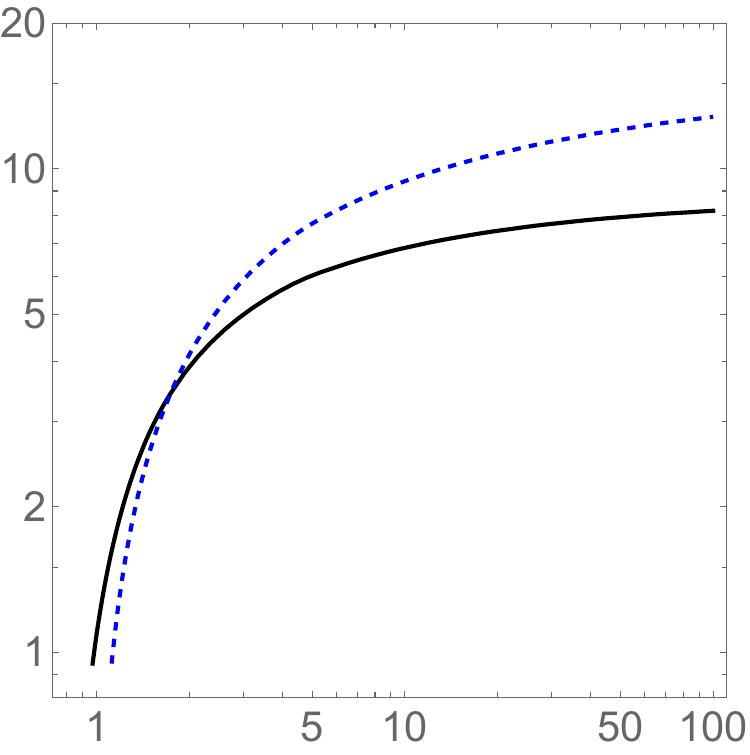}
    \caption{Plot of the gluon PDF $xg(x,Q)$ versus $Q (GeV)$  at $x=0.01$ (black solid) and  $x=0.005$ (blue dashed). 
    }
    \label{fig_gluon_evolution}
\end{figure}

\begin{figure}[h!]
    \centering
    \includegraphics[width=0.45\linewidth]{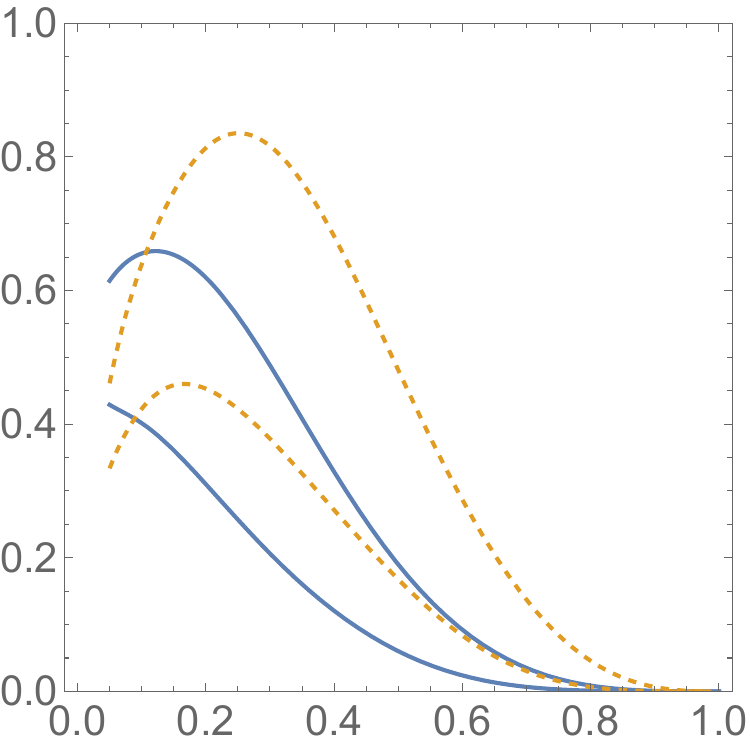}
        \includegraphics[width=0.45\linewidth]{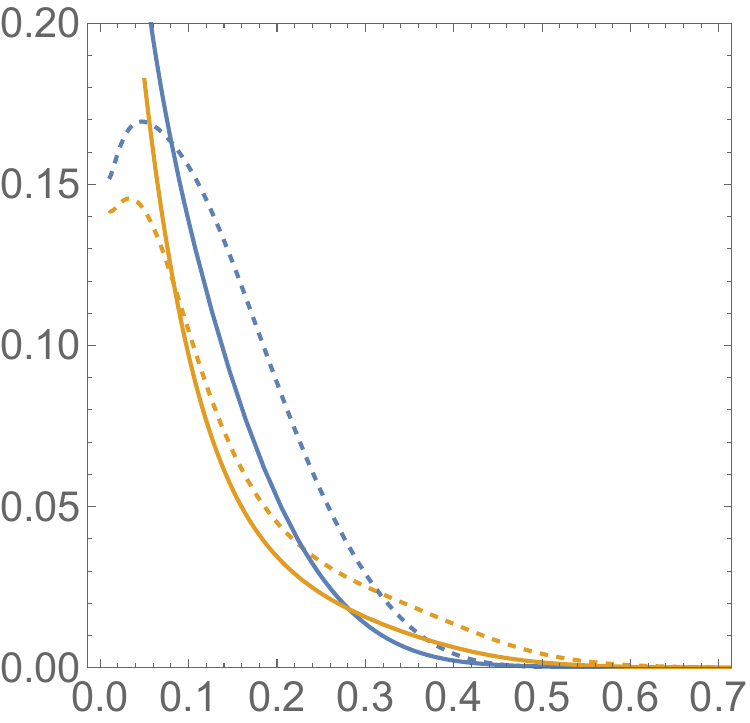}
    \caption{
    the $xu(x),xd(x)$ distributions, and the other of antiquarks $x\bar u(x), x\bar d(x)$, upper and lower curves~\cite{Miesch:2025wro}. Those are evolving from resolution $Q^2=50\, \rm GeV^2$ (solid lines) to $Q^2=1\, GeV^2$ (dashed). It is this lower one at which the bridging to
    spectroscopy is supposed to take place.}
    \label{fig_DGLAP}
\end{figure}

Two more plots in Fig.\ref{fig_DGLAP} show the evolution
of the $u,d$ distributions in the proton, as well as  that of the antiquark "sea" $\bar u, \bar d$. The low-resolution ones (dashed) lack perturbative small-$x$ partons and thus turn down at the left sides of the plots. The "evolved" ones (solid) 
move to the left and have a different form.

These PDFs are believed to be dominated by the $qqq$ and  the $qqqq\bar q$ Fock components
of the proton wave functions, which we discuss in the pentaquark chapters.
One may use the empirical PDF plots like this and compare them to the PDFs derived from the "theoretical wave functions", e.g.
for our $L=1$ pentaquark admixture to the nucleons. Those are supposed to be
defined at the boundary of DGLAP desciption, namely at the "boundary scale" $Q=1\, \rm GeV$ between the chiral and gluonic evolutions.

\begin{figure}
    \centering
    \includegraphics[width=0.55\linewidth]{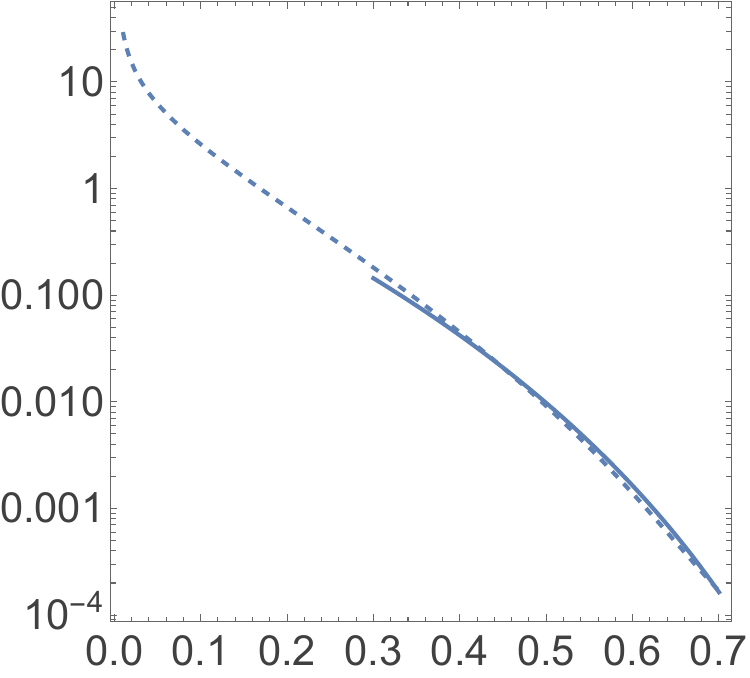}
    \caption{Left:The PDFs of antiquarks $u(x,1)+d(x,1)$ at scale $Q=1\, GeV$ (no $x$ prefactor), from ManeParse, compared to the fit $\sim (1-x)^8$ 
    shown by the solid line~\cite{Miesch:2025wro}}
    \label{fig_antiquark_PDF_Q1}
\end{figure}

\section{The force acting on a quark and twist-3 operators}

This section follows our paper \cite{Liu:2025ypg}.

Higher twist operators are of higher dimensions, including
  higher powers of quark and gluon fields. They carry information
about the partonic correlations. Their general theory has been carried out 
in the early 1980's in~\cite{Shuryak:1981kj} for unpolarized targets, and 
in~\cite{Shuryak:1981pi} for polarized ones. They provide the QCD corrections
to the partonic sum rules, such as the Bjorken and the Ellis-Jaffe sum rule. 

The most interesting  correlation between partons are those 
 stemming from a 
polarized target. While the structure function $g_1(x,Q^2)$ 
starts with the usual twist-2 operators, the structure function
 $g_2(x,Q^2)$ starts with twist-3. Since in experiments the structure functions can be  separated $kinematically$, this fact  offers the most direct access to the higher twist  physics. For a  transversely polarized nucleon, at $90^0$ to the incoming momentum, $g_1(x,Q^2)$ vanishes and the remaining DIS amplitude is purely twist-3. 
 The pertinent physics is related with the local operator\footnote{Although the operator carries three open indices $\rho,\mu,\nu$, the last pair is antisymmetrized, and the total spin is 2 and not 3.} 
\be 
\label{OFORCE}
O_{\bar q G q}= ig\big(\bar q \gamma^\rho  G^{\mu\nu} q \big)
\ee
given by the value of the gauge field strength on the struck quark. 
(The color indices  are not explicitly shown but assumed, here and below.)

The magnitude of this effect has  been  discussed in~\cite{Burkardt:2008ps}. It was pointed out that in the large-$N_c$ limit, the forces on the $u$ and $d$ quarks in the proton should be equal in magnitude but  opposite in sign. The suggested magnitude was of the order of
\be \label{eqn_Burkardt} F_{u}=-F_{d}\sim 0.1\, \rm GeV/fm \ee
Note that it is an order of magnitude smaller than  
the confining "string tension"  force
\be F_{\bar Q Q}=\sigma\approx 1\,\rm  GeV/fm\approx 0.2\, GeV^2 \ee 

\subsection{Twist-3 force in the Instanton Liquid Model (ILM)}
In the late 1970's to early 1980's  studies of semiclassical pseudoparticles - instantons and anti-instantons - 
had led to view the QCD vacuum as an "instanton liquid" \cite{Shuryak:1981ff}. The  vacuum is very inhomogeneous, with 4-quark operators of the type $LR+RL$ 
(here left-handed current means $L=\bar q (1+\gamma_5) q$, and right-handed with the opposite sine on front of $\gamma_5$)
strongly enhanced through fermionic zero modes, of a single pseudoparticle . Unfortunately, the operator (\ref{OFORCE}) is of different type, $LL+RR$, therefore it $not$ enhanced in a single pseudoparticle, and thus it does not appear in the
leading order in the pseudoparticle density.  The evaluation of matrix elements of such operators has been postponed in earlier literature, in favor of  discussion of the "most enhanced" effects.

A single instanton has a probability proportional to very small product of light quark "Lagrangian" (Higgs-induced)
masses $m_u m_d m_s/\Lambda_{QCD}^3\sim 10^{-4}$. Therefore an ideal gas
of independent pseudoparticles would be extremely dilute and thus irrelevant.
Fortunately \cite{Shuryak:1981ff}, 
the  QCD vacuum is not an ideal gas but rather  a "liquid", with strong correlations mediated by light quark exchanges. 
As a result, small "Lagrangian" masses are substituted by much larger
effective quark masses, the
   $SU(N_f)$ chiral symmetry is broken by a nonzero quark condensate. In the  ILM  the density and typical size of the pseudoparticles are
\be \label{eqn_ILM}
{N\over V_4}=
n_{I+\bar{I}}\approx 1\rm  fm^{-4},\,\,\,\, \rho\sim {1\over 3 } fm  \ee
In spite of  smallness of the diluteness parameter
\be {N \rho^4\over V_4 N_c}\approx ({1 \over 3})^5 \ll 1 \ee
 properties of instanton ensemble are not expandable in simple Taylor 
series in diluteness. For example, (for two light quark flavors)
a constituent quark mass scales as a $root$ of diluteness
\be
 M \rho\sim \sqrt {N \rho^4\over V_4}  
\ee
as it follows from Bethe-Salpeter equation summing infinite number of quark loop diagrams.

Derivation of effective action $S_{eff}$ for constituent quarks in the mean field approach \cite{Diakonov:1985eg} 
complemented this mass by a specific form factor $\mathcal{F}(k)$
\be S_{eff}=\int {d^4k \over (2\pi)^4} \psi^\dagger(k) [\slashed k - i M \mathcal{F}(k)] \psi(k)
\ee
related to quark zero mode. The form factor describes
dependence of the constituent quark mass on the 
momentum scale under investigation.
Further studies of quark
propagators 
were done using numerical simulations of the instanton ensembles,
see review \cite{Schafer:1996wv}.

To our knowledge, the first attempts to theoretically quantify twist-3 and twist-4 matrix
elements  have been carried in~\cite{Balla:1997hf}. They have used a version of  the
ILM  for vacuum structure, and chiral soliton version 
(following from it at large number of colors) for the nucleon. Without going into great details, their main conclusion is that the twist-3 matrix elements are non-zero but still suppressed by a power of the  diluteness parameter.
Let us also mention the recent study in~\cite{Hatta:2024otc} devoted to the twist-3
contributions  for other (e.g. momentum) sum rules.

\subsection{Lattice studies}
We would not go into the technicalities and long history of lattice gauge field simulations, and the number of key achievements. It is sufficient to note that current lattice simulations are able to work with fermions as light as the quarks in the real world,  reproducing major parts of hadronic phenomenology rather well. This includes the nucleon mass, form factors, PDFs and even GPDs. 

Recently, there have been new attempts to  quantify the values of higher twist operators. In particular 
the recent work by the Adelaide group \cite{Crawford:2024wzx}, where  numerical evaluation  of the twist-3 operators in the nucleon, were quoted well inside the (statistical) error bars. Remarkably, their results 
\be \label{eqn_lattice_Adelaide} 
F_{u} \approx 3\, \rm GeV/fm,\,\, \,\, F_{d}= 0. \pm \rm 0.05\, GeV/fm
\ee
are  much larger than the one suggested earlier in (\ref{eqn_Burkardt}).
The force on $u$ and $d$ are surprisingly different. Furthermore, as is clear from Fig.~5 of that paper, it is normal both to the directions of motion ($z$) and the nucleon's spin ($x$), being localized inside a small sphere of radius only $0.2-0.3\,\rm  fm$.
The vanishing force on the $d$ quark brings in a (perhaps simplistic)  explanation:
if $d$ quark is always locked into the spin-zero $ud$ "good diquark", there would be no polarization-sensitive effects associated with it.


In DIS scattering, the twis-3 PDF  measures the average Lorentz force in the nucleon. While it is  not amenable to partonic densities, it describes the colored force 
interaction on a struck parton. For a polarized nucleon along the x-direction  with light cone momentum $P^+$ along the z=direction, the average Lorentz force force 
\bea
\label{FY}
d_2=-\frac {\langle PS_x|\bar q (0)\gamma^+gG^{+y}(0)q(0)|P,S_x\rangle}
{2m_N(P^{+2}) S_x}
= \frac{F^y(0)}{\sqrt 2 m_N S_x}
\eea
is related to  the second Mellin moment of the quark intrinsic $\bar g_2(x) $
(after subtraction of the twis-2 contribution)~\cite{Aslan:2019jis}
\bea
\frac {d_2}3=\int_0^1dx\,x^2\, \bar g_2(x)
\eea
On the light front, the color valued field strength is
\bea
G^{+y}=\frac 1{\sqrt 2}(-E^y+B^x)=-\frac 1{\sqrt 2}(\vec E+\vec v\times \vec B)^y
\eea
for a struck parton along the negative z-direction.

At low resolution, the Lorentz force (\ref{FY}) can be evaluated using the ILM. 
For that we first note that in Euclidean signature,  the contribution from a 
single instanton or anti-instanton  vanishes because of self-duality, i.e. 
$E=\pm B$. Vacuum tunneling configurations carry zero energy-momentum tensor.
Second we also note that the force involves the quark charge which is chiral
even. In the ILM the contribution to (\ref{FY}) stems from IA molecules
which induce chirally even flips,
and is therefore of second order in the packing fraction. 

\begin{figure}
    \centering
\includegraphics[width=0.7\linewidth]{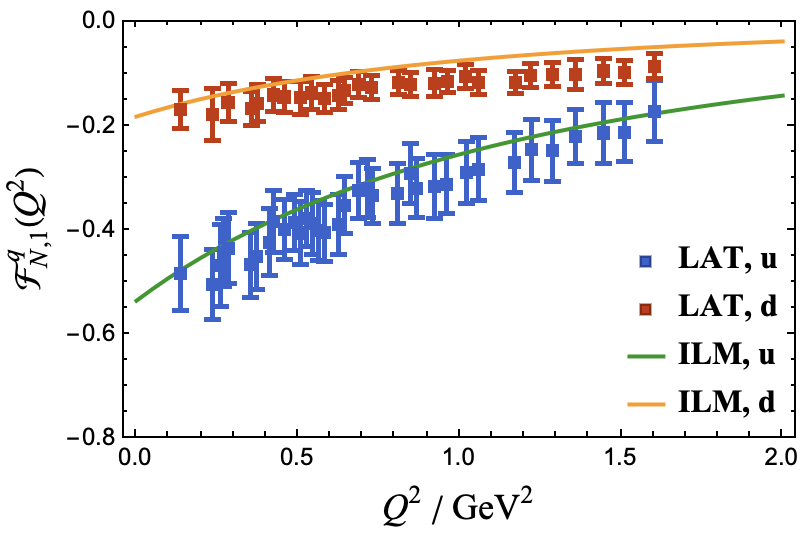}
\caption{Color force form factors $\mathcal{F}^q_{N,1}(Q^2)$ from ILM~\cite{Liu:2025ypg}  with parameters $n_{mol}=7.248$ fm$^{-4}$, $\gamma_{I\bar{I}}=24.22$ fm$^{4}$, and determinantal quark mass $m^*=82.6$ MeV evolved to 2 GeV compared to the lattice calculation with pion mass $450$ MeV in \cite{Crawford:2024wzx}.}
\label{fig:nuc_force1}
\end{figure}

\begin{figure}
    \centering
\includegraphics[width=0.7\linewidth]{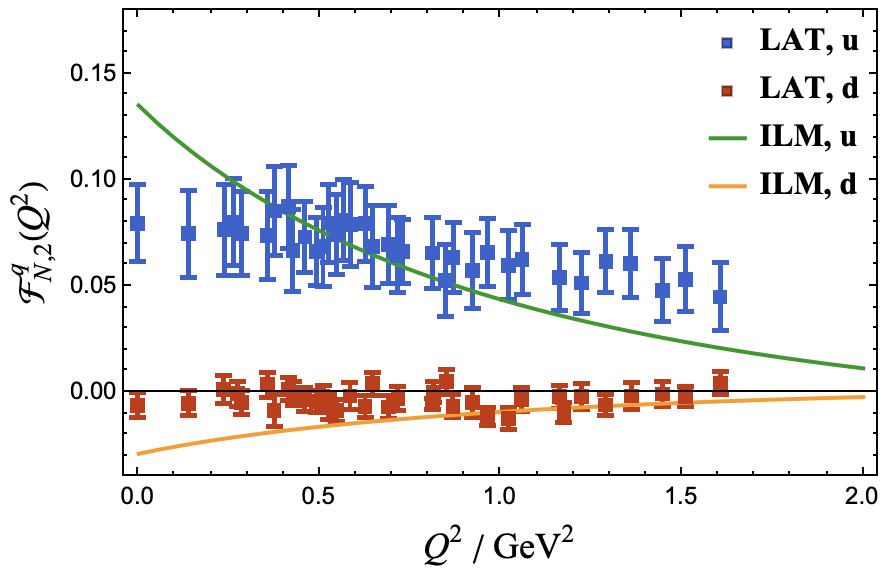}
\caption{Color  Lorentz force form factors $\mathcal{F}^q_{N,2}(Q^2)$ using the ILM~\cite{Liu:2025ypg}  with parameters $n_{\rm mol}=7.248$ fm$^{-4}$, $\gamma_{I\bar{I}}=24.22$ fm$^{4}$, and a determinantal quark mass $m^*=82.6$ MeV, evolved to 2 GeV and compared to the lattice calculation with pion mass $450$ MeV in~\cite{Crawford:2024wzx}.}
\label{fig:nuc_force2}
\end{figure}

\begin{figure}
    \centering
\includegraphics[width=0.7\linewidth]{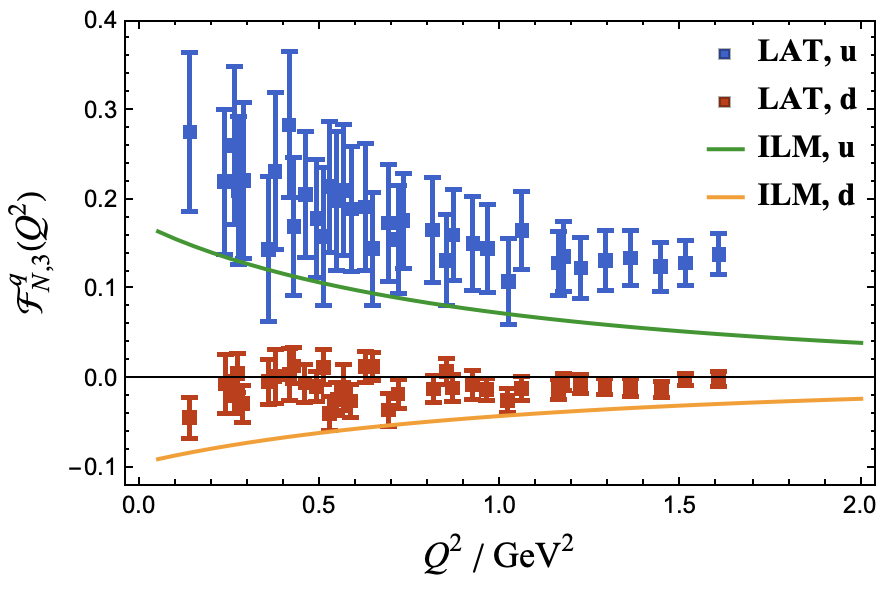}
\caption{Color Lorentz force form factors $\mathcal{F}^q_{N,3}(Q^2)$ from the ILM~\cite{Liu:2025ypg} with parameters $n_{\rm mol}=7.248$ fm$^{-4}$, $\gamma_{I\bar{I}}=24.22$ fm$^{4}$, and a determinantal quark mass $m^*=82.6$ MeV, evolved to 2 GeV and compared to the lattice calculation with pion mass $450$ MeV in~\cite{Crawford:2024wzx}.}
\label{fig:nuc_force3}
\end{figure}

Its evaluation is rather technical, so one should consult the original work \cite{Liu:2025ypg}. The force is different for different hadrons (pions, nucleons, rho mesons), which is not
surprising since even their sizes are different. As an example of 
the results see Figs.\ref{fig:nuc_force1},\ref{fig:nuc_force2},\ref{fig:nuc_force3} 

The color Lorentz force arises naturally from the non-perturbative QCD vacuum fluctuations of the gauge field, including  correlated instanton-anti-instanton  pairs.
By combining semiclassical strength and spatial structure of these fields  with the operator structure of higher-twist quark-gluon correlators, we have shown  that   molecular correlations provide an important mechanism behind the large higher-twist effects observed in polarized deep inelastic scattering, and confirmed in recent lattice simulations.
This agrees with our previous derivation of the confining potential
in quarkonia, from the same setting.

One new  result  of this study is the explicit construction and analysis of the gauge field structure of the molecular ensemble,
and its role in the quark-colored force. By employing the "ratio ansatz" for the correlated instanton-anti-instanton pair, we derived explicit expressions for the color-electric and color-magnetic field components and computed their spatial profiles. The overlap of the quark zero modes across the molecular pair was analyzed in detail, revealing a strong delocalization effect that enhances the color force.  Using Monte Carlo sampling over molecular orientations and separations, we obtained a robust estimate of the average color Lorentz force, $F\approx2$-$3\,\mathrm{GeV/fm}$, acting on a single quark. This result indicates that the nonperturbative color forces are comparable to, or can even exceed the magnitude of the confining string tension.

Our  analysis shows a  direct connection between these microscopic nonperturbative fields and the experimentally accessible twist-3 observables. The color Lorentz force operator $\bar{q}gG^{+i}\gamma^+q$ whose expectation value defines the twist-3 matrix element $d_2$
is shown to acquire  contributions from the molecular component of the instanton liquid. The induced  form factors  encode the nonlocal structure of this coupling and provide a unified description of the quark-gluon interaction over a broad range of momentum transfers. In particular, the molecular contributions dominate at low $Q^2$, leading to enhanced color forces consistent with phenomenological extractions from polarized structure functions.  These findings confirm that the twist-3 sector is direct manifestation of the underlying topological fields in the QCD vacuum.
Remarkably, the   emergent color Lorentz force  form factors are shown to be intimately related to the hadronic gravitational and transversity form factors, offering additional insights to the nature of the mass and force distributions in hadrons.


\chapter{The origin of hadronic mass and spin}

\section{Introduction}
In  general, one may relate hadronic masses $O(\Lambda_{QCD})$ to quantum breaking of the scale (conformal) symmetry, induced by the running coupling and captured by the trace anomaly.
Even in the chiral limit, the trace of the QCD energy-momentum tensor (EMT) is nonzero due to a quantum (scale) anomaly~\cite{Vainshtein:1981wh,Nowak:1996aj}. The nucleon mass $M_N$ can be related to the change of the gluon field (trace anomaly) in the nucleon state relative to the vacuum
\bea
M_N\sim \frac 1{2M_N}\bigg(\langle N|F^2|N\rangle-\langle 0|F^2|0\rangle\bigg)\approx \chi
\eea

As we already discussed, most 
hadronic masses emerge from "constituent" effective quark masses
 described within the \emph{instanton vacuum} picture. As we detailed earlier, the QCD vacuum at low resolution is a dilute ensemble of instantons and anti-instantons whose density and size set a nonperturbative hadronic scale. The delocalization of the quark zero modes across the ensemble generates spontaneous chiral symmetry breaking with the emergence of a constituent-like quark mass.
Quark hopping in the QCD vacuum, from one instanton to the next, induces a momentum-dependent dynamical mass $M(p)\!\approx\! M_0\,F^2(p\rho)$ with $M_0\approx 380\,\text{MeV}$, with $F(p\rho)$ a profiling of the zero mode. The gluon condensate measures the  instanton plus anti-instanton  density $n$ or 
$\langle 0| F^2/32\pi^2|0\rangle \approx   n$. 

The nucleon matrix element of $F^2$ in the trace anomaly, is  tied to the variance of instanton-number fluctuations in a grand-canonical treatment of the ensemble with
the instanton-liquid "compressibility'' $\chi=\partial^2{\rm ln}Z/{\partial n^2}$
($Z$ referring to the grand-canonical ensemble)~\cite{Zahed:2021fxk}.
This remarkable identity shows that most hadrons scoop their mass
from the non-perturbative glue in the QCD vacuum.
For the pion (as a Goldstone boson) this contribution vanishes in the chiral limit.

To quantify the budgeting of the mass at low resolution, Ji's  gauge-invariant energy decomposition is useful~\cite{Ji_95,Ji:2021}. In particular, at a hadronic ("soft'') renormalization scale tied to the instanton size, a \emph{hierarchy} of contributions emerge: a dominant  $M_q$ (quark kinetic/potential from zero-mode hopping) contribution, plus a sizeable anomalous piece $M_a$ stemming from the displaced gluon condensate ("epoxy''), with
a comparatively small valence-gluon piece $M_g$

\begin{figure}
\hfill
    \includegraphics[scale=0.30]{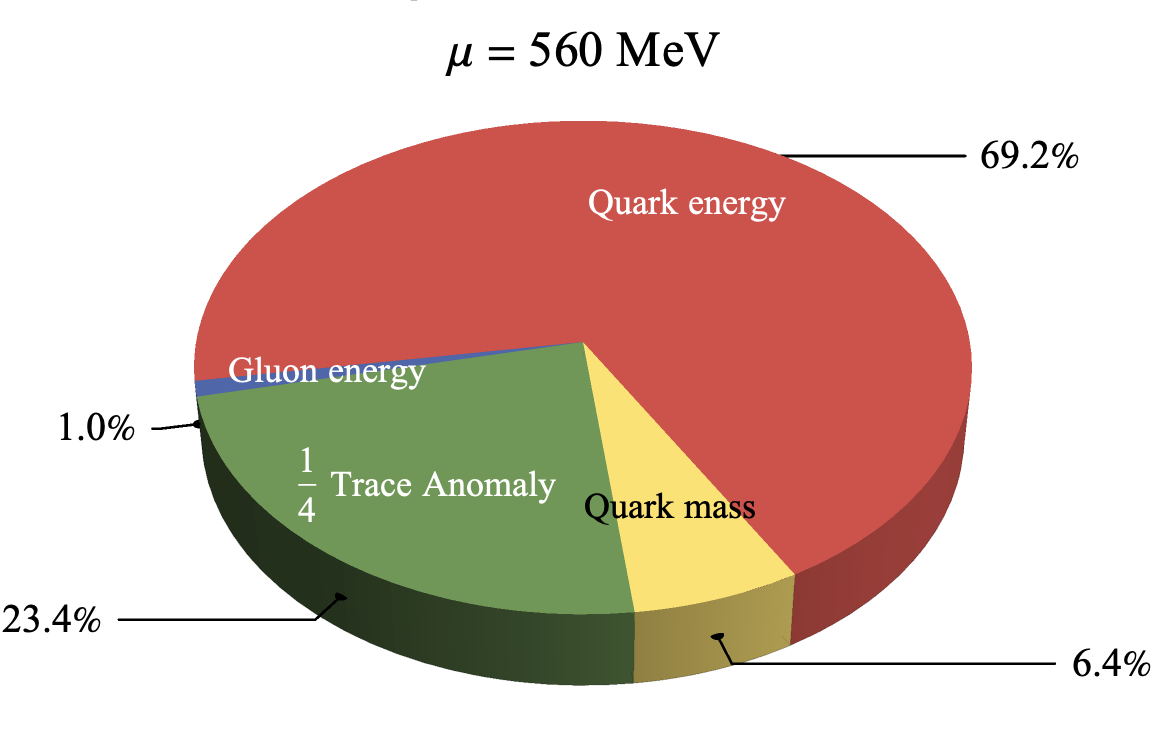}
\hfill
    \includegraphics[scale=0.30]{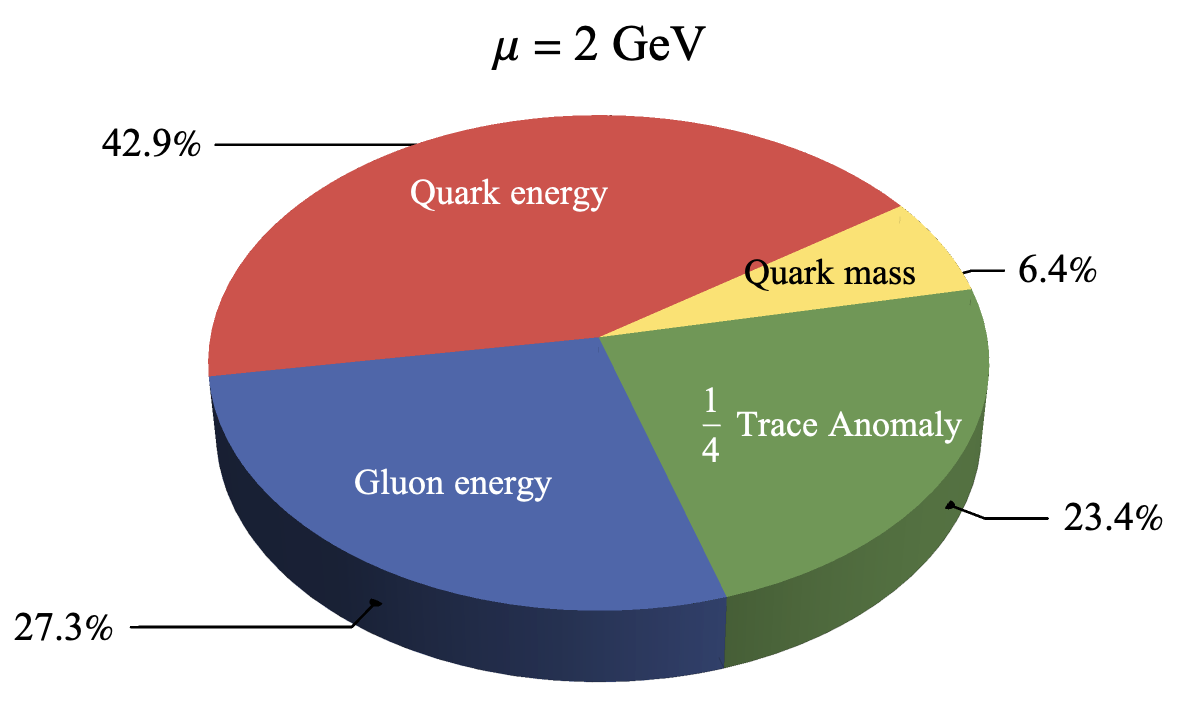}
\hfill
    \includegraphics[scale=0.30]{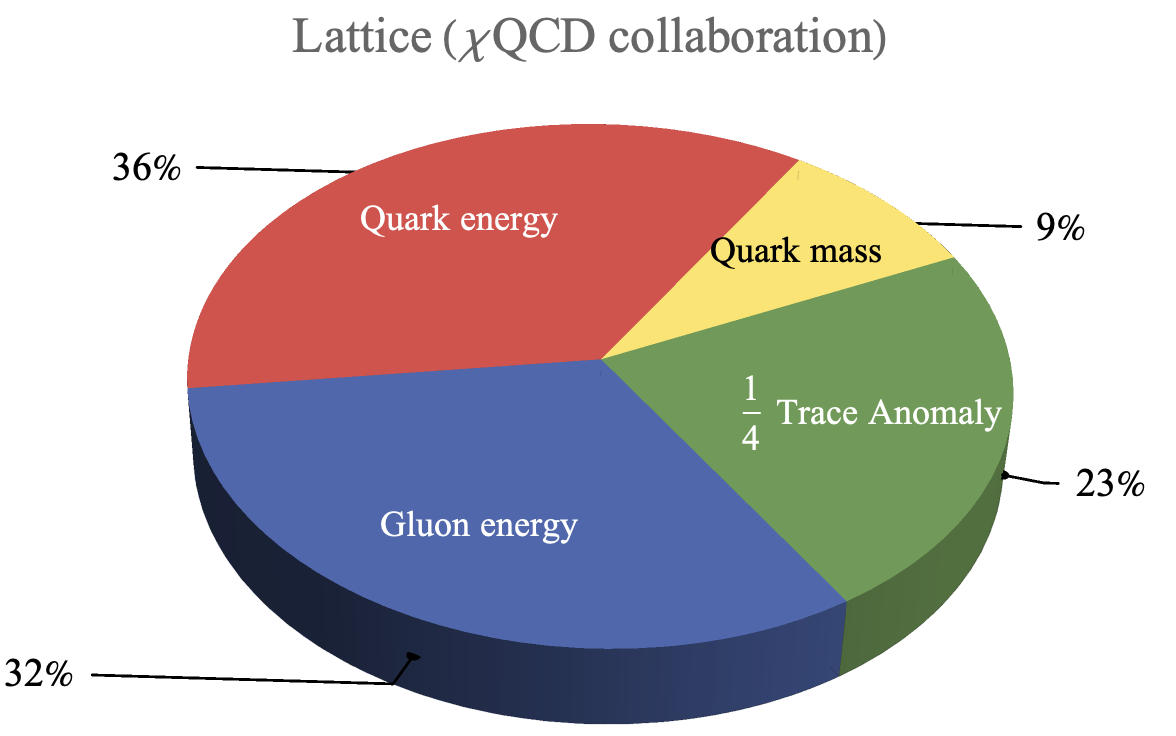}
\caption{ 
Mass decomposition using Ji's nucleon mass sum rule, in the QCD instanton vacuum at the resolution $\mu=560~\mathrm{MeV}
\sim1/\rho$~\cite{Liu:2024rdm} (a), and after DGLAP evolution at a resolution 
$\mu=2~\mathrm{GeV}$ (b); (c) Mass decomposition using Ji's nucleon mass sum rule, at a resolution of $\mu=2$ GeV from the lattice collaboration~\cite{Yang:2018nqn}.
}
\label{PIEVAC}
\end{figure}

More specifically, in QCD the stress tensor includes
 quark fields $q$ and gluon field strength $F^{a}_{\mu\nu}$, 
\begin{align}
T^{\mu\nu} &= - F^{a\mu\lambda} F^{a\nu}{}_{\lambda} + \frac{1}{4} g^{\mu\nu} F^{a}_{\alpha\beta}F^{a\,\alpha\beta}
+ \frac{i}{4}\,\bar q\,\gamma^{\{\mu}\,\overleftrightarrow{D^{\nu\}}}\, q, 
\end{align}
is "anomalous" \footnote{All heavy quark mass effects are contained in the running beta function},
with a nonzero trace
\begin{align}
T^\mu{}_\mu &= \frac{\beta(g)}{2g}\, F^{a}_{\alpha\beta}F^{a\,\alpha\beta} + (1+\gamma_m)\, m\, \bar qq,
\end{align}
so scale symmetry is broken by quantum effects (trace anomaly), even for $m\to 0$.
For a nucleon  state $\lvert N\rangle$ in its rest frame, Ji's mass decomposition is given by~\cite{Ji_95,Ji:2021}
\begin{equation}
M_N \;=\; \frac{1}{2M_N}\,\langle N\lvert T^{00}\rvert N\rangle
\;=\; M_q + M_g + M_m + M_a,
\end{equation}
with $M_q$ (quark kinetic/potential), $M_g$ (gluon field energy), $M_m$ (explicit quark-mass term), and $M_a$ (anomalous/trace part). At a scale $\mu\sim 1/\rho$ fixed by the instanton mean size~\cite{Zahed:2021fxk} 
\bea
M_q:M_a:M_g\approx 0.70:023:0.07
\eea
So about 70\% of the nucleon mass originates from the quark zero mode dynamics, 25\% from the displaced gluon condensate (anomaly), and 7\% from valence gluons. In contrast and since the pion is a Goldstone mode, its mass vanishes in the chiral limit, as per the Gell-Mann-Oakes-Renner relation 
\bea
f_\pi^2M_\pi^2=-2m\langle 0|\overline q q|0\rangle
\eea
The pion mass budget is still of similar structure as the nucleon but shifted at finite quark mass~\cite{Zahed:2021fxk}
\bea
M_\pi:M_q:M_g\approx 0.85: 0.12:0.03
\eea
Under RG evolution to higher $\mu$, the relative gluon contribution increases and the quark (kinetic/potential) piece decreases/mixes, bringing the composition closer to current lattice results quoted at $\mu=2$~GeV~\cite{Yang:2018nqn}
\begin{equation}
M_q : M_g : M_a : M_m \;=\; 0.33(4)(4) : 0.37(5)(4) : 0.23(1)(1) : 0.09(2)(1).
\end{equation}
A qualitative comparison between the lattice results at high resolution and the instanton vacuum model is shown in Table.

\begin{center}
\begin{tabular}{lcccc}
\hline
\textbf{Framework} & $M_q$ & $M_g$ & $M_a$ & $M_m$\\
\hline
Lattice QCD (PRL'18, $\mu=2$ GeV) & $0.33(4)(4)$ & $0.37(5)(4)$ & $0.23(1)(1)$ & $0.09(2)(1)$\\
Inst.~vacuum (soft scale, qualitative) & dominant & small & sizeable & small\\
\hline
\end{tabular}
\end{center}

In summary, the instanton-vacuum picture provides a clear physical mechanism for mass generation and a soft-scale hierarchy ($M_q$ dominant, $M_g$ small). The  current lattice results at $\mu=2$~GeV quantify the decomposition with a larger gluon share. The two viewpoints are consistent once renormalization-scale dependencce, operator mixing and evolution are taken into account.

\section{The quenching of the intrinsic spin}
The decomposition of the nucleon spin is a central question in hadronic physics. 
Deep inelastic scattering experiments (e.g., EMC, COMPASS, HERMES) show that the 
intrinsic quark spin accounts for only $\sim 30\%$ of the nucleon spin, leading 
to the so-called "proton spin puzzle.'' Lattice QCD calculations confirm this trend, 
but the precise mechanisms remain debated.

The QCD instanton vacuum offers a simple mechanism for why the nucleon 
intrinsic spin gets quenched. This mechanism  once combined with the observation that 
the non-perturbative glue in this vacuum carries  almost no angular momentum, a simple understanding of the nucleon spin decomposition emerges. 

A simple budgeting of the nucleon spin is given by Ji's decomposition~\cite{Ji_95}
\begin{equation}
\frac{1}{2} = S_q + L_q + J_g \,,
\end{equation}
where $S_q$ is the intrinsic quark spin contribution, $L_q$ is the quark orbital angular momentum, and  $J_g$ is the total gluon angular momentum (spin + orbital).
More specifically, the quark intrinsic spin operator is
\begin{equation}
\bm{S}_q = \int d^3x \; q^\dagger \frac{\bm{\Sigma}}{2}\, q \,,
\qquad \Sigma^i = \gamma^5 \gamma^0 \gamma^i \,,
\end{equation}
so that the flavor-singlet axial charge is
\begin{equation}
\Sigma = \frac{1}{2m_N}\langle P,S| \bar{q}\gamma^\mu\gamma_5 q|P,S\rangle \, s_\mu \,,
\end{equation}
where $s_\mu$ is the nucleon spin vector. The  gauge-invariant quark orbital angular momentum operator is
\begin{equation}
\bm{L}_q = \int d^3x \; q^\dagger \, (\bm{x}\times i\bm{D}) \, q \,,
\end{equation}
where $\bm{D} = \bm{\nabla} - ig \bm{A}$ is the covariant derivative.
The total gluon angular momentum operator is
\begin{equation}
\bm{J}_g = \int d^3x \; \bm{x}\times (\bm{E}^a \times \bm{B}^a) \,,
\end{equation}
with $\bm{E}^a$ and $\bm{B}^a$ the chromo-electric and chromo-magnetic fields.
This gauge-invariant form includes both gluon spin and orbital contributions together.
This spin decomposition is gauge invariant and provides a basis for connecting theory 
with experiment and lattice QCD.

The $U(1)$ axial anomaly provides a key relation between the divergence of the 
singlet axial current and the topological charge density
\begin{equation}
\partial^\mu J^5_\mu = 2i m \bar{\psi} \gamma_5 \psi 
+ \frac{g^2}{16\pi^2} F\tilde F \,,
\end{equation}
where $F\tilde F$ is the dual field strength tensor. In the chiral limit, the divergence of the singlet axial current is sourced completely by the topological charge density.
At low resolution, the major observation is that the quark intrinsic  spin is not conserved separately,  but mixes with  topological charge fluctuations, in the instanton vacuum
\begin{equation}
\Sigma \;\approx\; \frac {N_f}3\frac{1}{32\pi^2}\int d^4x\;
\frac{\langle 0|\, F\tilde F(x)\,F\tilde F(0)\,|0\rangle}{\langle 0|\,F^2(0)\,|0\rangle}\,,
\label{eq:A1}
\end{equation}
 This highlights the role of vacuum topology in reducing the 
observable quark spin contribution, through the scooping of the topological susceptibility. In pure Yang-Mills with no light quark screening,
the topological fluctuations per 4-volume,  are given by the Witten-Veneziano relation~\cite{Veneziano:1979ec,Witten:1979vv}
\begin{equation}
\bigg(\frac 1{32\pi^2}\bigg)^2\int \frac{d^4x}{V_4}\;
\langle 0|\, F\tilde F(x)\,F\tilde F(0)\,|0\rangle=\frac{f_\pi^2M_1^2}{2N_f}
\label{eq:A11}
\end{equation}
with $M_1$ related to the "bare" eta-prime mass. In QCD with light quarks, screening of the topological fluctuations maybe substantial, leading to a quenching of the intrinsic spin
\begin{equation}
\Sigma=\frac {N_f}3
\left[ \frac{M_1^2\,\langle\bar qq\rangle}{n}\left(\frac{2N_f\langle\bar qq\rangle}{f_\pi^2}\;-\;\frac{M_1^2}{N_f}\sum_{f=1}^{N_f}\frac{1}{m_f}\right)^{-1} \right]
\;\xrightarrow[m_f\to m]{}\;
\frac{1}{3}\Big[\frac{f_\pi^2\,M_1^2}{2\,n}\Big]\Big[1+\frac{M_1^2}{m_\pi^2}\Big]^{-1}\!.
\end{equation}
In the chiral limit one effectively trades $M_1\to m_\pi$, causing a total depletion of the intrinsic nucleon spin!

At low resolution or  "soft'' renormalization scale with
$\mu \sim 1/\rho$, the QCD instanton vacuum suggests the following partition of
the nucleon spin
\begin{equation}
S_q \approx 0.60,\qquad
L_q \approx 0.36,\qquad
J_g \approx 0.04 \,.
\end{equation}
Hence,  the majority of the spin arises from quarks, primarily through their 
intrinsic spin, while gluons contribute only a few percent. The chief reason is that at
low resolution the glue is dominated by tunneling self-dual gauge fields that carry
neither energy nor angular momentum  This contrasts 
with high-scale lattice QCD results, which show  a comparable  $S_q$ but a  larger $J_g$. 
The difference is attributed to QCD scale evolution. The  gluon contributions grow 
at higher resolution at the expense of the quark orbital contribution, since the intrinsic spin is fixed by the topological susceptibility which is RG invariant. 

In summary, the instanton vacuum explains how topological charge fluctuations "absorb'' part of the quark spin through the anomaly, leaving a reduced observable intrinsic spin $\Sigma$. The relation  to the topological susceptibility suggests that in the chiral limit, the light quark masses vanish and screening becomes perfect, an indication that $\Sigma$ could approach zero. At finite quark masses, the screening is partial, so a sizable quark intrinsic spin fraction survives, consistent with phenomenology.
The small gluon contribution at the instanton scale shows that the glue is "soft'', but evolves to larger values at higher renormalization scales. 

These observations 
provide a natural explanation  for the observed suppression of the quark intrinsic spin in the nucleon at low resolution. The decomposition~\cite{Zahed:2022wae} 
\begin{equation}
\frac{1}{2} \approx 0.60 + 0.36 + 0.04
\end{equation}
illustrates the dominance of the quark degrees of freedom at the non-perturbative scale, 
with gluonic contributions emerging at higher resolution. This framework 
connects the proton spin puzzle with the topological structure of the QCD vacuum.

\begin{figure}
    \includegraphics[scale=0.46]{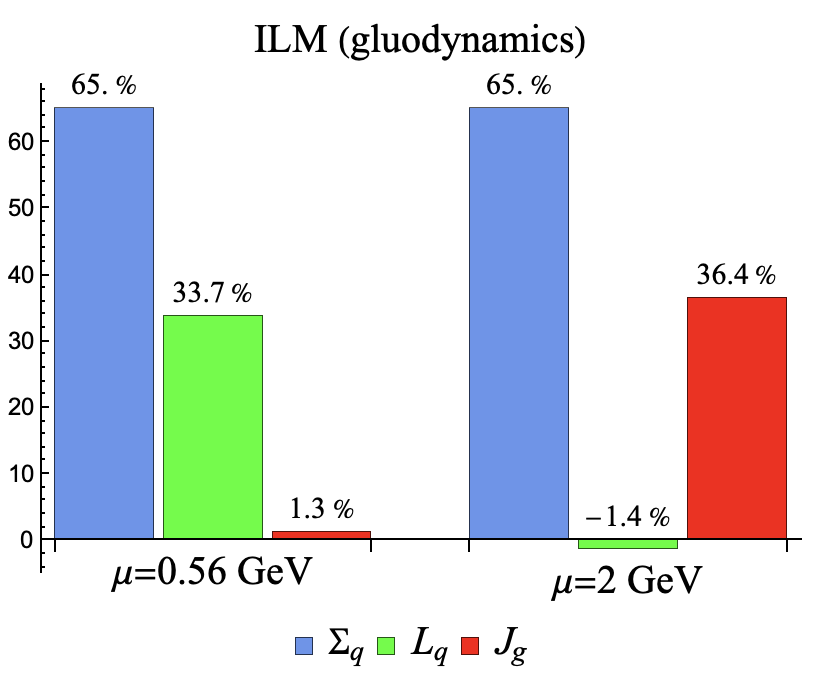}
    \hfill
    \includegraphics[scale=0.46]{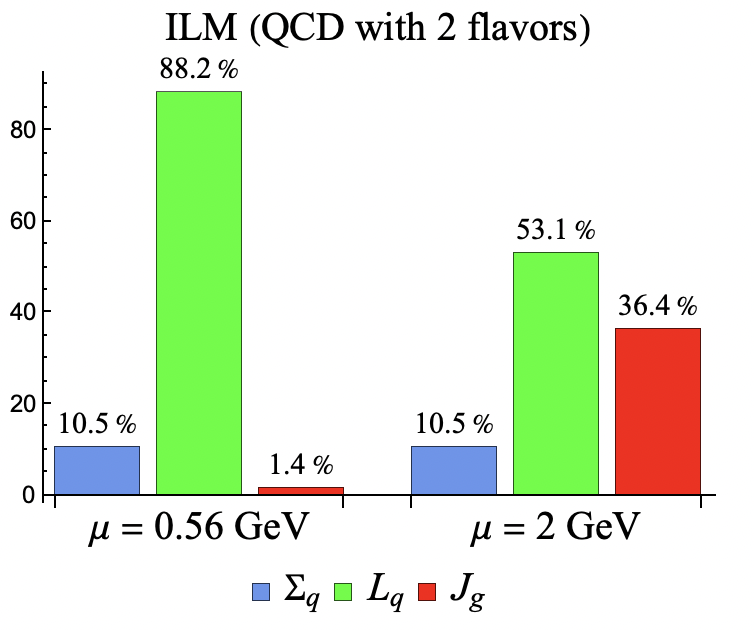}
    \hfill
    \includegraphics[scale=0.46]{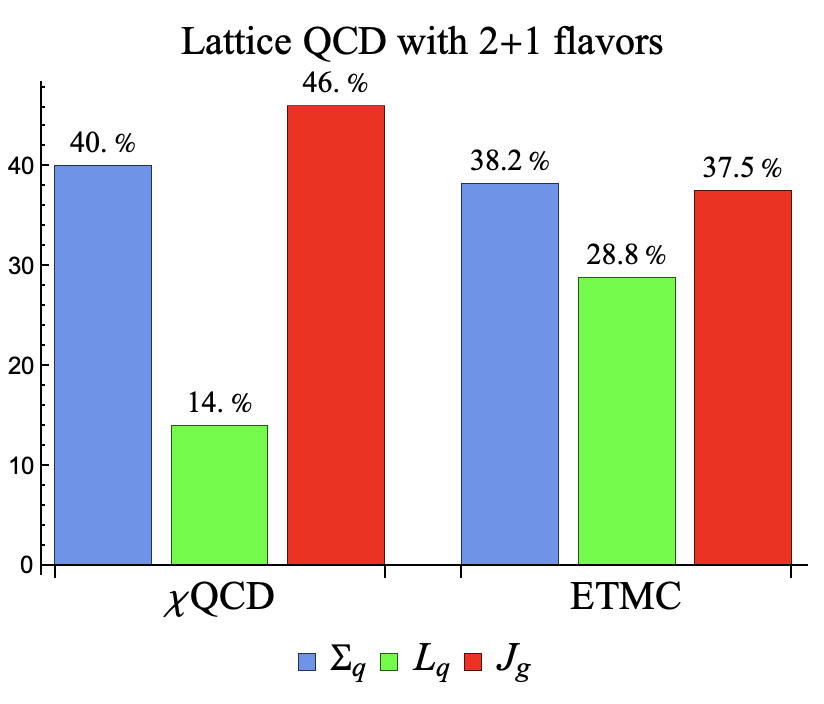}
    \caption{a: Spin decomposition using Ji's sum rule, in the quenched QCD instanton vacuum at $\mu=0.56\,{\rm GeV}$ (left),  and $\mu=2\,{\rm GeV}$ (right); b: Spin decomposition using Ji's sum rule, in the $N_f=2$ QCD instanton vacuum at $\mu=0.56\,{\rm GeV}$ (left) and $\mu=2\,{\rm GeV}$ (right); 
    c: Spin decomposition using Ji's sum rule at a resolution of $\mu=2\,{\rm GeV}$, from the lattice collaboration $\chi$QCD ~\cite{Wang:2021vqy} (left) and the ETMC~\cite{Alexandrou:2020sml} (right). All the results from the ILM are from~\cite{Liu:2024rdm}.}
    \label{fig:spin_sum}
\end{figure}

\chapter{ Generalized parton distributions (GPDs)}
\section{Introduction and Physical Motivation}
   Parton distributions (PDFs) were generalized to
 GPDs, which provide a unified light-front framework for describing the internal structure of hadrons~\cite{Ji:1996nm,Belitsky:2005qn}. They interpolate between parton distribution functions (PDFs) in the forward limit and electromagnetic as well as gravitational form factors when integrated over the parton momentum fraction. Because the defining matrix elements are off-forward, GPDs encode transverse spatial information in addition to longitudinal momentum flow, giving access to a genuine (2+1)-dimensional tomography of hadrons.

GPDs arise as off-forward matrix elements of twist-2 quark and gluon operators on the light front. Their phenomenological importance is underscored by deeply virtual Compton scattering (DVCS) and deeply virtual meson production~\cite{Ji:1996nm, Radyushkin:1997ki}, both of which probe specific combinations of quark GPDs over a controlled range of momentum transfer. Experimental programs at JLab, COMPASS, and future EIC facilities~\cite{AbdulKhalek:2021gbh, Anderle:2021wcy} aim to map these functions with increasing precision.

The Instanton Liquid Model (ILM)~\cite{Shuryak:2021hng, Shuryak:2021mlf} provides an analytically tractable description of nonperturbative QCD at the low-resolution scale $\mu_0\sim1$ GeV. Instantons induce nonlocal interactions that generate a momentum-dependent dynamical mass and nonlocal vertex functions. These structures naturally feed into the corresponding light-front wave functions (LFWFs), providing characteristic $k_\perp$-dependent shapes for GPDs that differ from purely phenomenological parametrizations or perturbative constructions.

This section presents a complete ILM-based construction of mesonic and baryonic GPDs in the DGLAP region $\xi \le x \le 1$. Since particle number is preserved in this interval, GPDs reduce to overlaps of initial and final LFWFs ( an approach particularly well suited to the nonlocal ILM structure). The resulting GPDs define low-scale boundary conditions that may be evolved perturbatively to higher scales relevant for lattice QCD or phenomenology.

\section{Twist-2 Definitions and Kinematics}

The unpolarized meson GPD is defined through the standard light-front bilinear~\cite{Ji:1996nm,Belitsky:2005qn}
\begin{equation}
H(x,\xi,t)
=\int \frac{dz^-}{4\pi}\,e^{i\frac{x}{2}P^+z^-}
\langle M(p^+)|\bar{q}(0)\gamma^+[0,z^-]\,q(z^-)|M(p^-)\rangle,
\label{eq:GPDmeson}
\end{equation} 
with Wilson line $[0,z^-]\!=\!1$ in light-cone gauge. The polarized analogue is obtained by replacing $\gamma^+ \to \gamma^+\gamma_5$.
The symmetric-frame kinematics follow the standard GPD conventions
\begin{equation}
p^\pm = P^\pm\pm\frac{\Delta^\pm}{2},\qquad
t=-\Delta^2,\qquad
x=\frac{k^+}{P^+},\qquad
\xi=-\frac{\Delta^+}{2P^+},
\label{eq:kinematics}
\end{equation}
as summarized in Eq.\,(3). We adopt the mostly-plus metric so that $p^2=-m^2$ and $t<0$.

Because ILM LFWFs contain only the lowest Fock sector, they cannot describe the ERBL domain $0<x<\xi$, which involves parton-number-changing transitions. As emphasized in Refs.~\cite{Ji:1996nm,Belitsky:2005qn}, full polynomiality requires both DGLAP and ERBL regions, so only DGLAP-sector polynomiality constraints apply here.

\section{Mesonic GPDs from ILM Light-Front Wave Functions}

Mesons in the ILM possess three leading LFWF components labeled by orbital projection $\Lambda=0,\pm1$~\cite{Ji:2003fw}. At zero skewness the unpolarized GPD reduces to the overlap integral
\begin{equation}
H(x,0,t)
=\sum_{\Lambda=0,\pm1}\!
\int \frac{d^2k_\perp}{(2\pi)^3}\,
\psi^{*}_{P,\Lambda}(x,k'_\perp)\,
\psi_{P,\Lambda}(x,k_\perp),
\label{eq:HmesonOverlap}
\end{equation}
with the momentum shift
\begin{equation}
k'_\perp=k_\perp+(1-x)\Delta_\perp.
\end{equation}

Earlier ILM studies demonstrate that mesonic LFWFs are well approximated by Gaussians~\cite{Shuryak:2021hng, Shuryak:2021mlf},
\begin{equation}
\psi(x,k_\perp)\sim e^{-A(x)k_\perp^2},
\end{equation}
leading to the analytic expression
\begin{equation}
H(x,0,-\Delta_\perp^2)
\sim e^{-A(x)\left(\frac34-x(1-x)\right)\Delta_\perp^2},
\label{eq:mesonGaussian}
\end{equation}
consistent with Eq.\,(9).  
This yields an $x$-dependent transverse radius,
\[
\langle b_\perp^2(x)\rangle=4A(x)\left(\frac34-x(1-x)\right),
\]
which narrows at large $x$ as expected from the localization of high-momentum constituents.

\section{Nucleon and $\Delta$ GPDs in the ILM}

The nucleon and $\Delta$ LFWFs follow from solving ILM-inspired constituent Hamiltonians~\cite{Shuryak:2022thi, Shuryak:2022sii}. For baryons the twist-2 quark GPDs are defined by~\cite{Ji:1996nm,Belitsky:2005qn}
\begin{align}
\int \frac{dz^-P^+}{4\pi}\,e^{\frac{i}{2}xP^+z^-}
\langle p',\Lambda'|\bar{q}(0)\gamma^+q(z^-)|p,\Lambda\rangle
&=
\bar{N}(p',\Lambda')\left[H(x,\xi,t)\gamma^+
+E(x,\xi,t)\frac{i\sigma^{+\!j}\Delta_j}{2m_N}\right]N(p,\Lambda),\\
\int \frac{dz^-P^+}{4\pi}\,e^{\frac{i}{2}xP^+z^-}
\langle p',\Lambda'|\bar{q}(0)\gamma^+\gamma_5 q(z^-)|p,\Lambda\rangle
&=
\bar{N}(p',\Lambda')\left[\tilde H(x,\xi,t)\gamma^+\gamma_5
+\tilde E(x,\xi,t)\frac{\Delta^+\gamma_5}{2m_N}\right]N(p,\Lambda),
\end{align}
with conventions matching Eq.\,(12).

At $\xi=0$ the unpolarized GPDs reduce to LFWF overlaps~\cite{Shuryak:2022thi}:
\begin{align}
H(x,0,t)
&=\int \mathcal{D}[x_i,k_{i\perp}]\,
\delta(x-x_1)\,
\psi_+^\ast(x_i,k'_{i\perp},\lambda_i)\psi_+(x_i,k_{i\perp},\lambda_i),\\
E(x,0,t)
&=-2m_N\,q_L
\!\!\int\!\mathcal{D}[x_i,k_{i\perp}]\,
\delta(x-x_1)\,
\psi_+^\ast\psi_-.
\end{align}

Numerical ILM LFWFs yield GPDs well described by the Gaussian ansatz~\cite{Guidal:2004nd}
\[
\mathrm{GPD}(x,Q^2)\approx f_1(x)\exp[-Q^2 f_2(x)],
\]
with a pronounced $x$-dependent width. As $Q^2$ increases the peak of $H(x,0,Q^2)$ shifts to larger $x$, confirming that high-$Q^2$ probes select compact, high-momentum configurations, in agreement with the trends shown in Fig.\,1 of~\cite{Shuryak:2022thi}.

\section{GPD Slopes and Transverse Tomography}

The slope
\[
-d(\ln H)/dQ^2
\]
evaluated near $Q^2\simeq1$ GeV$^2$ shows a maximum near $x\approx 1/3$~\cite{Shuryak:2022sii}, see Fig.\,2. The corresponding transverse radius
\[
R_{\mathrm{rms}}(x)=\sqrt{2f_2(x)}
\]
is typically around $0.6\,\mathrm{fm}$.  

Using the Gaussian form, the impact-parameter density at zero skewness is
\begin{equation}
q(x,b_\perp)=\int\frac{d^2\Delta_\perp}{(2\pi)^2}\,
e^{-i\Delta_\perp\cdot b_\perp}\,
H(x,0,-\Delta_\perp^2),
\label{eq:tomography}
\end{equation}
which yields the analytic expressions noted in~\cite{Shuryak:2022sii}. These describe a narrowing of the profile at large $x$ and a broadening at small $x$, completing the intuitive 3D tomographic picture.

\section{Electromagnetic and Gravitational Form Factors}

Integrating GPDs over $x$ at zero skewness gives the standard form factors~\cite{Belitsky:2005qn}
\begin{equation}
F_1(t)=\int dx\,H(x,0,t),\qquad
F_2(t)=\int dx\,E(x,0,t),\qquad
G_A(t)=\int dx\,\tilde H(x,0,t),
\label{eq:FFfromGPD}
\end{equation}
as stated in Eq.\,(21). ILM results for $Q^4F_1^d(Q^2)$ match dipole-like behavior at low $Q^2$ and approach the perturbative scaling limit at high $Q^2$~\cite{Shuryak:2022sii}.

The first moment yields the gravitational form factors~\cite{Polyakov:2018zvc}
\[
A(t)=\int dx\,xH(x,0,t),\qquad
B(t)=\int dx\,xE(x,0,t),
\]
Eq.\,(22). ILM predictions show $A(t)$ falling more slowly than $F_1(t)$, implying a more compact mass distribution than charge distribution, consistent with the trends in Fig.\,4 of~\cite{Shuryak:2022sii}.

\section{Comparison to Lattice QCD and Global Picture}

Lattice-determined ratios such as $A_2(0,|t|)/A_1(0,|t|)$ agree well in slope with ILM predictions~\cite{Hagler:2007xi}. Normalization differences reflect the lower ILM scale where valence quarks carry most of the momentum.

Recent LaMET-based extractions of quasi-GPDs~\cite{Bhattacharya:2022vvo} show good agreement with ILM results for $x>0.4$ at fixed $t=-0.69$ GeV$^2$, despite strong $t$-dependence. This reinforces the interpretation of ILM predictions as physically realistic low-scale inputs for DGLAP evolution.

\section{Summary}

GPDs in the Instanton Liquid Model arise from the nonlocal structure of instanton-induced interactions and retain full analytic control in the DGLAP region. ILM LFWFs produce transparent predictions for the $x$- and $t$-dependence of GPDs, transverse tomography, and associated form factors. Comparison with lattice QCD and LaMET reconstructions shows good agreement at moderate and large $x$, highlighting the physical relevance of instanton-induced constituent dynamics at low resolution.


\begin{subappendices}
\section{Operator Structure and Kinematic Conventions for GPDs}
\label{app:GPDops}

Generalized parton distributions are defined through nonlocal quark bilinears on the light front. For a quark field $q$ in a meson or baryon state, the leading twist-2 unpolarized operator reads
\begin{equation}
\mathcal{O}^+(z^-)=\bar{q}(0)\gamma^+[0,z^-]q(z^-),
\end{equation}
where $[0,z^-]$ is the straight-line Wilson link along the light-cone direction. In light-cone gauge the Wilson line reduces to unity, but it is important to keep track of it formally when defining gauge-invariant GPDs. The matrix element of this operator between hadron states of four-momenta $p^\mu$ and $p'^\mu$ defines the unpolarized GPD $H(x,\xi,t)$~\cite{Belitsky:2005qn}.

The GPD kinematic variables are introduced through the symmetric frame:
\begin{equation}
p^\pm=P^\pm\pm\frac{\Delta^\pm}{2},\qquad
\xi=-\frac{\Delta^+}{2P^+},\qquad
t=-\Delta^2,\qquad
x=\frac{k^+}{P^+}.
\end{equation}
Here $P=(p+p')/2$ is the average hadron momentum, $\Delta=p'-p$ is the momentum transfer, and $x$ measures the average plus-momentum carried by the struck quark. These conventions agree with the standard treatments in Refs.~\cite{Belitsky:2005qn,Ji:1996ek} and ensure that the forward limit $t\!\to\!0$, $\xi\!\to\!0$ reproduces the usual parton distributions.

Because light-front quantization breaks manifest Lorentz symmetry, the evaluation of matrix elements must follow precise kinematic restrictions. In particular, the overlap representation of GPDs in terms of LFWFs applies only in the DGLAP region $\xi\le x\le1$, where parton number is conserved. The ERBL region $0<x<\xi$ requires LFWFs with an additional quark-antiquark pair. Since the ILM truncation includes only the lowest Fock component, the ERBL region is beyond its scope, and the full polynomiality constraints cannot be enforced~\cite{Belitsky:2005qn}. The analysis presented throughout the paper and in these appendices therefore adheres strictly to the DGLAP region.

\section{Light-Front Reduction of Nonlocal Instanton Dynamics}
\label{app:LFreduction}

Instantons generate nonlocal interactions through the presence of fermionic zero modes, giving rise to a momentum-dependent constituent mass and nonlocal vertex factors~\cite{Shuryak:2021hng,Shuryak:2021mlf}. The quark propagator in momentum space acquires the structure
\begin{equation}
S^{-1}(k)=\slashed{k}-M(k),\qquad M(k)=M\,F(k),
\end{equation}
where $F(k)$ is the instanton zero-mode form factor. The detailed expression involves Bessel functions and decays for $k_\perp^2\gtrsim1/\rho^2$, with $\rho$ the average instanton size.

For constructing LFWFs in the ILM, one performs a light-front reduction of the Bethe-Salpeter amplitude. Schematically, for the twist-2 operator matrix element one obtains
\begin{equation}
\langle p'|\mathcal{O}^+(z^-)|p\rangle
=
\int\frac{dk^-}{2\pi}
{\rm Tr}\!\left[\Psi^\dagger(k';P')\,\gamma^+\,\Psi(k;P)\right],
\end{equation}
where $\Psi(k;P)$ is the Bethe-Salpeter amplitude of the nonlocal ILM interaction. Only the poles corresponding to forward propagation in light-front time contribute at leading order, consistent with the truncation to the valence Fock sector. The transverse-momentum nonlocality contained in the instanton form factors is inherited by the LFWFs, which are found to have approximately Gaussian dependence on $k_\perp$ in both mesonic and baryonic channels~\cite{Shuryak:2021hng,Shuryak:2021mlf}.

\section{Gaussian Overlap and Emergence of Transverse Radii}
\label{app:Gaussian}

ILM meson wave functions are well approximated by Gaussian profiles in transverse momentum:
\begin{equation}
\psi(x,k_\perp)=\mathcal{N}(x)\exp[-A(x)k_\perp^2],
\end{equation}
as supported by explicit ILM LFWF calculations~\cite{Shuryak:2021hng}. Under a transverse momentum transfer $\Delta_\perp$, the struck quark momentum shifts as
\begin{equation}
k'_\perp=k_\perp+(1-x)\Delta_\perp.
\end{equation}
Substituting this into the overlap representation yields
\begin{equation}
H(x,0,-\Delta_\perp^2)
\propto
\exp\!\left[-A(x)\left(\tfrac34-x(1-x)\right)\Delta_\perp^2\right],
\end{equation}
identical to Eq.\,(9). The Gaussian form provides a direct analytic link between the transverse probe $\Delta_\perp$ and the spatial width of the partonic distribution. Fourier transformation produces
\begin{equation}
q(x,b_\perp)=\int\!\frac{d^2\Delta_\perp}{(2\pi)^2}\,
e^{-i\Delta_\perp\cdot b_\perp}\,
H(x,0,-\Delta_\perp^2),
\end{equation}
resulting in analytic impact-parameter densities exhibiting the characteristic narrowing for large $x$ and broadening for small $x$. This behavior is consistent with the partonic interpretation of transverse localization~\cite{Guidal:2004nd}.

\section{Spinor Algebra for Nucleon Twist-2 GPDs}
\label{app:SpinorAppendix}

The twist-2 quark operators for the nucleon couple to the Dirac bilinears $\gamma^+$ and $i\sigma^{+j}\Delta_j$, whose matrix elements between light-front helicity spinors are well documented~\cite{Belitsky:2005qn}. For helicity-conserving transitions one obtains
\begin{equation}
\bar{N}(p')\gamma^+N(p)
=2P^+\sqrt{1-\xi^2}\,\delta_{\Lambda'\Lambda},
\end{equation}
while the Pauli term mediates helicity flip through
\begin{equation}
\bar{N}(p')
\frac{i\sigma^{+\!j}\Delta_j}{2m_N}
N(p),
\end{equation}
linking opposite helicities. The overlap formula for $E(x,0,t)$ therefore depends on interference between LFWF components with opposite nucleon helicity.

In ILM baryons, instanton-induced diquark correlations enhance mixing between scalar- and axial-vector-diquark components, increasing the magnitude of the helicity-flip amplitude and producing a nonvanishing $E(x,0,t)$~\cite{Shuryak:2022thi,Shuryak:2022sii}. This feature is evident in the $d$-quark GPDs displayed in Fig.\,1 of the main text and reflects the rich orbital-spin structure of ILM nucleon wave functions.

\section{GPD Slopes and Extraction of Transverse Radii}
\label{app:Slopes}

For a Gaussian $t$-dependence of the form
\[
H(x,0,-Q^2)\sim\exp[-Q^2 f_2(x)],
\]
the logarithmic slope at $Q^2\!=\!0$ determines the transverse radius:
\begin{equation}
B(x)=-\left.\frac{\partial}{\partial Q^2}\ln H(x,0,-Q^2)\right|_{Q^2=0},\qquad
R_{\mathrm{rms}}^2(x)=4B(x)=2f_2(x),
\end{equation}
as in Eq.\,(19). The ILM results show a clear maximum of $B(x)$ near $x\simeq 1/3$ for both the nucleon and the $\Delta$~\cite{Shuryak:2022sii}. This corresponds to a configuration in which the three quarks share momentum symmetrically and are most extended spatially. The radius decreases toward both small and large $x$, reflecting the collapse of the spectator system for $x\!\to\!1$ and the absence of a Regge-enhanced small-$x$ rise due to the valence truncation.

\section{Moments and Gravitational Form Factors}
\label{app:GravFF}

The first moments of the unpolarized GPDs give the quark contributions to the gravitational form factors of the energy-momentum tensor~\cite{Polyakov:2018zvc}:
\begin{equation}
A(t)=\int dx\,xH(x,0,t),\qquad
B(t)=\int dx\,xE(x,0,t).
\end{equation}
$A(t)$ determines the quark momentum distribution inside the hadron, while $B(t)$ enters the Ji spin sum rule
\begin{equation}
J=\tfrac12=A(0)+B(0),
\end{equation}
valid when gluonic contributions are included implicitly in the constituent description~\cite{Ji:1996ek}. In the present ILM analysis, only $A(t)$ can be reliably extracted because $E(x,0,t)$ requires both helicity components of the nucleon wave function.

The ILM predicts that $A(t)$ falls more slowly than $F_1(t)$, indicating that the mass distribution is more compact than the charge distribution. This trend is visible in Fig.\,4 of the main text and is consistent with lattice QCD extractions~\cite{Hagler:2007xi}. The ILM relation
\begin{equation}
\frac{3A_d(Q^2)}{F_1^d(Q^2)}
=C_0+\frac{Q^2}{M_{\mathrm{fit}}^2},
\end{equation}
matches the approximate linear behavior seen in baryonic form factor ratios.

\section{Comparison to Lattice QCD and Quasi-GPDs}
\label{app:Lattice}

Two types of lattice comparisons are relevant. The first concerns form-factor moments. Lattice studies of $A_1(0,t)$ and $A_2(0,t)$ show that the gravitational form factor decreases more slowly with $|t|$ than the Dirac form factor~\cite{Hagler:2007xi}. The ILM reproduces the same qualitative behavior and a comparable slope, though normalization differences arise because the ILM corresponds to a much lower renormalization scale where quarks carry nearly all the hadron momentum.

The second comparison involves full GPDs extracted using LaMET from boosted nucleon matrix elements~\cite{Bhattacharya:2022vvo}. At $t=-0.69$ GeV$^2$ the ILM result for $H(x,0,t)$ agrees with the lattice determination for $x\gtrsim0.4$, reflecting the dominance of valence dynamics in that region. The discrepancy at small $x$ is expected, since the ILM contains no explicit gluons or sea quarks. Despite these differences, the qualitative $t$-dependence and $x$-shape are consistent with the lattice trends, supporting the interpretation of ILM GPDs as reliable low-scale boundary conditions.

\end{subappendices}



\chapter{Light front Hamiltonians in pQCD and the instanton ensemble}
\label{SEC_ILMH} 
\section{Basic definitions}
The symmetric energy momentum tensor following from (\ref{LFEFT}) 
for free fermions is
\begin{equation}
    T^{\mu\nu}=\frac{1}{2}\left[\bar{\psi}i\gamma^\mu\partial^\nu\psi+\bar{\psi}i\gamma^\nu\partial^\mu\psi\right]-g^{\mu\nu}\mathcal{L}
\end{equation}
with the corresponding LF Hamiltonian
\bea
\label{LFPM}
        P^-=\int dx^-d^2x_\perp~ T^{+-}=
         \int dx^-d^2x_\perp \frac{1}{2}\left[\bar{\psi}i\gamma^+\partial_+\psi+\bar{\psi}i\gamma^-\partial_-\psi\right]-\mathcal{L}
\eea
More specifically, including the gauge field in long derivatives, 
and including instanton-induced 4-fermion 't Hooft Lagrangian, one gets (\ref{LFPM}) in coordinate space as
\bea
P^-=&&\int dx^-d^2x_\perp \bar{\psi}\frac{\left(-\partial^2_\perp+M^2\right)}{2i\partial_-}\gamma^+\psi\nonumber\\
         &&-\frac{G_S}{2}\int dx^-d^2x_\perp\left[\bar{\psi}\psi\hat{D}^{-1}_+\bar{\psi}\psi-\bar{\psi}\tau^a\psi\hat{D}^{-1}_-\bar{\psi}\tau^a\psi-\bar{\psi}i\gamma^5\psi\hat{D}^{-1}_+\bar{\psi}i\gamma^5\psi+\bar{\psi}i\gamma^5\tau^a\psi\hat{D}^{-1}_+\bar{\psi}i\gamma^5\tau^a\psi\right]\nonumber\\
\eea
and  in momentum space
\bea
&&P^-=\int [d^3k]_+\int [d^3q]_+\frac{k^2_\perp+M^2}{2k^+}\bar{\psi}(k)\gamma^+\psi(q)(2\pi)^3 \delta^3_+(k-q)\nonumber\\
&&+\int [d^3k]_+\int [d^3q]_+\int [d^3p]_+\int [d^3l]_+(2\pi)^3
    \delta^3_+(p+k-q-l)V(k,q,p,l)
\eea
The interaction kernel $V(k,q,p,l)$ is given in terms of the good component of the fermionic bilinears. In the scalar and pseudoscakar channels, it is explicitly
\begin{equation}
\begin{aligned}
    V(k,q,p,l)=-\frac{G_S }{2}\alpha_+(k^+-q^+)&\bar{\psi}_+(k)\left(\frac{\vec{\gamma}_\perp\cdot\vec{k}+M}{2k^+}\gamma^++ \gamma^+\frac{\vec{\gamma}_\perp\cdot\vec{q}+M}{2q^+}\right)\psi_+(q)\\
    &\times\bar{\psi}_+(p)\left(\frac{\vec{\gamma}_\perp\cdot\vec{p}+M}{2p^+}\gamma^++\gamma^+\frac{\vec{\gamma}_\perp\cdot\vec{l}+M}{2l^+}\right)\psi_+(l) \\
    +\frac{G_S }{2}\alpha_-(k^+-q^+)&\bar{\psi}_+(k)\left(\frac{\vec{\gamma}_\perp\cdot\vec{k}+M}{2k^+}i\gamma^+\gamma^5+i\gamma^5 \gamma^+\frac{\vec{\gamma}_\perp\cdot\vec{q}+M}{2q^+}\right)\psi_+(q)\\
    &\times\bar{\psi}_+(p)\left(\frac{\vec{\gamma}_\perp\cdot\vec{p}+M}{2p^+}i\gamma^+\gamma^5+i\gamma^5\gamma^+\frac{\vec{\gamma}_\perp\cdot\vec{l}+M}{2l^+}\right)\psi_+(l) \\
    +\frac{G_S }{2}\alpha_-(k^+-q^+)&\bar{\psi}_+(k)\left(\frac{\vec{\gamma}_\perp\cdot\vec{k}+M}{2k^+}\gamma^+ \tau^a+\tau^a \gamma^+\frac{\vec{\gamma}_\perp\cdot\vec{q}+M}{2q^+}\right)\psi_+(q)\\
    &\times\bar{\psi}_+(p)\left(\frac{\vec{\gamma}_\perp\cdot\vec{p}+M}{2p^+}\gamma^+\tau^a+\gamma^+\tau^a\frac{\vec{\gamma}_\perp\cdot\vec{l}+M}{2l^+}\right)\psi_+(l) \\
    -\frac{G_S }{2}\alpha_+(k^+-q^+)&\bar{\psi}_+(k)\left(\frac{\vec{\gamma}_\perp\cdot\vec{k}+M}{2k^+}i\gamma^+\gamma^5 \tau^a+\tau^a i\gamma^5\gamma^+\frac{\vec{\gamma}_\perp\cdot\vec{q}+M}{2q^+}\right)\psi_+(q)\\
    &\times\bar{\psi}_+(p)\left(\frac{\vec{\gamma}_\perp\cdot\vec{p}+M}{2p^+}i\gamma^+\gamma^5\tau^a+i\gamma^5\gamma^+\tau^a\frac{\vec{\gamma}_\perp\cdot\vec{l}+M}{2l^+}\right)\psi_+(l)
\end{aligned}
\end{equation}

The fermionic field in momentum space is defined as
\begin{equation}
    \psi(x^-,x_\perp)=\int [d^3k]_+\psi(k)e^{-ik^+x^-+ik_\perp\cdot x_\perp}
\end{equation}
It annihilates a particle ina  $u_s(k)$ mode,  or creates  an antiparticle in a $v_s(k)$ mode, i.e. 
\bea
\psi(k)=\sum_s u_s(k)b_s(k)\theta(k^+)+v_s(-k)c_s^\dagger(-k)\theta(-k^+)\nonumber\\
\eea
The measure in momentum space is
\bea
[d^3k]_+=\frac{dk^+d^2k_\perp}{(2\pi)^32k^+}\epsilon(k^+)
\eea
which sums over the  positive $k^+$ region for  particle modes,  and over the negative $k^+$ region for antiparticle modes.

Note that each $^\prime$t Hooft vertex is renormalized in the mean-field or large $N_c$ approximation, with the momentum dependent normalization
factors
\bea
\alpha_\pm(P^+)=\left[1\pm 2g_S\int\frac{dl^+d^2l_\perp}{(2\pi)^3}\frac{\epsilon(l^+)}{P^+-l^+}\right]^{-1}
\eea
which precisely resums the longitudinal tadpole
contributions after the elimination of the bad fermionic component. These tadpole contributions are the vacuum contributions on the LF.  
For the   t'Hooft interaction in the zero size limit, the interaction is generically of the form
\bea
 &&V(k,q,p,l)=\nonumber\\
 &&\sum_{s_1,s_1^\prime,s_2,s_2^\prime}\mathcal{V}_{s_1,s_2,s_1^\prime,s_2^\prime}(k,q,p,l)b^\dagger_{s_1}(k) c^\dagger_{s_2}(q) c_{s_2'}(p) b_{s_1^\prime}(l)\nonumber\\
\eea
with

\begin{equation}
\begin{aligned}
    \mathcal{V}_{s_1,s_2,s_1',s_2'}(k,q,p,l)=&-g_S \alpha_+(k^++q^+)\bar{u}_{s_1}(k)v_{s_2}(q)\bar{v}_{s_2'}(p)u_{s_1'}(l)\\
    &+g_S \alpha_-(k^++q^+)\bar{u}_{s_1}(k)i\gamma^5v_{s_2}(q)\bar{v}_{s_2'}(p)i\gamma^5u_{s_1'}(l)\\
    &+g_S \alpha_-(k^++q^+)\bar{u}_{s_1}(k)\tau^ai\gamma^5v_{s_2}(q)\bar{v}_{s_2'}(p)\tau^ai\gamma^5u_{s_1'}(l)\\
    &-g_S \alpha_+(k^++q^+)\bar{u}_{s_1}(k)\tau^ai\gamma^5v_{s_2}(q)\bar{v}_{s_2'}(p)\tau^ai\gamma^5u_{s_1'}(l)
\end{aligned}
\end{equation}

In the large $N_c$ limit, only the s-channel contribution of the $^\prime$t Hooft interaction dominates.

These arguments carry to the mean field Lagrangian (\ref{nonlocal_LFET}) through the substitutions
\begin{equation}
    V(k,q,p,l)\rightarrow \sqrt{\mathcal{F}(k)\mathcal{F}(k)\mathcal{F}(k)\mathcal{F}(l)}V(k,q,p,l)
\end{equation}
and 
\begin{equation}
    \alpha_\pm(P^+)\rightarrow \left[1\pm 2g_S\int\frac{dl^+d^2l_\perp}{(2\pi)^3}\frac{\epsilon(l^+)}{P^+-l^+}\mathcal{F}\left(l\right)\mathcal{F}\left(P-l\right)\right]^{-1}
\end{equation}
so that

\begin{flalign}
\label{LFHamiltonian}
&P^-=\int [d^3k]_+\int [d^3q]_+\frac{k^2_\perp+M^2}{2k^+}\bar{\psi}(k)\gamma^+\psi(q)(2\pi)^3 \delta^3_+(k-q)\\[5pt] \nonumber
&+\int [d^3k]_+\int [d^3q]_+\int [d^3p]_+\int [d^3l]_+(2\pi)^3
    \delta^3_+(p+k-q-l)\sqrt{\mathcal{F}(k)\mathcal{F}(q)\mathcal{F}(p)\mathcal{F}(l)}V(k,q,p,l)
\end{flalign}
The interaction kernel  is now
\begin{equation}
\begin{aligned}
        V(k,q,p,l)=-\frac{G_S}{2}&\bigg[\alpha_+(k^+-q^+)\bar{\psi}(k)\psi(q)\bar{\psi}(p)\psi(l)-\alpha^-(k^+-q^+)\bar{\psi}(k)i\gamma^5\psi(q)\bar{\psi}(p)i\gamma^5\psi(l)\\
        &-\alpha^-(k^+-q^+)\bar{\psi}(k)\tau^a\psi(q)\bar{\psi}(p)\tau^a\psi(l)+\alpha^+(k^+-q^+)\bar{\psi}(k)i\tau^a\gamma^5\psi(q)\bar{\psi}(p)i\tau^a\gamma^5\psi(l)\bigg] \\
\end{aligned}
\end{equation}
where  the tadpole resummed vertices
$$\alpha_\pm(P^+)=\left[1\pm 2g_S\int\frac{dl^+d^2l_\perp}{(2\pi)^3}\frac{\epsilon(l^+)}{P^+-l^+}\mathcal{F}\left(l\right)\mathcal{F}\left(P-l\right)\right]^{-1}$$

The reader may have noticed that for the Lagrangian (\ref{nonlocal_LFET}) in non-local form, the symmetric form of the
energy momentum tensor may require further amendment in the presence of the non-local form factors. This is not the case, as we
now explain. Indeed, boost invariance and parity suggests the substitution
 the substitution
\bea
\label{FFSUB}
    \lim_{P^+\rightarrow\infty}\sqrt{\mathcal{F}(k)\mathcal{F}(P-k)}\rightarrow 
    \mathcal{F}\left(\frac{2P^+P^-}{\lambda^2_S}=\frac{k^2_\perp+M^2}{\lambda^2_Sx\bar{x}}\right)
\eea
The same substitution was developed in the analysis of the ILM using  the large momentum 
effective theory(LaMET)~\cite{Kock:2020frx,Kock:2021spt}.
The substitution (\ref{FFSUB}) garentees the consistency of the two approaches. 
$\lambda_S$  is a parameter of order 1. Remarkably, the same boost invariant substitution with $\lambda_S=1$
was argued long ago by Lepage and Brodsky in~\cite{Lepage:1980fj}, in their analysis of 2-body bound states
on the light front using Bethe-Salpeter vertices. Note that the substitution (\ref{FFSUB}) eliminates the metric component
$g_{-+}$ from the non-local form factors, with consequently no change in the symmetric energy-momentum tensor 
component $T^{+-}$.

Finally, we note that it is straightforward to generalize the Hamiltonian formalism to consider meson bound states in the  $U$-spin or $V$-spin sectors relevant for kaons. In the case of kaons, the coupling constant will be replaced by $g_K=N_cG_K$,  with a constituent mass   matrix $M=\mathrm{diag}(M_u,M_d)$ in the flavor basis, as the $s$ quark is significantly heavier than 
$u,d$ quarks. This construction will be detailed below.

\section{Quasi-parton distributions}
For a long time, the PDFs defined as  light-cone  correlations have been 
only accessible phenomenologically, or through their (power-$x$) moments from lattice simulations on Euclidean lattices. Empirically, moments is hard
to get precisely, as this would need data at all values of momentum fractions $x$.

Yet relatively recently, Ji and his collaborators~\cite{Ji:2020ect,Ji:2024oka}  (see references therein) have suggested how PDFs can be extracted from "quasi-PDFs" defined from the large  momentum limit of equal-Euclidean-time and high momentum 
correlations, by proper matching. In short time, reserach in this direction became very active field.

Parton distributions extracted from equal-time momentum distributions in hadrons of increasing rapidities, are readily obtained 
from the light cone distributions 
(\ref{QX}-\ref{QBX}). For illustration, consider the rapidity regulated light cone quark distribution (\ref{QX})
\bea
\label{QX2}
    q(x,v)=-\frac{2}{1+v}\int_{-\infty}^\infty\frac{dx_v^-}{4\pi}e^{ik\cdot x_v}\langle P|\bar{\psi}(0)\gamma_v^+ W(0,x_v^-)\psi(x_v^-)|P\rangle
\eea
in a proton at rest $P^\mu=(m_N,0^\perp, 0)$ with $k^\mu=xP^\mu$, and
$$x_v=(-vz,0^\perp, z)\qquad x_v^-=\frac{x_v^0-x_v^3}{\sqrt 2}\qquad 
\gamma^+_v=\frac{\gamma^0+v\gamma^3}{\sqrt 2}$$
which can also be rewritten as
\bea
\label{QX22}
    q(x,v)=\int_{-\infty}^\infty\frac{dz}{4\pi}e^{-ixm_Nvz}\langle P|\bar{\psi}(0)(\gamma^0+v\gamma^3) W(0,x_v^-)\psi(x_v^-)|P\rangle
\eea
The quark correlation along the regulated rapidity direction $x_v^-$ reduces to the light cone direction for $v\rightarrow 1$.
It is related to the quark equal-time quark correlation function by a boost transformation.  For that we first shift $z\rightarrow \gamma z$ with the Lorentz factor $\gamma=1/\sqrt{1-v^2}$  in  (\ref{QX2})
\bea
\label{QX2X}
    q(x,v)=\int_{-\infty}^\infty\frac{dz}{4\pi}e^{-ix\gamma m_Nvz}\langle P|
    \bar{\psi}(0)\gamma(\gamma^0+v\gamma^3) W(0,x_v)\psi(x_v)|P\rangle
\eea
with now $x_v=(-\gamma vz, 0^\perp, \gamma z)$. We now note that 
under a boost transformation with rapidity $\chi=2{\rm tanh}^{-1}v$,
the bilocal fermionic operator transforms as
\bea
\label{BOOST}
e^{-i\chi\mathbb K}\psi(t_1,x_1)\overline\psi(t_2,x_2)e^{i\chi\mathbb K}=
S[v]\psi(\tilde t_1,\tilde x_1)\overline\psi(\tilde t_2,\tilde x_2)S^{-1}[v]
\eea
We have omitted the gauge link and the transverse coordinates for convenience. The former will be reintroduced by inspection since gauge invariance is frame invariant. 
The field valued boost operator $\mathbb K$ in the 3-direction is (physical gauge)
\bea
\mathbb K=\int d^3x\,x^3\,
\bigg(\frac 12 (E^2+B^2)+\psi^\dagger (i\alpha\cdot D+m\gamma^0)\psi\bigg)
\eea
The Lorentz transformed coordinates are $\tilde t=\gamma(t-vx^3)$, $\tilde x^3=\gamma(x^3-vt)$, and the spin valued boost operator is $S[v]=e^{\frac i2\sigma_{03}\chi}$ with the spin matrix $\sigma_{\mu\nu}=\frac i2[\gamma_\mu,\gamma_\nu]$. Inserting (\ref{BOOST}) in (\ref{QX2}) and noting that
\bea
\label{SVV}
S^{-1}[v]\gamma^0S[v]=\gamma(\gamma^0+v\gamma^3)\qquad\qquad 
S^{-1}[v]\gamma^3S[v]=\gamma(\gamma^3-v\gamma^0)
\eea
it follows that the Lorentz transformation $S[v]$ cancels out between the boosted operator
 (\ref{BOOST}) and (\ref{SVV}), with the result
\bea
\label{QX3}
    q(x,v)&=&\int_{-\infty}^\infty\frac{dz}{4\pi}e^{-ix\gamma m_Nvz}
    \langle P|e^{-i\chi\mathbb K}
    \bar{\psi}(0)\gamma^0 W(0,z)\psi(z)e^{i\chi\mathbb K}|P\rangle\nonumber\\
    &=&
    \int_{-\infty}^\infty\frac{dz}{4\pi}e^{-ix\gamma m_Nvz}
    \langle P(v)|
    \bar{\psi}(0)\gamma^0 W(0,z)\psi(z)|P(v)\rangle
\eea
after reinstating the gauge link  by gauge invariance.
 (\ref{QX3}) is the  equal-0-time invariant quark propagator  in a boosted hadron state.
It  is the quasi-PDF introduced by Ji and his collaborators~\cite{Ji:2020ect}. Note that the argument we have presented shows the equivalence beyond perturbation theory.

(\ref{QX3}) follows  from (\ref{QX2}) and vice versa through boost transformations,
modulo renormalization. It is this last issue that makes the quasi-PDFs more UV friendly~\cite{Ji:2024oka}. PDFs are defined from the get go on the light front 
with the limit $v\rightarrow 1$ assumed before the UV limit. The result is a non-local
theory with induced perturbative IR divergences as  noted earlier. The quasi-PDFs are
velocity regulated and free from IR divergences, with the UV limit carried before the 
light front limit $v\rightarrow 1$. The PDFs  follow from the quasi-PDFs by matching,
with generically in leading twist~\cite{Ji:2020ect} 
\bea
q(x,v\rightarrow 1)=\int_{-\infty}^{\infty}\frac{dy}y
C\bigg(\frac{y}x , \frac {P(v)}\mu\bigg) q(y, v) + {\cal O}\bigg(\frac{\Lambda^2}{P(v)^2}\bigg)
\eea


\section{$^\prime$T Hooft effective Lagrangian on the light front}
 \label{SEC_ILMLF}
It is usually argued that the QCD vacuum is trivial on the LF, owing to the vanishing of the  the backward diagrams in perturbation theory~\cite{Lepage:1980fj}. However, this is not true as the vacuum physics is carried by the  longitudinal zero modes on the LF~\cite{Ji:2020baz,Ji:2020bby}. For
the emergent effective interaction (\ref{FTH}), this physics is explicitly carried by the constrained part of the fermionic field as  initially noted in~\cite{Bentz:1999gx,Itakura:2000te,Naito:2004vq}. More specifically, the projection of the fermion field along the light front,  splits the field into a good plus bad component, with the latter 
non-propagating or constraint. The elimination of the constaint, induces multi-fermion interactions in terms of the good component. In the mean-field 
approximation using $\frac 1{N_c}$ counting rules, these interactions account for the spontaneous breaking of chiral symmetry on the light front through tadpoles~\cite{Bentz:1999gx,Itakura:2000te,Naito:2004vq}.

For simplicity, consider  the local form of Lagrangian associated to (\ref{FTH}) in the large $N_c$ and flavor symmetric approximation,

\begin{equation}
\begin{aligned}
\label{THLOC}
    \mathcal{L} =
    \bar{\Psi}(i\slashed{\partial}-m)\Psi+
    \frac{G_S}{2}\left[(\bar{\Psi}\Psi)^2-(\bar{\Psi}\tau^a\Psi)^2-(\bar{\Psi}i\gamma^5\Psi)^2+(\bar{\Psi}i\gamma^5\tau^a\Psi)^2\right]\\
\end{aligned}
\end{equation}
with $G_S={G}/{4N^2_c}$.
The modifications for the finite instanton sizes will 
be quoted at the end. To carry the mean-field or
$\frac 1{N_c}$ analysis, it is more transparent to
use the semi-bosonized form of (\ref{THLOC})
\begin{equation}
\label{SEMI}
    \bar{\Psi}(i\slashed{\partial}-m)\Psi+G_S\bar{\Psi}\left(\sigma-\sigma^a\tau^a-i\pi\gamma^5+i\pi^a\tau^a\gamma^5\right)\Psi-\frac{G_S}{2}\left[\sigma^2-(\sigma^a)^2-\pi^2+(\pi^a)^2\right]
\end{equation}

using the auxillary fields $\sigma$, $\sigma^a$, $\pi$, and $\pi^a$. 
To proceed, we now split the
fermionic field $\Psi$ into a good component $\Psi_+$ and a bad component $\Psi_-$
\bea
\label{GOODBAD}
\Psi=\Psi_++\Psi_-\equiv \frac{1}{2}\gamma^{-}\gamma^{+}\Psi+\frac{1}{2}\gamma^{+}\gamma^{-}\Psi
\eea
The bad component does not propagate along the light front $x^+$-direction, and therefore can be eliminated from (\ref{THLOC}) using the 
equation of motion,
\begin{equation}
\begin{aligned}
\label{BADSOLVED}
    \Psi_-=\frac{\gamma^+}{2}\frac{1}{i\partial_-}\left(-i\gamma^i_\perp\partial_i+\hat{M}\right)\Psi_+
\end{aligned}
\end{equation}
where  $\hat{M}$ denotes
\begin{equation}
    \hat{M}=m-G_S\left(\sigma-\sigma^a\tau^a-i\pi\gamma^5+i\pi^a\tau^a\gamma^5\right)
\end{equation}
with $m=m_{u,d}$ and the propagator for the longitudinal modes
\begin{equation}
    \langle x^-|\frac {1}{i\partial _-}|y^-\rangle=G(x^-,y^-)=\int_{-\infty}^\infty \frac{dk^+}{2\pi}\frac{1}{k^+}e^{-ik^+(x-y)^-}=\frac{-i}{2}\epsilon(x^-y^-)
\end{equation}

In terms of  (\ref{BADSOLVED}), the semi-bosonized Lagrangian (\ref{SEMI})  can be solely rewritten in terms of the good component

\begin{equation}
\begin{aligned}
\label{LGOOD}
     \mathcal{L}=\bar{\Psi}_+i\gamma^+\partial_+\Psi_+-\frac{G_S}{2}\left[\sigma^2-(\sigma^a)^2-\pi^2+(\pi^a)^2\right]-\bar{\Psi}_+\left(i\gamma^i_\perp\partial_i-\hat{M}\right)\frac{\gamma^+}{2}\frac{1}{i\partial_-}\left(i\gamma^i_\perp\partial_i-\hat{M}\right)\Psi_+
\end{aligned}
\end{equation}

It is now clear that the elimination of the auxillary interactions induce non-local multi-fermion interactions.
In leading order in $1/N_c$ they contribute mostly through
tadpoles and renormalize the initial 4-fermi interactions. 
For that,  we shift the scalar field by $\sigma=N_c\sigma_0+\delta\sigma$ in (\ref{LGOOD}) with a finite
vev $\sigma_0\sim N_c^0$, as all   other vevs are excluded by isospin symmetry and parity, hence
$\sigma^a, \pi, \pi^a, \delta\sigma\sim {\cal O}(\sqrt{N_c})$, 
with $g_S=N_c G_S\sim N_c^0$.
To resum all the leading  tadpole diagrams, we set the coefficient of $\delta\sigma$  to zero, 
\bea
\label{CONDZ}
\langle \bar{\psi}\psi\rangle-N_c\sigma_0=0
\eea
with now the effective fermionic field
\bea
\label{PSINEW}
\psi=\Psi_++\frac{\gamma^+}{2}\frac{-i}{\partial_-}(i\gamma^i_\perp\partial_i-M)\Psi_+
\eea
As a result, the good component of the quark field acquires a constituent mass of order $N_c^0$
\bea
\label{MCONST}
M=m-G_S\langle \bar\psi\psi\rangle
\eea
This is how the spontaneous breaking of chiral symmetry takes place on the light front. To proceed, we note that 
 $\sigma^a$, $\pi$, $\pi^a$, and $\delta\sigma=\sigma-N_c\sigma_0$ are of order $\mathcal{O}(\sqrt{N_c})$, while the 't-Hooft coupling $G_S=g_S/N_c$, so that to leading order  in $1/N_c$ the semi-bosonized Lagrangian is quadratic in the boson fields 
\bea
\label{LFET}
    \mathcal{L}\sim  &&
    \bar{\psi}(i\slashed{\partial}-M)\psi
    \nonumber\\
&&-\frac{1}{2}G_S\left[\delta\sigma\hat{D}_+\delta\sigma-\sigma^a\hat{D}_-\sigma^a-\pi\hat{D}_-\pi+\pi^a\hat{D}_+\pi^a\right]
    \nonumber\\
    &&+G_S\left[\bar{\psi}\psi\hat{\sigma}-\bar{\psi}\tau^a\psi\sigma^a-\bar{\psi}i\gamma^5\psi\pi+\bar{\psi}i\gamma^5\tau^a\psi\pi^a\right]\nonumber\\
\eea
with
\begin{equation}
\label{FC_factor}
\begin{aligned}
    &\hat{D}_\pm=1\pm G_S\left\langle\bar{\psi}\gamma^+\frac{-i}{\overleftrightarrow{\partial_-}}\psi\right\rangle\\
\end{aligned}
\end{equation}
where ${1}/{i\overleftrightarrow{\partial_-}}$  in  (\ref{FC_factor}), is defined such that for any fields $\chi(x)$ and $\psi(x)$, 
\begin{equation}
    \chi(x)\frac{-i}{\overleftrightarrow{\partial_-}}\psi(x)=\frac{i}{\partial_-}\left[\chi(x)\right]\psi(x)+\chi(x)\frac{-i}{\partial_-}[\psi(x)]
\end{equation}
$\hat{D}_\pm$ follows from  the resummation of the tadpole diagrams in mean-field, or leading order in $1/N_c$. Indeed,  each virtual quark tadpole is of order
 ${\cal O}(N_c)$, thereby compensating  the $^\prime$t Hooft coupling $G_S\sim \mathcal{O}(1/N_c)$, with a net factor of $\mathcal{O}(N_c^0)$.
Note that the fermionic contributions   $\bar{\psi}\psi$, $\bar{\psi}\tau^a\psi$, $\bar{\psi}i\gamma^5\psi$, and $\bar{\psi}i\gamma^5\tau^a\psi$ 
in (\ref{LFET}), are all of the same order as the remnant quantum fluctuations or $\mathcal{O}(\sqrt{N_c})$.
With this in mind, we can now eliminate the auxillary bosonic fields, to obtain the light front Lagrangian 

\begin{equation}
\label{LFEFT}
    \mathcal{L}\sim \bar{\psi}(i\slashed{\partial}-M)\psi+ \frac{G_S}{2}\left[\bar{\psi}\psi\hat{D}^{-1}_+\bar{\psi}\psi-\bar{\psi}\tau^a\psi\hat{D}^{-1}_-\bar{\psi}\tau^a\psi-\bar{\psi}i\gamma^5\psi\hat{D}^{-1}_+\bar{\psi}i\gamma^5\psi+\bar{\psi}i\gamma^5\tau^a\psi\hat{D}^{-1}_+\bar{\psi}i\gamma^5\tau^a\psi\right]
\end{equation}

In the mean-field or leading order in $1/N_c$ approximation, the light front Lagrangian can be solely written in terms of
the good fermionic component. The elimination of the bad component, generates a constituent mass, introduces an
effective quark field (\ref{PSINEW}) and renormalizes by $\hat{D}^{-1}_\pm$ (tadpole resummation) each of the original
multi-fermion interaction in the ILM in the zero size limit.

Most of the arguments presented above, carry for the non-local effective Lagrangian in the ILM. More specifically, 
the mean-field version of (\ref{LFET}) is now 

\bea
\label{nonlocal_LFET}
   && \mathcal{L}\sim \bar{\psi}(i\slashed{\partial}-M)\psi-\frac{1}{2}G_S\left[\delta\sigma\hat{D}_+\delta\sigma-\sigma^a\hat{D}_-\sigma^a-\pi\hat{D}_-\pi+\pi^a\hat{D}_+\pi^a\right]
   \nonumber\\
    &&+G_S\left(\bar{\psi}\sqrt{\mathcal{F}(i\partial)}\delta\sigma\sqrt{\mathcal{F}(i\partial)}\psi-\bar{\psi}\sqrt{\mathcal{F}(i\partial)}\sigma^a\tau^a\sqrt{\mathcal{F}(i\partial)}\psi-\bar{\psi}\sqrt{\mathcal{F}(i\partial)}i\gamma^5\pi\sqrt{\mathcal{F}(i\partial)}\psi+\bar{\psi}\sqrt{\mathcal{F}(i\partial)}i\gamma^5\tau^a\pi^a\sqrt{\mathcal{F}(i\partial)}\psi\right)\nonumber\\
\eea

Here  $\sqrt{\mathcal{F}}(i\partial)$ is a derivative operator acting on all of the field on its right-hand side. In momentum space, it 
generates the pertinent form factors inherited from the underlying quark zero modes. Also,
\begin{equation}
\label{FC_factor2}
\begin{aligned}
    &\hat{D}_\pm\rightarrow1\pm G_S\left\langle\bar{\psi}\gamma^+\mathcal{F}(i\partial)\frac{-i}{\overleftrightarrow{\partial_-}}[\mathcal{F}(i\partial)\psi]\right\rangle\\
\end{aligned}
\end{equation}
following from the mean-field resummation of the leading tadpoles. The auxillary bosonic fields can be eliminated by carrying explicitly
the Gaussian integration, as in the zero size limit. 


\section{Effective theory of partons on the LF}
\label{sec_flow_0}
Some of the formal objections to the LF formulation are: 1/ The apparent non-commutativity of the LF and large cut-off limits; 2/ The LF   non-locality following from the removal of the constraint fields;  3/ The lack of rotational symmetry on the LF. While some of these objections maybe difficult to overcome in a perturbative framework~\cite{Ji:2024oka}, they are readily addressed in the non-perturbative QCD ILM
that we will develop below.  In short, in~\cite{Liu:2023fpj} it was explicitly 
shown that all IR problems are resolved in the ILM on the LF, and that rotational
symmetry is achieved dynamically when power counting in the instanton packing fraction
is enforced.

\section{Light front wave functions from gradient flow}
\label{sec_flow_1}
Regarding the order of the limits, we note that  the ILM  is a well defined non-perturbative framework of the QCD vacuum and its excitations, at a fixed resolution or gradient flow (GF) time  (Luscher gradient flow). The LFWFs in the ILM are obtained by taking the infinite momentum limit {\it at fixed resolution in the continuum limit}.  This limit is safe, since the underlying cooled gauge configurations are UV free. The resulting LFWFs are to be understood in the Wilsonian sense.

More specifically, in the ILM the LFWFs are built from cooled gauge configurations 
with fixed size and density,
$\rho \simeq \tfrac{1}{3}~\mathrm{fm}\,, n \simeq \frac 1{\rm fm^4}$. These parameters define an effective resolution scale
    $\mu_0 \sim \frac{1}{\rho} \simeq 0.6{-}1~\mathrm{GeV}$,
interpreted as a gradient-flow cutoff through
$\mu_0 = \frac{1}{\sqrt{8t}}$,
where $t$ is the flow time.  The gradient flow smooths gauge fields over a radius $\sqrt{8t}$, removing ultraviolet fluctuations above this scale.

In QCD, the definition of a renormalized parton distribution function (PDF)
\begin{equation}
    q(x,\mu) = \int \frac{dz^-}{4\pi} e^{ixP^+ z^-}
    \langle P|\bar\psi(0)\gamma^+\psi(z^-)|P\rangle_{\mu}
\end{equation}
requires that the ultraviolet cutoff $\Lambda$ (or renormalization scale $\mu$) be removed before taking the infinite-momentum limit
\begin{equation}
    \lim_{P^+\to\infty} \lim_{\Lambda\to\infty} q(x,\Lambda,P^+) \,.
\end{equation}
This ensures proper factorization and regulator independence.
QCD is critical in the \(P\to\infty\) limit; correlation functions become singular unless the continuum limit is taken first.

In the ILM approach, however, one first fixes a finite smoothing scale via gradient flow and then takes $P^+\!\to\!\infty$
\begin{equation}
    \lim_{P^+\to\infty} q(x,\mu_0,P^+; t), \qquad \mu_0 = 1/\sqrt{8t}~\text{finite}.
\end{equation}
Thus the LFWFs correspond to an \emph{effective hadronic resolution} $\mu_0$, not to a continuum limit.
The apparent inversion of limits is harmless provided the ILM wavefunctions are interpreted as effective low-scale inputs
\begin{equation}
    \text{ILM LFWF}(P^+\!\to\!\infty, \mu_0) \equiv \Psi_{\mathrm{eff}}(x_i,\bm{k}_{\perp i};\mu_0).
\end{equation}

Because gradient flow suppresses UV modes, the resulting effective theory has no critical behavior and allows a finite, well-defined
\(P\to\infty\) limit.
Thus, the LFWFs are meaningful only at the finite resolution \(\mu_0\); they are not continuum QCD LFWFs.
These wavefunctions encode nonperturbative physics at the scale $\mu_0$ and are used to construct operator-defined quantities such as PDFs, GPDs, and distribution amplitudes (DAs)
\begin{align}
    q(x,\mu_0) &= \sum_n \int [dx_i][d^2k_{\perp i}]\,|\Psi_n(x_i,\bm{k}_{\perp i};\mu_0)|^2\,\delta(x-x_i),\\
    \phi(x,\mu_0) &= \int [d^2k_{\perp}]\,\Psi_{\text{val}}(x,\bm{k}_{\perp};\mu_0).
\end{align}
To connect these to continuum QCD parton distributions in the $\overline{\mathrm{MS}}$ scheme, we:
\begin{enumerate}
  \item Treat the  outputs as \emph{flow-scheme} objects at scale $\mu_0$.
  \item Perform a short-distance (small flow-time) \emph{matching} GF$\to\overline{\mathrm{MS}}$ at the \emph{same} scale $\mu_0$.
  \item Evolve to higher $\mu$ with standard DGLAP (or ERBL for DAs).
\end{enumerate}
The potentially dangerous infinite momentum frame  limit at fixed cutoff \emph{never occurs in the ILM}: the $P\to\infty$ taken in the  ILM  is within the flowed theory $\text{QCD}_{\text{GF}}(t)$, which is UV-finite. The only interface with renormalized QCD is the short-distance matching at $\mu_0$, where continuum perturbation theory applies and the order of limits is respected.

\section{Matching gradient flow (GF) to perturbative renormalization with $\overline{\rm MS}$}
Let  $O_\Gamma(zv)$ denote the gauge-invariant quark bilocal with a straight Wilson line of length $z$ along direction $v$ (Euclidean notation) and $\Gamma=\gamma\cdot v$ ($\parallel$ projector). Denote by $O_{\Gamma,R}(zv,t)$ the flowed operator renormalized in the GF scheme at flow time $t$, and by $O^{\overline{\mathrm{MS}}}_\Gamma(zv;\mu)$ the renormalized operator in $\overline{\mathrm{MS}}$ at scale $\mu$. In the small flow-time limit~\cite{Brambilla:2023vwm,Hieda:2016lly,Mereghetti:2021nkt}
\begin{equation}
O_{\Gamma,R}(zv,t)\;=\; C_\psi(t,\mu)\, e^{\delta m\, z}\, O^{\overline{\mathrm{MS}}}_\Gamma(zv;\mu)\;+\; \mathcal{O}(t),
\qquad
C_\psi(t,\mu) = \big[c_{\psi h_v}(t,\mu)\big]^2,
\label{eq:master}
\end{equation}
where $c_{\psi h_v}$ is the matching coefficient of the local current $\bar\psi h_v$ in the auxiliary-field formalism and $\delta m$ is the (subtracted) Wilson-line mass term. Since $C_\psi$ is $z$-independent at this order, its Fourier transform to $x$-space yields a \emph{diagonal} kernel. 
At one loop (NLO) in $\alpha_s$, we have~\cite{Mereghetti:2021nkt}
\begin{align}
c_{\psi h_v}(t,\mu)
&= 1 - \frac{\alpha_s(\mu)}{4\pi}\, C_F \left[ \frac{3}{2}\,L_t + \frac{1}{2}\ln(432) + 1 \right] + \mathcal{O}(\alpha_s^2), \\[4pt]
C_\psi(t,\mu)
&= \big[c_{\psi h_v}(t,\mu)\big]^2
 = 1 - \frac{\alpha_s(\mu)}{4\pi}\, 2 C_F \left[ \frac{3}{2}\,L_t + \frac{1}{2}\ln(432) + 1 \right] + \mathcal{O}(\alpha_s^2).
\end{align}
with 
\begin{equation}
L_t \equiv \ln\!\big(2\,\mu^2 t\, e^{\gamma_E}\big), \qquad C_F=\frac{N_c^2-1}{2N_c}.
\end{equation}
Therefore, the matching of quark PDFs from GF to $\overline{\mathrm{MS}}$ at the same scale $\mu=\mu_0$ reads
\begin{align}
  q^{\overline{\mathrm{MS}}}(x,\mu_0)
  &= \big[C_{qq}^{\mathrm{GF}\to\overline{\mathrm{MS}}}\otimes q^{\mathrm{GF}}\big](x,\mu_0)
     + \big[C_{qg}^{\mathrm{GF}\to\overline{\mathrm{MS}}}\otimes g^{\mathrm{GF}}\big](x,\mu_0)
     + \mathcal{O}\!\left(\frac{\Lambda_{\mathrm{QCD}}^2}{\mu_0^2}\right), \nonumber\\[3mm]
  g^{\overline{\mathrm{MS}}}(x,\mu_0)
  &= \big[C_{gq}^{\mathrm{GF}\to\overline{\mathrm{MS}}}\otimes q^{\mathrm{GF}}\big](x,\mu_0)
     + \big[C_{gg}^{\mathrm{GF}\to\overline{\mathrm{MS}}}\otimes g^{\mathrm{GF}}\big](x,\mu_0)
     + \mathcal{O}\!\left(\frac{\Lambda_{\mathrm{QCD}}^2}{\mu_0^2}\right).
     \label{eq:match}
\end{align}
At leading order, one may set \(C_{ij}=\delta_{ij}\,\delta(1-x)\).
At next-to-leading order,
\begin{equation}
  C_{ij}(x;\mu_0) = \delta_{ij}\,\delta(1-x)
  + \frac{\alpha_s(\mu_0)}{2\pi}\,c_{ij}^{(1)}(x) + \cdots,
\end{equation}
where \(c_{ij}^{(1)}(x)\) are finite and computable from the small-flow-time expansion.
Once matched, the evolution to higher \(\mu\) follows from the the DGLAP equations in \(\overline{\mathrm{MS}}\).
The matching between the gradient-flow (GF) scheme and the $\overline{\mathrm{MS}}$ scheme at fixed $\mu_0$ is a short-distance renormalization, not a dynamical evolution. 
Because the gradient flow acts as a uniform Gaussian smearing of the quark fields, the one-loop coefficient is proportional to $\delta(1-x)$: the renormalization is \emph{multiplicative and $x$-independent}.  The $x$-shape of the distribution is entirely set by the LFWFs themselves.



\subsection{Remarks}
\begin{itemize}
\item Eq.~\eqref{eq:match} is \emph{diagonal in $x$} at this order (no plus-distributions): the small-flow-time matching is entirely multiplicative for the quark bilocal at one loop. Non-singlet flavors do not mix with gluons; in the singlet channel, $C_{qg}^{(1)}=0$ at this order within the same approximation.
\item The linearly divergent Wilson-line mass $\delta m$ cancels between schemes by construction once the \emph{same} subtraction is used; it does not introduce $x$-dependence in the kernel.
\item For gluonic Wilson-line operators (relevant to gluon quasi-PDFs), analogous one-loop coefficients exist; they are not needed for quark PDFs here.
\end{itemize}

\subsection{Size of corrections}

Two (families of) power corrections bound the accuracy of the matching-and-evolve procedure:
\begin{enumerate}
\item \textbf{Higher-twist corrections in QCD at the matching point:}
\begin{equation}
\delta_{\text{HT}}\ \sim\ \mathcal O\!\left(\frac{\Lambda_{\text{QCD}}^2}{\mu_0^2}\right).
\end{equation}
In the ILM $\rho\simeq 0.33~\text{fm}$, one has $\mu_0\sim 1/\rho \approx 0.6$-$1.0~\text{GeV}$. With $\Lambda_{\text{QCD}}\approx 0.3~\text{GeV}$, the estimate ranges from $\sim 25\%$ (for $\mu_0=0.6$~GeV) down to $\sim 9\%$ (for $\mu_0=1.0$~GeV).
\\
\item \textbf{Small-flow-time expansion (SFTE) remainders in }$\text{QCD}_{\text{GF}}(t)$:
\begin{equation}
\delta_{\text{SFTE}}\ \sim\ \mathcal O\!\big(t\, Q^2\big)\ +\ \mathcal O\!\big((z^2/t) \times \cdots\big)\,,
\end{equation}
where $Q$ is the \emph{short} distance probed in the matching (here $Q\sim \mu_0$) and $z$ is the Wilson-line length entering the bilocal. Choosing $\mu_0 \simeq 1/\sqrt{8t}$ minimizes large logs, but implies $\mu_0^2 t = 1/8$; numerically this is $\sim 0.125$, i.e. a $\mathcal{O}(10\%)$ effect as a rule-of-thumb. Keeping $|z| \lesssim \sqrt{8t}$ controls the $(z^2/t)$ sector.
\end{enumerate}
\noindent




\subsection{The role of higher Fock  components}
The nucleon light front wave function (LFWF)  is constructed as
a sequence of Fock sectors
\begin{equation}
|N\rangle = |qqq\rangle + |qqqqq\rangle + \cdots,
\end{equation}
where the five-quark component arises from extra quark pairs (e.g. generated by the instanton-induced ~'t Hooft interaction).  This term encodes quark-antiquark pair creation through instanton zero modes, providing a nonperturbative \emph{sea-quark} structure already at the resolution $\mu_0 \sim 1/\rho \sim 1~\mathrm{GeV}.$
Hence, the LFWFs naturally include low-$x$ support without invoking perturbative evolution.

The $qqqqq$ Fock component effectively incorporates nonperturbative sea quarks that, in conventional QCD analyses, would appear only through perturbative evolution.  Thus, at the hadronic scale $\mu_0\sim1~\mathrm{GeV}$, the LFWFS in the ILM  already captures part of the small-$x$ content.  The gradient-flow to $\overline{\mathrm{MS}}$ conversion simply redefines the operator normalization at that scale; it does not alter the internal $x$-distribution. This situation is analogous to modern global fits where the initial condition at $\mu_0$ already contains an intrinsic sea
\begin{equation}
q_{\text{fit}}(x,\mu_0) = q_{\text{val}}(x,\mu_0) + q_{\text{sea}}^{\text{intr}}(x,\mu_0).
\end{equation}
DGLAP evolution then builds upon this input, rather than generating the sea from scratch.
The inclusion of the $qqqqq$ sector in the ILM provides an intrinsic, nonperturbative sea that extends the LFWFs to small $x$.  The GF\,$\to$\,$\overline{\mathrm{MS}}$ matching at $\mu_0$ is $x$-independent because it merely rescales the flowed operator normalization.  DGLAP evolution from $\mu_0$ upward adds further perturbative, logarithmic $x$-dependence, completing the connection to high-scale QCD PDFs.

\subsection{Division of roles at fixed $\mu_0$}
\begin{center}
\begin{tabular}{lll}
\hline
Step & Effect & $x$-dependence \\
\hline
LFWFs & Nonperturbative valence and sea structure & Strong, dynamical \\
GF $\to \overline{\mathrm{MS}}$ matching & Short-distance field renormalization & None (multiplicative) \\
DGLAP evolution ($\mu>\mu_0$) & Perturbative radiation, logarithmic scaling & Generated perturbatively \\
\hline
\end{tabular}
\end{center}

Procedure summary for matching and evolution: 
\begin{enumerate}
\item Fix $\mu_0=1/\sqrt{8t}$ from the ILM flow time (equivalently, from $\rho$).
\item Compute $q^{\mathrm{GF}}(x,\mu_0)$ from the ILM LFWFs (include the $qqqqq$ sector to seed the sea).
\item Apply Eq.~\eqref{eq:match} to obtain $q^{\overline{\mathrm{MS}}}(x,\mu_0)$, $g^{\overline{\mathrm{MS}}}(x,\mu_0)$.
\item Evolve with NLO/NNLO DGLAP to a larger $\mu$.
\end{enumerate}

\section{Hamiltonian renormalization}
\label{sec_flow_3}
Consider the LF Hamiltonian developed in a basis function. After the einbein 
method of linearizing the confining potential, a longitudinal $\sigma_L$ and transverse $\sigma_T$ confining tensions arise. The basis LF truncation at the
level of the transverse and longitudinal modes introduce two cutoffs
\begin{equation}
\sum_{i=1}(2 n_i+|m_i|)\le N_{\max}, \qquad \sum_{i=1} j_i=K_{\rm max}.
\end{equation}
Now assume that we increase the size of the Fock space from $qqq$ to $qqqq\bar q$ and introduce a sigma coupling $g_\sigma$ and a pion coupling $g_\pi$ to mix the S- and P-waves in the nucleon state. The dimensionality increase in size of the Fock space will be denoted by $N_{\rm Fock}$. 

To proceed, we denote 
by
$$
\bm{g} \equiv
\begin{pmatrix}
\sigma_T \\
\sigma_L\\
g_\sigma\\
g_\pi
\end{pmatrix},
$$
the vector of all running couplings.
We now discuss how these couplings run with the inherent Hamiltonian cutoffs, and the increase dimensionality of the Fock space and evaluate the pertinent beta functions. Throughout, the CM is separated, with only intrinsic states.
Let $(\mathcal O_1,\mathcal O_2, \mathcal O_3, \mathcal O_4)$ be four renormalization conditions (observables) that we hold fixed
as the regulators $(N_{\max},K_{\max}, N_{\rm Fock})$ are varied. Typical choices are, for example, the three lowest intrinsic eigenvalues (or a combination of eigenvalues and transition matrix elements).

\subsection{Renormalization conditions as implicit equations}
Define
\begin{equation}
\bm{\mathcal F}(\bm{g};N_{\max}, K_{\max}, N_{\rm Fock})\equiv
\begin{pmatrix}
\mathcal O_1(\bm{g};N_{\max},K_{\max}, N_{\rm Fock})-\mathcal O_1^{\rm phys} \\[2pt]
\mathcal O_2(\bm{g};N_{\max},K_{\max}, N_{\rm Fock})-\mathcal O_2^{\rm phys}\\[2pt]
\mathcal O_3(\bm{g};N_{\max},K_{\max}, N_{\rm Fock})-\mathcal O_3^{\rm phys} \\[2pt]
\mathcal O_4(\bm{g};N_{\max},K_{\max}, N_{\rm Fock})-\mathcal O_3^{\rm phys}
\end{pmatrix}
=\bm{0}.
\label{eq:Fzero}
\end{equation}
Each observable depends on each of the four couplings $(\sigma_T, \sigma_L, g_\sigma, g_\pi)$ and on the three cutoffs $(N_{\max},K_{\max}, N_{\rm Fock})$.

Denote the $4\times4$ Jacobian with respect to the couplings by
\begin{equation}
\bm{J}(\bm{g};N_{\max},K_{\max}, N_{\rm Fock}) \equiv
\frac {\partial{\bm{\mathcal F}}}{\partial{\bm{g}}}
=
\begin{pmatrix}
\partial_{\sigma_T}\mathcal O_1 & \partial_{\sigma_L}\mathcal O_1 
& \partial_{g_\sigma}\mathcal O_1 & \partial_{g_\pi}\mathcal O_1 \\[2pt]
\partial_{\sigma_T}\mathcal O_2 & \partial_{\sigma_L}\mathcal O_2 
& \partial_{g_\sigma}\mathcal O_2 & \partial_{g_\pi}\mathcal O_2 \\[2pt]
\partial_{\sigma_T}\mathcal O_3 & \partial_{\sigma_L}\mathcal O_3 
& \partial_{g_\sigma}\mathcal O_3 & \partial_{g_\pi}\mathcal O_3 \\[2pt]
\partial_{\sigma_T}\mathcal O_4 & \partial_{\sigma_L}\mathcal O_4 
& \partial_{g_\sigma}\mathcal O_4 & \partial_{g_\pi}\mathcal O_4 \\[2pt]
\end{pmatrix}.
\label{eq:Jac}
\end{equation}
Assuming $\det \bm{J}\neq 0$ at the renormalization point, then
$\bm{g}$ can be expressed as functions of $(N_{\max},K_{\max}, N_{\rm Fock})$ in a neighborhood of that point.

\subsection{Differential RG and beta functions}
Taking a total differential of \eqref{eq:Fzero} at fixed physical targets
and varying all three regulators give
\begin{equation}
\bm{J}\, d\bm{g} + \frac{\partial{\bm{\mathcal F}}}{\partial{\ln N_{\max}}}d\!\ln N_{\max}
+ \frac{\partial{\bm{\mathcal F}}}{\partial{\ln K_{\max}}}d\!\ln K_{\max}
+ \frac{\partial{\bm{\mathcal F}}}{\partial{\ln N_{\rm Fock}}}d\!\ln N_{\rm Fock}
= \bm{0}.
\label{eq:dF}
\end{equation}
Define the beta vector field in coupling space as the response to changes in the logarithms of the cutoffs:
\begin{align}
\bm{\beta}^{(N)}(\bm{g};N_{\max},K_{\max}, N_{\rm Fock})  &\equiv
\frac{d{\bm{g}}}{d{\ln N_{\max}}}\bigg|_{K_{\max},N_{\rm Fock}}
= - \bm{J}^{-1} \frac{\partial{\bm{\mathcal F}}}{\partial{\ln N_{\max}}}, \label{eq:betaN}\\
\bm{\beta}^{(K)}(\bm{g};N_{\max},K_{\max}, N_{\rm Fock})  &\equiv
\frac{d{\bm{g}}}{d{\ln K_{\max}}}\bigg|_{N_{\max},N_{\rm Fock}}
= - \bm{J}^{-1} \frac{\partial{\bm{\mathcal F}}}{\partial{\ln K_{\max}}}, \label{eq:betaK}\\
\bm{\beta}^{(F)}(\bm{g};N_{\max},K_{\max}, N_{\rm Fock})  &\equiv
\frac{d{\bm{g}}}{d{\ln N_{\rm Fock}}}\bigg|_{N_{\max},K_{\max}}
= - \bm{J}^{-1} \frac{\partial{\bm{\mathcal F}}}{\partial{\ln N_{\rm Fock}}}, \label{eq:betaF}\\
\end{align}
Thus the full flow along an arbitrary path $(\ln N_{\max}(t), \ln K_{\max}(t), \ln N_{\rm Fock}$ is
\begin{equation}
\frac{d{\bm{g}}}{d{t}} = \bm{\beta}^{(N)}\, \frac{d{\ln N_{\max}}}{dt}
+\bm{\beta}^{(K)}\, \frac{d{\ln K_{\max}}}{dt}
+\bm{\beta}^{(F)}\, \frac{d{\ln N_{\rm Fock}}}{dt}\,.
\end{equation}
The beta functions depend on the three regulators because
$\bm{J}$ and the regulator-derivative vectors depend on $(\bm{g};N_{\max},K_{\max}, N_{\rm Fock}) $, and running in one direction mixes into the other through $\bm{J}^{-1}$.

\subsection{Concrete asymptotics from basis LF truncation}

Asymptotically, regulator effects enter as inverse powers with possible cross terms. A generic local ansatz for the observables near a renormalization point is
\begin{align}
\mathcal O_a(\bm{g};N_{\max},K_{\max}, N_{\rm Fock}) 
= \mathcal O_a^{(\infty)}
&+\sum_{i\in{{\bf g}}} A_{a i}(\ln N_{\max},\ln K_{\max}), N_{\rm Fock})\, g_i\nonumber\\
&+ \frac{C_{a}}{N_{\max}^{p}} + \frac{D_{a}}{K_{\max}^{q}}
+ \frac{E_{a}}{N_{\rm Fock}^{r}}
+ \frac{F_{a}}{N_{\max}^{p} K_{\max}^{q}} + \cdots ,
\label{eq:Oansatz}
\end{align}
where $p,q,r>0$ are determined by basis selection rules and level spacing, and $A_{ai}$ may have mild logarithmic dependence.
Then
\begin{align}
\partial_{\sigma_T}\mathcal O_a &= A_{a,\sigma_T} + \mathcal O(g), \qquad
\partial_{\sigma_L}\mathcal O_a = A_{a,\sigma_L} + \mathcal O(g),\nonumber\\
\partial_{g_\sigma}\mathcal O_a &= A_{a,g_\sigma} + \mathcal O(g), \qquad
\partial_{g_\pi}\mathcal O_a = A_{a,g_\pi} + \mathcal O(g),\\
\partial_{\ln N_{\max}}\mathcal O_a &= \partial_{\ln N_{\max}}A_{a i}\, g_i
- p\, \frac{C_a}{N_{\max}^{p}} - p\, \frac{F_a}{N_{\max}^{p}K^{q}} + \cdots, \\
\partial_{\ln K_{\max}}\mathcal O_a &= \partial_{\ln K_{\max}}A_{a i}\, g_i
- q\, \frac{D_a}{K_{\max}^{q}} - q\, \frac{F_a}{N_{\max}^{p}K^{q}} + \cdots.\\
\cdots &= \cdots
\end{align}
Substituting into \eqref{eq:betaN}-\eqref{eq:betaK} yields explicit leading-order beta functions.

\subsection{Explicit example for a 2x2 evolution}
To illustrate explicitly how this works, let us switch the couplings $g_{\sigma, \pi}$ to the Fock states, and quantify the evolution in 
the three Fock space for a finite dimensional curt-off Hamiltonian.
With this in mind, and if $A_{ai}$ are treated as constants locally,
then
\bea
\bm{J}\approx 
\begin{pmatrix}
A_{1,\sigma_T}&A_{1,\sigma_L}\\ A_{2,\sigma_T}&A_{2,\sigma_L}
\end{pmatrix}
\eea
and
\begin{align}
\bm{\beta}^{(N)} &\simeq
\frac{1}{\det\bm{J}}
\begin{pmatrix}
 A_{2,\beta^2} & -A_{1,\beta^2}\\[2pt]
 -A_{2,\kappa^2} & A_{1,\kappa^2}
\end{pmatrix}
\begin{pmatrix}
p\, C_1 N_{\max}^{-p} + p\, F_1 (N_{\max}^{-p} K_{\max}^{-q})\\[2pt]
p\, C_2 N_{\max}^{-p} + p\, F_2 (N_{\max}^{-p} K_{\max}^{-q})
\end{pmatrix} + \cdots, \label{eq:LOBN}\\
\bm{\beta}^{(K)} &\simeq
\frac{1}{\det\bm{J}}
\begin{pmatrix}
 A_{2,\beta^2} & -A_{1,\beta^2}\\[2pt]
 -A_{2,\kappa^2} & A_{1,\kappa^2}
\end{pmatrix}
\begin{pmatrix}
q\, D_1 K_{\max}^{-q} + q\, F_1 (N_{\max}^{-p} K_{\max}^{-q})\\[2pt]
q\, D_2 K_{\max}^{-q} + q\, F_2 (N_{\max}^{-p} K_{\max}^{-q})
\end{pmatrix} + \cdots.
\label{eq:LOBK}
\end{align}
which show the power-law suppression in the regulators, and how the
transverse and longitudinal cutoffs $N_{\max}$ and $K_{\max}$ enter each component via $\bm{J}^{-1}$.

If only one renormalization condition is imposed in this simplified case, \eqref{eq:dF} gives one equation for two unknowns in $d\bm{g}$, so the flow is underdetermined. Two independent observables are required to determine the joint running of $(\sigma_T, \sigma_L)$. More specifically for this case, 
choose $\mathcal O_1$ as the lowest intrinsic eigenvalue dominated by transverse dynamics,  and $\mathcal O_2$ as the lowest longitudinal excitation energy. In general, we expect
\(
\partial_{\sigma_{T,L}}\mathcal O_{1,2}>0,
\)
(with magnitudes given by expectation values of the corresponding operators).
Compute these derivatives numerically at a reference point, estimate $(p,q)$ from convergence rates,
and plug into \eqref{eq:LOBN}-\eqref{eq:LOBK} to obtain numerical beta functions.

\subsection{Summary for 2x2 evolution}
\begin{itemize}
\item Each observable depends on both couplings $(\sigma_T, \sigma_L)$ and both regulators $(N_{\max},K_{\max})$.
\item Beta functions follow by differentiating the vector renormalization conditions and solving a $2\times2$ linear system with Jacobian $\bm{J}$; see Eqs.~\eqref{eq:betaN}-\eqref{eq:betaK}.
\item The resulting $\bm{\beta}$ depend on $(N_{\max},K_{\max})$ through explicit regulator terms and the inverse Jacobian.
\end{itemize}

\part{Hadronic spectroscopy on the light front}
\chapter{Quarkonia and "normal" mesons on light front}

Before we start this chapter, let us explain its title.
By {\em "normal hadrons"} we
mean those which can be described well by including only confining forces : examples include vector and tensor mesons, decuplet baryons, etc. Their typical
r.m.s. sizes are $\langle r \rangle \sim 1 \, fm$. Obviously,
there are spin-dependent forces, but those are subleading.

Quarkonia are
made of  quarks $c,b$ which are much heavier than effective light quark masses ($M_q\approx 0.35\, GeV, M_s\approx 0.5\, GeV$).
Respectively, their sizes and quark velocities are smaller, and perturbative Coulomb-like
potential $\sim 1/r$ needs to be included. 

Furthermore,  if these masses be {\em parametrically large}, they should be oulomb and be similar to positronium.  This is true, but however is not the case for real-world $c,b$ quarks. In fact the balance
between Coulombic and confining forces remains delicate. For example, {\em "the main spectral
gap"} between the 2s and the 1s states remains about the same, for quarkonia with $\bar s s, \bar c c$ and even  $\bar b b$ flavors. 

Yet there exist some "unusual" hadrons with sizes significantly smaller
than 1 fm. This happens because strong attraction between their
constituents can also neither be the  Coulombic nor confining forces, but
the instanton-induced ones responsible for chiral symmetry breaking.
The smallest one is the "scalar glueballs", and their unusually small size $\sim 0.2\, fm$ was predicted in \cite{Schafer:1994fd} and
later confirmed by lattice studies such as \cite{Abbott:2025irb}. However, glueballs
are not discussed in this work.

Examples of "unusual" mesons we do  discuss  are $\sigma$ and $\pi$ mesons: their properties are
defined mostly by instanton-induced 't Hooft Lagrangian rather than
confining forces \footnote{It would be interesting to work out their relatives - $\eta'$ and scalar isovector $a_1$ - in which t' Hooft Lagrangian
is $repulsive$, and therefore with unusually large masses and sizes,
but so far it is not done yet.}

\section{Boosting quarkonia WFs}

Heavy quark bound states are the pillars of hadronic spectroscopy. Already in 1970's, after $J/\psi$ and $Y$
families discoveries, it was shown that nonrelativistic approach, with a Cornell and spin-dependent potentials
provides quite accurate description of their properties. Naturally,
one would think that boosted versions of their WFs have a similar level of accuracy
on the LF. Before boosting, it is useful to transfer WFs from spherical to 
cylindrical coordinates (see e.g. \cite{Adhikari:2018umb}). 

There are some obvious differences between the description in the rest frame, and on the light front.
For instance, there are different symmetries:  3-dimensional  angular momenta $\vec S,\vec L,\vec J$ 
are reduced to their 2-dimensional transverse parts, 
with spin $\vec S$ and orbital momentum $\vec L$ projected onto 
longitudinal momentum $\vec P$. The projection of $J$ is denoted by  "meson helicity" $\Lambda$. Obviously,
hadron states with different  $\Lambda$ values,
are treated  differently: say $\rho(\Lambda=0)$ and $\rho(\Lambda=\pm 1)$ have different wave functions 
(even more than one: see below). While masses, magnetic and quadrupole moments, etc should turned out to be the same, the 3-dimensional
rotation is some complicated transformation,  involving all components of the wave functions, and we will not attempt to explicitly use it.

Our approach to quarkonia at LF were worked out in two papers, \cite{Shuryak:2021fsu,Shuryak:2021mlh}. The first question to be addressed is  whether the states of heavy quarkonia  can be
represented by linear Regge trajectories in {\em squared masses}, as would be natural for LF Hamiltonians?

\begin{figure}[h!]
\begin{center}
\includegraphics[width=7cm]{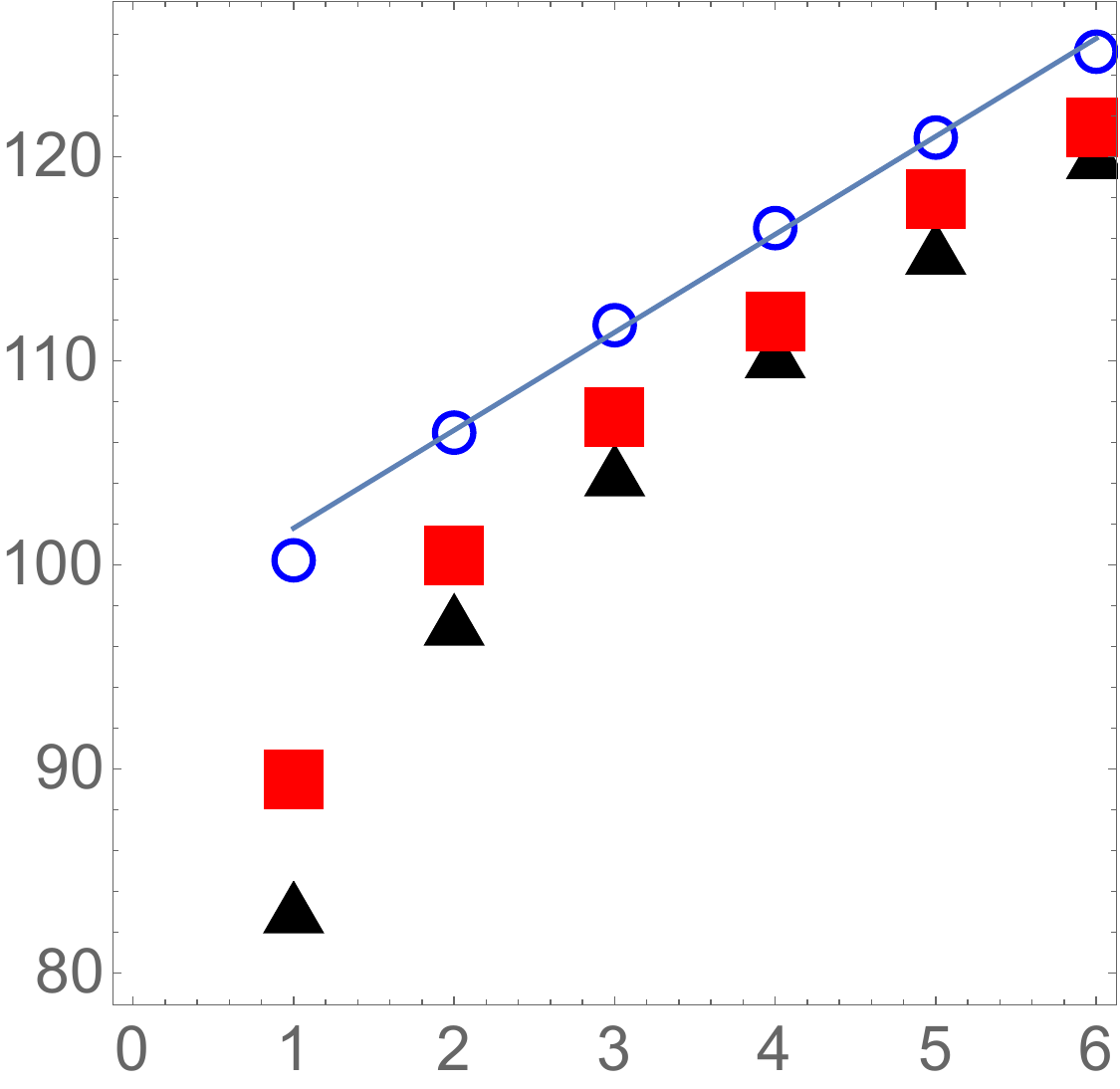}
\caption{$M_{n+1}^2\, (GeV)^2$ versus $n+1,n=0,..5$, for the six $S$ zero orbital momentum (L=0) states of  bottomonium.
The red squares correspond to the experimentally observed Upsilons. The black triangles 
show the masses obtained from the Schroedinger equation, with the Cornell potential (linear and Coulomb potentials, no spin forces). 
The blue circles show the masses if 
the Coulomb potential is switched off,  and only  the  linear potential is used. The straight line is shown for comparison. 
 }
\label{fig_6upsilons}
\end{center}
\end{figure}

In Fig.\ref{fig_6upsilons} we show the experimental masses of six $(nS), n=0-5$ Upsilon mesons, 
compared to the standard results from the Schroedinger equation, with the Cornell potential
(black triangles) and with only its linear part $V_{conf}=\sigma_T r$ (blue circles). The first 
observation 
 is that using a linear potential alone (blue circles), we 
 find a nearly-linear Regge trajectory. This observation will be important
 in the next subsection, as it shows that even for heavy bottomonia,  the light-front
 Hamiltonian can be approximated by an oscillator with good accuracy. Note 
however, that the slope of the straight line,  is here completely different 
from the $1/\alpha'$ slope of a similar trajectory for light mesons 
(e.g. for $\omega$ mesons we used in \cite{Shuryak:2021fsu}).

\begin{figure}[h!]
\begin{center}
\includegraphics[width=7cm]{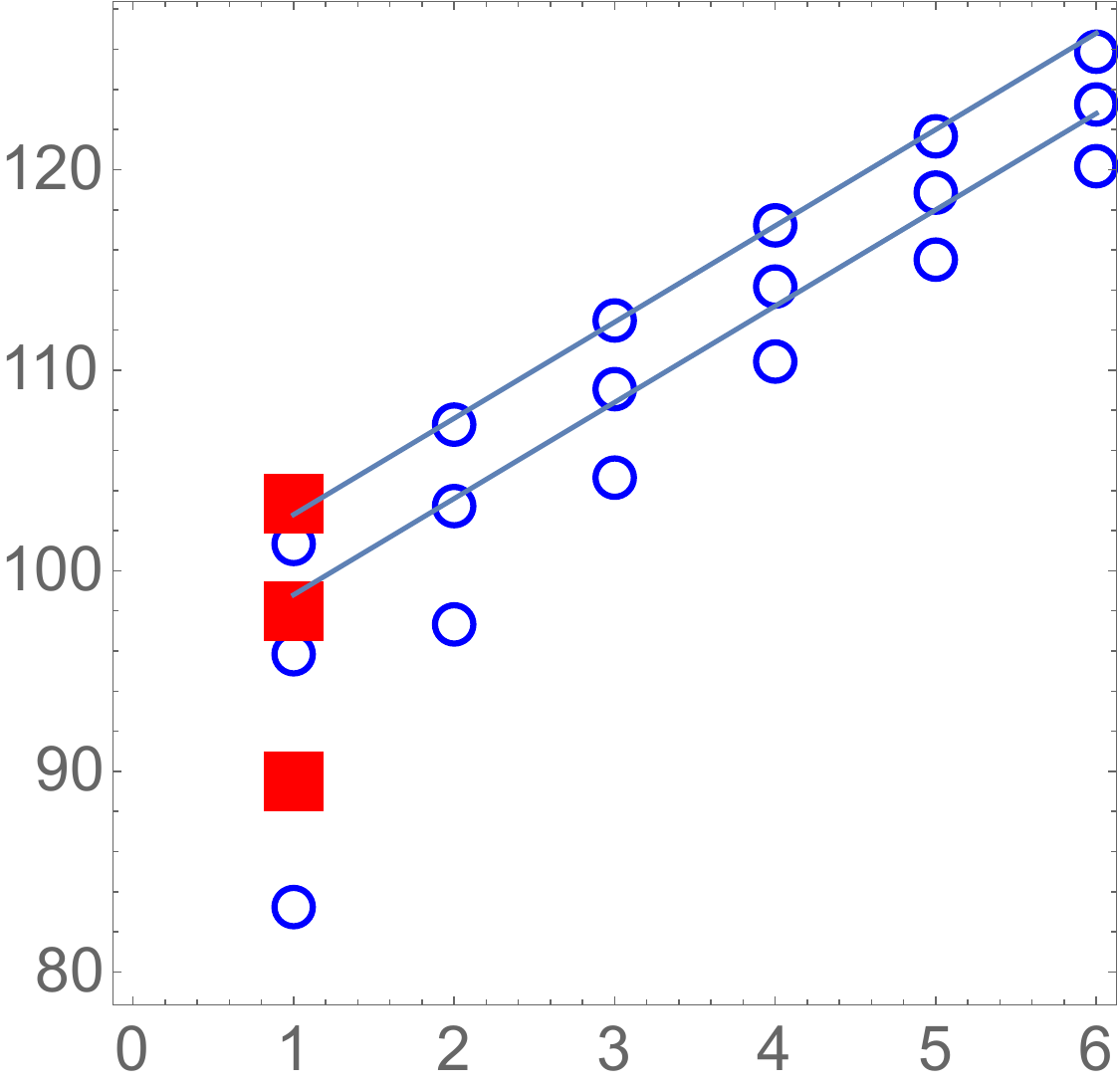}
\caption{$M_{n+1}^2\, (GeV)^2$ versus $n+1, n=0,1...$ for  three families  of  bottomonium
states, with orbital momentum $L=0,1,2$ (from bottom up). 
The red squares correspond to the experimentally observed $\Upsilon,h_b, \Upsilon_2$
mesons (from bottom up). The blue circles show
 masses obtained from the Schroedinger equation,
  with the Cornell potential (linear and Coulomb potentials, still without spin forces).  The straight lines are shown for comparison. 
 }
\label{fig_3L_bb}
\end{center}
\end{figure}

  The second observation is that the expected contribution from the  spin-dependent potential $V_{SS}$ (responsible for
splitting between squares and triangles),  is positive and decreases with $n$. The former is due to the 
positivity of the spin factor $\vec S_1\cdot\vec S_2=1/4$, and the second
to the fact that $V_{SS}(r)$ is rather short range, in comparison to the size
of the lowest Upsilon,  but much smaller than the sizes of  the excited ones.
Another way to anticipate the accuracy of an oscillator approximation in the light front
description (discussed in \cite{Shuryak:2021hng} and using $\omega_3$ mesons with $L=2$),  is to
study the mass dependence of bottomonium on its orbital momentum $L$. 
In Fig.\ref{fig_3L_bb} we show the calculated 18 squared masses for $n=0-5$ 
(left-to-right) and $L=0,1,2$ (bottom-to-top). While the Coulomb potential was included, it affects mostly and only $n=0,L=0$ Upsilon. The Regge trajectories for nonzero angular momentum
$L=1,2$,  show better linear dependence on $n$
 than $L=0$. The  corresponding wave function vanishes at the origin $r=0$, 
 and is less affected by short-range Coulomb and spin-dependent forces.
 
We further note, that  at larger $n$
(right side of the plot) the dependence on $L$ also becomes linear, as the $L=0,1,2$ 
points
become equidistant. This observation  encourages us to think that the oscillator
description of the light front Hamiltonian and LFWFs,  will need only relatively small corrections.  
 
In~\cite{Shuryak:2021hng}, we described how we may  include the  linear confining term in $H_{LF}$
(instanton induced at intermediate distances),
and make it  more user friendly, by eliminating the square root
using the well-known einbein $e=1/a$ trick, i.e. 

\bea
\label{EINBEIN} 2M \mathbb V_C(a,b,x,b_\perp)
\approx \sigma_T\bigg(\frac{|id/dx|^2+b\, b_\perp^2}a+a\bigg)
\eea
Here $a,b$ are variational parameters.
The minimization with respect to $a$ is assumed, followed by the substitution 
$b\rightarrow M^2\approx (2m_Q)^2$ for heavy mesons,  and most light ones.
(For the pion, this last substitution
is not valid, as we have shown in~\cite{Shuryak:2021hng}). 

For a numerical analysis of (\ref{EINBEIN}), we used in~\cite{Shuryak:2021hng}
a basis set of functions composed of a 2-dimensional transverse oscillator, times
longitudinal states  $sin(\pi n x)$ with odd $n$.
More specifically, the light front Hamiltonian can be re-arranged as follows

\be 
\label{HLFX}
H_{LF}=H_0+\tilde V + V_{perp}+V_{spin} 
\ee
with the spin-part including both the perturbative and non-perturbative instanton
contributions. As we noted earlier,  in the dense instanton vacuum, the central part is hardly
differentiable from the linear confining potential at  intermediate distances.

The first contribution  $H_0$ 

\begin{equation}
\label{H0X}
 H_0={\sigma_T \over a} \bigg( -{\partial^2  \over \partial x^2}-b{\partial^2  \over \partial  \vec k_\perp^2} \bigg) + \sigma_T  a + 4(m_Q^2+ k_\perp^2)
\end{equation} 
is diagonal in the functional basis used~\cite{Shuryak:2021hng}. 
In  this form, we make use of  the momentum representation, with $\vec k_\perp$ as variable. Similarly, one can
use the coordinate representation with $\vec b_\perp$ as a variable,  and $ \vec k_\perp=i \partial /\partial \vec b_\perp $.
The latter choice is much more convenient when discussing states with nonzero angular
momenta, in  relation  to the azimuthal angle coordinate $\phi$ in the transverse plane (see more on that in Appendix).

The second contribution $\tilde V$
\begin{equation}  
\tilde V(x,\vec k_\perp)\equiv (m_Q^2+k_\perp^2)\bigg({1 \over x \bar x} -4\bigg)
\end{equation}
 has nonzero matrix elements  $\langle n_1 |V(x,\vec k_\perp) | n_2\rangle $ for all $n_1,n_2$ pairs. 
The perturbative part $V_{perp}$ for heavy quarks is the Coulomb term, with 
running coupling and other radiative corrections. Finally, the last term $ V_{spin} $
contains matrices in spin variables and in orbital momenta, which we will consider later.
 

\begin{figure}[t]
\begin{center}
\includegraphics[width=6cm]{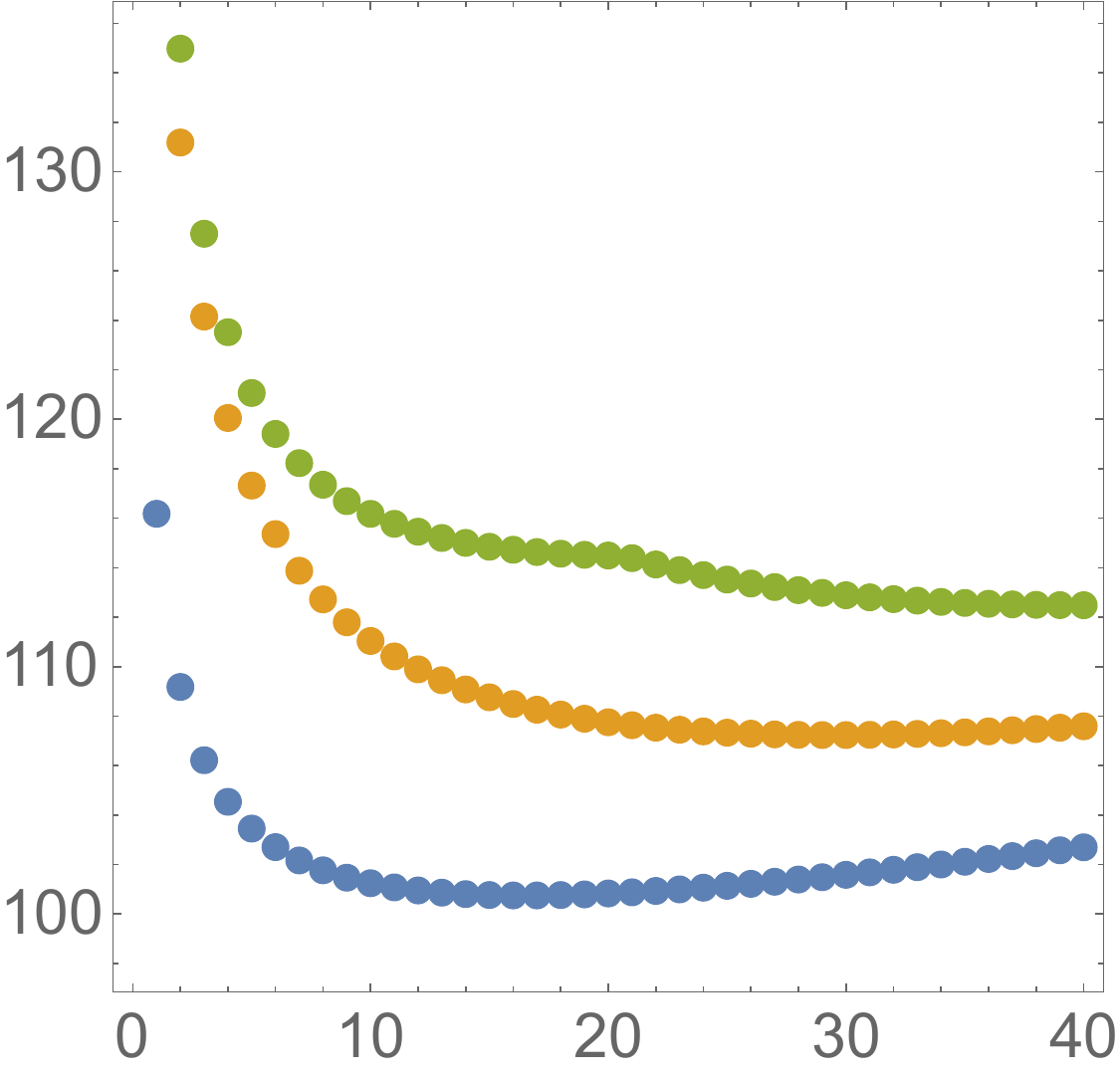}
\includegraphics[width=6cm]{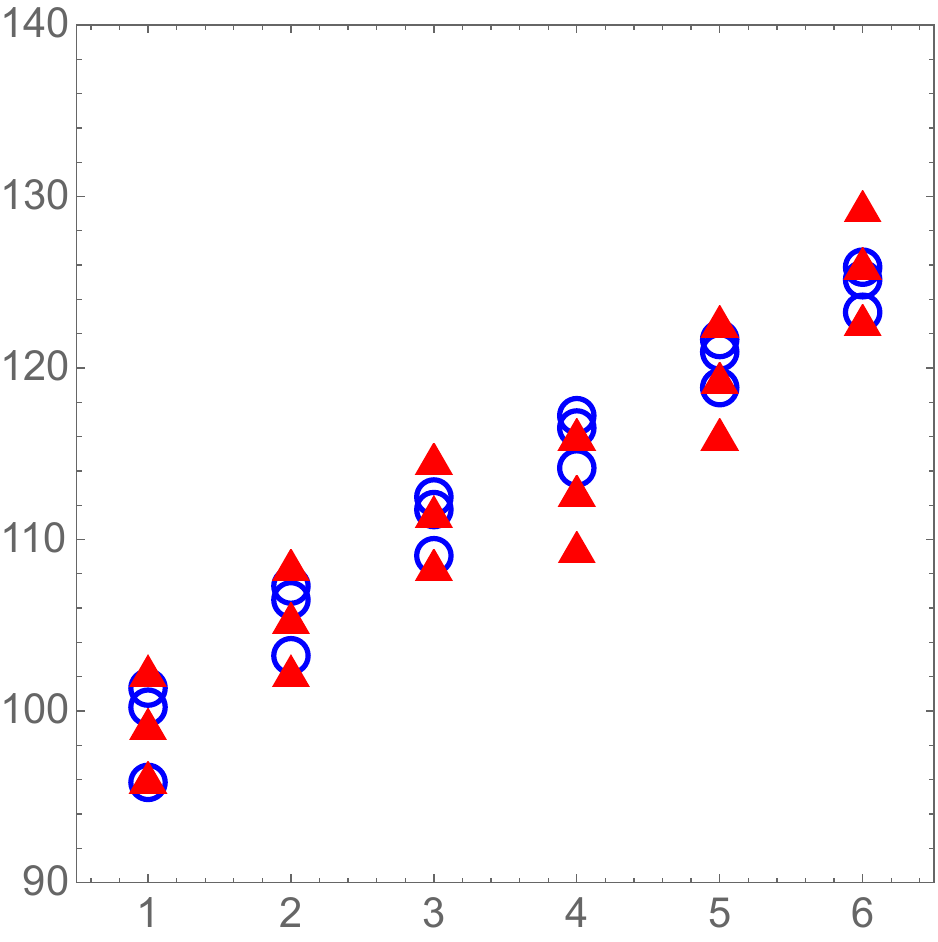}
\caption{Top: Squared masses $M^2_{n}$ for  $\bar b b$ mesons for $n=1,2,3$ versus the variational parameter $a$.
Bottom:  Squared masses  for $n=0..5$ (left to right) and orbital momentum 
$m=0,1,2$ (down to up), calculated from the light front Hamiltonian $H_{LF}$ (red triangles),
and shifted by a constant, $M^2_{n+1}-5\, GeV^2$. For comparison, the blue-circles show the squared masses $M^2_{n+1}$ calculated from the Schroedinger equation in the CM frame,  with only  linear plus centrifugal potentials.}
\label{fig_M2_vs_a}
\end{center}
\end{figure}

The  calculated masses (shifted by a constant  "mass renormalization", to make $n=0,m=0$ states the same) are shown in  Fig.\ref{fig_M2_vs_a}.
The bottom part shows  good agreement between the masses  obtained solving  the Shroedinger equation 
in the rest frame (blue circles),  and the masses following from the  light-front frame (red-triangles).
The slope is correct, and is determined by the same string tension $\sigma_T$. The splittings in orbital momentum are of the same scale, but not identical.
This is expected, as we compare the 2-dimensional  $m$-states on the light front,  with the 3-dimensional  $L$-states in the center of mass frame.

The  irregularity between the third and  fourth set of states, is due to our use of a 
 modest  basis set, with only three radial functions  (altogether 12 functions if one counts them with 4 longitudinal harmonics). 
 This can be eliminated using a larger set.

As a final note in this section, we recall that the chief goal of these calculations
is to generate the pertinent  LFWFs for all these states,  from the light front Hamiltonian  $H_{LF}$.  
The details about  the setting and some of these wavefunctions  can be found


\begin{figure}
    \centering
    \includegraphics[width=0.5
    \linewidth]{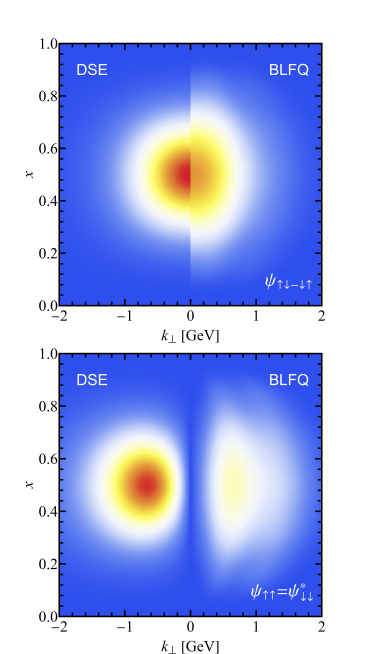}
    \caption{The LF wave function of $J/\psi$ and  $\eta_c$ mesons. The left side from Dyson-Schwinger equation, the right from Basis-light-front-quantization.}
    \label{fig_quarkonia_space}
\end{figure}
As a good picture is better than thousand words, let us present results of two modern approaches following \cite{Cao:2025bit}.
One of them is  Dyson-Schwinger (DSE) which is fully relativistic and
take interaction from averaged lattice configurations. 
Another is Basis light front quantization (BLFQ) by Vary et al in which LF Hamiltonian
is written as a matrix in some preselected basis, with kinetic energy and interactions. 
The results for LF WFs are shown in  Fig.\ref{fig_quarkonia_space}
for Dyson-Schwinger (DSE) and BLFQ
wave functions, of $J/\psi, \eta_c$ mesons.
It is  clear from it that while these two
wave functions are in qualitative agreement, they obviously are
not yet  derived quantitatively.
 So, unfortunately, the existing methods have not yet converged on say few-percent accuracy, even for the simplest hadrons, the heavy quarkonia.


\begin{subappendices}

\section{Heavy-quark reduction}

The heavy quark expansion of a Dirac fermion of mass $m_Q$ with fixed velocity, 
in an arbitrary gauge field is best achieved using the Foldy-Wuthuysen transformation on the relativistic
fermion propagator, 
\be
\label{EPOT}
e^{-i\frac{\slashed{D}_\perp}{2m_Q}}\,\frac 1{i\slashed{D}-m_Q}\,e^{-i\frac{\slashed{D}_\perp}{2m_Q}}
\ee
with $i\slashed{D} =i\slashed{\partial}+\slashed{A}$ and $\slashed{D}_\perp =\slashed{D}-\slashed{v}v\cdot D$ satisfying $[\slashed{D}_\perp, \slashed{v}]_+=0$.
We will refer to $v_\mu$ the 2-dimensional {\it light-cone-like} velocity along the 2-dimensional {\it light-cone-like} coordinate  $x_+$ in Euclidean signature, 
and to $v_{\perp \mu}$ its orthogonal  velocity along the 2-dimensional {\it light-cone-like} coordinate $x_-$ also in Euclidean signature,
\bea
v_\mu=({\bf 0}_\perp,  {\rm sin}\theta,{\rm cos}\theta),\qquad\qquad
v_{\perp \mu}=({\bf 0}_\perp,  -{\rm cos}\theta,{\rm sin}\theta),
\eea
with $x_+=v\cdot x$ and $x_-=v_\perp\cdot x$.
These light-cone-like Euclidean coordinations (lower indices) are not to be confused with the Minkowski light-cone coordinates $x^\pm=x^0\pm x^3$
(upper-indices).  With this in mind, and to order $1/m_Q^2$ the heavy quark propagator is

\begin{eqnarray}
\label{NR}
\frac 1{iv\cdot D}-\frac 1{iv\cdot D}\bigg(\frac 1{2m_Q}(i\slashed D_\perp)^2
-\frac 1{4m^2_Q}(i\slashed D_\perp)(iv\cdot D)(i\slashed D_\perp)\bigg) \frac 1{iv\cdot D} 
\end{eqnarray}
The bracket in (\ref{NR}) gives rise to a vertex insertion, which can be re-arranged

\be
\label{NR2}
\frac 1{2m_Q}\bigg((iD)^2-\frac 12\sigma_{\mu\nu}F_{\mu\nu}\bigg)
-\frac 1{4m^2_Q}\bigg(i\sigma_{\alpha\nu}iD_{\alpha} v_\mu  F_{\mu\nu}+ iD_{\nu} v_\mu F_{\mu\nu}\bigg)
\ee

with $\sigma_{\alpha\nu}=\frac 1{2i}[\gamma_\alpha, \gamma_\nu]$.  In (\ref{NR2})
we have dropped all terms that vanish on-shell, i.e $v\cdot D Q_v=0$
with $Q_v$ the heavy quark field.
When  inserted on a straight Wilson line, (\ref{NR2})  produces the spin corrections up to order $1/m_Q^2$.  
Note that in the Dirac representation $\sigma_{4i}$ is off-diagonal. The electric contribution  mixes  particles and anti-particles. It  does not contribute 
when inserted on a straight Wilson line defined as
\bea
{\bf W}(y,x)=\langle y_+|\frac 1{v\cdot D}|x_+\rangle \,\delta({\bf{x}}-{\bf{y}})
={\bf P}e^{i\int_{x_+}^{y_+}A\cdot dz}\,\theta(y_+-x_+)\delta({\bf{x}}-{\bf{y}})\nonumber\\
\eea
with the ordering along $x_+$ and the short hand notation 
$x_\mu=({\bf x}_\perp, x_-,x_+)\equiv ({\bf{x}}, x_+)$.

\subsection{Slated Wilson loop dressed with fields}

The undressed  Wilson loop  in the resummed instanton vacuum is

\bea
\label{W1}
\langle {\bf 1}_\theta\rangle =
\langle {\bf W}(\theta, 0_\perp)\,{\bf W}^\dagger(\theta, b_\perp)\rangle_C\approx e^{-Z_+V_C(\xi_\theta)}\nonumber\\
\eea
with $\xi_\theta=({\rm cos}^2\theta \,b_3^2+b_\perp^2)^{\frac 12}$,
and where $V_C(\xi_\theta)\rightarrow V_C(\xi_x)$ follows by analytical continuation $\theta\rightarrow -i\chi$. The spin dressed Wilson loop
to order $1/m_Q^2$ follows by inserting the corrections (\ref{NR}) on the Wilson lines

\bea
\label{W2}
&&\langle {\bf 1}_\theta\rangle \,\delta_{12}+\bigg(
+\frac{i}{4m_{Q1}^2}\int_{-\frac 12Z_+}^{+\frac 12Z_+} dz_+\bigg[\sigma_{1\alpha\nu}v_\mu \langle  F_{\mu\nu}(x_1, z_+)iD_{\alpha} (x_1,z_+) {\bf 1}_\theta\rangle + 1\leftrightarrow 2\bigg]\nonumber\\
&&+\frac{1}{4m_{Q1}^2}\int_{-\frac 12Z_+}^{+\frac 12Z_+} dz_+\,\int_{-\frac 12Z_+}^{+\frac 12Z_+}dz_+^\prime \bigg[\langle\sigma_{1\mu\nu}F_{\mu\nu}(x_1, z_+)(iD)^2(x_1, z_+^\prime) {\bf 1}_\theta\rangle + 1\leftrightarrow 2\bigg]\nonumber\\
&&+\frac{1}{8m_{Q1}m_{Q2}}\int_{-\frac 12Z_+}^{+\frac 12Z_+}dz_+\, \int_{-\frac 12Z_+}^{+\frac 12Z_+}dz_+^\prime 
\bigg[\langle\sigma_{1\mu\nu}F_{\mu\nu}(x_1, z_+)(iD)^2(x_2, z_+^\prime) {\bf 1}_\theta\rangle \nonumber\\
&&\qquad\qquad\qquad\qquad\qquad\qquad\qquad\qquad +\langle (iD)^2(x_1, z_+)\sigma_{2\mu\nu}F_{\mu\nu}(x_2, z^\prime_+) 
{\bf 1}_\theta\rangle\nonumber\\
&&\qquad\qquad\qquad\qquad\qquad\qquad\qquad\qquad-\frac 12\langle \sigma_{1\mu\nu}F_{\mu\nu}(x_1, z_+)\sigma_{2\alpha\beta}F_{\alpha\beta}(x_2, z^\prime_+) {\bf 1}_\theta\rangle
\bigg]
\bigg)\delta_{12}
\eea

after dropping the terms that vanish on-shell, the terms that vanish by parity after averaging in the presence of the undressed
Wilson loop, and those with no contribution to the spin-dependent potentials.   
In (\ref{W2}) we have labeled the quark masses for a general Wilson loop with unequal masses, and used the short hand notation
$$\delta_{12}=\delta({\bf x}_1-{\bf y}_1)\delta({\bf x}_2-{\bf y}_2)$$
Throughout, the affine integration parameters  $z_+, z_+^\prime$ in (\ref{W2}) are proper times. The conversion to ordinary times
$(z_+, z_+^\prime)\rightarrow (z_+, z_+^\prime)/\gamma_E$ amounts to extra Lorentz contraction factors of $1/\gamma_E=\sqrt{1+\dot{x}_E^2}$ 
in Euclidean signature, that will be added at the end by inspection.

\subsection{Identities}

To simplify (\ref{W2}) we use the identities  with slated Wilson lines (dropping the delta functions)

\be
\label{ID1}
{\bf W}(x_+,y_+){\bf W}(y_+,z_+)=\langle x_+|\frac 1{v\cdot D}|y_+\rangle\langle y_+|\frac 1{v\cdot D}|z_+\rangle
=\langle x_+|\frac 1{v\cdot D}|z_+\rangle={\bf W}(x_+, z_+)
\ee
 which is a property of the eikonalized and ordered  Wilson line. More impotantly, we have the identity

\bea
\label{ID2}
&&D_{ \nu}(x_+){\bf W}(x_+,y_+)-{\bf W}(x_+,y_+)D_{ \nu}(y_+)\nonumber\\
&&=\langle x_+|D_{ \nu}\frac 1{v\cdot D}-\frac 1{v\cdot D}D_{ \nu}|y_+\rangle=
\langle x_+|\frac 1{v\cdot D} [v\cdot D, D_{\nu}]\frac 1{v\cdot D}|y_+\rangle\nonumber\\
&&=\langle x_+|\frac 1{v\cdot D} (-iv_\mu F_{\mu \nu})\frac 1{v\cdot D}|y_+\rangle=
\int_{-\frac 12Z_+}^{+\frac 12Z_+} dz_+\,{\bf W}(x_+,z_+)(-iv_\mu F_{\mu\nu})(z_+){\bf W}(z_+,y_+)
\eea

The end-point derivative of  a Wilson line, amounts to an insertion of a pertinent field strength (plaquette in a lattice form) along the line,

\bea
\label{PLAQUETTE}
v_\mu F_{\mu \nu}=v_4F_{4\nu}+v_3 F_{3\nu}={\rm cos}\theta F_{4\nu} +{\rm sin}\theta F_{3\nu}
\eea
Finally, we have the large $|z_+|\rightarrow \infty$ identity 

\be
\label{ID3}
\,{\bf W}(y, z_+; x, z_+ )D_\alpha(x, z_+)\,{\bf W}(x, z_+; y, z_+)\rightarrow \partial_\alpha^y
\ee
as the fields are assumed to vanish at asymptotic $z_+$. 
A repeated use of (\ref{ID1}-\ref{ID3}) allows to simplify (\ref{W2}).



\subsection{Resulting spectrum}
The sum of the spin contributions  to the squared mass operator on the light front in the 
instanton vacuum, is now explicit and of the form

\bea
\label{LFHAMIL}
M^2_{SD,I}(\xi_x, b_\perp)=2MV_{SD,I}(\xi_x, b_\perp)=2M&&
\bigg(
\bigg[ \frac{\sigma_1\cdot ( {b}_{12}\times s_1\hat 3)}{4m_{Q1}}
-\frac{\sigma_2\cdot ( {b}_{21}\times s_2\hat 3)}{4m_{Q2}} \bigg]\,\frac 1{\xi_x}V_C^\prime(\xi_x)\nonumber\\
 &&+
\bigg[\frac{\sigma_1\cdot (b_{12}\times s_1\hat 3)}{2 m_{Q1}}-\frac{\sigma_2\cdot (b_{21}\times s_2\hat 3)}{2 m_{Q2}}\bigg]
\frac 1{\xi_x}\mathbb V_1^\prime(\xi_x)\nonumber\\
&&+
\bigg[\frac{\sigma_2\cdot(b_{12}\times s_1\hat 3)}{2 m_{Q2}}-\frac{\sigma_1\cdot(b_{21}\times s_2\hat 3)}{2 m_{Q1}}\bigg]\frac 1{\xi_x}
\mathbb V_2^\prime(\xi_x)\nonumber\\
&&+
\frac 1{4m_{Q1}m_{Q2}}\sigma_{1\perp i}\sigma_{2\perp j}\bigg[\bigg(\hat b_{21i}\hat b_{21j}-\frac 12\delta_{\perp ij}\bigg)\mathbb V_3(\xi_x)
\bigg]\bigg)\nonumber\\
\eea
with $b_{21}=-b_{12}=b_\perp$  and $s_{1,2}$ the signum of the velocity along the 3-direction (sign of the helicity).
All spin potentials $\mathbb V_{1,2,3}(\xi_x)$ are tied to the central potential $\mathbb V_C(\xi_x)$ in the instanton vacuum,
  \bea
\label{V12X}
\mathbb V_2(\xi_x)=\mathbb V_1(\xi_x)+V_C(\xi_x) \rightarrow \frac 12 V_C(\xi_x)\qquad\qquad
\mathbb V_3(\xi_x)=\frac{2b_\perp^2}{\xi_x^2}\mathbb V_C^{\prime\prime}(\xi_x)\qquad\qquad
\mathbb V_4 (\xi_x)=0
\eea
While on the light front $\mathbb V_{2,4}(\xi_x)$  are no longer tied by the Bianchi-identity (covariantized Lenz law),
we note that  the leading contributions  match the rest frame contributions at $\theta=0$. Therefore, the rest frame
relation $\mathbb V_2(R)=\frac 12 V_C(R)$ in the instanton vacuum~\cite{Eichten:1980mw} (note the sign convention difference), carries to the light front $\mathbb V_2(\xi_x)=\frac 12 V_C(\xi_x)$. This is not the case for $\mathbb V_{3,4}(\xi_x)$.

A key feature of the spin orbit contributions in (\ref{LFHAMIL}), is that a flip of a spin say $\sigma_1$,
can be compensated by a flip in the sign of the helicity say $s_1$ or $s_2$. This is reminiscent of the rest frame symmetry of the spin-orbit interactions,
that show that a flip in the spin can be compensated by a flip in the angular momentum, thereby preserving the total angular momentum.
The  light front Hamiltonian in the instanton vacuum,  is the squared mass operator for a $Q\bar Q\equiv Q_1Q_2$ pair, that includes the free  plus the central contribution (\ref{V12X}),  and the spin contributions (\ref{LFHAMIL}), 
\bea
\label{MASS2}
M^2=\sum_{i=1,2}\frac{k_\perp^2+m^2_{Qi}}{x_i}
+2M (V_{C}(\xi_x)+V_{SD,I}(\xi_x, b_\perp))
\eea
with Bjorken $x_{i=1,2}$ and satisfying  $x_1+x_2=1$. A detailed analyses of the spectrum and light front wavefunctions
following from (\ref{MASS2}) as applied to heavy and light mesons,
with comparative estimates from perturbative one-gluon exchange and confinement, will be detailed in a sequel.

\section{String induced scalar spin-orbit interaction}

Most of the spin-orbit interaction mediated by perturbative one-gluon exchange  and  non-perturbative instanton interactions, at short and intermediate distances respectively is vector-like. In the standard constituent quark model it yields substantial spin-orbit splitting (of the order of $100\,{\rm MeV}$ and must be omitted to account for the phenomenological success of the constituent quark model in the description of light and heavy-light hadrons. In contrast, the induced spin-orbit coupling from
a quantized NG string is scalar-like, with the same magnitude and
opposite sign. A large cancellation between the vector and scalar spin-orbit contributions, is usually argued to take place.

To understand the origin of the scalar spin-orbit interaction induced by a confining string, consider the standard NG string with end-points, carrying equal mass and spin,
\bea
\label{1}
S=\sigma_T\int d\tau\,d\sigma\, 
\sqrt{h}
+\int_B d\tau\, \bigg(M\sqrt{{\dot{X}}^2}
-\frac 12 \sigma_{\mu\nu}
\frac{{\dot{X}}^\mu {\ddot{X}}^\nu}{{\dot{X}}^2}
\bigg)
\eea
with the determinant
\bea
h={\rm det}\bigg(\eta_{\mu\nu}\partial_aX^\mu\partial_bX^\nu\bigg)
\eea
The boundary spin factor enforces Thomas precession along each of the end-point world lines~\cite{Strominger:1980xa,Polyakov:1988md}. Recall that 2 infinitesimal boosts are equivalent to 1 infinitesimal boost plus a Thomas precession. It is a Berry phase~\cite{Nowak:1990wu}.
Their contribution to the interaction potential is best seen by considering a specific solution to EOM following from (\ref{1}),
\bea
\partial_\alpha\bigg(\sqrt{h} h^{\alpha\beta}\partial_\beta X^\mu\bigg)=0
\eea
subject to the spinless boundary conditions
at $\sigma=\pm 1$
\bea
\sigma_T\sqrt{h}\partial_\sigma X^\mu\pm M\partial_\tau\bigg(\frac{\dot{X}^\mu}{\sqrt{\dot{X}}^2}\bigg)=0
\eea
The rotating string in the 12-plane is such a solution~\cite{Sonnenschein:2014bia} (and references therein), with the embedding
\bea
\label{2}
X^{\mu}=(\tau, 
r\sigma\, {\rm cos}(\omega\tau),
r\sigma\, {\rm sin}(\omega\tau), 0)
\eea
which solves the EOM, and for which the boundary condition reduces to
\bea
\label{SIGMAT}
\frac{\sigma_T}\gamma =\gamma M\frac{v^2}r
\eea
The end-point velocity is $v=\omega r$ and 
the Lorentz contraction factor is fixed by $1/\gamma^2=1-v^2$. In (\ref{SIGMAT}) the tension in the string balances the centrifugation, with the correct relativistic gamma factors.

From the Noether currents following from Poincare symmetry, the energy is
\bea
\label{P0}
P^0=2\gamma M+\frac{\sigma_T 2r}{v}\int_0^v\frac{dx}{\sqrt{1-x^2}}-\frac 12
(\sigma_1^z+\sigma_2^z)\gamma^2v^2\omega\nonumber\\
\eea
and angular  momentum
\bea
\label{JZ}
J^z=2\gamma Mr^2+\frac {2\sigma_T \omega}{v^3}
\int_0^v\frac {x^2\,dx}{\sqrt{1-x^2}}
+\frac 12 (\sigma_1^z+\sigma_2^z)\nonumber\\
\eea
 The spin contribution to the rotating string can be recast using the boundary constraint (\ref{SIGMAT})
 \bea
 \label{VTSL}
 V_{TSL}(r)=-\frac 12
(\sigma_1^z+\sigma_2^z)\gamma^2v^2\omega
=-\frac{1}2\frac {S^zL^z}{M^2 r} \frac{dV_C(2r)}{d2r}\nonumber\\
\eea
with $L^z=mvr$ and $S^z=\frac 12(\sigma_1^z+\sigma_2^z)$ and the
confining part of the central potential 
$V_C(r)=\sigma_T r$. 
It is the scalar spin-orbit coupling 
arising from Thomas precession, the resultant of
successive  boosts along the locus of the massive end-points with attached spins. This contribution from the confining string, was originally noted in~\cite{Buchmuller:1981fr,Gromes:1984ma,Pisarski:1987jb}.

We now note that the central contribution from the NG string, receives quantum corrections leading to the Arvis potential~\cite{Arvis:1983fp}
\bea
\label{ARVIS}
V_C(r)=\bigg(\sigma_T^2r^2-\frac{D_\perp}{24\alpha'}\bigg)^{\frac 12}=\sigma_T(r^2-r_c^2)^{\frac 12}
\approx\sigma_Tr-\frac{\pi D_\perp}{24 r}-\frac {\pi^2}{2\sigma_Tr^3}\bigg(\frac{D_\perp}{24}\bigg)^{\frac 12}
\eea
with $D_\perp=2$, $\alpha'=l_S^2=1/2\pi\sigma_T$ and the critical string length
$$r_c=\pi\bigg(\frac{\alpha'D_\perp}{6}\bigg)^{\frac 12}\rightarrow \frac{\pi l_S}{\sqrt 3}$$
The large distance expansion yields the linear confining part plus the universal Luscher term~\cite{Luscher:2010ik}. All the three contributions in (\ref{ARVIS}) are well reproduced by the high precision QCD analysis of the inter-quark potential. The Arvis potential shows that the string contribution no longer applies for $r<r_c\approx l_S$. 

If we regard the Arvis potential 
as a renormalized string tension 
$\sigma_T r\rightarrow \sigma_T(r)r$ that vanishes as $\sigma_T(r_c)=0$,
then its large distance suggests that the Thomas spin-orbit contribution in (\ref{VTSL}) receives a Luscher correction, 
 \bea
 \label{VTSLX}
 V_{TSL}(r)
\approx -\frac{1}2\frac {S\cdot L}{M^2 r }\bigg(\sigma_T+\frac {\pi D_\perp}{24r^2} \bigg)
\eea
which is significant at intermediate distances, and similar to the one following from one gluon exchange
\bea
\label{ONESL}
V_{GSL}(r)=\frac{1}2\frac {S\cdot L}{M^2 r }\frac{dV_{G}(r)}{dr}
\approx -\frac{1}2\frac {S\cdot L}{M^2 r }\frac{C_{12}\alpha_s}{r^2}
\nonumber\\
\eea
with the color factor 
$$C_{12}=\left<T_1^AT_2^A\right>= -\frac 43\, ({\rm meson}), 
-\frac 23\, (\rm baryon)$$

The remarkable phenomenological success of the Isgur and Karl analysis of the  even and odd parity baryonic spectra~\cite{Isgur:1978xj}, made use of only the one-gluon exchange spin-spin and tensor interactions. The spin-orbit interaction (\ref{ONESL}) was deemed too strong, and altogether omitted. We have recently confirmed their observations~\cite{Miesch:2023hvl}. This spin-orbit puzzle has been with us for decades.

We now suggest that the Luscher  contribution to the  spin-orbit emerging
from the string through Thomas precession (\ref{VTSLX}), may solve part of this puzzle. For $D_\perp=2$, it is about the right magnitude but opposite sign, to balance the vector spin-orbit interaction originating from one gluon exchange,
\bea
C_{12}\alpha_s +\frac{\pi}{12}\approx 0
\eea
for a Coulomb coupling 
$$\alpha_s\approx \frac \pi{16}\,\, \, ({\rm meson}),\,\,\, 
\frac \pi 8\,\,\, (\rm baryon)$$
The remaining long range
scalar spin-orbit interaction  in (\ref{VTSLX}),  proportional to the string tension, 
is also partly compensated by the 
vector spin-orbit interaction induced
by instantons in the intermediate range~\cite{Shuryak:2021hng}. These two observations explain why the spin-orbit interaction is absent in addressing
the hadronic spectroscopy~\cite{Isgur:1978xj,Miesch:2023hvl}, solving the puzzle.

\end{subappendices}

\subsection{QCD strings on the light front}


\section{Confinement in the basic relativistic meson problem}

Confinement-one of the defining nonperturbative features of QCD-must be handled relativistically, even for constituent quarks traditionally treated by semiclassical potential models.
Before turning to light-front quantization, it is useful to review the simplest relativistic model of meson confinement: two quarks of mass $m_Q$ connected by a classical string with tension $\sigma_T$. This setting captures the essential physics of linear confinement and the emergence of Regge trajectories.

\section{Confinement and rotating strings}
Experimentally, excited mesons lie on nearly linear Regge trajectories,
\[
J = a_M + \alpha' M_J^2,\qquad 
\alpha'=\frac{1}{2\pi\sigma_T},
\]
with small deviations for heavy quark systems. The classical rotating string with massive endpoints reproduces this structure accurately~\cite{Sonnenschein:2014jwa}.  The empirical linearity of Regge trajectories for both light and heavy mesons motivates the use of the Nambu-Goto string supplemented with massive quark endpoints.

Working in the rest frame, the quark-string system is represented by a classical Hamiltonian derived from the Nambu-Goto action. For equal-mass quarks separated by $r$ and rotating with angular momentum $J$, the Hamiltonian reads
\begin{equation}
H(r,p_r,J)
= 2\sqrt{p_r^2 + m_Q^2 + \frac{J^2}{4r^2}}
+ \sigma_T r ,
\label{H_rel}
\end{equation}
where $p_r$ is the radial momentum. The term $\sigma_T r$ represents the energy stored in the stretched flux tube between the quarks.
For a fixed total energy $E$, the radial momentum takes the form
\[
p_r^2(E,r)
=
\frac14(E - \sigma_T r)^2
- m_Q^2
- \frac{J^2}{4r^2},
\]
and the classical turning points $r_\pm$ are defined by $p_r(r_\pm)=0$.


The spectrum follows from the WKB quantization rule
\begin{equation}
\int_{r_-}^{r_+} dr\, p_r(E,J)
=
\pi\left(n + \frac12\right),
\label{WKB_condition}
\end{equation}
which encodes the relativistic dynamics exactly at the semiclassical level. Unlike the nonrelativistic case, the integral cannot be inverted analytically except in special limits. Numerical inversion yields the full radial spectrum $M_{n,J}$.
In the limit of light quarks, an analytic approximation emerges:
\[
M_{n,J}^2
\approx
4\pi\sigma_T \left(n + \frac{J}{2}\right),
\]
which demonstrates simultaneous linear radial and orbital Regge trajectories. The combination $\frac{2m_Q}/{E_n}$
controls mass corrections. These corrections distort the low-lying part of the trajectory while leaving the high-$n$ behavior essentially unchanged. This supports the conclusion that the rotating relativistic string with massive endpoints provides a realistic description of the observed meson spectrum.

\subsection{Confinement for radial strings}

For one half of the quark-string system the classical relativistic energy reads
\[
E = \sigma_T r + \sqrt{p^2 + m_Q^2},
\qquad (E-\sigma_T r)^2 = p^2 + m_Q^2 .
\]
Quantization of this Klein-Gordon-type relation is subtle: for strictly linear $V(r)=\sigma_T r$ the spectrum suffers from the usual Klein paradox and cannot be defined nonperturbatively in terms of fixed particle number. We therefore use the semiclassical WKB approximation, which correctly captures the high-lying spectrum.

The momentum vanishes at a turning point $r=r_*$, and in the massless limit ($m_Q=0$) $E=\sigma_T r_*$. For $r<r_*$ the motion is classically allowed, and the Bohr-Sommerfeld rule gives
\[
n+\frac12 = \frac{2}{\pi}\int_0^{r_*} dr\, p(r,E_n).
\]
For $J=0$ and $V=\sigma_T r$ the integral can be carried out analytically, yielding
\[
n = a_M + \alpha' E_n^2
\bigg[\sqrt{1-b^2}
+ b^2 \ln\!\left(\frac{1-\sqrt{1-b^2}}{b}\right)\bigg],
\qquad 
b=\frac{2m_Q}{E_n}.
\]
The quantum shift $a_M$ is not universal; following standard WKB practice we take $a_M=1/2$, which is known to reproduce the light vector meson system satisfactorily. Other channels experience small additional shifts, plausibly due to spin interactions discussed later.

For arbitrary $m_Q$, this relation cannot be inverted analytically; numerical inversion gives $M_n^2$ for given $n$. For light quarks with $m_Q\simeq 0.35$\,GeV and $\sigma_T=(0.42\text{ GeV})^2$, the resulting radial trajectory is nearly linear in $M_n^2$, closely following the asymptotic massless prediction
\[
M_n^2 \approx \frac{n+\frac12}{\alpha'}.
\]
Heavier quarks introduce curvature, as expected. Comparison with the experimental $\omega$ and $\omega_3$ trajectories shows good agreement: the extracted $\sigma_T$ matches that from heavy quarkonia, and the near-constant vertical shift between the two trajectories is consistent with two units of orbital excitation. The approximate relation $n+l=\text{const}$ often used phenomenologically works reasonably well for light quarks but is not expected to be exact for massive ones.

\section{Confinement on the light front}
\label{confinement}

Light-front quantization presents well-known challenges: boosting wave functions from the rest frame mixes Hamiltonian and momentum operators. Nevertheless, boost-invariant observables ( such as $M_n^2$ and transverse momentum scales)  can be compared directly between formulations.

\subsection{Confinement in 1+1 dimensions}

Chiral symmetry breaking generates an effective constituent mass $m_Q\approx0.35$\,GeV. In equal-time quantization the free Hamiltonian is
\[
P^0=\sqrt{\vec p_q^{\,2}+m_Q^2}+\sqrt{\vec p_{\bar q}^{\,2}+m_Q^2}.
\]
In light-front gauge $x^+=\tau$ the on-shell relation $2p^+p^-=m_Q^2$ yields the free light-front Hamiltonian,
\[
P^-=\frac{m_Q^2+p_{\perp q}^2}{2p_q^+}+
\frac{m_Q^2+p_{\perp\bar q}^2}{2p_{\bar q}^+}.
\]

Confinement arises from the QCD string with tension 
$\sigma_T\approx(420\text{ MeV})^2$. In $1+1$ dimensions the Nambu-Goto action reduces to~\cite{Bars:1976nk,Bars:1976re}
\[
P^-=\frac{m_Q^2}{2p_q^+}+\frac{m_Q^2}{2p_{\bar q}^+}
+\sigma_T |x_{\bar q}^- - x_q^-|.
\]
Introducing the CM and relative coordinates and momenta gives the meson mass operator
\[
M^2 = \frac{m_Q^2}{x(1-x)} + 2\sigma_T\,| i\,\partial_x |,
\]
with $x=p_q^+/P^+$ the Bjorken variable. Diagonalizing this operator yields the light-front wave functions $\varphi_n(x)$:
\[
\left(\frac{m_Q^2}{x(1-x)}+2\sigma_T|i\partial_x|\right)\varphi_n(x)=M_n^2\varphi_n(x).
\]

Using the Fourier representation of $|q|$, this becomes the 't~Hooft equation,
\[
M_n^2\varphi_n(x)=
\frac{m_Q^2}{x(1-x)}\varphi_n(x)
-\frac{2\sigma_T}{\pi}\, {\rm PV}\!\int_0^1 dy\,
\frac{\varphi_n(y)-\varphi_n(x)}{(x-y)^2}.
\]
The nonlocal kernel contains both a confining interaction and an induced self-energy term. Writing the equation as
\[
M_n^2\varphi_n(x)=
\frac{m_Q^2-2\sigma_T/\pi}{x(1-x)}\varphi_n(x)
-\frac{2\sigma_T}{\pi}\,{\rm PV}\!\int_0^1 dy\,
\frac{\varphi_n(y)}{(x-y)^2},
\]
one sees that the negative induced self-energy cancels the positive string term for $m_Q\to0$, giving a massless mode $\varphi_0(x)$ in the large-$N_c$ limit, analog of a Nambu-Goldstone boson.

A semiclassical WKB treatment of the 't~Hooft equation produces the Regge-like spectrum
\[
M_n^2 - m_Q^2\ln\!\left(\frac{x_+\bar x_-}{x_- \bar x_+}\right)
= 2\pi\sigma_T n,
\]
with turning points
\[
x_\pm=\frac12\left(1\pm\sqrt{1-\frac{4m_Q^2}{M_n^2}}\right),
\qquad M_n\ge2m_Q.
\]
For $m_Q\to0$ this reduces to the open-string form $M_n^2=n/\alpha'$, while for large $n$,
\[
\varphi_n(x)\approx \sqrt2\,\sin((n+1)\pi x),
\qquad 
M_n^2\simeq \frac{n}{\alpha'}+2m_Q^2\ln n .
\]

 \begin{figure}[h!]
	\centering
		\includegraphics[width=7cm]{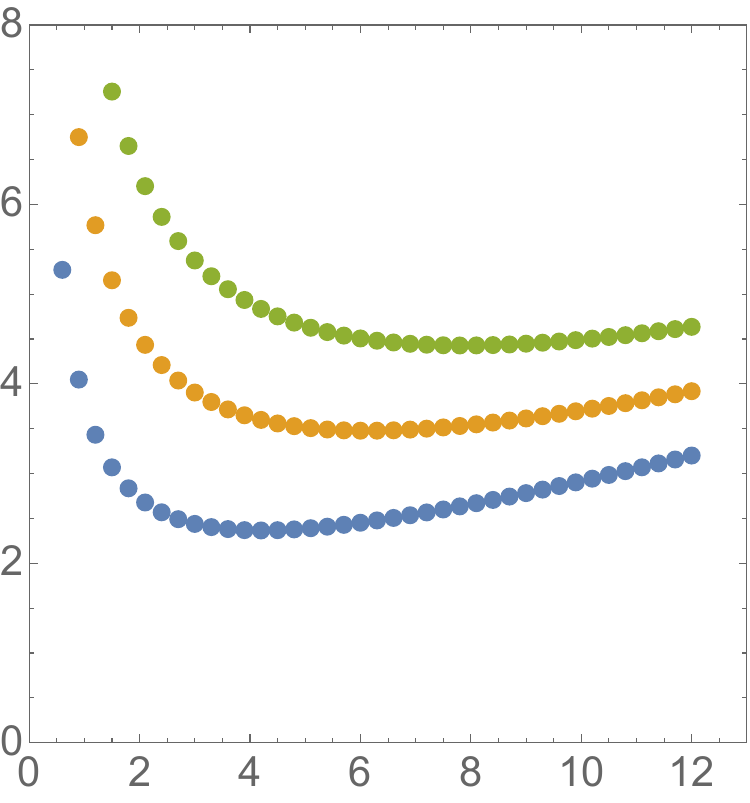}\\
	\includegraphics[width=7cm]{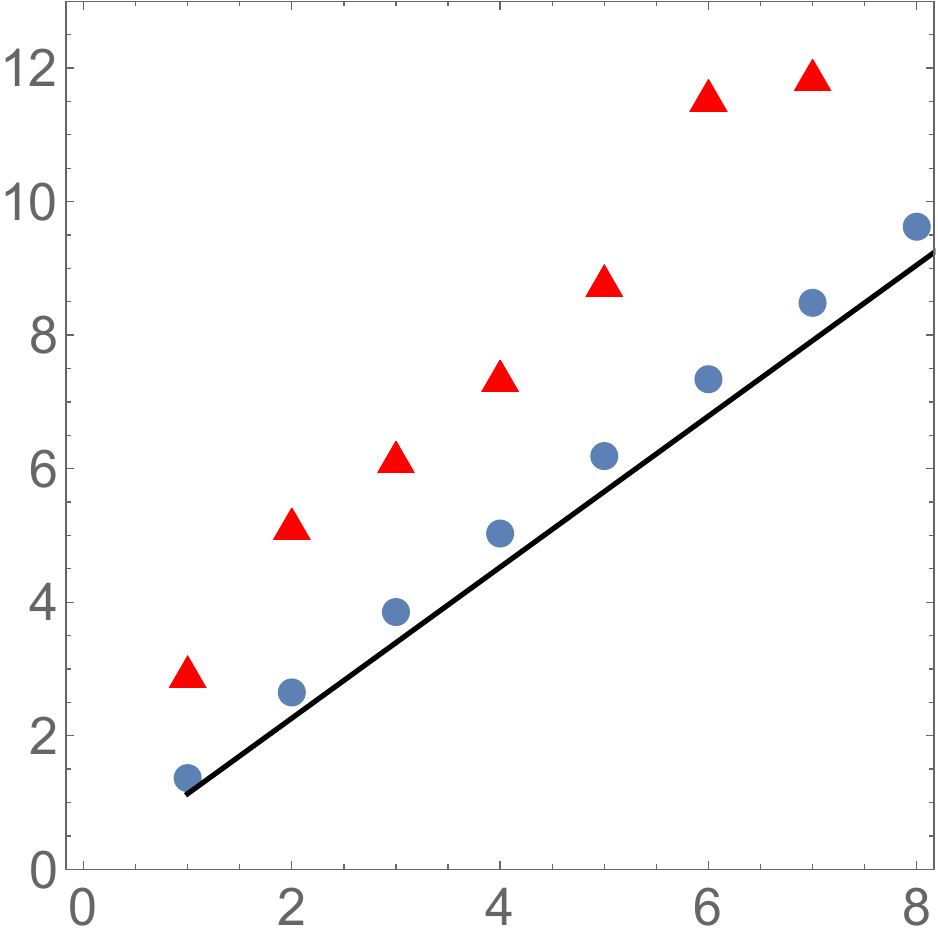}
	\caption{The upper plot shows the dependence of the three lowest eigenvalues $M^2_n$ on the parameter $a$,
	in  the region around their mimima. The lower plot shows  $M^2_{n+1}$ versus $n+1=1..7$.
	The results obtained from the Hamiltonian diagonalization with fixed $a=2.36$, are shown by red triangles. The blue disks are  the
	semiclassical results discussed previously, the line is a simple linear expression $M^2_n=n/\alpha'$, both shown for comparison. }
	\label{fig:comparem2}
\end{figure}

\begin{figure}[htbp]
\begin{center}
\includegraphics[width=6cm]{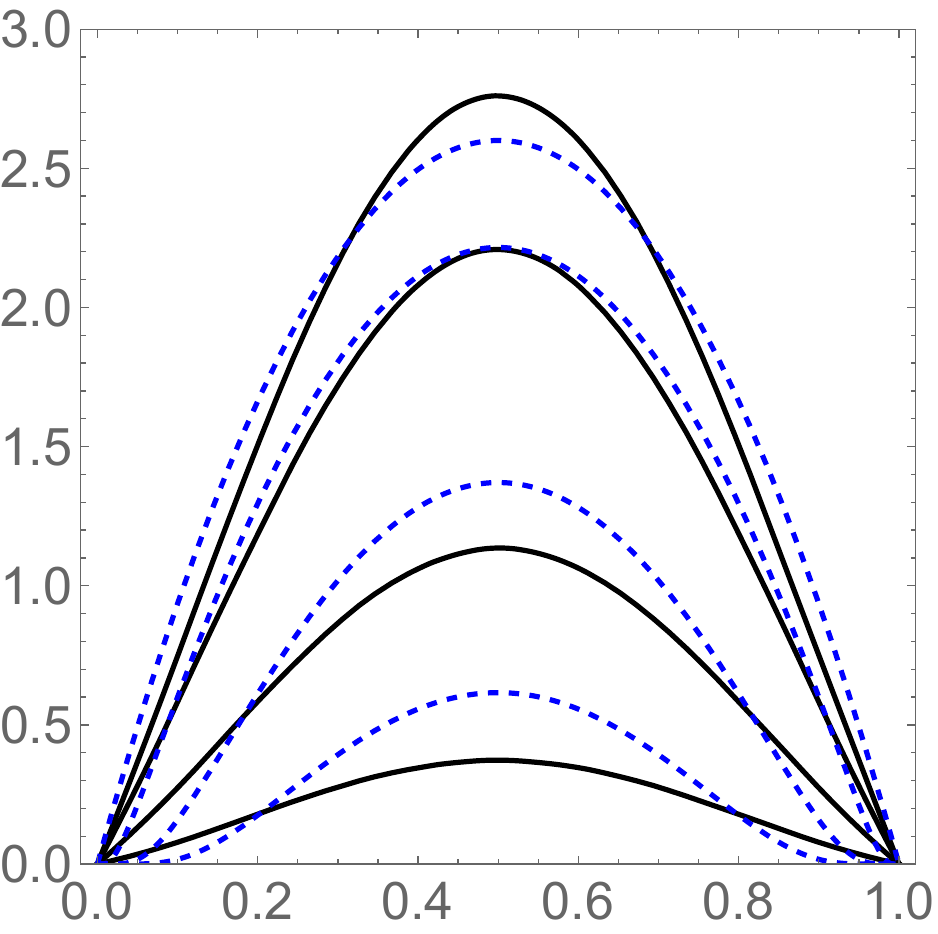}
\includegraphics[width=6cm]{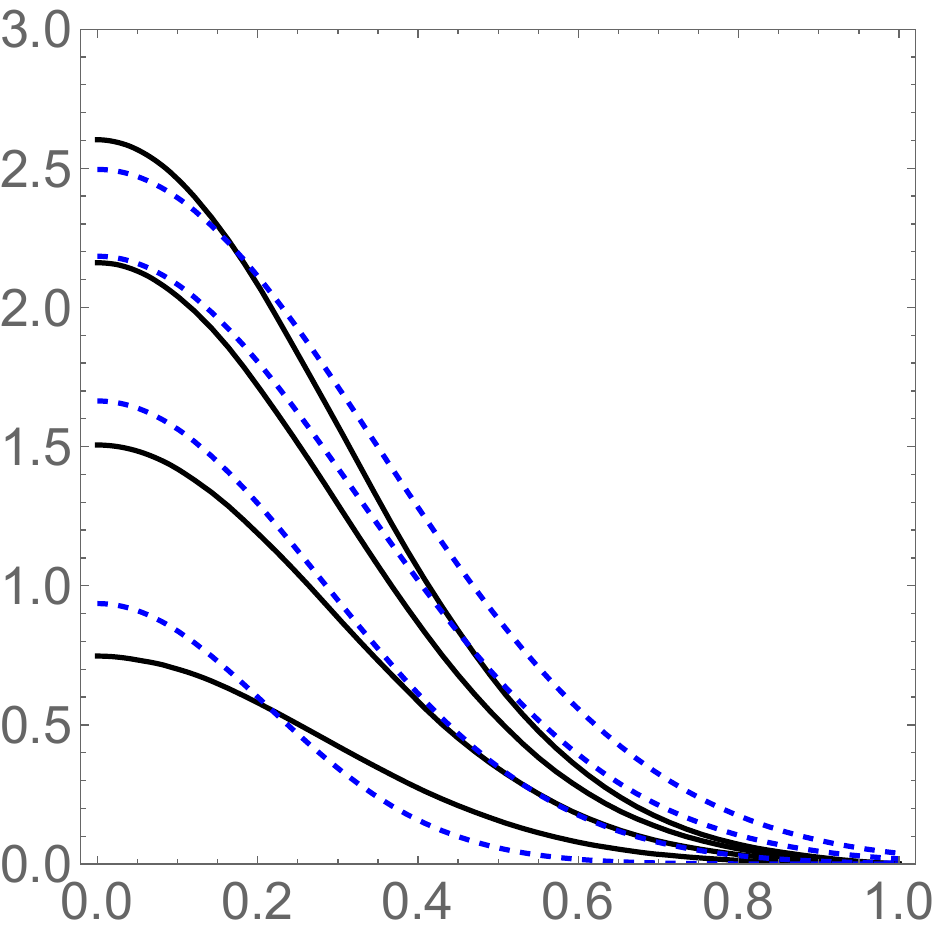}
\caption{The left plot shows the dependence of the ground state wave function $\Psi_0(x,p_\perp)$ versus
$x$ ar $p_\perp^2=0,0.2,0.4,0.6 \, GeV^2$, top to bottom curves. The right plot is versus $p_\perp$ at  $x=0.1,0,2,0.3,0,4$. All solid curves are from the
exact solution, while all dashed lines are for  the simplified  form 
}
\label{fig_ground}
\end{center}
\end{figure}

\subsection{Eliminating the Square Root in the Light-Front Hamiltonian}
To sidestep the nonlocal square root in the light-front Hamiltonian, we introduce an einbein parameter and define
\begin{equation}
M^2(a,b)=\frac{m_Q^2+k_\perp^2}{x(1-x)}+\sigma_T\!\left(\frac{|i\partial_x|^2+b\,x_\perp^2}{a}+a\right),
\end{equation}
with $a=1/e$ and $b$ tuned self-consistently so that $b\to M_{nl}^2$. Minimizing in $a$ after diagonalization recovers the original form.
To proceed, we decompose
\[
M^2 = H_0 + V,
\]
where
\begin{align}
H_0&= \frac{\sigma_T}{a}\left(-\partial_x^2 - b\,\partial_{k_\perp}^2\right) + \sigma_T a + 4(m_Q^2+k_\perp^2),\qquad
V = (m_Q^2+k_\perp^2)\left(\frac{1}{x(1-x)} -4 \right).
\end{align}
With $b=(2m_Q)^2$, diagonalization of a truncated basis reveals shallow minima of $M_n^2(a)$; In~\cite{Shuryak:2019zhv} the value  $a=2.36$ yields a spectrum consistent with the expected Regge slope.

Using these fixed parameters, the lowest LFWF is well approximated by
\begin{equation}
\psi_1(\rho,x)\approx e^{-\beta^2\rho^2/2}\,\sin(\pi x),\qquad
\beta=\left(\frac{4a}{\sigma_T b}\right)^{1/4},
\end{equation}
with higher harmonics suppressed at the percent level. The transverse width is
\[
\langle p_\perp^2\rangle = \frac{1}{\beta^2}=\sqrt{\frac{\sigma_T b}{4a}}.
\]
A full numerical solution in $(x,p_x,p_y)$ yields $M^2\simeq 2.58\,\text{GeV}^2$ and a wave function matching the analytic form to high precision. A comparison to the common ansatz
\[
\Psi_0(x,\xi)=4C_2\,x(1-x)\,e^{-C_1\xi^2},\qquad
\xi^2=\frac{p_\perp^2}{x(1-x)},
\]
shows qualitative agreement but endpoint deviations.

\chapter{Light quarks at the light front}

\section{Effective "constituent" quark mass and chiral condensate}
\label{SEC_ILMGAP}

Let us start this part, devoted to LF models, with discussion
of the  chiral condensate (\ref{CONDZ})  and constituent quark mass (\ref{MCONST}). The reader may think this discussion 
is unnecessary, as are both frame invariant scalars. We however think that  evaluating them in both frames
(in fact, in the mean-field or large $N_c$ approximation)
would be still instructive\footnote{Difficulties in deriving chiral symmetry breaking in LF formulation are well known in literature. 
An extreme point of view expressed in \cite{Brodsky:2009zd} was that QCD vacuum is $empty$ by itself, with nonperturbative phenomena ocuring only
inside hadrons. We, on the other hand, treat LF not as a
basis for fundamental theory, but simply a limit, from 
frame at rest to the large momentum limit. Thus, all Lorentz scalar must simply be preserved in that limit. }.

Discussion of the chiral condensate are needed, as they would directly lead to the LF description of the pion and sigma mesons. 

The light-front (LF) formulation of QCD provides a Hamiltonian description of hadronic structure in which observables are defined at fixed light-front time $x^+=x^0+x^3$~\cite{BrodskyPauliPinsky1998} (and references therein). This representation diagonalizes the boosts along the $z$-axis, making it particularly suited to relativistic bound-state problems. However, the apparent simplicity of the LF vacuum contrasts strongly with the complex instant-form picture, where the ground state hosts condensates, instantons, and other nonperturbative excitations.

The resolution, as discussed in Ref.~\cite{LiuShuryakZahed2023}, is that LF dynamics encodes the effects of the vacuum through {\it zero modes}, i.e.\ field components with longitudinal momentum $p^+=0$. These modes are responsible for reproducing chiral symmetry breaking, condensates, and anomalies, despite the vacuum apparent emptiness. The purpose of this chapter is to present a detailed  explanation of how these zero modes generate the effective vacuum condensate $\vev{\bar q q}_\LF$, and ensure equivalence with the instant-form theory.

\section{Constituent Quark Mass and Chiral Condensate on the Light Front}

In the preceding sections we have constructed the effective multi-fermion interactions emerging from the instanton liquid vacuum, projected them onto the light-front, eliminated the constrained ("bad'') fermion components, and derived a light-front Hamiltonian expressed solely in terms of the dynamical ("good'') fermion degrees of freedom. We now turn to show that this light-front Hamiltonian yields the same nonperturbative signatures of chiral symmetry breaking (namely a constituent quark mass and a nonvanishing chiral condensate) as found in the rest frame. In doing so, we demonstrate the frame-independence (boost invariance) of these emergent quantities, and thereby cement the physical equivalence between the instanton vacuum description in the rest frame and its light-front incarnation.


The starting point is the rest-frame ( Euclidean signature ) gap equation for the dynamical quark mass. Denoting by \(m\) the current (bare) quark mass, by \(g_S\) the effective 't Hooft coupling normalized by the number of colors, and by \(F(k)\) the form factor arising from the zero-mode structure of the instanton ensemble, one obtains \cite{Liu:2023feu,Liu:2023fpj} 
\[
M(k) = m + 2 \, g_S \, \int \frac{d^4 q}{(2\pi)^4} \; \frac{4 \, M(q)}{q^2 + M^2(q)} \; F(q) \,.
\]  
In the limit of low momenta (\(k \rho \ll 1\)), the running mass becomes approximately constant, \(M(k) \approx M(0) \equiv M\). This nonperturbative constituent mass represents the dynamical breaking of chiral symmetry, induced by the nonlocal multi-fermion interactions mediated by instantons. In that way, the vacuum generates a "mass gap'' even when the current mass \(m\) is small (or vanishing), characteristic of spontaneous chiral symmetry breaking.


On the light-front, after elimination of the bad components of the fermion fields in the mean-field approximation, only the good components, \(\psi_+\), remain dynamical. The elimination process renormalizes the multi-fermion interactions and yields a dynamical constituent mass function \(M(k^2)\) for the good-component quarks. The effective light-front quark propagator takes the form  
\[
S(k) \; \to\; \frac{i\bigl[\,\!\slashed{k} + M(k^2)\bigr]}{k^2 - M^2(k^2)} \; - \; \frac{i \, \gamma^+}{2 k^+} \,,
\]  
where the second term arises from the constrained structure of the original fermion field. Using this propagator, one computes the scalar condensate on the light-front as  
\[
\langle \bar \psi \psi \rangle \;=\; -2 \, N_c \, M \; \int \frac{d k^+ \, d^2 k_\perp}{(2\pi)^3} \; \frac{\epsilon(k^-)\,F(k^-)}{k^+}\bigg|_{k^-\frac {k_\perp^2+M^2}{2k^+}} \,,
\]  
where \(\epsilon(k^-)\) implements the light-front energy cutoff in the vacuum tadpole loops. In practice, with a suitable boost-invariant sharp cutoff, the integral yields a nonzero condensate, indicating that the vacuum structure on the light front is nontrivial, contrary to naive expectations of a "trivial LF vacuum.''


A non-trivial and central result of the analysis is that the values of both the constituent mass \(M\) and the chiral condensate computed on the light front are identical to those obtained in the rest frame, within the mean-field approximation the authors employ. This equivalence follows from the fact that both quantities are Lorentz scalars (frame-independent), and that the form factors arising from the instanton zero modes (which encode the nonlocality in the interactions) are handled carefully so as to respect boost invariance. Thus the breaking of chiral symmetry, manifested in a finite mass gap and a nonzero condensate in the rest-frame approach, survives intact in the light-front description.


This equivalence has important physical implications. First, it dispels the common lore that the light-front vacuum must be trivial: rather, the nonperturbative vacuum dynamics (here encoded in instanton-induced multi-fermion interactions) are still present and generate nonzero condensates via tadpole effects of the constrained modes. Second, it establishes that partonic descriptions of hadrons built on the light-front can faithfully capture essential vacuum physics such as spontaneous chiral symmetry breaking. Consequently, light-front wavefunctions and partonic distributions derived from this formalism inherently carry nonperturbative chiral dynamics; e.g., the resulting meson wavefunctions yield the correct constituent mass and condensate values, ensuring consistency with rest-frame phenomenology.


The demonstration that the constituent quark mass and chiral condensate are the same on the light front and in the rest frame underscores the universality of vacuum-induced chiral symmetry breaking in QCD, independent of the quantization frame. It validates the use of the light-front Hamiltonian  derived from the instanton liquid model (ILM),  as a legitimate starting point for nonperturbative hadron structure. In particular, since this construction is based only on the emergent 't Hooft interactions from the ILM (with form factors determined by the quark zero modes), it brings a concrete and controlled realization of chiral symmetry breaking into the light-front formalism. This sets the stage for subsequent analyses (as in later sections) of bound states such as pions and kaons, their light-front wavefunctions, distribution amplitudes and parton distributions --  all grounded in the same nonperturbative vacuum physics that yields the constituent mass and condensate.

\bigskip

\noindent In summary, the core message that the nonperturbative vacuum structure of QCD, realized via instantons, survives the light-front quantization: the emergent constituent quark mass and scalar condensate remain nonzero and coincide with their rest-frame values. This result justifies the use of the light-front formalism for realistic hadron phenomenology in a framework that fully accounts for spontaneous chiral symmetry breaking.

\begin{figure}
         \centering
         \includegraphics[scale=0.49]{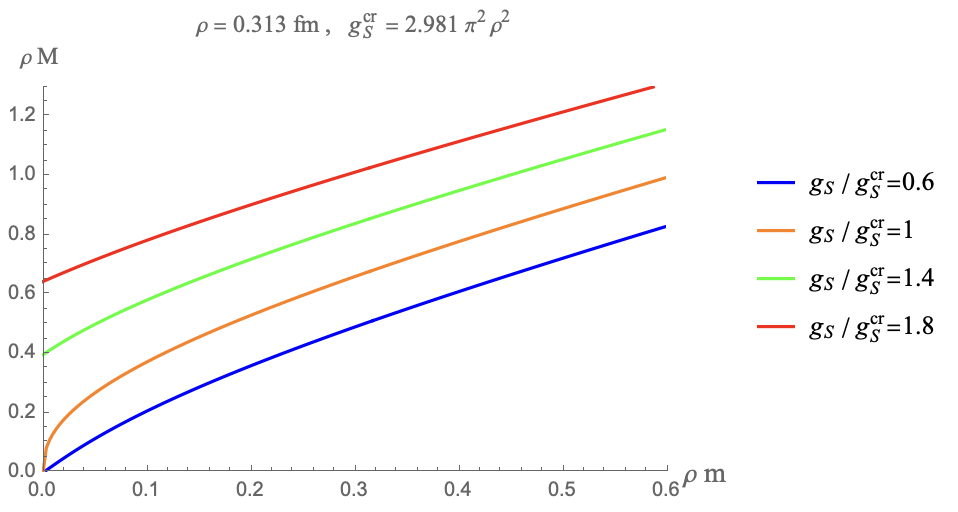}
         \caption{Constituent mass as a function of the current mass with different scalar couplings $g_S$, for a fixed instanton size $\rho=0.31$ fm.}
         \label{FIG1X}
\end{figure}
\begin{figure}
         \centering
         \includegraphics[scale=0.49]{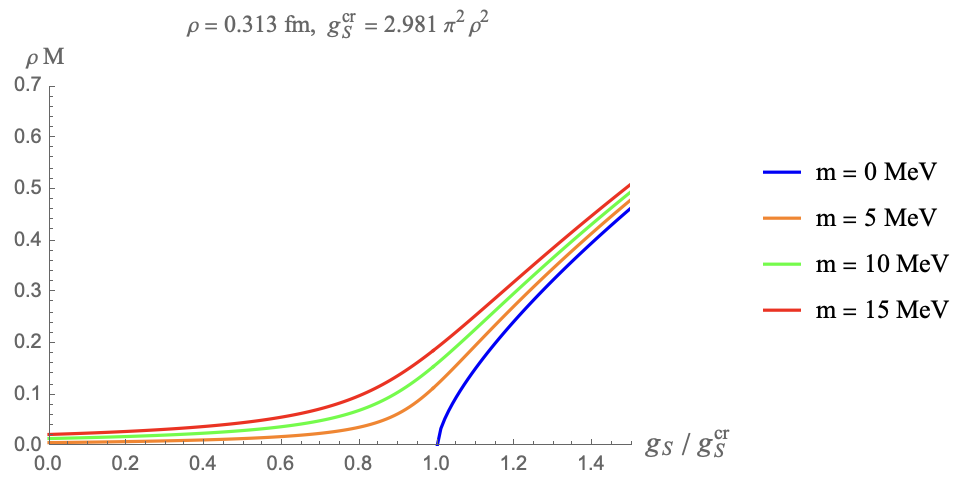}
          \caption{Constituent quark mass versus the scalar coupling $g_S$.}
          \label{FIG2X}
\end{figure}
\begin{figure}
    \centering
    \includegraphics[scale=0.48]{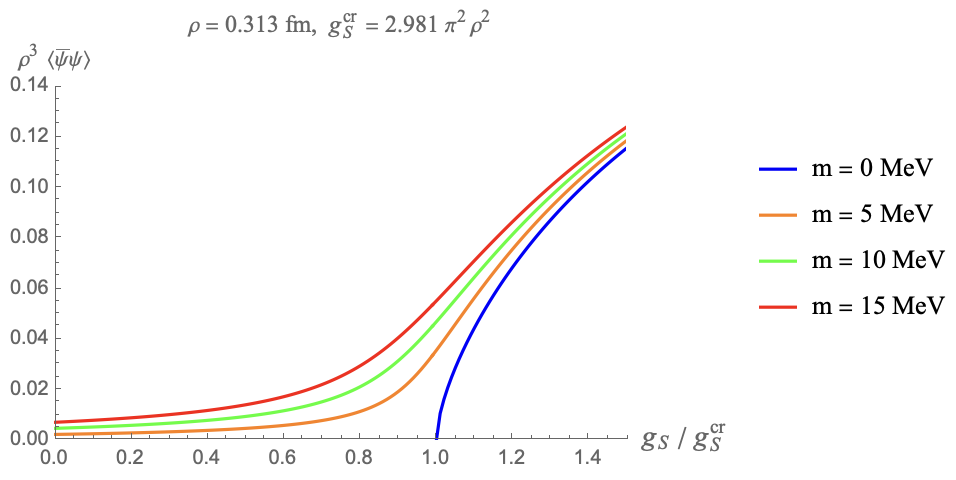}
    \caption{Quark condensate as a function of the scalar coupling $g_S$.}
    \label{FIG3X}
\end{figure}

 \section{LF sigma,pion wavefunctions vs those for "generic" mesons}
 \label{SEC_ILMLFS}
Now, we can use the light front Hamiltonian (\ref{LFHamiltonian}) to solve the eigenvalue equation for the LF meson wave functions, in the mean-field or large $N_c$ limit, 
\begin{equation}
\label{PMINUS}
    P^-|X,P\rangle=\frac{m_X^2}{2P^+}|X,P\rangle
\end{equation}
When restricted to the lowest Fock component, the meson LF wavefunctions are of the form 

\begin{equation}
\label{Meson_bound_state}
    |\mathrm{Meson} ~X,P\rangle=\frac{1}{\sqrt{N_c}}\int_0^1 \frac{dx}{\sqrt{2x\bar{x}}}\int\frac{d^2k_\perp}{(2\pi)^3}\sum_{s_1,s_2}\Phi_X(x,k_\perp,s_1,s_2)b^\dagger_{s_1}(k) c^\dagger_{s_2}(P-k)|0\rangle
\end{equation}
with the LF scattering normalization  $\langle P|P'\rangle=(2\pi)^32P^+\delta^{3}(P-P')$, or
\begin{equation}
\label{normal}
    \int_0^1dx\int\frac{d^2k_\perp}{(2\pi)^3}\sum_{s_1,s_2}\left|\Phi_X(x,k_\perp,s_1,s_2)\right|^2=1
\end{equation}
In the sigma and pion channels, with the valence quark assignments
\bea
    \sigma=\frac{1}{\sqrt{2}}(u\bar{u}+d\bar{d})\qquad 
    \pi^{\pm,0}=u\bar{d},\ d\bar{u},\ \frac{1}{\sqrt{2}}(u\bar{u}-d\bar{d})
\eea
the LF vertices  $\Phi_X$ (\ref{Meson_bound_state})  in the 2-particle Fock space take a simple form
\bea
\label{PHIX}
    \Phi_\sigma(x,k_\perp,s_1,s_2)&=&\phi_\sigma(x,k_\perp)\bar{u}_{s_1}(k)\frac{1}{\sqrt{2}}v_{s_2}(P-k)\nonumber\\
    \Phi_\pi^a(x,k_\perp,s_1,s_2)&=&\phi_\pi(x,k_\perp)\bar{u}_{s_1}(k)\frac{\tau^a}{\sqrt{2}}i\gamma^5v_{s_2}(P-k)
\eea
The  spin-flavor part is carried by the LF spinors  $u_s(k)$ (particle) 
and $v_s(k)$ (anti-particle), with $\phi_X$ a scalar-isoscalar wavefunction at the origin of 
the mesonic distribution amplitude (see below).
Inserting (\ref{PHIX}) in (\ref{Meson_bound_state}) and then in (\ref{PMINUS}), and unwinding the various contractions from the
4-Fermi interaction terms, yield the boost invariant eigenvalue equation
\begin{equation}
\label{eq:bound_state}
\begin{aligned}
        &m_X^2\Phi_X(x,k_\perp,s_1,s_2)=\frac{k_\perp^2+M^2}{x\bar{x}}\Phi_X(x,k_\perp,s_1,s_2)\\
        &+\frac{1}{\sqrt{2x\bar{x}}}\sqrt{\mathcal{F}(k)\mathcal{F}(P-k)}\int_0^1 \frac{dy}{\sqrt{2y\bar{y}}}\int\frac{d^2q_\perp}{(2\pi)^3}\sum_{s,s'}\mathcal{V}_{s_1,s_2,s,s'}(k,P-k,q,P-q)\Phi_X(y,q_\perp,s,s')\sqrt{\mathcal{F}(q)\mathcal{F}(P-q)}
\end{aligned}
\end{equation}
For the sigma and pion channels, the interaction kernels after contraction on the LF, are
\begin{equation}
\begin{aligned}
        &\sum_{s,s'}\mathcal{V}_{s,s',s_1,s_2}(q,q',k,k')\Phi_{\sigma}(y,q_\perp,s,s')\\
        =&-g_S\alpha_+(P^+)\mathrm{Tr}\left[(\slashed{q}+M)(\slashed{q'}-M)\right]\phi_{\sigma}(y,q_\perp)\bar{u}_{s_1}(k)\mathbf{1}\mathrm{tr}(\mathbf{1})v_{s_2}(k')\\
        =&-4g_S\alpha_+(P^+)\left(\frac{q_\perp^2+(y-\bar{y})^2M^2}{y\bar{y}}\right)\phi_{\sigma}(y,q_\perp)\bar{u}_{s_1}(k)v_{s_2}(k')
\end{aligned}
\end{equation}
\begin{equation}
\begin{aligned}
    &\sum_{s,s'}\mathcal{V}_{s,s',s_1,s_2}(q,q',k,k')\Phi^a_{\pi}(y,q_\perp,s,s')\\
    =&-g_S\alpha_+(P^+)\mathrm{Tr}\left[(\slashed{q}+M)(\slashed{q'}+M)\right]\phi_{\pi}(y,q_\perp)\bar{u}_{s_1}(k)i\gamma^5\tau^b\mathrm{tr}(\tau^a\tau^b)v_{s_2}(k')\\
    =&-4g_S\alpha_+(P^+)\left(\frac{q_\perp^2+M^2}{y\bar{y}}\right)\phi_{\pi}(y,q_\perp)\bar{u}_{s_1}(k)i\gamma^5\tau^av_{s_2}(k')
\end{aligned}
\end{equation}
The corresponding eigenvalue equations for the scalar-isoscalar wavefunctions $\phi_X$ are
\begin{equation}
\label{1X}
\begin{aligned}
        m_\sigma^2\phi_{\sigma}(x,k_\perp)=&\frac{k^2_\perp+M^2}{x\bar{x}}\phi_{\sigma}(x,k_\perp)\\
        &-\frac{2g_S\alpha_+(P^+)}{\sqrt{x\bar{x}}}\sqrt{\mathcal{F}(k)\mathcal{F}(P-k)}\int \frac{dy}{\sqrt{y\bar{y}}} \int\frac{d^2q_\perp}{(2\pi)^3}\left(\frac{q_\perp^2+(y-\bar{y})^2M^2}{y\bar{y}}\right)\phi_\sigma(y,q_\perp)\sqrt{\mathcal{F}(q)\mathcal{F}(P-q)}
\end{aligned}
\end{equation}
\begin{equation}
\label{4X}
\begin{aligned}
        m_{\pi}^2\phi_{\pi}(x,k_\perp)=&\frac{k^2_\perp+M^2}{x\bar{x}}\phi_{\pi}(x,k_\perp)\\
        &-\frac{2g_S\alpha_+(P^+)}{\sqrt{x\bar{x}}}\sqrt{\mathcal{F}(k)\mathcal{F}(P-k)}\int \frac{dy}{\sqrt{y\bar{y}}} \int\frac{d^2q_\perp}{(2\pi)^3} \left(\frac{q_\perp^2+M^2}{y\bar{y}}\right)\phi_{\pi}(y,q_\perp)\sqrt{\mathcal{F}(q)\mathcal{F}(P-q)}
\end{aligned}
\end{equation}

The solutions are generic, 

\begin{equation}
\label{GENX}
    \phi_{X}(x,k_\perp)=\frac{1}{\sqrt{2x\bar{x}}}\frac{C_{X}}{m^2_{X}-\frac{k_\perp^2+M^2}{x\bar{x}}}\sqrt{\mathcal{F}\left(k\right)\mathcal{F}\left(P-k\right)}
\end{equation}
with $C_X$ fixed by the normalization (\ref{normal}). It 
characterises the effective quark-meson coupling $g_X$ in in a covariant formulation, i.e.
$ C_X=-\sqrt{N_c}g_X$ (the minus sign enforces spectral positivity). The pion is found to be strongly bound, and the sigma a threshold state.

\section{Other light-front Hamiltonians}

Jia and Vary~\cite{Jia:2018ary} proposed a light-front Hamiltonian treatable via basis diagonalization, in analogy with atomic and nuclear structure methods. Their Hamiltonian contains effective quark masses, longitudinal and transverse confinement, and an NJL four-fermion term (not discussed here):
\[
H = H_M + H_{\parallel} + H_\perp + H_{NJL},
\qquad
H_M=\frac{m_Q^2}{x_1}+\frac{m_{\bar Q}^2}{x_2},
\]
\[
H_{\parallel}=\frac{\kappa^4}{(m_Q+m_{\bar Q})^2}\frac{1}{J(x)}\partial_xJ(x)\partial_x,
\qquad
H_\perp = k_\perp^2\Big(\frac{1}{x_1}+\frac{1}{x_2}\Big)+\kappa^4 x_1x_2 r_\perp^2 ,
\]
where $J(x)=x_1x_2$. With equal quark masses, $(1/x_1+1/x_2)=1/(x_1x_2)$.

Because both longitudinal and transverse confinements are quadratic, the Hamiltonian resembles coupled harmonic oscillators. In the basis constructed in~\cite{Jia:2018ary}, all terms except NJL become diagonal, yielding the analytic spectrum
\[
M_{nml}^2 = (m_Q+m_{\bar Q})^2
+2\kappa^2(2n+|m|+l_L+3/2)
+\frac{\kappa^4}{(m_Q+m_{\bar Q})^2}l_L(l_L+1),
\]
where $n$ is the transverse oscillator number, $m$ the helicity, and $l_L$ the longitudinal index.

The light-front kinetic structure naturally produces the term $k_\perp^2/(x\bar x)$, which couples transverse and longitudinal motion. The basis suppresses, but cannot fully remove, this mixing. Introducing the boost-invariant variable $\zeta = k_\perp/\sqrt{x(1-x)}$ can simplify some expressions, but if performed exactly it generates additional kinetic contributions absent from the model.

Qualitatively, the spectrum grows linearly with the quantum numbers $(n,m)$, consistent with Regge phenomenology. Quantitatively, however, the recommended confinement scale $\kappa=0.227\,\mathrm{GeV}$ yields slopes far too small. Matching the observed radial and orbital Regge trajectories requires significantly larger $\kappa$ (e.g. $\sim0.5\,\mathrm{GeV}$). Spin splittings, such as the $\omega_3-\omega$ mass gap, are also underestimated by nearly an order of magnitude.

Thus, while the Jia and Vary Hamiltonian captures the structural features of light-front confinement, its parameter tuning reflects the special low-lying $\pi$ and $\rho$ states (dominated by short-range spin and residual interactions) rather than the generic dynamics of excited mesons governed by confinement.


\section{Central potential from instantons on the Light Front}

Instantons provide localized semiclassical fluctuations of the gauge field.  
In the dense instanton liquid-including both dilute instantons and correlated \( I\bar I \) molecules-one can reproduce the central potential at intermediate distances \( r\sim 0.5\,\mathrm{fm} \) together with nonperturbative spin-independent forces.  
Their role on the light front follows from the dominance of instanton-induced magnetic fields in the relevant nonlocal correlators.

To extract the instanton-induced potential on the light front, we evaluate the connected part of the light-cone Wilson loop for a \(Q\bar Q\) pair separated by a transverse distance \( b_\perp \).  
As in Euclidean \(Q\bar Q\) scattering, we assign a relative angle \( \theta \) between Wilson lines, average over the instanton ensemble of density \(N/V_4\), and analytically continue the result to the light front by \( \theta \to -i\chi \) where \( \chi \) is the rapidity difference.

More specifically, the connected loop exponentiates~\cite{Shuryak:2021hng}
\[
\langle W(\theta,0_\perp)\,W^\dagger(\theta,b_\perp)\rangle_C
=
\exp\!\left[
-\frac{2N}{2N_c V_4}
\int d^4z\;\mathrm{Tr}_c
\left(1 - W_I(\theta,0_\perp)\,W_I^\dagger(\theta,b_\perp)\right)
\right],
\]
with \(W_I\) the Wilson line evaluated in a single instanton field.  
For a sloped trajectory in singular gauge,
\[
W_I=
\cos\!\left(\pi-\frac{\pi\gamma}{\sqrt{\gamma^2+\rho^2}}\right)
-i\,\hat n^a \tau^a 
\sin\!\left(\pi-\frac{\pi\gamma}{\sqrt{\gamma^2+\rho^2}}\right),
\]
where
\[
\gamma^2 = (z_4\sin\theta - z_3\cos\theta)^2 + (z_\perp - b_\perp)^2,
\qquad 
n^a=\eta^a_{\mu\nu}\dot x^\mu (z-b)^\nu .
\]
After integrating over the instanton position, the remaining dependence is encoded in a universal cylindrical integral,
\[
\mathbf I(\xi),\qquad \xi=\frac{b_\perp}{\rho},
\]
giving the Euclidean loop
\[
\langle W W^\dagger\rangle_C
=
\exp\!\left[
-\, Z_E^+\,
\frac{4\kappa}{N_c\rho}\,
\mathbf I(\xi)
\right],
\qquad 
\kappa=\frac{N\rho^{4}}{V_4}.
\]
Analytic continuation \(Z_E^+ \to i Z_M^+\), \( \theta\to -i\chi\) yields the LF Hamiltonian contribution
\[
P_I^-=
\frac{1}{\gamma_\beta}
\left(\frac{4\kappa}{N_c\rho}\right)
\mathbf I(\xi),
\qquad
\gamma_\beta=\cosh\chi.
\]

Thus, the induced contribution to the invariant mass of the \(Q\bar Q\) system is
\[
M^2
=
\frac{k_\perp^2}{x(1-x)} + 2P^+P_I^-
\approx
\frac{k_\perp^2}{x(1-x)}
+2M\left(\frac{4\kappa}{N_c\rho}\right)\mathbf I(\xi),
\]
with mass operator
\[
M=
\frac{|k_\perp|}{\sqrt{x(1-x)}}
+
\left(\frac{4\kappa}{N_c\rho}\right)\mathbf I(\xi)
+\mathcal O(\kappa^2).
\]
For small separations,
\[
\mathbf I(\xi)\approx \alpha \xi^2,\qquad \xi\ll 1,
\]
while for large \( \xi\),
\[
\mathbf I(\xi) \approx 2\Delta m_Q + \frac{C}{\xi^p},
\qquad p\ll 1.
\]


Allowing the separation to include a longitudinal component \( b_\mu=(0,b_\perp,b_3) \) modifies the geometry of the line-instanton interaction.  
After analytic continuation and conversion to dimensionless LF variables, the result is expressed through a generalized function
\[
\mathbf H(\tilde\xi_x),
\qquad
\tilde\xi_x^2
=
\frac{1}{(M\rho)^2}\Big(\frac{d}{dx}\Big)^2
+
\frac{b_\perp^2}{\rho^2},
\]
so that
\[
M^2
\approx
\frac{k_\perp^2+m_Q^2}{x(1-x)}
+
2M\left(\frac{4\kappa}{N_c\rho}\right)
\mathbf H(\tilde\xi_x)
\equiv
\frac{k_\perp^2+m_Q^2}{x(1-x)} + 2M V_C(\xi_x).
\]
Its limiting forms are
\[
\mathbf H(\tilde\xi_x)\approx 
c_2 \tilde\xi_x^2 + c_4 \tilde\xi_x^4 + \cdots,\qquad \tilde\xi_x\ll1,
\]
and at large distances,
\[
\mathbf H(\tilde\xi_x)\approx 
\mathrm{const} + \frac{C}{\tilde\xi_x^p},\qquad p\ll 1.
\]

\subsection{Spin-dependent forces}

In the instanton vacuum, spin-dependent interactions on the light front get a bit simpler, if one is using the self-duality of the gauge fields of instantons. Because $E$ and $B$ fields satisfy $BB=EE$ and $BE=\pm EE$, all spin potentials become linked to the central potential $V_C(\xi_x)$. Their construction parallels that detailed for the central potential.
Unfortunately, fields of $I\bar I$ molecules
are not selfdual, and their derivation should follow
generic definitions.

\subsection{Spin-spin interaction}

The transverse chromoelectric correlator controls the spin-spin force. Using the rotation relating $v_\mu F_{\mu i}$ to $E_i$ and its analytic continuation~\cite{Shuryak:2021yif}
\[
M^2_{SS}
=2M\,\frac{\sigma_{1\perp i}\sigma_{2\perp j}}{4m_{Q1}m_{Q2}}
R_{im}R_{jn}\,\partial_{1m}\partial_{1n}V_C(\xi_x),
\]
showing that instanton-induced spin-spin forces depend entirely on second derivatives of $V_C$. Thus
\[
\mathbb V_3(\xi_x)=\frac{2b_\perp^2}{\xi_x^2}V_C''(\xi_x),\qquad
\mathbb V_4(\xi_x)=0.
\]

\subsection{Spin-orbit structure}

The spin-orbit potentials retain the rest-frame relation
\[
\mathbb V_2(\xi_x)=\mathbb V_1(\xi_x)+V_C(\xi_x)\rightarrow \tfrac12 V_C(\xi_x),
\]
valid in the instanton background. Light-front kinematics modifies the tensor structure, but the dependence on $V_C$ persists. A spin flip may be compensated by a helicity flip, mirroring the rest-frame symmetry that preserves total angular momentum.

\subsection{Light-front Hamiltonian}

The squared-mass operator for a $Q\bar Q$ pair is
\[
M^2=\sum_{i=1}^2\frac{k_\perp^2+m_{Qi}^2}{x_i}
+2M\big(V_C(\xi_x)+V_{SD,I}(\xi_x,b_\perp)\big),
\]
with Bjorken fractions $x_1+x_2=1$. The instanton-induced spin contribution is
\[
V_{SD,I}=
\frac{\sigma\!\cdot\!(b_\perp\times s)}{\xi_x}
\left(V_C'+\mathbb V_1'+\mathbb V_2'\right)
+\frac{\sigma_{1\perp i}\sigma_{2\perp j}}{4m_{Q1}m_{Q2}}
\Big(\hat b_i\hat b_j-\tfrac12\delta_{ij}\Big)\mathbb V_3,
\]
with all $\mathbb V_{1,2,3}$ fixed by $V_C$. This compact structure encodes all instanton-mediated spin effects relevant for meson spectroscopy on the light front.

\section{Spin interaction from a string on the light front}
\label{I_spin_orbit}

Spin-dependent forces induced by the confining QCD string have been analyzed extensively in the rest frame~\cite{Buchmuller:1981fr,Pisarski:1987jb,Gromes:1984ma}. At large interquark separation $R$, the electric flux is localized within the string, so only the {\it self} spin-orbit (SO) terms survive. These originate mainly from Thomas precession and therefore carry a sign opposite to the familiar Coulombic SO force.

In the rest frame, and in the large-$R$ limit, the surviving SO potential is
\begin{equation}
\label{STRING1X_short}
V_{LS}^{\rm string}(R)\approx 
\left(\frac{\boldsymbol{\sigma}_1\!\cdot\!{\bf L}_1}{4m_{Q1}^2}
      -\frac{\boldsymbol{\sigma}_2\!\cdot\!{\bf L}_2}{4m_{Q2}^2}\right)
\left(\frac{1}{R}V_C'(R)+\frac{2}{R}V_1'(R)\right),
\end{equation}
with the convention ${\bf L}_1=-{\bf L}_2\equiv{\bf L}$.  
For the linear potential $V_C(R)=\sigma_T R$ and using $V_1(R)\approx -V_C(R)$ (the short-range cross-term $V_2$ vanishes at large $R$), (\ref{STRING1X_short}) reduces to
\[
V_{LS}^{\rm string}(R)\approx 
\left(\frac{\boldsymbol{\sigma}_1\!\cdot\!{\bf L}}{4m_{Q1}^2}
      -\frac{\boldsymbol{\sigma}_2\!\cdot\!{\bf L}}{4m_{Q2}^2}\right)
(1-2)\frac{\sigma_T}{R},
\]
which displays the characteristic negative sign of the string-induced SO interaction.  On the light front, this result translates to~\cite{Shuryak:2021yif}
\begin{equation}
\label{SLSTRING_short}
M_{LS,{\rm string}}^2 \approx 
2M\left[
\frac{\boldsymbol{\sigma}_1\!\cdot\!({\bf b}_{12}\!\times\! s_1\hat{3})}{4m_{Q1}}
-\frac{\boldsymbol{\sigma}_2\!\cdot\!({\bf b}_{21}\!\times\! s_2\hat{3})}{4m_{Q2}}
\right](1-2)\frac{\sigma_T}{\xi_x},
\end{equation}
where ${\bf b}_{ij}$ is the transverse separation and $\xi_x$ the longitudinal x-dependent measure of the string length on the light front.

As already noted in~\cite{Shuryak:2021yif} (Appendix B), the SO potential following the Eichten-Feinberg analysis~\cite{Eichten:1980mw} is larger by roughly a factor of two~\cite{Pineda:2000sz}. Importantly, the sign of (\ref{SLSTRING_short}) matches that of the string and of instanton effects in the dense regime, and is opposite to the perturbative Coulomb-exchange SO force (consistent with the original rest-frame findings~\cite{Buchmuller:1981fr,Pisarski:1987jb,Gromes:1984ma}.)

\chapter{Baryons at the light front}

\section{Confined N-quarks in 1+1 dimensions}
For simplicity, consider first the dimensionally reduced
"longitudinal" LF Hamiltonian for $N$ confined quarks of equal mass,
\bea
\label{HLF4}
H_{LF,L}=\sum_{i=1}^N\bigg(\frac{m_Q^2}{x_i}
+2\sigma_T  \bigg|\frac{i\partial}{\partial x_i}\bigg|\bigg)
\eea
The corresponding WFs for N-quarks are solutions to 
\bea
\label{HLF5}
\sum_{i=1}^N\bigg(\frac{m_Q^2}{x_i}
+2\sigma_T  \bigg|\frac{i\partial}{\partial x_i}\bigg|\bigg)\varphi_n[x]
=
M_n^2 \varphi_n[x]
\eea
giving mass squared of specific states.

Note that this Hamiltonian already includes two main
nonperturbative phenomena in QCD.
The first term is the kinetic energy, it includes
effective "constituent" quark masses stemming from chiral symmetry breaking.
The second, proportional to the string tension $\sigma_T$, describes quark confinement\footnote{For $N=3$ quarks,  it is similar to the baryonic
 equation derived in 2-dimensional QCD by
 \cite{Bars:1976nk,Durgut:1976bc}.} 
It will be solved in momentum representation, so
coordinates are present as a derivative over momenta.

\begin{figure}[h]
\begin{center}
\includegraphics[width=6cm]{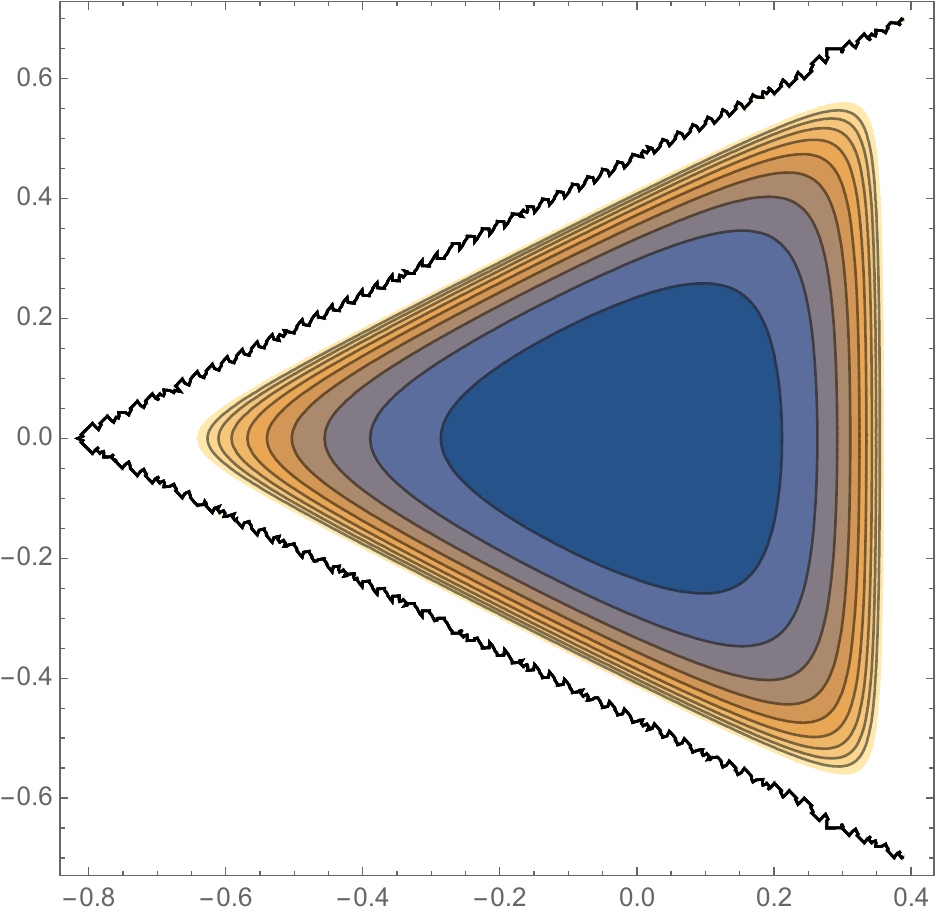}
\includegraphics[width=1cm]{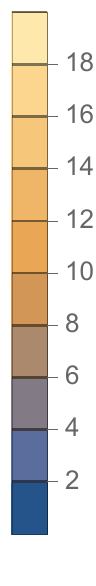}
\caption{Effective potential for equilateral triangular domain for $N=3$ (three quarks).}
\label{fig_V_in_Jacobi}
\end{center}
\end{figure}

 
 The variables $x_i$ are Bjorken-Feynman momentum fractions
 for  N identical particles moving  in
 a box  $0\leq x_i\leq 1$ and subject to the momentum constraint \bea
\label{SUMX2}
X=\sum_{i=1}^Nx_i=1
\eea
 The  condition (\ref{SUMX2}) makes physical support
 of the problem nontrivial. It is $\tilde N$-simplex with
one dimension lower $\tilde  N=N-1$. 
For baryons $N=3$, the domain is an equilateral triangles,
see Fig.\ref{fig_cube}, and our solution for quantum mechanics on it was developed in \cite{Shuryak:2021mlf}. 
For larger $N$
 the domain is the N-dimensional generalization of an equilateral triangle. For  $N=4$ it is a tetrahedron or 3-simplex, and for pentaquarks $N=5$ it is the 4-dimensional
 5-simplex with 5 corners, 10 edges and 10 triangular faces.
  The singular kinetic energy term forces the wavefunction to vanish at these boundaries (Dirichlet boundary condition). The shape of effective "cup potential" is shown in Fig.\ref{fig_V_in_Jacobi} for the baryons.

To solve this problem, we proceed in three steps. First, we unwind the 
square roots  by using the einbein trick 
\bea
\label{CONF2}
\sum_{i=1}^N\bigg|\frac{i\partial}{\partial x_i}\bigg|&=&\sum_{i=1}^N\frac 12\bigg(\frac 1{e_{iL}}+e_{iL}\bigg(\frac{i\partial}{\partial x_i}\bigg)^2\bigg)\nonumber\\
&\rightarrow&\frac 12\bigg(\frac 3{e_{L}}+e_{L}\sum_{i=1}^N\bigg(\frac{i\partial}{\partial x_i}\bigg)^2\bigg)
\eea
and  assume equal $e_{iL}=e_L$ at the extrema, in the steepest descent approximation.
Second, we diagonalize the free Laplacian after isolating the center of mass coordinate, using normal mode coordinates 
(analogue of Jacobi coordinates) see below. Third, we
minimize the energies with respect to $e_L$.

The Hamiltonian and wave functions use momentum representation and Jacobi coordinates (so, unlike
the group of Vary et al, we do not have issues with
center-of-CM motion). 

There are two methods to do quantum mechanics for such
Hamiltonian: \\ (i) One
is to work out complete set of eigenfunctions for 
Laplacian (the second term), and then representing the second
term as a matrix. In some truncated form it can be
diagonalized.\\
(ii) For a triangle, one can numerically solve Schrodinger equation, with laplacian and cup potential included, see \cite{Shuryak:2022thi}.

\section{LFWFs for  baryons, including Delta and Nucleons  }
\label{sec_Delta_Nucleon}
\begin{figure}[t!]
\begin{center}
\includegraphics[width=7cm]{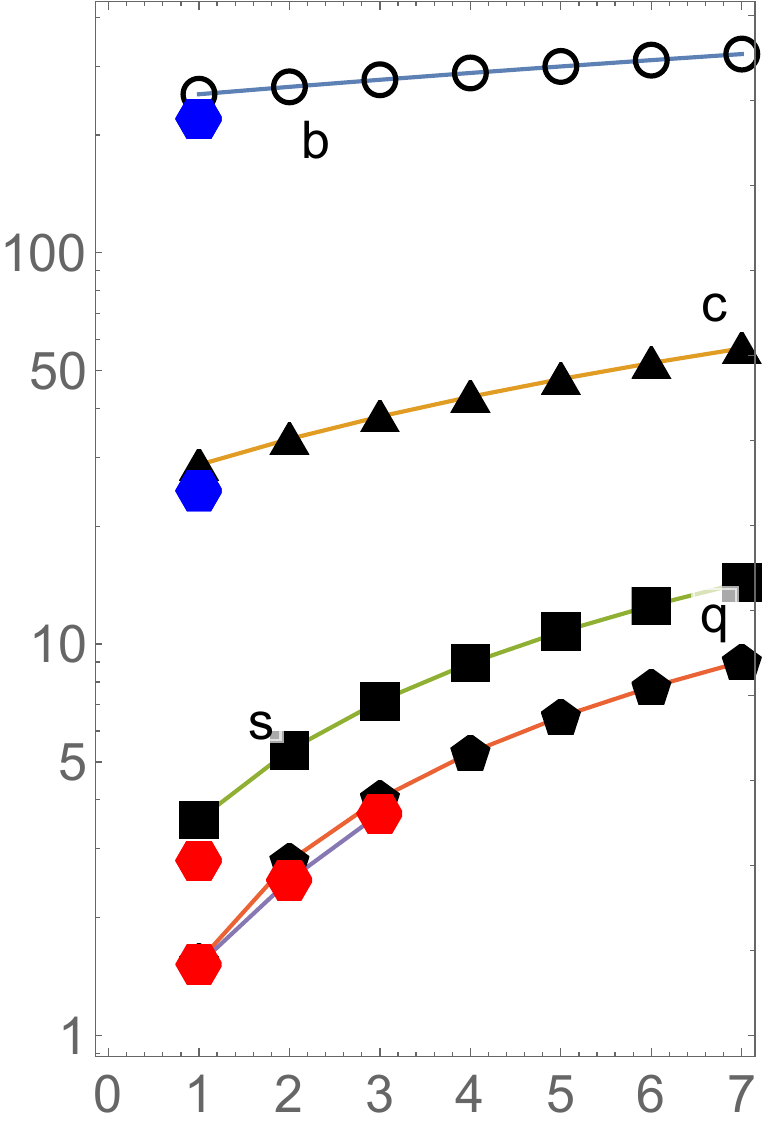}
\caption{Squared masses of baryons $M_{n+1}^2(Q,\frac 32)$ in $GeV^2$,  versus the principal quantum number $n+1=1..7$. The black  circles, triangles, squared and pentagons are results of our calculations for the  flavors  $b,c,s,q$. The red hexagons  are the experimental values of three $\Delta^{++}$ and one $\Omega^-$ masses, from PDG. 
The two blue  hexagons are  model predictions for masses of $ccc$ and $bbb$ baryons.}
\label{fig_all_masses}
\end{center}
\end{figure}

As an example of results derived in \cite{Shuryak:2022thi} let us show Fig.\ref{fig_all_masses} from it, comparing squared baryon masses calculated from LF Hamiltonian with experimental and lattice results,
for same flavor baryons $qqq,sss,ccc,bbb$. 
Many more details, including explicit wave functions can be found in this paper.

Further study of baryons was done in \cite{Shuryak:2022wtk}. The principle new element
introduced compared to the previous work was quark-quark (or diquark) correlations. Let us remind that for light quarks those appear not only due to spin-spin pQCD attraction, but 
due to instanton-induced 'tHooft Lagrangian as well. Needless to say, such effects are important
for nucleons, our main goal.

Diquark correlations of light quarks in nucleons and hadronic reactions, 
have been extensively discussed in the  literature in the past  decades, see e.g. \cite{Carroll:1968mlb,Jaffe:2003sg},
and more recently in the review~\cite{Barabanov:2020jvn}. 

In two-color QCD with $N_c=2$, diquarks are baryons. In the chiral limit,  QCD with two colors and flavors, admit 
Pauli-Gursey symmetry, an extended SU(4) symmetry that mixes massless baryons and mesons. 
In three-color QCD with $N_c=3$, diquarks play an important role in the light and heavy light baryons. 

The simplest way to understand diquark correlations in hadrons, is in single-heavy baryons where the heavy
spectator quark compensates for color, without altering the light diquark spin-flavor correlations.  A good example are
$Qud$ baryons,  with  $\Sigma_Q$ composed of a  light quark with a flavor symmetric assignment $I=1, J^P=1^+$,
and $\Lambda_Q$ composed of a light quark pair with a flavor asymmetric assignment $I=0, J^P=0^+$ state, the
so called {\it bad} and {\it good} diquark states. Note that the latter has no spin, and
thus no spin-dependent interaction with the heavy quark $Q$, while the former does. However, assuming that the
standard spin-spin interactions are of the form $(\vec \sigma_1 \vec \sigma_2)$, this spin interaction can be eliminated 
as follows
\bea  \label{eqn_binding} M( 1^+ud)-M( 0^+ud)
\approx \big((2M(\Sigma^*_Q)+M(\Sigma_Q)/3\big)-M(\Lambda_Q)  \approx 0.21\, {\rm GeV}  \nonumber
\eea 
 with the numerical value thus obtained from experimental masses of $cud,bud$ baryons yields the 
 binding of two types of light quark diquarks.  We note tha the 
 mass difference between heavy-light baryon and meson  iof $m(Qud)-m(Qu)\approx 329 \, MeV$,
 is close to a constituent quark mass.  With antisymmetric color and spin wave function, scalar diquarks must also be antisymmetric in flavor:
 so those can only be $ud, us, sd$ pairs. Those are called "good" diquarks in the literature, in contrast to  the
 "bad" ones made of same flavor $dd,uu,...bb$ and, by Fermi statistics, with a symmetric spin $S=1$ wave functions.   With antisymmetric color and spin wave function, scalar diquarks must also be antisymmetric in flavor:
 so those can only be $ud, us, sd$ pairs. Those are called "good" diquarks in the literature, in contrast to  the
 "bad" ones made of same flavor $dd,uu,...bb$ and, by Fermi statistics, with a symmetric spin $S=1$ wave functions\footnote{ Realization that diquarks  would turn to  Cooper pairs in dense quark matter, as pointed out in 
\cite{Rapp:1997zu,Alford:1997zt}. For a pedagogical review on "color superconductivity" see e.g.\cite{Schafer:2000et}.}

\begin{figure}[t]
\begin{center}
\includegraphics[width=8cm]{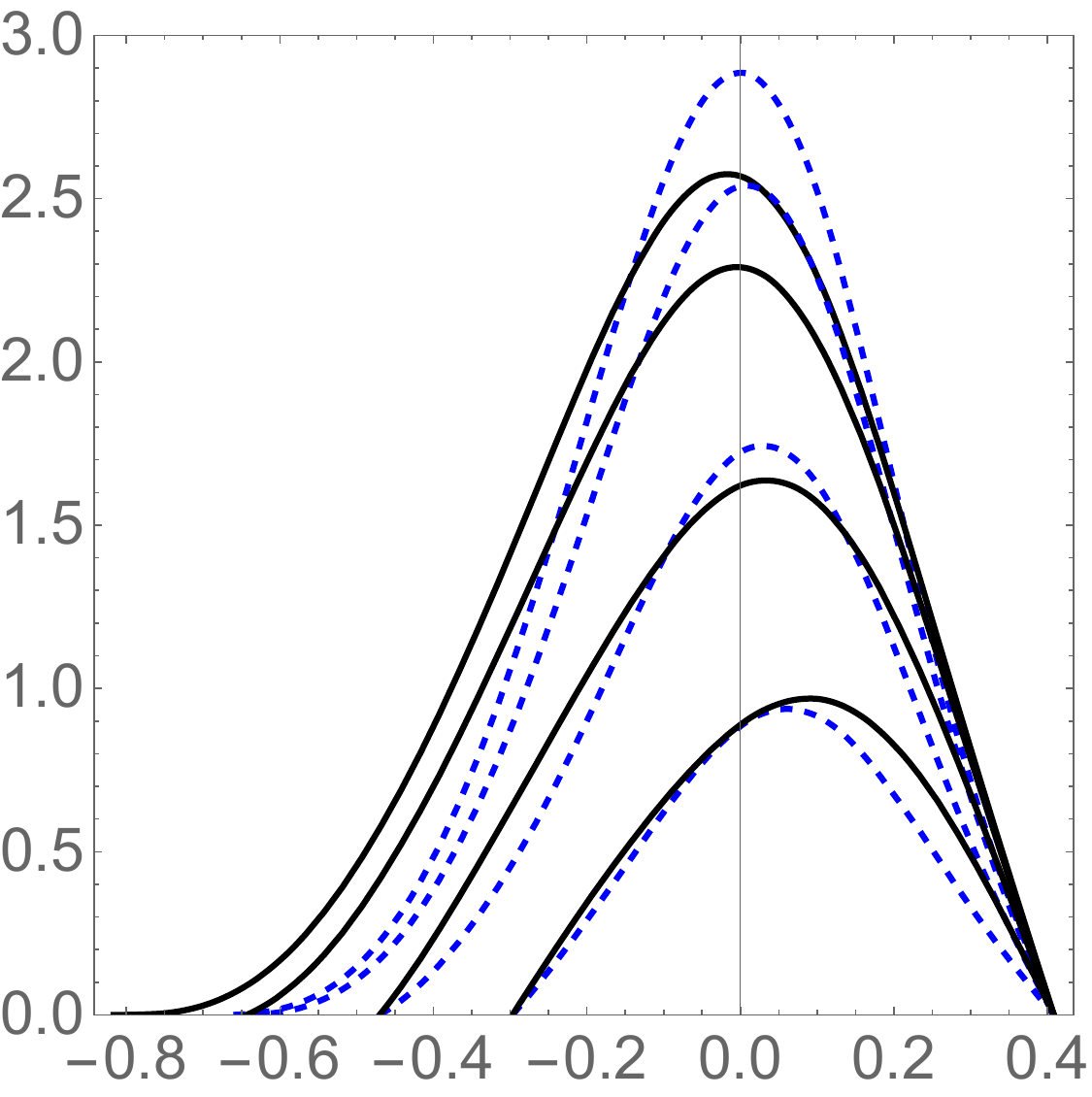}
\caption{
LFWFs for the lowest Delta (dashed lines) and N (solid lines). The plots are shown
versus the Jacobi coordinate $\lambda$, for  fixed $\rho=0,0.1,0.2,0.3 $, top to bottom.   }
\label{fig_Delta_N}
\end{center}
\end{figure}

We would not go in detail description of how diquark correlations in LF wave functions were taken into account in \cite{Shuryak:2022wtk}. Let us just mention that we started from $Qqq$ $\Lambda_c$-type baryons with only one light diquark, and proceed to nucleons posessing pairing in both $ud$ channels. The resulting 
effect of diquark pairing leads to difference between LF WFs for flavor-symmetric Delta baryons
and the nucleons, see Fig.\ref{fig_Delta_N}

\section{Formfactors and GPDs of Delta and Nucleons}
In this section we examplify our main point:
the LFWFs (e.g. those defined above) provide full description of the state, directly leading to all
kind of formfactors, and distributions (PDFs,GPDs etc). 

\begin{figure}[t]
\begin{center}
\includegraphics[width=6cm]{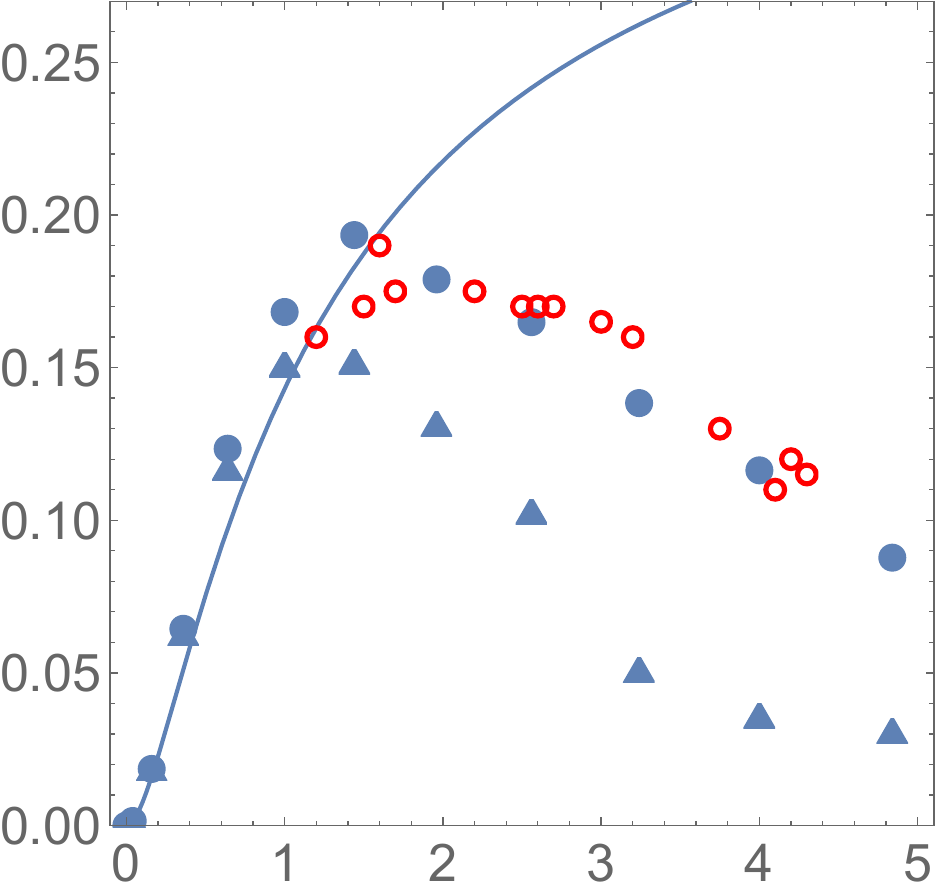}
\caption{$Q^4 F_1^d(Q^2),\, (GeV^4)$ versus the
momentum transfer $Q^2\, (GeV^2)$. The triangles and closed points correspond
to  the Delta and Proton LFWFs, respectively. The red circles are extraction from
the experimental  data on the $p$ and $n$ formfactors mentioned in the text.
The solid line shown for comparison, 
corresponds to the dipole form factor $Q^4/(1+ Q^2/m_\rho^2)^2$.}
\label{fig_Q4_F1d}
\end{center}
\end{figure}

We start with electromagnetic (virtual photon) 
formfactors, following from longitudinal wave functions
for the Delta and Proton  shown in Fig\ref{fig_Q4_F1d}. 
In order to emphaseise  better the most interesting region of intermediate $q^2$,  we plot
$Q^4 F_1^d(Q^2)$, see Fig.\ref{fig_Q4_F1d}.

Experiments are of course done with protons and neutrons, but using them one
can extract separate formfactors for $u$ or $d$ quarks. This was done e.g. in
\cite{1209.0683}, and the red circles in Fig.\ref{fig_Q4_F1d} are from Fig.8 of this work.
(For clarity we do not show the datapoints  in the range  $Q^2<1\, GeV^2$, as the error bars for these points
are $\pm 0.02$ on average.)  From the plot, we see that  this formfactor 
does not appear to reach a constant  at experimentally available $Q^2$,  and the measured points slowly decreasing
towards the right hand side.   Old  dipole parametrization 
$Q^4F=Q^4/(1+ Q^2/m_\rho^2)^2$ also asymptotes  a constant at large $Q^2$, but of a completely different magnitude.

Remarkably, our longitudinal proton wave functions 
reproduces such a trend, and  (with the parameter $A=4\, GeV^{-2}$  ) they follow the shape indicated by the data rather well. The calculated formfactor for the case when the struck quark is $u$ has a similar shape. Unfortunately, according to \cite{1209.0683}, the experimental trend is different, the constant at $Q^2\rightarrow \infty$ is approached from below. By the Drell-Yan relation this flavor difference is also seen in the PDFs of $u$ and
$d$ at $x\rightarrow 1$. Flavor asymmetry must be related with the asymmetry of the
spin-orbit part of the wave function, which in our approximations is so far ignored.

\begin{figure}[h!]
\begin{center}
\includegraphics[width=5.5cm]{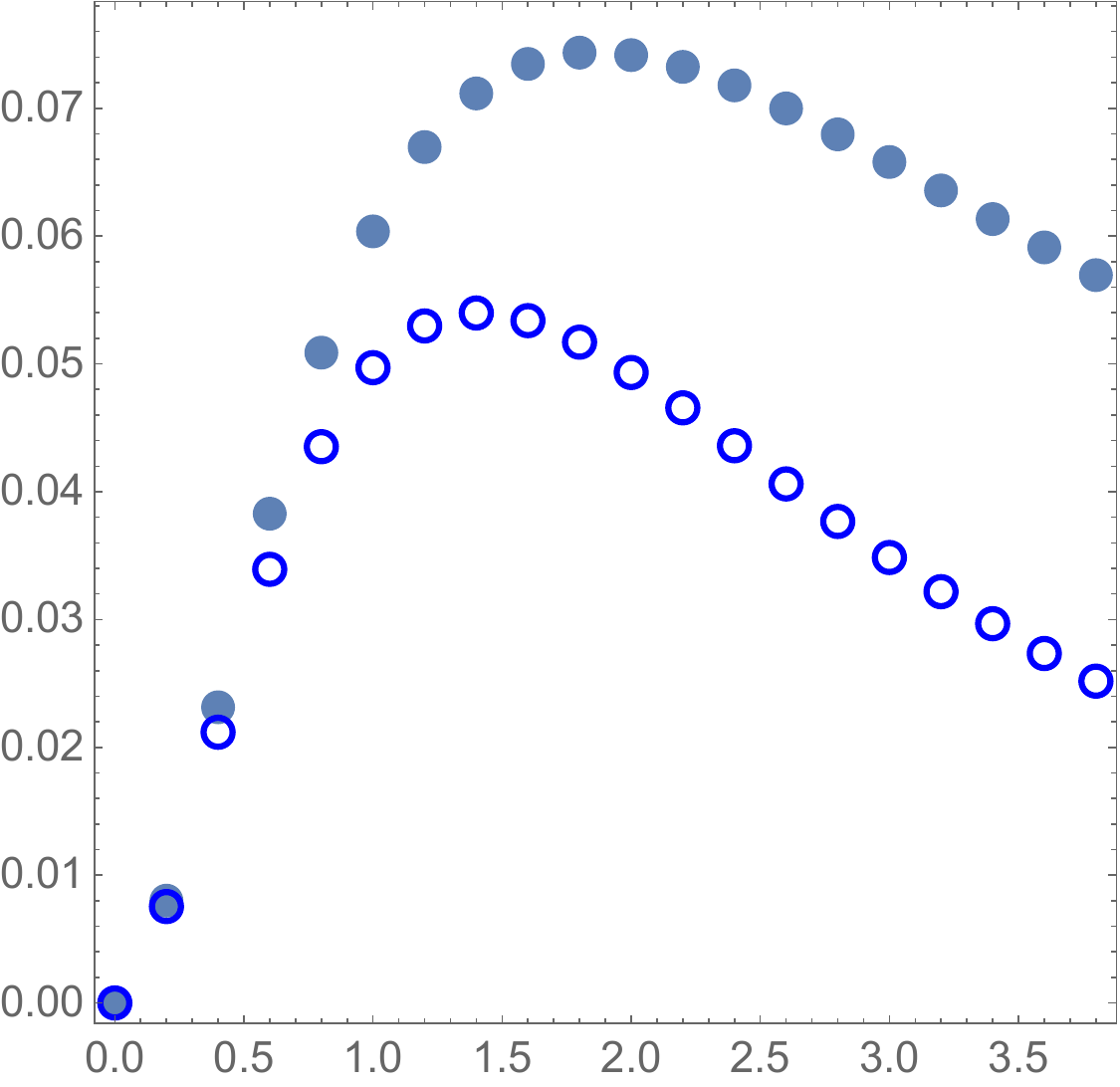} 
\includegraphics[width=5.5cm]{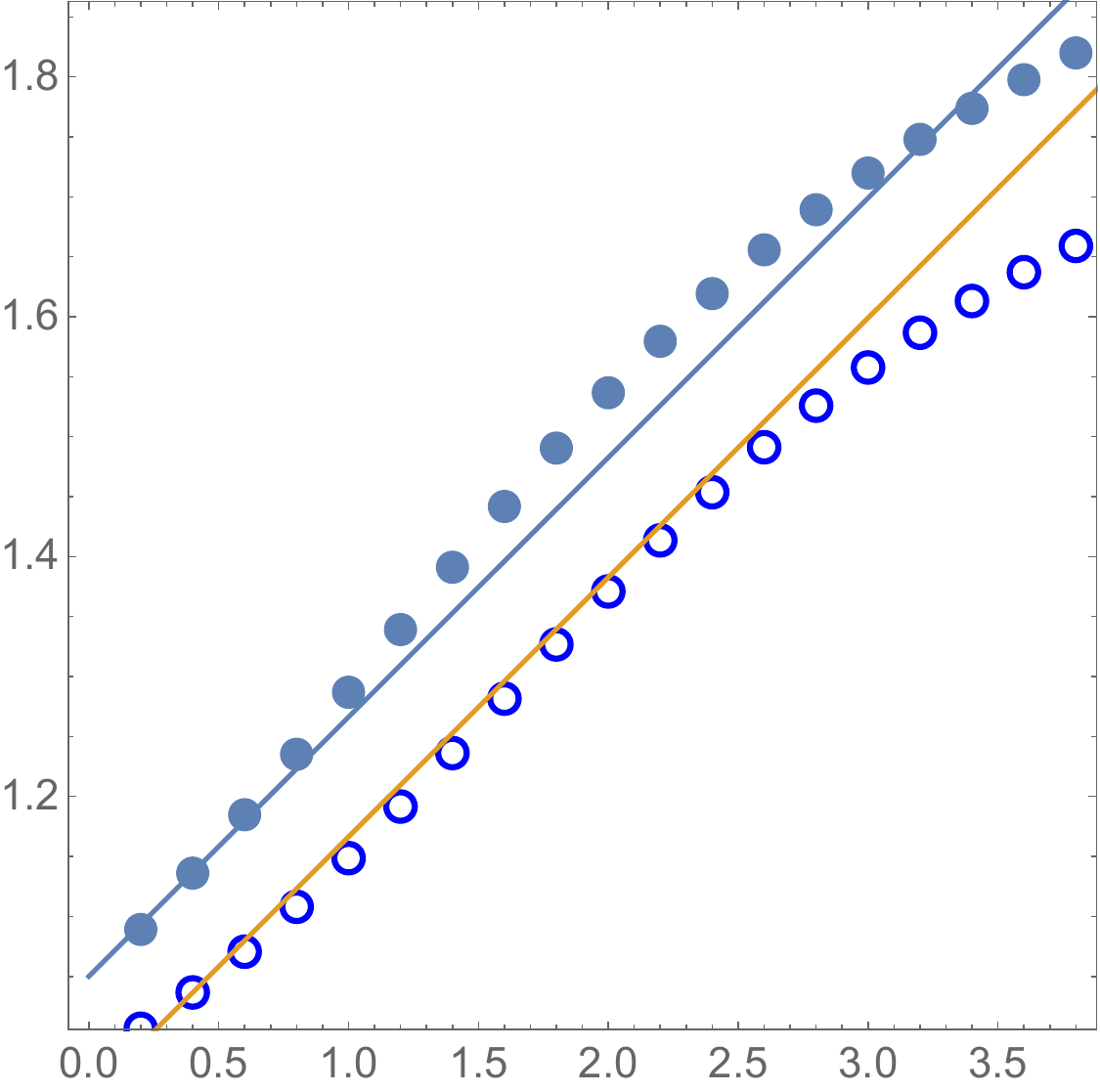} 
\caption{Scaled gravitational form factor $Q^4 A_d(Q^2)$ in GeV$^4$ versus $Q^2$ in GeV$^2$ (top), for the struck d-quark in a  nucleon (closed-points) and the isobar (open-points). Ratio of the gravitational to electromagnetic formfactors $3 A^d(Q^2)/F_1^d(Q^2)$ versus $Q^2$ in GeV$^2$ (bottom),  for the nucleon (closed-points) and the isobar (open points).
}
\label{fig_A_N_del}
\end{center}
\end{figure}

Note that the corresponding formfactor for Delta (triangles) is significantly softer, as one would expect from the size of the wave function.
Recall that the large difference between the Delta and Proton formfactors (so well seen
in this plot) is completely due to the
't Hooft quasi-local pairing $ud$ interaction. 
We also calculated gravitational formfactors from
the same LFWFs, see Fig.\ref{fig_A_N_del}

The same LFWFs were used  for 
evaluation of the "gravitational" formfactors.
While not experimentally available, we can 
substitute electromagnetic current to stress tensor $T^{\mu\nu}$\footnote{One more option sometimes discuss is the trace of the stress tensor, as if induced by a Higgs-like scalar. This formfactor gives distribution of the mass.}, as if it was scattering with a graviton. This formfactor described distribution of the energy/momentum inside baryons, rather than electric charge.
The first moment of the unpolarized GPDs at zero skewness,
is tied to the quark $A,B$ form factors of the energy-momentum tensor~\cite{Belitsky:2005qn} 
\be
A(t)=\int dx \,x H(x,0,t), \,\,\,
B(t)=\int dx \, xE(x,0,t)
\ee
They can be used to quantify the distribution of momentum, angular momentum and pressure-like stress,
inside the nucleon~\cite{Polyakov:2018zvc} (and references therein). More specifically, the total nucleon angular momentum at this low-resolution, is
given by Ji$^\prime$s sum rule~\cite{Ji:1996ek}
\bea
J=\frac 12=A(0)+B(0)
\eea
with the non-perturbative gluons implicit in the balance, as they enter implicitly  in the composition of the LF
Hamiltonian (mass, string tension, ...) for the constituent quarks. In  Fig.\ref{fig_A_N_del} (top) we show the numerical results for $Q^4A^d$ versus $Q^2$ in ${\rm GeV}^2$,  
for a struck d-quark, following from the integration of the  
unpolarized d-quark-A GPD, for the nucleon (filled-point) and $\Delta$-isobar (open-point). Again, we observe
that the isobar form factor falls faster than the mucleon form factor, an indication that the nucleon is more compact gravitationally
than the isobar, with a smaller gravitational radius. In Fig.\ref{fig_A_N_del} (top) we plot the ratio of the gravitational formfactor 
relative to the electromagnetic form factor, for the struck d-quark in the nucleon (filled-points) and isobar (open-points). The decrease in $Q^2$ of the gravitational form factor, is slower than the electromagnetic form factor  for both hadrons. This means that the spatial mass distribution of the struck d-quark, is more compact  than the spatial charge  distribution.

Generalized Parton Distributions (GPDs)
For the proton with quark assignment $uud$, we assume that the struck quark is $d$, with longitudinal 
momentum fraction $x_3$. In our (modified) Jacobi coordinates, this momentum fraction is directly related to the
longitudinal variable $\lambda$.  For the unpolarized d-quark GPD this amounts to integrating
the off-forward LFWFs over 5 variables,  the transverse momenta $\vec p_\rho,\vec p_\lambda$ and  
$\rho$  for the nucleon. To proceed, we approximate the dependence on the transverse momenta 
$\vec p_\rho,\vec p_\lambda$ by Gaussians, which is quite accurate, and carry the integrals analytically. 
The remaining integration over $\rho$ is performed numerically. We recall that the LFWFs are generated
at a  low renormalization scale, say $\mu_0=1$ GeV, with a nucleon composed of three constituent quarks, without
 constituent gluons.

\begin{figure}[htbp]
\begin{center}
\includegraphics[width=8cm]{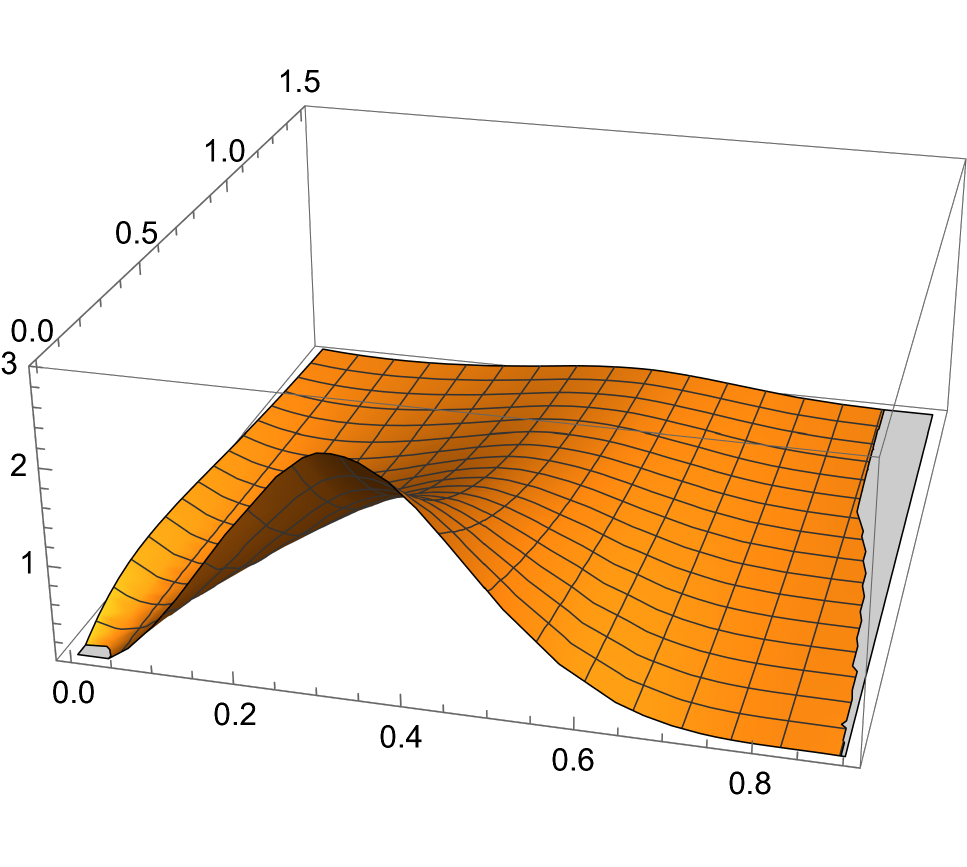}
\caption{The nucleon GPD function $H_d^N(x,\xi=0,Q^2)$ for a struck d-quark.}
\label{fig_Hd_N_3d}
\end{center}
\end{figure}

In Fig.\ref{fig_Hd_N_3d} we show the unpolarized nucleon GPD for the struck d-quark,
 as a function of $x,Q^2$ and zero skewness.  At small $Q^2$, the dependence
 on the longitunal parton momentum $x$, is that expected for a PDF with a maximum at 
 $x=\frac 13$. At  larger $Q^2$, the maximum of the  GPD clearly shifts towards
larger values of $x$. The GDP is not simply factorizable into the PDF times the form
factor, which are separable in $x$ and $Q^2$. The right shift in Fig.\ref{fig_Hd_N_3d}
shows that the nucleon shape changes with $x$. This is a key point of interest to us,
as we now proceed to detail these shape modifications.

Theoretical considerations~\cite{Guidal:2004nd} have suggested that the GPDs can be approximated
generically, by a  "Gaussian ansatz"  in the momentum transfer $Q^2$, with a width and a
pre-exponent that are x-dependent
\bea
 {\rm GPDs}(x,Q^2) \sim f_1(x)\,e^{-Q^2 f_2(x)}
 \label{GPD_ansatz}
\eea
We recall that  the standard nucleon formfactors are  dipole-like at low $Q^2$
(say lower than 10 GeV$^2$) with $F(Q^2)  \sim 1/(m^2+Q^2)^2$. At  very large $Q^2$
it asymptotes to a constant  $Q^4 F(Q^2) \rightarrow const$  which is fixed by the 
perturbative QCD scattering rule. In between, the $Q^2$ dependence remains an open
issue. 

Our LF formfactors were found to be consistent with these observations
(see below), and our GPDs at fixed $x$  are indeed numerically consistent with the exponent
$
exp[-{Q^2 f_s(x) \over \sqrt{1-\xi^2}}]
$
with no dependence of the pre-expont on the skewness, for $x>\xi$.
(We have checked that the x-integration of this exponent with $\xi=0$, returns the expected  formfactor.)

The r.m.s. size $R_{r.m.s.}(x)$ is maximal at $x\sim \frac 13$, and
is numerically about $0.6$ fm. It decreases sharply for $x\sim 1$, and
moderatly for $x\sim 0$. This can be explained by the fact that for $x\sim \frac 13$,
all three quarks carry about the same longitudinal momentum on the LH, which 
corresponds to three quarks at rest in the CM or rest frame. 
Semiclassically, this means a configuration in which a struck quark is near a "turning points" of
the wave function, with their QCD strings maximally streched. In contrast, $x$
away from $\frac 13$, corresponds to a struck quark rapidly moving in the CM frame,
which must happen near the hadron center, The corresponding size is therefore small.
 Note that for $x\rightarrow 1$, the distribution of the struck d-quark becomes nearly pointlike,
with slopes close to zero.  Our LFWFs show that the magnitude of this effect is 
different for nucleon and $\Delta$-isobar,   sensitive to the the diquark substructure of the former.

\section{Applications to negative parity baryons}


  The aim of this work is to $derive$ the light-front Hamiltonian and the corresponding wave functions
  (LFWFs), starting with the most basic meson settings and certain
  nonperturbative dynamics. In this methodical paper we will focus on "the main components" of the 
  wave functions with zero orbital momentum,  and ignore different  spin structures. Its generalization to full wave functions, with
  all allowed spin and angular momentum values, will be done in the next papers of the series.
  
   One Hamiltonian leads to an  infinite set of wave functions, any of which have infinitely many matrix elements. LFWFs are mutually orthogonal and can be properly normalized. The DAs are normalized
only  to certain empirical constants, like $f_\pi$ for a pion. The PDFs of baryons are traced
over all quarks but one: tracing mixes together all sectors of the wave function, with different
quantum numbers and even the number of partons.  Due to quantum entanglement, it leads to an entropy.


It is improved 
in  the LF  frame, as the motion of all quarks gets "frozen" (see the right sketch), 
 the distinctions between the heavy and the light quarks basically go away, as both
can be "eikonalized" and treated in fully relativistic formalism. If so, their interactions can  be deduced  from pertinent Wilson line correlators for any quarks.


\section{Wilson lines dressed by spin variables}
The perturbative contribution to the central potential on the light front,  induced by a  one-gluon exchange with an effective mass $m_G$ in the RIV, can be constructed  using the general technique of a sloped Wilson loop as we detailed in~\cite{Shuryak:2021hng}. In particular, the
one-gluon  interaction with spin effects,  follows by dressing the Wilson loop, with explicit spin factors

\bea
\label{ONEX}
\bigg<{\rm Tr}{\bf P}&&\bigg[{\rm exp}\bigg(+g\int d\tau_1(i\dot{x}(\tau_1)\cdot A(x(\tau_1)+\frac 1{4}\sigma_{1\mu\nu}F_{\mu\nu}(x(\tau_1))\bigg)
\nonumber\\
&&\times {\rm exp}\bigg(-g\int d\tau_2(i\dot{x}(\tau_2)\cdot A(x(\tau_2)+\frac 1{4}\sigma_{2\mu\nu}F_{\mu\nu}(x(\tau_2))\bigg)\bigg]\bigg>
\eea

with $\sigma_{\mu\nu}=\frac 1{2i}[\gamma_\mu, \gamma_\nu]$,  and $\sigma_{\mu\nu}=\eta_{a\mu\nu} \sigma^a$ using $^\prime$t Hooft symbol. 
The averaging is understood using the QCD action.
We have made explicit the gauge coupling $g$, for a perturbative treatment to follow.
For  massive quarks traveling on straight trajectories, the affine time
$\tau$ relates to the conventional time  $t$ through 
\bea
\label{SS2}
\mu=\frac {dt}{d\tau}=\frac{m_Q}{\sqrt{1+{\dot{\vec x}}^2}}\rightarrow \gamma m_Q
\eea
in Euclidean signature. We note that the holonomies tracing out the Wilson loop  are  unaffected by the exchange $\tau\rightarrow t$,
in contrast to the spin contributions which get rescaled by $1/\mu$. We now proceed to use (\ref{ONEX}) to derive the
spin interactions for one-gluon exchange, followed by the ones induced by instantons, all in Euclidean signature. The upshot of this construction will be the
derivation of the spin interactions on the light front
for both the perturbative and non-perturbative contributions.

\section{Central Coulomb interaction from one-gluon exchange}

The Coulomb interaction between a $Q\bar Q\equiv Q_1Q_2$ pair attached to the Wilson lines, can be obtained
in perturbation theory by expanding the holonomies, and averaging the $AA$ correlator in leading order. 
For that, we parametrize the world-lines by
\bea
x_\mu(t_1)=(0,0, {\rm sin}\theta\,t_1, {\rm cos}\theta\, t_1)\qquad\qquad
x_\mu(t_2)=(b_1,b_2, {\rm sin}\theta\, t_2+b_3, {\rm cos}\theta\, t_2)
\eea
The perturbative one-gluon contribution 
from (\ref{ONEX}) reads

\bea
\label{ONE1}
g^2 T_1^AT_2^B\int\,dt_1\int\,dt_2
\bigg({\rm cos^2}\theta \langle A_4^A(t_1)A^B_4(t_2)\rangle+{\rm sin^2}\theta \langle A_3^A(t_1)A^B_3(t_2)\rangle
+2{\rm sin}\theta {\rm cos}\theta  \langle A_4^A(t_1)A^B_3(t_2)\rangle\bigg)\nonumber\\
\eea

with the gluon correlator in Feynman gauge
\bea
\label{ONE2}
\langle A_\mu^A(t_1)A^B_\nu(t_2)\rangle=\frac 1{2\pi^2}\frac{\delta^{AB}\delta_{\mu\nu}}{|x(t_1)-x(t_2)|^2}
\eea
Inserting (\ref{ONE2}) into (\ref{ONE1}) and changing variables $T_E=t_1+t_2$ and $\tau=t_1-t_2$, yield
\bea
\label{ONE3}
\frac{g^2 T^A_1T_2^A}{2\pi^2}\int\,\frac {dT_E}2\int\,dt
\frac 1{t^2+{\rm cos}^2\theta\,b_3^2+b_\perp^2}
=\frac{g^2T_1^AT_2^A}{4\pi} \frac {T_E}{\sqrt{b_3^2{\rm cos}^2\,\theta+b_\perp^2}}
\eea
The analytical continuation $\theta\rightarrow -i\chi$ and $T_E\rightarrow iT_M$ of (\ref{ONE3}), re-exponentiates to the Coulomb contribution
\bea
{\rm exp}\bigg[-i\gamma T_M\bigg(-\frac{g^2T_1^AT_2^A}{4\pi} \frac {1/\gamma}{\sqrt{\gamma^2b_3^2+b_\perp^2}}\bigg)\bigg]
\eea
with $\gamma T_M$ the dilatated time along the light-like Wilson loop. The Coulomb contribution to the light front $Q\bar Q$ Hamiltonian $P_{Cg}^-$ follows, leading the squared invariant mass as
\bea
\label{ONE4}
2P^+P_{Cg}^-=
2P^+\bigg(-\frac{g^2T_1^AT_2^A}{4\pi} \frac {1/\gamma}{\sqrt{\gamma^2b_3^2+b_\perp^2}}\bigg)
\rightarrow 2M\bigg(-\frac{g^2T_1^AT_2^A}{4\pi} \frac{1}{\xi_x}\bigg)=2M\mathbb V_{Cg}(\xi_x)
\eea
with $P^+/M=\gamma$, and $\gamma b_3\rightarrow id/dx/M$ the conjugate of Bjorken-x.

 In the random instanton vacuum (RIV), the perturbative gluons acquire a momentum dependent mass
 from their rescattering through the instanton-anti-instanton ensemble~\cite{Musakhanov:2021gof} 
\bea
m_G(k\rho)=m_G\, \bigg(k\rho\,K_1(k\rho)\bigg)
\qquad\qquad
 m_G\rho &\approx& 2\bigg(\frac {6\kappa }{N_c^2-1}\bigg)^{\frac 12}\approx 0.55
\eea
using the estimate $\kappa=\pi^2\rho^4 n_{I+\bar I}$ in the right-most result. With this in mind, (\ref{ONE4}) is now

\bea
\label{ONEG}
\mathbb V_{Cg}(\xi_x)=-\frac{g^2 T_1^AT_2^A}{2\pi^2}  \frac 1{\xi_x}\int_0^\infty\frac {dx\,x{\rm sin}x}{x^2+(\xi_xm_G(x\rho/\xi_x))^2}         
\rightarrow -\frac{g^2 T_1^AT_2^A}{4\pi}\frac {e^{-m_G \xi_x}}{\xi_x}
\eea

with the right-most result following for a constant gluon mass.

\section{Spin-spin interaction from one-gluon exchange}

The perturbative spin-spin interaction follows from  the cross term  in (\ref{ONEX})
\bea
\label{SS1}
-\frac {g^2}{16}\int d\tau_1 d\tau_2 \langle \sigma_{1\mu\nu}F_{\mu\nu}(x(\tau_1))\sigma_{2\alpha\beta}F_{\alpha\beta}(x(\tau_2))\rangle
\eea
Note that the  perturbative electric field is purely imaginary in Euclidean signature, leading mostly to phases and not potentials in the long time
limit.  Also,  the Dirac representation $\sigma_{4i}$ is off-diagonal, an indication that the electric contribution mixes
particles and anti-particles, which is excluded by the use of straight Wilson lines on the light front.  With this in mind and using  (\ref{SS2}),  we can reduce (\ref{SS1})  to
\bea
\label{SS3}
-\frac{g^2}{4\mu^2}\int dt_1 dt_2 \langle \sigma_{1ij}F_{ij}(x(t_1))\sigma_{2kl}F_{kl}(x(t_2))
=-\frac{g^2}{4\mu^2}\sigma_1^a\sigma_2^b\int dt_1 dt_2 \langle B_a(x(t_1))B_b(x(t_2))\rangle
\eea
with
\bea
\label{SS4}
\big<B_a(x(t_1))B_b(x(t_2))\big> 
&=&T_1^AT_2^B\epsilon_{aij}\epsilon_{bkl}\partial_{1i}\partial_{2k}
\big<A^A_j(x(t_1)A_m^B(x(t_2))\big>\nonumber\\
&=&T^A_1T^A_2(\delta_{ab}\delta_{ik}-\delta_{ak}\delta_{bi})\partial_{1i}\partial_{2k}
\times\frac 1{2\pi^2}\frac 1{|x(t_1)-x(t_2)|^2} 
\eea

Inserting (\ref{SS4}) into (\ref{SS3}), and carrying the time integrations along the sloped Wilson loop, give~\cite{Shuryak:2021mlh}
\bea
\label{SS5}
-\frac{g^2T_1^AT_2^A}{4\pi} \frac {T_E}{4\mu^2}
\bigg[\sigma_{1\perp}\cdot \sigma_{2\perp}\,\bigg(-3 {\rm cos}^2\theta \frac{({\rm cos}\theta \,b_3)^2}{\xi_\theta^2}+{\rm cos}^2\theta\bigg)\frac 1{\xi_\theta^3}\bigg]
\eea
which is the dominant contribution under the analytical continuation $\theta\rightarrow -i\chi$, $\mu\rightarrow \gamma m_Q$ and
$T_E\rightarrow iT_M$, in the ultra-relativistic limit  $\gamma\gg 1$, and  in Minkowski signature. The final spin-spin contribution to the squared mass is in
general
\bea \label{eqn_VSS_pert}
H_{SS}=2M\bigg[\frac {\sigma_{1\perp}\cdot\sigma_{2\perp}}{4m_{Q1}m_{Q2}}
\bigg( \nabla_\perp^2 \mathbb V_{Cg}(\xi_x)\bigg)\bigg]
=2M\mathbb V_{SS}(\xi_x, b_\perp)
\eea

\section{Spin-orbit interaction from one-gluon exchange}

The  cross spin-orbit interaction is readily obtained   from the $12+21$ cross terms
\bea
-\frac{g^2 T_1^AT_2^B}{2}\,\sigma_2^a
\int d\tau_1 d\tau_2 \, i\dot{x}_i(\tau_1)\big<A_i^A(x(\tau_1)B_a^B(x(\tau_2))\big>+ 1\leftrightarrow  2
\eea
which can be reduced to
\bea
-\frac{ig^2T_1^AT_2^B{\rm sin}\theta}{2\mu}\sigma_2^a s_1
\int dt_1 dt_2\langle A^A_3(x(t_1))B^B_a(x(t_2))\rangle
+1\leftrightarrow 2
\eea
with  $s_{1,2}={\rm sgn}(v^3_{1,2})$  the signum of the 3-velocity of particle 1,2
(a more refined definition will be given below). 
After carrying the integrations, and the analytical continuations, the spin-orbit contribution to the squared mass is

\bea
\label{MLS12}
H_{SL,12}=2M\bigg[\bigg(\frac{\sigma_2\cdot (b_{12}\times s_1\hat 3)}{2m_{Q2}}
-\frac{\sigma_1\cdot (b_{21}\times s_2\hat 3)}{2m_{Q1}}\bigg)
\bigg(\frac 1{\xi_x}\mathbb V^\prime_{Cg}(\xi_x)\bigg)\bigg]
\eea

in general, with $b_{21}=-b_{12}\equiv b_\perp$.

The standard self spin-orbit interaction with Thomas precession is more subtle.  To unravel it,  we note that 
the insertion of a single spin contribution along the path-ordered Wilson loop amounts to expanding
the spin factors in (\ref{ONEX}) to first order,  and retaining the holonomies to all orders in ${\bf 1}_\theta$,
namely

\bea
\label{LS1}
 \frac 1{4\mu} \sigma_{1\mu\nu}\int dt_1 \langle gF_{\mu\nu}(x(t_1))\,{\bf 1}_\theta\rangle + 1\leftrightarrow 2\nonumber\\
\eea
with the path ordered color-spin  trace subsumed. Here ${\bf 1}_\theta$ refers to the slated Wilson loop 
without the spin dressing.
We now decompose
\bea
\label{LS2}
F_{\mu\nu}=v_{1\mu} v_{1\alpha} F_{\alpha\nu} +F_{\mu\nu}^\perp\equiv F_{\mu\nu}^{||}+F_{\mu\nu}^\perp
\eea
into a contribution parallel to $v_{1}=\dot{x}_1$ and a contribution orthogonal to $v_1$. The contribution 
parallel to the worldline when inserted in (\ref{LS1}) can be undone by the identity
(see Eq.71 in~\cite{Shuryak:2021hng})
\bea
\int dt_1 \langle gv_\alpha F_{\alpha\nu}(x(t_1))\,{\bf 1}_\theta\rangle
 =-i\partial_{1\nu}\langle {\bf 1}_\theta\rangle
\equiv -i\partial_{1\nu}\,e^{-T_E\mathbb V_C({\xi_\theta})}
\eea
with $\mathbb V_C(\xi_\theta)\approx \mathbb V_{Cg}(\xi_\theta)$ the central Coulomb potential in perturbation theory. The 
longitudinal contribution to (\ref{LS1}) 
\bea
\label{LS3}
- \frac 1{4\mu} \sigma_{1\mu\nu}v_{1\mu} i\partial_{1\nu}\,e^{-T_E\mathbb V_C({\xi_\theta})} + 1\leftrightarrow 2\nonumber\\
 \eea
is gauge-invariant. After carrying the analytical continuation, (\ref{LS3}) contributes both a real and imaginary part. The latter 
is an irrelevant  phase factor in Euclidean signature. The  {\it real } part contributes to the direct mass squared operator as

\bea
\label{MLS11}
H_{LS, 11}=
2M\bigg[\bigg(\frac{\sigma_1\cdot (b_{12}\times s_1\hat 3)}{4m_{Q1}}
-\frac{\sigma_2\cdot (b_{21}\times s_2\hat 3)}{4m_{Q2}}\bigg)
\bigg(\frac 1{\xi_x}\mathbb V^\prime_{Cg}(\xi_x)\bigg)\bigg]
\eea \label{eqn_pert_spin_orb}

in leading order in perturbation theory.
This is the standard spin-orbit contribution with the correct Thomson correction on the light front, familiar from atomic physics in the rest frame.
The total perturbative spin  contribution on the light front, is the sum of  (\ref{eqn_VSS_pert}), (\ref{MLS12}) and (\ref{MLS11}),

\bea
\label{PERP1}
H_{LS,g}=2M\bigg(
\frac{l_{1\perp}\cdot S_{1\perp}}{2m_{Q1}^2}-\frac{l_{2\perp}\cdot S_{2\perp}}{2m_{Q2}^2}
+\frac{l_{1\perp}\cdot S_{2\perp}}{m_{Q1}m_{Q2}} -\frac{l_{2\perp}\cdot S_{1\perp}}{m_{Q1}m_{Q2}}\bigg)
\frac 1{\xi_x}\mathbb V_{Cg}^\prime(\xi_x)
+2M\bigg(\frac {S_{1\perp}\cdot S_{2\perp}}{m_{Q1}m_{Q2}}\bigg)
 \nabla_\perp^2 \mathbb V_{Cg}(\xi_x)\nonumber\\
\eea
with the respective spins $\vec S_{1,2}=\vec \sigma_{1,2}/2$, and transverse  orbital momenta

\bea
\label{PERP2}
l_{1,2\perp}=\pm (b_\perp\times m_{Q1,2}s_{1,2}\hat 3)_\perp\qquad 
 s_{1,2}={\rm sgn}(v_{1,2}) \rightarrow \frac{Mx_{1,2}}{m_{Q1,2}}
 \eea

\section{Instanton-induced spin forces}

To construct the spin-dependent interactions in the ILM  
 on the light front,  we apply the general construction by Eichten and Feinberg~\cite{Eichten:1980mw},
to the slated Wilson loop  in Euclidean signature, followed by the analytical continuation $\theta\rightarrow -i\chi$ to Minkowski signature. This construction was detailed earlier.



\section{LF wave functions of baryons}

\begin{figure}[t!]
    \centering
    \includegraphics[width=6cm]{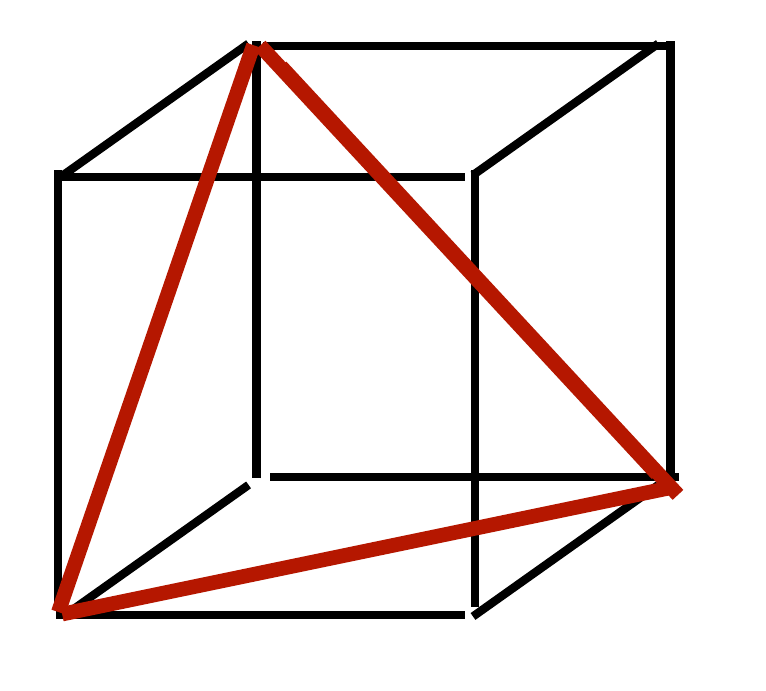}
    \caption{equilateral triangle formed by cube $x_1,x_2,x_3\in [0,1]^3$ cut by momentum normalization condition $x_1+x_2+x_3=1$}
    \label{fig_cube}
\end{figure}

 The variables $x_i$ are Bjorken-Feynman momentum fractions
 for  N identical particles moving  in
 a box  $0\leq x_i\leq 1$ and subject to the momentum constraint \bea
\label{SUMX}
X=\sum_{i=1}^Nx_i=1
\eea
 The  condition (\ref{SUMX}) makes physical support
 of the problem nontrivial. It is $\tilde N$-simplex with
one dimension lower $\tilde  N=N-1$. 
For baryons $N=3$, the domain is an equilateral triangles,
see Fig.\ref{fig_cube}, and our solution for quantum mechanics on it was developed in \cite{Shuryak:2021mlf}. 
For larger $N$
 the domain is the N-dimensional generalization of an equilateral triangle. For  $N=4$ it is a tetrahedron or 3-simplex, and for pentaquarks $N=5$ it is the 4-dimensional
 5-simplex with 5 corners, 10 edges and 10 triangular faces.
  The singular kinetic energy term forces the wavefunction to vanish at these boundaries (Dirichlet boundary condition). The shape of effective "cup potential" is shown in Fig.\ref{fig_V_in_Jacobi} for the baryons.

To solve this problem, we proceed in three steps. First, we unwind the 
square roots  by using the einbein trick 
\bea
\label{CONF22}
\sum_{i=1}^N\bigg|\frac{i\partial}{\partial x_i}\bigg|&=&\sum_{i=1}^N\frac 12\bigg(\frac 1{e_{iL}}+e_{iL}\bigg(\frac{i\partial}{\partial x_i}\bigg)^2\bigg)\nonumber\\
&\rightarrow&\frac 12\bigg(\frac 3{e_{L}}+e_{L}\sum_{i=1}^N\bigg(\frac{i\partial}{\partial x_i}\bigg)^2\bigg)
\eea
and  assume equal $e_{iL}=e_L$ at the extrema, in the steepest descent approximation.
Second, we diagonalize the free Laplacian after isolating the center of mass coordinate, using normal mode coordinates 
(analogue of Jacobi coordinates) see below. Third, we
minimize the energies with respect to $e_L$.


The Hamiltonian and wave functions use momentum representation and Jacobi coordinates\footnote{Therefore, unlike
the group of Vary et al, we do not have issues with
center-of-CM motion.}. 
There are two methods to do quantum mechanics for such
Hamiltonian: \\ (i) One, documented for triangle in \cite{Shuryak:2021mlf},
is to work out complete set of eigenfunctions for 
Laplacian (the second term), and then representing the second
term as a matrix. In some truncated form it can be
diagonalized.\\
(ii) For a triangle, one can also numerically solve Schrodinger equation, with laplacian and cup potential included\footnote{Unfortunately, this way of solving the problem for larger dimensions - tetra, pentaquarks - have not yet been done.}.



\chapter{Pentaquarks on the light front}

\section{The $A_4$ simplex and the forward wave functions}

The theory of hadrons on the light front largely parallels the discussion given above for mesons and baryons. 
The Hamiltonian is approximately a sum of (oscillator-like) Gaussian terms in transverse momenta, supplemented by a more complicated part depending on longitudinal momenta, or rather on Bjorken momentum fractions $x_i$, $i=1,\dots,5$. 

As usual, the first step is the elimination of the center-of-mass motion. As a result, the five momentum fractions are constrained to lie in the plane
\[
\sum_{i=1}^5 x_i = 1 .
\]
As in previous cases, this constraint is implemented using (modified) Jacobi coordinates $\alpha,\beta,\gamma,\delta$, which automatically incorporate this condition:
\ba \label{eqn_Jacobi_5}
x_1 &=& (6 + 15 \sqrt{2} \alpha+ 5 \sqrt{6} \beta+ 
    5 \sqrt{3} \gamma+ 3 \sqrt{5} \delta)/30  \nonumber \\
x_2 &=&  (6 - 15 \sqrt{2} \alpha+ 5 \sqrt{6} \beta+ 
    5 \sqrt{3} \gamma+ 3 \sqrt{5} \delta)/30, \nonumber \\
x_3 &=&  (6 - 10 \sqrt{} \beta+ 5 \sqrt{3} \gamma+ 
    3 \sqrt{5} \delta)/30, \nonumber \\
x_4 &=& 1/10 (2 - 5 \sqrt{3} \gamma+ \sqrt{5} \delta), \nonumber \\
x_5 &=& 1/5 - (2 \delta)/\sqrt{5}
\ea

The corresponding four-dimensional simplex (similar to Fig.~\ref{fig_cube}, but not plotted for obvious reasons) is known in mathematics as the $A_4$ simplex\footnote{Our colloquial name for it is a "starfish''. In \textit{Wolfram Mathematica} we use the command \texttt{starfish = Simplex[$a_1,\dots,a_5$]}, which constructs a simplex from a given set of points. This is particularly useful, for example, for defining integrals over this domain.}. 
It has five corners, corresponding, as usual, to configurations in which one of the $x_i=1$ and all others vanish. Their explicit locations in Jacobi coordinates are
\ba 
a_1 &=& \{1/\sqrt{2}, 1/\sqrt{6}, 1/(2 \sqrt{3}), 1/(2 \sqrt{5})\} \nonumber\\
a_2 &=& \{-(1/\sqrt{2}), 1/\sqrt{6}, 1/(2 \sqrt{3}), 1/(2 \sqrt{5})\} \nonumber\\
a_3 &=& \{0, -\sqrt{2/3}, 1/(2 \sqrt{3}), 1/(2 \sqrt{5}) \} \nonumber\\
a_4 &=& \{0, 0, -(\sqrt{3}/2), 1/(2 \sqrt{5}) \} \nonumber\\
a_5 &=& \{0, 0, 0, -(2/\sqrt{5}) \}
\ea
All corners have norm $4/5$, and all ten mutual scalar products are equal to $-1/5$. Consequently, all angles between pairs of vectors $\theta_{i>j}$ are equal, with $\cos(\theta_{i>j})=-1/4$.

There are ten line segments connecting pairs of corners $a_i,a_j$ with $i\neq j$, known as "edges''. For example, the above expressions show that the edge connecting $a_4$ and $a_5$ lies in the $\alpha=\beta=0$ plane. Associated with these edges are "faces'' defined by triples of corners $a_i,a_j,a_k$ ($i\neq j\neq k$), such as $(1,2,3)$. These faces are two-dimensional equilateral triangles. 
By "across'' we mean that the centers of edges, $(a_i+a_j)/2$, and the centers of complementary faces, $(a_i+a_j+a_k)/3$, are connected by straight lines passing through the global center $\alpha=\beta=\gamma=\delta=0$.

Our task is now to formulate quantum mechanics on this simplex. As in the baryon case, the confining part of the Hamiltonian becomes a quadratic form, which in momentum space turns into a Laplacian:
\be 
M^2=2P^+P^-=H_{conf}+V_{cup}+25\langle p_{\perp i}^2+m_i^2\rangle +H_{\lambda\lambda \sigma \sigma} .
\ee
The "cup potential'' is defined as
\be 
V_{cup}\sim \bigg(\sum_1^5{(p_{\perp i }^2+m_i^2) \over X_i}\bigg) -25\langle p_{\perp }^2+m^2\rangle .
\ee
Its name reflects its shape: it rises sharply near all boundaries of the simplex and is near zero in the middle, $\alpha\approx \beta\approx \gamma\approx\delta\approx 0$. The third term is a constant subtracted from the potential for convenience, making it approximately zero at the center; it is correspondingly added to the confining part of the Hamiltonian. The final color-spin term will be discussed later.

To avoid divergences at the boundaries of the simplex, the wave functions must vanish there. More precisely, we restrict ourselves to cases with finite $\psi V \psi$, which requires the wave function to vanish near a boundary as $\psi \sim (A-A_{\text{boundary}})^p$ with power $p>1/2$.

Using the Ritz variational approach, we start with the simple ansatz
\be \label{eqn_ansatz_faces}
\Psi_{faces}=\big(\prod_{i=1..5} x_i(\alpha,\beta,\gamma,\delta\in \text{"starfish''})\big) .
\ee
With the $x_i$ given by Eq.~(\ref{eqn_Jacobi_5}), this ansatz not only enforces Dirichlet boundary conditions but is in fact close to the true ground state. Using the Ritz method with a correction function consisting of several Gaussians, we found that corrections are at the level of a few percent, which we neglect in what follows.

An alternative approach, also developed earlier for baryons and the Laplacian on a triangle, is based on constructing an appropriate superposition of plane waves. These waves must vanish on all ten faces of the $A_4$ simplex and be eigenfunctions of the Laplacian. The required number of plane waves is determined by the size of the relevant permutation group $S_N$: it is $3!=6$ for baryons and $5!=120$ for the "starfish''.
The details of this construction are given in Appendix~\ref{sec_star}.

\begin{figure}[t!]
    \centering
        \includegraphics[width=0.75\linewidth]{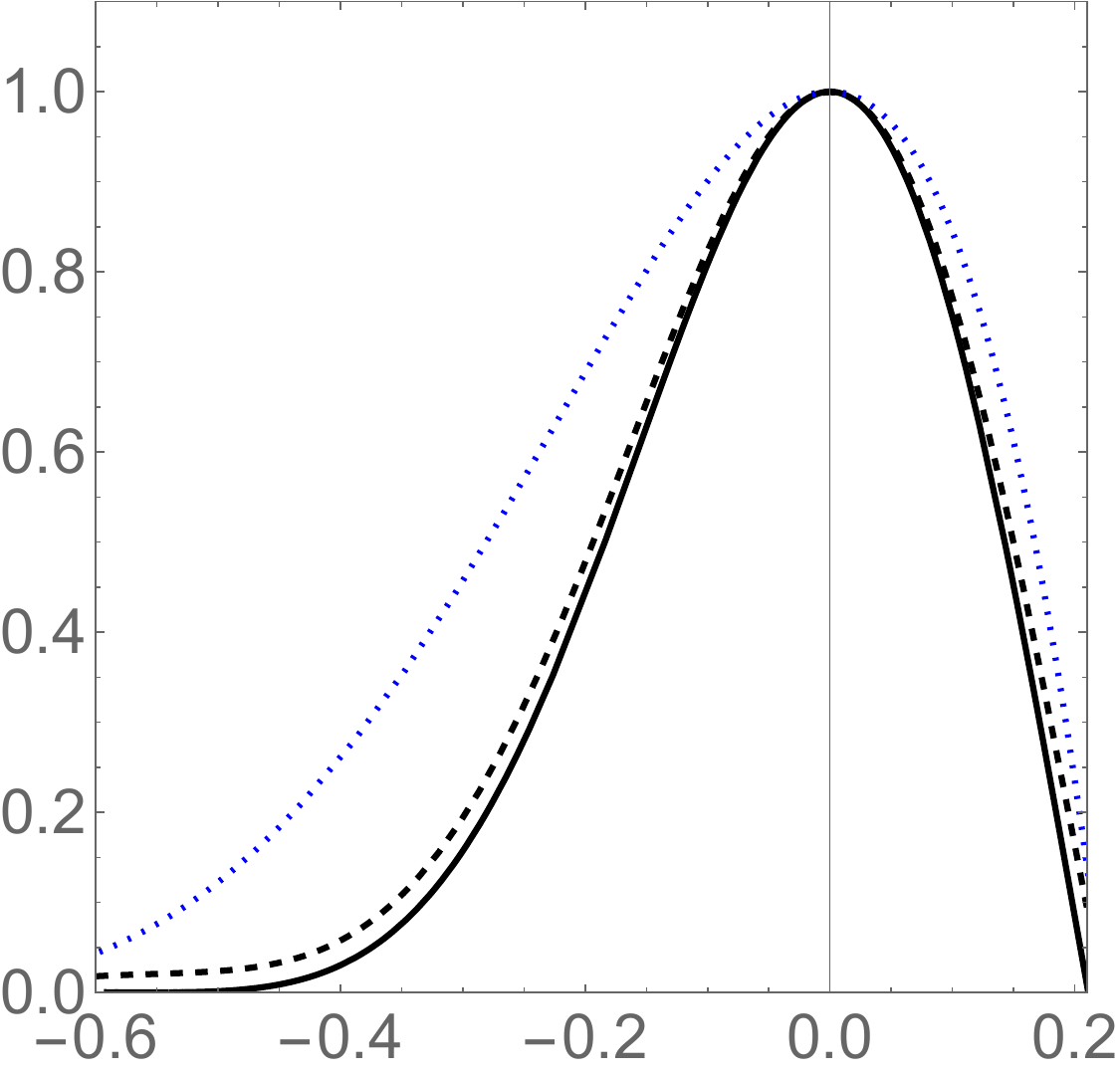}
    \caption{
    The lowest Laplacian eigenfunction $\psi[1,1,1,1]$ at $\alpha=\beta=\gamma=0$ versus $\delta$ (solid line) compared to our simplest  variational function (\ref{eqn_ansatz_faces}) (dotted line) and its improvement by Ritz variational procedure (dashed line). For a more clear comparison all are
    normalized to their value at the origin. }
    \label{fig_corr1}
\end{figure}

\begin{figure}[b!]
    \centering
    \includegraphics[width=0.75\linewidth]{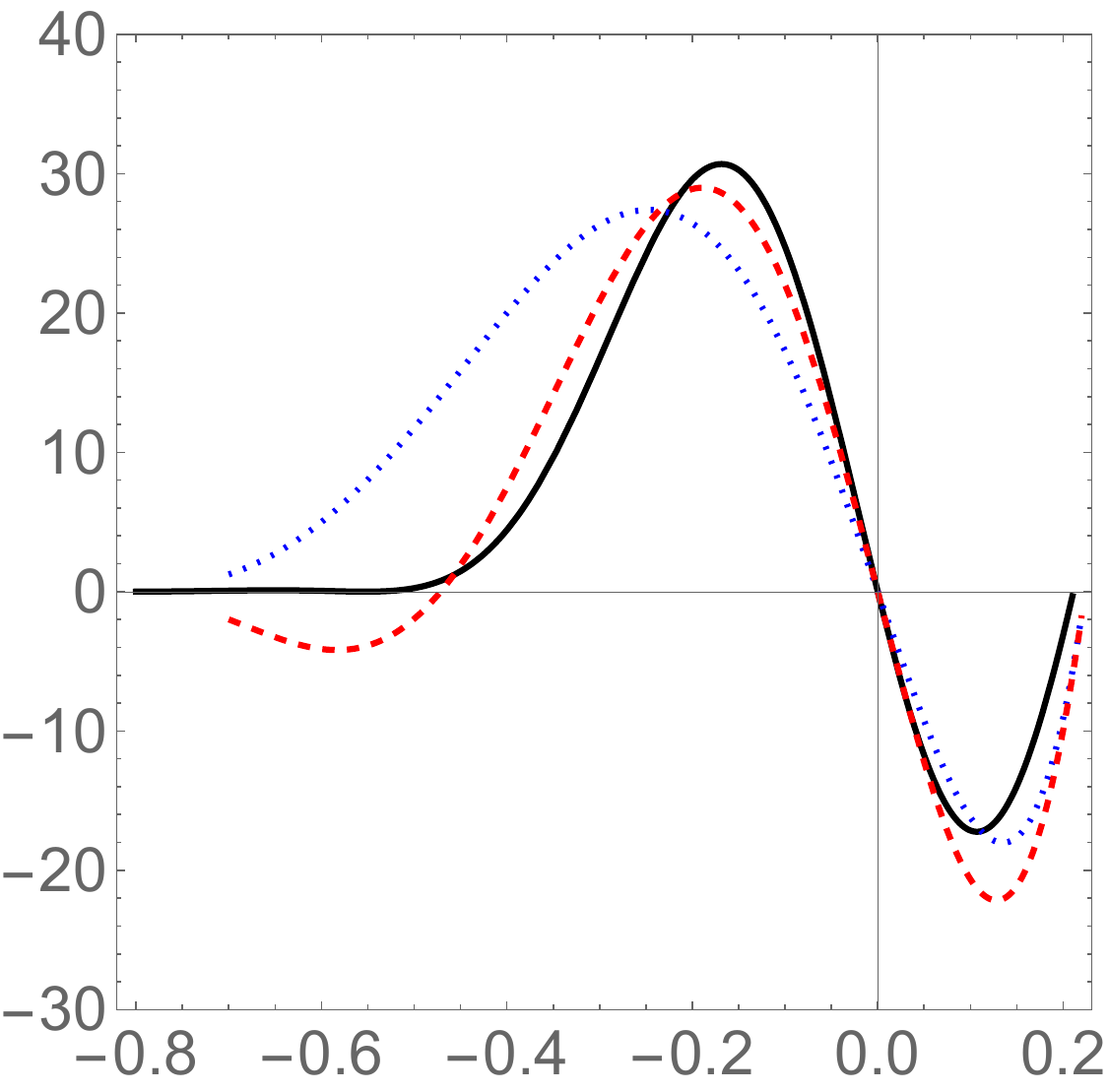}
    \caption{Comparison between the shapes of the P-shell ($L=1$) wave functions (arbitrary units) on 5-simplex $A_4$, at $\alpha=\beta=\gamma=0$ as a function of $\delta$.
  Black solid line is the exact Laplacian eigenstate with $\psi[\{2,1,1,1\}]$,
   blue dotted is initial trial function (\ref{eqn_psi1}) for $C=0$,  and red dashed is variational function with a single parameter, at value $C=4.5$. }  
    \label{fig_L1_var}
\end{figure}

The results of our study of wave functions on $A_4$ are summarized in two plots, corresponding to the ground $S$-shell and the first excited ($P$-shell) states. In Fig.~\ref{fig_corr1} we show the shape of the ground-state Laplacian eigenfunction on $A_4$, and in Fig.~\ref{fig_L1_var} that of the first excited states. While the variational Ritz method successfully reproduces the overall shapes of these wave functions (indeed, these calculations were performed before we discovered the exact eigensystem in the mathematical literature), it does not accurately describe their asymptotic tails.



The pentaquark wave function  provides a full description of the state, with all observables.
For example, it yields readily the probability momentum fraction $P(x_{\bar q})$. This is achieved by integration of the wave function squared over $\alpha,\beta,\gamma$ variables , as a function of  $\delta$. 
$$ PDF=\int_{\alpha,\beta,\gamma} |\Psi|^2 $$
(We recall  that the  4-th
Jacobi coordinate $\delta$ is 
directly related to the momentum fraction $x_5$, see (\ref{eqn_Jacobi_5}).)


\section{Pentaquark orbital-color-spin-flavor wave functions on the light front}

The nontrivial part of the construction concerns the wave functions (WFs)
depending on color, spin, and flavor variables.
The essential difference from traditional spectroscopy in the center-of-mass
(CM) frame is the symmetry structure: in the CM frame there is full spherical
$O(3)$ symmetry, while on the light front (LF) only axial $O(2)$ symmetry
remains.
As a result, the usual classification in terms of $J$, $L$, and $S$ is no longer
available; instead, only the projections $J_z$, $L_z$, and $S_z$ are good
quantum numbers.
LF wave functions of this type have been considered previously in the literature; see, for example \cite{Ji:2003yj,Belitsky:2005qn}.

Spectroscopy in the CM and on the LF therefore differs fundamentally due to the
absence of spherical symmetry in the latter case.
In the CM frame, a wave function with fixed total angular momentum and
projection $J,J_z$ (e.g.\ $J=1/2$, $J_z=1/2$) is constructed as a standard sum
over orbital projections using Clebsch-Gordon coefficients,
\[
\sum_{L_z}
\big(^{1/2,1/2}_{L,L_z,S,1/2-L_z}\big)
Y_{L,L_z}(\theta,\phi)\,|S,S_z\rangle ,
\]
which describes the addition of angular momenta $\vec L+\vec S=\vec J$.
On the LF, however, the quantum numbers $L$ and $S$ are not defined.
For example, in the $P$ shell the three possible values $L_z=-1,0,1$ and the six
possible values $S_z=-5/2,-3/2,\ldots,5/2$ can be combined freely, with the only
global constraint being their sum $J_z$.
Mixing of states with the same $J_z$ is determined by the spin-dependent part of
the Hamiltonian.

These mixing coefficients are to be fixed by diagonalizing the LF Hamiltonian.
Components with $L_z=\pm 1$ correspond to adding one quantum of the transverse
oscillator, while the $L_z=0$ component contains an extra linear coordinate
\[
z=-i\frac{\partial}{\partial p_z}\, .
\]

Since in our previous paper \cite{Miesch:2025wro} we derived pentaquark WFs in the
$S$ and $P$ shells, all that is required here is to repeat the same procedure on
the light front.
In the $S$ shell ($L=0$), the only modification concerns the spin
classification.
For five quarks there are six possible values of the spin projection,
$S_z=-5/2,-3/2,\ldots,5/2$.
The extreme case $S_z=5/2$ corresponds to the unique fully symmetric spin state
$\uparrow\uparrow\uparrow\uparrow\uparrow$.
In this case, Fermi statistics is enforced entirely by the (nontrivial)
color$\otimes$flavor wave functions already constructed in
\cite{Miesch:2025wro} for isospin $I=1/2$ and $I=3/2$.
(No states with $I=5/2$ exist.)

For convenience, we reproduce a shortened version of the table from that work,
which gives the number of antisymmetric states with fixed $I$ and $S_z$:

\begin{table}[h!]
    \centering
    \begin{tabular}{|c|c|c|c|} \hline
  $I/S_z$    & 1/2 & 3/2   & 5/2  \\  \hline
 -5/2  & 1 (15) & 1 (12) & 0\\
 -3/2 & 4 (75) & 4 (60) & 1 (15)\\
 -1/2 & 7 (150) & 7 (120)  & 2 (30)\\
  1/2 &  7 (150) & 7 (120)  & 2 (30) \\
  3/2 & 4 (75) & 4 (60) & 1 (15)    \\
  5/2 & 1 (15)  & 1 (12)  & 0 \\ \hline
    \end{tabular}
    \caption{Spin projection $S_z$ versus isospin $I$ for antisymmetric
    pentaquark states with $L=0$.
    Numbers in parentheses give the dimensions of the "good basis'' used for
    constructing and diagonalizing permutation matrices.}
    \label{tab_L0_LF}
\end{table}

With the new axially symmetric $(I,S_z)$ classification, both the number of
states and the permutation generators are organized differently.
Generating the analogous table in the $(I,S_z)$ basis, we find the same total
number of states, albeit distributed differently among the spin projections.

It is instructive to explain how this counting is related to the CM-frame
classification used in \cite{Miesch:2025wro}.
Consider, for example, the total number of states with $I=1/2$.
In the CM-frame table of \cite{Miesch:2025wro}, this number is
\[
2\times 3 + 4\times 3 + 6\times 1 = 24 ,
\]
where the first factors are the multiplicities $2S+1$ for fixed total spin $S$.
On the LF, spherical symmetry is absent and the total spin $S$ is not used.
Instead, the same states are counted as
\[
2(7+4+1)=24 ,
\]
where the factor of 2 accounts for the doubling due to negative values of $S_z$.
Thus the same physical states are present, but organized differently.
The different $2S+1$ orientations of a spin-$S$ multiplet, which are related by
rotations in the CM frame, are no longer related on the LF, where only axial
rotations preserving $S_z$ remain.

We now introduce the color-spin Hamiltonian,
\begin{equation}
\label{ean_H_lambdalambdasigmasigma}
H_{\lambda\lambda\sigma\sigma}
=
-\sum_{i>j}
V^{ij}_{\lambda\lambda\sigma\sigma}
(\vec\lambda_i\vec\lambda_j)
(\vec\sigma_i\vec\sigma_j) ,
\end{equation}
and compute the Hamiltonian matrix including all states.
Here the potential is replaced by its average value evaluated with the dominant
symmetric WF, and back-reaction on the spins is neglected.
Although the individual matrix elements are complicated $\mathcal{O}(1)$
numbers, diagonalization yields eigenvalues that reproduce exactly those found
in the alternative classification of \cite{Miesch:2025wro}.

In particular, the seven states with $S_z=1/2$ and $I=1/2$ have energies
\begin{align}
E_n &= 4.6667,\; 3.0000,\; 1.4415,\; -2.7749,\nonumber\\
&\phantom{=} -0.33333,\; -3.3333,\; -3.3333 ,
\end{align}
all of which were reported previously.
The four states with $S_z=3/2$ and $I=1/2$ have energies
\[
E_n = 3.000000,\; -0.3333333,\; -3.333333,\; -3.333333 .
\]
The repetition of eigenvalues between different $S_z$ sectors is a consequence
of the (now hidden) $O(3)$ symmetry of the Hamiltonian, which is not imposed
explicitly but remains present.

Turning to what were formerly called $P$-shell states, which involve linear
dependence on coordinates, one finds that the construction requires working in
a space of monomials of dimension $4\times 3^6\times 2^5\times 2^5$.
In the CM frame, one adds angular momenta $\vec L+\vec S=\vec J$ and classifies
states by $J$ and $J_z$ using Clebsch-Gordon coefficients.
On the LF, this is replaced by the much simpler condition $L_z+S_z=J_z$.

For $I=1/2$, $S_z=1/2$, and nominal $L=1$ states, the resulting energies are
\begin{align}
E_n &= 9.05,\; 9.05,\; -7.33,\; 4.67,\; 4.67,\; -3.33,\nonumber\\
&\phantom{=} -3.33,\; -3.33,\; -3.33,\; -3.33,\; -3.33,\; -3.33,\nonumber\\
&\phantom{=} 3.00,\; 3.00,\; 3.00,\; -2.77,\; -2.77,\; -2.77,\nonumber\\
&\phantom{=} 1.67,\; 1.67,\; 1.62,\; 1.62,\; 1.44,\; 1.44,\nonumber\\
&\phantom{=} 1.44,\; -0.333,\; -0.333 ,
\end{align}
with further repetitions for larger values of $S_z$ due to the presence of
multiple components with the same total $J$.

For baryon-pentaquark mixing, discussed below, one must use negative-parity
$P$-shell ($L=1$) pentaquark states.
In the CM-frame treatment, this was achieved by using wave functions linear in
the coordinates (on top of radial functions).
As reviewed in \cite{Miesch:2025wro}, enforcing Fermi statistics leads to
nontrivial orbital-color-spin-flavor structures; see Table~IV of that work.

On the light front, the same construction applies.
Linear dependence on coordinates is replaced by linear dependence on momenta.
This change does not affect permutation properties, so the
orbital-color-spin-flavor structure of the states remains unchanged.
There are again two classes of structures: those with $L_z=\pm 1$, involving
transverse momenta, and those with $L_z=0$, involving longitudinal momenta.
While Table~IV of \cite{Miesch:2025wro} lists antisymmetric $L=1$ structures with
all possible total spins $S$ and isospins $I$, here we present the analogous
classification in terms of spin projections $S_z$.

\begin{table}[h!]
    \centering
    \begin{tabular}{|c|c|c|c|} \hline
  $I/S_z$    & 1/2 & 3/2   & 5/2  \\  \hline
 -5/2  & 3 (60) & 3 (48) & 1 (12) \\
 -3/2 & 14 (300) & 13 (240) & 4 (60)\\
 -1/2 & 27 (600) & 24 (480)  & 7 (120)\\
  1/2 & 27 (600) & 24 (480)  & 7 (120) \\
  3/2 & 14 (300) & 13 (240) & 4 (60)    \\
  5/2 & 3 (60) & 3 (48) & 1 (12) \\ \hline
    \end{tabular}
    \caption{Spin projection $S_z$ versus isospin $I$ for antisymmetric
    pentaquark states with $L=1$.
    Numbers in parentheses denote the dimensions of the corresponding
    "good basis''.}
    \label{tab_L1_LF}
\end{table}

For mixing with the nucleon, one must select pentaquark states with $I=1/2$.
We therefore focus on the 27 states with $S_z=1/2$, corresponding to
$J_z=1/2$ and $L_z=0$.
These 27 $P$-shell wave functions are all fully antisymmetric under the quark
permutation group $S_4$.
The linear combinations of longitudinal momenta are then combined with the
color, spin, and isospin structures constructed in our previous work.


\section{Chiral evolution and flavor asymmetry of the sea}

Our {\em starting point} is the well-known traditional quark model used in hadronic spectroscopy.
The key phenomenon incorporated in this model is chiral symmetry breaking, which generates an
effective mass for the "constituent quarks.'' For light quarks this mass is
$m_q \sim 1/3 \, \mathrm{GeV}$. This scale is much smaller than the induced mass of gluons, and
therefore hadronic spectroscopy is traditionally described in terms of bound states of
constituent quarks, while gluonic states or excitations are treated as "exotica.''
The conventional states are two-quark mesons and three-quark baryons, although
tetraquarks $q^3 \bar q$ and pentaquarks $q^4 \bar q$ have also been observed recently,
primarily with heavy-quark content.

The {\em first ark of the bridge} (described in detail in this series of works) is the transfer
of such quark models from the center-of-mass frame to the light front.
For the simplest cases-such as heavy quarkonia-this amounts to a change from spherical
to cylindrical coordinates, followed by a transformation of longitudinal momenta into
the Bjorken-Feynman variable $x$. In general, however, it is more convenient to start directly
from light-front Hamiltonians $H_{LF}$ and perform their quantization.
One important advantage is that no nonrelativistic approximation is required, so that heavy
and light quarks are treated on the same footing.

The {\em second ark of the bridge} is constructed via {\em chiral dynamics}, which seeds the
quark sea through the production of additional quark-antiquark pairs.
Below we discuss how this can be implemented, both at first order in
the 't~Hooft effective action and via intermediate pions.

As the {\em third ark of the bridge}, we argue that one should employ the well-known DGLAP
evolution of parton distribution functions (possibly with modifications), evolving down to
a scale at which gluons are absent.
At that point, the $q\bar q$ sea should be reduced to the component generated purely by
chiral dynamics (step two).
The antiquark flavor asymmetry $\bar d-\bar u$ provides a key diagnostic, since it cannot
be generated by flavor-blind gluons and therefore cleanly separates gluonic and chiral
contributions.

\section{Properties of the "nucleon sea'' and baryon-pentaquark mixing}

\section{Baryon-pentaquark mixing on the light front}

In our paper \cite{Shuryak:2022wtk} we employed a "radiative'' (DGLAP-like) description,
in which a $\bar q q$ pair is produced from one of the quarks.
The mixing was attributed to the 't~Hooft four-fermion Lagrangian, which allows
the transition $u\rightarrow u d \bar d$ but not $u\rightarrow u u \bar u$.
Another mechanism considered was production via an intermediate pion propagator.
The probability, as a function of the momentum fraction of the initial quark,
was convoluted with its PDF in the nucleon, e.g.\ $u^{\mathrm{proton}}_v(x)$.
In this approach, any interference between the newly created quark and the spectator
quarks in the nucleon was neglected.
Finally, these calculations were performed at the level of density matrices (PDFs),
rather than using coherent wave functions.

It is worth noting that the observed flavor asymmetry of the antiquark sea, as well as
the presence of significant orbital angular momentum, can both be traced to
$\sigma$- and $\pi$-meson clouds in the nucleon.
The technical description in \cite{Miesch:2025wro} follows earlier literature and projects
the baryon wave function, multiplied by mesonic operators $T_\sigma$ and $T_\pi$,
onto pentaquark states.
The so-called meson-baryon-pentaquark overlap is given by an integral over all
coordinates $\vec x_i$, $i=1\ldots5$:
\be \label{eqn_Cn}
C_n=\int_x\Psi^{penta}_n(\vec x_i)\,\hat T(\vec x_4-\vec x_5)\,
B(\vec x_1,\vec x_2,\vec x_3) .
\ee
The antiquark is associated with the last coordinate $\vec x_5$.
All coordinates are expressed in terms of four Jacobi coordinates
$\vec \alpha,\ldots,\vec \delta$.
Each factor consists of a "main'' (radial) wave function, depending on the corresponding
hyperdistance ($Y_5$ for the pentaquark and $Y_3$ for the baryon), and the associated
color-spin-flavor structure, represented as vectors in the corresponding "monom'' spaces.

The present task is to reformulate all of these ingredients within the light-front framework.
The baryon and pentaquark wave functions are modified accordingly.

The nucleon color-spin-flavor structure remains as described in our previous work.
The radial wave function $B(Y)$, which depends on the hyperdistance
$Y_B^2=\vec \alpha^2+\vec \beta^2$, is split into a product of functions depending on
transverse and longitudinal momenta:
$$
B(\vec x_i) \rightarrow
B_\perp(\vec p_i^\perp)\, B_{\Delta 3}(\alpha,\beta) .
$$
The function $B_{\Delta 3}(\alpha,\beta)$ represents the ground-state baryon wave function
on the triangle and depends on two momentum fractions.\footnote{
For different baryons these fractions differ; for example, we discuss the difference
between the $\Delta$ and nucleon cases due to the "good diquark'' effect in the latter.}

The procedure for the pentaquark is analogous, with the wave function factorizing into
transverse and longitudinal components:
$$
\Psi(\vec x_i) \rightarrow
\Psi_\perp(\vec p_i^\perp)\, \Psi_{\Delta 5}(\alpha,\beta,\gamma,\delta) .
$$
This factorization reflects the independence of transverse and longitudinal variables
in the light-front Hamiltonian employed.
The longitudinal wave functions are defined on the 5-simplex $\Delta 5$, as introduced
earlier in this paper.

An additional complication arises from the fact that admixtures with nucleon-meson
configurations require P-shell pentaquarks.
This is a nontrivial issue, since part of the center-of-mass wave function contains
terms linear in the coordinates.
Because these terms have nontrivial transformation properties under quark permutations,
their treatment required extensive calculations of orbital-color-spin-flavor
wave functions obeying Fermi statistics \cite{Miesch:2025wro}.
Upon boosting to the light front, one finds that longitudinal momenta satisfy
$p_z \gg p_\perp$.
As a result, the components with $L_z=\pm1$, constructed from transverse momenta,
become subleading, and only the $L_z=0$ component survives.
This leads to a significant simplification, since $J_z=S_z$.
The longitudinal part is then generalized from the previously constructed
orbital-color-spin-flavor wave functions, with the coordinates replaced by
longitudinal momentum fractions $\alpha,\beta,\gamma,\delta$.

The mesonic operator $\hat T$ contains the $\bar q q$ vertex, which in coordinate space
is conventionally written as a Gaussian
$\exp[(\vec x_4-\vec x_5)^2/2\rho^2]$.
In momentum space this becomes the corresponding Gaussian form.
We assume that the boost to the light front compresses hadrons longitudinally, leaving
only the transverse-momentum dependence.

The operators providing the appropriate quantum numbers in $\hat T$ are
$(\vec S\cdot\vec L)$ for $T_\sigma$ and $(\vec S\cdot\vec p)$ for $T_\pi$.
Here $\vec S=\vec S_4+\vec S_5$ is the total spin of the quark-antiquark pair.
Since boosts leave only rotations in the transverse plane, we make the replacement
$$
(\vec S\cdot\vec L)\rightarrow S_z L_z .
$$
For the pion-related operator the situation is different: longitudinal momenta dominate,
and we therefore use
$$
(\vec S\cdot\vec p)\rightarrow (S^z_4 p_4+S^z_5 p_5) .
$$
The longitudinal momenta of the fourth quark and fifth antiquark are expressed in terms
of the Jacobi variables $\gamma$ and $\delta$.

Finally, we address isospin in the $N+\pi$ and $\Delta+\pi$ channels.
Unlike spin $S$, orbital angular momentum $L$, and total $J$, which are not good quantum
numbers in the light-front formulation, isospin remains well defined.
Adding a $\sigma$ meson presents no difficulty, but adding a pion with $I_\pi=1$ to
a nucleon or $\Delta$ requires that the total isospin match that of the proton,
$I=1/2$, $I_z=1/2$.
This is achieved using standard Clebsch-Gordan coefficients:
\ba \label{eqn_admixtures}
|N\pi\rangle &=& {1 \over \sqrt{3}} |p \pi^0\rangle-{\sqrt{2 \over 3}}|n \pi^+\rangle \\
|\Delta\pi\rangle &=& {1 \over \sqrt{2}} |\Delta^{++} \pi^-\rangle
-{\sqrt{1 \over 3}}| \Delta^+\pi^0\rangle
+{1\over \sqrt{6}}| \Delta^-\pi^+\rangle \nonumber
\ea
In general, there are three possible admixtures to the nucleon, with overlap coefficients
denoted by $C_{p\sigma}$, $C_{N\pi}$, and $C_{\Delta\pi}$.
Although their contributions will eventually be summed, we study the three admixtures
separately; see Table~\ref{tab_admixtures}.

Using the 27 $L=1$ pentaquark states defined above, we evaluated the projections of
$p+\sigma$, $N+\pi$, and $\Delta+\pi$ states onto these configurations.
The overlap coefficients $C_n$ in Eq.~(\ref{eqn_Cn}) were calculated as described in the
previous section, using appropriate combinations of coordinates in $|P_n\rangle$
together with the Ansatz function (\ref{eqn_ansatz_faces}).
In this calculation we included only functions normalized over momentum fractions
$\alpha,\beta,\gamma,\delta \in \mathrm{starfish}$.

\begin{figure}
    \centering
    \includegraphics[width=0.85\linewidth]{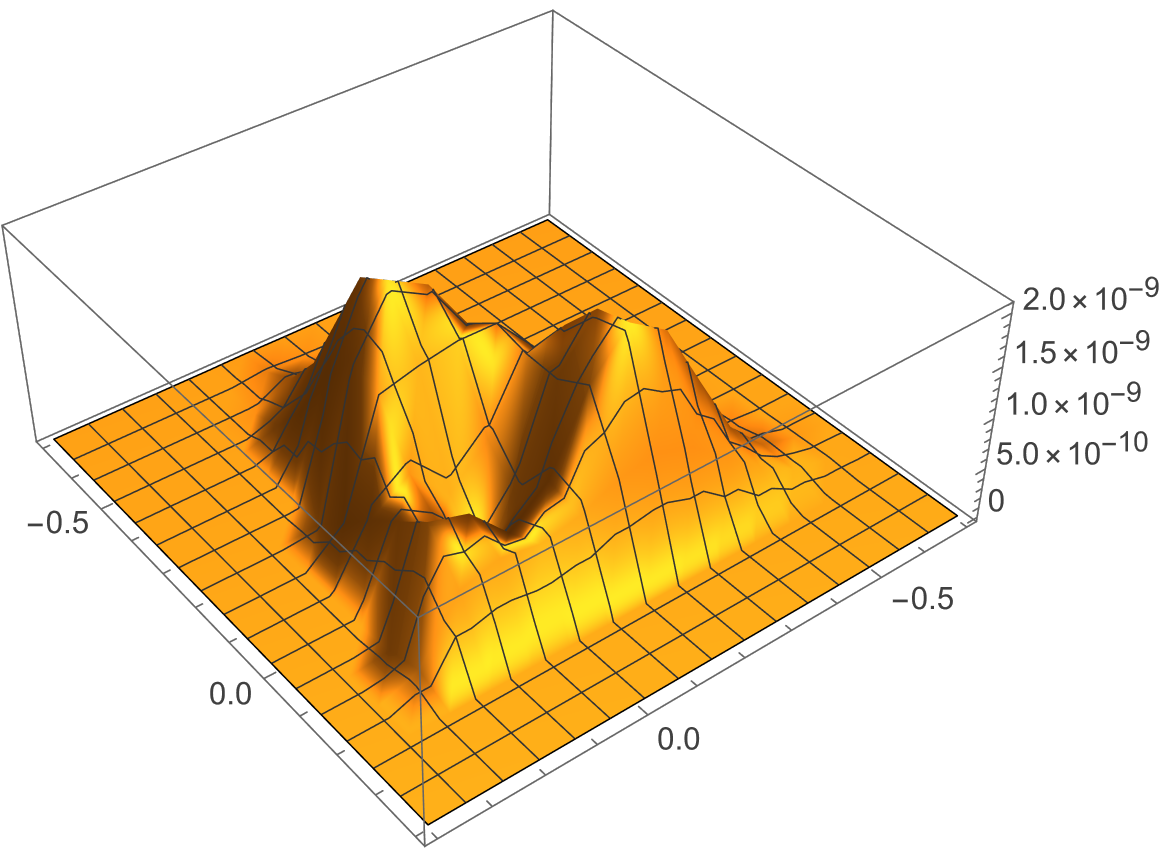}
    \caption{The shape of $\Delta \psi_{N\sigma}^2$ on the $\alpha=\beta=0$ and
    $\gamma,\delta$ plane.}
    \label{fig_Nsigma_shape}
\end{figure}

The individual 27 pentaquark states are not spherical, and we project onto
baryon-plus-meson wave functions in which particles 1-3 form the baryon and
particles 4-5 form the meson.
Nevertheless, one of the admixtures turns out to be spherical:
$$
\Delta \psi_{p\sigma}\cdot\Delta \psi_{p\sigma}
\sim (\alpha^2+\beta^2+\gamma^2+\delta^2) .
$$
This factor changes the shape of the wave function, which is shown in Fig.~\ref{fig_Nsigma_shape}.

\begin{figure}
    \centering
    \includegraphics[width=0.85\linewidth]{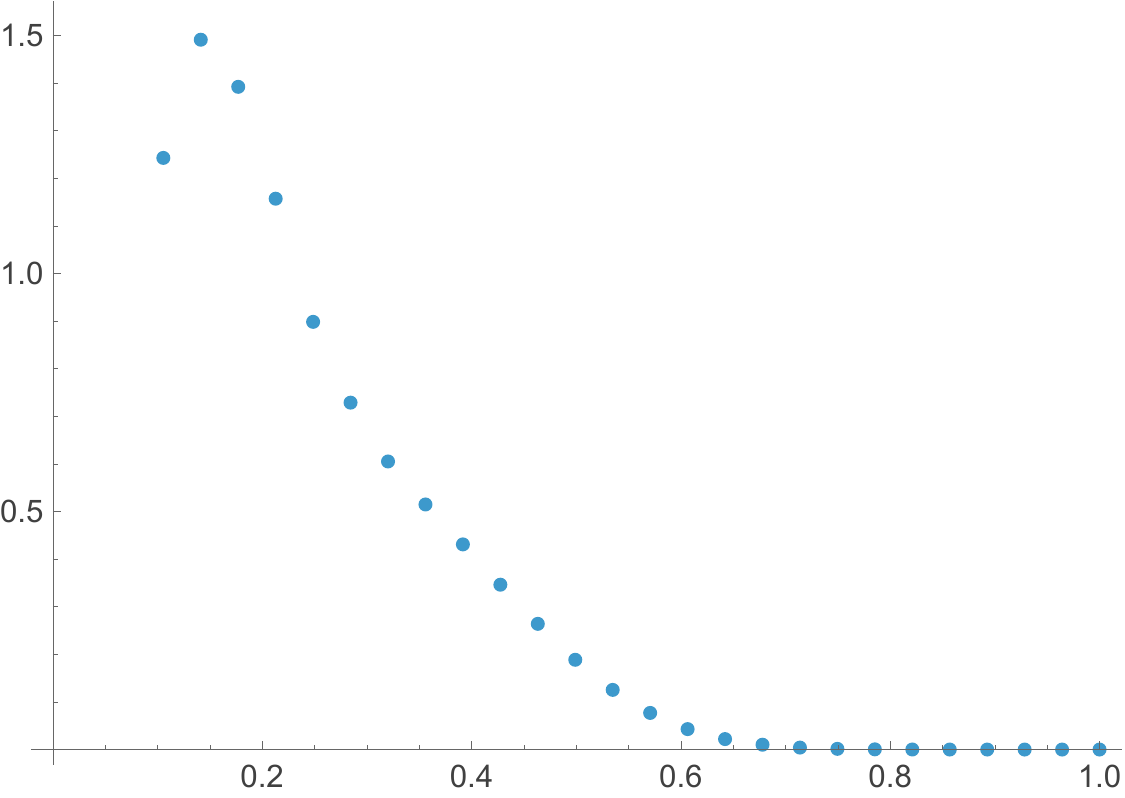}
    \caption{Predicted antiquark PDF from $\Delta\psi_{p\sigma}$ (arbitrary units)
    as a function of the Bjorken momentum fraction $x$.}
    \label{fig_Nsigma_PDF}
\end{figure}

Accordingly, the antiquark PDF (obtained by integrating $\Delta \psi^2$ over
$\alpha,\beta,\gamma$) is modified, as shown in Fig.~\ref{fig_Nsigma_PDF}.
Compared to the predictions for the ground ($L=0$) shell pentaquarks, the distribution becomes narrower and shifts toward
smaller $x$~\cite{Miesch:2025wro}.
Nevertheless, the large-$x$ behavior (more clearly visible in the lower panel)
exhibits the same $\sim (1-x)^8$ scaling.

Some properties of the three different five-quark admixtures to the proton are
listed in Table~\ref{tab_admixtures}.
We begin with the mean spin of the antiquark (particle~5),
\be
\langle S_5 \rangle =\Delta \psi_{A}\cdot S_5\cdot \Delta \psi_{p\sigma},
\ee
where $A=p\sigma,N\pi,\Delta\pi$.
All three admixtures yield negative contributions, which should be compared
to the total nucleon spin $S_z=+0.5$.

Another important observable is the isospin projection of the antiquarks.
We define the operators projecting onto $\bar d$ and $\bar u$ flavors as
\be
I_{\pm}= (\hat 1\pm \tau_z)/2 ,
\ee
applied to particle~5 (the antiquark).
Note that the isospin components for antiquarks are inverted.
In this case, the $p\sigma$ and $N\pi$ admixtures favor $\bar d$, while the
$\Delta\pi$ channel favors $\bar u$.

\begin{table}[]
    \centering
    \begin{tabular}{|c|c|c|c|} \hline
    operator & $p\sigma$ & $N\pi$ & $\Delta \pi$ \\
      $S_z(5)$   &  -0.177 & -0.112 & -0.0338 \\
      $\bar d=I_+(5)$ &  0.81  & 0.858 &  0.333 \\
      $\bar u=I_-(5)$ &  0.19  & 0.142 &  0.666 \\ \hline
    \end{tabular}
    \caption{Properties of the three five-quark admixtures to the nucleon
    [Eq.~(\ref{eqn_admixtures})], defined as expectation values of the operators
    listed in the first column.}
    \label{tab_admixtures}
\end{table}




\begin{subappendices}
\label{sec_star}
\section{Pentaquarks on LF: the forward wave function}
The longitudinal part of the LF Hamiltonian is not an oscillator, but rather a Laplacian (confining) operator supplemented by an irreducible potential $V$. For five quarks, the five momentum fractions are conveniently redefined through Jacobi combinations belonging to the pentachoron ("starfish'') or $A_4$ manifold of momentum fractions $\alpha,\beta,\gamma,\delta$. In this appendix we discuss several approaches to quantizing this system.

 \subsection{N fermions on a circle} 
 \label{sec_fermions_on_circle} 

Since fermions cannot pass each other in one dimension, their coordinates $s_i$ are ordered as
$s_1<s_2<\dots<s_N<s_1+1$. One may furthermore fix the center-of-mass coordinate by imposing $\sum_i s_i=0$, in which case the physical domain becomes a four-dimensional simplex. Geometrically, this simplex is similar (though not identical) to our symmetric simplex.

The fermions-on-a-circle problem admits known analytic solutions for the wave functions in the form of Slater determinants~\cite{Girardeau:1960cnk}, see Eq.~(\ref{eqn_slater_N}). These depend on four independent momenta $n_{21},n_{31},n_{41},n_{51}$, which must all be nonzero and distinct in order for the determinant not to vanish. For the ground state one naturally selects an "occupied Fermi sphere'' configuration, taking the lowest possible values $-2,-1,1,2$ (recall that zero momentum is already included). The corresponding Laplacian eigenvalue is proportional to the sum of squares
$0^2+2\times1^2+2\times2^2=10$.

Expanding the determinant yields a lengthy sum of plane waves. Remarkably, however, this expression can be simplified into a product of ten sinusoidal factors, one for each pair $i\neq j$,
\be \label{eqn_prod_sin}
\psi_{sin} \sim \prod_{i\neq j} \sin\!\left( {s_i-s_j \over 2}\right).
\ee
Unfortunately, the four-dimensional simplex relevant to this problem is not the same as the one required for pentaquarks. Although a mapping between the two domains exists, under this mapping the Laplacian transforms into a different elliptic operator. As a result, the elegant fermions-on-a-circle solution does not directly solve the problem of interest here.


We now observe that the standard simplex $\Delta_N[x_1,\ldots,x_N]$ is geometrically isomorphic to the ordered simplex $\Delta_N^\ast[s_1,\ldots,s_N]\subset\mathbb{R}^N$,
\bea
\label{SNORD}
 s_1\leq s_2\leq \dots\leq s_{N}\leq s_1+a,
\eea
where the parameter $a$ is fixed by the side length of the simplex. The faces of the ordered simplex correspond to configurations in which two successive coordinates coincide, $s_i=s_{i+1}$. The light-front wave functions defined on this domain are therefore equivalent to those of $N$ bosons interacting via hard-core repulsion on a circle of unit circumference,
\bea
\varphi[x]\rightarrow \varphi[s]\qquad {\rm with}\qquad
\varphi[s\,|\,s_i=s_j]=0.
\eea
Such impenetrable bosons are well known to map onto free fermions via the Pauli exclusion principle. The corresponding $N$-fermion wave functions are Slater determinants built from circular plane waves~\cite{Girardeau:1960cnk},
\bea
\label{SLAT1}
\varphi_n[s]=\frac {1}{N^{4/3}}
\begin{vmatrix}
    e^{i\tilde n_1s_1}& e^{i\tilde n_1s_2}  & \dots  & e^{i\tilde n_1s_N} \\
    e^{i\tilde n_2s_1}& e^{i\tilde n_2s_2}  & \dots  & e^{i\tilde n_2s_N} \\
    \vdots & \vdots & \ddots & \vdots \\
    e^{i\tilde n_Ns_1}& e^{i\tilde n_Ns_2}  & \dots  & e^{i\tilde n_Ns_N} 
\end{vmatrix},
\eea
with $\tilde n_i=2\pi n_i$. These wave functions have support on disjoint simplices related by permutation symmetry and possess eigen-energies
\bea
E[n]=\sum_{i=1}^N\tilde n_i^2.
\eea

The ordering $s_i\in[0,1]$ suggests an interpretation in terms of equidistant particles on a one-dimensional string of length one, undergoing longitudinal vibrations. To eliminate the center-of-mass coordinate, we perform the shift $s_j=\frac{S}{N}+\tilde s_j$. Substituting this into Eq.~(\ref{SLAT1}), and rearranging rows and columns of the Slater determinant, we obtain~\cite{He:2025dik}
\bea
\begin{vmatrix}
    e^{i\tilde n_1s_1}& e^{i\tilde n_1s_2}  & \dots  & e^{i\tilde n_1s_N} \\
    \vdots & \vdots & \ddots & \vdots \\
    e^{i\tilde n_Ns_1}& e^{i\tilde n_Ns_2}  & \dots  & e^{i\tilde n_Ns_N} 
\end{vmatrix}
=
\begin{vmatrix}
    e^{i\tilde n_1 \frac SN}e^{i\tilde n_1\tilde s_1}& \dots & e^{i\tilde n_1 \frac SN}e^{i\tilde n_1\tilde s_N} \\
    \vdots & \ddots & \vdots \\
    e^{i\tilde n_N \frac SN}e^{i\tilde n_1\tilde s_1}e^{i\tilde n_{N1}\tilde s_1}& \dots & e^{i\tilde n_N \frac SN}e^{i\tilde n_1\tilde s_N}e^{i\tilde n_{N1}\tilde s_N} 
\end{vmatrix}.
\eea
Extracting the common center-of-mass phase,
\bea \label{eqn_slater_N}
e^{i\frac SN\sum_{i=1}^N \tilde n_i},
\eea
yields
\bea
\label{SLAT4}
\varphi_n[s]=\frac {1}{N^{4/3}}
e^{i\frac SN\sum_{i=1}^N \tilde n_i}
\begin{vmatrix}
   1& 1  & \dots  & 1\\
   e^{i\tilde  n_{21}\tilde s_1}& \dots & e^{i\tilde  n_{21}\tilde s_N} \\
   \vdots & \ddots & \vdots \\
   e^{i\tilde  n_{N1}\tilde s_1}& \dots & e^{i\tilde  n_{N1}\tilde s_N} 
\end{vmatrix}
\equiv e^{i\frac SN{\sum_{i=1}^N \tilde n_i}}\varphi_{\tilde n}[\tilde s],
\eea
where $\tilde n_{i1}=\tilde n_i-\tilde n_1$ for $i=2,\ldots,N$. The coefficients $\tilde n_{i1}$ are ordered as $0<\tilde n_{21}<\tilde n_{31}<\tilde n_{41}<\dots$ to avoid redundant counting. Eliminating the center-of-mass degree of freedom yields reduced wave functions $\varphi_{\tilde n}[\tilde s]$ depending only on the $N-1$ relative coordinates, with energies
\bea
\label{SPECN}
\tilde E[\tilde n]&=&\frac{1}{\varphi_n[s]}\left[\left(\frac{\partial}{\partial s_i}\right)^2-N\left(\frac{\partial}{\partial S}\right)^2\right]\varphi_n[s]
\nonumber\\
&=&\sum_{i=1}^{N}\tilde n_i^2-\frac 1N\bigg(\sum_{i=1}^{N} \tilde n_i\bigg)^2 \nonumber\\
&=&\sum_{i=2}^{N}\tilde n_{i1}^2-\frac 1N\bigg(\sum_{i=2}^{N} \tilde n_{i1}\bigg)^2.
\eea
Since the set of Slater determinants is orthonormal and complete, the functions $\varphi_{\tilde n}[\tilde s]$ form a complete basis. The normal-mode coordinates may thus be identified with the Jacobi coordinates.

\subsection{Variational Ritz method for Laplacian eigenfunctions at $A_4$}
\label{sec_var}

For five bodies ($N=5$), the kinematical domain is a regular pentachoron, a four-dimensional simplex with five vertices, ten edges, ten equilateral faces, and five regular tetrahedra, which we refer to as a "starfish''. There are essentially two approaches to solving the Schrodinger equation in this geometry: (i) the variational Ritz method, employed in this work, and (ii) expansion in a complete basis defined on a different pentachoron arising in the $N$-fermion-on-a-circle problem, which will be discussed elsewhere~\cite{PENTAX}.

The Ritz method for elliptic partial differential equations is well established and relies on minimizing an energy functional over a finite set of parameters, typically the coefficients of a chosen basis. In the present case we explored several trial forms. For the ground state we ultimately settled on a relatively simple Ansatz involving Gaussian corrections,
\be \label{eqn_Ritz}
\Psi_{Ritz}= \Psi_{faces}\bigg[1+\sum_i^N C_i \exp(-Y^2/2R_i^2)\bigg],
\ee
where the first factor is the linear Ansatz (\ref{eqn_ansatz_faces}), and the bracketed term provides a correction function. The expectation value of the Laplacian (with or without the cup potential), normalized by the wave function norm,
$$
\langle - \nabla^2 \rangle ={\int_{simplex}\Psi_p(- \nabla^2) \Psi_p  \over \int_{simplex} \Psi_p^2},
$$
is minimized with respect to the coefficients $C_i$ using the \texttt{NMinimize} command. We tested various numbers of correction terms and choices of radii, ultimately returning to a set of three Gaussians. For the Laplacian-to-norm ratio we observed an improvement from the initial value $182$ down to $165$, corresponding to an improvement of roughly $10\%$.

We further found that including the cup potential,
$V_{cup}\sim (1/x_1+ 1/x_2+1/x_3+1/x_4+1/x_5-25)$,
reduces the magnitude of the correction function, bringing the ground-state wave function closer to the original linear Ansatz (\ref{eqn_ansatz_faces}). Consequently, this Ansatz was used in the calculations that follow.

The $P$-shell pentaquark states have negative parity, implemented through an additional factor linear in the coordinates. The simplest trial function we found that minimizes the Laplacian takes the form
\be 
\label{eqn_psi1}
\psi_1(C)\sim \Psi_{faces}\,\delta \,
\bigg(1-C(\alpha^2+\beta^2+\gamma^2+\delta^2)\bigg).
\ee
While the unmodified function $\psi_1(0)$ yields
$\langle - \nabla^2 \rangle / \langle 1\rangle\approx 312.0$,
the optimized choice $C=4.5$ reduces this value by approximately $10\%$. More elaborate trial functions lead to further, though modest, improvements. The resulting wave functions are compared in Fig.~\ref{fig_L1_var}. The change of sign near the origin is a robust feature, whereas a second zero near $\delta\approx -0.5$ in the variational solution is not robust and depends on the specific Ansatz. In the true $P_1$ state this second zero should be absent. We have verified that the variational function remains orthogonal to the $S$-shell ($L=0$) state within small numerical integration errors. We therefore conclude that the variational solution should not be trusted for $\delta<-0.5$.

\subsection{Exact Laplacian eigenfunctions at $A_4$}

The "equilateral'' (or regular) simplex on which we must solve the LF Schrodinger equation is known in the mathematical literature as the $A_4$ simplex. It is closely related to representations of the permutation group $S_4$. Problems involving tilings in arbitrary dimensions are intimately connected with the diagonalization of the Laplacian on such simplices, a connection that became clear to us only after most of this work had been completed.

We follow the derivation presented in Ref.~\cite{Doumerc_2008}. The so-called \emph{fundamental weights} in $\mathrm{dim}=5$ consist of four vectors $\omega_i$ embedded in five-dimensional space,
\ba 
&\omega&=\big\{ \{ 4/5, -1/5, -1/5, -1/5, -1/5\}, 
\{3/5, 3/5, -2/5, -2/5, -2/5\}, \nonumber \\
&&\{2/5, 2/5, 2/5, -3/5, -3/5\}, 
\{1/5,  1/5, 1/5, 1/5, -4/5\} \big\}.
\ea
Together with the origin $\{0,0,0,0,0\}$, these points form the five vertices of a simplex that is similar to, but distinct from, the "starfish'' simplex relevant for our problem. In Jacobi coordinates, the vertices of the latter are given by five different points $a_i$,
\ba 
&\{& 1/\sqrt{2}, 1/\sqrt{6}, 1/(2 \sqrt{3}), 1/(2 \sqrt{5})\}, \nonumber \\
&\{&-(1/\sqrt{2}), 1/\sqrt{6}, 1/(2 \sqrt{3}), 1/(2 \sqrt{5})\}, \nonumber \\
&\{&0, -\sqrt{2/3}, 1/(2 \sqrt{3}), 1/(2 \sqrt{5})\}, \nonumber \\
&\{&0, 0, -(\sqrt{3}/2), 1/(2 \sqrt{5})\}, \nonumber \\
&\{& 0, 0, 0, -(2/\sqrt{5})\}.
\ea

The affine transformation mapping one simplex onto the other was determined numerically using the \textit{Mathematica} command \texttt{FindGeometricTransform}. It consists of a rotation combined with translations,
\ba 
&\{& \{0.4 + 0.228825 \alpha + 0.213762 \beta + 0.237755 \gamma + 
  0.473607 \delta \}, \nonumber \\
&\{&0.2 - 0.59907 \alpha - 0.0816497 \beta + 0.0288675 \gamma + 
  0.111803 \delta \},\nonumber \\
&\{& -0.2  + 0.228825 \alpha - 0.559635 \beta - 
  0.11547 \gamma \}, \nonumber \\
&\{& 0.0707107 \alpha + 0.305049 \beta - 
  0.518007 \gamma - 0.111803 \delta \}, \nonumber \\
&\{& -0.4 + 0.0707107 \alpha + 
  0.122474 \beta + 0.366854 \gamma - 0.473607 \delta \} \}.
\ea

In arbitrary dimension, the eigenstates of the Laplacian on a simplex can be expressed as a sum over permutations of the full $S_{d-1}$ group. In \textit{Mathematica} notation this can be written compactly as
\be
\text{Sum[Signature[w] Exp[2 I*Pi*(Permute[p, w] . y)], \{w, Sn\}]},
\ee
or equivalently,
\be
\sum_{w\in S_{d}}\text{sign}[w]\, e^{2\pi i (wp)\cdot y}.
\ee
Here the "momenta'' are defined as $p=\sum_{i=1}^4 m_i\omega_i$, with $m_i$ being positive integers, $S_n$ denoting the $(d-1)!$ permutations of the permutation group, and $y$ representing the coordinates, either in the original simplex or after projection onto the "starfish'' simplex as described above. The permutations act on the components of $p$ (not on the $\omega_i$) and are required to enforce the correct boundary conditions on all faces of the simplex.

The ground state corresponds to $m=\{1,1,1,1\}$, or equivalently $p=\{-2,-1,0,1,2\}$, reminiscent of a five-element "Fermi sphere''. Owing to its complexity, the explicit expression for the real part of the wave function is given at the end of this section; its imaginary part vanishes identically.

The exact lowest Laplacian eigenvalue for $\psi[1,1,1,1]$ is
\be 
-\nabla^2\psi[1,1,1,1]/\psi[1,1,1,1]= 186.972,
\ee
which lies close to the values obtained from variational minimization. It should be emphasized, however, that while the exact value is uniform throughout the simplex, the variational estimates were obtained from ratios of integrals over the full domain.

The general expression for Laplacian eigenvalues on a unit $d$-simplex is
\be
E(m_1,m_2,\dots,m_{d-1})\propto\pi^2 |p|^2=
\pi^2\sum_{i=1}^{d-1}\sum_{j=1}^{d-1}\left(\min(i,j)-\frac{ij}{d}\right)m_i m_j.
\ee
For the present case with four integers, this reduces to
\ba 
E(m_1,m_2,m_3,m_4)\sim (&2m_1^2+3m_2^2+3m_3^2+3m_3m_4+2m_4^2 \nonumber \\
&+4m_2m_3+2m_2m_4+3m_1m_2+2m_1m_3+m_1m_4),
\ea
up to an overall scaling factor arising from the coordinate transformation from $x$ to $y$.

The predicted ratio of $P$-wave to $S$-wave eigenvalues is therefore
\be
{Pwave \over Swave}=(74/5)/10 =1.48.
\ee
This is smaller than the ratio obtained from the uncorrected variational Ansatz (\ref{eqn_ansatz_faces}),
\be
{Pwave \over Swave}={312./182.}=1.71.
\ee
As expected, variational corrections tend to lower this ratio toward the exact value, reaching approximately $1.6$.

The exact eigenfunction vanishes more rapidly than the variational one near the boundary corresponding to $x\rightarrow 1$. Computing the corresponding PDF (defined as the integral of the squared wave function over $\alpha,\beta,\gamma$) and plotting it on a logarithmic scale reveals a steeper falloff, consistent with a power-law behavior $\sim (1-x)^p$ with $p\approx 12$.

The expectation value of the cup potential $V_{cup}$ is significantly smaller for the exact eigenfunctions than for the variational ones, to the extent that it may be neglected at the $\sim 10\%$ level. One may wonder why $V_{cup}$ plays a more important role for baryons (defined on a triangle) than for pentaquarks (defined on a 5-simplex). This may reflect a more general trend as the number of constituents increases.

Finally, for completeness, we present the explicit form of the lowest Laplacian eigenstate $\psi(1,1,1,1)$. Owing to the requirement of Dirichlet boundary conditions on all ten faces of the $A_4$ simplex, a large number of plane waves is required:

\begin{small}
\ba 
& Re& \psi[1,1,1,1]= \\
&\cos(&2.5651 \beta +13.3389 \gamma -1.5708 \delta ) 
  -\cos (-10.4036 \alpha +4.85939 \beta +7.37354 \gamma -0.868315
   \delta +1.25664)
   \nonumber \\
-&\cos(&-3.2149 \alpha -12.7221 \beta +3.74594 \gamma -0.868315 \delta +1.25664)
   +\cos (-8.41672 \alpha
   -9.71879 \beta +4.65284 \gamma -0.165833 \delta +2.51327)\nonumber \\
+&\cos(&-9.41018 \alpha -0.573574 \beta +9.90275 \gamma -0.165833
   \delta +2.51327)
   +\cos (9.41018 \alpha +7.99807 \beta +5.65549 \gamma +1.5708 \delta )\nonumber \\
-&\cos(&-5.20182 \alpha +0.438171
   \beta -12.432 \gamma +2.27328 \delta +1.25664)
   -\cos (-6.19528 \alpha +9.58339 \beta -7.18206 \gamma +2.27328 \delta
   +1.25664)\nonumber \\
-&\cos(&0.993459 \alpha -7.99807 \beta -10.8097 \gamma +2.27328 \delta +1.25664)
   -\cos (4.20836 \alpha +11.0013
   \beta +6.56239 \gamma +2.27328 \delta +1.25664)\nonumber \\
 -&\cos(&10.4036 \alpha +2.5651 \beta +8.1847 \gamma +2.27328 \delta
   +1.25664)
   -\cos (11.3971 \alpha -6.58012 \beta +2.93479 \gamma +2.27328 \delta +1.25664)\nonumber \\
 -&\cos(&9.41018 \alpha +5.43297
   \beta -7.68338 \gamma +3.14159 \delta )
   +\cos (-9.41018 \alpha -1.99153 \beta -8.99585 \gamma +3.67824 \delta +3.76991)\nonumber \\
+ &\cos(&-10.4036 \alpha +7.15369 \beta -3.74594 \gamma +3.67824 \delta +3.76991)+\cos (-3.2149 \alpha -10.4278 \beta -7.37354
   \gamma +3.67824 \delta +3.76991)\nonumber \\
+&\cos(&8.57164 \beta +9.9985 \gamma +3.67824 \delta +3.76991)+\cos (7.18874 \alpha
   -9.00981 \beta +6.3709 \gamma +3.67824 \delta +3.76991)\nonumber \\
+&\cos(&6.19528 \alpha +0.135402 \beta +11.6208 \gamma +3.67824
   \delta +3.76991)-\cos (5.20182 \alpha -0.708976 \beta -12.0264 \gamma +3.84407 \delta +1.25664)\nonumber \\
-&\cos(&9.41018 \alpha
   +9.14522 \beta +0.0957439 \gamma +3.84407 \delta +1.25664)-\cos (-8.41672 \alpha -7.42449 \beta -6.46664 \gamma +4.38072
   \delta +5.02655)\nonumber \\
-&\cos(&0.993459 \alpha +3.13867 \beta +12.5277 \gamma +4.38072 \delta +5.02655)+\cos (-0.993459 \alpha
   +11.4395 \beta -5.86958 \gamma +4.54656 \delta +2.51327)\nonumber \\
+&\cos(&6.19528 \alpha -6.14194 \beta -9.49718 \gamma +4.54656
   \delta +2.51327)+\cos (4.20836 \alpha +12.1485 \beta +1.00264 \gamma +4.54656 \delta +2.51327)\nonumber \\
+&\cos(&11.3971 \alpha
   -5.43297 \beta -2.62496 \gamma +4.54656 \delta +2.51327)+\cos (-9.41018 \alpha +5.43297 \beta +6.56239 \gamma +5.0832
   \delta +6.28319)\nonumber \\
+&\cos(&-2.22144 \alpha -12.1485 \beta +2.93479 \gamma +5.0832 \delta +6.28319)+\cos (-5.20182 \alpha
   +5.29757 \beta -10.2126 \gamma +5.24904 \delta +3.76991)\nonumber \\
+&\cos(&11.3971 \alpha -1.72072 \beta +5.15417 \gamma +5.24904
   \delta +3.76991)-\cos (-7.42326 \alpha -9.14522 \beta +3.84169 \gamma +5.78569 \delta +7.53982)\nonumber \\
-&\cos(&-8.41672 \alpha
   +9.0916 \gamma +5.78569 \delta +7.53982) -\cos (-9.41018 \alpha +2.86787 \beta -6.77648 \gamma +6.654 \delta +6.28319)\nonumber \\
   -
   &\cos(&7.18874 \alpha -4.15042 \beta +8.59028 \gamma +6.654 \delta +6.28319)+\cos (5.20182 \alpha +4.15042 \beta -9.80701
   \gamma +6.81983 \delta +3.76991)\nonumber \\
+&\cos(&10.4036 \alpha +4.85939 \beta -2.93479 \gamma +6.81983 \delta +3.76991)-\cos
   (-9.41018 \alpha +6.58012 \beta +1.00264 \gamma +7.35648 \delta +7.53982)\nonumber \\
-&\cos(&-2.22144 \alpha -11.0013 \beta -2.62496
   \gamma +7.35648 \delta +7.53982)-\cos (-4.20836 \alpha +7.28909 \beta +7.87487 \gamma +7.35648 \delta +7.53982)\nonumber \\
-&\cos&
   (2.98038 \alpha -10.2924 \beta +4.24727 \gamma +7.35648 \delta +7.53982)-\cos (7.15369 \beta -8.90011 \gamma +7.52231
   \delta +5.02655)\nonumber \\
-&\cos(&11.3971 \alpha -0.573574 \beta -0.405578 \gamma +7.52231 \delta +5.02655)+\cos (-7.42326 \alpha
   -7.99807 \beta -1.71806 \gamma +8.05896 \delta +8.79646)\nonumber \\
+&\cos(&-3.2149 \alpha +1.85612 \beta +10.4041 \gamma +8.05896
   \delta +8.79646)-\cos (10.8659 \beta -1.12099 \gamma +8.2248 \delta +6.28319)\nonumber \\
-&\cos(&7.18874 \alpha -6.71552 \beta
   -4.74859 \gamma +8.2248 \delta +6.28319) 
+\cos (-7.42326 \alpha -4.28582 \beta +6.06107 \gamma +8.76144 \delta +10.0531)\nonumber \\
+
   &\cos(&-4.20836 \alpha +8.43624 \beta +2.31512 \gamma +9.62976 \delta +8.79646)+\cos (2.98038 \alpha -9.14522 \beta -1.31248
   \gamma +9.62976 \delta +8.79646)\nonumber \\
   +&\cos(&-8.41672 \alpha +2.29429 \beta -2.02789 \gamma +10.3322 \delta +10.0531)+\cos
   (2.98038 \alpha -5.43297 \beta +6.46664 \gamma +10.3322 \delta +10.0531)\nonumber \\
-&\cos(&6.19528 \alpha +3.57684 \beta -5.05842
   \gamma +10.4981 \delta +7.53982)-\cos (-7.42326 \alpha -3.13867 \beta +0.501322 \gamma +11.0347 \delta +11.3097)\nonumber \\
-&\cos&
   (-2.22144 \alpha -2.4297 \beta +7.37354 \gamma +11.0347 \delta +11.3097)+\cos (0.993459 \alpha +6.58012 \beta -4.15152
   \gamma +11.2006 \delta +8.79646)\nonumber \\
+&\cos(&7.18874 \alpha -1.85612 \beta -2.52921 \gamma +11.2006 \delta +8.79646)-\cos
   (-3.2149 \alpha +4.15042 \beta -0.715412 \gamma +12.6055 \delta +11.3097) \nonumber \\
   -&\cos(& 2.98038 \alpha -4.28582 \beta +0.9069
   \gamma +12.6055 \delta +11.3097) +\cos(-2.22144 \alpha -1.28255 \beta +1.8138 \gamma +13.308 \delta +12.5664)) \nonumber
\ea
\end{small}

\end{subappendices}

\chapter{Hard exclusive reactions,  formfactors}
\section{Pion PDF}

In this section we discuss  the pion parton
distribution function (PDF). 
At leading twist,  the partonic content of any
light hadron is encoded in bilocal light-front correlation functions, whose
Fourier transforms define the distribution functions. Since the pion is a
spin-zero Goldstone mode, its partonic structure at the twist-2 level reduces
to a single nontrivial distribution: the unpolarized quark PDF $q_\pi(x)$.
In particular,  for spin-0 hadrons the helicity and
transversity distributions do not contribute at leading twist, leaving the
unpolarized PDF as the unique partonic observable~\cite{Collins:2011zzd}.

The physical motivation for determining the pion PDF within the instanton-vacuum light-front
framework, follows directly from the role of the pion as the fundamental
Goldstone excitation of spontaneously broken chiral symmetry.  
Its partonic content is nonperturbative in origin, yet it must match perturbative
factorization descriptions at high resolution.  
In~\cite{Liu:2023feu,LiuShuryakZahed2023} parton distributions are obtained from
light-front matrix elements evaluated at low resolution, where the pion is
correctly dominated by its lowest quark-antiquark Fock component.
The light-front wave function used in the construction of the PDF is therefore
the same one obtained from the nonlocal, instanton-induced dynamics; its shape
encodes the full nonperturbative structure arising from the instanton vacuum.

The  leading-twist parton distributions follow from the
Fourier transform of a bilocal fermionic matrix element of the form
\begin{equation}
q_{\pi}(x)=\frac{1}{4\pi}
\int d\xi^{-}\,e^{-ixP^{+}\xi^{-}}
\langle \pi(P)\vert 
\bar{\psi}(0)\gamma^{+}W(0,\xi^{-})\psi(\xi^{-})
\vert\pi(P)\rangle ,
\label{eq:LFmatrix}
\end{equation}
where $W$ is the appropriate light-front Wilson line.
Section~VI further states that these matrix elements are to be computed using
the light-front wave functions determined earlier, and localized entirely in the
lowest $q\bar{q}$ component at the resolution scale of the instanton-vacuum
Hamiltonian. In the ILM at low resolution,  the pion is described solely by its $q\bar{q}$ component, with the unpolarized PDF given as
\begin{equation}
q_{\pi}(x)=
\int\frac{d^{2}k_{\perp}}
     {2(2\pi)^{3}}
\sum_{s_{1},s_{2}}
\bigl|\Phi_{\pi}(x,k_{\perp},s_{1},s_{2})\bigr|^{2}.
\label{eq:pionPDF}
\end{equation}
No additional higher-Fock corrections enter at the
resolution scale associated with the instanton vacuum.
The support is restricted to $0 < x < 1$ because 
$\Phi_{\pi}$ is defined only for $k^{+}=xP^{+}$ between
$0$ and $P^{+}$.
Normalization of the light-front state ensures the valence sum rule
\begin{equation}
\int_{0}^{1}\!dx\, q_{\pi}(x)=1,
\end{equation}
which follows immediately from the normalization condition of the
light-front wave function with no additional assumptions. 

In the chiral limit ($m_u=m_d\to 0$), the pion mass $M_\pi\to 0$ but $f_\pi$ remains finite, ensuring the Goldstone nature of the pion. The PDF is symmetric under $x\leftrightarrow 1-x$ in this limit, reflecting isospin symmetry. At large $x$, the counting rules predict $q_v^\pi(x)\sim (1-x)^{2+\gamma}$, where $\gamma$ encodes logarithmic QCD corrections~\cite{Brodsky:1973kr,Farrar:1975yb}. This behavior emerges naturally from the LFWF structure and provides a nontrivial check of consistency with perturbative QCD.

\begin{figure*}
  \includegraphics[height=5.5cm,width=.46\linewidth]{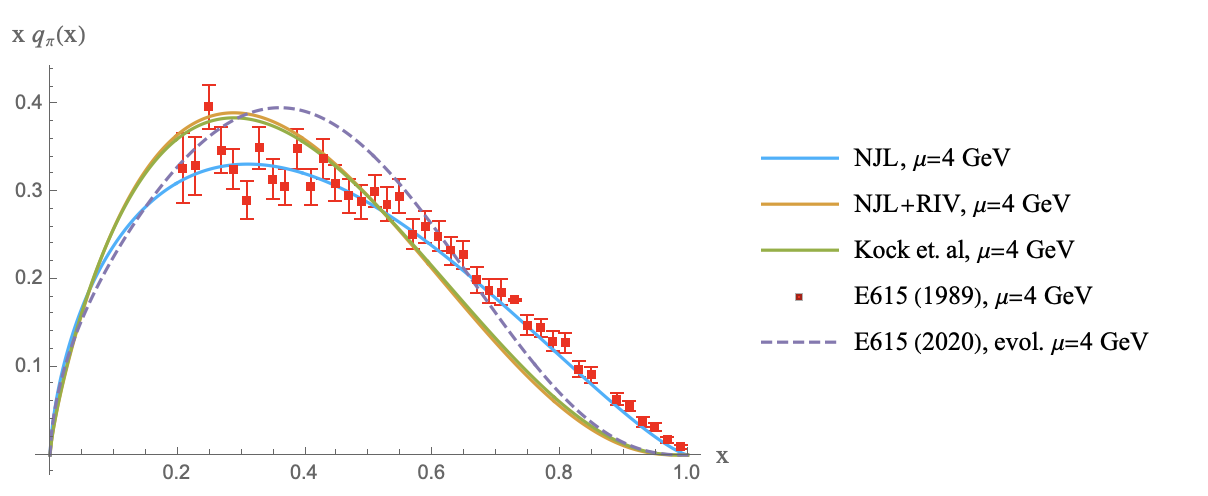}%
\hfill
  \includegraphics[height=5.5cm,width=.46\linewidth]{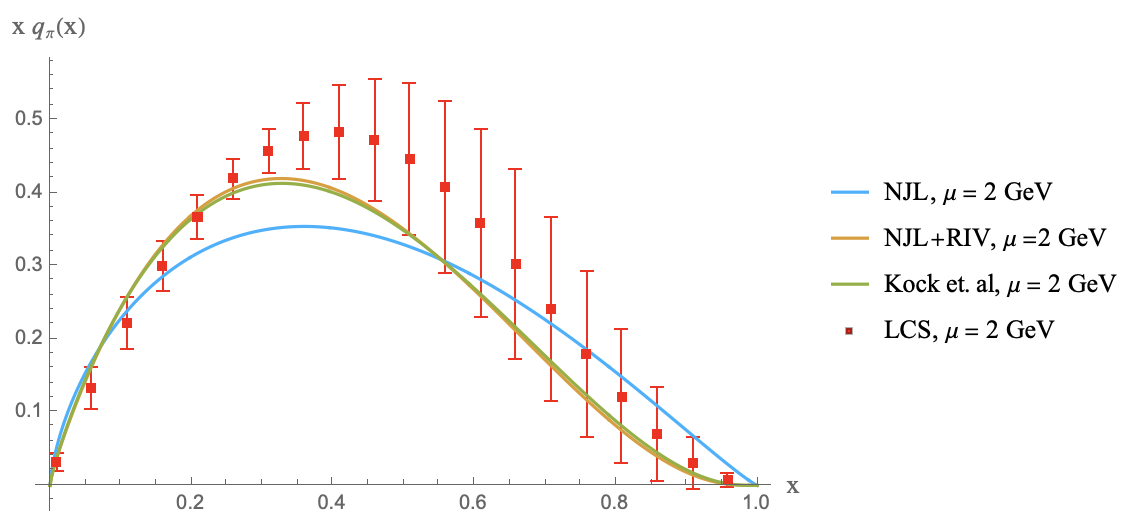}%
\caption{a: Pion parton momentum distribution function for zero instanton size (solid-blue) and finite instanton size 
of $\rho=0.317$ fm (solid-orange), both of which are evolved to $\mu=4$ GeV with a pion mass $m_\pi=135$ MeV.
The results are compared to the those extracted from the ILM using the LaMET (solid-green)~\cite{Kock:2020frx,Kock:2021spt},
also evolved to $\mu=4$ GeV.  The E615 data from 1989 (red) are from \cite{Conway:1989fs},  corresponding to a  fixed invariant muon pair mass $m_{\mu^+\mu^-}\geq 4.05$ GeV. The improved E615 data from 2010 (dashed-purple) are from  \cite{Aicher:2010cb}, using the original E615 experimental data from~\cite{Conway:1989fs}.\\
b: The same pion parton distribution functions as in a, but now evolved to $\mu=2$ GeV, for comparison with 
the lattice data using  cross sections (LCS) (red) from~\cite{Sufian:2020vzb}.
}
\label{fig_piLATDAT}
\end{figure*}


\section{Summary}

The leading-twist pion PDF in the instanton-vacuum light-front
approach is entirely determined by the valence light-front wave function through Eq.~(\ref{eq:pionPDF}).  
All nonperturbative features of the pion - including the momentum asymmetry,
endpoint behavior, and the normalization - originate from the instanton-induced
nonlocality encoded in the Bethe-Salpeter kernel and inherited by the projected light-front wave function.  

The nonlocal instanton model gives a realistic description of $f_\pi$, the shape of $\phi_\pi(x,\mu_0)$, and the pion PDF~\cite{Diakonov:1995ea,LiuShuryakZahed2023}. Quantitatively:
\begin{itemize}
\item $f_\pi \approx 92.4$~MeV (input or output of the model),
\item $\langle x\rangle_q^\pi \approx 0.5$ at low scale,
\item The DA is broad and nearly flat at $\mu_0\simeq1$~GeV, evolving toward $6x(1-x)$ at larger $\mu$.
\end{itemize}
Lattice QCD and experimental extractions (e.g.\ from $\pi\gamma$ transition form factors and LaMET computations) support this picture~\cite{Zhang2020,Shi2015,Kock:2020frx,Sufian:2020vzb}.

\section{Pion Distribution Amplitudes on the LF and pion formfactors} \label{sec_pion_DA}
The theory of {\em exclusive QCD reactions}  
\cite{Brodsky:1973kr,Chernyak:1977as,Radyushkin:1977gp}.
 is  based on two assumptions:\\
(i) the {\em separation of scales},  based on the assumption that the momentum transfer 
$Q$ (the scale in the
"hard block"), is large compared to the typical quark mass and transverse momenta inside hadrons;
(ii) that   the
"hard block" can be calculated {\em perturbatively using  gluon exchanges}.\\

However, the theory based on these two assumptions is insofar not  successful phenomenologically.
In particular, in the domain of $Q$  in which experimental and lattice results are available, the mesonic 
form factors, while approximately $\sim 1/Q^2$, are well above the one-gluon exchange predictions. 
This should not be surprising, as there is an important difference 
between the scales in DIS and jet physics on one hand, and exclusive processes on the other.
 The former operates  in the range $Q^2=10^2-10^4 \, {\rm GeV}^2$, while
 exclusive processes we study so far in  a different range, $Q^2=2-10\, {\rm GeV}^2$ (sometimes called semi-hard domain).
 
In this chapter (based on our paper \cite{Shuryak:2020ktq}) we will
accept the assumption (i) mentioned above: the $Q^2$ scale 
is  indeed large compared to the typical squared transverse momentum $\langle p_\perp^2 \rangle \sim 0.1\, {\rm GeV}^2$ within a hadron,
or the constituent quark mass $M^2\sim 0.1-0.15\, {\rm GeV}^2$. In the Breit frame description of form factors, conventional "collinear"
kinematics  should still hold.  So we still have a notion of a "hard block operator", sandwiched between two wave functions.
Yet we do not accept the second assumption (ii), showing that at such momentum transfer $Q$,  the nonperturbative 
quark interactions   are $not$ at all negligible in comparison  to gluon exchanges. Therefore a purely perturbative treatment of the 
"hard block"  needs to be appended by calculations of {\em leading nonperturbative contributions}, and  this paper makes the first steps in this direction.

Distribution Amplitudes (DAs) are defined by matrix elements
different from PDFs: unlike them they are matrix element
between vacuum and hadron state. In other words, the hadron
is produced by a given operator "out of nothing". Its usage
in hard processes (mainly elastic formfactors) is for incoming and outgoing hadrons in relativistic kinematics. Similar to 
PDFs, the moments of DAs can be related by OPE to local operators.
This relation allowed to evaluate DAs numerically, on the Euclidean lattices. 

The first pion DA  is   defined  using
 the forward matrix element of the twist-2 operator on the LF, 
\begin{equation}
\label{PITWIST2}
\begin{aligned}
        \phi_{\pi}(x)=&-i\int\frac{d\xi^-}{2\pi}e^{ixP^+\xi^-}\langle0|\bar{\psi}(0)\gamma^+\gamma^5\frac{\tau^a}{\sqrt{2}}W(0,\xi^-)\psi(\xi^-)|\pi^a(P)\rangle
\end{aligned}
\end{equation}
with the normalization  fixed by the pion weak decay constant
\begin{equation}
\label{PIWEAK}    \langle0|\bar{\psi}\gamma^+\gamma^5\frac{\tau^a}{\sqrt{2}}\psi|\pi^a(P)\rangle=if_\pi P^+
\end{equation}
In general, pion is characterized by three DAs
\begin{eqnarray}
	\label{WF1}
	&&\int_{-\infty}^{+\infty}\frac{p^+dz^-}{2\pi}e^{ixp\cdot z}\left<0\left|\overline{d}_\beta(0)[0,z]u_\alpha(z)\right|\pi^+(p)\right>\nonumber\\
	&&=
	\bigg(+\frac{if_\pi}4 \gamma^5\bigg(\slashed{p}\,\varphi^A_{\pi^+}(x)
	-\chi_\pi \varphi_{\pi}^P(x)-i \chi_\pi
	\sigma_{\mu\nu}\frac{p^\mu n^{\nu}}{p\cdot n}  {\varphi_{\pi}^{\prime\,T}(x)\over {6}}\bigg)\bigg)_{\alpha\beta}
\end{eqnarray}
Here $[x,y]$  is the shorthand notation for the gauge link between two points on the light front, and  $\sigma_{\mu\nu}=\frac 1{2i}[\gamma_\mu, \gamma_\nu]$. 
  In this example
 the first term contains momentum, while the other two do not. Therefore the axial $A$-DA is of leading twist, while the
 two others $P,T$-DA  are subleading (next twist) at large momentum $p\rightarrow \infty$.  
  
  The three functions $\varphi^i(x)$ have indices  $i=A,P,T$ standing for axial, pseudoscalar and tensor 
  gamma matrices in the operator.  They are all normalized to 1. Their explicit definition follows from (\ref{WF1}) by inversion
\begin{eqnarray}
\label{PIDIS}
	&&\varphi^A_{\pi^+}(x)=
	\frac  1{if_\pi}\int_{-\infty} ^{+\infty} \frac{dz^-}{2\pi}e^{ixp\cdot z}\left<0\left|\overline{d}(0)\gamma^+\gamma_5[0,z]u(z)\right|\pi^+(p)\right>\nonumber\\
	&&\varphi^P_{\pi^+}(x)=
	\frac  {1}{f_\pi\chi_\pi}\int_{-\infty} ^{+\infty}  \frac{p^+dz^-}{2\pi}e^{ixp\cdot z}\left<0\left|\overline{d}(0)i\gamma_5[0,z]u(z)\right|\pi^+(p)\right>\nonumber\\
	&&{\varphi^{T\prime}_{\pi^+}(x)}=
	\frac  {-6}{f_\pi\chi_{\pi}}\frac {p^\mu n^\nu}{p\cdot n}\int _{-\infty} ^{+\infty} \frac{p^+dz^-}{2\pi}e^{ixp\cdot z}\left<0\left|\overline{d}(0)\sigma_{\mu\nu}\gamma_5[0,z]u(z)\right|\pi^+(p)\right>
\end{eqnarray}
To relate the tensor pion DA amplitude in (\ref{PIDIS}) to the pseudoscalar light
front wavefunction with  helicity-1, we note that the dominant tensor
matrix element on the light cone reads
\bea
\langle 0|\overline{d}(0)i\sigma^{+i}\gamma_5u(z^-, z_\perp)|\pi^+(p)\rangle\nonumber\\
=2p^+\frac{\partial}{\partial z_\perp^i}\psi^P_1(z^-, z_\perp)
\eea
with ($x=k^+/p^+$)

\bea
\psi_1^P(x, k_\perp)=
\int dz^-dz_\perp e^{i(k^+z^-k_\perp \cdot z_\perp)}\psi_1^P(z^-, z_\perp)\nonumber\\
\eea
 Note the relation  $\psi^P_{\pm 1}(x, k_\perp)=k_\perp^\pm \psi_1^P(x, k_\perp)$ 
with $k_\perp^\pm =k_x\pm ik_y$. 
A comparison with (\ref{PIDIS}) shows that the twist-3 pion distribution amplitude
in $\varphi_T(x, k_\perp)$ matches the $m=1$ contribution through

\bea
\frac{\partial}{\partial k_\perp^i}\varphi_T(x, k_\perp)=\frac{6}{f_\pi\chi_\pi}k_\perp^i\psi^P_1(x, k_\perp)
\eea

Inserting the LF normalized pion state (\ref{Meson_bound_state}) with the corresponding solution (\ref{GENX}), the pion DA follows
\begin{equation}
\begin{aligned}
        \phi_{\pi}(x)&=\frac{\sqrt{N_c}M}{2\sqrt{2}\pi^2}\int_0^{\infty} dk^2_\perp \frac{C_{\pi}}{x\bar{x}m^2_{\pi}-(k_\perp^2+M^2)}\mathcal{F}\left(k\right)\mathcal{F}\left(P-k\right)\\
\end{aligned}
\end{equation}

 At  asymptotically increasing momentum transfer ${\rm log}(Q^2/\Lambda_{QCD}^2) \rightarrow  \infty$ one should include perturbative processes of gluon radiation. When 
the light-front  functions $\varphi_\pi(\xi) $ are decomposed into Gegenbauer polynomials, each of the polynomial carries  as a coefficient a (negative) power of the preceding log 
with a calculable {\it anomalous dimension}. Asymptotically, only the leading contribution survives  in the form (asymptotic wave function) 

    \begin{equation}
    \varphi_\pi \rightarrow  \varphi^{\rm asymptotic}_\pi(\xi)=\frac 34 (1-\xi^2) =6x\bar x
     \label{phi_asymptotic} 
     \end{equation} 
Needless to say that this limit is very far from the 
realistic kinematic range of interest in the present analysis.     
    Heavy nonrelativistic quarks should have values of quark momenta around $x\approx \bar x \approx 1/2$.
%
The opposite case of 
 light quarks generate much wider distributions.
 The ones used can be approximated by the form 

 \begin{equation}
 \varphi_\pi(\xi,p) = {\Gamma(3/2+p) \over \sqrt{\pi} \Gamma(1+p)}(1-\xi^2)^p   ={6^p\Gamma(3/2+p) \over \sqrt{\pi} \Gamma(1+p)}\,(x\bar x)^p
 \end{equation} 
The case $p=1$ is the "asymptotic" distribution,  while the case $p=0$ is called "flat". Several authors  have used an intermediate case $p=1/2$ called "semicircular".

\begin{figure}[h!]
	\begin{center}
		\includegraphics[width=8cm]{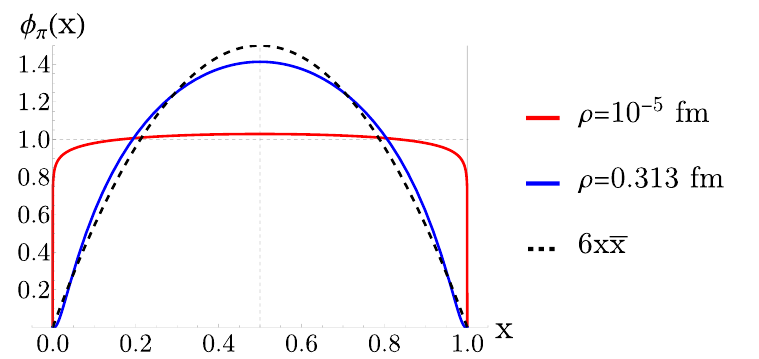}
		\includegraphics[width=8cm]{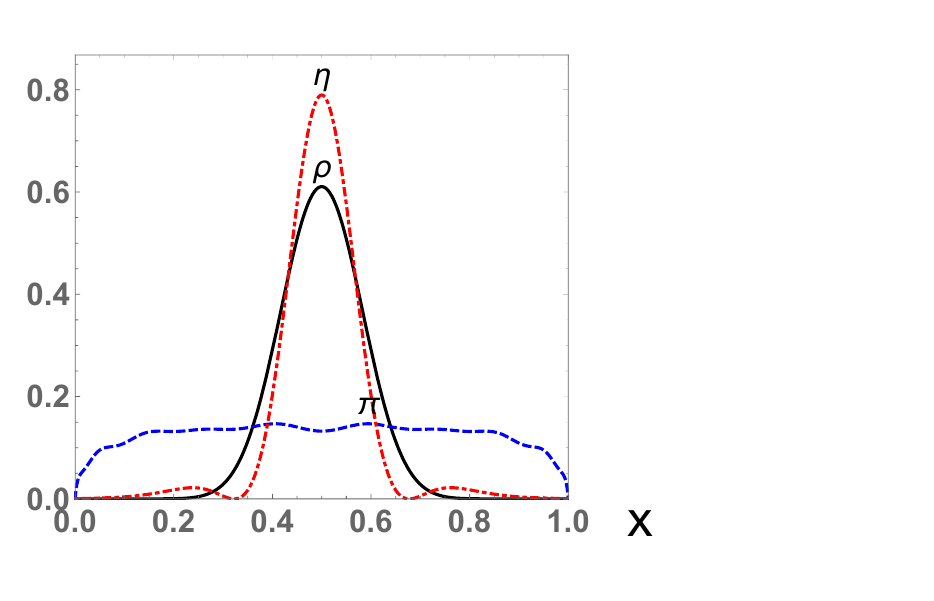}
		\caption{Left plot: the pion  DA  from \protect\cite{Kock:2020frx} sizes,
        for different instanton. 
		Right plot:  the light mesons - pion,rho and $\eta'$ light front distribution functions, from \protect \cite{Shuryak:2019zhv}.}
		\label{fig_pionwf}
	\end{center}
\end{figure}

 The pion is a particular particle, a Nambu-Goldstone mode,   and therefore
  its properties one can calculate in any theory in which  chiral symmetry gets spontaneously  broken. Historically  the NJL model
  and its nonlocal versions (some related with the  instanton liquid model) have been used to calculate the pion light-front wave function~\cite{Broniowski:2017wbr,Petrov:1997ve,Anikin:2000bn}. Before we briefly discuss the results of "realistic" models, related to larger set of hadronic wave functions,
  let us introduce some extreme cases.    
  For example, in \cite{hep-ph/0207266} a "flat"  pion wave function was
 used  as an "initial condition" for radiative evolution.   
Some typical  shapes of the pion and other light meson wave functions   are  shown in Fig.\ref{fig_pionwf}. Note that $\eta'$ is the channel in which 't Hooft Lagranigian is
repulsive, thus quarks are far from each other and have small
relative motion.

In  contrast, in 1980's Chernyak and collaborators \cite{Chernyak:1983ej} using the QCD sum rules arrived at pion wave function 

\begin{equation}
\varphi_{CZ}(x) = 30x(1-x)(2x-1)^2 \label{eqn_double_hump}
\end{equation}
known as  "the double-hump" one. 
But, since then no support for this shape has materialized, and it also does not
agree with lattice results on momentum fractions, so we will not discuss it. Let us state once again, that we 
see phenomenological failure of the pQCD expressions
not in the modified shape of the wave function, but in 
missing nonperturbative part of the hard blocks.

The light pseudoscalar mesons $P=\pi, K, \eta$ are related to chiral symmetry breaking and 
therefore they exist even in models without confinement, such as the NJL and ILM. 
Their wave functions and parton distribution amplitudes (PDA)
have been calculated in various approximations.  The question was addressed originally in the ILM framework
in~\cite{Petrov:1998kg}, and  more recently using the quasi-distribution proposal by one of us in~\cite{Kock:2020frx}.
The  distribution amplitude for the light  P-pseudoscalars was obtained to be~\cite{Kock:2020frx}

\begin{equation}
\label{PDFRIV}
\varphi_{P}(x)=\frac{2N_c}{f^2_P}\int\frac{d^2k_\perp}{(2\pi)^3}
\frac{\theta(x\bar x)}{(k_\perp^2+M^2(0, m_f)-x\bar xm_P^2)}\,M^2\bigg(\frac{\sqrt{k^2_\perp+M^2(0,m_f)}}{\lambda_P\sqrt{x\bar x}}\bigg)
\end{equation}
where the momentum-dependent quark mass is

\begin{equation}
M(k)=M(0)\bigg(\bigg|z\bigg(I_0K_0-I_1K_1\bigg)^\prime\bigg|^2\bigg)_{z=\frac 12 \rho k}
\end{equation}
Here $\lambda_P$ is a cut-off parameter of order 1, e.g. $\lambda_\pi=3.41$, and  $M(0)=386$ MeV  with an instanton size $\rho=0.3$ fm. As shown in~\cite{Kock:2020frx},
this momentum dependence has been confirmed by  lattice studies.
For a  light current quark mass $m_f$, the  running effective quark mass $M(k, m_f)\approx M(k)+m_f$. 
The corresponding  shape of the wave function is shown in the upper plot of  Fig.\ref{fig_pionwf}  as reproduced from \cite{Kock:2020frx},
is in agreement with \cite{Petrov:1998kg}. Both calculations show a wave functions rather close to the asymptotic one, and
very far from the "double-hump"  distribution (\ref{eqn_double_hump}).

Another approach to light front wave functions is based on some model-dependent Hamiltonians.
Jia and Vary \cite{Jia:2018ary} introduced a convenient form of it, including constituent quark masses
(that is, chiral symmetry breaking), plus some form of confinement, plus NJL-type residual interactions.
This approach was followed by one of us~\cite{Shuryak:2019zhv}, who calculated the
wave functions for the $\pi,\rho,\eta' $ mesons,  as shown in the lower plot of Fig.\ref{fig_pionwf}.  Also, the wave functions 
for  the $\Delta$ and nucleon were given in~\cite{Shuryak:2019zhv}, even with the inclusion of the 
 5-light-quark tail, providing the first results for the $anti$quark distribution inside the nucleon.

For a finite instanton size, the pion DA  is
\begin{equation}
\label{PIXX}
\begin{aligned}
        \phi_{\pi}(x)&=\frac{\sqrt{N_c}M}{\sqrt{2}\pi^2}C_{\pi}\int_{\frac{\rho M}{2\lambda\sqrt{x}\bar{x}}}^{\infty} dz  \frac{z^5}{\frac{\rho^2m^2_\pi}{4\lambda^2}-z^2} (F'(z))^4 \\
\end{aligned}
\end{equation}
In the zero instanton size approximation with a fixed transverse momentum cutoff $|k_\perp|<\Lambda$, the pion DA amplitude is a step function
\bea
        \phi_{\pi}(x)
        \xrightarrow{m_\pi\rightarrow0}\frac{\sqrt{N_c}M}{\sqrt{2}\pi}\left[\ln\left(1+\frac{\Lambda^2}{M^2}\right)\right]^{1/2}\theta(x\bar{x})
\eea
in agreement with a result established first in the local NJL model~\cite{Broniowski:2017wbr}.

\begin{figure}[h!]
\begin{center}
\includegraphics[width=8.cm]{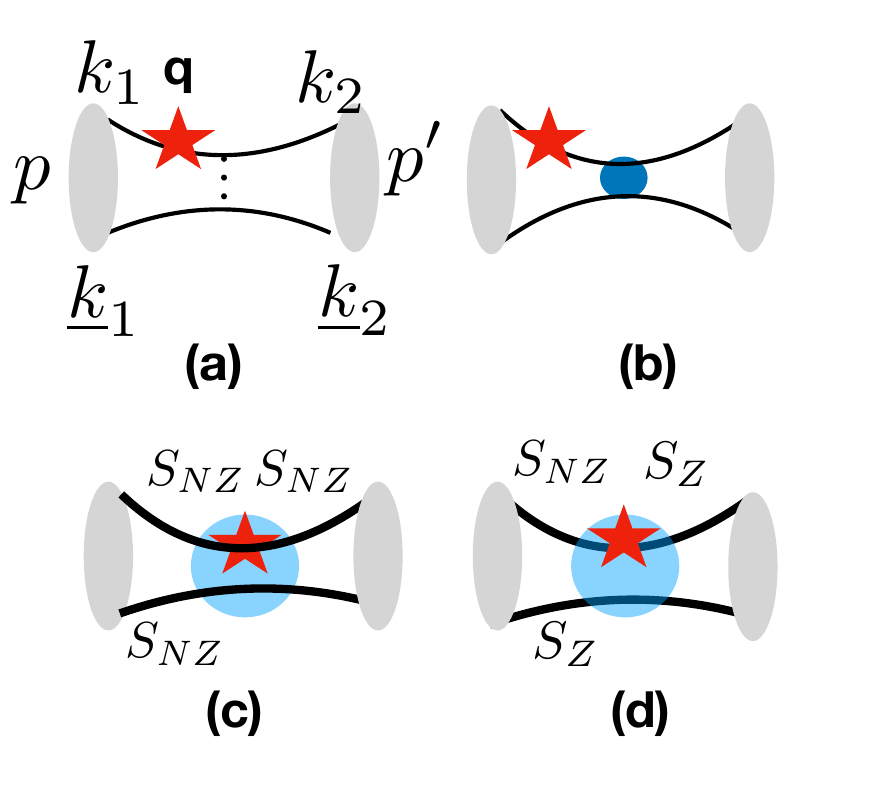}
\caption{The perturbative one-gluon exchange  diagrams: (a) explains our notations for momenta of quarks and mesons. The thin solid lines are free quark propagators, the red star indicates the virtual photon (or scalar) vertex, bringing in  large momentum $q^\mu$, and 
the	shaded ovals represent the (light-front) mesonic density matrices (distributions).
	The diagram (b) indicates a "Born-style contribution",
	in which the gluon propagator is substituted by
	the Fourier transform of the instanton field.
	The  diagrams (c,d) contain three propagators
	in the instanton background (thick lines). In (c) all of them are $S_{NZ}$, made of nonzero Dirac modes, while in (d) two of them are $S_{Z}$ made of quark zero modes. This last contribution we will refer to as  the "'t Hooft-style term".) 
}
\label{fig_diags}
\end{center}
\end{figure}

\section{Hard and semi-hard vertices of formfactors}
This section is based on paper \cite{Shuryak:2020ktq}, from which
our studies of the role of instanton-antiinstanton molecules 
began.

  Out of all formulae in this section,  the only part
 (to our knowledge)
 derived in literature long ago \cite{Brodsky:1973kr,Chernyak:1977as,Radyushkin:1977gp}  is the
 first term of (\ref{eqn_Vapi}), with the chirality-diagonal distribution functions.
 
  We will keep the notations of the contributions as explained in Fig.\ref{fig_diags}. For example,
the (photon-induced)  {\em vector scattering amplitude} on the pion, with
 perturbative one-gluon exchange  will be referred to as $V^\pi_a$, and reads explicitly

 \begin{eqnarray} \label{eqn_Vapi}
&&V^\pi_a(Q^2)=\epsilon_\mu(q)(p^\mu+p^{\prime\mu})\,(e_u+e_{\overline d})\,\bigg(\frac{2C_F\pi\alpha_sf_\pi^2}{N_cQ^2}\bigg)   \int  dx_1 dx_2  \nonumber\\
&&\times \bigg[ 
{ \varphi_\pi(x_1)\varphi_\pi(x_2)\over \bar x_1\bar x_2+m_{\rm gluon}^2/Q^2}
+
{\chi_\pi^2 \over Q^2} \bigg({\tilde\varphi_\pi(x_1)\tilde \varphi_\pi(x_2) \over 
	\bar x_1\bar x_2  + m_{\rm gluon}^2/Q^2}\bigg)
\bigg( {1\over \bar x_1 +E_\perp^2/Q^2}+{1\over \bar x_2 +E_\perp^2/Q^2}-2
\bigg)\bigg]\nonumber\\
\end{eqnarray}
Here we show explicitly the electromagnetic charges $e_u=2/3,e_{\overline d}=1/3$, although of course 
the total charge of a positive pion is $e_u+e_{\overline d}=1$. The color matrices give the factor
$C_F=(N_c^2-1)/2N_c=4/3$ with $N_c=3$ number of colors.  The large spacelike photon momentum is $q^\mu$ and  $q_\mu q^\mu=-Q^2<0$. The photon polarization
vector is $\epsilon_\mu(q)$, with $\epsilon_\mu q^\mu=0$.  The momenta of the initial and final mesons are called $p$ and $p'$. The pion decay constant is $f_\pi\approx 133\, {\rm MeV}$,  it 
characterizes 
the wave 
function at the origin in the transverse plane, $r_\perp=0$. 
For the pion distribution we use the expression (\ref{WF1}) 
	which includes not only the chirally diagonal part of the
 distribution  $\varphi_\pi(x)$ 
 but also the chirally non-diagonal one $\tilde\varphi_\pi(x)$. Both 
  depend only on the longitudinal momentum fraction $x$ of one of the quarks only. They are assumed to
  correspond to a 2-body sector of the full wave function.  
 Here the bar indicates that the momentum fractions are those of  antiquarks, $\bar x_i\equiv 1-x_i$. 
In terms of the  asymmetry parameters, these variables read as $x_i=(1+\xi_i)/2,\bar x_i=(1-\xi_i)/2$.
 The regulators are the gluon mass and quark "transverse energy".

\begin{figure}[h]
\begin{center}
\includegraphics[width=6cm]{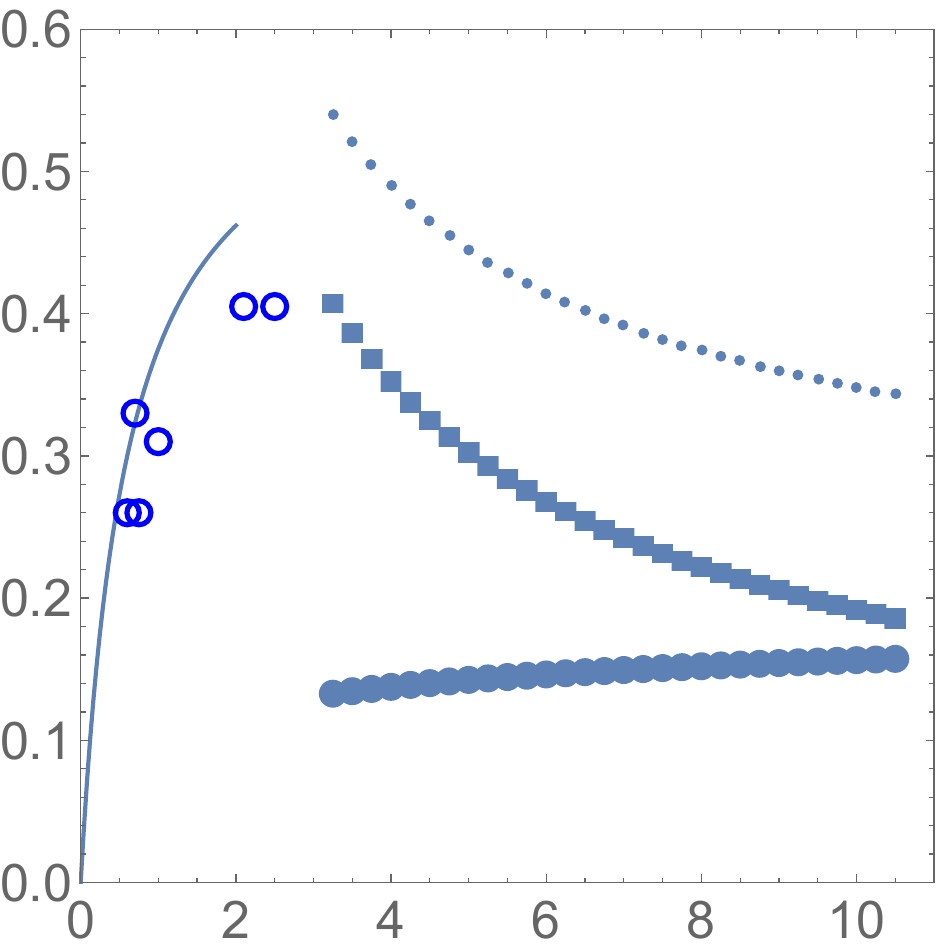}
\includegraphics[width=10cm]{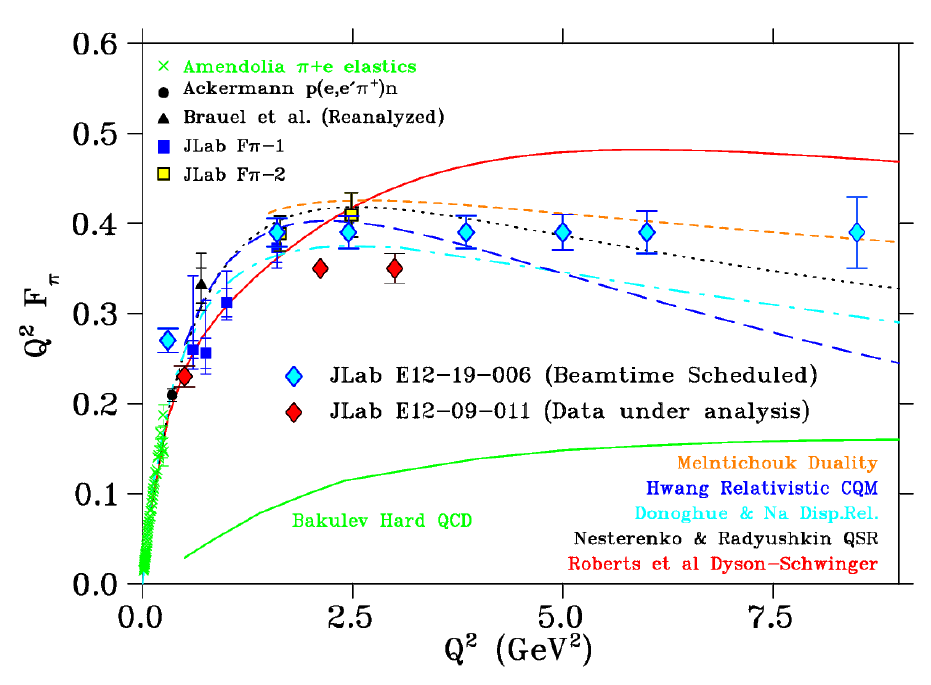}
\caption{The vector 
	 form factors of the pion times the squared momentum transfer, $Q^2 F_\pi(Q^2)\,\, (GeV^2)$ versus $Q^2 (GeV^2)$.
    {\bf Left}:  from \cite{Shuryak:2020ktq}.
The closed  discs show the  
perturbative gluon exchange contribution.  The squares correspond to the
instanton molecule contribution from  the nonzero mode propagators.  The dotted line above is their sum.
The  curve in the l.h.s. is the usual dipole formula, and the open points are some experimental data.\\
{\bf Right}: Points with symbols indicated in the upper left corner are experimental data, curves are various predictions of theoretical models mentioned in
the lower right corner. Light blue and red diamonds are $not$ data but location/accuracy projections for JLab E12 experiments (from their documents). 
}
\label{fig_pionV_all}
\end{center}
\end{figure}

 Note that the first term  in (\ref{eqn_Vapi}) is well known, 
but the second term $\sim \chi_\pi^2/Q^2$ is new, to our knowledge in the asymptotic analysis.
We will keep  it since $\chi_\pi$
is not small, unlike the masses and transverse momenta.
We note further that the last bracket in that term
averages to  a positive  contributions, as it will be shown in the
summary plot Fig.~\ref{fig_pionV_all}. 
 
The contribution we call 
the {\em Born-like instanton contribution} $V_b^\pi$ has the same traces and
 can be obtained by  substituting   in  $V_a^\pi$
 the Fourier transforms of the instanton gauge field 
  instead of  the  gluon propagator, with
 
 \begin{eqnarray} \label{eqn_substitution}
  \pi \alpha_s(Q/2) \rightarrow  \kappa \left<\mathbb G^2(Q\rho\sqrt{\overline{x}_1\overline{x}_2})
 \right> \end{eqnarray}
and is therefore

  \begin{eqnarray} \label{eqn_Vbpi}
&&V^\pi_b(Q^2)=\epsilon_\mu(q)(p^\mu+p^{\prime\mu})\,(e_u+e_{\overline d})\,
\bigg(\frac{2C_F\kappa f_\pi^2}{N_cQ^2}\bigg)   \int  dx_1 dx_2 \,\left<\mathbb G^2(Q\rho\sqrt{\overline{x}_1\overline{x}_2})\right> \nonumber\\
&&\times \bigg[ 
{ \varphi_\pi(x_1)\varphi_\pi(x_2)\over \bar x_1\bar x_2+m_{\rm gluon}^2/Q^2}
+
{\chi_\pi^2 \over Q^2} \bigg({\tilde\varphi_\pi(x_1)\tilde \varphi_\pi(x_2) \over 
	\bar x_1\bar x_2  + m_{\rm gluon}^2/Q^2}\bigg)
\bigg( {1\over \bar x_1 +E_\perp^2/Q^2}+{1\over \bar x_2 +E_\perp^2/Q^2}-2
\bigg)\bigg]\nonumber\\
\end{eqnarray}

The contribution $V_c^\pi$, with three $non$-zero mode propagators is
\begin{eqnarray} \label{eqn_Vcpi}
V_c^\pi= && \epsilon_\mu(q)(p^\mu+p^{\prime\mu})\,(e_u+e_{\overline d})\, 
{\kappa \pi^2 \rho^2 f_{\pi}^2 \chi_\pi^2 \over N_cM^2}
\langle  {\mathbb G}_V(Q \rho) \rangle 
\int dx\tilde\varphi_{\pi}(x) x 
\nonumber 
\end{eqnarray}
The  instanton induced formfactors  $\mathbb G$ and ${\mathbb G}_V(Q \rho)$ are derived explicitly in \cite{Shuryak:2020ktq}. The angular brackets indicate averaging over the instanton size.

The inclusion of instanton-antiinstanton molecules in the hard block
of the pion formfactor was the most important step made in \cite{Shuryak:2020ktq}, from which left plot of Fig.\ref{fig_pionV_all}
is taken. As one can see, their contribution (squares) with gluon exchange (disks) together (dotted line)  provides a realistic 
magnitude of the pion formfactor at intermediate momentum transfers.

For comparison, the right plot of Fig.\ref{fig_pionV_all}
summarizes current theoretical predictions and experimental projections\footnote{At the time of this writing we could not
locate any publications with their results, but perhaps showing
projected error bars and $Q^2$ range in this review is still worth it.
} for coming data of the JLab E12 experiments.

\begin{figure}[h!]
    \centering
    \includegraphics[width=0.45\linewidth]{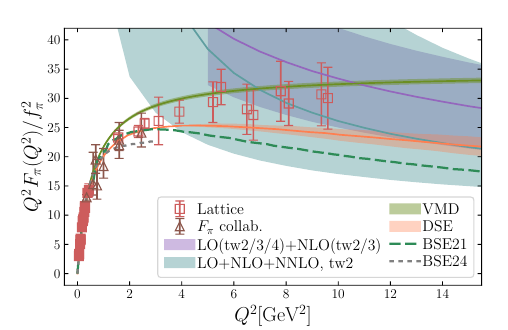}
        \includegraphics[width=0.45\linewidth]{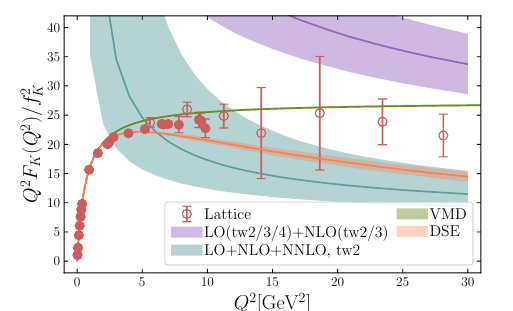}    
    \caption{Pion and kaon formfactors calculated on the lattice \cite{Ding:2024lfj} at larger momentum transfer values, shown by open squares.  }
    \label{fig_ff_lat}
\end{figure}

The pion and kaon formfactors were calculated on the lattice many times, for recent results at the largest momentum transfers see Fig.\ref{fig_ff_lat}.
according to them, $Q^2F(Q)\approx const$ up to rather high $Q^2$,
no expected downward transition to perturbative predictions are 
so far seen: this is one of the most striking puzzles of the QCD
theory. 

We will not give here lengthy expressions for other meson formfactors (scalar or Higgs-induced, gravitational or stress-tensor induced): all of them are obtained in the same general setting. Their relative size
and $Q^2$ dependence would perhaps provide better understanding of the
origin of the effective hard block. For example, scalar formfactor need extra chirality flip relative to vector ones:  perturbative
theory thus predict extra penalty in form of extra powers of  $1/Q$.
Contrary to that, lattice calculations reported by \cite{Davies:2019nut} observed scalar formfactor for $\bar c c$ channel to be several times
$larger$ than the vector one, also with flat $Q^2F(Q^2)$: it remains another open puzzle related with meson formfactors.

\begin{subappendices}
\section{Axial Ward Identity, GMOR Relation, and the Zero-Mode Condensate}
The axial Ward identity,
\begin{equation}
\partial_\mu J_5^\mu=2m\,\qb\ii\gamma_5 q+2\ii N_f\frac{g^2}{32\pi^2}F\tilde F,
\end{equation}
links explicit mass terms and topological effects. 
Solving for the constrained component $\psi_-$ in terms of $\psi_+$ gives
\begin{equation}
\psi_-=\frac{1}{2\ii\partial^+}(\bm{\gamma}_\perp\!\cdot\!\bm{D}_\perp+m)\gamma^+\psi_+.
\end{equation}
Substitution yields a zero-mode operator:
\begin{equation}
 \mathcal{Z}_{\text{zero}}(x)\equiv\frac{1}{2}\!\Big[\bar\psi_+(x)\frac{1}{\ii\partial^+}(\bm{\gamma}_\perp\!\cdot\!\Dlr_\perp+m)\gamma^+\psi_+(x)+\text{h.c.}\Big]_0\! .
\end{equation}
The projection onto $p^+=0$ reads
\begin{equation}
 [\mathcal{O}(x)]_0=\lim_{\eta\to0^+}\int\!\frac{\dd k^+}{2\pi}e^{-ik^+x^-}\tilde{\mathcal{O}}(k^+)\theta(\eta-|k^+|)=\frac{1}{L^-}\!\int_{-L^-/2}^{L^-/2}\!\dd y^-\,\mathcal{O}(y^-)\! .
\end{equation}

This operator $\mathcal{Z}_{\text{zero}}$ defines the LF vacuum condensate:
\begin{equation}
\vev{\qb q}_\LF\equiv\vev{0|\mathcal{Z}_{\text{zero}}|0}\neq0.
\end{equation}
Inserting into the axial Ward identity and taking vacuum-pion matrix elements yields the 
Gell-Mann-Oakes-Renner relation:
\begin{equation}
f_\pi^2m_\pi^2=-2\hat{m}\vev{\qb q}_\LF+\ord{\hat{m}^2},
\end{equation}
demonstrating that the pion mass and decay constant arise consistently from the zero-mode vacuum condensate. Physically, the LF zero-mode condensate plays the same role as the instant-form quark condensate in generating Goldstone bosons.

\section{Relation to Hadronic Matrix Elements}
Hadronic observables probe the underlying vacuum through their coupling to zero modes. Light-front wave functions describe Fock states with $p^+>0$, while the zero-mode contributions complete the symmetry relations and ensure correct normalization of chiral currents. For instance, the pion distribution amplitude and decay constant $f_\pi$ require inclusion of $\mathcal{Z}_{\text{zero}}$ to recover low-energy theorems. Thus, experimental quantities such as $m_\pi$, $f_\pi$, and form factors reflect the existence of a zero-mode vacuum condensate even though it is invisible in naive Fock space.


\section{Instanton Form Factors and Normalization}
The quark zero mode in an instanton background is
\begin{equation}
\psi_0(x)=\frac{\rho}{\pi}\frac{1}{(x^2+\rho^2)^{3/2}}\frac{1-\gamma_5\slashed{x}/\sqrt{x^2}}{2}U,
\end{equation}
where $\rho$ is the instanton size and $U$ a color rotation. Its Fourier transform gives the form factor
\begin{equation}
F(k\rho)=2z[I_0(z)K_1(z)-I_1(z)K_0(z)-I_1(z)K_1(z)/z],\quad z=k\rho/2.
\end{equation}
The nonlocal kernel is
\begin{equation}
\mathcal{K}(k_1,\ldots,k_{2N_f})=G\prod_iF(k_i\rho)(2\pi)^4\delta^{(4)}\!\Big(\sum k_i\Big),
\end{equation}
with $G$ determined by the instanton density $n$ and distribution $d(\rho)$. Using $n\simeq1~\text{fm}^{-4}$ and $\rho\simeq0.33~\text{fm}$ yields $G\sim5$-$10~\text{GeV}^{-2}$. These parameters quantitatively reproduce the observed condensate and $f_\pi$ values.

\section{From Covariant to Light-Front Gap Equation}
The Euclidean gap equation
\begin{equation}
M(p_E)=m+\int\!\frac{\dd^4\ell_E}{(2\pi)^4}K(p_E,\ell_E)\frac{M(\ell_E)}{\ell_E^2+M^2(\ell_E)}
\end{equation}
is transformed to LF variables via $\ell^\pm=\ell^0\pm\ell^3$:
\begin{equation}
M(x,\bm{\ell}_\perp)=m+\int_0^1\!\dd y\!\int\!\frac{\dd^2\bm{k}_\perp}{(2\pi)^3}K(x,\bm{\ell}_\perp;y,\bm{k}_\perp)\frac{M(y,\bm{k}_\perp)}{\bm{k}_\perp^2+M^2(y,\bm{k}_\perp)-y(1-y)P^{+2}}.
\end{equation}
Boundary terms at $x=0,1$ carry zero-mode contributions that generate $\vev{\qb q}_\LF$. Numerical solutions show the same critical coupling $G_{\text{crit}}$ as covariant approaches, confirming full equivalence when zero modes are retained.

\end{subappendices}

\chapter{Pion Distribution Amplitudes and formfactors }

\section{Why distribution amplitudes matter}
The pion's distribution amplitudes (DAs) encode how longitudinal momentum is shared by its valence quark and antiquark on the light cone.
They are the central nonperturbative inputs in factorized descriptions of hard exclusive processes and enter the hard kernels for the pion's electromagnetic and gravitational form factors (GFFs) \cite{Lepage:1980fj,Efremov:1980qeq,CollinsSoper1981,Diehl:2003ny,Belitsky:2005qn}.
Physically, $\varphi_\pi(x,\mu)$ gives the probability amplitude (at resolution scale $\mu$) to find the $u\bar d$ Fock state with the quark carrying the fraction $x$ of the pion's large $+$ momentum.

In this chapter we (i) define the twist-2 and twist-3 pion DAs with gauge links, (ii) summarize their evaluation in the instanton liquid model (ILM) of the QCD vacuum, and (iii) present their Efremov-Radyushkin-Brodsky-Lepage (ERBL) evolution~\cite{Efremov:1979qk,Lepage:1979zb}.

\section{Light-cone definitions}
Let $p^\mu$ denote the pion momentum and $z^\mu=(0,z^-,\bm{0}_\perp)$ a lightlike separation. With a straight Wilson line $W[0,z]=\mathcal{P}\exp\!\left[-ig\!\int_0^{z^-}\!A^+(t)\,dt\right]$, the twist-$2,3$ DAs are defined by~\cite{Shuryak:2020ktq,Kock:2020frx}
\begin{align}
\varphi_\pi^{A}(x,\mu) &= \frac{1}{iF_\pi}\!\int\!\frac{dz^-}{2\pi}\,e^{ixp^+z^-}
   \langle{0}|{\bar d(0)\gamma^+\gamma_5\,W[0,z]\,u(z)}|{\pi^+(p)}\rangle\bigg|_{\mu}, \label{eq:DAA}\\
\varphi_\pi^{P}(x,\mu) &= \frac{p^+}{F_\pi\chi_\pi(\mu)}\!\int\!\frac{dz^-}{2\pi}\,e^{ixp^+z^-}
   \langle{0}|{\bar d(0)i\gamma_5\,W[0,z]\,u(z)}|{\pi^+(p)}\rangle\bigg|_{\mu}, \label{eq:DAP}\\
\frac{\varphi_\pi^{T\,\prime}(x,\mu)}{6} &= \frac{p^\mu p^{\prime\nu}p^+}{F_\pi\chi_\pi(\mu)\,p\!\cdot\! p'}
\!\int\!\frac{dz^-}{2\pi}\,e^{ixp^+z^-}
\langle{0}|{\bar d(0)\sigma_{\mu\nu}\gamma_5\,W[0,z]\,u(z)}|{\pi^+(p)}\rangle\bigg|_{\mu}. \label{eq:DAT}
\end{align}
Here $F_\pi=\sqrt2\,f_\pi\approx 133\,\mathrm{MeV}$ and $\chi_\pi(\mu)=m_\pi^2/(\overline m_u(\mu)+\overline m_d(\mu))$ from PCAC.
The prime on $\varphi^{T}$ denotes differentiation with respect to $x$.
Eqs.~(\ref{eq:DAA})-(\ref{eq:DAT}) obey the normalization $\int_0^1\!dx\,\varphi^{A,P}(x,\mu)=1$ and $\int_0^1\!dx\,\varphi^{T\,\prime}(x,\mu)=0$.
We recall that twist counts suppression by powers of the hard scale $Q$ in collinear factorization. The leading-twist axial DA $\varphi^A$ dominates at asymptotic $Q$, while $\varphi^{P,T}$ are genuinely twist-3 and enter with $1/Q$ suppression-but can be numerically important at semi-hard scales \cite{Braun:2016wnp,Shuryak:2020ktq,Liu:2021xje,Liu:2023fpj}.


At the "instanton scale'' $\mu_0\equiv1/\bar\rho\approx0.63\,\mathrm{GeV}$, the ILM gives compact integral forms for the DAs~\cite{Liu:2023feu,Liu:2023fpj}
\begin{align}
\varphi_\pi^{A}(x;\mu_0) &= \frac{2N_c M^2}{f_\pi^2}\frac{1}{8\pi^2}\,\theta(x\bar x)
\int_0^\infty\!\!dk_\perp^2\,
\frac{{\cal F}^2\!\left(\frac{k_\perp}{\lambda_\pi^A\sqrt{x\bar x}}\right)}
{x\bar x\,m_\pi^2-k_\perp^2-M^2}, \\
\varphi_\pi^{P}(x;\mu_0) &= \frac{N_c M}{f_\pi^2\chi_\pi(\mu_0)}\frac{\theta(x\bar x)}{8\pi^2\,x\bar x}
\int_0^\infty\!\!dk_\perp^2\,
\frac{k_\perp^2+M^2}{x\bar x\,m_\pi^2-k_\perp^2-M^2}\,
{\cal F}\!\left(\frac{k_\perp}{\lambda_\pi^P\sqrt{x\bar x}}\right), \\
\frac{\varphi_\pi^{T\,\prime}(x;\mu_0)}{6} &=
\frac{N_c M}{f_\pi^2\chi_\pi(\mu_0)}\frac{(x-\bar x)\,\theta(x\bar x)}{8\pi^2\,x\bar x}
\int_0^\infty\!\!dk_\perp^2\,
\frac{k_\perp^2+M^2}{x\bar x\,m_\pi^2-k_\perp^2-M^2}\,
{\cal F}\!\left(\frac{k_\perp}{\lambda_\pi^T\sqrt{x\bar x}}\right),
\end{align}
with $\bar x\equiv1-x$.
The instantonic profile ${\cal F}$ is determined by the zero-mode wavefunction and can be expressed via modified Bessel functions; phenomenological scale parameters $(\lambda_\pi^A,\lambda_\pi^P,\lambda_\pi^T)$ fix the normalization at $\mu_0$.
Note that the  operator identity $$\partial_\nu\,\bar d(0)\sigma^{\mu\nu}\gamma^5 u(z)
= -\partial^\mu\,\bar d(0)i\gamma^5 u(z) + m\,\bar d(0)\gamma^\mu\gamma^5 u(z)$$ implies that $\varphi^P$ and $\varphi^T$ share the same low-energy coupling $\chi_\pi(\mu)$-and hence inherit the quark-mass anomalous dimension.

\begin{figure*}
    \centering
    \includegraphics[scale=0.7]{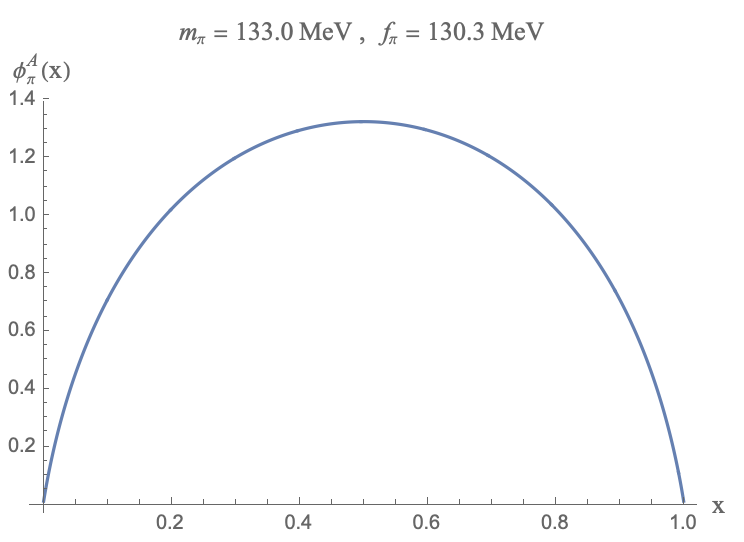}
    \caption{The un-evolved pion DA versus parton-$x$ at low resolution, in the the QCD instanton vacuum.}
    \label{PIDAX1}
\end{figure*}
\begin{figure*}
    \centering
    \includegraphics[scale=0.7]{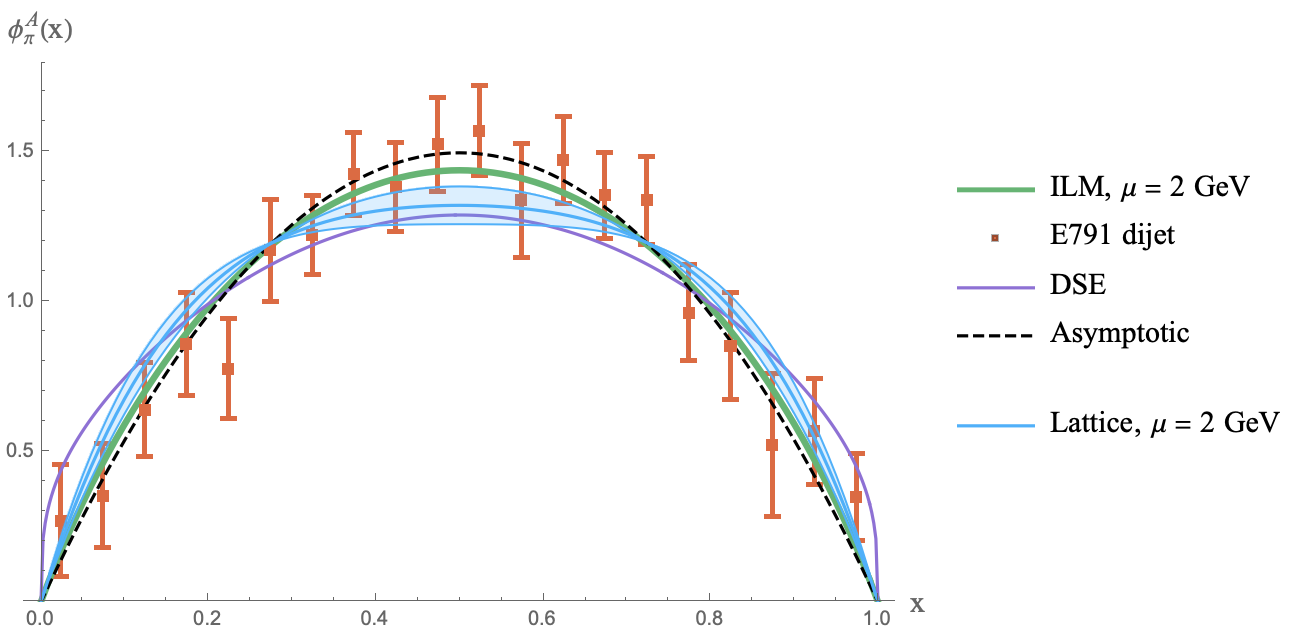}
    \caption{The evolved pion DA using the NLO ERBL equation to $\mu=2$ GeV (solid-green),
      compared with the lattice calculation (RQCD)~(shaded-blue)~ \cite{RQCD:2019osh} , Dyson-Schwinger result~\cite{Shi:2015esa}
      (solid-purple) and the asymptotic QCD result (dashed-black). The experimental data points (red squared points) are extracted from $\pi^-$ into di-jets via diffractive dissociation 
      with invariant dijet mass $6$ GeV~\cite{E791:2000xcx} and normalized in~\cite{Broniowski:2007si}.}
    \label{PIDAX2}
\end{figure*}

\section{Asymptotics and ERBL evolution}
At large $\mu$ the leading-twist DA approaches the conformal limit $\varphi_\pi^{A}(x,\mu)\to 6x\bar x$, while~\cite{Efremov:1980qeq}
\begin{align}
\varphi_\pi^{P}(x,\mu)&\to 1\times\Bigg[\frac{\alpha_s(\mu)}{\alpha_s(\mu_0)}\Bigg]^{-3C_F/b},\qquad
\varphi_\pi^{T}(x,\mu)\to 6x\bar x\times\Bigg[\frac{\alpha_s(\mu)}{\alpha_s(\mu_0)}\Bigg]^{C_F/b},
\end{align}
reflecting their distinct anomalous dimensions.
For practical evolution one expands in Gegenbauer polynomials~\cite{Diehl:2003ny} (and references therein)
\begin{align}
\varphi^{A,T}(x,\mu) &= 6x\bar x\sum_{n=0}^\infty a_n^{A,T}(\mu)\, C_n^{(3/2)}(x-\bar x),\qquad
\varphi^{P}(x,\mu) = \sum_{n=0}^\infty a_n^{P}(\mu)\, C_n^{(1/2)}(x-\bar x),
\end{align}
The polynomials are ortho-normal, and satisfy the following properties

\begin{align}
\int_0^1 dx\,[x(1-x)]^{\lambda-1/2}\,
C_n^{(\lambda)}(2x-1)C_m^{(\lambda)}(2x-1)
&= \delta_{mn}\frac{\pi 2^{1-2\lambda}\Gamma(n+2\lambda)}{n!\,(n+\lambda)\,\Gamma^2(\lambda)},\\
C_0^{(\lambda)}(z)&=1,\quad C_1^{(\lambda)}(z)=2\lambda z, \\
\frac{d}{dz}C_n^{(\lambda)}(z) &= 2\lambda\, C_{n-1}^{(\lambda+1)}(z).
\end{align}

With  this in mind and using the orthogonality,  yield  the moments
\begin{align}
a_n^{A,T}(\mu) &= \frac{2(2n+3)}{3(n+1)(n+2)}\int_0^1\!dy\, C_n^{(3/2)}(y-\bar y)\,\varphi^{A,T}(y,\mu),\\
a_n^{P}(\mu) &= (2n+1)\int_0^1\!dy\, C_n^{(1/2)}(y-\bar y)\,\varphi^{P}(y,\mu).
\end{align}
The leading-log ERBL evolution is diagonal in this basis~\cite{Efremov:1980qeq,Lepage:1979zb}
\begin{align}
a_n^{X}(\mu)=a_n^{X}(\mu_0)\left[\frac{\alpha_s(\mu)}{\alpha_s(\mu_0)}\right]^{\gamma_n^{X}/b},\qquad
X\in\{A,P,T\},
\end{align}
with one-loop anomalous dimensions
\begin{align}
\gamma_n^{A} &= C_F\!\left[-3 + 4\sum_{j=1}^{n+1}\frac{1}{j} - \frac{2}{(n+1)(n+2)}\right],\;
\gamma_n^{P} = C_F\!\left[-3 + 4\sum_{j=1}^{n+1}\frac{1}{j} - \frac{8}{(n+1)(n+2)}\right],\;
\gamma_n^{T} = C_F\!\left[-3 + 4\sum_{j=1}^{n+1}\frac{1}{j}\right],
\end{align}
and $b=\frac{11}{3}N_c-\frac{2}{3}N_f$. 
For the ERBL evolution at leading log it is sufficient to truncate at modest $n$ ($n=2,4$) for phenomenology, since higher moments evolve away with increasing $\mu$. In general, the evolution  is obtained by setting $\mu\sim \mathcal{O}(Q)$ for hard processes, while  for semi-hard kinematics, a matching to ILM at $\mu_0\simeq 1/\bar\rho$ captures large nonperturbative end-point effects.
The ILM naturally suppresses $x\to0,1$ via the zero-mode transverse profile ${\cal F}$, stabilizing convolution integrals in hard kernels.
Although $1/Q$ suppressed, $\varphi^{P,T}$ contributions can be sizable at a few GeV.


\section{Pion gravitational formfactors}
As we already mentioned many times, 
 pion occupies a unique position in the spectrum of strongly interacting particles. As the lightest hadron, its properties reflect the Nambu-Goldstone nature of chiral symmetry breaking in QCD. As a composite quark bound state, it simultaneously encodes nonperturbative dynamics of confinement and semiclassical vacuum structure. This dual character makes the pion an exceptionally sensitive probe of the short and long-distance structure of QCD as well as the structure of the QCD vacuum itself \cite{Schafer:1996wv,Diakonov:1995ea,Nowak:1996aj}.

Among all local operators in quantum field theory, the stress (energy-momentum) tensor $T_{\mu\nu}$ holds a privileged status. It governs the response of the theory to spacetime deformations, defines conserved charges associated with translations, and encodes the flow of energy, momentum, and stress. Its matrix elements between hadronic states define the gravitational form factors (GFFs), quantities that play the role of form factors in hypothetical couplings to gravity. Even though gravitational interactions are far too weak to be used experimentally, these form factors appear in hard exclusive processes, generalized parton distributions, deeply virtual Compton scattering, and dispersion relations connecting spacelike and timelike amplitudes \cite{Polyakov:2018}. They reveal fundamental aspects of hadron structure such as the distribution of mass, pressure, and shear forces, the decomposition of hadron mass into quark and gluon contributions, and the mechanical stability of strongly interacting systems \cite{Polyakov:2018}.

A profound feature of the QCD energy-momentum tensor is its anomalous trace,
\begin{equation}
T^\mu_{\ \mu} = \frac{\beta(g^2)}{4g^4} F^a_{\mu\nu} F^{a\mu\nu} + m \bar\psi\psi,
\end{equation}
which manifests the quantum breaking of classical scale invariance. The running of the strong coupling introduces a dynamical scale $\Lambda_{\rm QCD}$, and the expectation value of the trace anomaly contributes significantly to hadron masses \cite{Pagels:1978dd,Vainshtein:1981wh,Shifman:1988zk}. Understanding how this anomaly is realized in the pion provides deep insight into the interplay between chiral dynamics and conformal symmetry breaking.

The gravitational form factors (GFFs) of the pion encode the response of the system to a probe that couples to the energy-momentum tensor\footnote{Also known as the stress tensor.} (EMT)~\cite{Polyakov:2018zvc} . Although gravitons are not really available for experiments, the same matrix elements can be accessed through moments of generalized parton distributions (GPDs)~\cite{Diehl:2003ny} and by lattice QCD~\cite{Hackett:2023nkr}. The two scalar functions that parameterize the pion EMT, \(A_\pi(Q^2)\) and \(D_\pi(Q^2)\), summarize, respectively, how momentum is carried by constituents and how internal forces (pressure and shear) are arranged.

We define the pion EMT matrix element by
\begin{equation}
\label{eq:EMTdecomp}
\langle \pi(p') | T^{\mu\nu} | \pi(p) \rangle =
2 \bar p^\mu \bar p^\nu\, A_\pi(Q^2)
+ \frac{1}{2} (q^\mu q^\nu - g^{\mu\nu} q^2)\, D_\pi(Q^2)
+ 2 m_\pi^2 g^{\mu\nu} \bar C(Q^2),
\end{equation}
where \(q = p' - p\), \(\bar p = (p' + p)/2\), and \(Q^2 = -q^2\). Energy-momentum conservation enforces \(A_\pi(0) = 1\), while the mechanical stability of a bound state implies a negative \(D\)-term at the origin, \(D_\pi(0)<0\).

\section{The Energy-Momentum Tensor and the Trace Anomaly}

The QCD EMT in a symmetric, gauge-invariant form is
\begin{equation}
\label{eq:Tmunu}
T^{\mu\nu} =
- F^{a\,\mu\lambda}F^{a\,\nu}{}_{\lambda}
+ \frac{1}{4} g^{\mu\nu} F^a_{\rho\sigma}F^{a\,\rho\sigma}
+ \bar\psi \gamma^{(\mu} i\overleftrightarrow{D}^{\nu)} \psi.
\end{equation}
Classically, \(T^\mu{}_\mu=0\) in the massless limit. Quantum corrections break scale invariance, producing the trace anomaly
\begin{equation}
\label{eq:trace}
T^\mu{}_\mu = \frac{\beta(g)}{2g} F^a_{\rho\sigma}F^{a\,\rho\sigma}+ m\,\bar\psi\psi,
\end{equation}
with the QCD beta function
\begin{equation}
\label{eq:beta}
\beta(g) = \mu \frac{dg}{d\mu}
= -\frac{\beta_0}{16\pi^2} g^3 - \frac{\beta_1}{(16\pi^2)^2} g^5 + \cdots,
\qquad
\beta_0 = 11 - \frac{2}{3}N_f,
\quad
\beta_1 = 102 - \frac{38}{3}N_f.
\end{equation}
The anomaly implies that scale breaking is driven by gluon dynamics even in the chiral limit. In the vacuum, this is reflected in the nonzero gluon condensate \(\langle F^2\rangle\), while for hadrons it controls the scalar channel of the EMT and contributes to their mass budget.

\section{Decomposition and Physical Interpretation}

Projecting Eq.~\eqref{eq:EMTdecomp} in the Breit frame, one finds~\cite{Liu:2024vkj}
\begin{align}
\label{eq:T00}
T^{00}(Q^2) &= 2\!\left(m_\pi^2 + \frac{Q^2}{4}\right) A_\pi(Q^2) + \frac{Q^2}{2} D_\pi(Q^2), \\
\label{eq:Tii}
T^{ii}(Q^2) &= 2\bar p_i^2\, A_\pi(Q^2) - \frac{1}{6} Q^2 D_\pi(Q^2),
\end{align}
with no sum over \(i\). The time component tracks energy flow and normalizes \(A_\pi(0)=1\). The spatial trace is directly tied to internal stress. Eliminating \(A_\pi\) or \(D_\pi\) gives convenient relations:
\begin{align}
\label{eq:AfromT}
A_\pi(Q^2) &= \frac{3 T^{00}(Q^2) - T^\mu{}_\mu(Q^2)}{Q^2 + 4m_\pi^2}, \\
\label{eq:DfromT}
D_\pi(Q^2) &= \frac{T^\mu{}_\mu(Q^2) - T^{00}(Q^2)}{Q^2}.
\end{align}
Negative \(D_\pi(0)\) ensures that repulsive core pressure is balanced by attractive surface tension, as required by mechanical stability.

\section{Connection to GPDs and Energy Flow}

The GFFs are the second Mellin moments of pion GPDs. For the unpolarized GPD \(H_\pi(x,\xi,t)\) one has~\cite{Diehl:2003ny}
\begin{equation}
\label{eq:GPDmoment}
\int_{-1}^{1} dx\, x\, H_\pi(x,\xi,t) = A_\pi(t) + \xi^2 D_\pi(t),
\end{equation}
with \(t = q^2 = -Q^2\). This identity links the GFFs to experimentally accessible observables in deeply virtual Compton scattering (DVCS) and timelike Compton scattering (TCS). Physically, the integral weights partonic momentum flow by \(x\), while the \(\xi^2\) term encodes the stress redistribution characterized by \(D_\pi\).

\section{Perturbative Regime and Twist Expansion}

At large \(Q^2\), the EMT matrix element factorizes into~\cite{Lepage:1980fj}
\begin{equation}
\label{eq:fact}
\langle \pi | T^{\mu\nu} | \pi \rangle
= \int_0^1 dx_1 dx_2\, \Phi_\pi(x_1,\mu)\, T_H^{\mu\nu}(x_1,x_2,Q^2,\mu)\, \Phi_\pi(x_2,\mu) + \mathcal{O}\!\left(\frac{1}{Q^4}\right),
\end{equation}
with the leading-order hard kernel scaling like~\cite{Shuryak:2020vzl,Tong:2021ctu,Tong:2022zax,Liu:2024vkj}
\begin{equation}
\label{eq:hardkernel}
T_H \sim \frac{16\pi \alpha_s C_F}{Q^2} \frac{1}{x_1 x_2} \times \mathcal{P}^{\mu\nu}(x_1,x_2),
\end{equation}
where \(\mathcal{P}^{\mu\nu}\) encodes the EMT vertex structure. The distribution amplitude evolves according to the ERBL equation~\cite{Efremov:1979qk,Lepage:1979zb}
\begin{equation}
\label{eq:ERBL}
\mu \frac{d}{d\mu} \varphi_\pi(x,\mu) = \int_0^1 dy\, V(x,y)\, \varphi_\pi(y,\mu),
\end{equation}
driving \(\varphi_\pi(x,\mu)\) towards the asymptotic form \(6x(1-x)\). Twist-3 DAs \(\varphi_\pi^{P,T}\) contribute at \(\mathcal{O}(1/Q)\) and are numerically important in the few-GeV range, where they correct the leading-twist picture and feed the scalar channel~\cite{Shuryak:2020ktq}.


The QCD instanton vacuum offers a semiclassical, microscopic description of the nonperturbative ground state of QCD \cite{Shuryak:1982,Schafer:1996wv,Nowak:1996aj}. Instantons and anti-instantons are topological tunneling configurations that interpolate between vacua of different winding number. Their distribution in size and density can be extracted from semiclassical arguments, phenomenology, and lattice QCD simulations \cite{Schafer:1996wv,Diakonov:1995ea,Nowak:1996aj}. The resulting instanton liquid model (ILM) successfully explains the dynamical generation of the constituent quark mass, spontaneous breaking of chiral symmetry, and numerous hadronic observables \cite{Diakonov:1995ea,Nowak:1996aj,Kacir:1996qn}. It also naturally accommodates the gluonic scale (stress tensor trace) anomaly \cite{Novikov:1981}.

Modern high-statistics lattice QCD computations of pion gravitational form factors allow for quantitative comparison with theoretical predictions \cite{Hackett:2023,Wang:2024}. This motivates a fully detailed, mathematically expanded account of the ILM computation of pion gravitational form factors, including both trace and traceless components, quark and gluon contributions, and their evolution to higher resolution scales. This chapter provides exactly such an expanded, pedagogical exposition, synthesizing semiclassical QCD, chiral dynamics, and light-front quantization \cite{Liu:2024jno}.

\section{ Breaking of Scale Invariance in QCD and Instanton Vacuum }

The QCD vacuum is a medium populated by fluctuating gauge fields whose properties are far from trivial. Instantons are Euclidean classical solutions of the Yang-Mills equations with finite action,
\begin{equation}
S_{\text{inst}} = \frac{8\pi^2}{g^2},
\end{equation}
and nonzero topological charge \cite{Schafer:1996wv}. Their size $\rho$ and collective coordinates parametrize a manifold of degenerate configurations. Semiclassical analysis yields a size distribution \cite{Schafer:1996wv}
\begin{equation}
n(\rho) \sim \rho^{-5} \left( \rho \Lambda_{\rm QCD} \right)^{b} \exp\left(-C \rho^2/\bar R^2 \right),
\end{equation}
where $b = 11N_c/3 - 2N_f/3$ is the one-loop coefficient of the QCD beta-function and $C$ is a constant of order unity. Perturbation theory controls the small-$\rho$ tail \cite{Schafer:1996wv}, while interactions among instantons regulate the infrared \cite{Diakonov:1995ea,Nowak:1996aj}. Lattice cooling studies support the picture of a dilute but correlated ensemble with characteristic size $\bar\rho \approx 0.3\,\text{fm}$ and average separation $\bar R \approx 1\,\text{fm}$ \cite{Schafer:1996wv,Athenodorou:2018jwu}.

Instantons induce fermionic zero modes of definite chirality, whose delocalization across the ensemble gives rise to the quark condensate and a constituent quark mass $M(p)$ \cite{Diakonov:1995ea}. This mechanism provides a microscopic understanding of spontaneous chiral symmetry breaking. Conformal symmetry is broken by ultraviolet fluctuations encoded in the trace anomaly,
\begin{equation}
T^\mu_{\ \mu} = \frac{\beta(g^2)}{4g^4} F^a_{\mu\nu}F^{a\mu\nu} + m\bar\psi\psi,
\end{equation}
with $F^2$ counting, in the instanton ensemble, the number of pseudoparticles in a given volume modulo subtractions of perturbative contributions \cite{Novikov:1981}.

The instanton number fluctuates around its mean $\bar N$ according to a grand-canonical distribution,
\begin{equation}
P(N) = e^{bN/4}\left( \frac{\bar N}{N} \right)^{bN/4},
\end{equation}
resulting in a topological compressibility
\begin{equation}
\sigma_T = \frac{\langle (N - \bar N)^2\rangle}{\bar N} = \frac{4}{b}.
\end{equation}
This compressibility reflects the quantum nature of the vacuum and plays an essential role in enabling finite pion matrix elements of gluonic operators \cite{Novikov:1981,Nowak:1996aj}.

\section{The Pion Mass Identity from the Trace Anomaly}

For a one-pion state normalized as
\begin{equation}
\langle p'|p\rangle = 2E(p)(2\pi)^3 \delta^{(3)}(p - p'),
\end{equation}
the forward EMT matrix element is completely fixed by Lorentz invariance,
\begin{equation}
\langle p | T_{\mu\nu} | p\rangle = 2 p_\mu p_\nu,
\end{equation}
and therefore the trace satisfies
\begin{equation}
\langle p | T^\mu_{\ \mu} | p \rangle = 2 m_\pi^2.
\end{equation}
Using the anomalous expression for the trace yields the pion mass identity \cite{Liu:2024jno}
\begin{equation}
m_\pi = -\frac{b}{32\pi^2}\frac{\langle \pi | F^2 | \pi\rangle}{2m_\pi} + \frac{\langle \pi | m\bar\psi\psi | \pi \rangle}{2 m_\pi}.
\end{equation}

In the instanton vacuum the forward matrix element of $F^2$ is obtained from the sensitivity of the Euclidean pion two-point function to fluctuations in the instanton number. After isolating the connected part of the three-point correlator and employing the Feynman-Hellmann theorem, one finds
\begin{equation}
-\frac{b}{32\pi^2} \frac{\langle \pi | F^2 | \pi\rangle}{2m_\pi} = \frac{m_\pi}{2},
\end{equation}
while differentiation of the Euclidean pion energy with respect to the bare quark mass gives
\begin{equation}
\frac{\langle \pi | m\bar\psi\psi | \pi\rangle}{2 m_\pi} = \frac{m_\pi}{2}.
\end{equation}

Thus the pion mass decomposes symmetrically:
\begin{equation}
m_\pi = \frac{m_\pi}{2} \bigg|_{\rm gluon} + \frac{m_\pi}{2} \bigg|_{\rm quark}.
\end{equation}
This equal partition is a special consequence of the Goldstone nature of the pion; in other hadrons the mass decomposition is markedly different. In the chiral limit, the scalar gluonic matrix element vanishes, reflecting the decoupling of massive scalar glueball backgrounds from the massless Goldstone mode \cite{Liu:2024jno}.

\section{Gravitational Form Factors and General Structure}

The matrix element of the energy-momentum tensor between off-forward pion states takes the most general Lorentz- and parity-invariant form
\begin{equation}
\langle p'|T_{\mu\nu}|p\rangle = 2\bar p_\mu \bar p_\nu A(q^2) + \frac{1}{2}(q_\mu q_\nu - g_{\mu\nu} q^2) D(q^2),
\end{equation}
where $q=p'-p$ and $\bar p = (p+p')/2$. The scalar function $A(q^2)$ describes the distribution of momentum and energy, while $D(q^2)$ encodes mechanical properties such as pressure and shear forces \cite{Polyakov:2018}.

Separating the trace and traceless parts,
\begin{equation}
T_{\mu\nu} = \left(T_{\mu\nu} - \frac{1}{4} g_{\mu\nu} T^\alpha_{\ \alpha}\right) + \frac{1}{4} g_{\mu\nu} T^\alpha_{\ \alpha},
\end{equation}
one introduces the trace form factor
\begin{equation}
T_\pi(q^2) = \frac{\langle p'|T^\mu_{\ \mu}|p\rangle}{2m_\pi^2}
= G_\pi(q^2) + \sigma_\pi(q^2),
\end{equation}
where $G_\pi(q^2)$ and $\sigma_\pi(q^2)$ represent gluonic and quark contributions respectively. Chiral symmetry implies the low-energy theorem
\begin{equation}
D(0) = -1,
\end{equation}
reflecting the long-range nature of the pion \cite{Polyakov:2018,Liu:2024jno}.

\section{The stress tensor trace Form Factor in the Instanton Vacuum}

The ILM provides a microscopic computation of $G_\pi(q^2)$ by relating it to the Euclidean $F^2$-$F^2$ vacuum correlator. In the soft regime, $Q^2 \ll 1/\bar\rho^2$, the dynamics is dominated by the lightest scalar glueball of mass $m_{0^{++}} \approx 1.25\,\text{GeV}$. One obtains an approximately monopole form
\begin{equation}
G_\pi(Q^2) \simeq \left(1-\sigma_\pi(0)\right) \frac{1}{1 + Q^2/m_{0^{++}}^2}.
\end{equation}

In the semi-hard regime, $Q\bar\rho \sim 1$, nonlocal form factors emerge from Fourier-transforming the instanton gauge profile. These take the form of integrals over Bessel functions $J_n$ and modified Bessel functions $K_n$, which are well approximated by analytic expressions. Together with the quark sigma-term contribution,
\begin{equation}
\sigma_\pi(Q^2) = \frac{\sigma_\pi(0)}{1+Q^2/m_\sigma^2},
\end{equation}
with $m_\sigma \approx 683\,\text{MeV}$, the ILM yields a complete picture of the trace form factor across all relevant scales \cite{Liu:2024jno}.

\section{Traceless Gravitational Form Factors from Light-Front Dynamics}

The traceless quark contribution to the EMT dominates at leading order in instanton density and can be computed using the pion light-front wave function obtained by diagonalizing the ILM light-front Hamiltonian. Working in a Drell-Yan frame ($q^+ =0$), one evaluates matrix elements of the twist-two operator $T^{++}$, which avoids contributions from higher Fock states.

The gravitational form factors follow from
\begin{equation}
A(Q^2)=\frac{1}{2P^{+2}} \langle p'|T^{++}|p\rangle,
\end{equation}
and
\begin{equation}
D(Q^2)=\frac{2}{3Q^2} \left[\langle p'|T^\mu_{\ \mu}|p\rangle
-\frac{Q^2+4m_\pi^2}{4P^{+2}}\langle p'|T^{++}|p\rangle
\right].
\end{equation}
Gluonic traceless contributions arise only at next-to-leading order in the instanton density and are suppressed by $\kappa^2$. Nonetheless they can be computed and expressed in terms of quark GFFs, with coefficients given by overlap integrals of instanton zero modes \cite{Liu:2024jno}.

To compare with lattice QCD results, ILM predictions are evolved from the resolution scale $\mu_0 = 1/\bar\rho \approx 0.6\,{\rm GeV}$ to $\mu = 2\,{\rm GeV}$ using leading-order DGLAP evolution of twist-two operators \cite{Hackett:2023,Wang:2024}.

\section{Radii and Spatial Interpretation}

The spatial distribution of energy and pressure within the pion is encoded in the slopes of gravitational form factors at zero momentum transfer. The tensor and scalar radii are defined by
\begin{equation}
\langle r_A^2 \rangle = -6\frac{A'(0)}{A(0)}, \qquad
\langle r_D^2 \rangle = -6\frac{D'(0)}{D(0)}.
\end{equation}
For the pion these radii illustrate the contrast between short-range twist-two dynamics and long-range scalar interactions \cite{Polyakov:2018,Liu:2024jno}.

Additional radii follow from Breit-frame matrix elements of $T_{00}$ and $T^\mu_{\ \mu}$,
\begin{equation}
\langle r_M^2 \rangle = -6\frac{dT_{\pi}^{00}(Q^2)/dQ^2}{T_\pi^{00}(0)},\qquad
\langle r_S^2 \rangle = -6\frac{dT_{\pi}^{\mu\mu}(Q^2)/dQ^2}{T_\pi^{\mu\mu}(0)},
\end{equation}
revealing a large scalar radius controlled by chiral dynamics in agreement with chiral perturbation theory \cite{Polyakov:2018}.


In summary, the instanton liquid model provides a remarkably coherent and quantitatively accurate description of the pion's gravitational structure \cite{Schafer:1996wv,Liu:2024jno}. Semiclassical gauge configurations generate both the dynamical breaking of scale invariance and the spontaneous breaking of chiral symmetry. The resulting zero-mode structure yields a constituent quark picture consistent with the Goldstone nature of the pion. Gravitational form factors computed at the ILM resolution scale evolve naturally to lattice QCD scales and agree closely with numerical data \cite{Hackett:2023,Wang:2024}. The gluonic trace anomaly contributes half of the pion mass, a striking hallmark of conformal symmetry breaking in QCD. The traceless gravitational form factors reveal a mechanically stable pion whose distributions of pressure and shear are governed by chiral and semiclassical dynamics \cite{Polyakov:2018,Liu:2024jno}.

\begin{figure*}
\centering
\includegraphics[height=5.15cm,width=.485\linewidth]{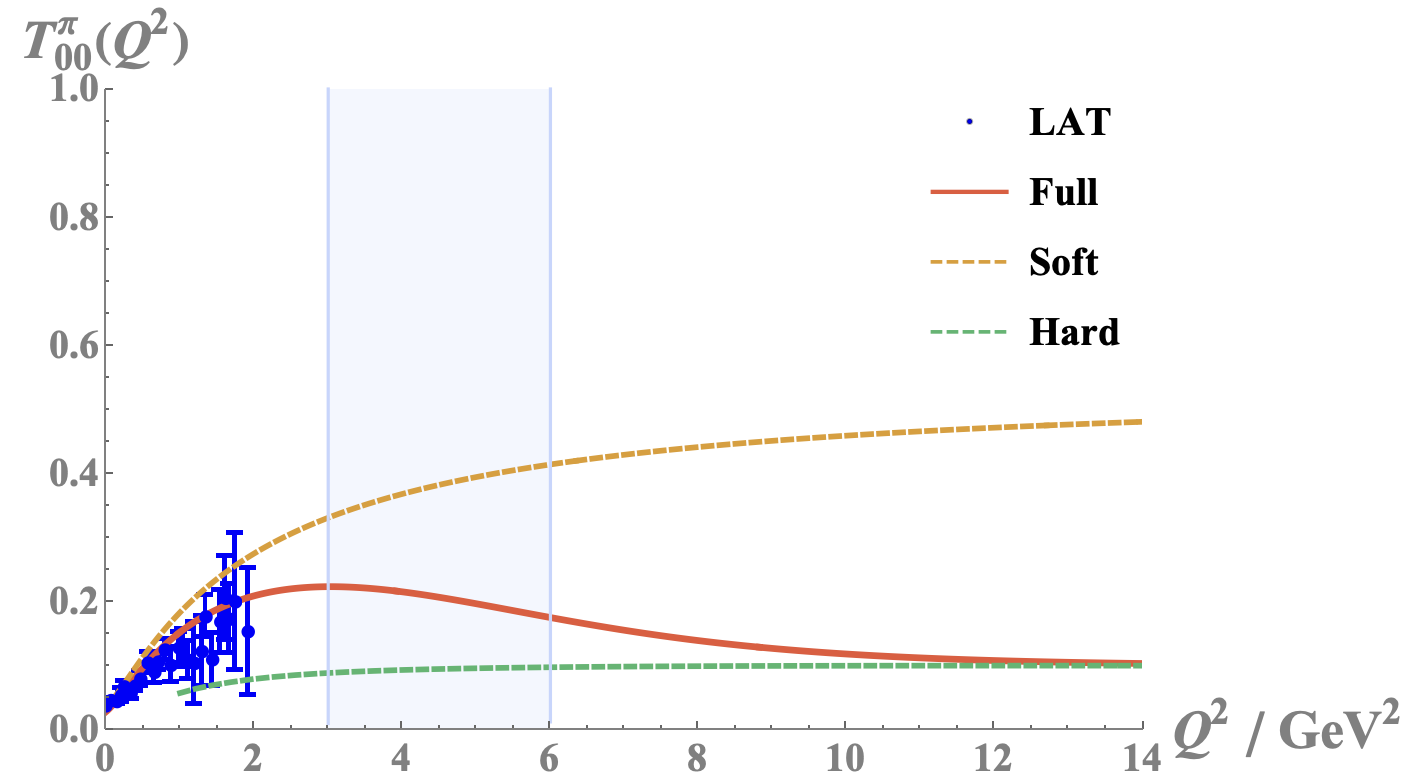}%
\includegraphics[height=5.15cm,width=.485\linewidth]{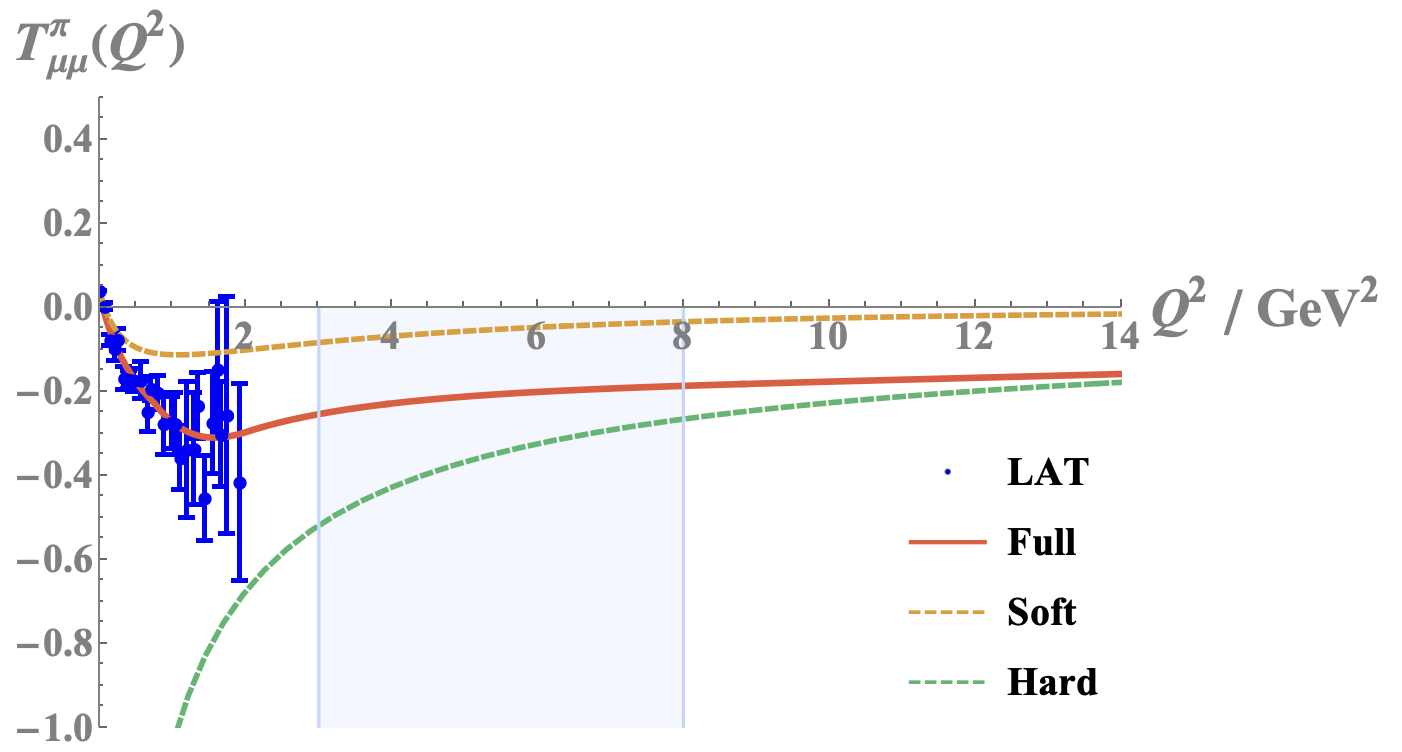}%
\caption{The pion 00-EMT (a) and pion trace-EMT (b) versus $Q^2$, with the  soft contribution from the ILM (dashed-orange line) from~\cite{Liu:2024jno},
the hard plus semi-hard contribution (dashed-green line) 
and the interpolation full sum (solid-red line) are from~\cite{Liu:2024vkj}, compared to the lattice results from~\cite{Hackett:2023nkr}.}
\label{fig:T00TMUMU}
\end{figure*}

\begin{figure*}
\centering
\includegraphics[height=5.15cm,width=.485\linewidth]{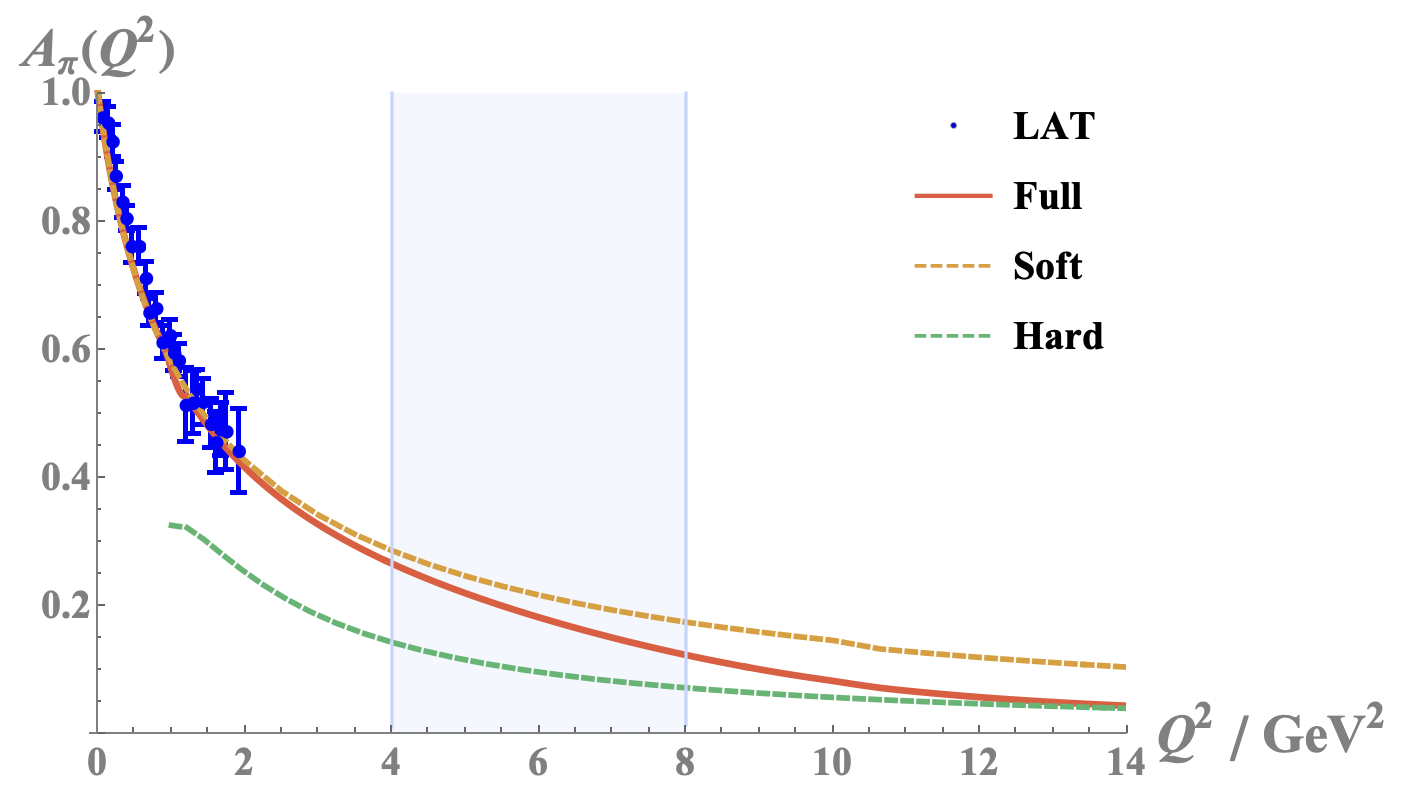}%
\includegraphics[height=5.15cm,width=.485\linewidth]{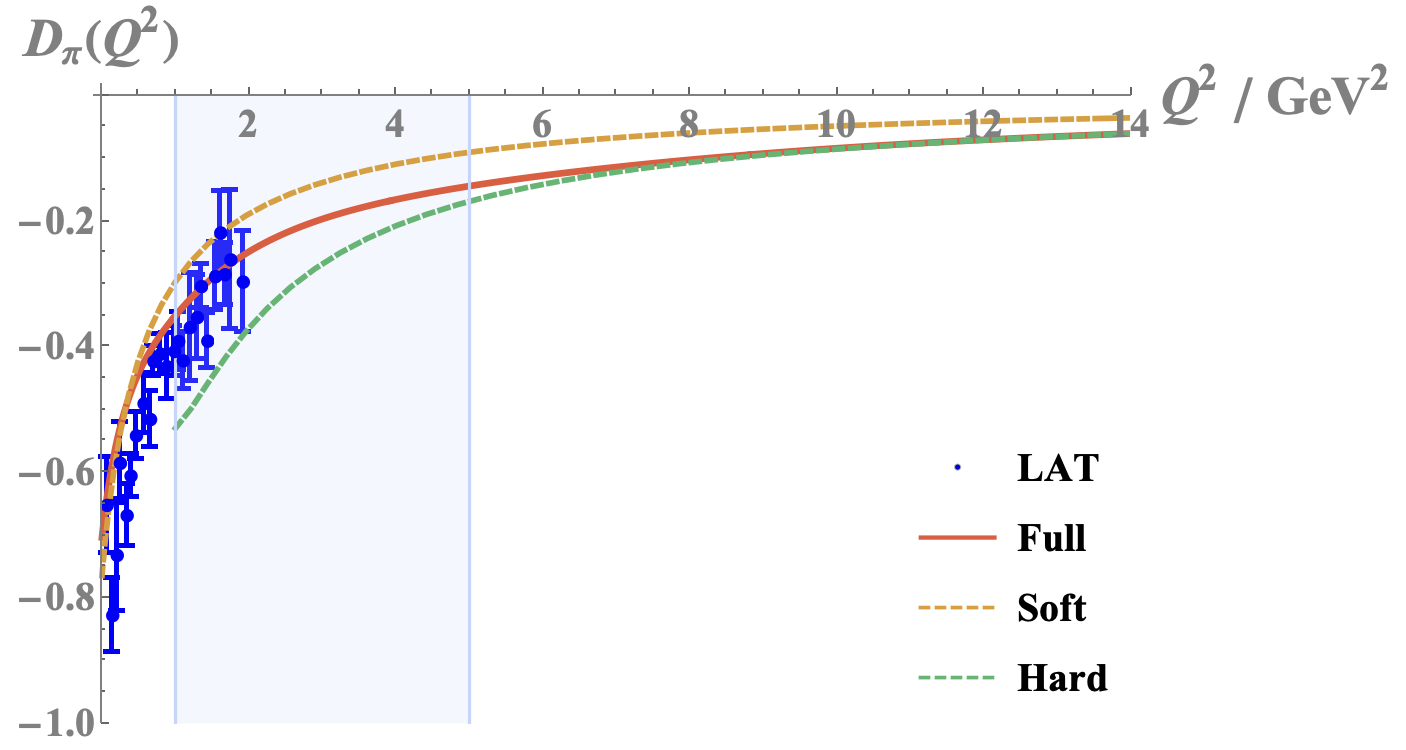}%
\caption{The pion A-GFF (a) and pion D-GFF (b) versus $Q^2$, with the  soft contribution from the ILM (dashed-orange line) from~\cite{Liu:2024jno},
the hard plus semi-hard contribution (dashed-green line)  and the interpolation full sum (solid-red line) are from~\cite{Liu:2024vkj}, compared to the lattice results from~\cite{Hackett:2023nkr}.}
\label{fig:ADALL}
\end{figure*}


\section{Pion Light-Front Wave Function in the ILM}

The pion state on the light front is written as
\begin{equation}
|P\rangle = \int \frac{dx}{\sqrt{2x(1-x)}} \int \frac{d^2 k_\perp}{(2\pi)^3} \sum_{s_1,s_2,a} \Phi_\pi^a(x,k_\perp,s_1,s_2)\; b_{s_1}^{a\dagger}(k)\, d_{s_2}^{a\dagger}(P-k)\,|0\rangle,
\end{equation}
where the wave function factorizes into a momentum profile and a Dirac-isospin factor,
\begin{equation}
\Phi_\pi^a(x,k_\perp) = \frac{1}{\sqrt{N_c}}\, \varphi_\pi(x,k_\perp)\, \bar u(k)\, i\gamma_5 \tau^a\, v(P-k).
\end{equation}
The scalar profile contains the instanton zero-mode form factor,
\begin{equation}
F(k) = \left[z(I_0 K_0 - I_1 K_1)'\right]^2_{z=k\rho/2},
\end{equation}
where modified Bessel functions encode the nonlocal structure of the zero mode \cite{Diakonov:1995ea,Schafer:1996wv,Nowak:1996aj}.

\section{Scalar Form Factor from Chiral and Instanton Dynamics}

Chiral Ward identities imply a representation for the scalar form factor in terms of the pion decay constant and the sigma field correlator,
\begin{equation}
F_S(q) = -\frac{1}{f_\pi} + \frac{\langle\sigma\rangle}{f_\pi^2}
- \frac{m_\pi^2}{f_\pi} \int d^4x\; e^{iq\cdot x}\; \langle T\,\sigma(x)\sigma(0)\rangle + \cdots.
\end{equation}
To one loop in chiral perturbation theory one finds
\begin{equation}
F_S(Q^2) = -\frac{1}{f_\pi} + \frac{1}{f_\pi^3}\left(Q^2 + \frac{1}{2}m_\pi^2\right)J(Q),
\end{equation}
where $J(Q)$ is the standard two-pion loop function \cite{Gasser:1983yg}. The sigma-term form factor follows from
\begin{equation}
\sigma_\pi(Q^2) = -\frac{f_\pi}{2} F_S(Q^2).
\end{equation}
Within the ILM the scalar channel is dominated by an emergent sigma meson of mass $m_\sigma \approx 683\,{\rm MeV}$, giving the monopole form
\begin{equation}
\sigma_\pi(Q^2) = \frac{1/2}{1 + Q^2/m_\sigma^2},
\end{equation}
in agreement with the semiclassical analysis \cite{Liu:2024jno}.







\section{Semi-Hard Instanton Effects}

At scales \(Q\rho \sim 1\) the instanton size \(\rho\) becomes relevant. The quark propagator in an instanton background can be schematically written as
\begin{equation}
\label{eq:Sinst}
S_I(x,y) = S_0(x-y) + S_{\text{zero}}(x,y) + S_{\text{NZM}}(x,y),
\end{equation}
where the zero-mode (chiral) and non-zero-mode (NZM) parts capture distinct physics. The NZM term induces an effective vector form factor \(\mathbb{G}_V(\rho Q)\) in the EMT, modifying the twist-3 contributions and producing semi-hard corrections:
\begin{equation}
\label{eq:NZM}
T_{00,\mathrm{NZM}}^\pi(Q^2) \sim \rho Q\, \mathbb{G}_V(\rho Q)
\int dx_1 dx_2\, \mathcal{K}(x_1,x_2)\, \varphi_\pi^P(x_1)\varphi_\pi^P(x_2) + \cdots,
\end{equation}
with \(\mathcal{K}\) a calculable kernel. This mechanism implements the transition between soft chiral dynamics and hard partonic scattering, and is essential for a realistic description of the trace channel at moderate \(Q^2\).

\section{Soft Sector and Matching Across Scales}

For \(Q \ll \rho^{-1}\), the zero modes dominate and the EMT is governed by chiral symmetry breaking. A practical description across all \(Q\) combines soft, NZM, and hard pieces:
\begin{equation}
\label{eq:matching}
\mathcal{O}(Q^2) = c_{\mathrm{soft}}(Q^2)\, \mathcal{O}_{\mathrm{soft}}(Q^2)
+ c_{\mathrm{NZM}}(Q^2)\, \mathcal{O}_{\mathrm{NZM}}(Q^2)
+ c_{\mathrm{hard}}(Q^2)\, \mathcal{O}_{\mathrm{hard}}(Q^2),
\qquad
\sum_i c_i(Q^2)=1,
\end{equation}
with a smooth switch in the few-GeV regime. This interpolation reproduces \(A_\pi(0)=1\), yields a negative \(D_\pi(0)\), and matches lattice trends for slopes and radii.

\section{Mechanical Densities and Internal Forces}

The spatial structure of internal forces follows from the Fourier transforms of \(D_\pi(Q^2)\)~\cite{Polyakov:2018zvc}
\begin{align}
\label{eq:pressure}
p_\pi(r) &= -\frac{1}{6\pi^2 r}
\int_0^\infty dQ\, \frac{Q^3 \sin(Qr)}{2E_\pi(Q)}\, D_\pi(Q^2), \\
\label{eq:shear}
s_\pi(r) &= -\frac{3}{8\pi^2}
\int_0^\infty dQ\, \frac{Q^4 j_2(Qr)}{2E_\pi(Q)}\, D_\pi(Q^2),
\end{align}
with \(E_\pi(Q) = \sqrt{m_\pi^2 + Q^2/4}\). For a dipole fit,
\begin{equation}
\label{eq:Ddipole}
D_\pi(Q^2) = \frac{D_\pi(0)}{(1+Q^2/M_D^2)^2},
\end{equation}
the integrals in Eqs.~\eqref{eq:pressure}-\eqref{eq:shear} can be evaluated analytically in terms of exponential integrals, yielding a positive core pressure and a negative surface pressure. The stability (von Laue) condition
\begin{equation}
\label{eq:Laue}
\int_0^\infty 4\pi r^2 p_\pi(r)\, dr = 0
\end{equation}
is automatically satisfied by any sufficiently fast-falling \(D_\pi(Q^2)\). The mechanical radius follows from the slope at the origin,
\begin{equation}
\label{eq:rm}
\langle r^2\rangle_D = \frac{6}{D_\pi(0)} \frac{dD_\pi(Q^2)}{dQ^2}\bigg|_{Q^2=0} = \frac{12}{M_D^2}.
\end{equation}
Empirically one finds \(\sqrt{\langle r^2\rangle_D}\approx 0.45\text{-}0.55~\mathrm{fm}\)~\cite{Liu:2024vkj}, smaller than the pion charge radius, indicating that confining forces act at shorter distances than electric screening.

\section{Summary and Outlook}

The pion GFFs provide a unifying view of QCD dynamics across scales. The function \(A_\pi(Q^2)\) traces how energy and momentum flow through the hadron, while \(D_\pi(Q^2)\) encodes the mechanical landscape of pressure and shear that stabilizes it. Perturbative QCD governs the asymptotic region through factorization and ERBL evolution; instanton non-zero modes supply semi-hard corrections that connect to the trace anomaly; and zero modes control the soft domain set by chiral symmetry breaking. The negative \(D\)-term emerges as a robust signature of confinement. Future lattice and experimental studies, especially at the Electron-Ion Collider, will refine the gluon contribution and test these mechanisms with higher precision.

\begin{figure*}
\centering
\includegraphics[height=5.1cm,width=.48\linewidth]{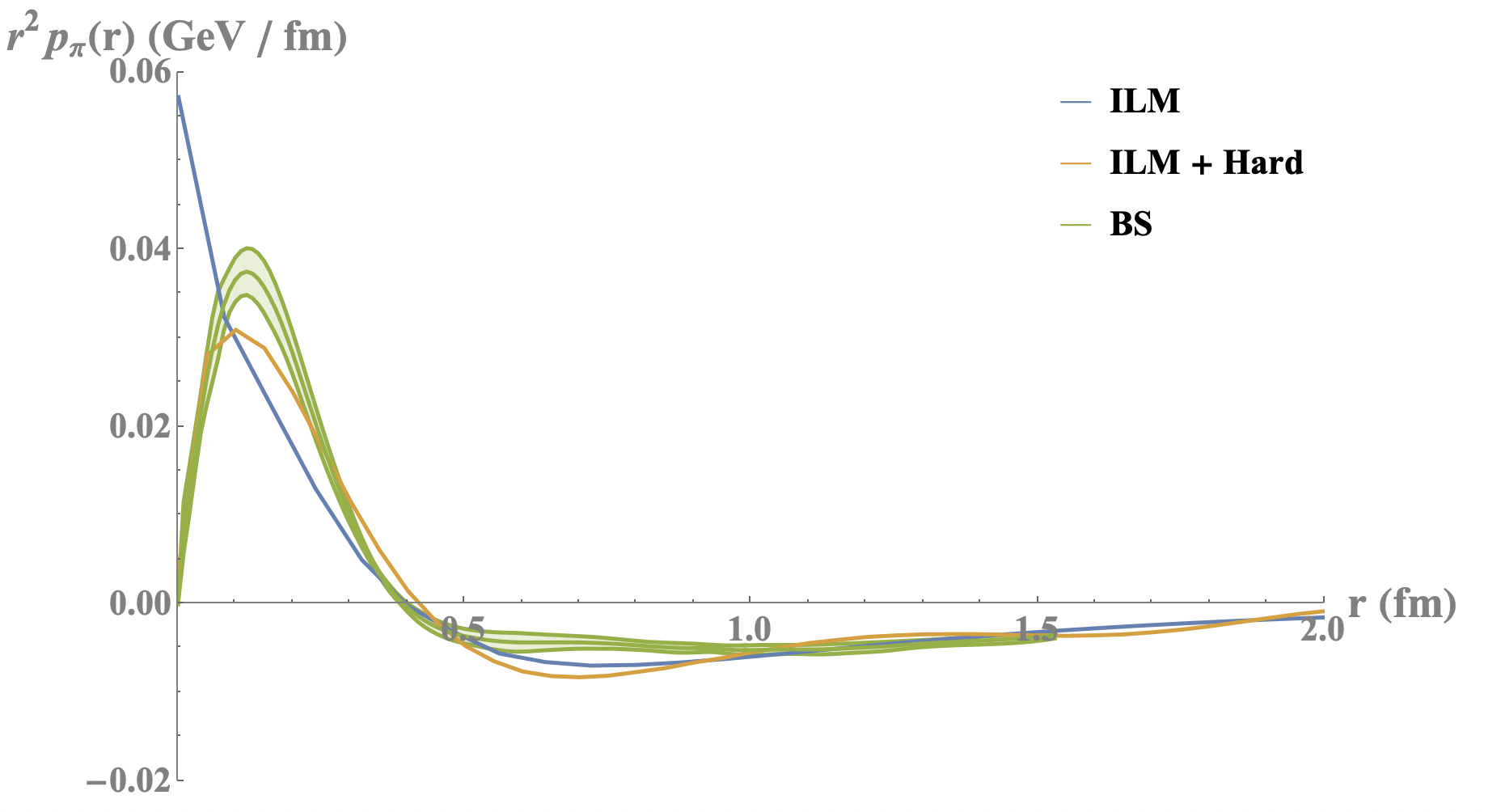}%
\includegraphics[height=5.1cm,width=.47\linewidth]{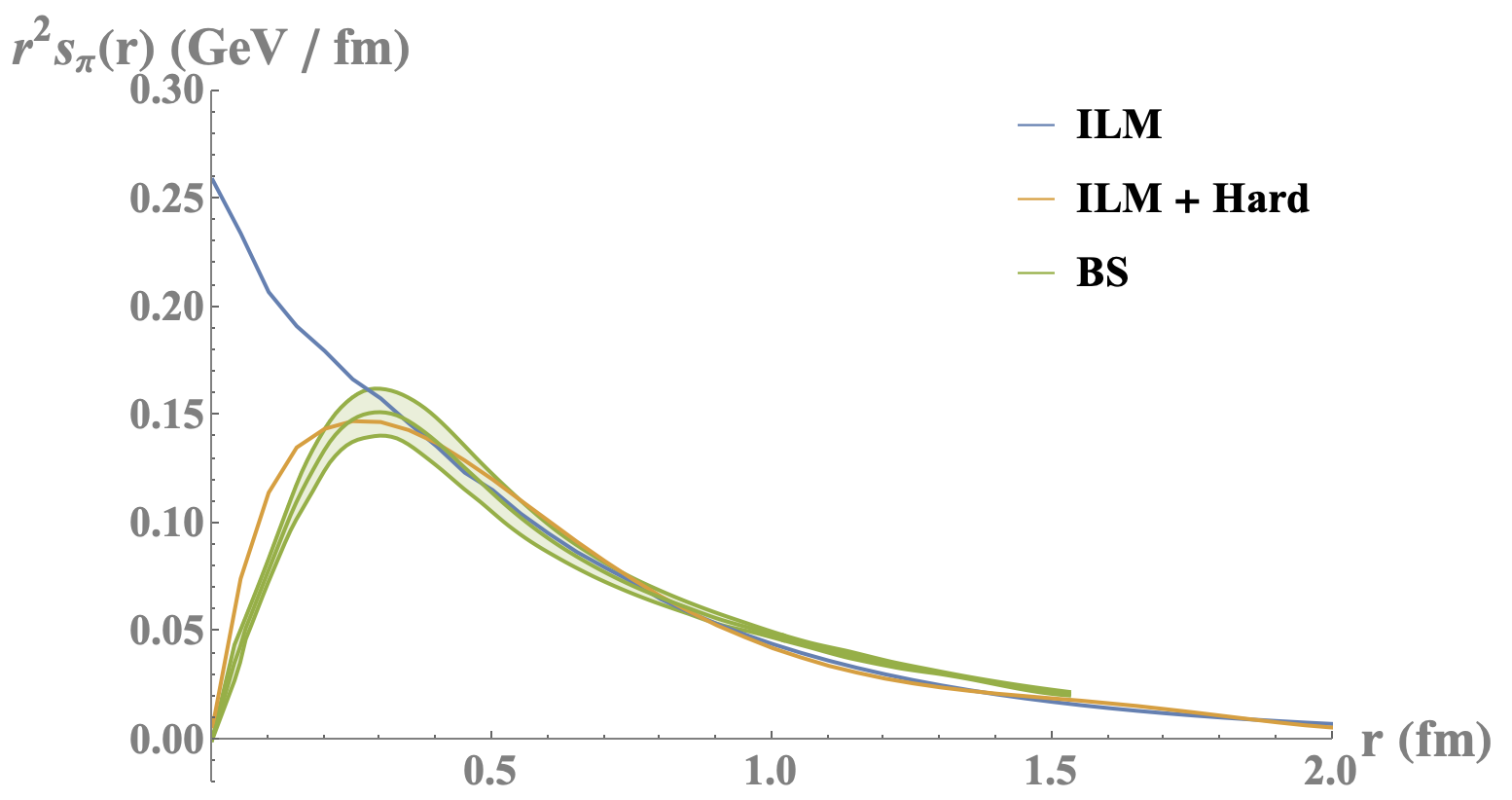}%
\caption{The pressure (a) and shear (b) in the pion calculated using the ILM (blue-solid line) and the ILM plus pQCD (orange solid line). The comparison is to the recent results from the
Bethe-Salpeter resummation~\cite{Xu:2023izo}.}
\label{fig:PS}
\end{figure*}

\begin{subappendices}

\section{Energy-Momentum (stress) Tensor and related vertices}

The EMT inserts the following vertices into quark and gluon lines. For quarks,
\begin{equation}
V_q^{\mu\nu}(k',k) = \frac{i}{4}
[\gamma^\mu (k' + k)^\nu + \gamma^\nu (k' + k)^\mu]
- \frac{i}{2} g^{\mu\nu} (\slashed{k}' + \slashed{k} - 2m).
\end{equation}
For the quark-gluon insertion,
\begin{equation}
V_{qg}^{\mu\nu,a} = i g_s T^a [\gamma^{(\mu} A^{\nu)} - g^{\mu\nu} \slashed{A}],
\end{equation}
and for gluons,
\begin{equation}
V_g^{\mu\nu,\alpha\beta} = -i g_s f^{abc}
\big[g^{\mu\alpha}(k_1 - k_2)^\nu + g^{\nu\alpha}(k_1 - k_2)^\mu\big] + \cdots.
\end{equation}
These rules generate the hard kernels that appear in Eq.~\eqref{eq:hardkernel}.


In the static limit, the spatial components of the EMT decompose as
\begin{equation}
T^{ij}(\bm{r}) = 
\left( \frac{r^i r^j}{r^2} - \frac{1}{3}\delta^{ij} \right) s_\pi(r)
+ \delta^{ij} p_\pi(r).
\end{equation}
Taking the divergence and imposing conservation \(\partial_i T^{ij}=0\) leads to coupled differential constraints that the transforms in Eqs.~\eqref{eq:pressure}-\eqref{eq:shear} satisfy. The von Laue condition in Eq.~\eqref{eq:Laue} expresses mechanical equilibrium, while the integrated shear relates to the \(D\)-term via
\begin{equation}
-\frac{32\pi m_\pi}{45} \int_0^\infty dr\, r^4 s_\pi(r) = D_\pi(0).
\end{equation}


The instanton liquid model is specified by an average instanton size \(\bar\rho \simeq 0.33~\mathrm{fm}\), a constituent mass \(M \simeq 350~\mathrm{MeV}\), and a packing fraction \(\kappa \sim 0.1\). These values set the crossover between soft (\(Q \ll \rho^{-1}\)), semi-hard (\(Q\rho \sim 1\)), and hard (\(Q \gg \rho^{-1}\)) dynamics. Simple parametrizations such as Eq.~\eqref{eq:Ddipole} provide useful analytic control and allow direct interpretation of radii and densities.





\section{Light-front Hamiltonian in terms of the good component}
\label{sec_good}
In the previous section we have seen how the instanton-induced 't~Hooft interaction in the instanton liquid model can be rewritten on the light front in terms of the good fermionic component only, after eliminating the bad component in the mean-field approximation and resumming the leading tadpole diagrams. The outcome is an effective light-front Lagrangian in which the good component field \(\psi\) carries a dynamically generated constituent mass \(M\), and interacts through nonlocal four-fermion operators in the scalar, pseudoscalar and isovector channels. In this section the authors carry out the canonical Legendre transformation of this Lagrangian to derive the light-front Hamiltonian \(P^-\), first in the zero-size limit of the instantons and then in the finite-size case with explicit nonlocal form factors. The resulting Hamiltonian is expressed entirely in terms of the good component and its associated creation and annihilation operators, and provides the starting point for the bound-state and partonic analysis that follows.

The mean-field Lagrangian in the zero-size limit can be written as
\begin{eqnarray}
L \;=&\; \bar\psi\,(i\slashed\partial - M)\,\psi
\;-\; \frac{1}{2}\,G_S\,
\Bigl[\,
\delta\sigma\,\hat D_+\,\delta\sigma
- \sigma^a\,\hat D_-\,\sigma^a
- \pi\,\hat D_-\,\pi
+ \pi^a\,\hat D_+\,\pi^a
\Bigr]\nonumber\\
&+ G_S\,
\Bigl[\,
\bar\psi\psi\,\sigma
- \bar\psi\tau^a\psi\,\sigma^a
- \bar\psi i\gamma_5\psi\,\pi
+ \bar\psi i\gamma_5\tau^a\psi\,\pi^a
\Bigr] ,
\label{eq:L-mf-zero}
\end{eqnarray}
where the bosonic fields \(\sigma,\sigma^a,\pi,\pi^a\) describe scalar-isoscalar, scalar-isovector, pseudoscalar-isoscalar and pseudoscalar-isovector channels, respectively, and \(\delta\sigma = \sigma - N_c\sigma_0\) denotes the fluctuation around the nonzero scalar vacuum expectation value that encapsulates spontaneous chiral symmetry breaking. The operators \(\hat D_\pm\) are nonlocal kernels generated by the tadpole resummation, schematically of the form
\begin{equation}
\hat D_\pm \;=\; 1\pm {G_S}\left<
\bar\psi\,\gamma^+ \,\frac{-i}{{\overleftrightarrow{\partial_-}}}\,\psi\right>
\,+\,\cdots,
\label{eq:Dhat-schematic}
\end{equation}
where \(\partial_- = \partial/\partial x^-\) is the longitudinal light-front derivative, and the dots indicate channel-dependent details that are not essential for the Hamiltonian structure. The key point is that integrating out the Gaussian bosonic fields in this mean-field Lagrangian generates strings of four-fermion operators with kernels proportional to \(\hat D_\pm^{-1}\).

Carrying out the Gaussian integration over \(\delta\sigma,\sigma^a,\pi,\pi^a\) yields an effective Lagrangian purely in terms of \(\psi\),
\begin{equation}
L \;=\; \bar\psi\,(i\slashed\partial - M)\,\psi
\;+\; \frac{G_S}{2}
\Bigl[
\bar\psi\psi \,\hat D^{-1}_+\,\bar\psi\psi
- \bar\psi\tau^a\psi \,\hat D^{-1}_-\,\bar\psi\tau^a\psi
- \bar\psi i\gamma_5\psi \,\hat D^{-1}_-\,\bar\psi i\gamma_5\psi
+ \bar\psi i\gamma_5\tau^a\psi \,\hat D^{-1}_+\,\bar\psi i\gamma_5\tau^a\psi
\Bigr] .
\label{eq:L-eff-zero}
\end{equation}
This is the mean-field light-front Lagrangian associated with the instanton-induced interaction in the zero-size limit. It contains a free part for a constituent quark of mass \(M\), and a set of nonlocal four-fermion operators that act in the scalar and pseudoscalar isosinglet and isotriplet channels.

The promotion of this Lagrangian to a Hamiltonian proceeds through the symmetric energy-momentum tensor,
\begin{equation}
T^{\mu\nu}
\;=\;
\frac{1}{2}\,\Bigl[\,
\bar\psi\, i\gamma^\mu \partial^\nu \psi
+ \bar\psi\, i\gamma^\nu \partial^\mu \psi
\Bigr]
- g^{\mu\nu} L,
\label{eq:Tem}
\end{equation}
from which the light-front Hamiltonian is obtained as the plus-minus component integrated over the light-front spatial coordinates,
\begin{equation}
P^- \;=\; \int dx^- d^2 x_\perp\, T^{+-}
\;=\; \int dx^- d^2 x_\perp\,\frac{1}{2}
\Bigl[\,
\bar\psi\,i\gamma^+ \partial^+ \psi
+ \bar\psi\,i\gamma^- \partial^- \psi
\Bigr]
- \int dx^- d^2 x_\perp\,L .
\label{eq:Pm-def}
\end{equation}
Here \(x^+ = (x^0 + x^3)/\sqrt{2}\) is the light-front "time'' and \(x^- = (x^0 - x^3)/\sqrt{2}\) its conjugate coordinate; \(\gamma^\pm = (\gamma^0 \pm \gamma^3)/\sqrt{2}\) are the associated Dirac matrices. Inserting the effective Lagrangian \eqref{eq:L-eff-zero} into \eqref{eq:Pm-def}, and using the equations of motion to eliminate time derivatives in favor of spatial ones, one arrives at the Hamiltonian density purely in terms of spatial operators.

Performing these steps, one obtains the Hamiltonian in coordinate space as
\begin{eqnarray}
P^-
\;=&&\;
\int dx^- d^2 x_\perp\,
\bar\psi(x)\,
\frac{-\partial_\perp^2 + M^2}{2\,i\partial_-}\,
\gamma^+ \psi(x)\nonumber\\
&&-
\frac{G_S}{2}
\int dx^- d^2 x_\perp\,
\Bigl[
\bar\psi\psi\,\hat D_+^{-1}\,\bar\psi\psi
- \bar\psi\tau^a\psi\,\hat D_-^{-1}\,\bar\psi\tau^a\psi
- \bar\psi i\gamma_5\psi\,\hat D_-^{-1}\,\bar\psi i\gamma_5\psi
+ \bar\psi i\gamma_5\tau^a\psi\,\hat D_+^{-1}\,\bar\psi i\gamma_5\tau^a\psi
\Bigr] .\nonumber\\
\label{eq:Pm-coord}
\end{eqnarray}
The first term is the familiar kinetic light-front Hamiltonian for a fermion of mass \(M\), in which the longitudinal derivative \(\partial_-\) is inverted by the same Green's function that appeared in the elimination of the bad component. The second term is the interaction part, expressed as nonlocal four-fermion operators whose kernels are precisely the tadpole-resummed \(\hat D_\pm^{-1}\).

For applications to bound states it is convenient to rewrite the Hamiltonian in momentum space. Introducing the measure
\begin{equation}
[d^3k]_+ \;=\; \frac{dk^+\,d^2k_\perp}{(2\pi)^3\,2k^+\,\epsilon(k^+)},
\label{eq:measure}
\end{equation}
where \(\epsilon(k^+) = \theta(k^+) - \theta(-k^+)\) distinguishes particle and antiparticle regions, and the Fourier transform of the good field,
\begin{equation}
\psi(x^-,x_\perp)
\;=\;
\int [d^3k]_+\, \psi(k)\,e^{-ik^+ x^- + i\vec k_\perp\cdot\vec x_\perp} ,
\label{eq:psi-Fourier}
\end{equation}
the Hamiltonian becomes
\begin{eqnarray}
P^- \;=&&\;
\int [d^3k]_+ \int [d^3q]_+\,
\frac{k_\perp^2 + M^2}{2k^+}\,
\bar\psi(k)\,\gamma^+\,\psi(q)\,
(2\pi)^3\,\delta^{(3)}_+(k-q)
\nonumber\\
&&+
\int [d^3k]_+ [d^3q]_+ [d^3p]_+ [d^3l]_+\,
(2\pi)^3\,\delta^{(3)}_+(p+k-q-l)\,
V(k,q;p,l),
\label{eq:Pm-mom-general}
\end{eqnarray}
where the delta function
\(\delta^{(3)}_+(k-q) = \delta(k^+ - q^+)\,\delta^{(2)}(\vec k_\perp - \vec q_\perp)\)
enforces momentum conservation, and \(V(k,q;p,l)\) is the interaction kernel generated by the four-fermion part of \eqref{eq:Pm-coord}. The first term in \eqref{eq:Pm-mom-general} is the kinetic part, while the second term describes the effective two-body interaction in all channels connected by the 't~Hooft interaction.

The explicit form of the kernel in the scalar and pseudoscalar channels can be displayed in terms of the good-component spinors \(\psi_+\). In the zero-size limit one can write
\begin{small}
\begin{align}
&V(k,q;p,l)
\;=\;
-\frac{G_S}{2}\,\alpha_+(k^+ - q^+)\,
\bar\psi_+(k)
\Bigl(
\frac{\vec\gamma_\perp\cdot\vec k_\perp + M}{2k^+}\,\gamma^+
+ \gamma^+\,\frac{\vec\gamma_\perp\cdot\vec q_\perp + M}{2q^+}
\Bigr)
\psi_+(q)\,
\bar\psi_+(p)
\Bigl(
\frac{\vec\gamma_\perp\cdot\vec p_\perp + M}{2p^+}\,\gamma^+
+ \gamma^+\,\frac{\vec\gamma_\perp\cdot\vec l_\perp + M}{2l^+}
\Bigr)
\psi_+(l)
\nonumber\\
&\;+\;
\frac{G_S}{2}\,\alpha_-(k^+ - q^+)\,
\bar\psi_+(k)
\Bigl(
\frac{\vec\gamma_\perp\cdot\vec k_\perp + M}{2k^+}\,i\gamma^+\gamma_5
+ i\gamma_5\gamma^+\,\frac{\vec\gamma_\perp\cdot\vec q_\perp + M}{2q^+}
\Bigr)
\psi_+(q)\,
\bar\psi_+(p)
\Bigl(
\frac{\vec\gamma_\perp\cdot\vec p_\perp + M}{2p^+}\,i\gamma^+\gamma_5
+ i\gamma_5\gamma^+\,\frac{\vec\gamma_\perp\cdot\vec l_\perp + M}{2l^+}
\Bigr)
\psi_+(l)
\nonumber\\
&\;+\;
\frac{G_S}{2}\,\alpha_-(k^+ - q^+)\,
\bar\psi_+(k)
\Bigl(
\frac{\vec\gamma_\perp\cdot\vec k_\perp + M}{2k^+}\,\gamma^+\tau^a
+ \tau^a\gamma^+\,\frac{\vec\gamma_\perp\cdot\vec q_\perp + M}{2q^+}
\Bigr)
\psi_+(q)\,
\bar\psi_+(p)
\Bigl(
\frac{\vec\gamma_\perp\cdot\vec p_\perp + M}{2p^+}\,\gamma^+\tau^a
+ \gamma^+\tau^a\,\frac{\vec\gamma_\perp\cdot\vec l_\perp + M}{2l^+}
\Bigr)
\psi_+(l)
\nonumber\\
&\;-\;
\frac{G_S}{2}\,\alpha_+(k^+ - q^+)\,
\bar\psi_+(k)
\Bigl(
\frac{\vec\gamma_\perp\cdot\vec k_\perp + M}{2k^+}\,i\gamma^+\gamma_5\tau^a
+ \tau^a i\gamma_5\gamma^+\,\frac{\vec\gamma_\perp\cdot\vec q_\perp + M}{2q^+}
\Bigr)
\psi_+(q)\,
\bar\psi_+(p)
\Bigl(
\frac{\vec\gamma_\perp\cdot\vec p_\perp + M}{2p^+}\,i\gamma^+\gamma_5\tau^a
+ i\gamma_5\gamma^+\tau^a\,\frac{\vec\gamma_\perp\cdot\vec l_\perp + M}{2l^+}
\Bigr)
\psi_+(l).
\label{eq:V-gamma-structure}
\end{align}
\end{small}
Although lengthy, this expression makes explicit how the Hamiltonian couples scalar, pseudoscalar and isovector channels through combinations of \(\gamma^+\), \(\gamma_5\), \(\vec\gamma_\perp\), and the Pauli matrices \(\tau^a\). The functions \(\alpha_\pm\) are tadpole-resummed vertex functions that encode the nontrivial vacuum structure through longitudinal zero modes; they depend only on the total longitudinal momentum flowing through the interaction.

For many purposes a compact form of the kernel is more transparent. In terms of the effective field \(\psi\) (rather than explicitly the projected \(\psi_+\)), and suppressing explicit Dirac structures, the interaction kernel in the zero-size limit can be summarized as
\begin{eqnarray}
V(k,q;p,l) \;=\;&&
-\frac{G_S}{2}\,\Bigl[
\alpha_+(k^+ - q^+)\,\bar\psi(k)\psi(q)\,\bar\psi(p)\psi(l)
-\alpha_-(k^+ - q^+)\,\bar\psi(k)i\gamma_5\psi(q)\,\bar\psi(p)i\gamma_5\psi(l)
\nonumber\\&&-\alpha_-(k^+ - q^+)\,\bar\psi(k)\tau^a\psi(q)\,\bar\psi(p)\tau^a\psi(l)
+\alpha_+(k^+ - q^+)\,\bar\psi(k)i\tau^a\gamma_5\psi(q)\,\bar\psi(p)i\tau^a\gamma_5\psi(l)
\Bigr].\nonumber\\
\label{eq:V-compact}
\end{eqnarray}
In this representation one clearly sees the decomposition into scalar-isoscalar, pseudoscalar-isoscalar, scalar-isovector, and pseudoscalar-isovector channels, with the vertex functions \(\alpha_\pm\) distinguishing attractive and repulsive combinations after tadpole resummation.

The vertex functions themselves follow from the resummation of quark tadpoles and take the form
\begin{equation}
\alpha_\pm(P^+)
\;=\;
\left\{
1
\;-\;
2g_S
\int dl^+ d^2l_\perp\,
\frac{1}{(2\pi)^3}
\frac{\epsilon(l^+)}{P^+ - l^+}
\right\}^{-1},
\label{eq:alpha-zero}
\end{equation}
in the zero-size limit, where \(g_S = N_c G_S\) is the 't~Hooft coupling scaled with the number of colors. The integral runs over the longitudinal momenta of intermediate tadpole loops, and the denominator encodes the propagation along the light-front direction. The appearance of a nontrivial \(\alpha_\pm\) expresses the impact of the instanton-induced vacuum on the light-front Hamiltonian: the vacuum is not trivial, but contributes through these resummed vertex factors.

The above construction extends naturally to the finite-size instanton case, where the nonlocal form factor \(F(k)\) associated with the quark zero modes must be retained. In the Lagrangian, this amounts to replacing each quark field by the smeared field \(\sqrt{F(i\partial)}\,\psi\), so that each four-fermion vertex is dressed by a product of form factors. In momentum space, this is implemented by inserting factors of \(F(k)\) for each external leg. As a result, the Hamiltonian becomes
\begin{eqnarray}
P^- \;=\;&&
\int [d^3k]_+ [d^3q]_+\,
\frac{k_\perp^2 + M^2}{2k^+}\,
\bar\psi(k)\gamma^+\psi(q)
(2\pi)^3\,\delta^{(3)}_+(k-q)
\nonumber\\&&\;+\;
\int [d^3k]_+ [d^3q]_+ [d^3p]_+ [d^3l]_+\,
(2\pi)^3\,\delta^{(3)}_+(p+k-q-l)\,
\sqrt{F(k)F(q)F(p)F(l)}\,V(k,q;p,l),\nonumber\\
\label{eq:Pm-mom-F}
\end{eqnarray}
where the kernel \(V(k,q;p,l)\) is structurally the same as in \eqref{eq:V-compact}, but with the vertex functions modified to
\begin{equation}
\alpha_\pm(P^+)
\;\longrightarrow\;
\left\{
1
\;-\;
2g_S
\int dl^+ d^2l_\perp\,
\frac{1}{(2\pi)^3}
\frac{\epsilon(l^+)}{P^+ - l^+}\,
F(l)\,F(P-l)
\right\}^{-1}.
\label{eq:alpha-F}
\end{equation}
These replacements reflect the finite size of instantons in the ILM and the corresponding momentum-dependent suppression of high-virtuality contributions. The factors \(\sqrt{F}\) in \eqref{eq:Pm-mom-F} appear for each external quark leg, while the combination \(F(l)F(P-l)\) enters the resummed tadpole loop in \eqref{eq:alpha-F}. Importantly, because the form factor \(F(k)\) falls rapidly for large \(|k|\), the resulting nonlocal Hamiltonian is well behaved and does not lead to pathological propagation.

To summarize, starting from the mean-field Lagrangian for the instanton-induced 't~Hooft interaction on the light front, the canonical construction yields a light-front Hamiltonian of the form
\begin{equation}
P^- \;=\; P^-_{\rm kin} \;+\; P^-_{\rm int},
\label{eq:Pm-split}
\end{equation}
where \(P^-_{\rm kin}\) describes a constituent quark of mass \(M\) with the standard light-front kinetic operator, and \(P^-_{\rm int}\) encodes nonlocal four-fermion interactions in the scalar and pseudoscalar channels, dressed by vertex functions \(\alpha_\pm\) that resum vacuum tadpoles and by form factors \(F(k)\) inherited from the instanton zero modes. All of this is formulated entirely in terms of the good component of the quark field; the bad component has been eliminated as a constraint. The Hamiltonian thus obtained is boost invariant and provides a concrete, frame-independent realization of the nonperturbative instanton vacuum on the light front. It is this Hamiltonian that will be diagonalized in subsequent sections in the quark-antiquark sector to obtain the light-front wave functions of pions and kaons, and from them their distribution amplitudes and parton distribution functions.

\section{Relation to the Wilsonian picture and light-front criticality}

At first sight, the light-front Hamiltonian treatment presented above  may appear to be in tension with the Wilsonian perspective emphasized earlier, in particular with the discussion of light-cone criticality in Sec.~\ref{sec_wilsonian}. The latter stresses that a formulation defined strictly at the light-front critical surface generically exhibits infrared pathologies associated with longitudinal zero modes and the absence of an intrinsic resolution scale.

The present construction should not, however, be interpreted as a canonical quantization performed directly at that critical point. Rather, the infrared sensitivity characteristic of light-front dynamics manifests itself here through a well-identified, power-counting-enhanced class of diagrams, namely tadpole chains associated with longitudinal zero-mode exchanges. These contributions are not treated perturbatively; instead, they are resummed into effective vertex functions $\alpha_\pm(P^+)$, which encode the vacuum structure and the dominant infrared physics of the theory.

From this viewpoint, the resummation procedure plays a role analogous to a Wilsonian matching step: degrees of freedom responsible for the strongest infrared enhancements are effectively integrated out and absorbed into renormalized operators of the light-front Hamiltonian. Subsequent calculations are then performed with these dressed interactions, which remain finite and well defined in the light-cone limit.

Thus, while the technical implementation differs from a conventional Wilsonian flow, the underlying logic is the same: potentially singular contributions associated with unresolved modes are reorganized into effective couplings before physical observables are computed. In this sense, the analysis above should be viewed as a concrete realization of the Wilsonian philosophy within a light-front Hamiltonian framework, rather than an exception to it.

\end{subappendices}




\chapter{Pions on the LF}
The pion plays a central role in QCD as both the pseudo-Goldstone boson of spontaneous chiral symmetry breaking (S$\chi$SB) and the lightest hadronic bound state. Understanding its structure within the light-front (LF) framework provides an illuminating example of how nonperturbative phenomena-traditionally associated with the vacuum-emerge from hadronic wave functions. In this chapter we will derive, in pedagogical fashion, the pion's decay constant, distribution amplitude (DA), and parton distribution function (PDF) using the nonlocal instanton-induced kernel of Ref.~\cite{LiuShuryakZahed2023}. We will also interpret these results in terms of chiral symmetry realization, angular momentum composition, and QCD evolution.

\section{Light-Front pion bound-state}

In the preceding sections we have constructed an effective light-front (LF) Hamiltonian for quarks derived from the induced multi-flavor interactions of the instanton liquid model (ILM)~\cite{Diakonov:1985eg,Schafer:1996wv,Diakonov:2002fq,Nowak:1996aj}, and shown that (within the mean-field (large \(N_c\)) approximation and after elimination of the  bad) light-cone components. The resulting constituent quark mass and chiral condensate coincide with their rest-frame values.  We now use that Hamiltonian to diagonalize the two-body sector (valence \(q\bar q\)) in the light-front and extract the meson spectrum.  In particular, for two light quark flavors (u, d) with equal current mass, we examine scalar and pseudoscalar mesons.  The key result is that only the pseudoscalar isosinglet and isotriplet modes (the pions) emerge as strongly bound (Goldstone) states, while all other scalar or pseudoscalar modes remain unbound on the light front, in the mean-field approximation and away from the chiral limit~\cite{Diakonov:1995ea,Schafer:1996wv,Nowak:1996aj,Kacir:1996qn}.  We moreover extract the corresponding light-cone wave function for the pion, which yields a nonperturbative pion distribution amplitude (DA).  


The LF Hamiltonian derived can be cast, in the two-body valence approximation, into an eigenvalue equation  
\begin{equation}
   P^- \bigl|X, P_i\bigr\rangle = \frac{m_X^2}{2 P^+} \bigl|X, P_i\bigr\rangle,
\end{equation}
or equivalently (multiplying by \(2 P^+\))  
\begin{equation}
   \hat H_{\rm LF} \bigl|X, P_i\bigr\rangle = m_X^2 \bigl|X, P_i\bigr\rangle.  \label{eq:LF_eigen}
\end{equation}
In the valence (leading \(1/N_c\)) sector the meson state is dominated by a constituent-quark and antiquark, and may be written as  
\begin{equation}
  \bigl|{\rm Meson}\; X, P_i\bigr\rangle = \frac{1}{\sqrt{N_c}} \! \int_0^1\! \frac{d x}{\sqrt{2 x \bar x}} \! \int \! \frac{d^2 k_\perp}{(2\pi)^3} \sum_{s_1, s_2} \Phi_X(x, k_\perp, s_1, s_2) \; b^\dagger_{s_1}(k)\; d^\dagger_{s_2}(P-k) \; |0\rangle, \label{eq:meson_state}
\end{equation}
with normalization  
\begin{equation}
  \int_0^1 d x \int \frac{d^2 k_\perp}{(2\pi)^3} \sum_{s_1, s_2} \bigl|\Phi_X(x, k_\perp, s_1, s_2)\bigr|^2 = 1.
\end{equation}
Here \(x = k^+/P^+\), \(\bar x = 1-x\), and \(M\) denotes the constituent quark mass generated by the ILM instanton-induced dynamics.  The light-front Hamiltonian matrix elements between these two-body states lead to bound-state equations in momentum space.  


For two light flavors (u, d) with identical current mass \(m\), the relevant mesonic channels are the isosinglet scalar \(\sigma\), the isosinglet pseudoscalar \(\eta_0\) (denoted \(\sigma_5\) in the original paper), the isotriplet pseudoscalar \(\pi^{\pm,0}\), and the corresponding non-singlet scalar partners.  The flavor structure in the valence sector is  
\[
  \sigma = \frac{1}{\sqrt 2} (u \bar u + d \bar d), \qquad 
  \pi = \frac{1}{\sqrt 2}(u \bar u - d \bar d), \qquad
  \pi^\pm = u \bar d,\; d \bar u.
\]  
By construction of the instanton-induced interaction in the ILM, the scalar-isoscalar (\(\sigma\)) channel and the pseudoscalar isotriplet (\(\pi\)) channel are strongly attractive, while the pseudoscalar isosinglet (\(\eta_0\)) channel and non-singlet scalar/pseudoscalar partners are strongly repulsive~\cite{tHooft:1976rip,Shifman:1979uw,Schafer:1996wv,Kacir:1996qn}.  As a result, only the \(\pi\) emerges as a bound state, while the others remain unbound or appear only at threshold.


Denoting by \(\Phi_X\) the light-cone wave function (LCWF) of meson \(X\), the Hamiltonian eigenvalue equation reduces to an integral equation of the general form  
\[
   \bigl[k_\perp^2 + M^2\bigr]\Phi_X(x,k_\perp) \;+\; \int d x' d^2 q_\perp\; V_X(x,k_\perp; x', q_\perp)\; \Phi_X(x', q_\perp) = m_X^2\, \Phi_X(x,k_\perp).
\]
Here \(V_X\) is the effective two-body kernel arising from the instanton-induced four-fermion interaction (after eliminating the bad components), projected in the scalar or pseudoscalar channel.  In the pseudoscalar (\(\pi\)) channel the kernel is attractive and strong enough to bind.

In  the ILM, the interaction kernel (in the two-body light-front Hamiltonian) takes the form (suppressing spinor and color structure for clarity)  
\begin{equation}
  V_\pi(k, q) \; \sim\; -\, g_S \; F(k)\, F(q)\; \bar u(k)\,\Gamma_\pi\, u(P-k)\; \bar v(q)\,\Gamma_\pi\, v(P-q),
  \label{eq:kernel_formfactor}
\end{equation}
where \(g_S\) is the effective coupling associated with the 't Hooft interaction, and \(\Gamma_\pi\) denotes the spin-flavor Dirac structure appropriate for the pseudoscalar isotriplet channel (e.g. \(\gamma^5 \tau^a\) in flavor space, plus the usual LF spinor projectors).  The factors \(F(k)\), \(F(q)\) stem from the zero-mode profile, thus making the effective interaction non-local in momentum space and strongly suppressing large momenta (\(\gtrsim 1/\rho\)).  This non-locality (finite instanton size) is essential for rendering the integrals finite and preserving the boost-invariance of the LF Hamiltonian~\cite{Liu:2023feu,Liu:2023fpj}. More specifically,
in Euclidean momentum space the form factor is given by  
\begin{equation}
  F(k) = \bigl(z\, F_0(z)\bigr)^2 , 
  \qquad z = \tfrac12\, k \rho, 
  \qquad F_0(z) = I_0(z) K_0(z) - I_1(z) K_1(z), 
  \label{eq:F_of_k}
\end{equation}
where \(\rho\) is the (average) instanton size, and \(I_{\nu}, K_{\nu}\) are modified Bessel functions.   
This form factor ensures that the induced interactions are  smeared over a distance scale \(\sim \rho\),  avoiding the divergences associated with a strictly local four-fermion interaction. 

\begin{figure*}
  \includegraphics[height=5.5cm,width=.46\linewidth]{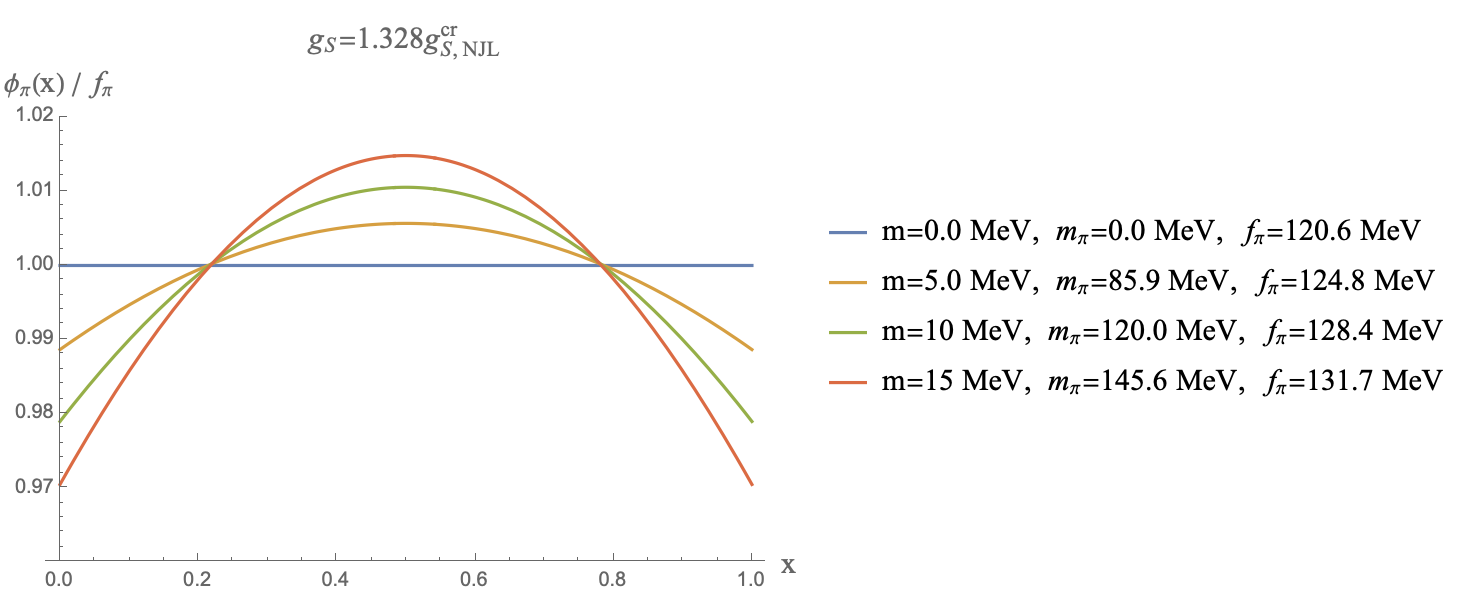}%
  \includegraphics[height=5.5cm,width=.46\linewidth]{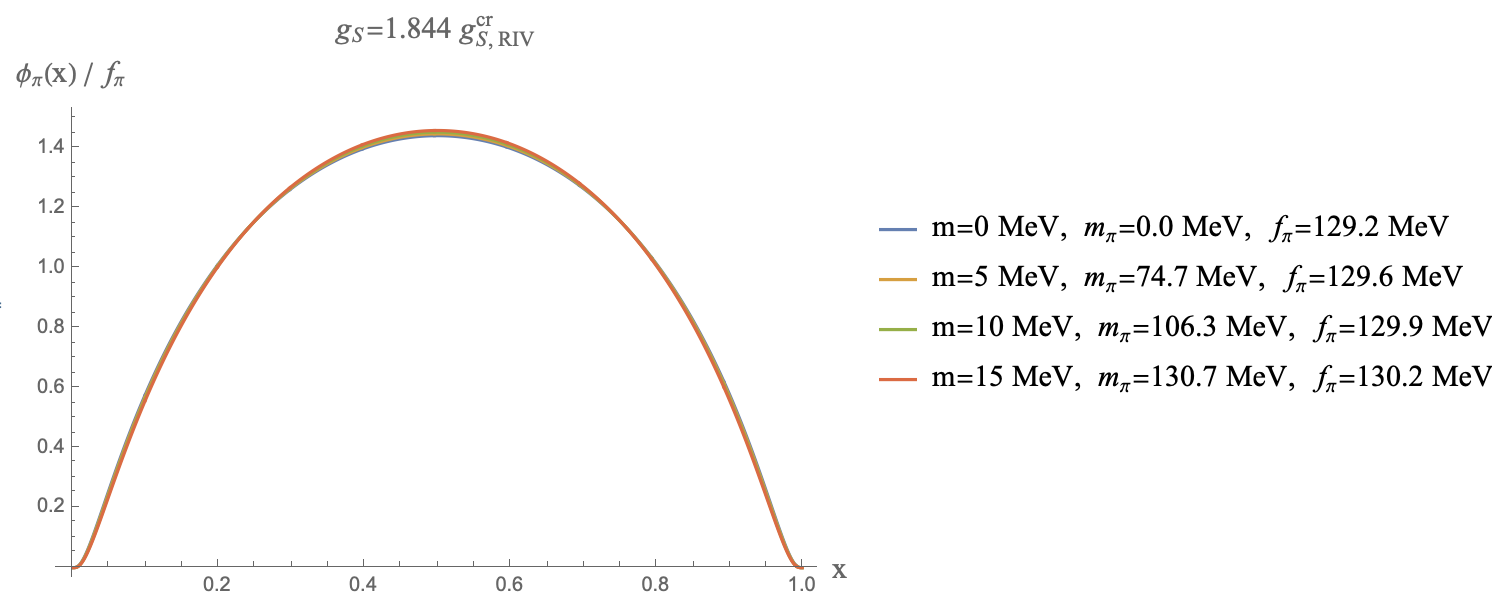}%
  \includegraphics[height=5.5cm,width=.46\linewidth]{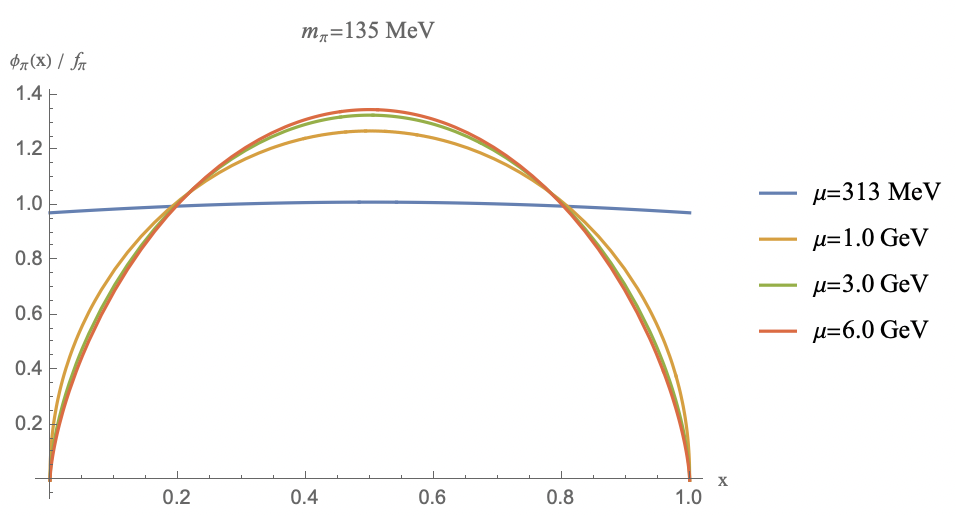}%
  \includegraphics[height=5.5cm,width=.46\linewidth]{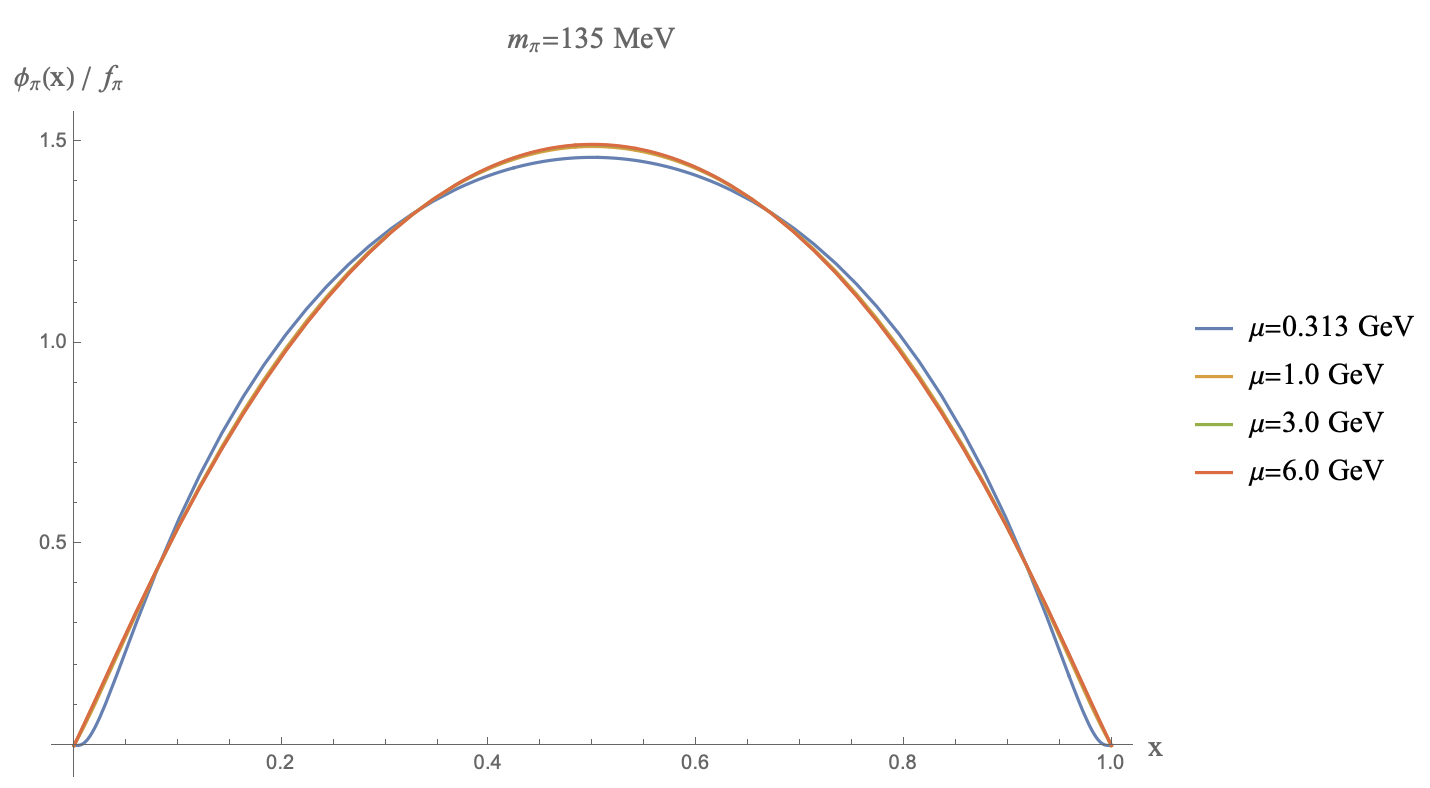}%
\caption{
a: The pion DA with different pion mass $m_\pi$ in the zero instanton size limit;
b: The pion DA with different pion mass $m_\pi$ in the ILM with finite instanton size $\rho=(630\mathrm{MeV})^{-1}$;
c: The ERBL evolution of pion DA in zero instanton size limit;
d: The ERBL evolution of pion DA in RIV with finite instanton size $\rho=(630\mathrm{MeV})^{-1}$.
}
\label{fig_pionERBL}
\end{figure*}

\section{Pion DA}

The good-component light-front spinors for quarks and antiquarks are given by 
\begin{equation}
  u_s(p) = \frac{1}{\sqrt{2 p^+}} \left( \slashed p + M \right) \begin{pmatrix} \chi_s \\ \chi_s \end{pmatrix}, 
  \qquad 
  v_s(p) = \frac{1}{\sqrt{2 p^+}} \left( \slashed p - M \right) \begin{pmatrix} \chi_s \\ -\chi_s \end{pmatrix},
  \label{eq:LF_spinors}
\end{equation}
with \(\chi_s\) a two-component spinor (helicity).
Thus the pseudoscalar Dirac structure for the pion, in the valence Fock sector, can be written as  
\begin{equation}
  \Phi_\pi(x, k_\perp, s_1, s_2) = \varphi_\pi(x, k_\perp)\; \bar u_{s_1}(k)\, \gamma^5 \tau^a\, v_{s_2}(P-k),
  \label{eq:LCWF_spinor_structure}
\end{equation}
where \(\varphi_\pi(x, k_\perp)\) is a scalar function encoding the momentum dependence (effectively the radial wave function in LF variables), and \(\tau^a\) is a Pauli matrix in flavor space selecting the appropriate isospin component of the pion.  The normalization of \(\varphi_\pi\) then follows from eq. \eqref{eq:LCWF_spinor_structure} after squaring/summing/integrating over spin and momentum~\cite{Liu:2023feu,Liu:2023fpj}.

 \begin{figure*}
  \includegraphics[height=5.5cm,width=.46\linewidth]{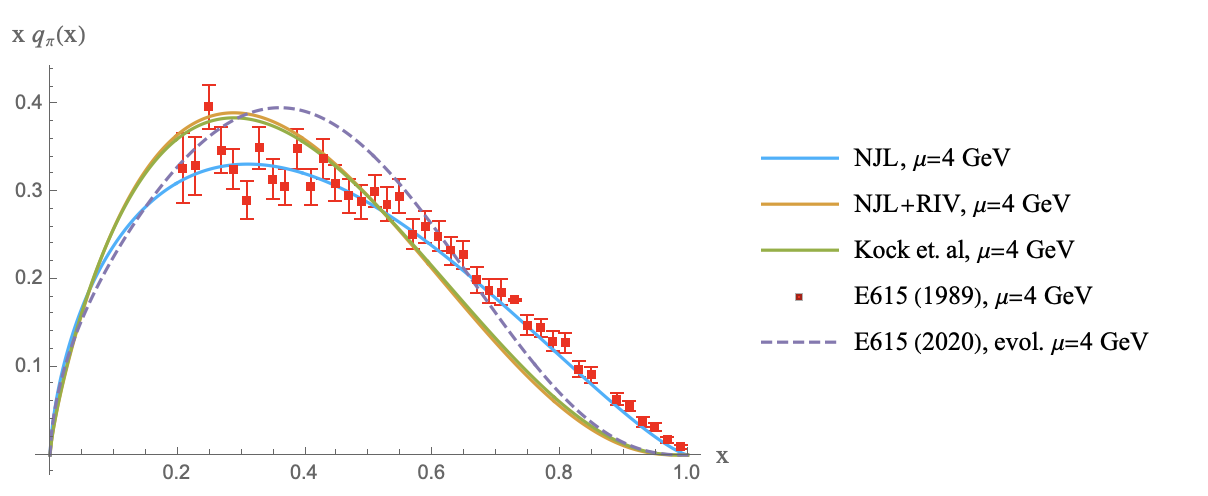}%
  \includegraphics[height=5.5cm,width=.46\linewidth]{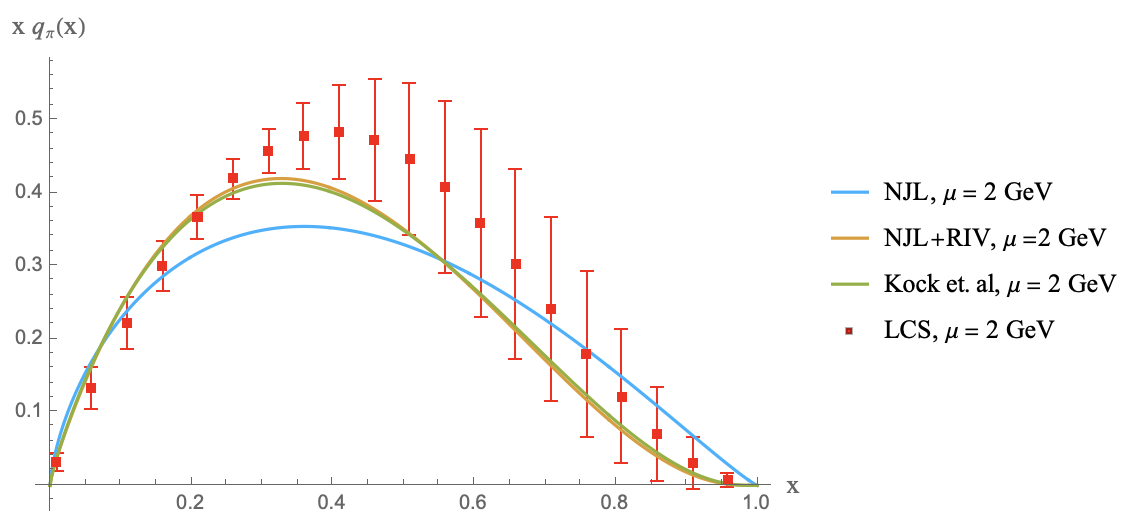}%
\caption{a: Pion parton momentum distribution function for zero instanton size (solid-blue) and finite instanton size 
of $\rho=0.317$ fm (solid-orange), both of which are evolved to $\mu=4$ GeV with a pion mass $m_\pi=135$ MeV.
The results are compared to the those extracted from the ILM using the LaMET (solid-green)~\cite{Kock:2020frx,Kock:2021spt},
also evolved to $\mu=4$ GeV.  The E615 data from 1989 (red) are from \cite{Conway:1989fs},  corresponding to a  fixed invariant muon pair mass $m_{\mu^+\mu^-}\geq 4.05$ GeV. The improved E615 data from 2020 (dashed-purple) are from  \cite{Aicher:2010cb}, using the original E615 experimental data from~\cite{Conway:1989fs}.\\
b: The same pion parton distribution functions as in a, but now evolved to $\mu=2$ GeV, for comparison with 
the lattice data using  cross sections (LCS) (red) from~\cite{Sufian:2020vzb}.
}
\label{fig_piLATDAT2}
\end{figure*}

In practice, one inserts the explicit zero-mode profile (eq. \eqref{eq:F_of_k}) when computing the overlap integrals.  For equal-mass quarks the kernel simplifies to a function depending only on the relative momenta \(k_\perp, q_\perp\) and the longitudinal momentum fractions \(x, x'\).  One obtains an explicit expression 
(in the so-called  "zero-instanton-size limit" \(F \to 1\)) for \(V_\pi\), but in the finite-\(\rho\) ILM the factors \(F(k) F(q)\) regulate the integrals and modify quantitatively the binding condition.
In the zero-instanton-size limit (pointlike interaction), and with a boost-invariant momentum cutoff \(\Lambda\), the bound-state equation simplifies and yields a relation for the pion mass \(m_\pi\).  In that limit the bound-state condition reads (for \(\pi\))  
\[
   m_\pi^2 \;-\; \frac{m}{M} \;=\; \frac{g_S}{4\pi^2} \, m_\pi^2 \int_0^1 d y \int_0^{\Lambda^2} d q_\perp^2 \; \frac{1}{q_\perp^2 + M^2 - y \bar y \, m_\pi^2},
\]
which upon evaluation yields a pion mass that vanishes in the chiral limit \(m \to 0\), thus realizing the Goldstone theorem on the light-front.  More precisely one finds  
\[
  m_\pi^2 = \frac{2m}{f_\pi^2}\, |\langle\bar \psi \psi \rangle| + \mathcal O(m^2),
\]
with \(f_\pi\) the pion decay constant and \(\langle\bar \psi \psi \rangle\) the chiral condensate --  the same relation as in the rest frame~\cite{GellMann:1968rz,Donoghue:1992dd}.  This demonstrates that the spontaneous breaking of chiral symmetry by instanton effects in the ILM is correctly carried over in the light-front formalism, and that the pion emerges as a genuine light-front Goldstone mode~\cite{Diakonov:1997sj,Schafer:1996wv,Nowak:1996aj}.  


In all other scalar and pseudoscalar channels (e.g., \(\sigma_5\), \(\eta_0\), non-singlet scalar/pseudoscalar), the instanton-induced kernel is repulsive or insufficiently attractive; correspondingly the integral equations admit no bound-state solution below the two-constituent quark threshold.  In particular the would-be scalar \(\sigma\) is at threshold (mass \(\simeq 2M\) in the chiral limit), and the \(\eta_0\) remains unbound.  This result aligns with the expectation from the ILM that only the pseudoscalar-isotriplet modes are strongly bound due to the 't Hooft interaction~\cite{tHooft:1976rip}.


The pion decay constant $f_\pi$ quantifies the overlap between the vacuum axial current and the one-pion state
\begin{equation}
\langle 0| \bar{q}(0)\gamma^\mu \gamma_5 q(0) | \pi(P)\rangle = i f_\pi P^\mu.
\end{equation}
On the light front, we use the "plus'' component ($\mu=+$) to avoid instantaneous terms. In terms of the valence LFWF, the pion decay constant reads 
\begin{equation}
f_\pi = \sqrt{\Nc}\int_0^1\!\dd x\!\int\!\frac{\dd^2\bm{k}_\perp}{(2\pi)^3}\, \frac{\phi_\pi(x,\bm{k}_\perp)\, \mathcal{S}(x,\bm{k}_\perp)}{\sqrt{x(1-x)}},
\end{equation}
where $\mathcal{S}$ contains spinor traces involving $\gamma^+\gamma_5$ and the nonlocal form factors. The factor $\sqrt{x(1-x)}$ arises from the Jacobian of the LF normalization.

Physically, $f_\pi$ measures how much of the pion internal structure is "aligned'' with the axial current. In the instanton framework, $f_\pi$ is directly proportional to the dynamical quark mass $M(k)$ and to the instanton density, establishing a link between the microscopic (instanton) and macroscopic (hadronic) scales.
At high scales, $\phi_\pi(x)\rightarrow \phi_\pi(x,\mu)$ evolves according to the Efremov-Radyushkin-Brodsky-Lepage (ERBL) equation~\cite{Efremov:1979qk,Lepage:1980fj}, which governs its approach to the asymptotic form $\phi_\pi^{\text{as}}(x)=6x(1-x)$. The nonlocal instanton model provides an initial condition at a low scale $\mu_0\simeq 0.6$-$1$~GeV that is broader than the asymptotic shape, consistent with lattice and experimental constraints~\cite{Shi2015,Zhang2020}.


\begin{subappendices}

\section{Melosh rotations and spinor algebra}

In this appendix we summarize the role of Melosh rotations in connecting canonical spin to light-front helicity. This is essential for understanding how the pion, although a spin-zero bound state, contains nontrivial orbital and spin correlations encoded in its light-front wave function~\cite{Melosh:1974cu,Keister:1991sb,Bakker:2013cea}. The relativistic nature of LF dynamics implies that the naive nonrelativistic spin-addition rules do not directly apply; Melosh rotations correct for this mismatch.

On the light front, the helicity basis differs from the instant form due to boosts along the $x^-$ direction. The spin wave function is obtained by applying the Melosh rotation:
\begin{equation}
R_M(x,\bm{k}_\perp) = \frac{m + xM_0 - i\sigma\cdot(\hat{n}\times \bm{k}_\perp)}{\sqrt{(m+xM_0)^2 + \bm{k}_\perp^2}},
\end{equation}
where $\hat{n}$ is the unit vector along the $z$-axis and $M_0$ is the invariant mass of the $q\bar q$ pair. This rotation connects canonical spin to LF helicity.
Using these rotations, the spinor factor in the axial current matrix element is
\begin{equation}
\mathcal{S}(x,\bm{k}_\perp) = \frac{m^2 + \bm{k}_\perp^2 - x(1-x)M_\pi^2}{\sqrt{(m^2+\bm{k}_\perp^2)/[x(1-x)]}},
\end{equation}
up to normalization. This factor controls the relativistic suppression of helicity nonconserving components and explains why the pion DA is broader at low scales than the asymptotic form.

\section{From Bethe-Salpeter to LFWFs}

Here we outline how the covariant Bethe-Salpeter (BS) description of the pion in the instanton vacuum~\cite{Liu:2023feu,Liu:2023fpj} is mapped to a light-front wave function. The key step is to integrate over the LF energy $k^-$, which enforces on-shell propagation of one of the constituents and produces the familiar LF energy denominator.

To obtain the LF wave function, integrate the BS amplitude over $k^-$:
\begin{equation}
\Psi_\pi(x,\bm{k}_\perp) = \int\!\frac{\dd k^-}{2\pi}\, \Gamma_\pi(k;P)\, S(k+\tfrac{P}{2})S(k-\tfrac{P}{2}).
\end{equation}
Closing the contour in the lower half-plane picks the pole of $S(k^-)$ corresponding to the on-shell quark. This leads to
\begin{equation}
S(k) \approx \frac{\slashed{k}+M(k)}{k^2-M^2(k)+i\epsilon} \to \frac{1}{2k^+}\frac{\gamma^+}{k^- - (M^2+\bm{k}_\perp^2)/k^+}.
\end{equation}
Performing the $k^-$ integral yields the LF energy denominator $M_\pi^2-\mathcal{M}^2$. The nonlocality of $\Gamma_\pi$ introduces form factors $F((k-\ell)^2)$ that regularize UV behavior and encode the instanton profile. This procedure ensures that chiral Ward identities and the GMOR relation remain valid in the LF representation.

\section{Instanton Form Factors and Normalization}

The quark zero mode in an instanton background follows the BPST form~\cite{Belavin:1975fg} and is a central building block of the ILM. In coordinate space (regular gauge)
\begin{equation}
\psi_0(x)=\frac{\rho}{\pi}\frac{1}{(x^2+\rho^2)^{3/2}}\frac{1-\gamma_5\slashed{x}/\sqrt{x^2}}{2}U,
\end{equation}
where $\rho$ is the instanton size and $U$ a color rotation. Its Fourier transform gives the form factor
\begin{equation}
F(k\rho)=2z[I_0(z)K_1(z)-I_1(z)K_0(z)-I_1(z)K_1(z)/z],\quad z=k\rho/2,
\end{equation}
which is closely related to the function in Eq.~\eqref{eq:F_of_k}. The nonlocal kernel is
\begin{equation}
\mathcal{K}(k_1,\ldots,k_{2N_f})=G\prod_iF(k_i\rho)(2\pi)^4\delta^{(4)}\!\Big(\sum k_i\Big),
\end{equation}
with $G$ determined by the instanton density $n$ and distribution $d(\rho)$. Using $n\simeq1~\text{fm}^{-4}$ and $\rho\simeq0.33~\text{fm}$ yields $G\sim5$-$10~\text{GeV}^{-2}$~\cite{Shuryak:1981ff,Schafer:1996wv,Nowak:1996aj}. These parameters quantitatively reproduce the observed condensate and $f_\pi$ values and set the model scale for the DA and PDF.

\section{From Covariant to Light-Front Gap Equation}

Finally, we comment on the relation between the covariant gap equation and its LF counterpart. The Euclidean gap equation
\begin{equation}
M(p_E)=m+\int\!\frac{\dd^4\ell_E}{(2\pi)^4}K(p_E,\ell_E)\frac{M(\ell_E)}{\ell_E^2+M^2(\ell_E)}
\end{equation}
is transformed to LF variables via $\ell^\pm=\ell^0\pm\ell^3$:
\begin{equation}
M(x,\bm{\ell}_\perp)=m+\int_0^1\!\dd y\!\int\!\frac{\dd^2\bm{k}_\perp}{(2\pi)^3}K(x,\bm{\ell}_\perp;y,\bm{k}_\perp)\frac{M(y,\bm{k}_\perp)}{\bm{k}_\perp^2+M^2(y,\bm{k}_\perp)-y(1-y)P^{+2}}.
\end{equation}
Boundary terms at $x=0,1$ carry zero-mode contributions that generate $\vev{\bar q q}_\LF$. Numerical solutions show the same critical coupling $G_{\text{crit}}$ as in covariant approaches, confirming full equivalence when zero modes are retained~\cite{Chang:2013pq,Liu:2023feu,Liu:2023fpj}. This supports the consistency of implementing S$\chi$SB and Goldstone dynamics directly on the light front.

\end{subappendices}


\chapter{Kaons on the LF}
\label{sec:Kaon}

The kaon provides a central bridge between the dynamics of spontaneous chiral symmetry breaking and the explicit breaking introduced by the strange quark mass~\cite{GellMann:1968rz,Donoghue:1992dd}.  Within the light-front formulation~\cite{Brodsky:1997de}, its description becomes especially transparent because the wave function is expressed directly in terms of boost-invariant momentum fractions.  The appearance of an asymmetric mass distribution between the constituent quarks becomes immediately reflected in the internal kinematic structure, the bound-state dynamics, and ultimately the partonic content of the state.

In the instanton liquid model, the nonperturbative vacuum induces a multi-fermion interaction with the correct chiral and flavor structure to generate Goldstone modes~\cite{Diakonov:1995ea,Schafer:1996wv,Nowak:1996aj}.  The $u$, $d$, and $s$ quarks enter symmetrically through the 't Hooft determinant~\cite{tHooft:1976rip}, but the heavier strange quark produces explicit SU(3) flavor breaking.  The kaon thus emerges from the same vacuum structure as the pion, yet with a built-in mass asymmetry that becomes encoded directly into its light-front wave function.

In what follows, the dynamics is organized in the SU(3) subgroups known as U-spin and V-spin, which interchange nonstrange and strange quarks~\cite{Liu:2023feu}.  Because the $u$ and $d$ quarks remain nearly degenerate, both sectors exhibit identical light-front dynamics.  It suffices to write the analysis for a single kaon wave function $\Phi_K$, with the understanding that the charged and neutral kaons follow by trivial flavor replacement.


\section{Flavor structure and the light-front kaon wave function}
\label{sec:LFWF}

The relevant quark-antiquark content of the kaon can be formulated in either the U-spin or V-spin representation, both of which rotate nonstrange quarks into strange quarks~\cite{Liu:2023feu}.  For the charged kaons, the $u\bar{s}$ and $s\bar{u}$ configurations define the U-spin doublet, while the neutral $d\bar{s}$ and $s\bar{d}$ belong to the V-spin sector.  In either case, isospin symmetry ensures that $u$ and $d$ can be treated equivalently, so that the light-front wave functions obey
\[
\Phi_{K^+}=\Phi_{K^-}=\Phi_{K^0}=\Phi_{\bar{K}^0}\equiv\Phi_K.
\]

The kaon contains both a pseudoscalar and a scalar channel due to the chirality structure of the instanton-induced interaction.  These channels are represented by the light-front wave functions
\begin{align}
\Phi_{K_5}(x,k_\perp,s_1,s_2)
&=
\phi_{K_5}(x,k_\perp)\,
\bar{u}_{s_1}(k)\tau^\pm v_{s_2}(P-k),
\\
\Phi_{K}(x,k_\perp,s_1,s_2)
&=
\phi_K(x,k_\perp)\,
\bar{u}_{s_1}(k)i\gamma^5\tau^\pm v_{s_2}(P-k),
\end{align}
where the scalar functions $\phi_{K_5}$ and $\phi_K$ encode the nonperturbative momentum structure and $\tau^\pm$ denote the U- or V-spin raising and lowering operators.

The unequal quark masses $M_u$ and $M_s$ introduce an inherent asymmetry in the longitudinal momentum distribution~\cite{Keister:1991sb}.  The light quark tends to carry a larger momentum fraction because the kinetic term $(k_\perp^2+M^2)/x$ penalizes the heavier strange quark more strongly as $x$ decreases.  This UV-finite, boost-invariant structure is preserved by the instanton form factor $\mathcal{F}$, which depends only on the combination
\[
\frac{k_\perp^2+M_u^2}{x}
+
\frac{k_\perp^2+M_s^2}{\bar{x}},
\]
and therefore ensures that the nonlocal interaction is consistent with light-front symmetry~\cite{Liu:2023feu,Liu:2023fpj}.

\section{Physical content of the kaon bound-state equations}
\label{sec:BoundStatePhysics}

The instanton-induced interaction contains the same essential physics that produces the pion as a Goldstone boson, but extended to three flavors~\cite{Nowak:1988bh,Kacir:1996qn,Liu:2023feu}.  Its determinantal structure explicitly couples left- and right-handed quarks and mixes the pseudoscalar and scalar channels.  When restricted to the valence two-body sector, the effective Hamiltonian on the light front contains the free kinematic term and a nonlocal kernel determined by the instanton zero-mode profile~\cite{Liu:2023feu}.

The resulting bound-state equation takes the form of an inhomogeneous integral equation.  The kernels include the functions $\alpha_{K\pm}$ that incorporate quark-loop dressing of the 't Hooft vertex; these encode the effects of mass generation and the interplay between $u$ and $s$ constituents~\cite{tHooft:1976rip}.  The scalar and pseudoscalar channels are repulsive and attractive, respectively, producing an unbound scalar partner but a deeply bound pseudoscalar kaon.

All mathematical expressions for the kernels and their spin traces appear in Appendix~\ref{app:BoundStateMath}.  Here we emphasize their physical meaning.  The pseudoscalar interaction is enhanced relative to the scalar channel because instantons induce chirality flips that match the pseudoscalar quantum numbers~\cite{Schafer:1996wv,Kacir:1996qn}.  The SU(3)-breaking mass difference $M_s-M_u$ tilts the kernel and shifts the momentum distribution in favor of the lighter quark.  The resulting light-front wave function is therefore asymmetric even before evolution to higher scales~\cite{Shi2015}.

\section{Kaon mass, the scalar channel, and the GOR relation}
\label{sec:MassSpectrum}

The mass eigenvalues follow from a spectral equation derived from the bound-state kernel.  The pseudoscalar channel yields an eigenvalue $m_K$ in the expected range between $(M_s-M_u)$ and $(M_s+M_u)$, while the scalar kaon partner $K_5$ is pushed above the cutoff and remains unbound.  The physical kaon mass arises from the interplay between chiral symmetry breaking and explicit SU(3) violation~\cite{GellMann:1968rz,Donoghue:1992dd,Kacir:1996qn}.

When expanded in the limit of small current masses, the mass equation reduces to the Gell-Mann-Oakes-Renner relation,
\[
m_K^2
=
\frac{m_u+m_s}{f_K^2}
|\langle\bar{u}u\rangle+\langle\bar{s}s\rangle|
+\cdots,
\]
which shows that the kaon remains a pseudo-Goldstone boson despite the heavier strange quark.  The condensate receives contributions from both flavors weighted symmetrically, reflecting the structure of the 't Hooft interaction.

The decay constant $f_K$ follows from the normalization of the axial current.  In the instanton liquid model it inherits a logarithmic sensitivity to the instanton size and matches the chiral expansion when the constituent masses and form factor are used consistently~\cite{Donoghue:1992dd,Kacir:1996qn}.  The detailed derivation is presented in Appendix~\ref{app:MassMath}.

\begin{figure*}
  \includegraphics[height=5.5cm,width=.46\linewidth]{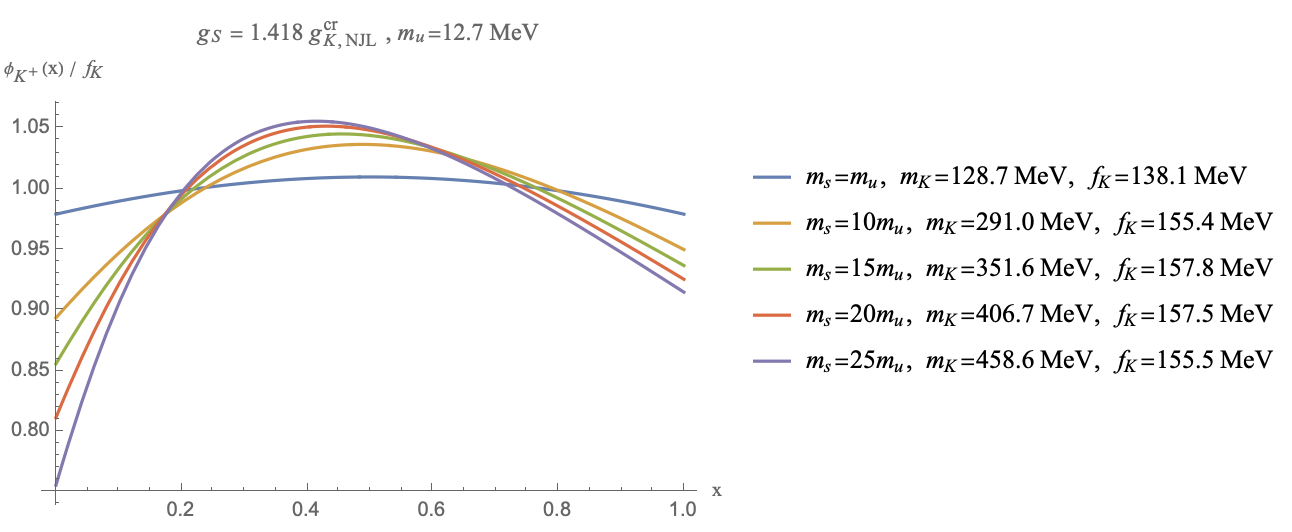}%
  \includegraphics[height=5.5cm,width=.46\linewidth]{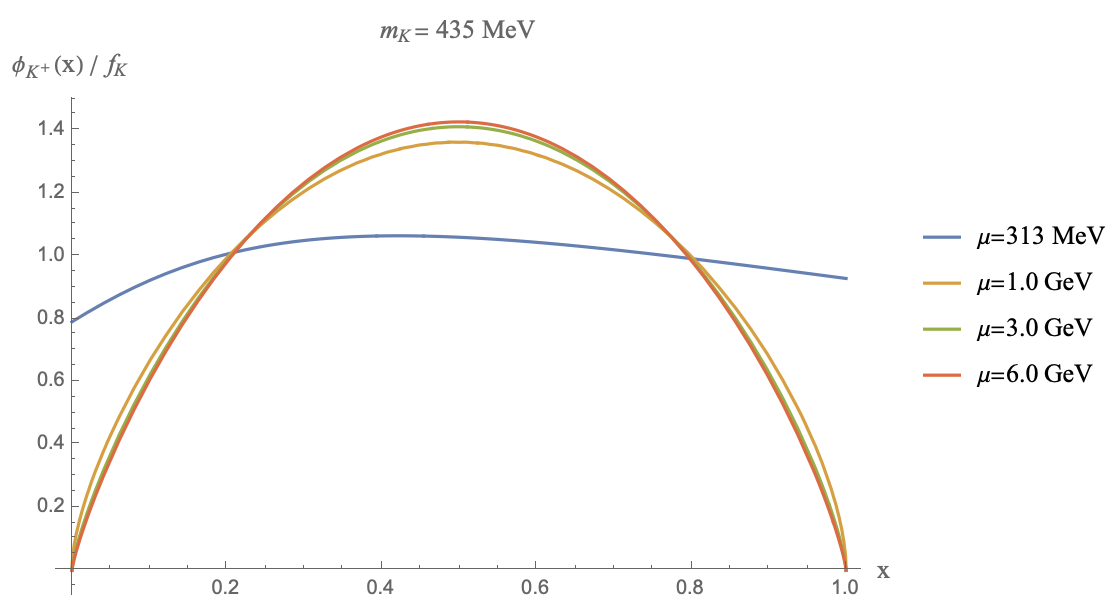}%
  \includegraphics[height=5.5cm,width=.46\linewidth]{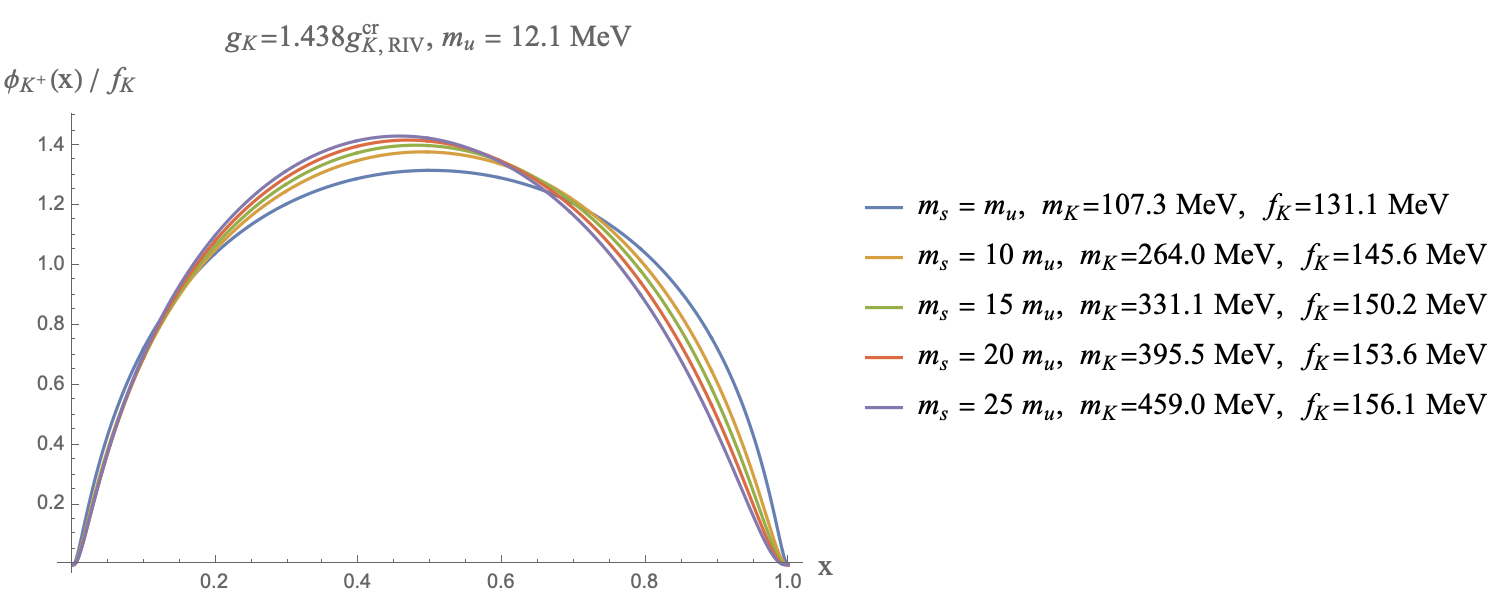}%
  \includegraphics[height=5.5cm,width=.46\linewidth]{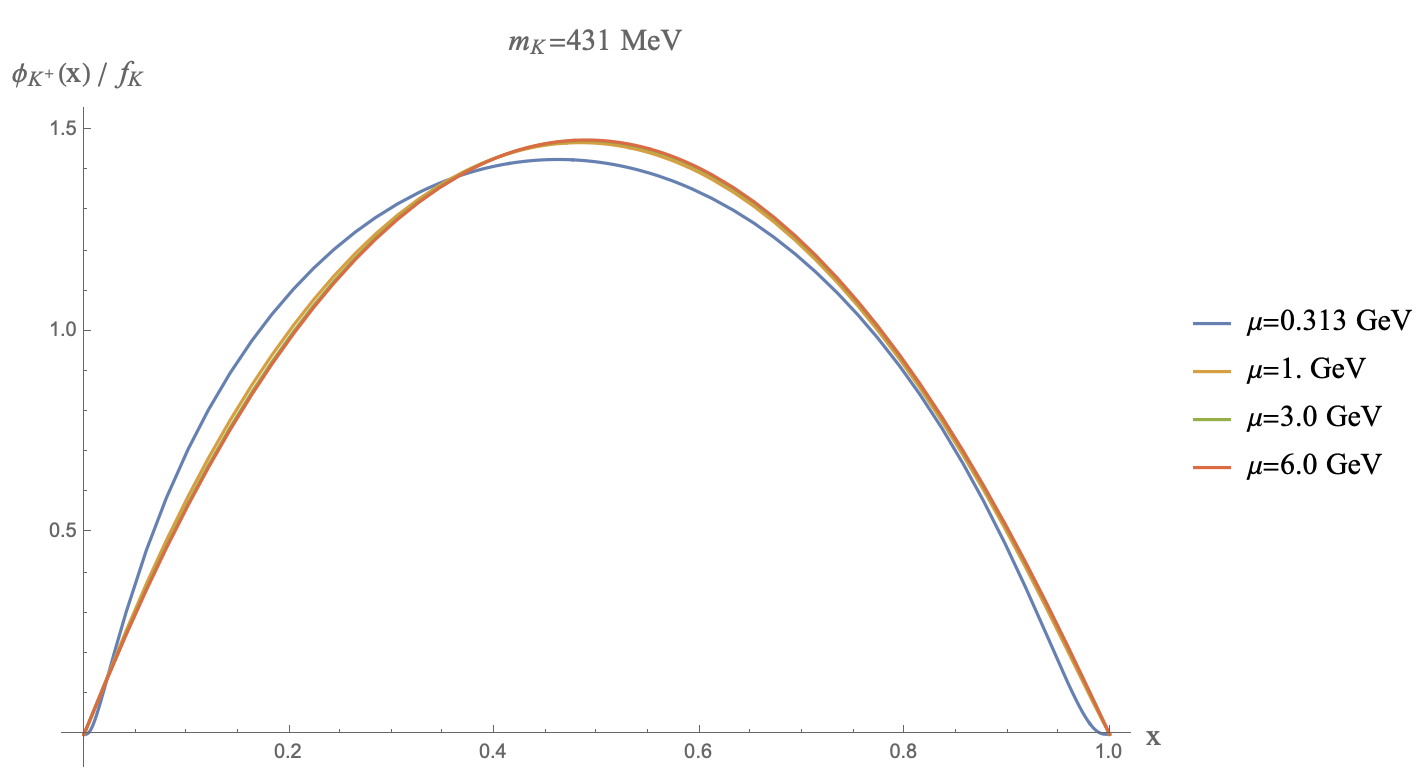}%
\caption{
a: The positive charged kaon DA versus parton $x$, for fixed coupling $g_S/g_{K}^{cr}=1.418$ and fixed $m_u=12.7$ MeV,
but different ratios $m_s/m_u$, in the zero instanton size limit;
b: The positive charged kaon DA versus parton $x$, for fixed $m_K=435$ MeV using the ERBL evolution from the initial 
$\mu=0.313$ GeV to $\mu=6$ GeV;
c,d: Same as a,b but in the ILM, for a fixed instanton size $\rho=0.313$ fm.
}
\label{fig_DGLAPKDAX}
\end{figure*}

\begin{figure*}
  \includegraphics[height=5.5cm,width=.46\linewidth]{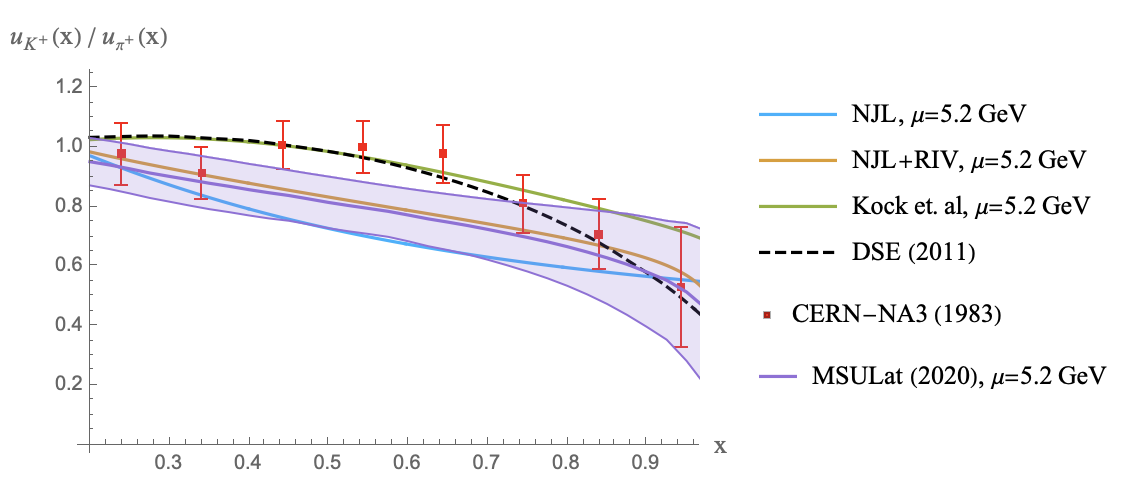}%
  \includegraphics[height=5.5cm,width=.46\linewidth]{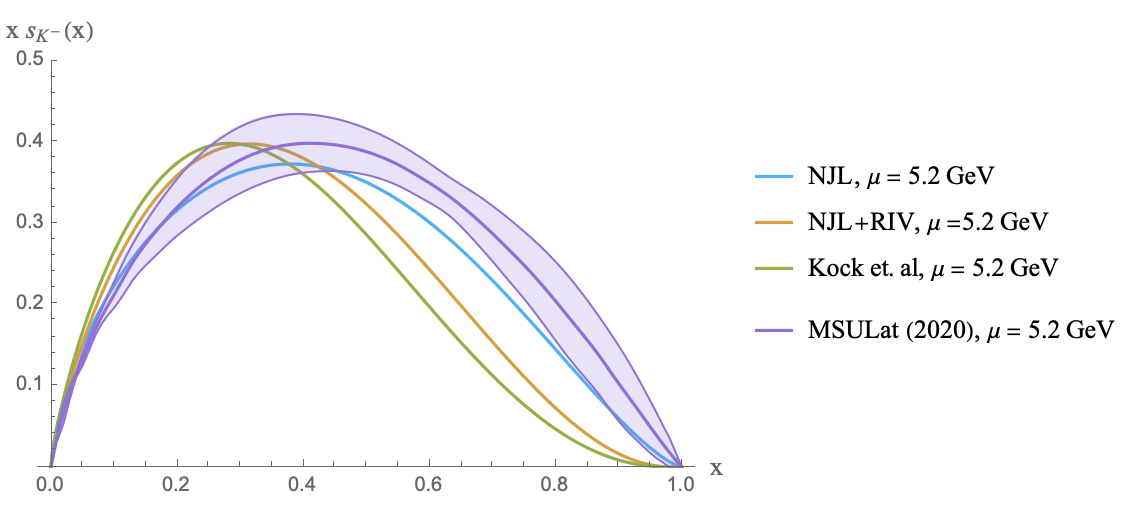}%
\caption{
a: The ratio of positive 
charge kaon to pion PDF versus parton $x$, in the zero instanton size limit, with $m_K=435$ MeV and evolved to $\mu=5.2$ GeV (solid-blue),
in the ILM with finite instanton size $\rho=0.318$ fm with $m_K=431$ MeV and evolved to $\mu=5.2$ GeV (solid-orange). The comparison is to the results of
the ILM using the LaMET extraction in~\cite{Kock:2020frx,Kock:2021spt} and evolved to $\mu=5.2$ GeV (solid-green), and the 
results from  the Dyson-Schwinger equation with full Bethe-Salpeter wavefunction~\cite{Bashir:2012fs} (dashed-black). The data for the measured ratio (red)
are from~\cite{Saclay-CERN-CollegedeFrance-EcolePoly-Orsay:1980fhh} using  muon pair production, using the invariant mass cuts between $4.1$ GeV and $8.5$ GeV to eliminate the meson production on resonance. The recent lattice data MSULat (band-purple) are from~\cite{Lin:2020ssv}, using the LaMET construction, at $\mu=5.2$ GeV.
b: The same as in b, but for the strange quark momentum distribution for the negative kaon $x s_{K^-}(x)$ versus $x$.
}
\label{fig_DGLAPK}
\end{figure*}

\section{Distribution amplitudes and parton distribution functions}
\label{sec:PDFDAphysics}

The light-front wave function directly determines both the distribution amplitude (DA) and the parton distribution function (PDF).  Because the kaon contains a heavier strange quark, the momentum distribution is skewed toward the nonstrange quark.  This phenomenon is already present at the model scale, well before perturbative evolution is applied~\cite{Shi2015,Liu:2023feu}.

The DA measures the projection of the wave function onto zero transverse separation.  Its asymmetry is a clear signature of SU(3) breaking and is consistent with lattice calculations and Dyson-Schwinger results~\cite{Shi2015,Zhang2020}.  Under ERBL evolution the DA slowly approaches the asymptotic form, but retains a noticeable skewness at experimentally relevant scales~\cite{Liu:2023feu}.

The PDF is obtained by integrating the absolute square of the wave function over transverse momentum.  The strange antiquark distribution peaks at larger momentum fractions than the light quark distribution.  This asymmetry has been observed in Drell-Yan experiments and lattice extractions and is a robust prediction of the instanton framework~\cite{Liu:2023feu}.

All explicit integral expressions for the DA and PDF are given in Appendix~\ref{app:MassMath}.  Their forms depend on the instanton form factor or on the transverse cutoff in the zero-size limit, but their qualitative behavior follows directly from the constituent mass hierarchy~\cite{Liu:2023feu}.

\begin{figure*}
    \centering
   \includegraphics[height=5.5cm,width=.56\linewidth]{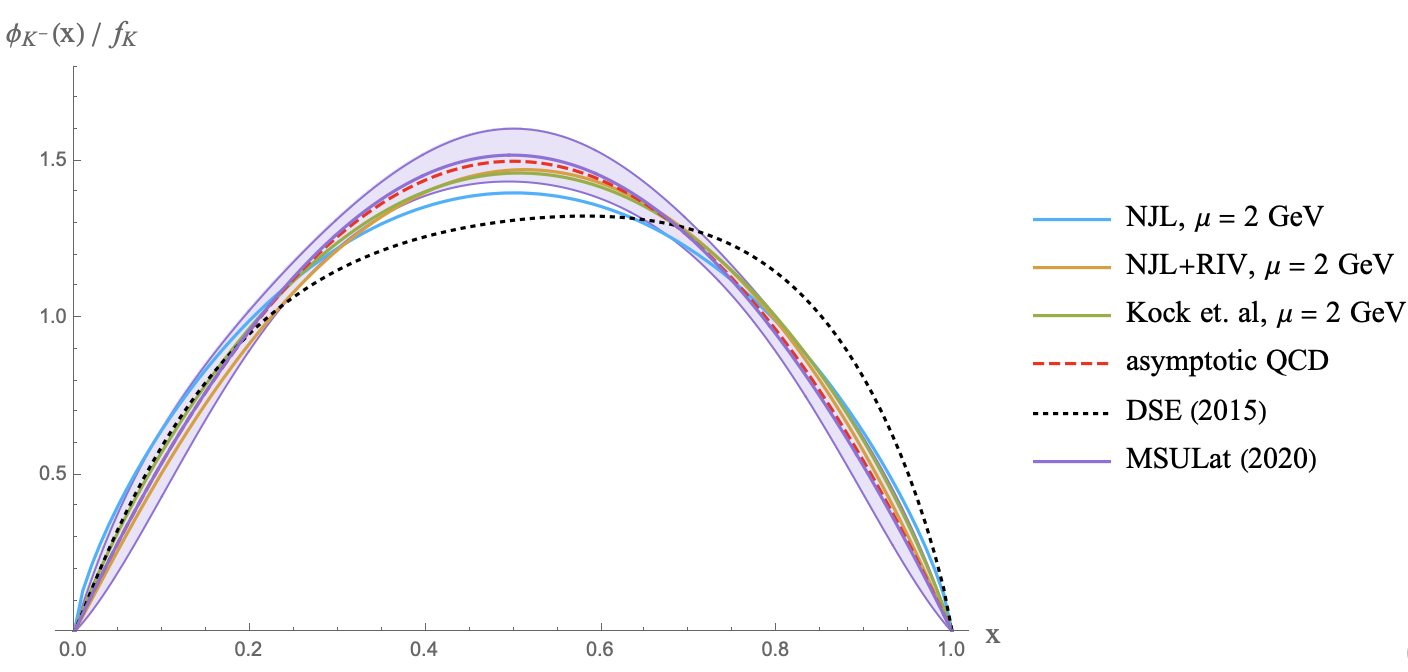}
    \caption{Evolution of the negative kaon DA  with $m_K=435$ MeV,  from $\mu=0.313$ GeV to $\mu=2$ GeV  in the zero instanton size limit (solid-blue), and in the ILM with a finite instanton size with $m_K=431$ MeV (solid-orange)~\cite{Liu:2023feu}. 
    The results are compared to those obtained also from the ILM (solid-green), using the
    LaMET construction in~\cite{Kock:2020frx,Kock:2021spt}, also evolved to $\mu=2$ GeV. The results using the 
    Dyson-Schwinger equation with Bethe-Salpeter wavefunctions (dashed-black) are  from~\cite{Bashir:2012fs}.
    The QCD asymptotic result of $6x\bar x$ (dashed-red) is from~\cite{Lepage:1980fj,Efremov:1980qeq}. The recent lattice data MSULat(2020) (purple-band) are from 
    \cite{Lin:2020ssv} using the LaMET construction.}
\label{KDA_EXPTXX}
\end{figure*}

\begin{subappendices}

\section{ Full Bound-State Equation and Instanton Kernel}
\label{app:BoundStateMath}

In this appendix we collect the explicit light-front bound-state equations and the corresponding instanton-induced kernels that underlie the qualitative discussion in Sec.~\ref{sec:BoundStatePhysics}.  The goal is to make clear how the nonlocal 't~Hooft interaction, when projected on the valence sector and expressed in light-front variables, leads to the coupled scalar and pseudoscalar equations for the kaon.  Throughout, we keep the notation as close as possible to the main text, so that the physical interpretation of each ingredient remains transparent~\cite{Liu:2024lfv}.

The starting point is the effective ILM interaction in the three-flavor sector.  After bosonization in the scalar and pseudoscalar channels and restriction to the $q\bar q$ Fock sector, one arrives at a light-front Hamiltonian consisting of a free kinetic term plus a nonlocal kernel that encodes instanton zero-mode overlap.  The structure of the kernel is constrained by chiral symmetry and SU(3) flavor, and the explicit expressions given below show how this information feeds into the kaon bound-state problem.

\subsection{ Light-front wave functions and Dirac structure}

The kaon wave functions in the scalar and pseudoscalar channels are
\begin{align}
\Phi_{K_5}(x,k_\perp,s_1,s_2)
&=
\phi_{K_5}(x,k_\perp)\bar{u}_{s_1}(k)\tau^{\pm}v_{s_2}(P-k),\\
\Phi_{K}(x,k_\perp,s_1,s_2)
&=
\phi_{K}(x,k_\perp)\bar{u}_{s_1}(k)i\gamma^5\tau^{\pm}v_{s_2}(P-k).
\end{align}
The spinor factors implement the scalar and pseudoscalar Dirac structures, while the scalar functions $\phi_{K_5}$ and $\phi_K$ capture the nonperturbative momentum dependence.  The use of U- or V-spin raising and lowering operators $\tau^\pm$ ensures that the same formalism applies to both charged and neutral kaons, differing only by flavor labels.

\subsection{ Boost-invariant instanton form factor}

The nonlocality of the instanton zero-mode implies
\begin{equation}
\lim_{P^+\rightarrow\infty}\sqrt{\mathcal{F}_u(k)\mathcal{F}_s(P-k)}\rightarrow 
\mathcal{F}\!\left(
\frac{k^2_\perp+M_u^2}{x}+\frac{k^2_\perp+M_s^2}{\bar{x}}
\right).
\end{equation}
This shows explicitly that, in the infinite-momentum frame, the nonlocal form factor depends only on boost-invariant combinations of the transverse momentum and longitudinal fractions.  The same functional form appears in the pion sector, but with $M_u=M_d$, so that the kaon case illustrates how SU(3) breaking deforms an otherwise universal instanton profile~\cite{Liu:2023feu}.

\subsection{Complete bound-state equation}

The kaon bound-state equation is
\begin{equation}
\begin{aligned}
m_K^2\Phi_K(x,k_\perp,s_1,s_2)
&=
\left[\frac{k_\perp^2+M_u^2}{x}+\frac{k_\perp^2+M_{s}^2}{\bar{x}}\right]\Phi_K(x,k_\perp,s_1,s_2)
\\
&+
\frac{1}{\sqrt{2x\bar{x}}}\sqrt{\mathcal{F}_u(k)\mathcal{F}_s(P-k)}
\int_0^1 \frac{dy}{\sqrt{2y\bar{y}}}
\int\frac{d^2q_\perp}{(2\pi)^3}
\sum_{s,s'}
\mathcal{V}_{s,s',s_1,s_2}
\Phi_K(y,q_\perp,s,s')
\sqrt{\mathcal{F}_u(q)\mathcal{F}_s(P-q)}.
\end{aligned}
\end{equation}
The first line encodes the kinetic energy of quark and antiquark with unequal constituent masses, while the second line represents the nonlocal interaction mediated by the ILM kernel.  The structure is entirely analogous to the pion case, but with modified kinematics and flavor factors appropriate to the $u\bar s$ system.

\subsection{Instanton-induced spin kernel}

The interaction kernel is
\begin{equation}
\begin{aligned}
\mathcal{V}_{s,s',s_1,s_2}
&=
-g_K\Big[
\alpha_{K+}\,\bar{u}_{s_1}(k)\tau^+i\gamma^5v_{s_2}(k')
\bar{v}_{s'}(q')\tau^-i\gamma^5u_{s}(q)
-
\alpha_{K-}\,\bar{u}_{s_1}(k)\tau^+v_{s_2}(k')
\bar{v}_{s'}(q')\tau^-u_{s}(q)
\\
&\qquad
+\alpha_{K+}\,\bar{u}_{s_1}(k)\tau^-i\gamma^5v_{s_2}(k')
\bar{v}_{s'}(q')\tau^+i\gamma^5u_{s}(q)
-
\alpha_{K-}\,\bar{u}_{s_1}(k)\tau^-v_{s_2}(k')
\bar{v}_{s'}(q')\tau^+u_{s}(q)
\Big].
\end{aligned}
\end{equation}
The structure of $\mathcal{V}$ displays explicitly how the ILM couples pseudoscalar and scalar channels: the terms proportional to $\alpha_{K+}$ involve $i\gamma^5$ and drive the attractive pseudoscalar interaction, while those proportional to $\alpha_{K-}$ generate the repulsive scalar contribution.  The dressing functions $\alpha_{K\pm}$ resum tadpole-like quark loops and therefore encode the response of the vacuum to the presence of the kaon state~\cite{Liu:2023feu}.

\subsection{Dressing functions \texorpdfstring{$\alpha_{K\pm}$}{alphaKpm}}

\begin{equation}
\alpha_{K\pm}=
\left[
1
\pm
g_K\int\frac{dk^+d^2k_\perp}{(2\pi)^3}
\frac{\epsilon(k^+)}{P^+-k^+}
\left(
\mathcal{F}_u(k)\mathcal{F}_s(P-k)
+
\mathcal{F}_s(k)\mathcal{F}_u(P-k)
\right)
\right]^{-1}.
\end{equation}
These functions summarize the strength of the effective interaction in each channel.  In the chiral limit and in the pion sector, the analogous functions reduce to simple constants fixed by the gap equation; here, the presence of the strange quark mass and the asymmetry between $M_u$ and $M_s$ makes the dressing more intricate and feeds directly into the different binding patterns of $K$ and $K_5$.

\subsection{Scalar and pseudoscalar bound-state equations}

\begin{align}
m_{K_5}^2\phi_{K_5}
&=
\left[\frac{k^2_\perp+M_{u}^2}{x}+\frac{k^2_\perp+M_{s}^2}{\bar{x}}\right]\phi_{K_5}
+
\frac{2g_K\alpha_{K-}}{\sqrt{x\bar{x}}}\sqrt{\mathcal{F}_u(k)\mathcal{F}_s(P-k)}
\int_0^1 \frac{dy}{\sqrt{y\bar{y}}}\int\frac{d^2q_\perp}{(2\pi)^3}
\!\left(\frac{q_\perp^2+y^2M_{s}^2-2y\bar{y}M_{u}M_{s}+\bar{y}^2M_{u}^2}{y\bar{y}}\right)
\phi_{K_5}
\sqrt{\mathcal{F}_u(q)\mathcal{F}_s(P-q)},
\\
m_{K}^2\phi_{K}
&=
\left[\frac{k^2_\perp+M_{u}^2}{x}+\frac{k^2_\perp+M_{s}^2}{\bar{x}}\right]\phi_K
-
\frac{2g_K\alpha_{K+}}{\sqrt{x\bar{x}}}\sqrt{\mathcal{F}_u(k)\mathcal{F}_s(P-k)}
\int_0^1 \frac{dy}{\sqrt{y\bar{y}}}\int\frac{d^2q_\perp}{(2\pi)^3}
\!\left(\frac{q_\perp^2+y^2M_{s}^2+2y\bar{y}M_{u}M_{s}+\bar{y}^2M_{u}^2}{y\bar{y}}\right)
\phi_{K}
\sqrt{\mathcal{F}_u(q)\mathcal{F}_s(P-q)}.
\end{align}
The two equations differ only in the sign in front of the $2y\bar y M_u M_s$ term and in the corresponding dressing factors $\alpha_{K\pm}$.  This sign difference is ultimately responsible for the attractive versus repulsive character of the pseudoscalar and scalar channels, respectively, and explains why the kaon emerges as a pseudo-Goldstone boson while its scalar partner is unbound.

\subsection{Auxiliary integrals}

The integral
\begin{equation}
\int_0^1 dx\int d^2k_\perp\frac{1}{x}
\left[
\mathcal{F}_u^2(k)+\mathcal{F}_s^2(k)
-\mathcal{F}_u(k)\mathcal{F}_s(P-k)
-\mathcal{F}_s(k)\mathcal{F}_u(P-k)
\right]
\end{equation}
vanishes for large cutoff.  This identity plays the same role as in the pion sector: it guarantees that, after appropriate subtractions consistent with the gap equation, the bound-state integral equations remain finite and independent of the UV regularization, with the instanton size $\rho$ providing the only physical scale.

\section{Kaon Mass Spectrum, GOR Limit, DA and PDF}
\label{app:MassMath}

In this appendix we present the spectral equations for the kaon mass, together with the explicit expressions for the DA and PDF in both the ILM and in the zero-size limit.  These formulas make concrete the qualitative statements in Secs.~\ref{sec:MassSpectrum} and~\ref{sec:PDFDAphysics}, and show how the interplay between dynamical mass generation and explicit SU(3) breaking controls the kaon structure~\cite{Liu:2023feu}.

The starting point is the resummed kaon propagator in the pseudoscalar channel, obtained by summing quark-loop diagrams with nonlocal ILM form factors.  The pole condition for this propagator yields the kaon mass equation, whose small-mass expansion reproduces the Gell-Mann-Oakes-Renner relation.  The same Bethe-Salpeter amplitude, when projected on the light front and integrated over transverse momentum, yields the DA and PDF.

\subsection{Spectral equations for the masses}

The mass eigenvalues satisfy
\begin{equation}
1=
\int_0^1dy\int d^2q_\perp
\frac{V_X(y,q_\perp)}{y\bar{y}m_X^2-(q_\perp^2+M^2)}
\mathcal{F}_u(q)\mathcal{F}_s(P-q),
\end{equation}
with potentials
\begin{equation}
V_X=
\begin{cases}
\displaystyle
+\frac{2g_K}{(2\pi)^3}\alpha_{K-}\,\frac{q_\perp^2+y^2M_s^2-2y\bar{y}M_uM_s+\bar{y}^2M_u^2}{y\bar{y}},
\\[12pt]
\displaystyle
-\frac{2g_K}{(2\pi)^3}\alpha_{K+}\,\frac{q_\perp^2+y^2M_s^2+2y\bar{y}M_uM_s+\bar{y}^2M_u^2}{y\bar{y}}.
\end{cases}
\end{equation}
Here $X=K_5$ or $K$ respectively.  The structure of $V_X$ encodes the attractive or repulsive nature of each channel and depends sensitively on the constituent masses $M_u$ and $M_s$.  In the limit $M_s\to M_u$ the potentials reduce smoothly to their pion counterparts, confirming that the kaon sector interpolates continuously between SU(2) and SU(3) chiral dynamics.

\subsection{Gell-Mann-Oakes-Renner relation}

Expanding the pseudoscalar kaon mass equation yields
\begin{equation}
m^2_K=
\frac{m_u+m_s}{f_K^2}
|\langle \bar{u}u\rangle+\langle \bar{s}s\rangle|
+\mathcal{O}(m_u^2,m_s^2,m_um_s),
\end{equation}
with condensate
\[
|\langle \bar{u}u\rangle+\langle \bar{s}s\rangle|
=\frac{N_c}{2g_K}(M_u+M_s-m_u-m_s).
\]
This relation makes explicit how the ILM saturates the kaonic GMOR relation: the difference between constituent and current masses provides a measure of the quark condensates, while the coupling $g_K$ encapsulates the strength of the underlying instanton-induced interaction~\cite{Liu:2023feu}.

\subsection{Light-front wave function form}

\begin{equation}
\phi_{K^+}(x,k_\perp)=
\frac{1}{\sqrt{2x\bar{x}}}
\frac{C_K}{m^2_K-\frac{k_\perp^2+xM_s^2+\bar{x}M_u^2}{x\bar{x}}}
\sqrt{\mathcal{F}_u(k)\mathcal{F}_s(P-k)}.
\end{equation}
This form illustrates that the light-front wave function is dominated by configurations where the invariant mass of the $u\bar s$ pair is close to the physical $m_K^2$, with the instanton form factors suppressing large virtualities.  The overall normalization constant $C_K$ is fixed by requiring that the probability carried by the valence Fock sector be unity or by matching the decay constant $f_K$.

\subsection{Parton distribution function}

\begin{align}
u_{K^+}(x)
&=
\frac{C_K^2}{4\pi^2}
\int_0^{\infty} dk^2_\perp
\frac{
k_\perp^2+x^2M_s^2+2x\bar{x}M_sM_u+\bar{x}^2M_u^2
}{
[x\bar{x}m^2_K-(k_\perp^2+xM_s^2+\bar{x}M_u^2)]^2
}
\mathcal{F}_u(k)\mathcal{F}_s(P-k),
\end{align}
and in ILM variables
\begin{equation}
u_{K^+}(x)=
\frac{C_K^2}{2\pi^2}
\int_{ \frac{\rho\sqrt{\bar{x}M_u^2+xM_s^2}}{2\lambda_K\sqrt{x\bar{x}}}}^{\infty}
dz \,
\frac{
z^2-\frac{\rho^2}{4\lambda^2_K}(M_s-M_u)^2
}{\left(\frac{\rho^2m_K^2}{4\lambda^2_K}-z^2\right)^2}
z^5(F'(z))^4.
\end{equation}
These expressions show explicitly that the strange-quark mass shifts the distribution toward larger $x$ for the strange antiquark and toward smaller $x$ for the light quark.  The ILM representation in terms of the dimensionless variable $z$ highlights the role of the instanton size $\rho$ as the natural UV regulator.

Zero-size limit:
\begin{equation}
u_{K^+}(x)=
\frac{C_K^2}{4\pi^2}\theta(x\bar{x})
\left[
\frac{(m^2_K-(M_s-M_u)^2)x\bar{x}\Lambda^2}{(xM_s^2+\bar{x}M_u^2-x\bar{x}m_K^2)(xM_s^2+\bar{x}M_u^2+\Lambda^2-x\bar{x}m_K^2)}
+
\ln\left(1+\frac{\Lambda^2}{xM_s^2+\bar{x}M_u^2-x\bar{x}m_K^2}\right)
\right].
\end{equation}
In this limit the instanton profile is effectively replaced by a sharp transverse cutoff $\Lambda$, and the logarithmic dependence reflects the familiar structure of light-front constituent models~\cite{Keister:1991sb}.  The ILM can thus be viewed as providing a physically motivated smearing of this cutoff in terms of a finite-size instanton form factor.

\subsection{Distribution amplitude}

\begin{align}
\phi_{K}(x)
&=
\frac{\sqrt{N_c}}{\sqrt{2}\pi^2}
\int dk^2_\perp
\frac{\phi_K(x,k_\perp)}{\sqrt{2x\bar{x}}}
[xM_s\mathcal{F}_s(P-k)+\bar{x}M_u\mathcal{F}_u(k)].
\end{align}
In the ILM:
\begin{equation}
\phi_{K^+}(x)=
\frac{\sqrt{N_c}(xM_s+\bar{x}M_u)}{\sqrt{2}\pi^2}
C_{K}\int_{ \frac{\rho\sqrt{\bar{x}M_u^2+xM_s^2}}{2\lambda_K\sqrt{x\bar{x}}}}^{\infty}
dz\,
\frac{z^5}{\frac{\rho^2m_K^2}{4\lambda_K^2}-z^2}
(F'(z))^4.
\end{equation}
The numerator $(xM_s+\bar{x}M_u)$ makes explicit the skewness of the DA: even before evolution, the heavier strange quark biases the momentum distribution, so that the DA is no longer symmetric around $x=\tfrac12$ as in the pion case~\cite{Kock:2020frx,Kock:2021spt,Liu:2023feu}.

Zero-size limit:
\begin{align}
\phi_{K^+}(x)
&=
\frac{\sqrt{N_c}(xM_s+\bar{x}M_u)}{2\sqrt{2}\pi^2}
C_{K}\,
\theta(x\bar{x})
\ln\left(1+\frac{\Lambda^2}{xM_s^2+\bar{x}M_u^2-x\bar{x}m_K^2}\right).
\end{align}
The decay constant is
\begin{align}
f_K
&=
C_{K}\frac{\sqrt{N_c}M}{2\sqrt{2}\pi^2}
\int_0^1dx\frac{xM_s+\bar{x}M_u}{M}
\ln\left(1+\frac{\Lambda^2}{xM_s^2+\bar{x}M_u^2-x\bar{x}m_K^2}\right).
\end{align}
These relations show how the same underlying wave function controls both the DA and the decay constant.  In practice, $f_K$ is used to fix $C_K$, after which the ILM predictions for the DA and PDF follow without additional free parameters, up to the choice of instanton size~\cite{Liu:2023feu}.

\end{subappendices}



\chapter{Scalar mesons  on the LF}
\label{sec:scalarILM}

\section{Physical background and role of scalar mesons}

The Instanton Liquid Model (ILM) provides a semiclassical description of the QCD vacuum in terms of an ensemble of instantons and anti-instantons with characteristic size \(\rho\) and density \(n_{I+\bar I}\)~\cite{Diakonov:1995ea,Schafer:1996wv,Nowak:1996aj}.  
Quarks propagating in this background acquire exact zero modes of definite chirality in the field of a single (anti-)instanton~\cite{tHooft:1976rip}.  
By integrating out the gauge fields, one obtains an effective nonlocal multi-fermion interaction - the nonlocal 't~Hooft vertex - which encodes both the spontaneous breaking of chiral symmetry and the explicit breaking of the axial \(U_{A}(1)\) symmetry.

At low resolution, the quark propagation is dominated by these zero modes and one can parameterize the quark self-energy in terms of a momentum dependent constituent mass \(M(k)\), which is nonzero even in the chiral limit once the scalar 't~Hooft interaction exceeds a critical strength~\cite{Diakonov:1995ea,Nowak:1988bh}.  
The same interaction generates strong correlations in the scalar channels, giving rise to scalar mesons as quark-antiquark bound states.

From the point of view of chiral symmetry, the scalar-isoscalar \(\sigma\) field plays a special role.  
In a purely linear sigma model, the \(\sigma\) is the chiral partner of the pion; in the ILM, a similar phenomenon occurs dynamically: the same nonlocal interaction that generates a nonzero quark condensate and hence the pion as a pseudo-Goldstone boson, also binds a scalar \(\bar q q\) state whose mass tracks the dynamical threshold \(2M\)~\cite{Diakonov:1995ea,Nowak:1988bh}.  
The isovector scalar \(a_{0}\) is affected by both single-instanton and instanton-molecule contributions and thus provides a probe of the interplay between these two sectors of the ILM.

Working on the light front adds two crucial elements~\cite{Brodsky:1998hn}.  
First, the hadronic states are expanded in Fock components with a clear partonic interpretation: probabilities for finding quarks with given longitudinal momentum fraction \(x\) and transverse momentum \(k_\perp\).  
Second, the constrained ("bad'') component of the quark field is eliminated, inducing additional nonlocal interactions that are essential to realize spontaneous chiral symmetry breaking on the light front~\cite{Liu:2024lfv,Liu:2023fpj}.


\section{Instanton-induced scalar interactions and the gap equation}

In the ILM, the effective action for two light flavors can be written schematically as
\begin{equation}
\label{THOOFT1}
\begin{aligned}
    \mathcal{L}_I=&\frac{G_I}{8(N^2_c-1)}\left\{\frac{2N_c-1}{2N_c}\left[(\bar{\psi}\psi)^2-(\bar{\psi}\tau^a\psi)^2-(\bar{\psi}i\gamma^5\psi)^2+(\bar{\psi}i\gamma^5\tau^a\psi)^2\right]+\frac{1}{4N_c}\left[\left(\bar{\psi}\sigma_{\mu\nu}\psi\right)^2-\left(\bar{\psi}\sigma_{\mu\nu}\tau^a\psi\right)^2\right]\right\}\\
\end{aligned}
\end{equation}
in the ultra-local limit, where \(m\) is the current quark mass~\cite{Nowak:1988bh}.  The  effective coupling 
\begin{equation}
    G_I=\int d\rho n(\rho)\rho^{N_f}(2\pi\rho)^{2N_f}=\frac{n_{I+\bar{I}}}{2}(4\pi^2\rho^3)^{N_f}\left(\frac{1}{m^*\rho}\right)^{N_f}
\end{equation}
is fixed by the mean-instanton density 
\begin{equation}
    \frac{n_{I+\bar{I}}}{2}=\int d\rho n(\rho) \prod_{f=1}^{N_f}(m^*_f\rho)
\end{equation}
with $m^*_f$ the induced determinantal mass~\cite{Schafer:1996wv,Liu:2023fpj}. At low resolution, the instanton distribution is sharply peaked around the average instanton size $\rho\approx0.31$ fm, with a mean density $n_{I+\bar{I}}\sim 1$ $\mathrm{fm^{-4}}$.

However, the instanton size is not small in comparison to the size of the light hadrons,
e.g. pion, kaon, nucleon ..., and cannot be ignored in general. More importantly, this size fixes 
the UV scale and provides for a natural cut-off both in Euclidean or light front signature (see below). Finite size instantons yield finite
size zero modes, and therefore non-local effective interactions between the light quarks. The net effect is captured by the
substitution
\begin{equation}
\label{SUBX}
    \psi(x)\rightarrow\sqrt{{F}(i\partial)}\psi(x)
\end{equation}
in the local approximation. Here ${{F}(i\partial)}$ is the zero mode profile, that acts as a form factor. In singular gauge
its form is more user friendly  in momentum space
\begin{equation}
\label{M_cut_off}
    {F}(k)=\left[(zF_0'(z))^2\right]\bigg|_{z=\frac{k\rho}{2}}
\end{equation}
where $F_0(z)=I_0(z)K_0(z)-I_1(z)K_1(z)$ are spherical Bessel functions, and $k=\sqrt{k^2}$ is the Euclidean $4$-momentum.
Inserting (\ref{SUBX}) into (\ref{THOOFT1}) yields the non-local form of the effective action in the QCD instanton vacuum in leading order
in $1/N_c$
\begin{equation}
\label{TNLX}
\begin{aligned}
\mathcal{L}=&\bar{\psi}[i\slashed{\partial}-m]\psi+\frac{G_S}{2}(\bar{\psi}\sqrt{{F}(i\partial)}\sqrt{{F}(i\partial)}\psi)^2-\frac{G_S}{2}(\bar{\psi}\sqrt{{F}(i\partial)}\tau^a\sqrt{{F}(i\partial)}\psi)^2\\
&-\frac{G_S}{2}(\bar{\psi}\sqrt{{F}(i\partial)}i\gamma^5\sqrt{{F}(i\partial)}\psi)^2+\frac{G_S}{2}(\bar{\psi}\sqrt{{F}(i\partial)}i\gamma^5\tau^a\sqrt{{F}(i\partial)}\psi)^2 
\end{aligned}
\end{equation}
with $G_S=G_I/4N_c^2$.
Instanton-anti-instanton molecular configurations contribute additional nonlocal interactions.
As we will detail below, these interactions operate chiefly in the vector channels, but are subleading in the $1/N_c$ book-keeping.

The light quarks through zero modes acquire a  momentum dependent mass, in leading order in 
$1/N_c$ (mean-field approximation)
\begin{equation}
M(k) \approx M\,F(k)\,,
\end{equation}
The constituent quark mass \(M\equiv M(0)\) satisfies the gap-equation~\cite{Diakonov:1995ea,Nowak:1996aj}.
\begin{equation}
\frac{m}{M}
=
1 - 8g_S
\int\!\frac{d^4k}{(2\pi)^4}\,
\frac{F^2(k)}{k^2+M^2}\,,
\label{eq:gap4d}
\end{equation}
where \(g_S=N_cG_S\) is the scalar 't~Hooft coupling. 
The nonlocal form factor plays the role of a smooth ultraviolet cutoff of order \(1/\rho\).  
In dimensionless variables \(z=k\rho/2\), \eqref{eq:gap4d} can be written more explicitly as
\begin{equation}
\frac{m}{M}
=
1 - \frac{4g_S}{\pi^2\rho^2}
\int_0^\infty\!dz\,
\frac{z^3}{z^2+\frac{\rho^2M^2}{4}}\,
F^2(z).
\label{eq:gapz}
\end{equation}

In the chiral limit \(m\to 0\), a nontrivial solution \(M\neq 0\) exists only if the scalar coupling exceeds a critical value \(g_S^{\rm cr}\), determined from the condition that the coefficient of \(M\) in the linearized gap equation vanishes:
\begin{equation}
g_S^{\rm cr}
=
2\pi^2\rho^2
\left[
 8\int_0^\infty\!dz\,
 z\,F^2(z)
\right]^{-1}
\simeq 2.981\,\pi^2\rho^2.
\label{eq:gScr}
\end{equation}
For the standard two-flavor ILM parameters \(\rho \approx (636\text{ MeV})^{-1}\), this yields \(g_S^{\rm cr}\approx 72.6~\text{GeV}^{-2}\)~\cite{Liu:2023feu,Liu:2023fpj}.  
Once \(g_S>g_S^{\rm cr}\), the constituent mass \(M\) grows rapidly with \(g_S\), and the quark condensate
\begin{equation}
\rho^3\langle\bar\psi\psi\rangle
=
-\frac{4N_c}{\pi^2}\,\rho M
\int_0^\infty\!dz\,
\frac{z^3}{z^2+\frac{\rho^2M^2}{4}}\,
z\,F^2(z)
\label{eq:condensate}
\end{equation}
develops in magnitude, signaling the spontaneous breaking of chiral symmetry in the ILM.

\section{Effective scalar channels in the ILM}

The scalar channels are obtained by projecting the nonlocal 't~Hooft interaction onto the color-singlet, spin-0 bilinears.  
For two flavors, the relevant isoscalar and isovector scalar operators are
\begin{equation}
\sigma(x) \sim \bar\psi(x)\psi(x), 
\qquad
a_0^a(x) \sim \bar\psi(x)\tau^a\psi(x),
\end{equation}
where \(\tau^a\) are Pauli matrices in flavor space.

At leading order in \(1/N_c\), the scalar sector of the effective action may be written as
\begin{equation}
\mathcal{L}_{S}
=
\bar\psi(i\!\not\!\partial - M)\psi
+ \frac{G_\sigma}{2}
\bigl(\bar\psi \sqrt{F}\sqrt{F}\psi\bigr)^2
+ \frac{G_{a_0}}{2}
\bigl(\bar\psi \sqrt{F}\,\tau^a\sqrt{F}\psi\bigr)^2,
\label{eq:Lscalar}
\end{equation}
with $G_\sigma=-G_{a_0}=G_S$.
In the isoscalar scalar channel \(\sigma\), the net interaction is purely attractive and directly tied to the same scalar vertex that generates the quark condensate.  
In the isovector channel \(a_0\), the molecular contribution competes with the single-instanton term, resulting in a weaker net attraction.  
This competition is one of the main reasons that the \(a_0\) is heavier and less bound than the \(\sigma\) in the ILM~\cite{Schafer:1996wv,Kacir:1996qn,Liu:2023fpj}.

\section{Light-front representation of scalar mesons}

We now turn to the light-front Hamiltonian formulation~\cite{Brodsky:1997de,Brodsky:1998hn}.  
On the light front, quark fields are decomposed as
\begin{equation}
\psi = \psi_+ + \psi_-,
\qquad
\psi_\pm = \Lambda_\pm \psi,
\qquad
\Lambda_\pm = \frac{\gamma^0\gamma^\pm}{2},
\end{equation}
where \(\psi_+\) is the dynamical "good'' component and \(\psi_-\) is the constrained "bad'' component.  
Eliminating \(\psi_-\) using its equation of motion generates additional multi-fermion interactions, which in leading order reduce to tadpole contributions that renormalize the couplings in \eqref{eq:Lscalar}.  
The corresponding scalar kernels on the light front carry a characteristic denominator of the form
\begin{equation}
\frac{1}{1+2g_X w_+(P^+)},
\end{equation}
where \(w_+(P^+)\) is an even function of \(P^+\) arising from tadpole resummation and \(g_X\) is the channel-dependent effective coupling (\(g_\sigma, g_{a_0}\))~\cite{Liu:2024lfv}.

A scalar meson state \(|X(P)\rangle\) (\(X=\sigma, a_0\)) is expanded in the lowest Fock sector as
\begin{equation}
|X(P)\rangle
=
\int_0^1\frac{dx}{\sqrt{2x\bar x}}
\int\frac{d^2k_\perp}{(2\pi)^3}
\sum_{s_1,s_2}
\Phi_X(x,k_\perp,s_1,s_2)\,
b_{s_1}^\dagger(k)\,
d_{s_2}^\dagger(P-k)\,|0\rangle,
\label{eq:scalarFock}
\end{equation}
where \(x=k^+/P^+\), \(\bar x=1-x\), and \(b^\dagger, d^\dagger\) are quark and antiquark creation operators.

Because the scalar has spin zero, the spin structure of the valence wave function is particularly simple.  
One may factorize the light-front wave function as
\begin{equation}
\Phi_X(x,k_\perp,s_1,s_2)
=
\frac{C_X}{\sqrt{N_c}}
\sqrt{2x\bar x}\,
\phi_X(x,k_\perp)\,
\bar u_{s_1}(k)\,v_{s_2}(P-k),
\label{eq:PhiFactorized}
\end{equation}
where \(\phi_X(x,k_\perp)\) is a scalar function and \(C_X\) is a normalization constant.  
The latter is fixed by the probabilistic normalization condition
\begin{equation}
\int_0^1\!dx
\int\!\frac{d^2k_\perp}{(2\pi)^3}
\sum_{s_1,s_2}
\bigl|\Phi_X(x,k_\perp,s_1,s_2)\bigr|^2
= 1.
\label{eq:normPhi}
\end{equation}

\section{Light-front bound-state equation for scalar mesons}

Using the effective interaction \eqref{eq:Lscalar} and projecting the light-front Hamiltonian onto the two-body Fock sector, one derives the bound-state equation for \(\phi_X(x,k_\perp)\).  
The result can be written in the compact form
\begin{align}
m_X^2\,\phi_X(x,k_\perp)
&=
\frac{k_\perp^2+M^2}{x\bar x}\,\phi_X(x,k_\perp)
\nonumber\\
&\quad
-\frac{4\sqrt{F(k)F(P-k)}}{\sqrt{2x\bar x}}
\frac{g_X}{1+2g_Xw_+(P^+)}
\int_0^1\!\frac{dy}{\sqrt{2y\bar y}}
\int\!\frac{d^2q_\perp}{(2\pi)^3}
\left[
\frac{q_\perp^2+(y-\bar y)^2M^2}{y\bar y}
\right]
\nonumber\\
&\qquad\qquad
\times \phi_X(y,q_\perp)\,\sqrt{F(q)F(P-q)}\,.
\label{eq:LFscalarBound}
\end{align}
This equation holds for \(X=\sigma,a_0\) with the appropriate channel-dependent coupling \(g_X\).  
The kernel is entirely determined by the constituent mass \(M\), the instanton form factor \(F(k)\), and the tadpole function \(w_+(P^+)\).  
The nonlocality of the interaction ensures a finite result without the need of ad hoc cutoffs: the instanton size \(\rho\) provides the physical ultraviolet scale~\cite{Liu:2023feu,Liu:2023fpj}.

\begin{figure*}
  \includegraphics[height=5.5cm,width=.46\linewidth]{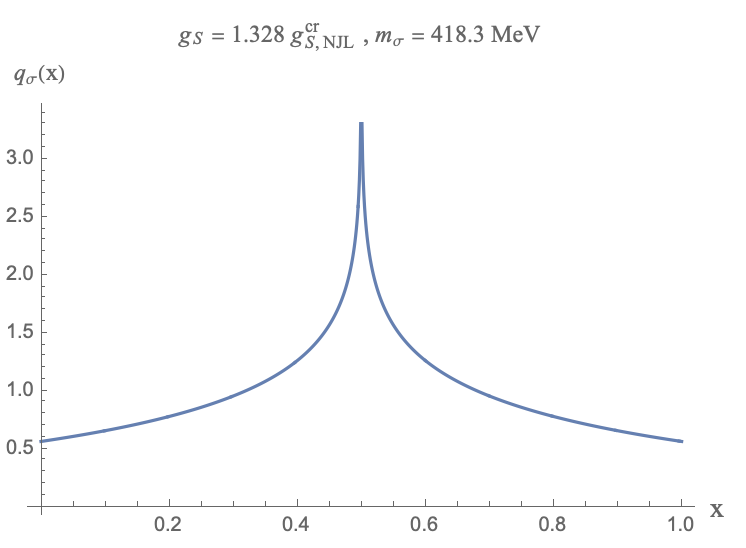}%
  \includegraphics[height=5.5cm,width=.46\linewidth]{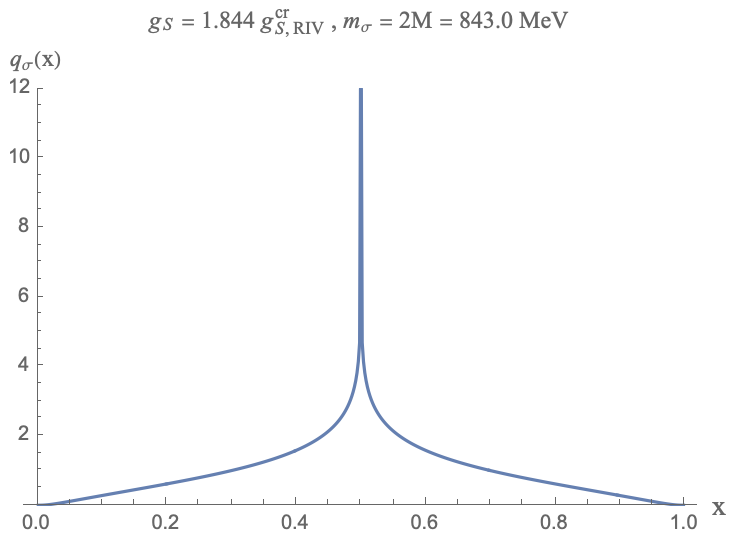}%
\caption{a: The sigma PDF in zero instanton size and in the chiral limit, with a  treshold sigma  mass $2M=418.9$ MeV;
b: The sigma PDF with the finite instanton size $1/\rho=630$ MeV, with a  treshold sigma  mass $2M=418.9$ MeV in the chiral limit;}
\label{fig_sigpi}
\end{figure*}

\begin{figure*}
  \includegraphics[height=3.cm,width=.99\linewidth]{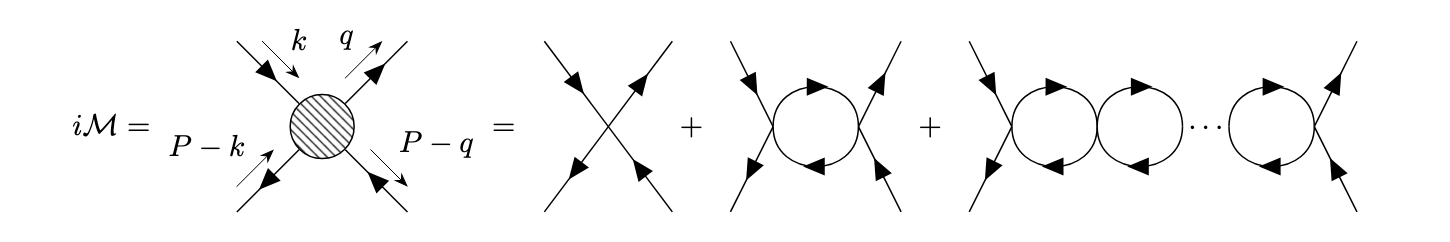}%
\caption{Contributions to the vacuum polarization function in the mean-field (bubble) approximation.}
\label{fig_bubble}
\end{figure*}

The same dynamics can be recast in a covariant language by resumming the scalar polarization bubble. 
In leading order (LO), the vacuum polarization function contributes to the 4-point function through the bubble-chain
shown in~FIg.~\ref{fig_bubble}.

The scalar propagator is
\begin{equation}
D_X(P^2)
=
\frac{iG_X}{1-G_X\,\Pi_{SS}(P^2)},
\label{eq:Dscalar}
\end{equation}
with scalar polarization
\begin{equation}
\Pi_{SS}(P^2)
=
-2iN_c\int\!\frac{d^4k}{(2\pi)^4}\,
\frac{4k\!\cdot\!(P-k)-4M^2}
{\bigl(k^2-M^2\bigr)\bigl((P-k)^2-M^2\bigr)}
F(k)F(P-k).
\label{eq:PiSS}
\end{equation}
The scalar mass follows from the pole condition
\begin{equation}
1
=G_X\,\Pi_{SS}(m_X^2).
\label{eq:poleCondition}
\end{equation}
Integrating over the light-front energy and using the gap equation, one arrives at a light-front form of the mass equation
\begin{equation}
1 - \frac{g_X}{g_S}\left(1-\frac{m}{M}\right)
=
-2g_X (m_X^2 - 4M^2)
\int_0^1\!dx
\int\!\frac{d^2k_\perp}{(2\pi)^3}\,
\frac{F(k)F(P-k)}
{x\bar x m_X^2 - (k_\perp^2+M^2)}.
\label{eq:massEquationLF}
\end{equation}
This equation is particularly useful to study how the scalar masses depend on the current quark mass \(m\) and on the coupling ratios \(g_X/g_S\).
Numerical solutions of \eqref{eq:LFscalarBound} and \eqref{eq:massEquationLF} exhibit a characteristic behavior.  
In the chiral limit \(m\to 0\), the \(\sigma\) mass satisfies
\begin{equation}
m_\sigma \approx 2M,
\end{equation}
so it lies very close to the constituent quark threshold.  
The \(\sigma\) is therefore a "threshold'' bound state whose mass tracks the order parameter of chiral symmetry breaking, in line with ILM expectations~\cite{Diakonov:1997sj,Nowak:1996aj}.  
As the current mass \(m\) increases, the constituent mass \(M\) changes only mildly, but the pole equation shifts so that \(m_\sigma\) moves above threshold and the state gradually becomes unbound.  
This illustrates explicitly how chiral symmetry breaking controls the existence of a light \(\sigma\).   The isovector scalar \(a_0\) is governed by the coupling
\begin{equation}
g_{a_0} = -g_S + 8g_V.
\end{equation}
where $g_V$ arises from a Fierzing of the vector interactions induced by instanton-anti-instanton molecules (see below). 
Since \(g_V\) is parametrically smaller than \(g_S\), the net attraction in the \(a_0\) channel is significantly reduced.  
As a result, the \(a_0\) mass lies well above the \(\sigma\) and is only weakly bound or even unbound, depending on the precise choice of ILM parameters~\cite{Liu:2023feu,Liu:2023fpj}.

This pattern is qualitatively consistent with phenomenology, where the light isoscalar scalar meson is anomalously light and broad, while the isovector scalar lies at a substantially higher mass.

\begin{figure}
    \includegraphics[scale=0.5]{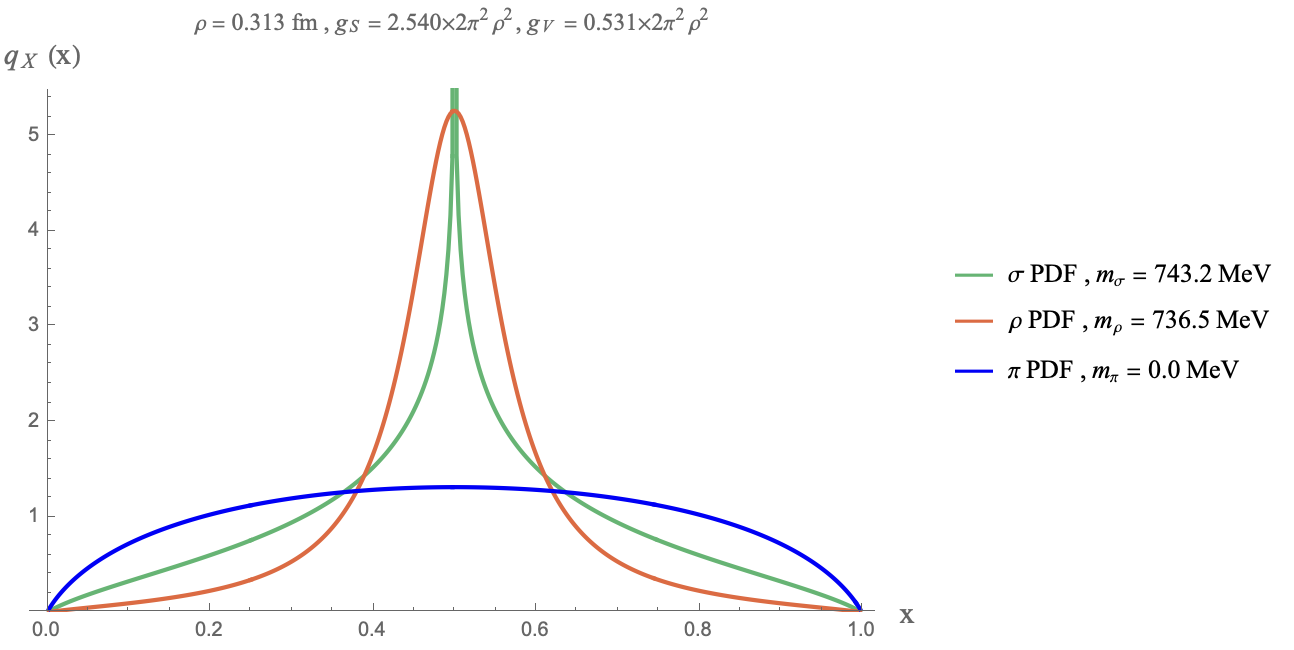}
    \caption{Parton density functions in the chiral limit~\cite{Liu:2023fpj}.}
    \label{TLXX2}
\end{figure}

\section{Scalar parton distributions on the light front}

One of the advantages of the light-front formulation is the direct access it provides to the partonic structure of hadrons.  
For a spin-0 meson \(X\) (such as \(\pi\) or a scalar meson), the leading-twist quark distribution function is defined by~\cite{Brodsky:1997de,Belitsky:2005qn}
\begin{equation}
q_X(x)
=
\int_{-\infty}^{\infty}\!\frac{d\xi^-}{4\pi}\,
e^{ixP^+\xi^-}\,
\langle X(P)|
 \bar\psi(0)\gamma^+W(0,\xi^-)\psi(\xi^-)
|X(P)\rangle,
\end{equation}
with \(W\) a Wilson line ensuring gauge invariance.  
In terms of the light-front wave function \(\Phi_X\) in \eqref{eq:scalarFock}, one finds the familiar relation
\begin{equation}
q_X(x)
=
\int\!\frac{d^2k_\perp}{(2\pi)^3}
\sum_{s_1,s_2}
\bigl|\Phi_X(x,k_\perp,s_1,s_2)\bigr|^2.
\label{eq:qScalar}
\end{equation}
Due to charge symmetry, the antiquark distribution satisfies
\begin{equation}
\bar q_X(x)
=
q_X(1-x),
\end{equation}
so that the total valence number and momentum sum rules are manifestly satisfied:
\begin{align}
\int_0^1\!dx\,
\bigl[q_X(x)-\bar q_X(x)\bigr]
&= 1,
\\
\int_0^1\!dx\,x\,
\bigl[q_X(x)+\bar q_X(x)\bigr]
&= 1
\end{align}
in the pure two-body truncation.

For scalar mesons in the ILM, \(q_X(x)\) is obtained by inserting the solution of \eqref{eq:LFscalarBound} into \eqref{eq:qScalar}.  
The resulting distributions are typically broad in \(x\), reflecting the strong binding and the nonlocality of the instanton-induced kernel.  
In the chiral limit, the \(\sigma\) distribution is particularly delocalized, consistent with its role as a collective mode tied to the quark condensate~\cite{Liu:2024lfv}.

\section{Summary of the scalar sector in the ILM}

To summarize, the ILM, through its nonlocal 't~Hooft interaction, provides a unified mechanism for spontaneous chiral symmetry breaking and for binding scalar mesons~\cite{Diakonov:1995ea,Schafer:1996wv,Nowak:1996aj}.  
The constituent quark mass \(M\) is generated dynamically once the scalar coupling exceeds a critical value, and the same coupling controls the attraction in the \(\sigma\) channel, leading to \(m_\sigma\approx 2M\) in the chiral limit.  
Instanon-anti-instanton molecular configurations introduce additional vector-like interactions which weaken the attraction in the isovector scalar channel, explaining naturally why \(m_{a_0}\gg m_\sigma\) in typical ILM parameter sets.  
The light-front formulation yields a well-defined bound-state equation for scalar mesons, fully consistent with the covariant Bethe-Salpeter resummation and directly related to their partonic structure via the light-front wave functions~\cite{Liu:2023feu,Liu:2023fpj}.  
More formal aspects of the gap equation and of the covariant Bethe-Salpeter amplitudes in the scalar channels are collected in the following appendices.


\begin{subappendices}

\section{Bethe-Salpeter wave functions and quark-meson couplings}

After summing the geometric series of scalar polarization bubbles, the full scalar propagator reads
\begin{equation}
D_X(P^2)
=
\frac{iG_X}{1-G_X\Pi_{SS}(P^2)},
\qquad X=\sigma,a_0,
\end{equation}
where \(\Pi_{SS}\) is given by Eq.~\eqref{eq:PiSS}.  
Near the pole at \(P^2=m_X^2\), the propagator can be written as
\begin{equation}
D_X(P^2)
\simeq
\frac{Z_X}{P^2-m_X^2+i\epsilon},
\end{equation}
with residue \(Z_X=g_{Xqq}^2\), where \(g_{Xqq}\) is the effective quark-meson coupling.  
Differentiating with respect to \(P^2\) gives the standard relation
\begin{equation}
g_{Xqq}^2
=
\left.
\left(
 \frac{\partial\Pi_{SS}(P^2)}{\partial P^2}
\right)^{-1}
\right|_{P^2=m_X^2},
\end{equation}
for \(X=\sigma,a_0\).  
Analogous expressions hold for pseudoscalar and vector mesons with their respective polarization functions~\cite{Liu:2023fpj}.

The corresponding Bethe-Salpeter wave functions are
\begin{align}
\Psi_\sigma(k;P)
&=
g_{\sigma qq}\,S(k)\,
\sqrt{F(k)}\,\sqrt{F(P-k)}\,S(k-P),
\\
\Psi_{a_0}(k;P)
&=
g_{a_0 qq}\,S(k)\,
\sqrt{F(k)}\,\tau^a\sqrt{F(P-k)}\,S(k-P),
\end{align}
where \(S(k)\) is the dressed quark propagator
\begin{equation}
S(k) = \frac{1}{\not\!k-M(k)}\approx \frac{1}{\not\!k-M}.
\end{equation}
These scalar Bethe-Salpeter amplitudes have the same nonlocal structure as the original 't~Hooft interaction: each quark leg is dressed with a form factor \(\sqrt{F}\) reflecting the instanton zero mode.

\section{From Bethe-Salpeter amplitudes to light-front wave functions}

The light-front wave functions \(\Phi_X\) that appear in \eqref{eq:scalarFock} can be obtained by integrating the Bethe-Salpeter amplitudes over the light-front energy \(k^-\) and projecting on appropriate spin states.  
One finds
\begin{equation}
\frac{1}{\sqrt{2x\bar x}}\,
\Phi_X(x,k_\perp,s_1,s_2)
=
iP^+
\int_{-\infty}^{\infty}\!\frac{dk^-}{2\pi}\,
\bar u_{s_1}(k)\,
\frac{\gamma^+}{2k^+}\,
\Psi_X(k;P)\,
\frac{\gamma^+}{2(P^+-k^+)}\,
v_{s_2}(P-k),
\qquad (k^+=xP^+).
\end{equation}
Performing the \(k^-\) integration by residues and comparing with the factorized form \eqref{eq:PhiFactorized} yields a relation between the normalization constant \(C_X\) and the covariant coupling \(g_{Xqq}\).  
For scalar and pseudoscalar channels one finds the simple identity
\begin{equation}
g_{Xqq}
=
-\frac{C_X}{\sqrt{N_c}},
\qquad X=\sigma,a_0,\pi,\eta_0,
\end{equation}
so that the normalization determined from the covariant Bethe-Salpeter equation matches exactly the probabilistic normalization of the light-front wave function \eqref{eq:normPhi}.  

This provides a powerful consistency check: the scalar meson properties (masses, couplings, and wave-function normalizations) computed in the covariant ILM framework can be translated directly into the light-front language without ambiguity.  
All observables that can be expressed in terms of the light-front wave functions - such as parton distribution functions and, in principle, distribution amplitudes specialized to scalar channels - are therefore fully determined once the ILM parameters are fixed in the vacuum sector~\cite{Liu:2023feu,Liu:2023fpj}.

\end{subappendices}




\chapter{Vectors on the LF}
\label{sec:vectorILM}

\section{Physical Background}
Vector mesons, notably the $\rho$ and $\omega$, represent the lowest-lying $J^{PC}=1^{-}$ excitations of QCD and provide a crucial testing ground for understanding nonperturbative dynamics. Because they couple directly to the electromagnetic current through $\bar{q}\gamma_\mu q$, they dominate hadronic contributions in many low-energy processes such as $e^+e^-$ annihilation and hadronic $\tau$ decay, forming the empirical basis of vector meson dominance \cite{Sakurai:1969ss}. Their internal quark structure and dynamical mass generation relate directly to the mechanisms of confinement and chiral symmetry breaking.

Within the ILM described  earlier, spontaneous chiral symmetry breaking is driven primarily by near-zero quark modes of single instantons. These modes generate a momentum-dependent constituent mass through the gap equation and produce strong attraction in pseudoscalar and scalar channels. In contrast, vector mesons receive only weak contributions from single instantons, since the 't~Hooft interaction flips helicity and therefore does not contribute directly to chirality-preserving vector bilinears. Instead, vector channels are governed by instanton-anti-instanton molecules \cite{Schafer:1994nv,Shuryak:2021fsu}, which give rise to nonlocal quark interactions preserving chirality. These molecular interactions are suppressed in the vacuum, but dominate the formation of vector-meson bound states, rendering the vector sector a sensitive probe of molecular correlations.

Light-front (LF) Hamiltonian quantization provides a powerful framework for analyzing such nonlocal interactions. The decomposition of quark fields into dynamical ($\psi_+$) and nondynamical ($\psi_-$) components reveals how the constrained field elimination generates additional multi-fermion operators. At leading order in $1/N_c$, these reduce to tadpole contributions that renormalize the effective vertices. The resulting LF Hamiltonian encodes the same nonlocalities appearing in the Euclidean ILM but rephrased in the language of LF wave functions, making explicit the partonic structure of vector mesons. In the LF formulation, the longitudinal and transverse polarizations obey distinct integral equations due to the lack of manifest Lorentz symmetry as noted earlier \cite{BrodskyPauliPinsky1998}. Nevertheless, it is possible to show that the physical masses coincide, and that Lorentz covariance can be restored by properly removing spurious terms induced by instantaneous interactions \cite{Liu:2023fpj}.

From a phenomenological perspective, the ILM's description of vector mesons is appealing because the instanton size $\rho\approx 1/3$ fm sets the characteristic nonlocality scale, which naturally produces extended vector bound states with charge radii. Furthermore, the ILM predicts broad distribution amplitudes at the intrinsic resolution scale $\mu\sim 1/(2\rho)\approx 0.6$ GeV, a feature consistent with lattice QCD extractions. The consistency between light-front results, covariant Bethe-Salpeter structures, and Euclidean path-integral derivations provides a cross-validated and physically transparent description of mesons in QCD.

\section{Physical details}
The lightest isovector vector meson is the $\rho(770)$ and the lightest isoscalar is the $\omega(782)$, each carrying quantum numbers $J^{PC}=1^{-}$. These states are created from the QCD vacuum by the local vector currents 
\[
J_\mu^a(x)=\bar{\psi}(x)\gamma_\mu \frac{\tau^a}{2}\psi(x), 
\qquad
J_\mu^{(0)}(x)=\bar{\psi}(x)\gamma_\mu\psi(x),
\]
and their dominance in hadronic electromagnetic processes forms the foundation of vector meson dominance \cite{Sakurai:1969ss}. 

In the Instanton Liquid Model (ILM), the fermionic zero modes of single instantons lead to a momentum-dependent constituent mass $M(k)\approx MF^2(k)$ as we discussed earlier, which  is the key to spontaneous chiral symmetry breaking, with the light pions and sigmas.

However, vector mesons cannot be generated by single instantons alone. The 't~Hooft interaction violates chirality and contributes only to channels built from bilinears such as $\bar{\psi}_L\psi_R$. Vector bilinears $\bar{\psi}\gamma_\mu\psi$ are chirality preserving and therefore require instanton-anti-instanton molecular correlations. Such molecules behave as tightly correlated pairs of instantons with approximate dipole structure, and their fermionic zero modes overlap in a manner that produces a nonlocal vector vertex \cite{Shuryak:2021fsu,Liu:2023fpj}.

The resulting effective interaction in the vector channel is suppressed by the small probability for molecular formation, typically a few percent of the total instanton density. This explains the phenomenological fact that vector mesons are heavier than the pion: their binding arises from weaker interactions than those responsible for chiral symmetry breaking.

To describe vector mesons in a partonic language, light-front (LF) quantization is  useful. The 
However, the main conceptual challenge is to ensure that Lorentz covariance of the vector channel, manifest in the covariant Bethe-Salpeter amplitude, is correctly reproduced in the LF Hamiltonian framework. A central result, derived in \cite{Liu:2023fpj}, is that all spurious Lorentz-violating LF terms can be cancelled by enforcing an identity between LF energy denominators and the covariant ILM propagators. This restores the equality of longitudinal and transverse vector-meson masses.

\section{Nonlocal Vector Interactions from Instanton Molecules}
In the local approximation, the maximally locked instanton-anti-instanton configurations or molecules,
induce flavor mixing interactions of the form~\cite{Schafer:1994nv}
\begin{equation}
\label{THHOFT2}
 \begin{aligned}
\mathcal{L}_{I\bar{I}}=G_{I\bar{I}}\bigg\{&\frac{1}{N_c(N_c-1)}\left[(\bar{\psi}\gamma^\mu\psi)^2+(\bar{\psi}\gamma^\mu\gamma^5\psi)^2\right]-\frac{N_c-2}{N_c(N_c^2-1)}\left[(\bar{\psi}\gamma^\mu\psi)^2- (\bar{\psi}\gamma^\mu\gamma^5\psi)^2\right]\\
    &+\frac{2N_c-1}{N_c(N_c^2-1)}\left[(\bar{\psi}\psi)^2+(\bar{\psi}\tau^a\psi)^2+(\bar{\psi}i\gamma^5\psi)^2+(\bar{\psi}i\gamma^5\tau^a\psi)^2\right]\\
     &-\frac{1}{2N_c(N_c-1)}\left[(\bar{\psi}\gamma^\mu\psi)^2+(\bar{\psi}\tau^a\gamma^\mu\psi)^2+(\bar{\psi}\gamma^\mu\gamma^5\psi)^2+(\bar{\psi}\tau^a\gamma^\mu\gamma^5\psi)^2\right]\bigg\}  
\end{aligned}
\end{equation}
which are $LL$ and $RR$ chirality preserving.
 The effective molecule-induced coupling is defined as
 \begin{equation}
 \label{MOLX}
     G_{I\bar{I}}=\int d\rho_I d\rho_{\bar{I}}\int dud^4R ~\frac{1}{8T_{I\bar{I}}^2} (4\pi^2\rho^2_I)(4\pi^2\rho^2_{\bar{I}})n(\rho_I)n(\rho_{\bar{I}})T_{I\bar{I}}(u,R)^{2N_f}\rho_I^{N_f}\rho_{\bar{I}}^{N_f}
 \end{equation}
Here  $R=z_I-z_{\bar{I}}$ is the relative molecular separation, $u_\mu=\frac{1}{2i}\mathrm{tr}(U_{\bar{I}}\tau^+_\mu U^\dagger_I)$ is the relative
molecular orientation with the locked color with $\tau_\mu^+=(\vec{\tau},-i)$, and $T_{I\bar I}$ is the hopping quark matrix.  (\ref{MOLX}) is readily 
understood as the unquenched tunneling density for a molecular configuration, whereby a pair of quark lines is removed by the division $T_{I\bar I}^2$
to account for the induced 4-Fermi interaction. The strength of the induced molecular coupling $G_{I\bar I}$ to the single coupling $G_I$ is
\begin{equation}
\label{hopping}
G_{I\bar{I}}=\frac{G_I^2}{128\pi^4\rho^2} \xi
\end{equation}
where the dimensionless and positive hopping parameter is defined as 
\bea
\label{HOPPING}
\xi=\frac{1}{\rho^4}\int dud^4R\left[\rho T_{I\bar{I}}(u,R)\right]^{2N_f-2}
\eea

Thanks to the smallness of the instanton plus anti-instanton density $n(\rho)$, the complex many-body dynamics of the ILM can be  organized around the dilute limit. For that the  $1/N_c$ 
book-keeping is useful, with $n(\rho)\sim N_c$ and both $G_I$ and $G_{I,\bar I}$ of the same order in $1/N_c$, but with a parametrically
small ratio $G_{I\bar I}/G_I$ from the diluteness. With this in mind, the leading contributions in $1/N_c$ from single and molecular instantons and anti-instantons are respectively
\bea
\label{THOOFT_Nc}
\mathcal{L}_{I}&=&\frac{G_I}{8N_c^2}\left[(\bar{\psi}\psi)^2-(\bar{\psi}\tau^a\psi)^2-(\bar{\psi}i\gamma^5\psi)^2+(\bar{\psi}i\gamma^5\tau^a\psi)^2\right]\nonumber\\
\mathcal{L}_{I\bar{I}}&=&\frac{G_{I\bar{I}}}{2N_c^2}\bigg[4\left[(\bar{\psi}\psi)^2+(\bar{\psi}\tau^a\psi)^2+(\bar{\psi}i\gamma^5\psi)^2+(\bar{\psi}i\gamma^5\tau^a\psi)^2\right]\nonumber\\
&&\qquad -\left[(\bar{\psi}\gamma^\mu\psi)^2+(\bar{\psi}\tau^a\gamma^\mu\psi)^2-3(\bar{\psi}\gamma^\mu\gamma^5\psi)^2+(\bar{\psi}\tau^a\gamma^\mu\gamma^5\psi)^2\right]\bigg] 
\eea
The induced $^\prime$t Hooft interaction ${\cal L}_I$ does not operate in the light vector channels, but the molecular induced interaction ${\cal L}_{I\bar I}$ does. 
The molecular interaction is equally attractive in the scalar $\sigma, a_0$ and pseudoscalar $\pi,\eta'$ channels.
Since the instanton molecules are topologically neutral, the molecular interaction are $U(1)_A$ symmetric. 
 Note that this Lagrangian predicts no splitting between the isoscalar ($\omega$) and isovector ($\rho$) vector channels. To illustrate how these interactions operate in various mesonic channels, 
 a Fierzing yields
\begin{equation}
\begin{aligned}
\mathcal{L}=&\bar{\psi}(i\slashed{\partial}-M)\psi+\frac{G_\sigma}{2}(\bar{\psi}\psi)^2+\frac{G_{a_0}}{2}(\bar{\psi}\tau^a\psi)^2+\frac{G_{\eta'}}{2}(\bar{\psi}i\gamma^5\psi)^2+\frac{G_\pi}{2}(\bar{\psi}i\gamma^5\tau^a\psi)^2\\
 &-\frac{G_\omega}{2}(\bar{\psi}\gamma_\mu\psi)^2-\frac{G_\rho}{2}(\bar{\psi}\gamma_\mu\tau^a\psi)^2-\frac{G_{f_1}}{2}(\bar{\psi}\gamma_\mu\gamma^5\psi)^2-\frac{G_{a_1}}{2}(\bar{\psi}\gamma_\mu\gamma^5\tau^a\psi)^2 
\end{aligned}
\end{equation}
with the effective couplings
\begin{align*}
G_\sigma&=G_S  &  G_{a_0}&=-G_S+8G_V   & 
G_\pi &=G_S    &  G_{\eta'}&=-G_S+8G_V\\
G_\omega&=G_V   &  G_\rho&=G_V        & G_{a_1}&=G_V    &  G_{f_1}&=-3G_V
\end{align*}
where $G_S=\frac{G_{I}}{4N_c^2}+\frac{4G_{I\bar{I}}}{N_c^2}$ and  $G_V=\frac{G_{I\bar{I}}}{N_c^2}$.
As we noted earlier, the non-local form is readily obtained through the substitution
\begin{equation}
\label{SUBX2}
    \psi(x)\rightarrow\sqrt{{F}(i\partial)}\psi(x)
\end{equation}
with the leading order $1/N_c$ 
\begin{equation}
\label{TNLX2}
\begin{aligned}
\mathcal{L}=&\bar{\psi}[i\slashed{\partial}-M(k)]\psi+\frac{G_S}{2}(\bar{\psi}\sqrt{{F}(i\partial)}\sqrt{{F}(i\partial)}\psi)^2-\frac{G_S}{2}(\bar{\psi}\sqrt{{F}(i\partial)}\tau^a\sqrt{{F}(i\partial)}\psi)^2\\
&-\frac{G_S}{2}(\bar{\psi}\sqrt{{F}(i\partial)}i\gamma^5\sqrt{{F}(i\partial)}\psi)^2+\frac{G_S}{2}(\bar{\psi}\sqrt{{F}(i\partial)}i\gamma^5\tau^a\sqrt{{F}(i\partial)}\psi)^2-\frac{G_V}{2}(\bar{\psi}\sqrt{{F}(i\partial)}\gamma_\mu\sqrt{{F}(i\partial)}\psi)^2\\
&-\frac{G_V}{2}(\bar{\psi}\sqrt{{F}(i\partial)}\gamma_\mu\tau^a\sqrt{{F}(i\partial)}\psi)^2+\frac{3G_V}{2}(\bar{\psi}\sqrt{{F}(i\partial)}\gamma_\mu\gamma^5\sqrt{{F}(i\partial)}\psi)^2-\frac{G_V}{2}(\bar{\psi}\sqrt{{F}(i\partial)}\gamma_\mu\gamma^5\tau^a\sqrt{{F}(i\partial)}\psi)^2 
\end{aligned}
\end{equation}

\section{Covariant analysis of vector mesons}
In the covariant formulation, the starting point is the dressed quark propagator in the ILM is
\[
S(k)=\frac{\slashed{k}+M(k)}{k^2-M^2(k)+i\epsilon},
\]
In $1/N_c$ book-keeping in the ILM, the Bethe-Salpeter equations in the vector channels amount
to resumming the bubble graphs as for the scalar with vector sources (mean-field approximation). More specifically,  the polarization tensor in the vector channel arises from the one-loop bubble using the dressedquark propagator,
\begin{equation}
\Pi_{\mu\nu}(P)
=
-i\!\int\!\frac{d^4k}{(2\pi)^4}
\mathrm{tr}\bigl[\gamma_\mu(\slashed{k}+M)\gamma_\nu(\slashed{P}-\slashed{k}+M)\bigr]
\frac{F(k)F(P-k)}{(k^2-M^2)[(P-k)^2-M^2]}.
\label{eq:PiVector}
\end{equation}
The trace yields
\[
\mathrm{tr}\bigl[\gamma_\mu(\slashed{k}+M)\gamma_\nu(\slashed{P}-\slashed{k}+M)\bigr]
=
4\bigl[k_\mu(P-k)_\nu + k_\nu(P-k)_\mu - g_{\mu\nu}(k\cdot(P-k)-M^2)\bigr].
\]
Using Feynman parameters and shifting $k\to k-\alpha P$, one obtains the standard transverse structure
\[
\Pi_{\mu\nu}(P)
=
\left(g_{\mu\nu}-\frac{P_\mu P_\nu}{P^2}\right)\Pi_T(P^2),
\qquad
P^\mu\Pi_{\mu\nu}=0.
\]
The bound-state mass follows from the pole condition
\[
1-g_V\Pi_T(m_V^2)=0.
\]

While in the covariant formulation Lorentz symmetry symmetry is manifest, this is not the case 
for the light fron formulation. However, the two formulations should yield the same observables,
in the $1/N_c$ book-keeping approximation of the ILM. The simplest way to see how this arizes
in the leading order or mean-field approximation is as follows: In the driving bubble kernel,
 the measure can be split through  $d^4k\rightarrow dk^-dk^+dk_\perp$ , and then the  $k^-$-integration carried first, 
\begin{equation}
\label{LF_integral}
    \int_{-\infty}^{\infty
    }\frac{dk^-}{2\pi}\frac{i}{(k^2-M^2)[(P-k)^2-M^2]}\mathcal{F}\left(k\right)\mathcal{F}\left(P-k\right)\bigg|_{P^2=m_X^2}\rightarrow\frac{\theta(x\bar{x})}{2x\bar{x}P^+}\frac{1}{m_X^2-\frac{k_\perp^2+M^2}{x\bar{x}}}\left[zF'(z)\right]^4\bigg|_{z=\frac{\rho k_\perp}{2\lambda_X\sqrt{x\bar{x}}}}
    \end{equation}
This can be justified by doing the contour integral along $k'^4=\frac{-k^3+ik^4}{\sqrt{2}}$ in Euclidean space. The parameter $\lambda_X$ is of order $1$, and can be determined by matching the integrals on both sides.  Its origin goes back to the removing of the  spurious poles in the two-body non-local form factor, in the analytical continuation from Euclidean to Minkowski signature~\cite{Kock:2020frx,Kock:2021spt}. Effectively, $\lambda_X$ is a measure of the non-locality related to the finite-sized instanton vacuum with effective size cut-off $\rho/\lambda_X$ which depends on the bound state mass $m_X$, constituent mass $M$ and instanton size $\rho$. with typivally~\cite{Kock:2020frx,Kock:2021spt}, 
\bea
\label{LAMBDASV}
  \lambda_S=2.464  \qquad  \lambda_V=3.542
\eea

\section{Light-Front Representation of Vector Mesons}

A physical vector meson is described on the light front by its lowest Fock component~\cite{Liu:2023fpj}
\[
|V(P,\lambda)\rangle
=
\int_0^1\!\frac{dx}{\sqrt{2x\bar{x}}}
\int\!\frac{d^2k_\perp}{(2\pi)^3}
\sum_{s_1s_2}
\Phi^\lambda_V(x,k_\perp;s_1,s_2)\,
b^\dagger_{s_1}(k)\,d^\dagger_{s_2}(P-k)\,|0\rangle.
\]
with the  LF momentum variables are
\[
k^+=x P^+,
\qquad
(P-k)^+=\bar{x}P^+,
\qquad
k_\perp,\qquad \bar{x}=1-x.
\]
In this framework, the vector meson light front   wave function factorizes
\[
\Phi_V^\lambda(x,k_\perp;s_1,s_2)
=
\frac{C_{V,\lambda}}{\sqrt{N_c}}
\sqrt{2x\bar{x}}\,
\phi_V^\lambda(x,k_\perp)\,
\bar{u}_{s_1}(k)\gamma^\mu \epsilon_\mu^{(\lambda)}v_{s_2}(P-k).
\]
with the spinor bilinears splitting into transverse and longitudinal contributions
\[
\bar{u}(k)\gamma^\mu\epsilon_\mu^{(\lambda)}v(P-k)
=
\bar{u}(k)
\left[
\gamma^+\epsilon_{(\lambda^-)}
+\gamma^-\epsilon_{(\lambda^+)}
-\gamma^i\epsilon_{(\lambda^i)}
\right]v(P-k).
\]
Because 
$\epsilon^{+}_{A}=0$
and $\epsilon^{+}_{(0)}\neq 0$, the longitudinal polarization couples to the instantaneous part of the Dirac propagator, producing the LF asymmetry between $\lambda=0$ and $\lambda=\pm 1$ which must ultimately be removed.
The LF wavefunction normalization follows from
\[
\int dx \int \frac{d^2k_\perp}{(2\pi)^3}
\sum_{s_1s_2}|\Phi_V^\lambda|^2=1.
\]

\section{Transverse Vector Bound-State Equation}

The study of light-front (LF) bound states provides a powerful nonperturbative framework for describing relativistic hadrons. In the ILM, quarks acquire a dynamical mass together with nonlocal, chirally symmetric interactions that can be recast in a light-front Hamiltonian formalism. In this formulation, only the so-called good component of the fermion field is dynamical, while the bad component is constrained and must be eliminated. This elimination gives rise to instantaneous fermion-exchange kernels that contribute to the two-body interaction. 

The LF Hamiltonian for the $\rho$/$\omega$ mesons contains a kinetic term, and a nonlocal interaction arising from ILM molecular contributions. For the vector mesons with transverse polarization, the instantaneous fermion exchange through the bad component does not contribute. Hence, the bound state equation follows solely from the use of the good component, giving~\cite{Liu:2023fpj}
\begin{equation}
m_V^2 \phi_V^T(x,k_\perp)
=
\frac{k_\perp^2+M^2}{x\bar{x}}\,\phi_V^T(x,k_\perp)
-
4g_V \sqrt{F(k)F(P-k)}
\int_0^1 dy \int\frac{d^2q_\perp}{(2\pi)^3}
\frac{4(q_\perp^2+M^2)}{\sqrt{2x\bar{x}2y\bar{y}}}\,
\phi_V^T(y,q_\perp)\sqrt{F(q)F(P-q)}.
\label{eq:TransMain}
\end{equation}
Its derivation from the LF Hamiltonian in the mean-field approximation, uses the reduction
\[
\gamma^i\slashed{q}\gamma^i=-\slashed{q}+2q^i\gamma^i,
\]
after summing over $i=1,2$, which yields
\[
\bar{u}(k)\gamma^i(\slashed{q}+M)\gamma^i v(P-q)
=
4(q_\perp^2+M^2)\,\bar{u}(k)v(P-q),
\]
producing the kernel in Eq.~\eqref{eq:TransMain}.
The equation is an eigenvalue problem for $m_V^2$, symmetric in $x\leftrightarrow\bar{x}$, with attractive kernel due to positive $g_V$ from instanton molecules. Its solution determines $\phi_V^T$ and the vector-meson mass.

\begin{figure}
    \includegraphics[scale=0.5]{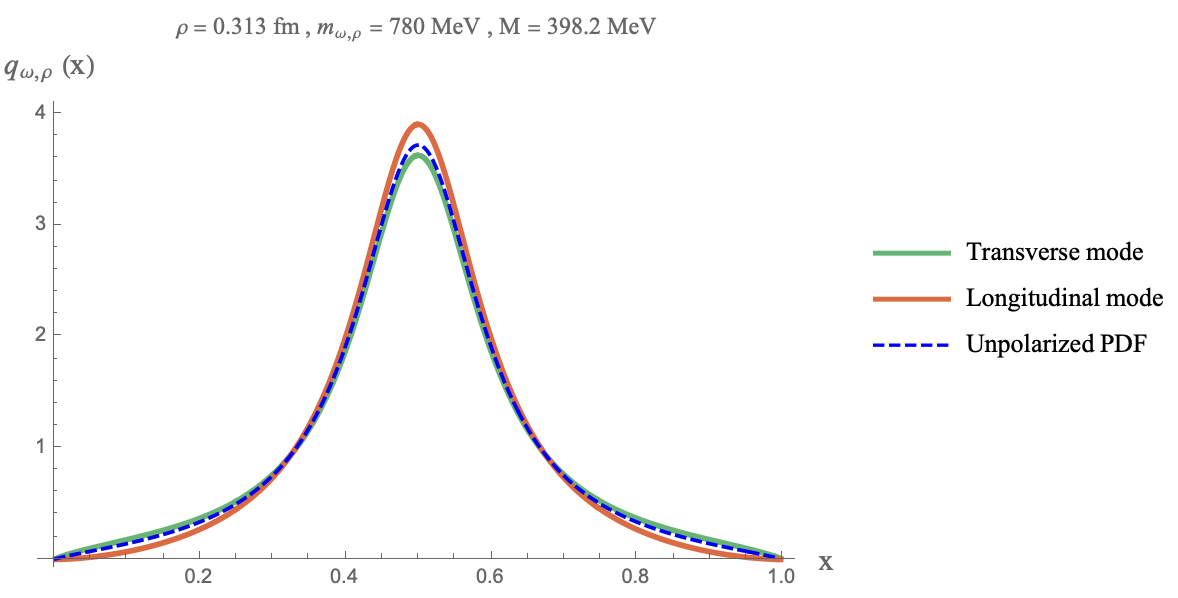}
    \caption{Vector mesons PDFs versus parton-$x$: transverse polarization (solid-green), longitudinal polarization (solid-red) and unpolarized (dashed-blue)~\cite{Liu:2023fpj}.}
    \label{TLXX1}
     \end{figure}

\section{Longitudinal Vector Bound-State Equation on the Light Front}

For vector mesons, the longitudinal polarization contains additional instantaneous contributions not present in the transverse channel. A covariant treatment in the ILM, however, requires that the longitudinal and transverse channels ultimately coincide once nonlocal form factors and Bethe-Salpeter constraints are enforced. 

Here, we derive the longitudinal light-front bound-state equation for a neutral vector meson in a quark-antiquark model with an effective vector interaction inspired by the ILM.  The goal is to make explicit the role of the instantaneous fermion contribution in the longitudinal channel and to show how, after enforcing a covariant structure for the non-local interaction, the longitudinal equation becomes identical to the transverse one, leading to a unique vector mass and a common light-front wave function. The exact ILM derivation can be found in~\cite{Liu:2023fpj} (Appendix B).

\subsection{Light-front kinematics and quark fields}
 For completeness, we recall the light front coordinates
\begin{equation}
x^\pm = x^0 \pm x^3 , 
\qquad
\bm{x}_\perp = (x^1,x^2),
\end{equation}
and similarly for momenta.  The total meson momentum is chosen as
\begin{equation}
P^\mu = \bigl(P^+,\,\tfrac{m_V^2}{P^+},\,\bm{0}_\perp\bigr),
\end{equation}
with $m_V$ the physical vector meson mass.  In the valence $q\bar q$ Fock sector the quark carries longitudinal momentum fraction $x$ and transverse momentum $\bm{k}_\perp$, while the antiquark carries $\bar x = 1-x$ and $-\bm{k}_\perp$.

The Dirac field is decomposed into "good'' and "bad'' components,
\begin{equation}
\psi_\pm = \Lambda_\pm \psi, 
\qquad 
\Lambda_\pm = \frac{1}{2}\gamma^\mp\gamma^\pm,
\end{equation}
with $\psi_+$ dynamical and $\psi_-$ constrained.  In the presence of an effective non-local interaction $V_{\rm ILM}$ and a constituent quark mass $M$, the light-front Dirac equation yields the constraint
\begin{equation}
i\partial^+ \psi_- 
= \left(i\gamma^\perp\partial_\perp + M + V_{\rm ILM}\right)\psi_+,
\end{equation}
so that
\begin{equation}
\psi_- 
= \frac{1}{i\partial^+}\left(i\gamma^\perp\partial_\perp + M + V_{\rm ILM}\right)\psi_+.
\label{eq:psi_minus_constraint}
\end{equation}
Eliminating $\psi_-$ in favor of $\psi_+$ generates a {\it bare} instantaneous fermion interactions in the light-front Hamiltonian, which will play a crucial role in the longitudinal channel. For clarity of the argument, we will ignore the dressing of the bare interactions in the ILM. The full treatement
including the dressing in the ILM, is given in~\cite{Liu:2023fpj}.

\subsection{Effective vector interaction and non-local form factor}

We consider an effective four-fermion interaction in the vector channel,
\begin{equation}
\mathcal{L}_{\rm int}
= -\frac{g_V}{2}\,\bigl[\bar\psi(x)\gamma^\mu \mathcal{F}(i\partial)\psi(x)\bigr]\,
   \bigl[\bar\psi(x)\gamma_\mu \mathcal{F}(i\partial)\psi(x)\bigr],
\end{equation}
where $g_V$ is a coupling and $\mathcal{F}(i\partial)$ is a non-local operator encoding the momentum-dependent ILM form factor.  In momentum space the corresponding vertex attaches a form factor $F(k)$ to each quark leg, with $k$ the off-shell quark momentum.  For a quark carrying $(xP^+,\bm{k}_\perp)$ and an antiquark carrying $(\bar x P^+, -\bm{k}_\perp)$ we will denote these factors as
\begin{equation}
F(k) \equiv F\bigl(x,\bm{k}_\perp\bigr),
\qquad
F(P-k) \equiv F\bigl(\bar x,-\bm{k}_\perp\bigr).
\end{equation}
A convenient shorthand is
\begin{equation}
\mathcal{F}_{x,\bm{k}_\perp} \equiv \sqrt{F(k)F(P-k)}.
\end{equation}

\subsection{Vector meson wave function on the light front}

The light-front valence state of a neutral vector meson with polarization $\lambda$ is expanded as
\begin{equation}
\bigl| V(P,\lambda) \bigr\rangle_{\rm val}
= \sum_{s_1,s_2}
\int_0^1 \frac{dx}{\sqrt{x\bar x}}
\int \frac{d^2\bm{k}_\perp}{(2\pi)^3}\;
\psi^\lambda_{s_1 s_2}(x,\bm{k}_\perp)\,
b^\dagger_{s_1}(xP^+,\bm{k}_\perp)\,
d^\dagger_{s_2}(\bar x P^+,-\bm{k}_\perp)\,|0\rangle,
\end{equation}
where $s_1,s_2$ denote light-front helicities and $b^\dagger,d^\dagger$ are quark and antiquark creation operators.

The spinor structure is encoded in a light-front spin projector $S^\lambda_{s_1 s_2}(x,\bm{k}_\perp)$, while the dynamical information is in a scalar wave function $\phi_V(x,\bm{k}_\perp)$.  For a longitudinally polarized state ($\lambda = 0$) one may write
\begin{equation}
\psi^L_{s_1 s_2}(x,\bm{k}_\perp)
= S^L_{s_1 s_2}(x,\bm{k}_\perp)\,\phi_V^L(x,\bm{k}_\perp),
\end{equation}
where $S^L_{s_1 s_2}$ is constructed from the polarization vector $\varepsilon_L^\mu(P)$ and the Dirac structure $\gamma_\mu$ coupled to the spinors.

The free $q\bar q$ invariant mass squared is
\begin{equation}
M_0^2(x,\bm{k}_\perp)
= \frac{\bm{k}_\perp^2 + M^2}{x\bar x}.
\label{eq:M0_def}
\end{equation}
The interacting system satisfies a light-front Hamiltonian eigenvalue problem,
\begin{equation}
\bigl( P^-_{\rm free} + P^-_{\rm int} \bigr)\, \bigl| V(P,\lambda)\bigr\rangle
= \frac{m_V^2}{P^+}\,\bigl| V(P,\lambda)\bigr\rangle,
\end{equation}
which reduces in the valence sector to an integral equation for $\phi_V^\lambda(x,\bm{k}_\perp)$.

\subsection{Transverse reference equation}

It is convenient to recall first the transverse ($\lambda=\pm 1$) equation, which is free of instantaneous fermion contributions once $\psi_-$ is eliminated.  For a transverse polarization one obtains a homogeneous integral equation of the form
\begin{equation}
\bigl[m_V^2 - M_0^2(x,\bm{k}_\perp)\bigr]\,
\phi_V^T(x,\bm{k}_\perp)
= \int_0^1 dy \int \frac{d^2\bm{q}_\perp}{(2\pi)^3}\;
K_T\bigl(x,\bm{k}_\perp; y,\bm{q}_\perp\bigr)\,
\phi_V^T(y,\bm{q}_\perp),
\label{eq:transverse_eq_general}
\end{equation}
where $K_T$ is the transverse kernel generated by the effective vector interaction.  In the ILM, a typical structure is
\begin{equation}
K_T\bigl(x,\bm{k}_\perp; y,\bm{q}_\perp\bigr)
= -\,\frac{g_V N_c}{\sqrt{2x\bar x\,2y\bar y}}\;
\mathcal{F}_{x,\bm{k}_\perp}\,
\mathcal{F}_{y,\bm{q}_\perp}\,
\mathcal{N}_T\bigl(x,\bm{k}_\perp; y,\bm{q}_\perp\bigr),
\label{eq:KT_def}
\end{equation}
where $\mathcal{N}_T$ is the Dirac trace projected onto the transverse polarization.  For a pure vector current-current interaction this trace is dominated by terms proportional to $4\bigl(\bm{q}_\perp^2 + M^2\bigr)$; more precisely,
\begin{equation}
\mathcal{N}_T\bigl(x,\bm{k}_\perp; y,\bm{q}_\perp\bigr)
= 4\bigl(\bm{q}_\perp^2 + M^2\bigr)
+ \text{(terms that vanish after symmetric integration)}.
\end{equation}
After discarding the odd terms in $\bm{q}_\perp$ by symmetric integration, the transverse kernel may be represented schematically as
\begin{equation}
K_T\bigl(x,\bm{k}_\perp; y,\bm{q}_\perp\bigr)
\simeq 
-\,\frac{4g_V N_c}{\sqrt{2x\bar x\,2y\bar y}}\;
\mathcal{F}_{x,\bm{k}_\perp}\,
\bigl(\bm{q}_\perp^2 + M^2\bigr)\,
\mathcal{F}_{y,\bm{q}_\perp}.
\label{eq:KT_schematic}
\end{equation}
The precise numerical normalization is not essential for the discussion of the difference between transverse and longitudinal channels; what matters is the dependence on $\bm{q}_\perp^2+M^2$ and the overall non-local factors.

\subsection{Instantaneous fermion exchange and the longitudinal kernel}

In contrast, the longitudinal polarization is sensitive to the instantaneous fermion term originating from the constraint~\eqref{eq:psi_minus_constraint}.  When the light-front Hamiltonian is reexpressed solely in terms of the good component $\psi_+$, the effective interaction in the vector channel acquires an additional "seagull'' piece, which in the valence sector appears as an instantaneous quark exchange.  Diagrammatically, this corresponds to the propagation of a quark line with the instantaneous light-front propagator $\gamma^+/(2p^+)$.

The longitudinal bound-state equation may be written as
\begin{equation}
\bigl[m_V^2 - M_0^2(x,\bm{k}_\perp)\bigr]\,
\phi_V^L(x,\bm{k}_\perp)
= \int_0^1 dy \int \frac{d^2\bm{q}_\perp}{(2\pi)^3}\;
K_L\bigl(x,\bm{k}_\perp; y,\bm{q}_\perp\bigr)\,
\phi_V^L(y,\bm{q}_\perp).
\label{eq:longitudinal_eq_general}
\end{equation}
The kernel $K_L$ naturally splits into a regular (ladder-type) contribution $K_{\rm reg}$, identical to that appearing in the transverse channel, and an instantaneous contribution $K_{\rm inst}$:
\begin{equation}
K_L = K_{\rm reg} + K_{\rm inst},
\qquad
K_{\rm reg} \equiv K_T.
\end{equation}

Carrying out the Dirac algebra for the longitudinal polarization vector $\varepsilon_L^\mu(P)$ and using the instantaneous propagator for the internal quark line, one finds a kernel of the form
\begin{align}
K_L\bigl(x,\bm{k}_\perp; y,\bm{q}_\perp\bigr)
&= -\,\frac{g_V N_c}{\sqrt{2x\bar x\,2y\bar y}}\;
\mathcal{F}_{x,\bm{k}_\perp}\,
\mathcal{F}_{y,\bm{q}_\perp}\,
\Bigl[
\mathcal{N}_T\bigl(x,\bm{k}_\perp; y,\bm{q}_\perp\bigr)
+ \mathcal{N}_{\rm inst}\bigl(x,\bm{k}_\perp; y,\bm{q}_\perp\bigr)
\Bigr],
\label{eq:KL_def}
\end{align}
with $\mathcal{N}_T$ as in the transverse case and $\mathcal{N}_{\rm inst}$ encoding the instantaneous contribution.  After performing the light-front energy integration and projecting onto the longitudinal channel, the instantaneous trace can be arranged as
\begin{equation}
\mathcal{N}_{\rm inst}\bigl(x,\bm{k}_\perp; y,\bm{q}_\perp\bigr)
= -\,\bigl[2x\bar x\,m_V^2 + \bm{k}_\perp^2 + M^2\bigr]
+ \text{(terms that vanish after symmetric integration)}.
\label{eq:Ninst_form}
\end{equation}
Here the dependence on $m_V^2$ reflects the coupling of the longitudinal polarization vector to the total momentum $P^\mu$, while the combination $\bm{k}_\perp^2+M^2$ arises from the numerator structure of the instantaneous propagator and the spinor projectors.

Substituting~\eqref{eq:KT_schematic} and~\eqref{eq:Ninst_form} into~\eqref{eq:KL_def}, and dropping odd terms in $\bm{q}_\perp$ by symmetric integration, the longitudinal kernel can be written in a schematic but physically transparent form:
\begin{align}
K_L\bigl(x,\bm{k}_\perp; y,\bm{q}_\perp\bigr)
&\simeq 
-\,\frac{4g_V N_c}{\sqrt{2x\bar x\,2y\bar y}}\;
\mathcal{F}_{x,\bm{k}_\perp}\,
\bigl(\bm{q}_\perp^2 + M^2\bigr)\,
\mathcal{F}_{y,\bm{q}_\perp}
\nonumber\\
&\quad
-\,\frac{g_V N_c}{\sqrt{2x\bar x\,2y\bar y}}\;
\mathcal{F}_{x,\bm{k}_\perp}\,
\bigl[2x\bar x\,m_V^2 + \bm{k}_\perp^2 + M^2\bigr]\,
\mathcal{F}_{y,\bm{q}_\perp}.
\label{eq:KL_schematic}
\end{align}
The first line is identical to the transverse kernel, while the second line is the characteristic instantaneous contribution.

\subsection{Difference between longitudinal and transverse kernels}

Equations~\eqref{eq:KT_schematic} and~\eqref{eq:KL_schematic} make the origin of the difference between longitudinal and transverse kernels explicit.  For fixed $(x,\bm{k}_\perp)$ one finds
\begin{align}
K_L\bigl(x,\bm{k}_\perp; y,\bm{q}_\perp\bigr)
- K_T\bigl(x,\bm{k}_\perp; y,\bm{q}_\perp\bigr)
&\simeq
-\,\frac{g_V N_c}{\sqrt{2x\bar x\,2y\bar y}}\;
\mathcal{F}_{x,\bm{k}_\perp}\,
\bigl[2x\bar x\,m_V^2 + \bm{k}_\perp^2 + M^2\bigr]\,
\mathcal{F}_{y,\bm{q}_\perp}.
\label{eq:kernel_difference}
\end{align}
Thus the \emph{difference} between the longitudinal and transverse kernels is proportional to the combination
\begin{equation}
2x\bar x\,m_V^2 + \bm{k}_\perp^2 + M^2,
\label{eq:difference_struct}
\end{equation}
modulated by the non-local form factors.  Left untreated, this difference would imply distinct integral equations and potentially different eigenvalues $m_{V,L}^2$ and $m_{V,T}^2$, in apparent conflict with rotational covariance.

\subsection{Covariant ILM prescription and restoration of equality}

In the covariant ILM, the underlying Bethe-Salpeter equation for the vector channel has a Dirac structure proportional to
\begin{equation}
\Gamma^\mu(k;P) \sim \gamma^\mu \,F(k)F(P-k),
\end{equation}
with the form factors chosen such that the full amplitude is rotationally invariant in the meson rest frame.  When this covariant structure is projected onto the light front, the combination appearing in~\eqref{eq:difference_struct} must match the invariant $k^2 + M^2$ evaluated in an instantaneous approximation with the same non-local dressing.

This requirement motivates the replacement
\begin{equation}
x\bar x\,m_V^2 + \bm{k}_\perp^2 + M^2
\;\longrightarrow\;
2\bigl(\bm{k}_\perp^2 + M^2\bigr)\,F^2(k),
\label{eq:ILM_replacement}
\end{equation}
where $F(k)$ is the ILM quark form factor evaluated at the off-shell momentum corresponding to $(x,\bm{k}_\perp)$.  The numerical factor $2$ reflects the two dressed quark legs in the covariant amplitude.  A similar replacement-like  is shown to be exact in the full treatement in the ILM, in leading order in $1/N_c$. With this in mind, the instantaneous contribution in~\eqref{eq:KL_schematic} is reshaped into the same functional form as the transverse kernel.  More precisely, after~\eqref{eq:ILM_replacement} the longitudinal kernel becomes
\begin{equation}
K_L\bigl(x,\bm{k}_\perp; y,\bm{q}_\perp\bigr)
\simeq K_T\bigl(x,\bm{k}_\perp; y,\bm{q}_\perp\bigr),
\end{equation}
up to terms that vanish after symmetric integration over $\bm{q}_\perp$ and higher-order corrections in the ILM expansion.

Consequently, the longitudinal and transverse bound-state equations,
\begin{align}
\bigl[m_V^2 - M_0^2(x,\bm{k}_\perp)\bigr]\,
\phi_V^T(x,\bm{k}_\perp)
&= \int_0^1 dy \int \frac{d^2\bm{q}_\perp}{(2\pi)^3}\;
K_T\bigl(x,\bm{k}_\perp; y,\bm{q}_\perp\bigr)\,
\phi_V^T(y,\bm{q}_\perp),
\\[0.5em]
\bigl[m_V^2 - M_0^2(x,\bm{k}_\perp)\bigr]\,
\phi_V^L(x,\bm{k}_\perp)
&= \int_0^1 dy \int \frac{d^2\bm{q}_\perp}{(2\pi)^3}\;
K_L\bigl(x,\bm{k}_\perp; y,\bm{q}_\perp\bigr)\,
\phi_V^L(y,\bm{q}_\perp),
\end{align}
collapse onto a single integral equation,
\begin{equation}
\bigl[m_V^2 - M_0^2(x,\bm{k}_\perp)\bigr]\,
\phi_V(x,\bm{k}_\perp)
= \int_0^1 dy \int \frac{d^2\bm{q}_\perp}{(2\pi)^3}\;
K\bigl(x,\bm{k}_\perp; y,\bm{q}_\perp\bigr)\,
\phi_V(y,\bm{q}_\perp),
\label{eq:unified_vector_eq}
\end{equation}
with a common kernel $K \equiv K_T \simeq K_L$ and a unique scalar light-front wave function $\phi_V$.

The physical consequence is that the vector meson mass and wave function are independent of the polarization:
\begin{equation}
\phi_V^L(x,\bm{k}_\perp) = \phi_V^T(x,\bm{k}_\perp) \equiv \phi_V(x,\bm{k}_\perp),
\qquad
m_{V,L}^2 = m_{V,T}^2 \equiv m_V^2.
\end{equation}
This equality signals the restoration of rotational covariance (within the approximations of the ILM) at the level of the light-front bound-state equation.  In practice, this construction provides a consistent starting point for computing longitudinal and transverse distribution amplitudes, parton distributions, and form factors of light vector mesons within the ILM framework.

\begin{figure}
    \centering
    \includegraphics[scale=0.6]{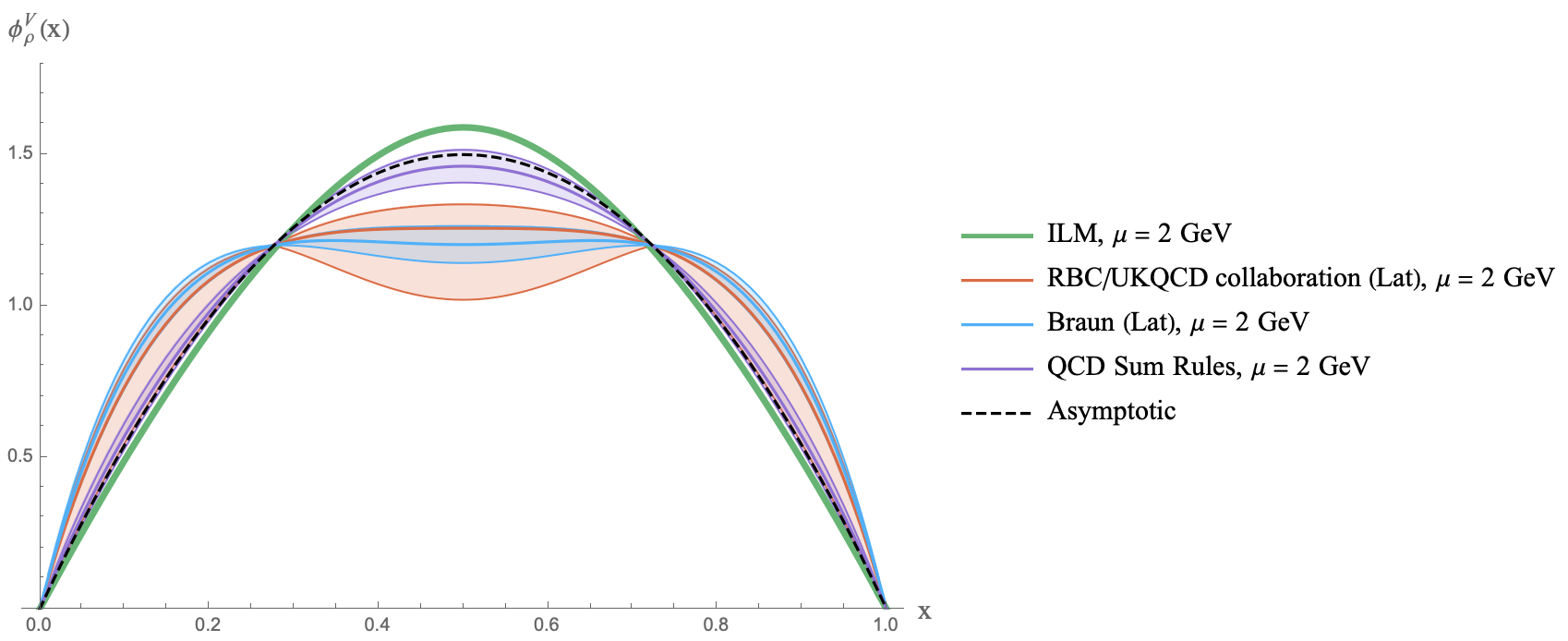}
    \caption{Evolved longitudinal  $\rho$ DA to $\mu=2$ GeV in solid-green~\cite{Liu:2023fpj},
    compared with the lattice in RBC/UKQCD collaboration~\cite{Boyle:2008nj} in filled-red and the lattice \cite{Braun:2016wnx} in filled-blue. 
    The QCD asymptotic result of $6x\bar x$ is in dashed-black and the QCD sum result~\cite{Stefanis:2015qha} is in filled-purple.}
    \label{RHOX2}
\end{figure}

\begin{figure}
    \centering
    \includegraphics[scale=0.7]{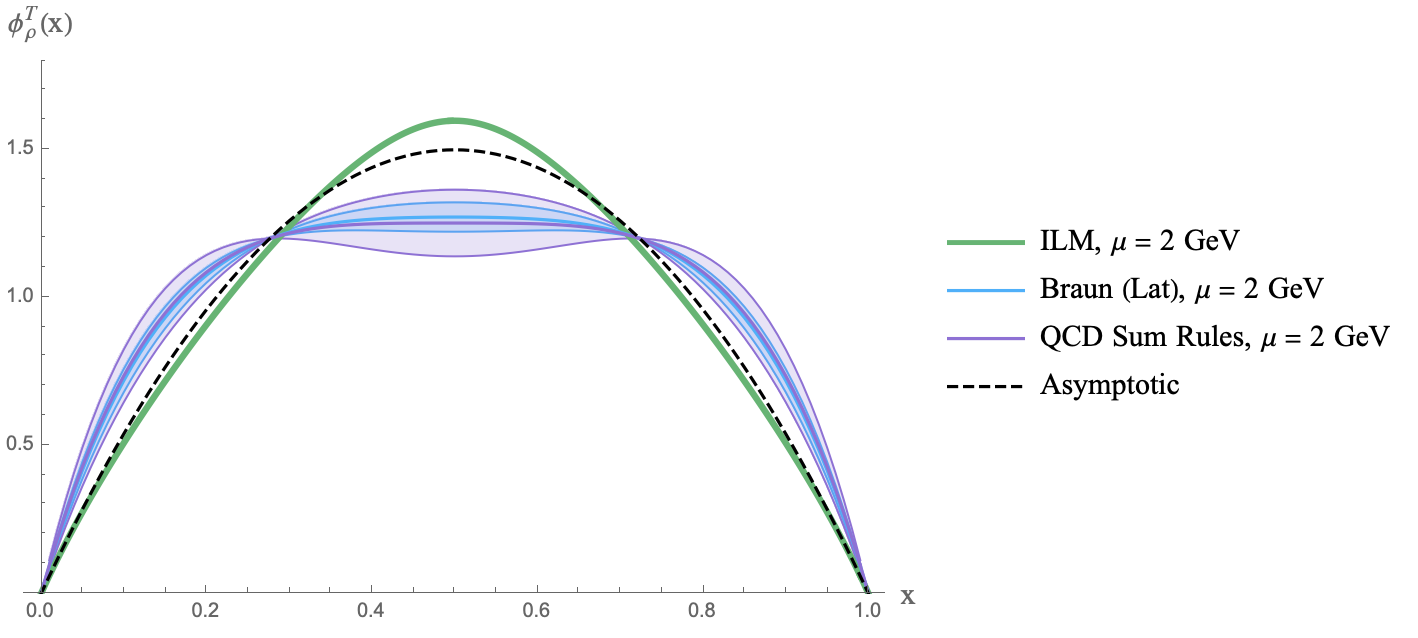}
    \caption{Evolved  transverse DA for the rho meson  at $\mu=2$ GeV  in solid-green~\cite{Liu:2023fpj}, and
    compared with the lattice result at $\mu=2$ GeV~\cite{Braun:2016wnx} in filled-blue and the QCD sum rule result also at $\mu=2$ GeV~\cite{Ball:1996tb} 
    in filled-purple.}
    \label{TENSORX2}
\end{figure}

\section{Decay Constants and Distribution Amplitudes}

The decay constant is defined via
\[
\langle 0|\bar{\psi}(0)\gamma^\mu \psi(0)|V(P,\lambda)\rangle
=
f_V m_V \epsilon^\mu_{(\lambda)}.
\]
Contracting with $\epsilon^{(0)}_\mu$ and using LF quantization gives
\[
f_V
=
-\frac{2}{\sqrt{2N_c}m_V}
\int dx\,d^2k_\perp\,
\frac{\phi_V^L(x,k_\perp)}{\sqrt{2x\bar{x}}}
\bigl[x\bar{x}m_V^2+k_\perp^2+M^2\bigr]F(k)F(P-k).
\]

Using the covariance replacement
\[
x\bar{x}m_V^2+k_\perp^2+M^2
\to
2(k_\perp^2+M^2)F^2(k),
\]
transforms this into the ILM-canonical form involving Bessel functions.

The longitudinal DA is defined by projecting the LF wave function onto the minimal-spin structure and integrating over $\mathbf{k}_\perp$:
\[
\phi_V(x)
=
\frac{\sqrt{2N_c}}{f_V m_V}
\int\!\frac{d^2k_\perp}{16\pi^3}
(k_\perp^2+M^2)F^2(k)\,
\phi_V^L(x,k_\perp).
\]
The transverse DA follows from replacing $\phi_V^L$ by $\phi_V^T$ after covariance restoration ensures equality.

\begin{center}
\begin{tabular}{|l|c|c|c|c|}
    \hline
    & $f_\pi$ (MeV) & $f_{\rho}$ (MeV) & $f^{T}_{\rho}$ (MeV) & $f^T_\rho/f_\rho$ \\
    \hline
    ILM~\cite{Liu:2023fpj} & $130.3$ & $203.97$ & $92.48$ & $0.453$ \\
    Lattice ($2$GeV) \cite{Braun:2016wnx} & - & 199(4)(1) & 124(4)(1) & $0.629(8)$ \\
    PDG (exp) \cite{ParticleDataGroup:2018ovx} & $130.3\pm0.3$ & $210\pm4$ & - & -\\
    \hline
\end{tabular}
\end{center}

\begin{figure}
    \centering
    \includegraphics[scale=0.55]{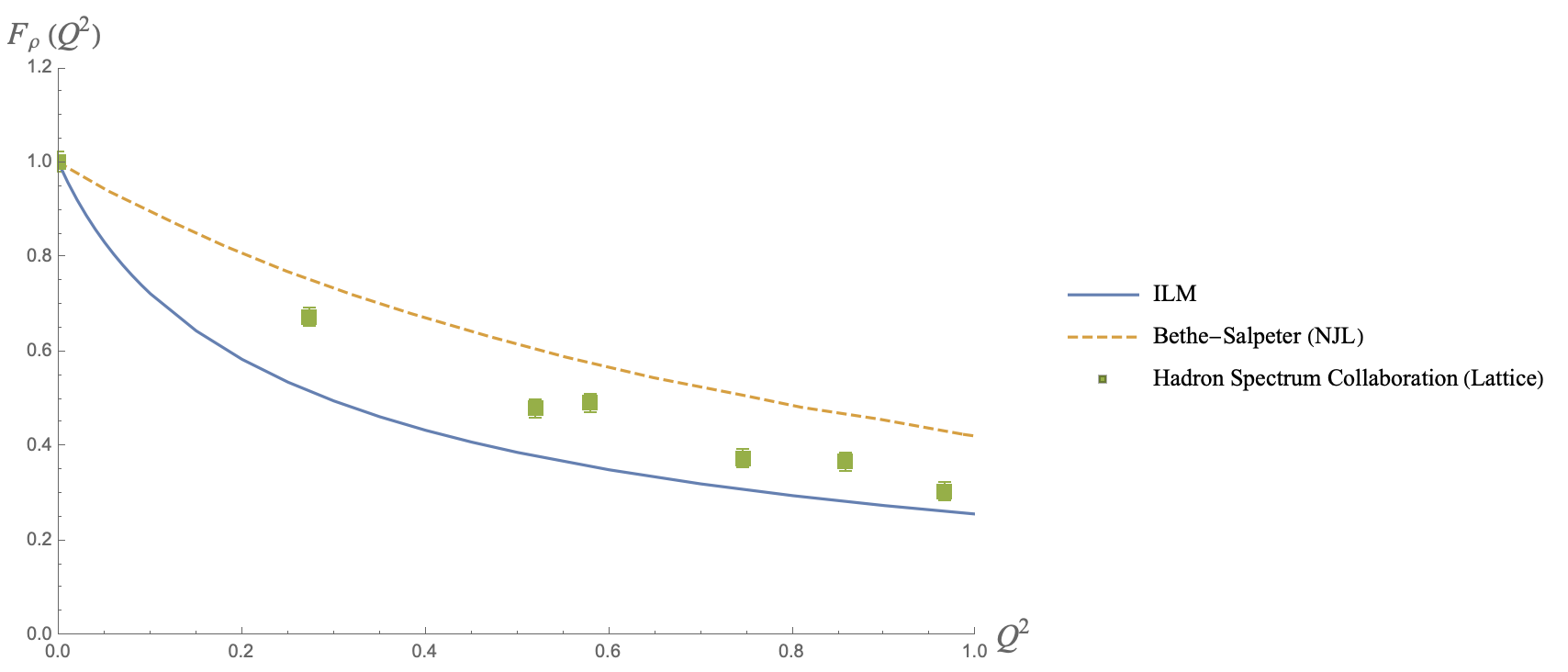}
    \caption{ILM calculations~\cite{Liu:2023fpj} are compared with the recent lattice calculation \cite{Shultz:2015pfa}, and the model analysis by Bethe-Salpeter equation using the random-phase approximation in the NJL model \cite{Carrillo-Serrano:2015uca}.}
    \label{EMRHO}
\end{figure}

\section{Electromagnetic Structure and Charge Radii}

In the Drell-Yan-West frame $q^+=0$, the vector meson electromagnetic form factor is given entirely by the valence wave function:
\[
F_V(Q^2)
=
\int_0^1 dx \int\!\frac{d^2k_\perp}{16\pi^3}
\frac{\phi^\ast(x,k_\perp+(1-x)q_\perp)\phi(x,k_\perp)}{x\bar{x}}
\,\mathcal{S}_{V}(x,k_\perp,q_\perp),
\]
where $\mathcal{S}_V$ is a spin-overlap factor calculable from Dirac algebra. Expanding for small $Q^2$ gives
\[
F_V(Q^2)=1-\frac{Q^2}{6}r_V^2+\cdots.
\]
Numerical evaluation in the ILM yields 
\[
r_V\approx 0.997\,\mathrm{fm},
\]
consistent with the expectation that vector mesons have spatial extent $\sim 2\rho$.

\begin{center}
\begin{tabular}{|l|c|}
\hline
Reference & $r_\rho$ (fm) \\
\hline
 ILM~\cite{Liu:2023fpj} & $0.997$ \\
 de Melo~\cite{deMelo:1997hh} & $0.608$ \\
 Bhagwat~\cite{Bhagwat:2006pu}  &  $0.735$ \\
   Krutov~\cite{Krutov:2016uhy}  &  $0.748$ \\
   Carrillo-Serrano\cite{Carrillo-Serrano:2015uca} & $0.819$ \\
Owen~\cite{Owen:2015gva} & $0.819$ \\
\hline
\end{tabular}    
\end{center}

\section{Mass Spectrum and Covariance Restoration}

Once the longitudinal and transverse kernels are made identical by enforcing the ILM-covariant replacement, the mass satisfies
\[
m_{V,L}^2=m_{V,T}^2 \equiv m_V^2.
\]
The solution of the eigenvalue equation produces a value close to the physical $\rho$ and $\omega$ masses when standard ILM parameters ($\rho\approx 1/3\,\mathrm{fm}$, $M\approx 350\,\mathrm{MeV}$) are used. The separation between $\rho$ and $\omega$ arises from small isospin-breaking effects not included in the leading ILM analysis.

\section{Summary}

The ILM provides a unified dynamical mechanism for constituent mass generation and vector meson binding. Instanton molecules generate a nonlocal chirality-preserving interaction that produces vector bound states. On the light front, longitudinal and transverse integral equations differ due to instantaneous interactions, but enforcing covariance conditions restores their equality. The resulting LF wave functions allow computation of decay constants, distribution amplitudes, and charge radii in a fully nonperturbative and physically transparent framework.

\begin{subappendices}

\section{Covariant Bethe-Salpeter Structure of Vector Mesons}
\label{app:BSvector}

The covariant formulation of vector mesons in the Instanton Liquid Model follows from the nonlocal effective action induced by instantons and instanton-anti-instanton molecules. The basic object is the color-singlet vector correlator,
\[
\Pi_{\mu\nu}(x-y)
=
\langle\,0\,|\,T\{\bar{\psi}(x)\gamma_\mu\psi(x)\,
\bar{\psi}(y)\gamma_\nu\psi(y)\}\,|\,0\,\rangle.
\]
In the ILM, this correlator is evaluated by expanding the quark determinant around the instanton ensemble and keeping the lowest nontrivial contributions from zero-mode physics. The dressed quark propagator has the form
\[
S(k)=\frac{\slashed{k}+M(k)}{k^2-M^2(k)+i\epsilon},
\]
where $M(k)$ is the constituent mass generated by the single-instanton sector. The nonlocal form factor $F(k)$ encodes the Fourier transform of the instanton zero-mode profile and appears at the quark-meson vertex.

The leading contribution to the vector correlator is obtained by evaluating the quark loop with two vector insertions and nonlocal dressings,
\begin{equation}
\Pi_{\alpha\beta}(P)
=
-i\!\int\!\frac{d^4k}{(2\pi)^4}\,
\mathrm{tr}
\left[
\Gamma_\alpha(\slashed{k}+M(k))\,
\Gamma_\beta(\slashed{P}-\slashed{k}+M(P-k))
\right]
\frac{F(k)F(P-k)}
{[k^2-M^2(k)][(P-k)^2-M^2(P-k)]}.
\label{eq:VecPolAppendixExpanded}
\end{equation}

The Dirac trace decomposes into two pieces,
\[
\mathrm{tr}\big[\gamma_\alpha \slashed{k}\gamma_\beta(\slashed{P}-\slashed{k})\big]
+
M(k)M(P-k)\,\mathrm{tr}[\gamma_\alpha\gamma_\beta],
\]
and after use of the identity
\[
\mathrm{tr}[\gamma_\alpha\gamma_\rho\gamma_\beta\gamma_\sigma]
=
4(g_{\alpha\rho}g_{\beta\sigma}-g_{\alpha\beta}g_{\rho\sigma}+g_{\alpha\sigma}g_{\rho\beta}),
\]
one obtains a tensor proportional to
\[
g_{\alpha\beta},
\qquad
P_\alpha P_\beta,
\]
as expected from Lorentz symmetry. The nonlocal factors $F(k)$ and $F(P-k)$ soften the ultraviolet behavior and reflect the instanton size $\rho$.

The correlator is decomposed as
\[
\Pi_{\mu\nu}(P)
=
\left(g_{\mu\nu}-\frac{P_\mu P_\nu}{P^2}\right)\Pi_T(P^2)
+
\frac{P_\mu P_\nu}{P^2}\Pi_L(P^2).
\]
Because the vector current is conserved and the ILM interactions preserve chirality in the molecular sector, the longitudinal component vanishes at the physical point,
\[
P^\mu \Pi_{\mu\nu}(P)=0,
\]
ensuring $\Pi_L(P^2)=0$ for on-shell vector mesons. The vector meson mass emerges from the zero of the Dyson-resummed propagator,
\[
D_{\mu\nu}(P)=
\frac{-i}{1-g_V\Pi_T(P^2)}
\left(g_{\mu\nu}-\frac{P_\mu P_\nu}{P^2}\right),
\]
giving the bound-state condition
\[
1-g_V\Pi_T(m_V^2)=0.
\]

Near this pole, the propagator assumes the standard form
\[
D_{\mu\nu}(P)
=
\frac{-iZ_V}{P^2-m_V^2}
\left(g_{\mu\nu}-\frac{P_\mu P_\nu}{m_V^2}\right)
+
\cdots,
\]
with residue
\[
Z_V^{-1}
=
\left.
\frac{\partial\Pi_T(P^2)}{\partial P^2}
\right|_{P^2=m_V^2}.
\]

This allows identification of the covariant Bethe-Salpeter amplitude,
\begin{equation}
\Psi_{\mu}(k;P)
=
g_{Vqq}\,
S(k)\gamma_\mu S(k-P)\sqrt{F(k)F(P-k)},
\label{eq:BSampAppendix}
\end{equation}
which enters the LSZ reduction of three-point functions and forms the starting point for LF reduction.

\section{Reduction to Light-Front Dynamics}
\label{app:LFvector}

The transformation from the covariant Bethe-Salpeter amplitude \eqref{eq:BSampAppendix} to a light-front wave function proceeds by integrating over the LF energy $k^-$ and projecting onto the dynamical spinor components. Explicitly, the LF projection is
\[
\Phi_V^\lambda(x,k_\perp)
=
iP^+
\int\!\frac{dk^-}{2\pi}\,
\bar{u}(k)\frac{\gamma^+}{2k^+}\,
\Psi_\mu(k;P)\epsilon^\mu_{(\lambda)}
\frac{\gamma^+}{2(P^+-k^+)}\,v(P-k).
\]

The propagator has poles at
\[
k^-=\frac{k_\perp^2+M^2(k)}{k^+}-i\epsilon,\qquad
(P-k)^-=\frac{(P_\perp-k_\perp)^2+M^2(P-k)}{P^+-k^+}-i\epsilon,
\]
and the contour integral projects the quark line on-shell. Because the nonlocal form factors $F(k)$ are functions of spatial momentum only in the instanton model, they do not contribute poles in $k^-$. This simplifies the residue calculation, yielding
\[
\Phi_V^\lambda(x,k_\perp)
=
\frac{C_{V,\lambda}}{\sqrt{N_c}}\,\sqrt{2x\bar{x}}\,
\phi_V^\lambda(x,k_\perp)\,
\bar{u}(k)\gamma^\mu\epsilon_\mu^{(\lambda)}v(P-k),
\]
with the scalar LF wave function $\phi_V^\lambda$ defined by the on-shell value of the Bethe-Salpeter kernel.

The elimination of $\psi_-$ in the LF Hamiltonian generates additional terms,
\[
\psi_- = \frac{1}{i\partial^+}\left[i\gamma^\perp\partial_\perp + M + \text{nonlocal ILM vertex}\right]\psi_+,
\]
leading to instantaneous four-fermion interactions. These interactions contribute to longitudinal polarization because the polarization vector satisfies $\epsilon_{(0)}^+\neq 0$, causing longitudinal spin projections to couple directly to instantaneous quark exchange. The induced interaction is governed by a tadpole term,
\[
w_+(P^+)
=
\int dx\,d^2k_\perp\,2x\,F(k)F(P-k)
-
\frac{1}{2g_S}\left(1-\frac{m}{M}\right),
\]
which mirrors the structure of the scalar ILM gap equation. This connection highlights the continuity between the nonlocal interactions that generate $M(k)$ and those that produce vector binding.

\section{Longitudinal Bound-State Equation and Covariance Restoration}
\label{app:LongitudinalAppendix}

The longitudinal LF kernel differs from the transverse one because of the presence of the instantaneous interaction. The raw equation takes the form
\[
m_V^2\,\phi_V^L(x,k_\perp)
=
\frac{k_\perp^2+M^2}{x\bar{x}}\phi_V^L
-
K_V\phi_V^L
+\Delta_{\text{LF}}(x,k_\perp),
\]
where $K_V$ is the same nonlocal kernel appearing in the transverse channel. The additional term $\Delta_{\text{LF}}$ may be written schematically as
\begin{equation}
\Delta_{\text{LF}}(x,k_\perp)
=
g_V\,\sqrt{F(k)F(P-k)}
\frac{
x\bar{x}m_V^2 + k_\perp^2 + M(k)M(P-k)
}
{x\bar{x}}\,
\Xi(x,k_\perp),
\label{eq:DeltLFExpanded}
\end{equation}
where $\Xi$ represents a smooth function associated with the overlap of spinor numerators.

This term has no analog in the covariant Bethe-Salpeter equation and must therefore be removed or reinterpreted. The comparison of LF and covariant amplitudes shows that the combination
\[
x\bar{x}m_V^2 + k_\perp^2 + M(k)M(P-k)
\]
naturally appears in LF energy denominators but should be replaced by the covariant denominator arising from quark propagators,
\[
k^2-M^2(k)
\qquad\text{and}\qquad
(P-k)^2-M^2(P-k).
\]

The ILM approximation $M(k)\approx M$ at low momentum allows one to identify
\[
k^2-M^2 = k_\perp^2 - x\bar{x}P^2,
\]
leading to the replacement
\[
x\bar{x}m_V^2 + k_\perp^2 + M(k)M(P-k)
\approx
2(k_\perp^2+M^2)F^2(k),
\]
derived by matching the structure of the covariant quark loop with the LF energy denominator.

Substituting this relation into \eqref{eq:DeltLFExpanded} eliminates $\Delta_{\mathrm{LF}}$ and renders the longitudinal equation identical to the transverse one. The equality
\[
m_{V,L}=m_{V,T}
\]
is thus a consequence of enforcing the correct covariant structure inside the LF Hamiltonian.

\section{Decay Constants and Distribution Amplitudes from Light-Front Wave Functions}
\label{app:DAvector}

The longitudinal decay constant is extracted from the matrix element
\[
\langle 0|\bar{\psi}(0)\gamma^+\psi(0)|V(P,\lambda=0)\rangle
=
f_V\,m_V\,\epsilon^+_{(0)}.
\]
Using the LF wave function yields
\[
f_V
=
-\frac{2}{\sqrt{2N_c}\,m_V}
\int dx \int\!\frac{d^2k_\perp}{(2\pi)^3}
\frac{\phi_V^L}{\sqrt{2x\bar{x}}}
\bigl[x\bar{x}m_V^2 + k_\perp^2 + M^2\bigr]
F(k)F(P-k).
\]

Applying the covariance-restoring identity gives the covariant expression involving the Bessel-profile kernel
\[
(zF_0(z))^6,
\qquad
z=k\rho,
\]
which arises from the Fourier transform of the instanton zero mode. The integration measure changes accordingly,
\[
d^2k_\perp \to \frac{dz\,z}{\rho^2},
\qquad
M\to M\,F_0(z).
\]

The longitudinal distribution amplitude follows from integration over transverse momenta of the minimal spin component of the LF wave function,
\[
\phi_V(x)
=
\int \frac{d^2k_\perp}{16\pi^3}
\phi_V^L(x,k_\perp)\,
\mathcal{P}_{\text{min}}(x,k_\perp),
\]
where $\mathcal{P}_{\text{min}}$ denotes the projector onto the leading-twist component. After substitution of the Bessel representation and elimination of spurious LF terms, one obtains the compact form given in the main text.

The transverse DA is computed similarly, using the transverse LF wave function $\phi_V^T$. At the ILM scale, both DAs are broad, reflecting the extended nonlocality of instanton zero modes.

\section{Tadpole Structures and Gap-Like Mass Equations}
\label{app:TadpoleVector}

The tadpole term $w_+(P^+)$ entering the LF Hamiltonian plays a central role in renormalizing the effective vector kernel. Its explicit form is
\[
w_+(P^+)
=
\int_0^1 dx
\int\!\frac{d^2k_\perp}{(2\pi)^3}
2x\,F(k)F(P-k)
-
\frac{1}{2g_S}\left(1-\frac{m}{M}\right).
\]
The first term originates from the instantaneous contraction of two nonlocal vector vertices in the LF Hamiltonian, while the second term derives from the ILM gap equation,
\[
M = m + g_S \int\!dk\, F^2(k),
\]
which expresses the dynamical constituent mass in terms of the scalar coupling.

Using this relation, one may rewrite the vector bound-state equation in a form analogous to the scalar gap equation. Combining tadpole and dynamical terms yields
\[
1 - \frac{g_V}{g_S}\left(1-\frac{m}{M}\right)
=
-2g_V(m_V^2-4M^2)
\int\!dx\,d^2k_\perp\,
\frac{F(k)F(P-k)}{x\bar{x}m_V^2-(k_\perp^2+M^2)}.
\]
This relation highlights the weaker binding in vector channels due to the suppressed ratio $g_V/g_S$, reflecting the molecular rather than single-instanton origin of vector interactions. In particular, the vector mass satisfies
\[
m_V > 2M,
\]
in contrast to the pion, whose mass remains anomalously small because of Goldstone dynamics.

\end{subappendices}


\chapter{Transverse-momentum-dependent parton distributions (TMDs)}

\section{Soft functions}

The transverse-momentum-dependent parton distributions (TMDs) have become indispensable tools in the theoretical description of high-energy scattering involving hadrons. They encode much more than the longitudinal momentum structure familiar from collinear parton distribution functions; they also reveal the intrinsic transverse motion of partons inside a fast-moving hadron. The dependence on the transverse separation $b_T$, defined as the Fourier conjugate to the transverse momentum $k_T$, allows TMDs to probe the three-dimensional momentum structure of hadronic matter \cite{Collins:1981uk,Collins:1981va,Collins:1984kg}. 

The TMDs depend on several scales, including the ultraviolet renormalization scale $\mu$ and the rapidity scale $\zeta$ that arises from the need to regularize lightlike Wilson lines. This rapidity scale has no analogue in collinear factorization and is intimately connected with the geometry of Wilson lines that extend near the light cone. The logarithmic sensitivity of TMDs to this scale is governed by the Collins-Soper (CS) kernel \cite{Collins:1981uk}, a central nonperturbative function encoding long-distance correlations that appear in soft-gluon exchanges. The CS kernel determines how TMDs evolve with rapidity and is a central building block in the TMD evolution equations. Although the CS kernel can be computed perturbatively for small separations $b_T$, its behavior at large $b_T$ is intrinsically nonperturbative.

Modern global analyses of TMDs require a reliable parameterization of the nonperturbative CS kernel \cite{Davies:1984sp,Landry:2002ix,Konychev:2005iy,Scimemi:2017etj,Scimemi:2019cmh,Bacchetta:2019sam,Bury:2022czx}. Several groups have proposed forms based on phenomenological considerations, but a compelling nonperturbative calculation rooted directly in QCD dynamics has been difficult to obtain. Recently, the development of large-momentum effective theory (LaMET) \cite{Ji:2013dva,Ji:2014gla,Ji:2020ect} has permitted the extraction of the CS kernel from lattice QCD through quasi-TMDs \cite{Shanahan:2020zxr,Shanahan:2021tst,Avkhadiev:2023poz}. These calculations are advancing rapidly, but the underlying physics at long distances remains challenging to interpret.

A parallel approach is to compute the nonperturbative soft functions directly in the ILM~\cite{Liu:2024sqj}. Since soft functions are Wilson loops built from Wilson lines, and since Wilson loops are deeply sensitive to long-distance gauge fields, the ILM offers a natural platform for evaluating them. The ability to compute Wilson loops in Euclidean space using instanton ensembles provides a particularly powerful way to extract soft functions that depend on an Euclidean angle $\theta$, which can then be analytically continued to the physical Minkowski rapidity $\chi=i\theta$. This procedure opens a new, fully Euclidean, first-principles route to the CS kernel. For completeness, we note the original attempt to study scattering amplitudes using this approach~\cite{Shuryak:2000df,Shuryak:2003rb}.

This chapter provides an expanded account of this construction. We begin with an overview of Wilson lines and their role in TMD factorization, including a detailed discussion of cusp divergences and the structure of soft functions. We then turn to a comprehensive description of the ILM calculation of Wilson loops with cusp angles, explaining how the instanton profile enters and how the angular dependence factorizes. Both weak-field and strong-field limits are examined, and the resulting CS kernel is analyzed in detail, including its asymptotic form at large transverse separation. Finally, we discuss the matching to perturbation theory at small $b_T$, the interpretation of the ILM at different gradient-flow cooling levels, and comparisons with lattice and phenomenological results.

\subsection{Wilson Lines, Wilson Loops, and the Geometry of Soft Functions}

Wilson lines arise naturally in QCD whenever one seeks to maintain gauge invariance in matrix elements involving colored fields. If a quark is created at point $y$ and annihilated at $x$, gauge covariance requires that the operator insert a parallel-transport factor connecting these two points along some path. The straight Wilson line connecting $x$ and $y$ is 
\begin{equation}
W[x;y]=\mathcal{P}\exp\left( ig\int_x^y dz^\mu A_\mu(z)\right),
\end{equation}
where $\mathcal{P}$ enforces ordering along the path. In TMD correlators, the path is typically chosen to be nearly lightlike, as required by the kinematics of fast-moving partons.

When two or more Wilson lines are combined to form a closed contour, one obtains a Wilson loop 
\begin{equation}
W(C)=\frac{1}{N_c}\mathrm{Tr}\,\mathcal{P}\exp\left( ig \oint_C dx^\mu A_\mu(x)\right).
\end{equation}
Wilson loops are highly sensitive to nonperturbative gauge dynamics. They probe confinement, the static quark potential~\cite{Diakonov:2009jq,Simonov:1996ati}, and the color interactions underlying hadronic scattering~\cite{Shuryak:2000df}.

A central geometric feature arises when the contour $C$ contains one or more cusps. A cusp occurs when the tangent direction of the path changes discontinuously. If two straight segments meet at an angle $\chi$, then the loop has a cusp of that angle. In Minkowski signature, the cusp angle may be infinite if the lines are exactly lightlike. Such cusps generate additional divergences beyond the usual ultraviolet divergences associated with self-energy of the lines. The resulting cusp anomalous dimension $\Gamma_{\rm cusp}(\chi,\alpha_s)$ governs the dependence of the loop on the angle~\cite{Polyakov:1980ca,Dotsenko:1979wb,Brandt:1982gz,Korchemskaya:1992je}.

Soft functions in TMD factorization are Wilson loops whose geometry is determined by the directions of initial-state and final-state partons~\cite{Collins:2011zzd}. For instance, in the Drell-Yan process, two quarks approach one another from opposite directions along the light cone, producing soft gluon exchange encoded in Wilson lines pointing to the past. Semi-inclusive deep-inelastic scattering (SIDIS) involves a mixture of past- and future-oriented lines, reflecting the different time orderings of the soft exchanges~\cite{Vladimirov:2014hla}. In all cases, the transverse separation $b_T$ between the lines captures the partonic impact parameter dependence.

\begin{figure}
    \centering
    \includegraphics[width=0.3\linewidth]{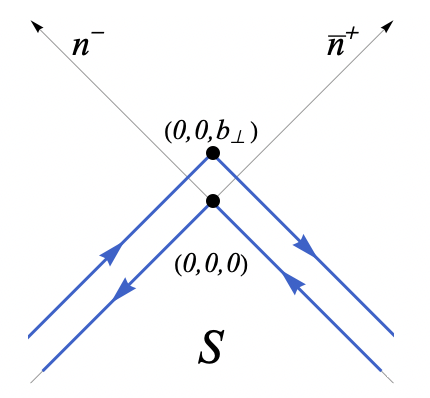} 
\includegraphics[width=0.3\linewidth]{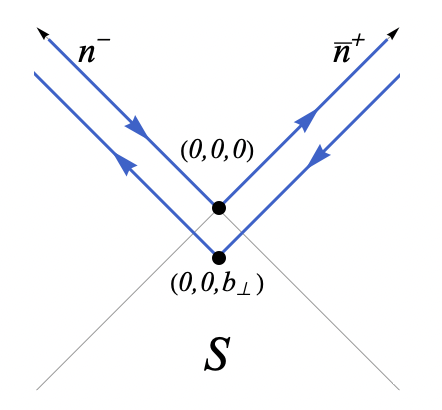}
\includegraphics[width=0.3\linewidth]{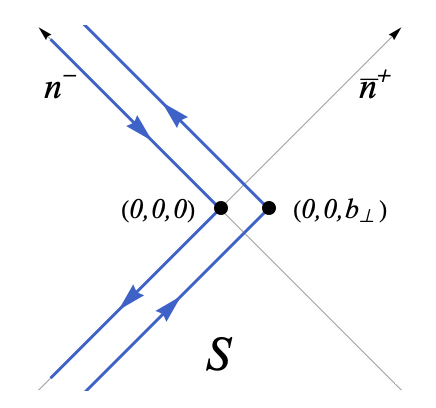}
    \caption{The Contours of the Wilson lines for the soft function of the cross section in (a) Drell-Yan,  (b) electron-positron annihilation,
    and (c) SIDIS. Hadron $A$ is moving near the light-cone direction $n$,
    and hadron $B$ is moving near the  light-cone direction $\bar{n}$ \protect\cite{Vladimirov:2014hla}.}
    \label{fig:wl}
\end{figure}

An important aspect is that soft functions contain both ultraviolet divergences and rapidity divergences. The former are renormalized in the usual manner, while the latter arise because lightlike Wilson lines extend to infinite rapidity. To regularize these divergences, one can tilt the Wilson lines slightly off the light cone by giving them large but finite rapidity parameters. In Euclidean space, these become finite real angles $\theta$ between the Wilson lines. This is the key to the analytic continuation used throughout the ILM analysis.

\subsection{Analytic Structure of Cusp Singularities and Rapidity Evolution}

Renormalization of soft functions leads to a multiplicative structure in which the dependence on the cusp angle is entirely encoded in a function $K(\mu,b_T,\chi)$ plus an angle-independent term $P(\mu,b_T)$~\cite{Vladimirov:2017ksc}
\begin{equation}
S(\mu,b_T,\chi)=\exp\left[K(\mu,b_T,\chi)+P(\mu,b_T)\right].
\end{equation}
The evolution with respect to $\mu$ follows the renormalization-group equation 
\begin{equation}
\frac{dK}{d\ln\mu^2}=-\gamma_K(\chi,\alpha_s),
\end{equation}
where $\gamma_K$ is the cusp anomalous dimension.

There are two limiting regimes of particular importance. The first is the small-angle limit $\chi\to0$, which corresponds physically to nearly parallel Wilson lines. In this limit the kernel behaves quadratically in the angle~\cite{Korchemsky:1991zp,Mitev:2015oty}
\begin{equation}
K(\chi\to0)\sim K_B(b_T)\chi^2,
\end{equation}
and the corresponding cusp anomalous dimension reduces to the bremsstrahlung function. 

The second regime is the large-angle limit $\chi\to\infty$, relevant for TMD factorization where the Wilson lines are effectively lightlike. In this limit one finds 
\begin{equation}
K(\chi\to\infty)\sim K_{\rm CS}(b_T)\chi,
\end{equation}
which defines the CS kernel $K_{\rm CS}(b_T)$~\cite{Collins:1981uk,Collins:1981va}. This kernel is nonperturbative at large $b_T$ but becomes perturbatively calculable when $b_T$ is small. It plays a central role in controlling the rapidity evolution of TMDs through the Collins-Soper equation.

The ILM analysis hinges on computing $K(\theta)$ for real Euclidean angles $\theta$ and analytically continuing to $K(i\chi)$ for Minkowski rapidity $\chi$~\cite{Liu:2024sqj}. A key simplification arises because the angle dependence factorizes in both the weak and strong field instanton regimes, allowing the extraction of the CS kernel from the limit of large imaginary angle.

\subsection{Soft Functions in the Instanton Liquid Model}

The ILM describes the QCD vacuum as a statistical ensemble of instantons and anti-instantons, each represented by a classical gauge configuration localized in space-time. The instanton field is characterized by its size $\rho$, center $z$, and color orientation, though for the soft-function calculation the hedgehog structure ensures that path ordering is not needed. The contribution of a single instanton to a Wilson loop is~\cite{Shuryak:2000df,Liu:2024sqj}
\begin{equation}
W_I(\rho,z)=\exp\left(i\tau^a\phi_I^a(\rho,z)\right),
\end{equation}
where $\phi_I^a$ encodes the projection of the instanton field onto the Wilson line. The explicit form is 
\begin{equation}
\phi_I^a(\rho,z)=\int_C dx^\mu \,\bar\eta^a_{\mu\nu}\frac{(x-z)^\nu\rho^2}{(x-z)^2[(x-z)^2+\rho^2]},
\end{equation}
where $\bar\eta^a_{\mu\nu}$ is a 't~Hooft symbol.
Since instantons form a dilute ensemble, the contributions exponentiate,
\begin{equation}
W(C)=\exp\left[ \frac{n_{I+A}}{N_c}\int d^4z \left(\mathrm{Tr}_c\, W_I(\rho,z)-1\right)\right].
\end{equation}
This structure holds for any contour, including those relevant for soft functions.

\begin{figure}
    \centering
    \includegraphics[width=0.6\linewidth]{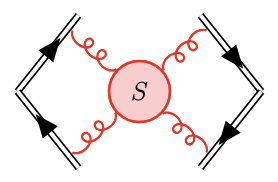}
    \caption{Soft function attached to the pertinent Wilson lines in Fig.~\ref{fig:wl}.}
    \label{fig:soft_wilson}
\end{figure}

The geometry of the contour for the soft function involves four half-infinite Wilson lines that form two cusps separated by a transverse displacement $b_T$. Each cusp is characterized by an angle $\theta$ in Euclidean space. The evaluation of the instanton contribution requires integrating over the instanton center position $z$ and using explicit parametrizations for the instanton profile along each of the four half-infinite lines.
After subtracting the angle-independent self-energy, the resulting expression defines the cusp kernel~\cite{Liu:2024sqj} 
\begin{equation}
K(\rho,b_T,\theta)=\ln\frac{W(b_T,\theta)}{W(b_T,0)},
\end{equation}
which vanishes for $\theta\to0$.

\subsection{Weak-Field Limit and Factorization of Angular Dependence}

In the weak-field regime where the instanton profile is small, the expansion of the loop exponent yields
\begin{equation}
\label{eq:weak-theta}
\ln W(\rho,b_T,\theta)\approx -\frac{n_{I+A}}{N_c}\int d^4z\,\phi^2(z,\rho,b_T,\theta).
\end{equation}
The instanton squared profile function at zero angle reads
\begin{equation}
\begin{aligned}
\phi^2(z,\rho,b_\perp,\theta=0)=&-2\pi^2\frac{z_3^2+z_\perp^2-\frac14b^2_\perp}{\sqrt{z_3^2+\left(z_\perp+\frac12b_\perp\right)^2}\sqrt{z_3^2+\left(z-\frac12b_\perp\right)^2}}\\
        &\times\left(1-\frac{\sqrt{z_3^2+\left(z_\perp+\frac12b_\perp\right)^2}}{\sqrt{z_3^2+\left(z_\perp+\frac12b_\perp\right)^2+\rho^2}}\right)\left(1-\frac{\sqrt{z_3^2+\left(z_\perp-\frac12b_\perp\right)^2}}{\sqrt{z_3^2+\left(z_\perp-\frac12b_\perp\right)^2+\rho^2}}\right)\\
        &+\pi^2\left(1-\frac{\sqrt{z_3^2+(z_\perp+\frac12b_\perp)^2}}{\sqrt{z^2_3+(z_\perp+\frac12b_\perp)^2+\rho^2}}\right)^2+\pi^2\left(1-\frac{\sqrt{z_3^2+(z_\perp-\frac12b_\perp)^2}}{\sqrt{z^2_3+(z_\perp-\frac12b_\perp)^2+\rho^2}}\right)^2\,.
\end{aligned}
\end{equation}
The UV divergence as $z\rightarrow0$ is regulated by the instanton size $\rho$, and the IR divergence $z\rightarrow\infty$ cancels out between the two-anti-parallel Wilson lines. The absence of the $z_4$ dependence is due to rotational symmetry in Euclidean space.

At finite angle, the expression is manageable but more involved. 
The integration in \eqref{eq:weak-theta} can be reorganized using a change of variables that separates the angular dependence from the remaining geometry. The result is~\cite{Liu:2024sqj} 
\begin{equation}
K^{(1)}(\rho,b_T,\theta)=K^{(1)}_{\rm CS}(b_T/\rho)\,h(\theta),
\end{equation}
where 
\begin{equation}
h(\theta)=\theta\cot\theta-1,
\end{equation}
and $K^{(1)}_{\rm CS}$ is a purely radial integral independent of the angle. This factorization is remarkable because it greatly simplifies the analytic continuation and because numerical analysis shows that it persists even in the strong-field regime.
The explicit expression for the weak-field kernel is 
\begin{equation}
\label{eq:weak-CS}
K^{(1)}_{\rm CS}(b_T)=
\frac{4\pi^2 n_{I+A}\rho^4}{N_c}
\int_0^\infty dk \frac{F_g^2(\rho k)}{k}\left(J_0(k b_T)-1\right),
\end{equation}
where $F_g$ is the instanton form factor
\begin{equation}
    \mathcal{F}_g(k)=\frac4{k^2}-2K_2(k)\,.
\end{equation}
For $b_T\gg\rho$, the oscillatory behavior of the Bessel function ensures that the integral yields a logarithmic dependence,
\begin{equation}
K^{(1)}_{\rm CS}(b_T)\sim
-\frac{4\pi^2 n_{I+A}\rho^4}{N_c}\left[\ln\left(\frac{b_T}{\rho}\right)+1\right].
\end{equation}
This logarithmic growth is in excellent agreement with phenomenological models that have long assumed a similar behavior, highlighting the physical relevance of the ILM.

\begin{figure*}
    \centering
\includegraphics[width=.5\linewidth]{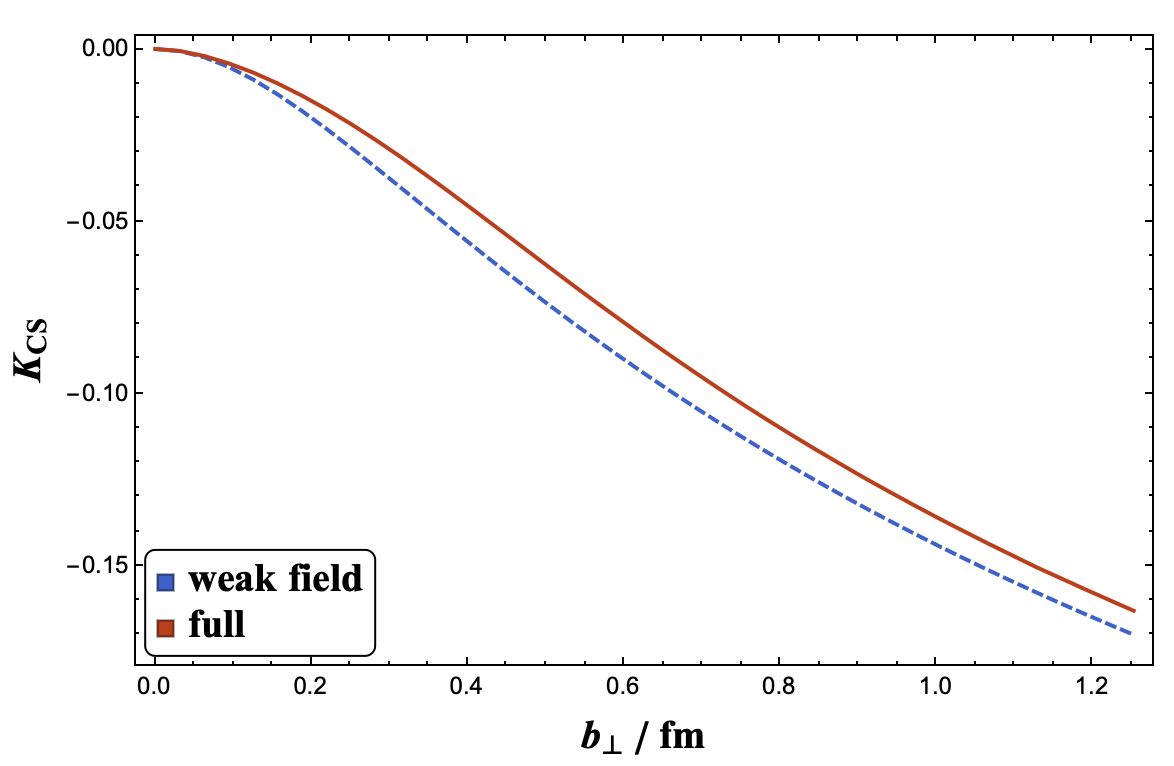}
\includegraphics[width=.46\linewidth]{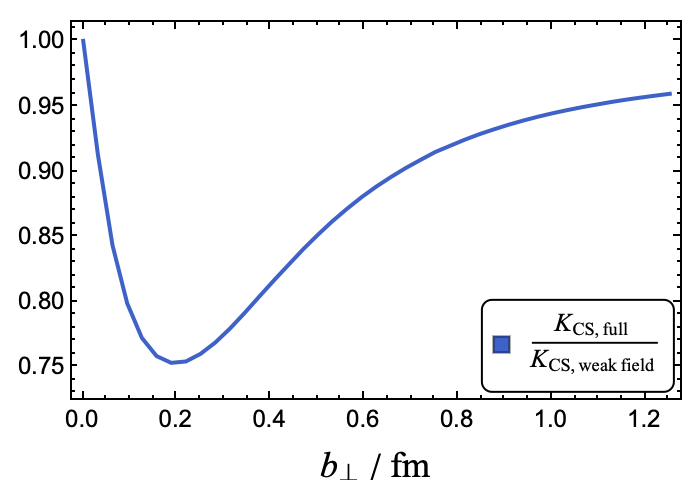}
    \caption{(a) Instanton liquid estimation on CS kernel with the full calculation~\eqref{eq:strong-CS}, and the weak field approximation \eqref{eq:weak-CS}. We also plot their ratio in (b).}
    \label{fig:KCSX}
\end{figure*}

\subsection{Strong Field Evaluation and the Full Instanton Contribution}

Beyond the weak-field approximation, the instanton profile must be treated exactly. Numerical integration over $z$ shows that the factorization of angular dependence remains extremely accurate even when $\phi$ is not small. Thus the full kernel can still be written in the form
\begin{equation}
\label{eq:strong-CS}
K(\rho,b_T,\theta)\approx K_{\rm CS}(b_T/\rho)\,h(\theta),
\end{equation}
where $K_{\rm CS}$ is the fully nonperturbative ILM result~\cite{Liu:2024sqj}.

The ILM prediction for $K_{\rm CS}$ has several noteworthy properties. At small separations $b_T\ll\rho$, the instanton field appears nearly uniform over the region probed by the Wilson lines, leading to a suppressed contribution. At larger separations $b_T\sim\rho$, the sensitivity to the instanton core becomes pronounced, and the kernel grows in magnitude. At asymptotically large $b_T$, the kernel approaches the logarithmic form predicted by the weak-field analysis. These features match smoothly onto perturbative expectations at small $b_T$ and phenomenological behavior at large $b_T$.

\begin{figure*}
    \centering
\includegraphics[width=.5\linewidth]{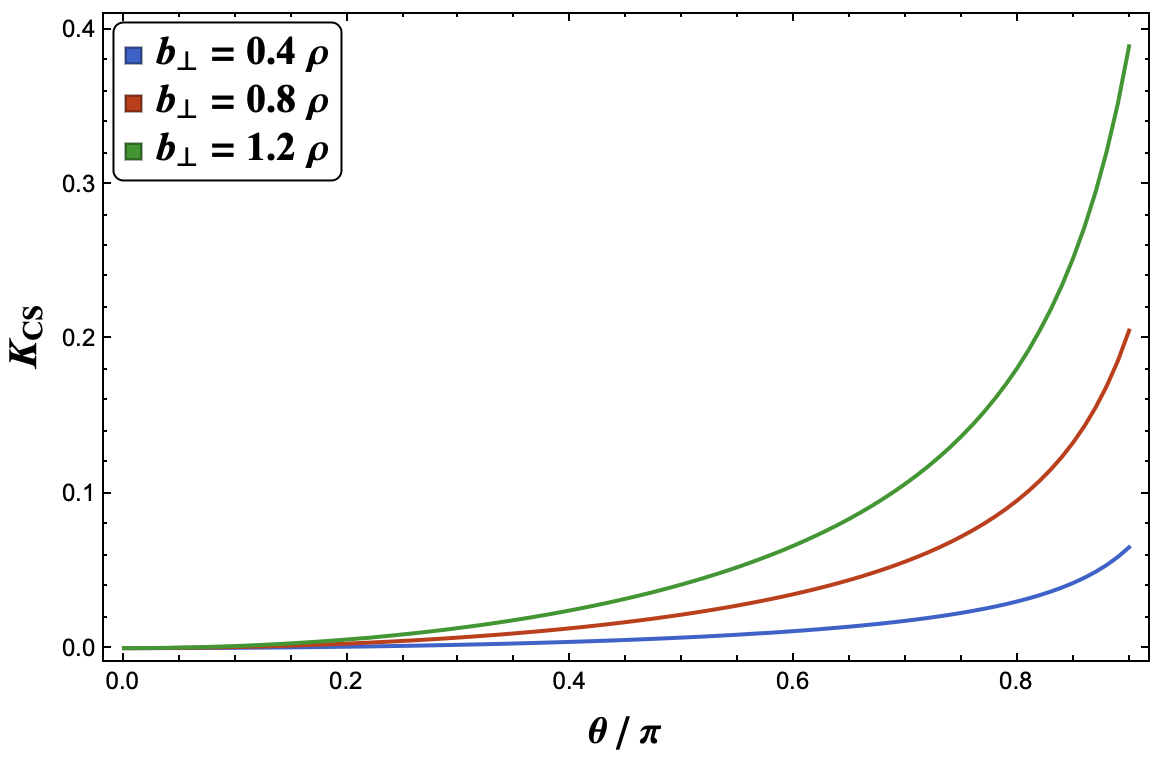}
\includegraphics[width=.5\linewidth]{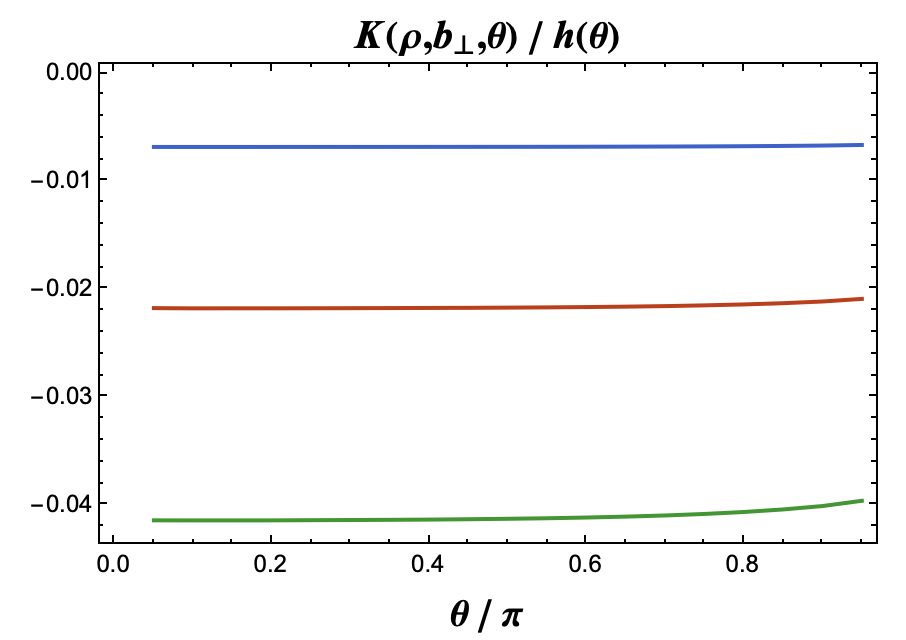}
    \caption{The angular dependence of $K$ in \eqref{eq:strong-CS}. }
    \label{fig:theta}
\end{figure*}

\subsection{Collins-Soper Evolution and Matching to Perturbation Theory}

The ILM describes the nonperturbative regime of QCD corresponding to low renormalization scale $\mu$. In gradient flow studies, the flow time $t$ serves as an inverse resolution scale $\mu\sim 1/\sqrt{8t}$. At long flow times, most instanton-anti-instanton pairs dissociate, leaving a dilute ensemble with density $n\approx 1\,\mathrm{fm}^{-4}$. At shorter flow times corresponding to higher resolution $\mu\sim2\,\mathrm{GeV}$, the effective instanton density increases substantially, consistent with the picture of instanton molecules or tightly bound instanton-antiinstanton pairs.

To construct a full CS kernel valid across all $b_T$, one must combine the ILM result with perturbation theory. The matching uses the standard $b^\ast$ prescription of TMD factorization. The perturbative kernel satisfies an RG equation involving $\Gamma_{\rm cusp}$, which is now known up to four loop order. The complete expression becomes
\begin{equation}
\label{eq:max-CS}
K_{\rm CS}(b_T,\mu)=
K_{\rm ILM}(b_T,\mu)
+K_{\rm CS}^{\rm pert}(b^\ast,\mu_b)
-2\int_{\mu_b}^{\mu}\frac{d\mu'}{\mu'}\,\Gamma_{\rm cusp}(\alpha_s(\mu')),
\end{equation}
where $b^\ast$ and $\mu_b$ are defined in terms of a parameter $b_{\max}$ chosen to optimize the transition region
\begin{equation}
    \mu_b=\frac{2e^{-\gamma_E}}{b^*(b_\perp)}\xrightarrow{b_\perp\rightarrow\infty}\frac{2e^{-\gamma_E}}{b_{\rm max}}
\end{equation}
with the asymptotic saturation following from the conventional $b^*$-parameterization~\cite{Collins:2011zzd},
\begin{equation}
\label{b_star}
    b^*(b_\perp)=\frac{b_\perp}{\sqrt{1+b_\perp^2/b_{\rm max}^2}}\,,
\end{equation}
which reduces to $b_\perp$ in the limit $b_\perp\to0$.
Here $\gamma_E$ is  Euler constant, and the optimal non-perturbative distance is  $b_{\rm max}=0.56$ fm, which is chosen to optimize the interpolation between non-perturbative and perturbative contributions in the lattice results of ASWZ24~\cite{Avkhadiev:2023poz}.
The resulting CS kernel agrees well with lattice calculations and modern phenomenological fits.

\begin{figure*}
    \centering
\includegraphics[width=0.5\linewidth]{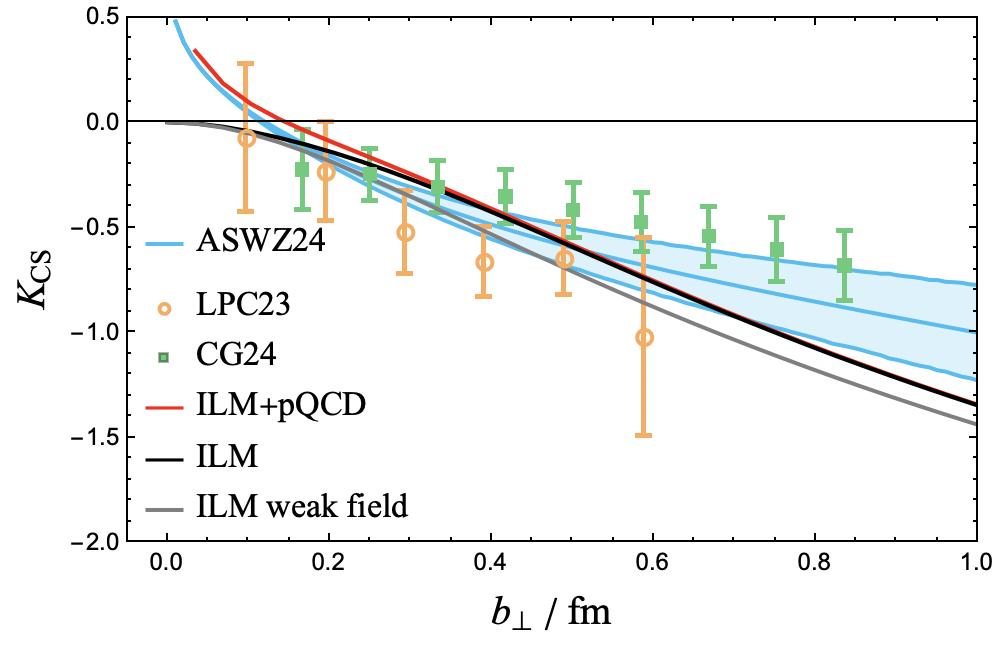}
\includegraphics[width=0.5\linewidth]{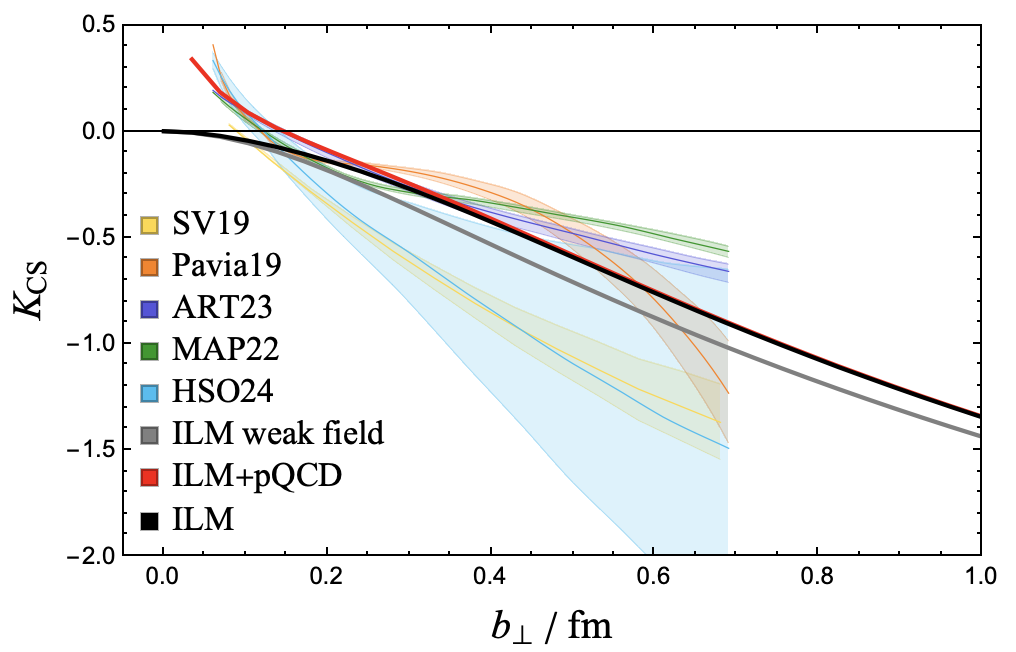}
    \caption{(a) The black-solid curve is the  ILM result for the CS kernel with $\rho=0.343$ fm, $n_{I+A}=7.46$ fm$^{-4}$~\cite{Liu:2024sqj}, compared to  the weak field approximation gray-solid curve and the optimized combination 
    with the short-distance perturbative contribution red-solid curve . The results are compared to the recent lattice calculation by ASWZ24~\cite{Avkhadiev:2023poz},  a new lattice calculation using Coulomb-gauge TMD correlators~\cite{Zhao:2023ptv} by CG24~\cite{Bollweg:2024zet} and results from Lattice Parton Collaboration LPC23~\cite{LatticePartonLPC:2023pdv}. (b) shows the comparison of our results with the phenomenologically  extracted CS kernel  by SV19~\cite{Scimemi:2019cmh}, Pavia19~\cite{Bacchetta:2019sam}, MAP22~\cite{Bacchetta:2022awv}, ART23~\cite{Moos:2023yfa}, and HSO24~\cite{Gonzalez-Hernandez:2022ifv}.}
    \label{fig:CS}
\end{figure*}

\subsection{Conclusions}

The instanton liquid model provides a natural and physically transparent framework for computing the nonperturbative structure of soft functions and their rapidity evolution. Its success relies on the strong support from gradient-flow lattice studies, which confirm that instantons dominate the long-distance, low-resolution sector of QCD. The ILM captures the essential topological fluctuations of the vacuum and thus provides the correct gauge-field background in which soft Wilson loops can be evaluated.

The resulting CS kernel displays a smooth transition from perturbative to nonperturbative behavior, matching onto perturbative results at small transverse separation and reproducing a logarithmic growth at large separation. The angular dependence factorizes in a universal manner across weak- and strong-field regimes, enabling a straightforward analytic continuation from Euclidean to Minkowski space and offering a new prescription for computing soft functions on the lattice. Comparisons with lattice calculations and phenomenological extractions show very good agreement, highlighting the utility of the ILM in the study of TMD evolution.

The combination of analytical control, numerical tractability, and physical grounding makes the ILM an attractive tool for continued exploration of nonperturbative soft physics in QCD. Its application to the CS kernel sets the stage for future studies of TMD observables in Euclidean frameworks, including potential extensions to generalized TMDs, jet soft functions, and small-$x$ dipole amplitudes.

\section{TMDs of Pion and Kaon}

The transverse-momentum-dependent parton distribution functions (TMDPDFs) of the pion and kaon encode the multidimensional structure of these Nambu-Goldstone bosons in QCD. Whereas standard collinear parton distribution functions (PDFs) characterize the distribution of longitudinal momentum at a fixed renormalization scale, TMDPDFs resolve the simultaneous dependence on the light-cone momentum fraction $x$ and the intrinsic transverse momentum $\vec k_\perp$ (or equivalently the transverse position $\vec b_\perp$), thus providing a genuine three-dimensional tomography of hadronic structure~\cite{Boussarie:2023izj}. They play a central role in the description of low-$q_T$ Drell-Yan (DY) production, semi-inclusive deep inelastic scattering (SIDIS), and, prospectively, electron-ion collider (EIC) measurements.

The theoretical description of TMDs is complicated by the appearance of light-like Wilson lines, which are essential for gauge invariance but introduce rapidity divergences that do not appear in collinear factorization. A consistent factorization and renormalization scheme therefore requires soft subtraction and controlled evolution in both the renormalization scale~$\mu$ and the rapidity scale~$\zeta$~\cite{Collins:2011zzd,Rogers:2015sqa}. The appropriate framework for these tasks is the Collins-Soper-Sterman (CSS) formalism~\cite{CollinsSoper:1981,CSS:1985}.

At low resolution, where perturbative methods become unreliable, a first-principles description of nonperturbative dynamics becomes essential. In this domain, the QCD instanton liquid model (ILM)~\cite{Shuryak:1981ff,Schafer:1996wv,Nowak:1996aj} provides a concrete semiclassical picture of the QCD vacuum as an ensemble of instantons and anti-instantons with characteristic size $\rho\simeq 1/3\,$fm. The ILM naturally generates a momentum-dependent dynamical mass for quarks, nonlocal effective interactions, and nontrivial gauge-field configurations that directly influence Wilson-line structures. This framework has been successfully applied to pion and kaon structure~\cite{Liu:2023feu,Liu:2023fpj}, and in particular the complete TMD analysis~\cite{Liu:2024sqj} which we now review.

\subsection{Definition of Meson TMDPDFs}

The starting point for the TMD description is the gauge-invariant quark correlation function at fixed light-cone momentum fraction $x$ and transverse momentum $\vec k_\perp$. For a pion with momentum $p^\mu$, the leading-twist TMDPDFs are defined from matrix elements of quark bilinears connected by staple-shaped Wilson lines, whose direction distinguishes SIDIS from DY~\cite{Collins:2011zzd}. Explicitly,
\begin{align}
q_\pi(x,k_\perp) &= \int\!\frac{dz^-}{2\pi}\!\int\!\frac{d^2 b_\perp}{(2\pi)^2}
\,e^{ixp^+ z^- - i \vec k_\perp\!\cdot\! \vec b_\perp}
\langle \pi | \bar\psi(0)\gamma^+ W^{(\pm)}[0;z^-,b_\perp]\psi(z^-,b_\perp) | \pi \rangle, \\
\Delta q_\pi(x,k_\perp) &= \int\!\frac{dz^-}{2\pi}\!\int\!\frac{d^2 b_\perp}{(2\pi)^2}
\,e^{ixp^+ z^- - i \vec k_\perp\!\cdot\! \vec b_\perp}
\langle \pi | \bar\psi(0)\gamma^+\gamma_5 W^{(\pm)}\psi(z^-,b_\perp) | \pi \rangle,\\
\delta q_\pi(x,k_\perp) &= \int\!\frac{dz^-}{2\pi}\!\int\!\frac{d^2 b_\perp}{(2\pi)^2}
\,e^{ixp^+ z^- - i \vec k_\perp\!\cdot\! \vec b_\perp}
\langle \pi | \bar\psi(0)i\sigma^{\alpha+}\gamma_{5} W^{(\pm)}\psi(z^-,b_\perp) | \pi \rangle.
\end{align}
for the unpolarized, spin and transversity respectively. 
The Wilson line staples run space-like for SIDIS, and time-like for DY~\cite{Kumano:2020ijt}, as illustrated in Fig.~\ref{fig:wl}. 
The resummation of the collinear gluons from the final state in the SIDIS process, yields $W^{(+)}$. Similarly,  the resummation of the collinear gluons 
from the initial state in the Drell-Yan process, yields $W^{(-)}$~\cite{Angeles-Martinez:2015sea,GrossePerdekamp:2015xdx,Aidala:2012mv,Barone:2010zz,DAlesio:2007bjf}.

The transverse coordinate dependence can be obtained by using the Fourier transform
\bea
\tilde{q}_{\pi}(x,b_\perp)=\int\frac{d^2k_\perp}{(2\pi)^2}e^{ik_\perp\cdot b_\perp} q_{\pi}(x,k_\perp)
\eea
The pion is on-shell with energy $E_{\vec{p}}=\sqrt{m_\pi^2+\vec{p}^2}$ and $3$-momentum $\vec{p}$,
\begin{equation}
\begin{aligned}
    p^\mu=&p^+\bar n^\mu-\frac {m^2_\pi}{2p^+} n^\mu+p^\mu_\perp
\end{aligned}
\end{equation}
with hadron mass $m_\pi$. Here $n=(1,0,0_\perp)$ and $\bar n=(0,1,0_\perp)$ are the longitudinal and transverse  light cone vectors. Similar definitions hold for other hadrons.
For a spin-zero target such as the pion or kaon, only two leading-twist TMDs survive
\begin{align}
f_1^{q/\pi}(x,k_\perp), \qquad 
h_{1}^{\perp\,q/\pi}(x,k_\perp).
\end{align}
The latter, the Boer-Mulders function~\cite{Boer:1997nt}, describes a correlation between quark transverse momentum and quark transverse spin inside an unpolarized hadron, and is thus naively $T$-odd. As shown in the ILM construction, $h_1^\perp$ vanishes at lowest Fock order~\cite{Liu:2025mbl} and is generated only through Wilson-line effects or higher Fock-state components.

\subsection{Rapidity Divergences and Instanton-Induced Soft Separation}

The Wilson lines $W^{(\pm)}$ run along nearly light-like directions and engender \emph{rapidity divergences} that cannot be renormalized by ordinary ultraviolet counterterms. The modern treatment of these divergences introduces an independent rapidity scale $\zeta$, soft subtraction factors, and a renormalized TMD~\cite{Collins:2011zzd}. 

Within the ILM, the nonperturbative gauge-field configurations of instantons induce a distinctive structure for Wilson-line correlators. A remarkable consequence is the "ILM soft separation,'' whereby the rapidity dependence factorizes from the constituent-quark distribution~\cite{Liu:2025mbl}
\begin{equation}
\tilde f_{1}^{q/\pi}(x,b_\perp;\mu\!\sim\!1/\rho,y_q-y_n)
\approx
\tilde f^{q/\pi}_1(x,b_\perp)\,
\exp\!\left[ K_{\mathrm{CS}}^{(\mathrm{inst})}(b_\perp/\rho)\,(y_q-y_n)\right].
\end{equation}
Here $K_{\mathrm{CS}}^{(\mathrm{inst})}$ is the instanton-induced contribution to the Collins-Soper kernel, encoding the cusp behavior of Wilson lines in the semiclassical vacuum. Because the instanton ensemble is dilute, the nonperturbative rapidity dependence is dominated by single-instanton physics.

\begin{figure*}
    \centering
    \includegraphics[width=1\linewidth]{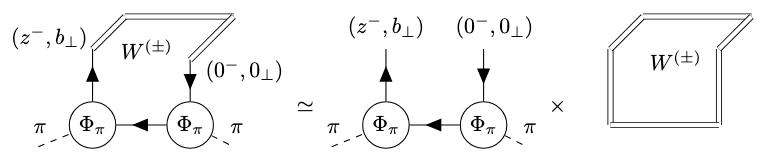}
    \caption{Soft separation for the quark pion TMD, which is approximated into a constituent quark distribution without rapidity dependence and a rapidity dependent factor generated by the stapled Wilson line~\cite{Liu:2025mbl}.}
    \label{fig:tmd_sub}
\end{figure*}

After soft subtraction in the Collins scheme~\cite{Collins:2011zzd}, the renormalized TMD at the initial ILM scale is~\cite{Liu:2024sqj}
\begin{equation}
\tilde F_{1}^{q/\pi}(x,b_\perp;\mu_0,\zeta_0)
=
\tilde f_{1}^{q/\pi}(x,b_\perp)
\exp\!\left[
K^{(\mathrm{inst})}_{\rm CS}\!\left(\frac{b_\perp}{\rho}\right)
\ln\!\left(\frac{\zeta_0}{1/\rho^{2}}\right)
\right].
\end{equation}

This expression is a crucial bridge between the nonperturbative ILM description and the perturbative CSS evolution that connects to experimental scales.

\subsection{Constituent-Quark TMDPDFs from ILM Light-Front Wave Functions}

In the ILM, the pion and kaon are treated as bound $q\bar q$ states formed through instanton-induced nonlocal interactions. This produces nonlocal light-front wave functions with momentum-dependent constituent quark masses. The pion LFWF in momentum space is~\cite{Liu:2023feu}
\begin{equation}
\Phi_\pi(x,k_\perp,s_1,s_2)
=
\frac{1}{\sqrt{N_c}}
\left[
C_\pi\sqrt{x\bar x}\left(
m_\pi^2 - \frac{k_\perp^2+M^2}{x\bar x}
\right)
F\!\left(\frac{k_\perp}{\lambda_\pi\sqrt{x\bar x}}\right)
\right]
\bar u_{s_1}(k_1)i\gamma_5 \tau^a v_{s_2}(k_2),
\end{equation}
with the ILM nonlocal form factor defined earlier.
After performing the spin contractions of the LFWF overlap, one finds the unpolarized constituent-quark TMD of the pion at $\mu_0=1/\rho$,
\begin{equation}
\label{eq:pion_tmd}
f_{1}^{q/\pi}(x,k_\perp)
=
\frac{C_\pi^2}{(2\pi)^3}\,
\frac{2(k_\perp^2+M^2)}{\left(x\bar x m_\pi^2 - k_\perp^2 - M^2\right)^2}
F^2\!\left(\frac{k_\perp}{\lambda_\pi\sqrt{x\bar x}}\right).
\end{equation}
This distribution is valence-dominated, symmetric under $x\leftrightarrow \bar x$, and sharply peaked around $x\simeq 0.5$, reflecting the dominance of lowest Fock components at low resolution.

For the kaon, $SU(3)$ flavor symmetry breaking leads to distinct constituent masses $M_u$ and $M_s$~\cite{Liu:2025mbl}, producing an asymmetric LFWF
\begin{equation}
\label{eq:kaon_tmd}
f_{1}^{u/K^+}(x,k_\perp)
=
\frac{C_K^2}{(2\pi)^3}
\frac{2\!\left(k_\perp^2+\bar x^{2}M_u^{2}
+x^{2}M_s^{2}
+2x\bar x M_uM_s\right)}{\left(x\bar x m_K^2-k_\perp^2-x M_s^2-\bar x M_u^2\right)^{2}}
F^2\!\left(\frac{k_\perp}{\lambda_K\sqrt{x\bar x}}\right).
\end{equation}
The heavier strange quark biases the kaon distribution toward smaller $x$, in agreement with phenomenology and lattice-QCD results.

\begin{figure*}
\centering
%
%
\includegraphics[width=0.325\linewidth]{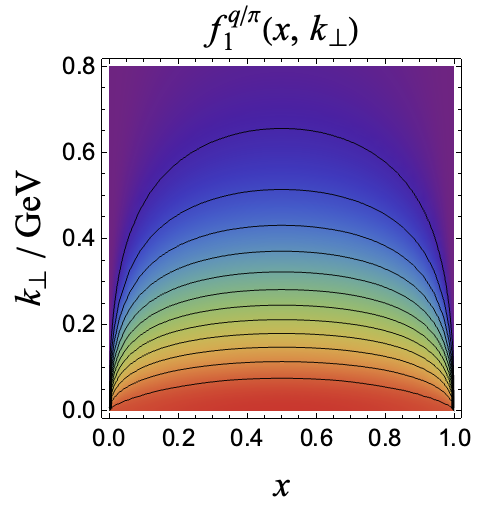}
 \includegraphics[width=0.325\linewidth]{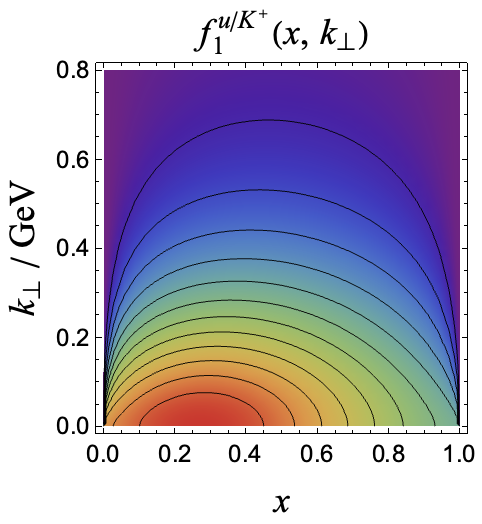}
\includegraphics[width=0.355\linewidth]{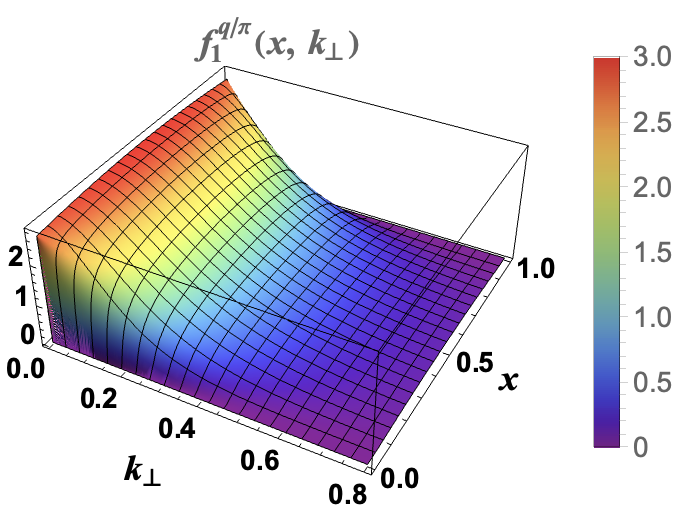}
\includegraphics[width=0.355\linewidth]{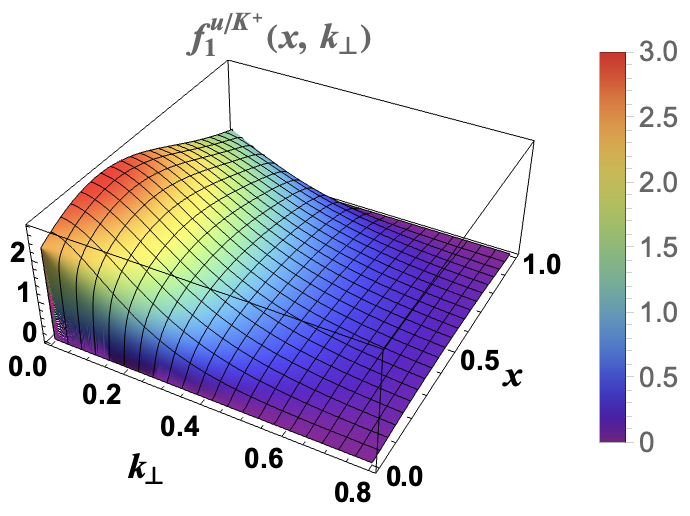}
\includegraphics[width=0.38\linewidth]{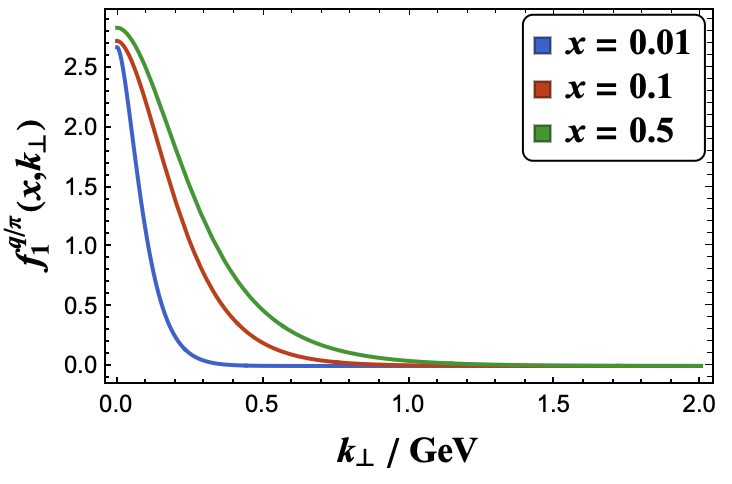}
\includegraphics[width=0.38\linewidth]{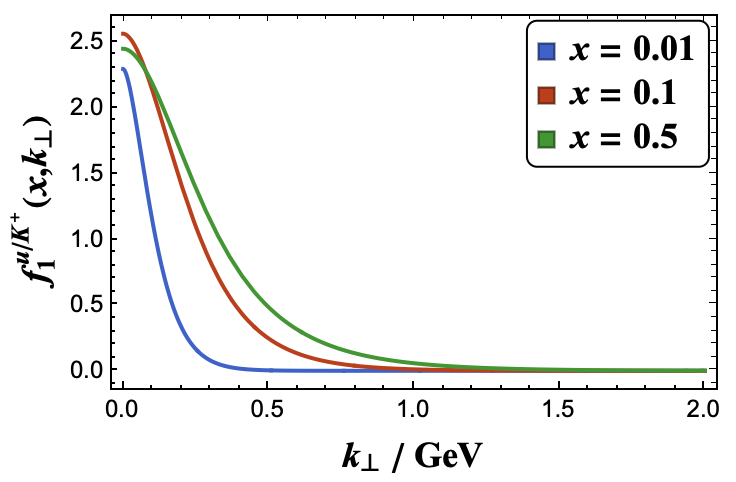}
\includegraphics[width=0.38\linewidth]{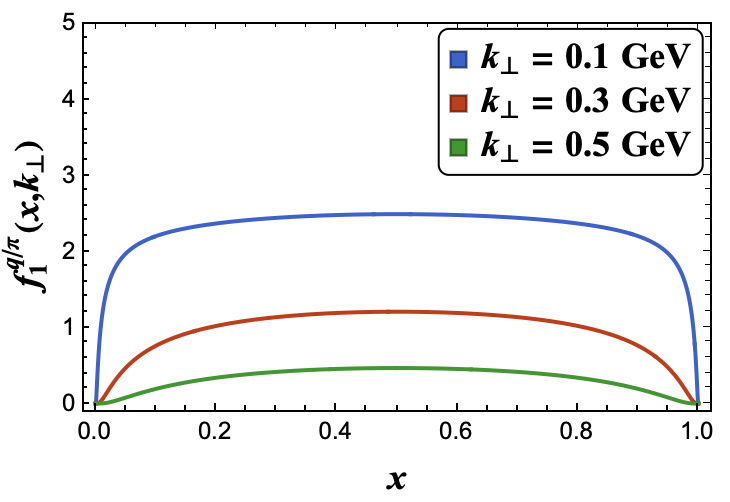}
\includegraphics[width=0.38\linewidth]{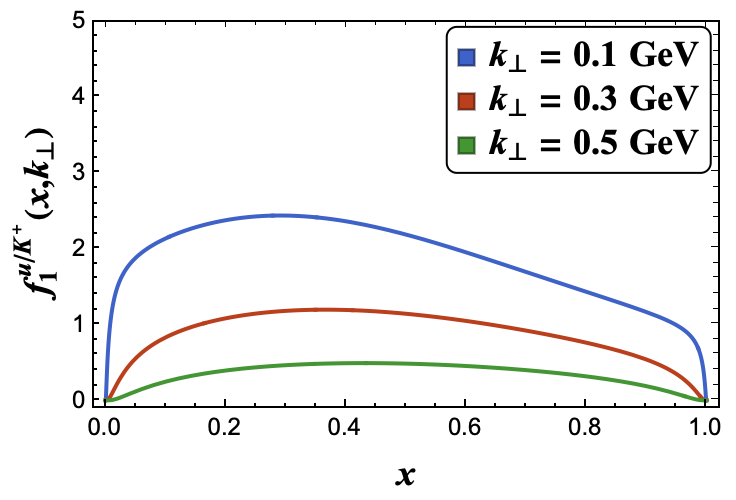}
\caption{Constituent quark TMD distribution in pion (left) \eqref{eq:pion_tmd} and kaon (right) 
\eqref{eq:kaon_tmd} at low resolution $\mu=1/\rho$
with $\rho=0.313$ fm: (a,b) are the density plots, (c,d) the 3D plots,
(e,f) the transverse momentum dependent plots for fixed $x$, and (g,h) the longitudinal momentum dependence for fixed $k_\perp$. The pion parameters are
$C_\pi=7.240$, $m_\pi=139.0$ MeV, $M=398.17$ MeV.
The kaon parameters are $C_K=6.60$, $m_K=458.0$ MeV, $M_u=394.4$ MeV, $M_s=556.5$ MeV.}
\label{fig:pion_TMD}
\end{figure*}

\subsection{TMD Evolution from the ILM Scale to Experimental Scales}

The soft-subtracted TMDs obtained above must be evolved to high momentum scales relevant for experiment. Their scale dependence is governed by the CSS equations~\cite{CollinsSoper:1981,CSS:1985},
\begin{align}
\frac{d}{d\ln\sqrt{\zeta}} \ln F &= K_{\rm CS}(b_\perp,\mu),\\
\frac{d}{d\ln\mu}\ln F &= \gamma_F(\alpha_s(\mu)) - \Gamma_{\rm cusp}(\alpha_s(\mu))\ln\frac{\zeta}{\mu^2}.
\end{align}

At small impact parameters $b_\perp\ll 1/\Lambda_{\mathrm{QCD}}$, the TMD factorizes into a convolution of perturbative Wilson coefficients and collinear PDFs,
\begin{equation}
\tilde F_{1}^{q/h}(x,b_\perp;\mu,\mu^2)
=
\sum_{i=q,\bar q,g}
\int_x^1\frac{dx'}{x'}
C_{q/i}\!\left(\frac{x}{x'},b_\perp,\mu_b\right)
f_{i/h}(x',\mu_b)
e^{-S_{\mathrm{Sud}}(\mu_b,\mu)},
\end{equation}
where $\mu_b=2e^{-\gamma_E}/b_\perp$, and the Sudakov factor $S_{\mathrm{Sud}}$ resums large logarithms. Perturbative expressions for the anomalous dimensions and coefficients are known through N$^3$LO~\cite{Braun:2017cih}. 

In the large-$b_\perp$ region, perturbation theory fails, and the ILM provides the correct nonperturbative input. A smooth matching between perturbative and ILM contributions is essential~\cite{Liu:2024sqj,Liu:2025mbl}, implemented through a matching function centered near $b_{\mathrm{NP}}\sim\rho$.

The combined ILM + CSS evolution predicts the migration of probability density from valence $x\approx 0.5$ at low scales to small $x$ at higher scales, and a compression of the transverse width as gluon and sea radiation are generated. This qualitative behavior is consistent with extractions from Drell-Yan measurements.

\subsection{Conclusion}

The ILM provides a snapshot of the pion and kaon at their natural low-resolution scale: valence-dominated, with transverse extent of order the instanton size. CSS evolution then builds the experimentally observed small-$x$ and small-$b_\perp$ structure from perturbative gluon radiation. The resulting meson tomography therefore combines the semiclassical picture of the QCD vacuum with perturbative QCD dynamics.

The emerging picture for the pion is especially striking: at $\mu_0=1/\rho$, its TMD is concentrated at $x\simeq 0.5$ in a region of transverse size $b_\perp\simeq \rho$, but at $\mu=2\,\mathrm{GeV}$ the distribution shifts toward low $x$ and becomes sharply peaked near the origin in impact-parameter space. The kaon exhibits the same behavior, with additional asymmetry reflecting the heavier strange quark.

These results highlight the importance of incorporating both nonperturbative vacuum structure and perturbative evolution in a consistent framework. They also underscore the need for future lattice calculations, Drell-Yan analyses, and EIC measurements to fully resolve the multidimensional structure of pseudoscalar mesons.



\chapter{Summary and Discussion}

\section{From vacuum fields to potentials}

In this review we have argued that the starting point for any unified description of hadrons must be better understanding of the 
QCD vacuum itself. While being the true ground state of this theory, it is not an empty  but a densely populated medium filled with 
nonperturbative gluonic and quark configurations, on top of the usual zero point perturbative fluctuations (gluons).

Fluctuations with nontrivial topology - instantons and anti-instantons  -dominate the nonperturbative  structure of this medium, 
producing zero modes for light quarks and giving rise to the spontaneous breakdown of the $SU(N_f)$ chiral symmetry. 
Their impact is well quantified, as they provide a calculable a dynamical mechanism for generating (momentum-dependent) 
constituent quark masses, pion properties etc. This mechanism of quark effective mass generation, together with the chirality-changing
 interactions inherent in the instanton ensemble, becomes an essential building block of all effective low-energy Hamiltonians.

The instanton fields affect quark propagation in a specific and channel-dependent way. Correlation functions reveal strong attraction 
in pseudoscalar and scalar $\bar q q$ channels, moderate effects in vector and axial channels, and repulsion in some others. 
Thus they  directly determine which hadrons are light, which are heavy, and which channels host strong dynamical correlations or diquark clustering. 

At the same time, confinement-the long-distance property encoded in the area law for Wilson loops- is based on a distinct and 
complementary piece of the quark-gluon interaction. As Wilson loops are sensitive to gauge field strength,
they also obtain large contribution from (topologically neutral) instanton-antiinstanton pairs (molecules). Important novel element of this review
is that a combination of instanton-induced topological and non-topological effects produce effective potentials applicable across the full hadronic spectrum, from pions to quarkonia.



The underlying structure of the QCD vacuum is revealed from
lattice configurations via $gradient-flow$ cooling.  After subsequently smoothing away ultraviolet noise (gluons)  
one finds that the vacuum is permeated by an intricate network of long, thin, string-like center vortices. 
As the lattice evolves in Euclidean time, these vortices trace out two-dimensional worldsheets:  extended, fluctuating surfaces that weave throughout spacetime.
A key feature of this vortex network is the presence of numerous branching and recombination points where 
several vortex sheets meet or split. In center-projected descriptions, these branch points naturally act as 
sources and sinks of Abelian magnetic flux, and thus manifest as monopoles and anti-monopoles. 
Their appearance is not an artifact of projection but reflects genuine geometric singularities inherent to the vortex surfaces.

What is particularly striking is that these same branching regions are strongly correlated with the positions of instantons 
and anti-instantons once the configuration is gradient-flowed. This correlation is not accidental. Both instantons and vortex branch points arise from localized regions where the 
gauge field attains nontrivial topology due to singular or rapidly varying geometry. In simple terms, the vortex sheets provide 
a scaffold that  "guides"  the topological charge: when sheets twist, intersect, or fold over themselves, they naturally create
 the localized lumps of topological charge identified as instantons.

This geometric viewpoint offers a unifying physical intuition. Confinement in the vortex picture stems from the percolation 
of these extended center flux surfaces, which disorder Wilson loops and enforce an area law. Chiral symmetry breaking, 
on the other hand, is driven by the presence of instantons, whose zero modes form a near-zero Dirac spectrum. If both
 monopole worldlines and instanton-like topological lumps originate from the very same vortex geometry, then confinement 
 and chiral symmetry breaking cease to be independent phenomena. Instead, they emerge as two tightly linked consequences 
 of the underlying vortex structure of the vacuum.
From this perspective, the apparent duality between monopoles (the topological sources of confinement) and instantons 
(the topological agents of chiral symmetry breaking) is not mysterious. It reflects the fact that the non-Abelian vacuum 
organizes itself through a common geometric mechanism: the fluctuating, branching, and knotted center vortices whose 
topology encodes both magnetic monopole worldlines and topological charge density. Thus, the long-distance physics of 
QCD  string tension, mass gap, and chiral condensate, may all be different faces of the same underlying vortex-driven 
topological order, as noted formally by the authors~\cite{Ramamurti:2018evz}.

Returning to $gradient-flow$ cooling, let us once again emphasize its important role, as a modern form of Renormalization Group. Its modern version reveals
relatively dense medium made of strongly correlated instanton-antiinstanton pairs (or molecules). Further cooling converges on dilute "instanton liquid"
made of separate instantons. Nontrivial topology of those make them stable under
cooling. Only uncovering those subsequent layers of vacuum structure, one can
return to physical vacuum (zero or minimal cooling) with confidence that
nonperturbative dynamics is fully included.

\section{Wave functions of multiquark hadrons}

The expansion of hadronic spectroscopy in the last decade has revealed that multiquark states are not isolated curiosities but a very important components 
of the QCD dynamics. Tetraquarks, pentaquarks, hexaquarks, and even twelve-quark configurations arise naturally once the underlying forces and 
symmetries are examined in a unified framework. Constructing the wave functions of these states requires a careful synthesis of geometric, algebraic, and dynamical inputs.
And, last but not least, admixture with them (e.g. baryons and pentaquarks) do not turn out to be small, and offer resolutions of some important puzzles.

The formalism developed in this review rests on three pillars. First, Jacobi coordinates are used to decouple center-of-mass motion from internal dynamics, an 
essential step when treating systems with more than two constituents. Second, when quarks share equal masses, hyperspherical symmetry offers a powerful simplification: 
all relative coordinates combine into a single hyperradius and a set of hyperangles. This hypercentral approximation captures dominant features of multiquark dynamics 
and replaces complicated many-body kinematics by a manageable set of radial and angular functions defined on high-dimensional spheres. 
Third, the enforcement of Fermi statistics is achieved through the representation theory of permutation groups~$S_n$. The orbital, color, spin, 
and flavor components of the wave function must combine into overall antisymmetric states; the $S_n$ representations provide a systematic language
 for achieving and classifying these combinations.

Within this structure, multiquark wave functions can be built explicitly, not only for ground states but also for excited configurations with nontrivial orbital angular momentum. 
The technical developments discussed in this review (including the use of "monom"  bases, optimized coordinate systems, and Mathematica-assisted
 realizations of $S_n$ representations)  enable the solution of multiquark Schrodinger equations with a clarity and control not previously available. In the tetraquark sector, 
diquark-antidiquark configurations and meson-meson molecular components can be treated within one and the same framework; pentaquark states 
require a careful analysis of color-spin couplings among four quarks and one antiquark; hexaquarks and larger clusters reveal
 high-dimensional analogues of nuclear cluster dynamics, including tendencies toward color-singlet or diquark substructures.

A particularly illuminating outcome of this unified construction is the natural appearance of \emph{mixing} between conventional hadrons 
and their multiquark counterparts. Such mixing arises because states with the same global quantum numbers inevitably couple through 
the QCD Hamiltonian. Even if the multiquark configurations do not manifest themselves as sharp experimental resonances, their virtual 
admixtures significantly influence hadronic properties.

One striking example occurs in the charmonium sector. A nominally "pure"  $c\bar c$ state such as the $\chi_{c1}$ or the excited $\psi$ 
family generically mixes with four-quark configurations $c\bar c q\bar q$. In the hyperspherical formalism, these tetraquark components
 appear naturally when solving the coupled Schrodinger equations: the heavy $c\bar c$ pair interacts not only via the central potential 
 but also through instanton-induced forces that are sensitive to the presence of light quark pairs. The resulting wave functions show that 
 conventional charmonium states acquire a nontrivial $c\bar c q\bar q$ admixture whose probability may reach several percent, precisely 
 the level needed to account for the observed deviations from naive potential-model predictions. This mixing also helps explain 
 the anomalous decay patterns and near-threshold structures observed in the charmonium-like $X$, $Y$, and $Z$ states, many of
  which sit close to open-charm thresholds where tetraquark configurations play an enhanced role.

An even more consequential example is provided by the nucleon. In the rest frame the $qqq$ component dominates, but the 
full wave function necessarily includes pentaquark configurations $qqqq\bar q$. The color, spin, and flavor couplings of four 
quarks and an antiquark can be organized systematically through $S_4$ and $S_5$ representations, revealing a small but 
robust admixture of configurations such as $[ud][ud]\bar d$ or $[ud][us]\bar s$. These admixtures provide a natural microscopic 
explanation for the observed flavor asymmetry $\bar d > \bar u$ in the proton sea. The mechanism is straightforward in the present 
framework: once the $ud$ diquark is recognized as the most attractive channel in the instanton-induced interaction, the lowest-energy 
pentaquark configurations contain $\bar d$ more readily than $\bar u$, producing the empirically observed excess. The same pentaquark 
components also impact spin observables by modifying the balance between quark spin and orbital angular momentum inside the nucleon. 

These examples illustrate a general principle: multiquark configurations, once included consistently, are not exotic decorations but essential 
pieces of the full hadronic wave function. In charm and bottom sectors they correct the shortcomings of simple quarkonium models; in the 
light-quark sector they resolve longstanding puzzles in proton structure; in nuclear systems they suggest hidden correlations that link 
quark-level dynamics to few-body nuclear forces. In all cases, the unified construction based on Jacobi coordinates, hyperspherical methods, 
and permutation-group representations provides a mathematically coherent way to build, classify, and analyze these multiquark wave functions.

Crucially, multiquark states need not appear as visible resonances in order to influence hadronic properties. 
Their virtual admixtures modify the structure of ordinary hadrons: the nucleon contains pentaquark components 
that explain the flavor asymmetry of the sea; heavy quarkonia receive contributions from virtual four-quark states;
 and nuclear systems show evidence of underlying multiquark correlations. In this unified framework, wave functions
  for multiquark clusters are constructed and handled on an equal footing with those of traditional mesons and baryons. 
  This broadens hadronic spectroscopy into a genuinely multi-body domain, where the rich internal structure of QCD becomes manifest.

\section{Bridging spectroscopy and parton observables}
The mainstream of partonic physics has long been driven by experimental data on hard processes such as deep-inelastic scattering, 
elastic lepton-hadron scattering, and Drell-Yan production. Over several decades, global fits combined with perturbative DGLAP
 evolution have produced increasingly accurate parton distribution functions for quarks, antiquarks, and gluons. 
 This phenomenology has been remarkably successful, yet it has also exposed a number of persistent puzzles: 
 the decomposition of the nucleon spin and the unexpectedly large orbital component, the flavor asymmetry of 
 the antiquark sea, and the unusual properties of the Roper resonance. A deeper conceptual issue is that form 
 factors, PDFs, GPDs and related observables come from different experiments or lattice simulations and, in practice, 
 lack a unifying theoretical origin. Spectroscopy and partonic observables have thus historically formed largely disconnected research domains.

The guiding idea of the present program is to reverse this logic and to construct partonic observables directly 
from the underlying nonperturbative structure that also governs hadronic spectroscopy. Rather than beginning with 
experimental PDFs and evolving them upward in scale, the approach begins with first-principles dynamics in the rest frame,
 proceeds through a systematic construction of wave functions, and finally boosts them to the light front. Symbolically, the theoretical chain of development reads
\[
({\rm Jacobi\ coordinates})\ \longrightarrow\ {\rm LF\ Hamiltonian}\ \longrightarrow\ {\rm LF\ wave\ functions}\ \longrightarrow\ ({\rm formfactors,\ PDFs,\ GPDs}\ldots).
\]
In this formulation, all observables arise from the same approximations and share a common dependence on all parameters.

The Hamiltonians used throughout the analysis include two dominant nonperturbative phenomena of QCD: effective constituent-like 
quark masses arising from chiral symmetry breaking and the long-range confining potential. Even with only these ingredients retained, 
the Schrodinger equations for mesons, baryons, and multiquark states can be solved to an impressive degree of realism. 
While quark-model spectroscopy of the simplest hadrons has been developed for half a century, the present work extends 
these methods decisively to multi-quark systems, tetraquarks, pentaquarks, hexaquarks, and even higher clusters. 
This systematic construction is made possible by the use of Jacobi coordinates to remove center-of-mass motion, 
hyperspherical methods to exploit approximate dynamical symmetries, and a novel application of representation theory 
of the permutation groups~$S_n$ to enforce Fermi statistics in orbital-color-spin-flavor space. The notions of "monom"  and "good coordinate"  
wave-function representations, aided by symbolic computation tools, help manage the complexity of these multiquark configurations.

Even if such multiquark states are not individually identifiable as resonances in scattering experiments, their theoretical wave functions 
allow one to quantify their virtual admixtures inside ordinary hadrons. These admixtures naturally modify hadronic properties and help
 explain several long-standing puzzles, such as the flavor asymmetry of the nucleon sea and the importance of higher-Fock components
  in spin decompositions. The nucleon, in particular, acquires naturally an admixture of pentaquark configurations, while hybrid $\bar q g q$ 
  components provide the light-front analogue of including extra gluons at the Hamiltonian level.

The transition from rest-frame spectroscopy to partonic observables is achieved by boosting the wave functions to the light front. 
In momentum space the natural variables are the Bjorken fractions~$x_i$, which satisfy the normalization $\sum_i x_i=1$. Together 
with modified Jacobi coordinates, these variables define quantum mechanics on simplices such as $A_2$ for baryons and $A_4$ for pentaquarks. 
Constructing complete bases of Laplacian eigenstates on these manifolds uncovers deep mathematical structures and provides a consistent 
set of tools for light-front quantization. Using either these bases or optimized low-lying approximations, one obtains light-front wave functions
 that encode the full nonperturbative structure of hadrons at a low instanton-defined scale~$\mu_0$. From these, without external 
 phenomenological input, one can compute electromagnetic form factors, PDFs, DAs, GPDs, quasi-distributions, and even gravitational 
 form factors describing the internal pressure and shear forces.

Perhaps the most important conceptual advance is the establishment of a direct bridge between spectroscopy and partonic physics. 
Rest-frame potentials, instanton-induced interactions, and multiquark admixtures leave characteristic imprints on the boosted light-front wave functions. 
Twist-three quark-gluon correlations, semi-hard instanton contributions, and the structure of the energy-momentum tensor all reflect 
the same nonperturbative mechanisms that control the spectroscopy of mesons and baryons. In this way, phenomena that once appeared 
unrelated: mass splittings, orbital excitations, sea-quark asymmetries, mechanical forces inside hadrons-- emerge as different manifestations of a single underlying dynamical picture.

The final step in connecting these results to experimental observables is scale evolution. Traditional DGLAP evolution offers a perturbative, 
probabilistic description of how PDFs change with the resolution scale~$\mu$, but this description is incomplete because PDFs are not
 the fundamental objects,  they correspond to reduced density matrices obtained by integrating out degrees of freedom. A more coherent 
 evolution framework should incorporate additional Fock components directly at the Hamiltonian level, treating extra gluons in the same 
 manner as extra $q\bar q$ pairs. The recent realization, particularly in the lattice community, that gradient flow represents a form of 
 renormalization-group evolution opens new possibilities. As the flow time~$\tau$ increases, hard gluon modes are suppressed, 
 revealing first a dense instanton-antiinstanton ensemble and eventually the dilute instanton liquid. Understanding this evolution 
 semiclassically may ultimately yield a more microscopic connection between vacuum structure, Hamiltonians, and the scale dependence of partonic observables.

In summary, the program described here merges hadronic spectroscopy and partonic physics into a unified theoretical framework. 
Both domains draw from the same vacuum-derived Hamiltonians, the same multiquark wave functions, and the same light-front 
quantization procedures. Further developments,  including more complete treatments of spin-dependent interactions, 
improved renormalization of light-front Hamiltonians, and systematic inclusion of higher Fock sectors,  promise to strengthen 
this bridge and bring us closer to a genuinely unified and predictive theory of hadron structure.





\section{What lies ahead}

Needless to say, the program outlined in many papers and in this review is only partially completed, and certain theoretical 
and phenomenological challenges still are  ahead. Significant progress has been made, yet many of the essential ingredients 
required for a fully unified description of hadronic structure remains to be incorporated.

We started this review with modern semiclassical theory applied to tunneling
in quantum mechanics: as Fig.\ref{fig_DWP_gap_semi} shows, with two and three
loop corrections to instanton density one indeed get good quantitative 
description of the gap between lowest energy levels. Unfortunately, for gauge theory instantons we are still lacking those: 
the last semiclassical result here remains the classic paper by \cite{'tHooft:1977hy} fixing the one-loop corrections.
 That is why the instanton physics is still based on empirical constants 
(e.g. quark effective mass) or lattice data, rather than straightforward semiclassics.

The quark-model description of the basic mesons and baryons has now been extended to {\em multiquark hadrons} 
containing four, five, six, or even nine quarks. Technical difficulties related to incorporation of Fermi statistics
have been overcomed.
In this work we largely focused on the most symmetric sectors, such as systems composed of quarks of identical flavor or 
those dominated by light quarks. These cases offer valuable theoretical clarity and reduce computational complexity, but 
they also represent only a narrow subset of the full multiquark spectrum. The rapidly developing experimental program in
 heavy-flavor spectroscopy (covering states with charm, bottom, and mixed flavors)  poses new challenges and opportunities.
  Constructing a systematic theoretical framework that can treat arbitrary flavor combinations, account for the interplay 
  between heavy-quark symmetry and light-quark dynamics, and connect smoothly to known quark-model limits remains an important open direction.

Our "bridge'' toward partonic physics on the light front (LF) began with the simplest mesons and baryons and has already 
yielded new insights into quantization methods, renormalization issues, and the interpretation of wave functions as parton amplitudes. 
The inclusion of higher Fock sectors has been successfully carried out for the $\bar qq$ systems, notably in detailed studies of the pion, 
sigma, and related mesons. For baryons, the baryon-pentaquark mixture has been developed extensively, both in the conventional CM 
frame and on the LF. This sector has shown particular promise: it provides access to the long-standing puzzles of the nucleon spin 
decomposition, the flavor asymmetry of the sea, and the emergence of nontrivial correlations in higher Fock components.

Yet a great deal remains to be accomplished before truly complete picture emerges. For example, we have provided spectroscopic model fitting lattice glueballs,
 but not yet for $hybrid$ sectors, such as $\bar q g q$ configurations. Beyond their intrinsic spectroscopic interest, these states are 
 essential for understanding how gluonic degrees of freedom manifest themselves in hadron structure and how they participate in the t
 ransition between constituent-like and partonic descriptions. Incorporating these sectors consistently into a Hamiltonian approach on the 
 LF presents both conceptual and computational challenges, particularly in relation to gauge invariance, the counting of independent 
 degrees of freedom, and the scaling of the Hilbert space with increasing Fock number. The spectroscopically fitted "effective gluon mass" of 
 about 0.9 GeV qualitatively agree with DGLAP-based disappearence of gluon PDF at $\mu< 1 \, GeV$, but this important part of the "bridge"
still needs to be quantified.

Equally important is the task of extending DGLAP evolution from its traditional probabilistic formulation for parton distribution 
functions (PDFs) into a fully Hamiltonian description of the pertinent wave functions. Such a program would allow the evolution 
of PDFs, generalized parton distributions, and other correlation functions to be understood in terms of transitions among an 
ever-increasing tower of Fock components. Realizing this vision requires combining nonperturbative LF Hamiltonians with 
controlled truncation schemes, renormalization procedures, and matching to perturbative renormalization group transformations in the appropriate kinematic regimes.

In summary, while the foundations have been laid and several key mechanisms have been clarified, the long-term goal 
of a unified, multiscale Hamiltonian description of hadrons-- capable of linking spectroscopy, partonic observables, and 
the full complexity of QCD dynamics, remains a rich and demanding frontier for future work.

\chapter{Epilogue}

The central theme of this review has been the construction of a bridge between two complementary but historically 
disconnected frameworks for describing hadrons: the spectroscopy-oriented rest-frame approach and the partonic light-front description. 
The success of this endeavor rests on a unified treatment of the QCD vacuum, confinement, and chiral symmetry breaking on the 
one hand, and the scale evolution, operator structure, and factorization properties of high-energy processes on the other. The resulting
 synthesis offers a  continuous view of hadron structure across all relevant distance scales.

Another outcome of this work is the demonstration that instanton physics provides a natural mechanism for embedding nonperturbative
 QCD dynamics into the light-front formalism. The Instanton Liquid Model, long known to reproduce a broad spectrum of phenomenological 
 signatures in low-energy QCD, proves equally well-suited to generating the nonlocal interactions and dynamical mass functions required to 
 describe hadrons at low resolution. Once these structures are implemented in the light-front Hamiltonian, the resulting wave functions 
 automatically incorporate essential features of the QCD vacuum, including spontaneous chiral symmetry breaking and strong correlations 
 among quark flavors and spins. Important novel features are related with
contributions of $I\bar I$ $molecules$, 
and incorporation of {\em gradient flow} version of RG on the lattice.

By constructing light-front wave functions for a wide variety of systems - quarkonia, light mesons, baryons, pentaquarks, tetraquarks, 
and heavier multiquark states - we have shown that hadronic spectroscopy and partonic observables share a common dynamical infrastructure.
 The same interactions that reproduce the observed mass spectra and spin splittings also determine the shapes of DAs, PDFs, TMDs and GPDs, 
 as well as the pressure, shear, and energy distributions encoded in the gravitational form factors. This coherence strengthens the central thesis: 
 the spectroscopy of hadrons and the momentum-space structure revealed in scattering experiments emerge from the same underlying dynamics.

Another important contribution of this review is the explicit mapping of low-energy wave functions onto high-energy partonic observables through the 
use of perturbative evolution equations. The DGLAP and ERBL formalisms ensure that distribution functions computed at the low nonperturbative 
scale defined by the ILM evolve consistently into the region probed by experiments. This multiscale evolution encapsulates the transition from the
 nonperturbative "constituent'' regime to the perturbative "partonic'' regime and shows how the effects of confinement and chiral symmetry 
 breaking are gradually diluted as the resolution increases.

The exploration of gravitational form factors and the energy-momentum tensor further enhances the scope of the program.
 These quantities, which are becoming increasingly accessible through both lattice simulations and experimental proxy processes, 
 reveal profound aspects of hadron structure that go beyond charge and momentum distributions. The instanton-based light-front 
 framework naturally accommodates these observables, leading to predictions for pressure profiles, mass distributions, and mechanical 
 stability conditions. These results highlight the deep connection between internal forces in hadrons and the dynamics of their underlying quark and gluon degrees of freedom.

Looking ahead, several exciting avenues present themselves. Extending the light-front Hamiltonian to include dynamical gluons will 
be essential for capturing the full structure of higher-twist observables, gluon distributions, and the gluonic contributions to the 
energy-momentum tensor. Incorporating higher Fock sectors more systematically will allow for a refined understanding of 
sea-quark asymmetries and the role of quark-antiquark pairs in baryon structure. The emergent Electron-Ion Collider will 
provide unprecedented data on GPDs, TMDs, and exclusive processes, offering stringent tests of the framework developed here. 
Furthermore, improvements in lattice QCD calculations of quasi-PDFs and gravitational form factors will supply additional inputs and benchmarks.

The overarching conclusion of this review is that the boundary between hadronic spectroscopy and partonic descriptions can be bridged  
with carefully constructed theoretical tools. By embedding the essential nonperturbative physics of instantons and confinement directly into 
the light-front framework, we obtain a comprehensive theory of hadrons that is simultaneously sensitive to their spectroscopy, their internal 
forces, and their high-energy scattering signatures. This integrated picture reflects the spirit of QCD itself: a theory whose diverse phenomena,
 though often studied in isolation, ultimately arise from the same fundamental dynamics.

The hope is that the synthesis presented here will serve not only as a coherent summary of a substantial body of previous work, but as a 
foundation for future advances. The interplay between vacuum structure, hadronic wave functions, and partonic observables is a fertile 
ground for new discoveries. As experimental capabilities expand and theoretical techniques evolve, the program outlined in this review 
will continue to provide a powerful framework for unraveling the multifaceted nature of the strong interaction.

\section*{Acknowledgements}
This work is supported by the Office of Science, U.S. Department of Energy under Contract  No. DE-FG-88ER40388.
This research is also supported in part within the framework of the Quark-Gluon Tomography (QGT) Topical Collaboration, under contract no. DE-SC0023646.

\newpage
\bibliography{all,big,big2,reference,revnew,ref,refx,MAIN_FINALNotes}
\bibliographystyle{apalike}

\end{document}